\newif\iffigs\figstrue
\documentclass[a4paper,11pt]{article}
\usepackage{latexsym,amssymb,lscape,graphics}
\usepackage{amsmath}
\usepackage{graphicx}        
\usepackage{longtable}
\usepackage{youngtab}
\usepackage{multirow}
\usepackage{color}
\usepackage{slashed,epsfig}
\usepackage{amsfonts}

\newcommand{\mathsym}[1]{{}}

\newtheorem{definizione}{Definition}[section]
\newtheorem{teorema}{Theorem}[section]

\newcommand{\bd}{\begin{definizione}}
\newcommand{\ed}{\end{definizione}}

\def\IC{\relax\,\hbox{$\inbar\kern-.3em{\rm C}$}}
\def\IG{\relax\,\hbox{$\inbar\kern-.3em{\rm G}$}}
\def\IB{\relax{\rm I\kern-.18em B}}
\def\ID{\relax{\rm I\kern-.18em D}}
\def\IL{\relax{\rm I\kern-.18em L}}
\def\IF{\relax{\rm I\kern-.18em F}}
\def\IH{\relax{\rm I\kern-.18em H}}
\def\II{\relax{\rm I\kern-.17em I}}
\def\IN{\relax{\rm I\kern-.18em N}}
\def\IP{\relax{\rm I\kern-.18em P}}
\def\IQ{\relax\,\hbox{$\inbar\kern-.3em{\rm Q}$}}
\def\bfzero{\relax\,\hbox{$\inbar\kern-.3em{\rm 0}$}}
\def\IK{\relax{\rm I\kern-.18em K}}
\def\IG{\relax\,\hbox{$\inbar\kern-.3em{\rm G}$}}
 \font\cmss=cmss10 \font\cmsss=cmss10 at 7pt
\def\IR{\relax{\rm I\kern-.18em R}}
\def\ZZ{\relax\ifmmode\mathchoice
{\hbox{\cmss Z\kern-.4em Z}}{\hbox{\cmss Z\kern-.4em Z}}
{\lower.9pt\hbox{\cmsss Z\kern-.4em Z}} {\lower1.2pt\hbox{\cmsss
Z\kern-.4em Z}}\else{\cmss Z\kern-.4em Z}\fi}
\def\bfone{\relax{\rm 1\kern-.35em 1}}

\def\inbar{\vrule height1.5ex width.4pt depth0pt}
\def\bfzero{\relax{\rm I\kern-.18em 0}}
\def\bfone{\relax{\rm 1\kern-.35em 1}}

\DeclareFontFamily{U}{rsf}{} \DeclareFontShape{U}{rsf}{m}{n}{
  <5> <6> rsfs5 <7> <8> <9> rsfs7 <10-> rsfs10}{}
\DeclareMathAlphabet\Scr{U}{rsf}{m}{n}

\renewcommand{\baselinestretch}{1.0}
\newcommand{\ft}[2]{{\textstyle\frac{#1}{#2}}}
\def\tilde{\widetilde}

\def\1bar{1\hskip -.275cm -}
\def\2bar{2\hskip -.275cm -}
\def\3bar{3\hskip -.275cm -}

\newsavebox{\uuunit}
\sbox{\uuunit}
                 {\setlength{\unitlength}{0.825em}
                      \begin{picture}(0.6,0.7)
                                      \thinlines
                                      \put(0,0){\line(1,0){0.5}}
                                      \put(0.15,0){\line(0,1){0.7}}
                                      \put(0.35,0){\line(0,1){0.8}}
                                     \multiput(0.3,0.8)(-0.04,-0.02){10}{\rule{0.5pt}{0.5pt}}
                      \end {picture}}

\makeatletter \@addtoreset{equation}{section} \makeatother

\def\bfone{\relax{\rm 1\kern-.35em 1}}

\def\bfone{\relax{\rm 1\kern-.35em 1}}
\font\cmss=cmss10 \font\cmsss=cmss10 at 7pt

\begin{document}
\begin{titlepage}
\begin{center}
\vskip 0.2cm
\vskip 0.2cm
{\Large\sc  Classification of  Arnold-Beltrami Flows \\
\vskip 0.2cm
and their Hidden Symmetries
  }\\[1cm]
{\sc P.~Fr\'e${}^{\; a,c,}$\footnote{Prof. Fr\'e is presently fulfilling the duties of Scientific Counselor of the Italian Embassy in the Russian Federation, Denezhnij pereulok, 5, 121002 Moscow, Russia. \emph{e-mail:} \quad {\small {\tt pietro.fre@esteri.it}}} and A.S.~Sorin$^{\; b,c}$ \\[10pt]
{${}^a$\sl\small Dipartimento di Fisica, Universit\'a di Torino\\INFN -- Sezione di Torino \\
via P. Giuria 1, \ 10125 Torino \ Italy\\}
\emph{e-mail:} \quad {\small {\tt fre@to.infn.it}}\\
\vspace{5pt}
{{\em $^{b}$\sl\small Bogoliubov Laboratory of Theoretical Physics $\&$}}\\
{{\em Veksler and Baldin Laboratory of High Energy Physics}}\\
{{\em Joint Institute for Nuclear Research,}}\\
{\em 141980 Dubna, Moscow Region, Russia}~\quad\\
\emph{e-mail:}\quad {\small {\tt sorin@theor.jinr.ru}}\\
\vspace{5pt}
{{\em $^{c}$\sl\small  National Research Nuclear University MEPhI\\ (Moscow Engineering Physics Institute),}}\\
{\em Kashirskoye shosse 31, 115409 Moscow, Russia}~\quad\\
\quad \\
\quad \vspace{6pt}}
\vspace{8pt}
\begin{abstract}
In the context of mathematical hydrodynamics,  we consider the group theory structure which underlies the so named ABC flows introduced by Beltrami, Arnold and Childress. Main reference points are Arnold's theorem stating that, for flows taking place on  compact three manifolds $\mathcal{M}_3$, the only velocity fields able to produce chaotic streamlines are those satisfying Beltrami equation and the modern topological conception of contact structures, each of which admits a representative contact one-form also satisfying Beltrami equation.We advocate that Beltrami equation is nothing else but the eigenstate equation for the first order Laplace-Beltrami operator $\star_g \mathrm{d}$, which can be solved by using time-honored harmonic analysis. Taking for $\mathcal{M}_3$ a torus $T^3$ constructed as $\mathbb{R}^3/\Lambda$, where $\Lambda$ is a crystallographic lattice, we present a general algorithm to construct solutions of the Beltrami equation which utilizes as main ingredient the orbits under the action of the point group   $\mathfrak{P}_\Lambda$ of three-vectors in the momentum lattice $^\star\Lambda$. Inspired by the crystallographic construction of space groups, we introduce the new  notion of a \textit{Universal Classifying Group $\mathfrak{GU}_\Lambda$} which contains all space groups as proper subgroups. We show that the $\star_g \mathrm{d}$ eigenfunctions are naturally arranged into irreducible representations of $\mathfrak{GU}_\Lambda$ and by means of a systematic use of the branching rules with respect to various possible subgroups $ \mathrm{H}_i \subset \mathfrak{GU}_\Lambda$ we search and find Beltrami fields with non trivial hidden symmetries. In the case of the cubic lattice the point group is the proper octahedral  group $\mathrm{O_{24}}$ and the Universal Classifying Group $\mathfrak{GU}_{cubic}$ is a finite group $\mathrm{\mathrm{G_{1536}}}$ of order $|\mathrm{\mathrm{G_{1536}}}| \, = \, 1536$ which we study in full detail deriving all of its 37 irreducible representations and the associated character table. We show that the $\mathrm{O_{24}}$ orbits in the cubic lattice are arranged into $48$ equivalence classes, the parameters of the corresponding Beltrami vector fields  filling all the 37 irreducible representations of $\mathrm{\mathrm{G_{1536}}}$. In this way we obtain an exhaustive classification of all \textit{generalized $\mathrm{ABC}$-flows} and of their hidden symmetries. We make several conceptual comments about the need of a field-theory yielding Beltrami equation as a field equation and/or an instanton equation and on the possible relation of Arnold-Beltrami flows with (supersymmetric) Chern-Simons gauge theories. We also suggest linear generalizations of Beltrami equation to higher odd-dimensions that are different from the non-linear one proposed by Arnold and possibly make contact with M-theory and the geometry of flux-compactifications.
\end{abstract}
\end{center}
\end{titlepage}
\tableofcontents
\noindent {}
\newpage
\part{The Article}
\section{Introduction}
\label{introduczia}
Classical hydrodynamics of ideal, incompressible, inviscid fluids subject to no external forces is described by Euler equation in three dimensional Euclidian space $\mathbb{R}^3$, namely by:
\begin{equation}\label{EulerusEqua}
    \frac{\partial}{\partial t}\mathbf{u}\, + \, \mathbf{u}\cdot \nabla \mathbf{u} \, = \, - \, \nabla p \quad ; \quad \nabla \cdot \mathbf{u} \, = \, 0
\end{equation}
where $\mathbf{u} \, = \, \mathbf{u}\left( x \, , \,t\right)$ denotes the local velocity field and $p(\mathbf{x})$ denotes the local pressure\footnote{Note that we have put the density $\rho = 1$.}. In vector notation eq. (\ref{EulerusEqua}) reads as follows:
\begin{equation}\label{EulerusEquaVec}
    \frac{\partial}{\partial t} u^i\, + \, u^j \, \partial_j \, u^i \, = \, - \, \partial^i \, p \quad ; \quad \partial^\ell  \, u_\ell\, = \, 0
\end{equation}
and admits some straightforward rewriting that, notwithstanding the kinder-garden arithmetic involved in its derivation, is at the basis of several profound and momentous theoretical  developments which have kept  the community of  dynamical system theorists busy for already fifty years \cite{arnoldus,Childress,Henon,Dombre,ArnoldBook,ZaslBook,Bogoyav,arnoldorussopapero,FFMF,Dynamo,Gilbert,Etnyre2000,Ghrist2007}.
\par
With the present paper we aim at introducing into the classical  field of mathematical fluid-mechanics a new group-theoretical approach that allows for a more systematic classification and algorithmic construction of the so named Beltrami flows, hopefully providing new insight into their properties.
\subsection{Beltrami Flows and Arnold Theorem}
Let us then begin with the rewriting of eq.(\ref{EulerusEquaVec}) which is the starting point of the entire adventure. The first step to be taken in our raising conceptual ladder is that of promoting the fluid trajectories, defined as the solutions of the following first order differential system\footnote{In mathematical hydrodynamics people distinguish two notions, that of trajectories, which are the solutions of the differential equations (\ref{streamlines}) and that of streamlines. Streamlines are the instantaneous curves that at any time $t=t_0$ admit the velocity field $u^{i}(x,t_0)$ as tangent vector. Introducing a new parameter $\tau$, streamlines at time $t_0$, are the solutions of the differential system:
\begin{equation}\label{strimotti}
    \frac{d}{d\tau}x^i(\tau) \, = \, u^i(x(\tau),t_0)
\end{equation}
In the case of steady flows where the velocity field is independent from time, trajectories and streamlines coincide. Since we are mainly concerned with steady flows, in the present paper we often use the word streamlines and trajectories indifferently.}:
\begin{equation}\label{streamlines}
    \frac{d}{dt}x^i(t) \, = \, u^i(x(t),t)
\end{equation}
to smooth maps:
\begin{equation}\label{mappini}
    \mathcal{S} \, : \, \mathbb{R}_t \, \rightarrow \, \mathcal{M}_g
\end{equation}
from the time real line $\mathbb{R}_t $ to a smooth Riemannian manifold $\mathcal{M}_g$ endowed with a metric $g$. The classical case corresponds to $\mathcal{M} \, = \, \mathbb{R}^3$, $g_{ij}(x) \, = \, \delta_{ij}$, but any other Riemannian three-manifold might be used and there exist generalization also to higher dimensions. Adopting this point of view, the velocity field $\mathbf{u}\left( x \, , \,t\right)$ is turned into a time evolving vector field on $\mathcal{M}$ namely into a smooth family of sections of the tangent bundle $T\mathcal{M}$:
\begin{equation}\label{Ufildo}
    \forall \, t \, \in \, \mathbb{R} \, : \, u^i(x,t) \, \partial_i \, \equiv \, {\mathrm{U}}(t) \, \in \, \Gamma\left(T\mathcal{M}\, , \, \mathcal{M}\right)
\end{equation}
Next, using the Riemannian metric, which allows to raise and lower tensor indices, to any ${\mathrm{U}}(t)$ we can associate a family of sections of the cotangent bundle $CT\mathcal{M}$ defined by the following time evolving one-form:
\begin{equation}\label{Omfildo}
  \forall \, t \, \in \, \mathbb{R} \, : \,    \Omega^{[\mathrm{U}]}(t)  \, \equiv \, g_{ij} \, u^i(x,t) \, dx^j \, \in \, \Gamma\left(CT\mathcal{M}\, , \, \mathcal{M}\right)
\end{equation}
Utilizing the exterior differential and the contraction operator acting on differential forms, we can evaluate the Lie-derivative of the one-form $\Omega^{[\mathrm{U}]}(t)$ along the vector field $\mathrm{U}$. Applying definitions (see for instance \cite{pietrobook}, chapter five, page 120 of volume two) we obtain:
\begin{eqnarray}\label{trivialone}
    \mathcal{L}_\mathrm{U} \Omega^{[\mathrm{U}]}(t) & \equiv & {\rm i}_\mathrm{U}  \cdot {\rm d}  \Omega^{[\mathrm{U}]} \, + \, {\rm d} \left({\rm i}_\mathrm{U} \cdot \Omega^{[\mathrm{U}]} \right) \nonumber\\
                                      & = &\left ( u^\ell \partial_\ell \, u^i  \, + \, g^{i k} \partial_{k} \, \underbrace{\parallel \mathrm{U} \parallel^2}_{g_{mn} \, u^m \, u^n} \right)\, g_{ij} \, dx^j
\end{eqnarray}
and Euler equation can be rewritten in either one of the following two equivalent index-free reformulations:
\begin{eqnarray}
    - \, {\rm d } \left( p \, - \, \ft 12 \,\parallel \mathrm{U} \parallel^2\right ) & = & \partial_t \Omega^{[\mathrm{U}]} \, + \,  \mathcal{L}_\mathrm{U} \Omega^{[\mathrm{U}]} \label{EulerBezIndex}\\
    &\mbox{or}& \nonumber\\
    - \, {\rm d } \left( p \, + \, \ft 12 \,\parallel \mathrm{U} \parallel^2\right ) & = & \partial_t \Omega^{[\mathrm{U}]} \, + \,  {\rm i}_\mathrm{U}  \cdot {\rm d}  \Omega^{[\mathrm{U}]} \, \label{bernullini}
\end{eqnarray}
Eq.(\ref{bernullini}) is one of the possible formulations of the classical Bernoulli theorem. Indeed from eq.(\ref{bernullini}) we immediately conclude that
\begin{equation}\label{finocchionabis}
    H \, = \, p \, + \, \ft 12 \,\parallel \mathrm{U} \parallel^2
\end{equation}
is constant along the trajectories defined by eq. (\ref{streamlines}). Turning matters around we can say that in steady flows, where $\partial_t \mathrm{U} \, = \,0$, the fluid trajectories necessarily lay on the level surfaces  $H(\mathrm{x})\, = \, h \, \in \, \mathbb{R}$ of the function:
\begin{equation}\label{ignobilia}
    H \, : \, \mathcal{M} \, \rightarrow \, \mathbb{R}
\end{equation}
defined by (\ref{finocchionabis}). Then if $H(\mathbf{x})$ has a non trivial $x$-dependence it defines a natural foliation of the $n$-dimensional manifold $\mathcal{M}$ into a smooth family of $(n-1)$-manifolds (all diffeomorphic among themselves) corresponding to the level surfaces. Furthermore, as already advocated, the trajectories, \textit{i.e.} the solutions of eq.(\ref{streamlines}), lay on these surfaces. In other words the dynamical system encoded in eq.(\ref{streamlines}) is effectively $(n-1)$-dimensional admitting $H$ as an additional conserved hamiltonian. In the classical  case $n\, = \, 3$ this means that the differential system (\ref{streamlines}) is actually two-dimensional, namely non chaotic and in some instances even integrable\footnote{Here we rely on a general result established by the theorem of Poincar\'e-Bendixson  \cite{PoincareNonchaos} on the limiting orbits of planar differential systems  whose corolloray is generally accepted to establish that two-dimensional continuous systems cannot be chaotic.}. Consequently we reach the conclusion that no chaotic trajectories (or streamlines) can exist if $H(x)$ has a non trivial $x$-dependence: the only window open for lagrangian chaos occurs when $H$ is a constant function. Looking at eq.(\ref{bernullini}) we realize that in steady flows where $\partial_t \Omega^{[\mathrm{U}]} \, =\,0$, the only open window for chaotic trajectories is provided by velocity fields that satisfy the  condition:
\begin{equation}\label{cinesinotulipano}
    {\rm i}_\mathrm{U}  \cdot {\rm d}  \Omega^{[\mathrm{U}]} \, = \, 0
\end{equation}
This weak condition (\ref{cinesinotulipano}) is certainly satisfied if the velocity field $\mathrm{U}$ satisfies the strong condition:
\begin{equation}
    {\rm d}  \Omega^{[\mathrm{U}]} \, = \, \lambda \, \star_g \, \Omega^{[\mathrm{U}]} \quad \Leftrightarrow \quad
   \star_g \, {\rm d}  \Omega^{[\mathrm{U}]} \, = \, \lambda \,  \Omega^{[\mathrm{U}]} \label{Beltrami}
\end{equation}
where $\star_g$ denotes the Hodge duality operator in the metric $g$:
\begin{eqnarray}\label{gongo}
    \star_g \, \Omega^{[\mathrm{U}]} &  = & \epsilon_{\ell m n} \, g^{\ell k}\, \Omega^{[\mathrm{U}]}_k \, dx^m \, \wedge \,dx^n
    \, =\,  u^\ell \, dx^m \, \wedge \,dx^n \, \epsilon_{\ell m n}\label{gruscia1}\\
    \star_g \, \mathrm{d} \, \Omega^{[\mathrm{U}]} &  = & \epsilon_{\ell m n} \, g^{m p}\,g^{n q}  \partial_p \left( g_{qr} u^r\right)  \, dx^\ell \label{gruscia2}
\end{eqnarray}
The heuristic argument which leads to  consider velocity fields that satisfy  \textit{Beltrami condition} (\ref{Beltrami}) as the unique steady candidates compatible with chaotic trajectories was transformed by Arnold  into a rigorous theorem \cite{arnoldus,ArnoldBook} which, under the strong hypothesis that $(\mathcal{M},g)$ is a closed, compact Riemannian three-manifold, states the following:
\begin{teorema} (\textbf{Arnold})
There are only two possibilities:
\begin{description}
  \item[a)] Either the form $\Omega^{[\mathrm{U}]}$ is an eigenstate of the Beltrami operator $\star_g \, \mathrm{d}$ with a non vanishing eigenvalue $\lambda \ne 0$
  \item[b)] or the manifold $\mathcal{M}$ is subdivided into a finite collection of cells, each of which admits a foliation diffeomorphic to $T^2 \times \mathbb{R}$ and every two-torus $T^2$ is an invariant set with respect to the action of the velocity field $\mathrm{U}$: in other words, all trajectories lay on some $T^2$  immersed in the manifold $\mathcal{M}$.
\end{description}
\end{teorema}
Henceforth the desire to investigate the on-set of chaotic trajectories in steady flows of incompressible fluids motivated the interest of the  dynamical system community  in Beltrami vector fields defined by  the condition (\ref{Beltrami}). Furthermore, in view of the above powerful theorem proved by Arnold, the focus of attention concentrated on the rather unphysical, yet mathematically very interesting case  of compact three-manifolds. Within this class, the most easily treatable case is that of flat compact manifolds without boundary, so that the most popular playground turned out to be the three torus $T^3$. Reporting literally the words of Robert Ghrist in his very nice review \cite{Ghrist2007}: \textit{on those occasions when compactness is desired and the complexities of boundary conditions are not, the fluid domain is usually taken to be a Euclidian $T^3$ torus given by quotienting out Euclidian space $\mathbb{R}^3$ by the action of three mutually orthogonal translations}. These slightly ironical words are meant to emphasize the main point which is outspokenly put forward by the same author few lines below: \textit{Since so little is known about the rigorous behavior of fluid flows, any methods which can be brought to bear to prove theorems about their behavior are of interest and potential use}. Certainly most physical contexts for fluid dynamics do not correspond to the idealized situation of a motion in a compact manifold or, said differently, periodic boundary conditions are not the most appropriate to be applied either in a river, or in the atmosphere or in the charged plasmas environing a compact star, yet the message conveyed by Arnold theorem that Beltrami vector fields play a distinguished role in chaotic behavior is to be taken seriously into account and gives an important hint. Moreover, although boundary terms usually encode relevant physical phenomena, yet the history of periodic boundary conditions is a very rich and noble one in Quantum Mechanics, Classical Field Theory and also in Quantum Field Theory. It suffices to recall that periodic boundary conditions of quantized fields provide a formulation of finite temperature  quantum field theory.
\par
In our opinion such arguments are  a sufficient justification for the fifty year  long efforts devoted by dozens of authors to  the study of  steady flows generated by Beltrami vector fields. On the other hand, what  is  somewhat surprising is that an overwhelming part of such efforts focused on a  single example constructed on the $T^3$ torus. The following vector field:
\begin{equation}\label{bagcigaluppi}
  \mathbf{u}(x,y,z) \, = \, \mathbf{V}^{(ABC)}(x,y,z) \, \equiv \, \left(
\begin{array}{l}
 C \cos (2 \pi  y)+A \sin (2 \pi  z) \\
 A \cos (2 \pi  z)+B \sin (2 \pi  x) \\
 B \cos (2 \pi  x)+C \sin (2 \pi  y)
\end{array}
\right)
\end{equation}
which satisfies the Beltrami condition with eigenvalue $\lambda \, = \, 1$ and which contains three real parameters $A,B,C$ defines  what is known in the literature by the name of an ABC-flow (Arnold-Beltrami-Childress) \cite{arnoldus,Childress,beltramus} and during the last half century it was the target of fantastically numerous investigations.
\par
Main aim of our work was to understand the  principles underlying the construction  of the ABC-flows,  use systematically such principles to construct and classify all other Arnold-like Beltrami flows, deriving also, as a bonus,  their hidden discrete symmetries.
\par
The issue of symmetries happens to be quite relevant at least in  two respects. On one hand, in the case of the ABC model, it occurs that the choice of parameters ($A:B:C=1$) which leads to the Beltrami vector field with the largest group of automorphisms leads also to the most extended distribution of chaotic trajectories. On the other hand symmetries of Beltrami flows have proved to be crucial in connection with their use in modeling \textit{magneto-hydrodynamic fast dynamos} \cite{zeldus,Dynamo,Gilbert}. By this words it is understood the mechanism that in a steady flow of charged particles generates a large scale magnetic field whose magnitude might be exponentially increasing with time. No analytic results do exist on fast dynamos and all studies have been so far numerical, yet while dealing with these latter, crucial simplifications occur and optimization algorithms become available if the steady flow possesses a large enough group $\mathcal{G}$ of symmetries. In this case the magnetic field can be developed into irreducible representations of  $\mathcal{G}$ and this facilitates the numerical determination of growing rates of different modes. It is important to stress that the linearized dynamo equations for the magnetic field $\mathbf{B}$ coincide with the linearized equations for perturbations around a steady flow. Therefore the same development of perturbations into irreps of $\mathcal{G}$ is of great relevance also for the study of fluid instabilities. In plasma physics Beltrami Flows are known under the name of Force--Free Magnetic Fields \cite{FFMF}.
\subsection{The conception of contact structures}
Last but not least let us mention that Beltrami vector fields are intimately related with the mathematical conception of \textit{contact topology}. This latter, vigorously developed in the last two decades starting from classical results of analysis that date back  to Darboux, Goursat and other XIX century maitres, is a mathematical theory aiming at providing an intrinsic geometrical-topological characterization of \textit{non integrability}, namely of the issues discussed above. As we have seen from our sketch of Arnold Theorem, the main obstacle to the onset of chaotic trajectories has a distinctive geometrical flavor: trajectories are necessarily ordered and non chaotic if the manifold where they take place has a foliated structure $\Sigma_h \times \mathbb{R}_h$, the two dimensional level sets $\Sigma_h$ being invariant under the action of the velocity vector field $U$. In this case each streamline lays on some surface $\Sigma_h$.  Equally adverse to chaotic trajectories is the case of \textit{gradient flows} where there is  a foliation provided by the level sets of some function $H(x)$ and the velocity field $\mathrm{U}=\nabla H$ is just the gradient of $H$. In this case all trajectories are orthogonal to the leaves $\Sigma_h$ of the foliation and their well aligned tangent vectors are parallel to its normal vector.
\par
\begin{figure}[!hbt]
\begin{center}
\iffigs
\includegraphics[height=80mm]{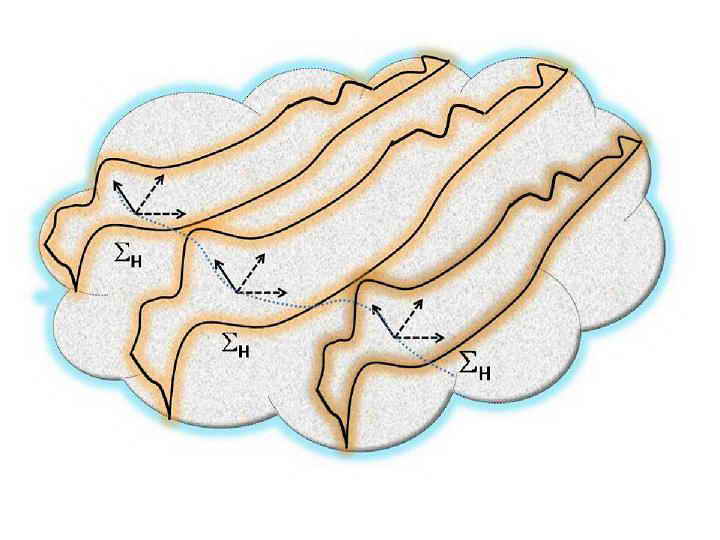}
\else
\end{center}
 \fi
\caption{\it  Schematic view of the foliation of a three dimensional manifold $\mathcal{M}$. The family of two-dimensional surfaces $\Sigma_h$ are typically the level sets $H(\mathbf{x}) =h$ of some function $H\, : \, \mathcal{M} \, \rightarrow\, \mathbb{R}$. At each point of $p\, \in \, \Sigma_h \subset \mathcal{M}$ the dashed vectors span the tangent space $T_p^\parallel\Sigma_h$, while the solid vector span the normal space to the surface $T_p^\perp\Sigma_h$. Equally adverse to chaotic trajectories is the case where the velocity field $\mathrm{U}$  lies in $T_p^\perp\Sigma_h$ (gradient flow) or in $T_p^\parallel\Sigma_h$ }
\label{fogliattone}
 \iffigs
 \hskip 1cm \unitlength=1.1mm
 \end{center}
  \fi
\end{figure}
 In conclusion in presence of a foliation we have the following decomposition of the tangent space to the manifold $\mathcal{M}$ at any point $p \in \mathcal{M}$
 \begin{equation}\label{finchius}
    T_p\mathcal{M} \, = \, T_p^\perp\Sigma_h \, \oplus \, T_p^\parallel\Sigma_h
\end{equation}
and no chaotic trajectories are possible in a region $\mathfrak{S} \subset \mathcal{M}$ where $\mathrm{U}(p) \, \in \, T_p^\perp\Sigma_h $ or $\mathrm{U}(p) \, \in \, T_p^\perp\Sigma_h$ for $\forall p \, \in \, \mathfrak{S} $ (see fig.\ref{fogliattone}).
\par
This matter of fact motivates an attempt to capture the geometry  of the bundle of subspaces orthogonal to the lines of flow by introducing  an intrinsic topological indicator that distinguishes necessarily non chaotic flows from possibly chaotic ones.
Let us first consider the extreme case of a gradient flow where $\Omega^{[\mathrm{U}]} \, = \, \mathrm{d} H$ is an exact form. For such flows we have:
\begin{equation}\label{canedellascala}
\Omega^{[\mathrm{U}]} \, \wedge \, \mathrm{d}\Omega^{[\mathrm{U}]} \, = \, \Omega^{[\mathrm{U}]} \, \wedge \, \underbrace{\mathrm{d} \, \mathrm{d}H}_{= \, 0} \, = \, 0
\end{equation}
Secondly let us consider the opposite case where the velocity field $\mathrm{U}$ is orthogonal to a gradient vector field $\nabla H$ so that the integral curves of $\mathrm{U}$  lay on the level surfaces $\Sigma_h$. Furthermore let us assume that $\mathrm{U}$ is self similar on neighboring level surfaces. We can characterize this situation in a Riemannian manifold $(\mathcal{M},g)$ by the following two conditions:
\begin{equation}\label{gattoascensore}
{\rm i}_{\nabla H}\Omega^{[\mathrm{U}]} \, \Leftrightarrow \,  g\left(\mathrm{U}\, ,\, \nabla H \right) \, = \, 0 \quad ; \quad \left[ U\, , \, \nabla H\right] \, = \, 0
\end{equation}
The first of eq.s(\ref{gattoascensore}) is obvious. To grasp the second it is sufficient to introduce, in the neighborhood of any point $p\, \in \, \mathcal{M}$ a local coordinate system composed by $(h,x^\parallel )$ where $h$ is the value of the function $H$ and $x^\parallel $ denotes some local coordinate system on the level set $\Sigma_h$.  The situation we have described corresponds to assuming that:
\begin{equation}\label{caruccio}
    \mathrm{U} \, \simeq \, U^{\parallel} (x^\parallel) \, \partial_\parallel  \quad ; \quad \partial_h \, U^{\parallel} (x^\parallel) \, = \, 0
\end{equation}
Under the conditions spelled out in eq.(\ref{gattoascensore}) we can easily prove that:
\begin{equation}\label{suschione}
    {\rm i}_{\nabla H} \, \mathrm{d}\Omega^{[\mathrm{U}]} \, = \,0
\end{equation}
Indeed from the definition of the Lie derivative we obtain:
\begin{eqnarray}\label{cincischiando}
    {\rm i}_{\nabla H} \, \mathrm{d}\Omega^{[\mathrm{U}]} \, = \, \underbrace{\mathcal{L}_{\nabla H} \,\Omega^{[\mathrm{U}]}}_{= \, \Omega^{\left[[U\, ,\, \nabla H]\right]} \, = \, 0} \, - \, \mathrm{d}\left(\underbrace{{\rm i}_{\nabla H}\Omega^{[\mathrm{U}]}}_{= 0}\right)
\end{eqnarray}
Since we have both ${\rm i}_{\nabla H}\Omega^{[\mathrm{U}]} \, = \, 0$ and $\left[U\, ,\, \nabla H\right ] \, = \, 0$ it follows that also in this case:
\begin{equation}\label{franceschiello}
    \Omega^{[\mathrm{U}]} \, \wedge \, \mathrm{d}\Omega^{[\mathrm{U}]} \, = \, 0
\end{equation}
Indeed the three-form $\Omega^{[\mathrm{U}]} \, \wedge \, \mathrm{d}\Omega^{[\mathrm{U}]} $ has no projection in the direction $\nabla H$ and therefore it lives on the two dimensional surfaces $\Sigma_h$: but in two dimensions any three-form necessarily vanishes.
\begin{figure}[!hbt]
\begin{center}
\iffigs
\includegraphics[height=80mm]{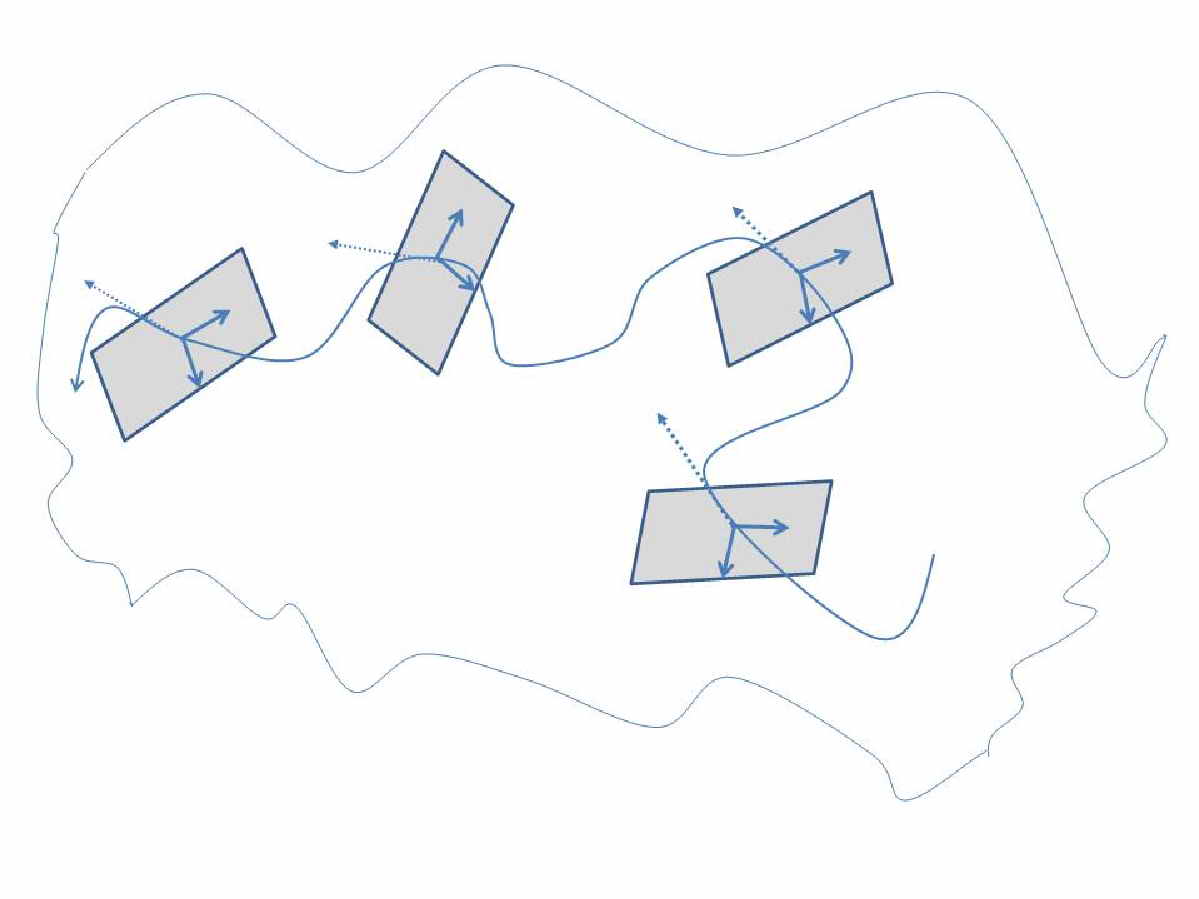}
\else
\end{center}
 \fi
\caption{\it  A schematic picture of a contact structure. Given the Reeb field $\mathrm{U}$, associated with a contact one-form $\alpha$, we can consider the family of hyperplanes orthogonal, at each point of the manifold, to the vector field $\mathrm{U}$. These hyperplanes are defined as the kernel of the contact form $\alpha$ and constitute the contact structure. }
\label{contattone}
 \iffigs
 \hskip 1cm \unitlength=1.1mm
 \end{center}
  \fi
\end{figure}
This heuristic arguments motivate the notion of \textit{contact form} and \textit{contact structure} that capture the non integrability of a vector field in a \textit{topological, metric independent way}.
\begin{definizione}
Let $\mathcal{M}$ be smooth three-manifold. A contact form $\alpha \, \in \, \Gamma\left( CT\mathcal{M},\mathcal{M}\right)$ is a one-form such that
\begin{equation}\label{contattusforma}
    \alpha \, \wedge \, \mathrm{d}\alpha \, \ne \, 0
\end{equation}
\end{definizione}
\begin{definizione}
Let $\mathcal{M}$ be smooth three-manifold and $\alpha$ a contact form on it. The rank two vector-bundle  of all vector fields $X$ which satisfy the condition:
\begin{equation}\label{strucontattus}
    {\rm i}_X \, \alpha \, = \, 0
\end{equation}
is named the \textbf{contact structure} $\mathcal{CS}_\alpha$ defined by $\alpha$.
\end{definizione}
In view of what we discussed above it is clear that the definition of contact structures captures the notion of maximal \textit{non-integrability}. At each point $p \, \in \, \mathcal{M}$ the contact structure is a two-dimensional plane singled out in the tangent space $T_p\mathcal{M}$ by the condition (\ref{strucontattus}). This smooth family of planes, however, cannot be considered as the tangent plane of the level surfaces of any foliation. This is ruled out by the condition (\ref{contattusforma}).
\par
Let us next introduce the notion of \textit{Reeb-like field}
\begin{definizione}(\textbf{Reeb-like field}).
Let $\mathcal{M}$ be a smooth three-manifold and $\alpha$ a contact one-form defining a contact-structure. A  Reeb--like  field for $\alpha$ is a vector field  $\mathrm{U}$ satisfying  the following two conditions:
\begin{equation}\label{ribba}
    {\rm i}_\mathrm{U} \, \alpha \, > \, 0 \quad ; \quad {\rm i}_\mathrm{U} \, \mathrm{d} \alpha \, = \, 0
\end{equation}
\end{definizione}
Let us stress that this definition makes sense only in view of eq.(\ref{contattusforma}); this latter imposes that $\mathrm{d} \alpha$ has support on the orthogonal complement $W^\perp\, \subset \, T\mathcal{M}$  of the one-dimensional sub-bundle $W^\parallel \, \subset \, T\mathcal{M}$  forming the support of $\alpha$. The second of eq.s (\ref{ribba}) imposes that the Reeb-like  vector field should have no component along $W^\perp$:  see fig.\ref{contattone}.
\par
As wee see the main reason to introduce the contact form  conception is that, so doing one liberates the notion of a vector field capable to generate chaotic trajectories from the use of any  metric structure. A vector field $U$ is potentially interesting for chaotic regimes if it is a Reeb-like field for at least one contact form $\alpha$. In this way the mathematical theorems about the classification of contact structures modulo diffeomorphisms (theorems that are metric-free and of topological nature) provide new global methods to capture the topology of hydro-flows.
\par
Instead if we work in a Riemannian manifold endowed with a metric $(\mathcal{M},g)$ we can always invert the procedure and define the contact form $\alpha$ that can admit $\mathrm{U}$ as a Reeb-like field by identifying
\begin{equation}\label{gominato}
    \alpha \, = \, \Omega^{[\mathrm{U}]}
\end{equation}
In this way the first of the two conditions (\ref{ribba}) is automatically satisfied: ${\rm i}_\mathrm{U}\Omega^{[\mathrm{U}]} \, = \, \parallel \mathrm{U}\parallel^2 \, > 0$. It remains to be seen whether $\Omega^{[\mathrm{U}]}$ is indeed a contact form, namely whether $\Omega^{[\mathrm{U}]} \, \wedge \, \mathrm{d} \Omega^{[\mathrm{U}]} \, \ne \,0$ and whether the second condition ${\rm i}_\mathrm{U} \, \mathrm{d} \Omega^{[\mathrm{U}]} \, = \, 0$ is also satisfied. Both conditions are automatically fulfilled if $\mathrm{U}$ is a Beltrami field, namely if it is an eigenstate of the operator $\star_g \, \mathrm{d}$ as advocated in eq.(\ref{Beltrami}). Indeed the implication ${\rm i}_\mathrm{U} \, \mathrm{d} \Omega^{[\mathrm{U}]} \, = \, 0$ of Beltrami equation was shown  in eq. (\ref{cinesinotulipano}), while from the Beltrami condition it also follows:
\begin{eqnarray}\label{cirimellaG}
 \Omega^{[\mathrm{U}]} \, \wedge \, \mathrm{d} \Omega^{[\mathrm{U}]} & = &\Omega^{[\mathrm{U}]} \, \wedge \, \star_g \Omega^{[\mathrm{U}]} \, = \, \parallel \mathrm{U}\parallel^2 \, \mathrm{Vol}  \, \ne \, 0 \nonumber\\
 \mathrm{Vol} & \equiv & \frac{1}{3!} \, \times \, \epsilon_{ijk} \, dx^i \, \wedge \, dx^j \, \wedge \, dx^k
\end{eqnarray}
In this way the conceptual circle closes and we see that all Beltrami vector fields can be regarded as Reeb-like fields for a bona-fide contact form. Since the same contact structure (in the topological sense) can be described by different contact forms,  once Beltrami fields have been classified
it remains the task to discover how many inequivalent contact structures they actually describe. Yet it reasonable to assume that every contact structure has a contact form representative that is derived from a Beltrami Reeb-like field. Indeed a precise correspondence is established by a theorem proved in \cite{Etnyre2000,Ghrist2007}:
\begin{teorema}
\label{teoremuccio}
Any rotational Beltrami vector field on a Riemannian $3$-manifold is a Reeb-like field for some contact form. Conversely any Reeb-like field associated to a contact form on a $3$-manifold is a rotational Beltrami field for some Riemannian metric. Rotational Beltrami field means an eigenfunction of the $\star_g \mathrm{d}$ operator corresponding to a non vanishing eigenvalue $\lambda$.
\end{teorema}
\subsection{Beltrami equation  at large and harmonic analysis}
All the arguments presented in previous sections have been instrumental to enlighten the role of Beltrami vector fields from various viewpoints
related with hydrodynamics and other mathematical-physical issues. Let us now consider from a more general point of view Beltrami equation (\ref{Beltrami}) which constitutes the main topic of the present paper. The one here at stake  is the case $p=1$ of an eigenvalue equation that can be written in any $(2\, p + 1)$-dimensional Riemannian manifold $\left(\mathcal{M}_p \, , \, g\right)$, namely:
\begin{equation}\label{choucrut}
    \star_g  \mathrm{d}\omega^{(p)} \, = \, \lambda \, \omega^{(p)}
\end{equation}
The eigenfunctions of the $\star_g  \mathrm{d}$ operator are 1-forms for $p=1$, namely in three-dimensions, but they are higher differential forms in higher odd dimensions. A particularly interesting case is that of $7$-manifolds where the eigenfunctions of $\star_g  \mathrm{d}$ are three-forms and can be related with a $\mathrm{G_2}$-structure of the manifold. An important general observation is that the relation encoded in theorem \ref{teoremuccio} between eq.(\ref{choucrut}) and contact structures, as they are defined in current mathematical literature, is true only for $p=1$ and it is lost for higher $p$. Indeed contact structures are always defined in terms of a contact one-form and eq.(\ref{contattusforma}) is replaced by:
\begin{equation}\label{salsicciafresca}
    \alpha \, \wedge \, \underbrace{\mathrm{d}\alpha \, \wedge \,\mathrm{d}\alpha \, \dots \, \mathrm{d}\alpha}_{p-\mbox{times}} \, \ne \, 0
\end{equation}
Hence the problem of determining the spectrum and the eigenfunctions of the operator $\star_g  \mathrm{d}\omega^{(p)}$  is a general one and can be addressed in the same way in all odd-dimensions, yet its relation with flows and contact-structures is peculiar to $d=3$ and has not a general significance. Whether the contact-structure viewpoint or the pure geometrical view point encoded in eq.(\ref{choucrut}) is more fundamental is certainly a matter of debate and bears  also on personal scientific tastes, yet it is absolutely  clear that once the correspondence of theorem \ref{teoremuccio} has been established, the classification of Beltrami fields is reduced to a classical problem of differential geometry whose solution can be derived within a time honored framework which makes no reference to trajectories, dynamical systems, contact structures and all the rest of the conceptions debated in previous subsections of this introduction.
\par
The framework we refer to is that of \textit{harmonic analysis} on compact Riemannian manifolds $\left( M,g\right)$  and its application to the spectral analysis of Laplace-Beltrami  operators (for reviews see the book \cite{castdauriafre} and the articles \cite{armoniepietro}). As thoroughly discussed in the quoted references there are, on a Riemann manifold $\left( M,g\right)$, several invariant differential operators, generically named Laplace-Beltrami some of which are of second order, some other of first order. They act on the sections of vector bundles $E \rightarrow \mathcal{M}$  of different rank, for instance the tangent bundle, the bundle of $p$-forms, the bundle of symmetric two tensors, the spinor bundle etc. Among the first order operators the most important ones are the Dirac operator acting on sections of the spinor bundle and the $\star_g \mathrm{d}$-operator acting on $p$-forms in a $(2 \, p +1)$-dimensional manifold. The spectrum of all Laplace-Beltrami operators is sensitive both to the topology and to the metric of the underlying manifold. Each eigenspace is organized into irreducible representations of the isometry group $\mathrm{G}$ of the metric $g$ and the eigenfunctions assigned to a particular representation are generically named \textit{harmonics}.
\par
Here comes an important distinction in relation with the nature of the group $\mathrm{G}$. If $\mathrm{G}$  is a Lie group and if the manifold
$\mathcal{M}$ is homogeneous under its action, than  $\mathcal{M} \sim \mathrm{G/H}$ where $\mathrm{H}\subset \mathrm{G}$ is the stability subgroup of some reference point $p_0 \, \in \, \mathcal{M}$. In this case harmonic analysis reduces completely to group-theory and the spectrum of any Laplace-Beltrami operator can be derived in pure algebraic terms without ever using any differential operations. In the case $\mathrm{G}$ is not a Lie group and/or $\mathcal{M}$ is not homogeneous under its action, then matters become more complicated and ad hoc techniques have to be utilized case by case to analyze the spectrum of invariant operators.
\subsection{Harmonic analysis on the $T^3$ torus and the Universal Classifying Group}
\label{introT3}
The reasons to compactify Arnold-Beltrami flows on a $T^3$ have already been discussed and we do not resume the issue. We just observe that $\mathbb{R}^3$ is a non-compact coset manifold so that harmonic analysis over $\mathbb{R}^3$ is a complicated matter of functional analysis. After compactification, namely after imposing periodic boundary conditions, things drastically simplify. Firstly, as we explain in a detailed way in section \ref{elementi}, the compactification is obtained by quotienting $\mathbb{R}^3$ with respect to a discrete subgroup of the translation group which constitutes a lattice:
\begin{equation}\label{quazientus}
    T^3 \, = \, \frac{\mathbb{R}^3}{\Lambda}
\end{equation}
Secondly we implement the programme of harmonic analysis by presenting a general algorithm to construct solutions of the Beltrami equation which utilizes as main ingredient the orbits under the action of the point group   $\mathfrak{P}_\Lambda$ of three-vectors in the momentum lattice $^\star\Lambda$ which is just the dual of the lattice $\Lambda$. In the language of crystallography the point group  is just the discrete subgroup $\mathfrak{P}_\Lambda \subset  \mathrm{SO(3)}$ of the rotation group which maps the lattice $\Lambda$ and its dual $^\star\Lambda$ into themselves:
\begin{equation}\label{puntarellona}
    \mathfrak{P}_\Lambda \, \Lambda \, = \, \Lambda \quad ; \quad \mathfrak{P}_\Lambda \, ^\star\Lambda\, = \, ^\star\Lambda
\end{equation}
In the case of the cubic lattice, that is the main example studied in this paper we have  $\mathrm{G_{cubic}}\, = \, \mathrm{O_{24}}$ where $\mathrm{O_{24}} \sim \mathrm{S_4}$ is the proper octahedral  group of order $|\mathrm{O_{24}}|\, = \, 24$.
In the case of the hexagonal lattice which we also briefly analyze, the point group  is the dihedral group $\mathrm{{\cal D}_6}$
of order $|\mathrm{D_{6}}|\, = \, 12$.
\par
Thirdly, as we explain in detail in section \ref{maingruppo}, which   constitutes the hard-core of the present paper, a general argument, inspired by the logic that crystallographers use to derive and classify space groups, leads us to introduce a large finite group $\mathfrak{GU}_\Lambda$, named by us the \textit{Universal Classifying Group for the Lattice $\Lambda$}, made out of discretized rotations and translations that are defined by the structure of $\Lambda$. All eigenfunctions of the $\star_g \mathrm{d}$-operator can be organized into a finite number of classes and each class decomposes in a specific unique way into the irreducible representations of $\mathfrak{GU}_\Lambda$. Hence all Arnold-Beltrami vector fields are in  correspondence with the irreps of $\mathfrak{GU}_\Lambda$. Knowing the branching rules of such irreps with respect to its various subgroups $\mathrm{H_i} \subset \mathfrak{GU}_\Lambda$ and selecting the identity representation one obtains Arnold-Beltrami vector fields invariant with respect to those $\mathrm{H_i}$ for which we were able to find an identity irrep $D_1$ in the branching rules. In this way we can classify all Arnold Beltrami flows and also uncover their  \textit{hidden symmetries}.
\par
In this paper we consider in an extensive way the case of the cubic lattice and we construct the corresponding Universal Classifying Group $\mathfrak{GU}_{cubic} \, = \, \mathrm{\mathrm{G_{1536}}}$. This latter is
a finite group  of order $|\mathrm{\mathrm{G_{1536}}}| \, = \, 1536$ which we study in full detail deriving all of its 37 irreducible representations and the associated character table. We also analyze a large class of its subgroups $\mathrm{H_i }\subset \mathrm{\mathrm{G_{1536}}}$ systematically constructing their irreps and character tables.
This allows the derivation of all the branching rules of the 37  $\mathrm{\mathrm{G_{1536}}}$ irreps with respect to the considered subgroups which are displayed in dedicated tables in the appendices.
We show that the $\mathrm{O_{24}}$ orbits in the cubic lattice arrange into $48$ equivalence classes, the parameters of the corresponding Beltrami vector fields  filling all the 37 irreducible representations of $\mathrm{\mathrm{G_{1536}}}$. In this way we obtain an exhaustive classification of all \textit{generalized $\mathrm{ABC}$-flows} and of their hidden symmetries.
 In this way we fulfill the task of classifying and constructing all possible generalizations of the ABC-flows.
\par
From our analysis emerges the following pattern. The Universal Classifying Group contains at least
two\footnote{It is known \cite{bravaislat} that there are 4 different Space-Groups $\Gamma^{I}_{24}$ ($I=1,\dots,4$) of order $24$, isomorphic to the point group  $\mathrm{O_{24}}$ but not conjugate one to the other under the action of the continuous translation group. One of them is the point group  itself $\Gamma^{1}_{24}\, = \, \mathrm{O_{24}}$ which is a subgroup of the first of the two groups of order 192  identified by us: $\mathrm{O_{24}} \subset \mathrm{G_{192}}$. Another of the four mentioned groups is $\Gamma^{2}_{24}\, = \, \mathrm{GS_{24}}$ which is a subgroup of the second group of order 192 identified by us: $\mathrm{GS_{24}} \subset \mathrm{GF_{192}}$. It remains  to see  whether  $\Gamma^{3}_{24} $ and $ \Gamma^{4}_{24}$ are contained in the two already identified subgroups $\mathrm{G_{192}}$ and $\mathrm{GF_{192}}$ or if there exists other two such non conjugate subgroups of order 192 that respectively contain $\Gamma^{3}_{24} $ and $\Gamma^{3}_{24} $. We do not know the answer to such a question. Extensive but lengthy calculation can resolve the issue. } isomorphic but not conjugate subgroups of order 192, namely $\mathrm{G_{192}}$ and $\mathrm{GF_{192}}$ in our nomenclature. The classical ABC-flows are obtained from the lowest lying momentum orbit of length 6 which produces an irreducible $6$-dimensional representation of the Universal Classifying Group: $D_{23}\left[\mathrm{\mathrm{G_{1536}}},6\right] $. The three parameter ABC-flow is just the irreducible $3$-dimensional representation $D_{12}\left[\mathrm{GF_{192}},3\right]$ in the split $D_{23}\left[\mathrm{\mathrm{G_{1536}}},6\right] \, = \, D_{12}\left[\mathrm{GF_{192}},3\right]\oplus D_{15}\left[\mathrm{GF_{192}},3\right]$. With respect to the isomorphic but not conjugate subgroup $\mathrm{G_{192}}$ the representation $D_{23}\left[\mathrm{\mathrm{G_{1536}}},6\right] $ remains instead irreducible: $D_{23}\left[\mathrm{\mathrm{G_{1536}}},6\right] \, = \, D_{20}\left[\mathrm{G_{192}},6\right]$, so that there is no proper way of reducing the six parameters to three. The most symmetric case $A:A:A=1$ simply corresponds to the identity representation of the subgroup  $\mathrm{GS_{24}}\subset \mathrm{GF_{192}}$ which occurs in the splitting  of  the $3$-dimensional representation $D_{12}\left[\mathrm{GF_{192}},3\right]\, = \, D_{1}\left[\mathrm{GS_{24}},1\right]\oplus D_{3}\left[\mathrm{GS_{24}},2\right] $
\par
All other Beltrami flows arising from different instances of the $48$ classes of momentum vectors have similar structures. The result of the construction algorithm produces a representation of the Universal Classifying Group that can be either reducible or irreducible. This latter can be split into irreps of either $\mathrm{G_{192}}$ or $\mathrm{GF_{192}}$ and apparently all cases of invariant Beltrami vector fields have invariance groups that are subgroups of one of the two groups $\mathrm{G_{192}}$ or $\mathrm{GF_{192}}$. It would be interesting to transform this observation into a theorem. At the moment we have not found an obvious proof.
\par
A much shorter sketch of the Hexagonal Lattice is also discussed, to emphasize the generality of our methods, but we do not address the construction of the Universal Classifying Group for this case, which might be performed along the same lines.
\subsection{Organization of the paper}
This very long paper is organized into three parts: the Article, the Appendices and the Bibliography.
The appendices that fill almost 100 pages are  tables whose content is boring,  yet it constitutes an essential and indispensable part  of the presented results:
\begin{enumerate}
  \item Appendix \ref{descriptio} contains the definition of all the relevant groups and subgroups by explicit enumeration of their conjugacy classes of elements.
  \item Appendix \ref{carateropoli} contains all the character tables of the relevant groups and subgroups that we have explicitly constructed, since most of them are not available in the literature.
  \item Appendix \ref{salamicrudi} contains the classification of momentum vectors in the cubic lattice and reports the irreducible representations of the classifying group $\mathrm{\mathrm{G_{1536}}}$ to which each momentum class leads when solving the Beltrami equation.
  \item Appendix \ref{brancicardo} contains all the branching rules of all the irreps of $\mathrm{\mathrm{G_{1536}}}$ with respect to all considered subgroups. This information is essential to spot all Beltrami vector fields that are invariant with respect to some subgroup $\mathcal{H}_i \subset \mathrm{\mathrm{G_{1536}}}$.
  \item Appendix \ref{ABCsubgroups} contains the description and the list of conjugacy classes of some additional subgroups that play a role in understanding all cases and subcase of the classical ABC-flows.
  \item Appendix \ref{largone} contains some formulae too large for the main text that had to be displayed in landscape format.
\end{enumerate}
As for the Article it is divided into the following ten sections:
\begin{enumerate}
  \item Section \ref{introduczia} is the present conceptual introduction.
  \item Section \ref{elementi} presents in a brief way all the elements of lattice theory and point group  theory that are needed in our constructions.
  \item Section \ref{fantasmabeltrami} presents the algorithm for the construction of solutions of the Beltrami equation that has been systematically implemented on a computer by means of a purposely written MATHEMATICA code.
  \item Section \ref{reticolocubico} provides the definition of the cubic lattice and of its  octahedral  point group  that constitute the main example dealt with in the present paper.
  \item Section \ref{maingruppo} contains the definition and the construction, in the case of the cubic lattice, of the Universal Classifying Group. The same section contains also a detailed description of the induction algorithm utilized to construct all the irreps and the character tables of the relevant groups and subgroups.
  \item Section \ref{triplettoni} contains the classification of the $48$ momentum classes in the cubic lattice and the description of their organization into point group  orbits.
  \item Section \ref{beltramotti} contains a detailed discussion of several examples of Beltrami fields on the cubic lattice with an in depth analysis of their hidden symmetries.
  \item Section \ref{hexareticolo} contains a brief description of the hexagonal lattice and of its point group  ${\cal D}_6$. In this case we do not construct the Universal Classifying Group and we just use the example to illustrate the new features that appear when the lattice is not self-dual as in the case of the cubic one.
  \item Section \ref{HexaFildi} briefly presents some examples of Beltrami vector field on the hexagonal lattice for illustrative  purposes.
    \item Section \ref{soklucio}, named the Conclusions, contains a wide conceptual discussion of the obtained results and of the entire field of ABC-flows from the perspective of authors and readers that do not belong to the community of experts in this field of mathematical hydrodynamics.
\end{enumerate}
To the reader who has no time to follow the technical developments and is rather interested in obtaining a conceptual assessment of the matters dealt with in the article we suggest the reading of the introduction and immediately after of the conclusions. He can come back to the other sections at another  time.
\section{Basic Elements of Lattice and  Finite Group Theory needed in our construction}
\label{elementi}
In this section we summarize the main definitions and we fix our conventions for all those items in Lattice Theory and in  Finite Group  Theory that we are going to utilize in the sequel and which are essential in our construction.
\subsection{Lattices}
\label{reticoli}
We begin by fixing  our notations for space and momentum lattices that define a three torus $\mathrm{T}^3$ endowed with a flat metric structure.
\par
Let us consider the standard $\mathbb{R}^3$ manifold and introduce a basis of three  linearly independent 3-vectors that are not necessarily orthogonal to each other and of equal length:
\begin{equation}\label{sospirone}
    \vec{\mathbf{w}}_\mu \, \in \, \mathbb{R}^3 \quad \mu \, = \, 1, \dots \, 3
\end{equation}
Any vector in $\mathbb{R}$ can be decomposed along such a basis and we have:
\begin{equation}\label{xvec}
    \vec{\mathbf{r}} \, = \,  r^\mu\vec{\mathbf{w}}_\mu
\end{equation}
The flat (constant) metric on $\mathbb{R}^3$ is defined by:
\begin{equation}\label{gmunu}
    g_{\mu\nu} \, = \, \langle  \vec{\mathbf{w}}_\mu \, , \,  \vec{\mathbf{w}}_\nu \rangle
\end{equation}
where $\langle \, ,\, \rangle$ denotes the standard Euclidian scalar product.
The space lattice $\Lambda$ consistent with the metric (\ref{gmunu}) is the free abelian group (with respect to sum) generated
by the three basis vectors (\ref{sospirone}), namely:
\begin{equation}\label{reticoloLa}
 \mathbb{R}^3 \, \ni \,   \vec{\mathbf{q}}  \, \in \, \Lambda \, \Leftrightarrow \, \vec{\mathbf{q}} \, = \, q^\mu \, \vec{\mathbf{w}}_\mu \quad \mbox{where} \quad q^\mu \, \in \, \mathbb{Z}
\end{equation}
The momentum lattice is the dual lattice $\Lambda^\star$ defined by the property:
\begin{equation}\label{reticoloLastar}
    \mathbb{R}^3 \, \ni \,   \vec{\mathbf{p}}  \, \in \, \Lambda^\star \, \Leftrightarrow \, \langle \vec{\mathbf{p}} \, , \, \vec{\mathbf{q}}\rangle \, \in \, \mathbb{Z} \quad \forall \, \vec{\mathbf{q}}\, \in \, \Lambda
\end{equation}
A basis for the dual lattice is provided by a set of three \textit{dual vectors} $\vec{\mathbf{e}}^\mu$ defined by the relations\footnote{In the sequel for the scalar product of two vectors we utilize also the equivalent shorter notation $\vec{\mathbf{a}}\, \cdot \vec{\mathbf{b}} \, = \, \langle \vec{\mathbf{a}}\, \cdot \vec{\mathbf{b}}\rangle $.}:
\begin{equation}\label{dualvecti}
    \langle \vec{\mathbf{w}}_\mu \, , \, \vec{\mathbf{e}}^\nu \rangle \, = \, \delta^\nu_\mu
\end{equation}
so that
\begin{equation}\label{pcompi}
    \forall \, \vec{\mathbf{p}} \, \in \, \Lambda^\star \quad \vec{\mathbf{p}} \, = \, p_\mu \, \vec{\mathbf{e}}^\mu \quad \mbox{where } \quad p_\mu \, \in \, \mathbb{Z}
\end{equation}
\subsection{The three torus $\mathrm{T}^3$}
\label{torello}
The three torus is topologically defined as the product of three circles, namely:
\begin{equation}\label{tritorustop}
    \mathrm{T}^3 \, \equiv \, \mathbb{S}^1 \, \times \, \mathbb{S}^1  \, \times \, \mathbb{S}^1  \, \equiv \, \frac{\mathbb{R}}{\mathbb{Z}} \, \times \, \frac{\mathbb{R}}{\mathbb{Z}}\, \times \, \frac{\mathbb{R}}{\mathbb{Z}}
\end{equation}
Alternatively we can define the three-torus by modding $\mathbb{R}^3$ with respect to a three dimensional lattice. In this case the three-torus comes automatically equipped with a flat constant metric:
\begin{equation}\label{metricT3}
     \mathrm{T}^3_g \, = \, \frac{\mathbb{R}^3}{\Lambda}
\end{equation}
According to (\ref{metricT3}) the  flat Riemannian space $\mathrm{T}^3_g$ is defined as the set of equivalence classes with respect to the following equivalence relation:
\begin{equation}\label{equivalenza}
    \vec{\mathbf{r}}^\prime \, \sim \, \vec{\mathbf{r}} \quad \mbox{iff} \quad \vec{\mathbf{r}}^\prime \, - \, \vec{\mathbf{r}} \, \in \, \Lambda
\end{equation}
The metric (\ref{gmunu}) defined on $\mathbb{R}^3$ is inherited by the quotient space and therefore it endows  the topological torus (\ref{tritorustop}) with a flat Riemannian structure. Seen from another point of view the space of flat metrics on  $\mathrm{T}^3$ is just the coset manifold $\mathrm{SL(3,\mathbb{R})}/\mathrm{O(3)}$ encoding all possible symmetric matrices, alternatively all possible space lattices, each lattice being spanned by an arbitrary triplet of basis vectors (\ref{sospirone}).
\subsection{Bravais Lattices}
\label{bravireticoli}
Every lattice $\Lambda$ yields a metric $g$ and every metric $g$ singles out an isomorphic copy $\mathrm{SO_g(3)}$ of the continuous rotation group $\mathrm{SO(3)}$, which leaves it invariant:
\begin{equation}\label{copiaso3}
    M \, \in \, \mathrm{SO_g(3)} \quad \Leftrightarrow \quad M^T \, g \, M \, = \, g
\end{equation}
By definition $\mathrm{SO_g(3)}$ is  the conjugate of the standard $\mathrm{SO(3)}$ in $\mathrm{GL(3,\mathbb{R})}$:
\begin{equation}\label{Sconiugo}
    \mathrm{SO_g(3)} \, = \, \mathcal{S} \, \mathrm{SO(3)} \, \mathcal{S}^{-1}
\end{equation}
with respect to the matrix $\mathcal{S}\,  \in \,\mathrm{GL(3,\mathbb{R})}$ which reduces the metric $g$ to the Kronecker delta:
\begin{equation}\label{Smatrucca}
    \mathcal{S}^T \, g \, \mathcal{S} \, = \, \mathbf{1}
\end{equation}
Notwithstanding this a generic lattice $\Lambda$  is not invariant with respect to any proper subgroup of the rotation group $\mathrm{G} \, \subset \, \mathrm{SO_g(3)}\, \equiv \, \mathrm{SO(3)}$. Indeed by invariance of the lattice one understands the following condition:
\begin{equation}\label{GruppoReticolo}
    \forall \, \gamma \, \in \, \mathrm{G} \quad \mbox{and} \quad \forall \, \vec{\mathbf{q}}\, \in \, \Lambda \,\, : \quad \quad \gamma\,\cdot \, \vec{\mathbf{q}} \, \in \, \Lambda
\end{equation}
Lattices that have a non trivial symmetry group $\mathrm{G} \subset \mathrm{SO(3)}$ are those relevant to Solid State Physics and Crystallography. There are 14 of them grouped in 7 classes  that  were already classified in the XIX century
by Bravais \cite{bravaislat}. The symmetry group $\mathrm{G}$ of each of these Bravais lattices $\Lambda_B$ is necessarily one of the well known  finite  subgroups of the three-dimensional rotation group $\mathrm{O(3)}$. In the language universally adopted by Chemistry and Crystallography for each Bravais lattice $\Lambda_B$ the corresponding invariance group $\mathrm{G_B}$ is named the \textit{Point Group}. For purposes different from our present one, the point group  can be taken as the lattice invariance subgroup within $\mathrm{O(3)}$ that, besides rotations, contains also improper rotations and reflections. Since we are interested in Beltrami equation, which is covariant only under proper rotations, of interest to us are only those point groups  that are subgroups of $\mathrm{SO(3)}$.
\par
According to a standard nomenclature the $7$ classes of Bravais lattices are respectively named \textit{Triclinic, Monoclinic, Orthorombic, Tetragonal, Rhombohedral, Hexagonal and Cubic}. Such classes are specified by giving the lengths of the basis vectors $\vec{\mathbf{w}}_\mu$ and the three angles between them, in other words, by specifying the 6 components of the metric (\ref{gmunu}).
\subsection{The proper Point Groups}
\label{puntigruppi}
Restricting one's attention to proper rotations the proper point groups  that appear in the $7$ lattice classes are either the cyclic groups
$\mathbb{Z}_h$ with $h=2,3,4$ or the dihedral groups ${\cal D}_h$ with $h=3,4,6$ or the tetrahedral group $\mathrm{T}$ or the octahedral  group $\mathrm{O_{24}}$. In this paper we restrict our attention to the two lattices with the largest possible point groups, namely the Hexagonal lattice with $\mathrm{D}_6$ symmetry and the cubic lattice with $\mathrm{O_{24}}$ symmetry. We think that these two examples suffice to clarify the principles of the construction we want to present and furthermore provide the potentially more interesting Beltrami flows  to be analyzed in connection with the problem of the origin of chaotic trajectories.
\subsection{Point Group Characters}
\label{gruppicaratteri}
Another fundamental ingredient in our construction are the characters of the point group  and of other classifying groups that will emerge in our construction.
\par
Given a finite group $\mathrm{G}$, according to standard theory and notations \cite{mieibukki} one defines its order and the order of its conjugacy classes as follows:
\begin{eqnarray}\label{ordinari}
    g & = & \left|\mathrm{G} \right| \, = \, \mbox{$\#$ of group elements} \nonumber\\
    g_i & = & \left|\mathcal{C}_i \right| \, = \, \mbox{$\#$ of group elements in the conjugacy class $\mathcal{C}_i$} \quad i\, = \, i,\dots,r
\end{eqnarray}
If there are $r$ conjugacy classes one knows from first principles that there are exactly $r$ inequivalent irreducible representation $D^{\mu}$ of dimensions $n_\mu \, = \, \mbox{dim} \, D^\mu$, such that:
\begin{equation}\label{dimensiali}
    \sum_{\mu \, = \, 1}^r \, n_\mu^2 \, = \, g
\end{equation}
For any reducible or irreducible representation of dimension $d$:
\begin{equation}\label{RRepra}
    \forall \, \gamma \, \in \, \mathrm{G} \quad : \gamma \, \rightarrow \, \quad \mathfrak{R}\left[\gamma\right]  \, \in \, \mbox{Hom}\left[\mathbb{R}^d \, ,\, \mathbb{R}^d\right]
\end{equation}
the character vector is defined as:
\begin{equation}\label{caratdefi}
    \chi^{\mathfrak{R}}\, = \, \left\{ \mbox{Tr}\, \left(\mathfrak{R}\left[\gamma_1\right] \right) \, , \, \mbox{Tr}\, \left(\mathfrak{R}\left[\gamma_2\right] \right)\, , \,\dots \, \mbox{Tr}\, \left(\mathfrak{R}\left[\gamma_r\right] \right)\right \} \, , \quad \gamma_i \, \in \, \mathcal{C}_i
\end{equation}
The choice of a representative $\gamma_i$ within each conjugacy class $\mathcal{C}_i$ is irrelevant since all representatives have the same trace.
In particular one can calculate the characters of the irreducible representations:
\begin{equation}\label{fundamentalcarat}
   \chi^\mu \, = \,  \chi\left[D^\mu\right]\, \, = \, \left\{ \mbox{Tr}\, \left(D^\mu\left[\gamma_1\right] \right) \, , \, \mbox{Tr}\, \left(D^\mu\left[\gamma_2\right] \right)\, , \,\dots \, , \,\mbox{Tr}\, \left(D^\mu\left[\gamma_r\right] \right)\right \} \, , \quad \gamma_i \, \in \, \mathcal{C}_i
\end{equation}
that are named \textit{fundamental characters} and constitute the \textit{character table}. We stick to the widely adopted convention that the first conjugacy class is that of the identity element $\mathcal{C}_1 \, = \, \left\{\mathbf{e}\right\}$, containing only one member. In this
way the first entry of the character vector is always the dimension $d$ of the considered representation. In the same way we order the irreducible representation starting always with the identity one dimensional representation which associates to each group element simply the number $1$.
\par
It is well known that for any finite group $\mathrm{G}$, the character vectors satisfy the following two fundamental relations:
\begin{equation}\label{idechar1}
    \sum_{\mu \, = \, 1}^r \, \chi_i ^\mu \, \chi_j^\mu \, = \, \frac{g}{g_i} \, \delta_{ij}
\end{equation}
and
\begin{equation}\label{idechar2}
    \sum_{i \, = \, 1}^r \, g_i \,\chi_i ^\mu \, \chi_i^\nu \, = \, g \, \delta^{\mu\nu}
\end{equation}
Utilizing these identities one can immediately retrieve the decomposition of any given reducible representation $\mathcal{R}$ into its irreducible components. Suppose that the considered representation is the following direct sum of irreducible ones:
\begin{equation}\label{multippi}
    \mathfrak{R} \, = \, \oplus_{\mu =1}^r \, a_\mu \, D^\mu
\end{equation}
where $a_\mu$ denotes the number of times the irrep $D^\mu$ is contained in the direct sum and it is named the \textit{multiplicity}.
Given the character vector of any considered representation $\mathfrak{R}$ the vector of its multiplicities is immediately obtained by use of (\ref{idechar2}):
\begin{equation}\label{multipvector}
    a_\mu \, = \, \frac{1}{g} \, \sum_{i}^r \, g_i \chi_i^{\mathfrak{R}} \, \chi_i^\mu
\end{equation}
Furthermore one can construct the projectors onto the invariant subspaces $a_\mu \, D^\mu$ by means of another classical formula that we will extensively use in the sequel\footnote{We recall that according to standard conventions the first conjugacy class is always the class of the identity, so that the first component $\chi^\mu$ of any character is just the dimension of that irrep $D_\mu$. Hence in formula (\ref{proiettori})
$\mbox{dim}D_\mu \, = \,D_\mu$.}:
\begin{equation}\label{proiettori}
    \Pi^\mu_{\mathfrak{R}} \, = \, \frac{
    \mbox{dim}D_\mu}{g} \, \sum_{k=1}^r \, \chi^\mu_k \, \sum_{\ell \,=\, 1}^{g_k} \, \underbrace{\mathfrak{R}\left[{\gamma_\ell}\right]}_{\gamma_\ell \in \, \mathcal{C}_k}
\end{equation}
\section{The spectrum of the $\star \mathbf{d}$ operator on $\mathrm{T}^3$ and Beltrami equation}
\label{fantasmabeltrami}
The main issues of the present paper is the construction of vector fields defined over the three-torus $\mathrm{T}^3$ that are eigenstates of the
$\star_g \mathrm{d}$ operator, namely of solutions of the following equation:
\begin{eqnarray}
    \star_g \mathrm{d}\Omega^{(n;I)} &=& \, m_{(n)} \, \Omega^{(n;I)} \nonumber\\
    \Omega^{(n;I)}\left[V_{(m;J)} \right] &=& \delta^n_m \, \delta^I_J \label{formaduale}
\end{eqnarray}
where $\mathrm{d}$ is the exterior differential, and $\star_g$ is the Hodge-duality operator which, differently from the exterior differential, can be defined only with reference to a given metric $g$. By $\Omega^{(n;i)}$ we denote a one-form:
\begin{equation}\label{omegas}
    \Omega^{(n;I)} \, = \, \Omega^{(n;I)}_\mu \,dx^\mu
\end{equation}
which is declared to be dual to the vector field we are interested in:
\begin{eqnarray}\label{vettocampo}
V_{(m;J)} & = & V_{(m;J)}^\mu \, \partial_\mu \nonumber\\
    \Omega^{(n;I)}\left[V_{(m;J)} \right] & \equiv &  \Omega^{(n;I)}_\mu \, V_{(m;J)}^\mu \, = \, \delta^n_m \, \delta^I_J
\end{eqnarray}
and by means of the composite index $(n;I)$ we make reference to the quantized eigenvalues $m_{(n)}$ of the $\star_g \mathrm{d}$ operator
(ordered in increasing magnitude $|m_{(n)}|$) and to a basis of the corresponding eigenspaces
\begin{equation}\label{superpongomega}
    \star_g \mathrm{d}\Omega^{(n)} \, = \, \, m_{(n)} \, \Omega^{(n)} \quad \Rightarrow \quad \Omega^{(n)}\, = \,
    \sum_{I=1}^{d_n} \,c_I \, \Omega^{(n;I)}
\end{equation}
the symbol $d_n$ denoting the degeneracy of $|m_{(n)}|$ and $c_I$ being constant coefficients.
\par
Indeed, since $T^3$ is a compact manifold, the eigenvalues $m_{(n)}$ form a discrete set.
Their values and their degeneracies are a property of the metric $g$ introduced on it.
Here we outline the general procedure to construct the eigenfunctions of $\star_g \mathrm{d}$, to calculate the
eigenvalues and to determine  their degeneracies.
What follows is an elementary and straightforward exercise in harmonic analysis.
\par
In tensor notation, equation (\ref{formaduale}) has the following appearance:
\begin{equation}\label{tensoBeltra}
    \frac{1}{2} \,\sqrt{|\mbox{det}g|} \, g_{\mu\nu} \, \epsilon^{\nu\rho\sigma} \partial_\rho \Omega_\sigma \, = \, m \, \Omega_\mu
\end{equation}
The equation written above is named Beltrami equation since it was  already considered by the great Italian mathematician Eugenio Beltrami in 1881 \cite{beltramus}, who presented one of its periodic solutions previously constructed by Gromeka in 1881. Such a solution was inherited by Arnold and it is essentially the basis of his Hydrodynamical Model. Here we will see that Arnold Model just corresponds to the lowest eigenfunction of the $\star_g \, \mathrm{d}$-operator in the case of the cubic lattice.
Many more similar models can be constructed choosing higher eigenvalues, choosing irreducible representation of the point group  in their eigenspaces or changing the lattice.
\par
Introducing the basis vectors of the dual lattice $\Lambda^\star$ we can write:
\begin{equation}\label{bardacco1}
    \Omega \, = \, \Omega_\mu \, dr^\mu  \, = \, \Omega_\mu \, e^\mu_i \, dx^i \, = \, \Omega_i \, dx^i
\end{equation}
where $e^\mu_i$ are the components of the vectors $\vec{\mathbf{e}}^\mu$ in a standard orthogonal basis of $\mathbb{R}^3$ and
\begin{equation}\label{bardacco2}
    x^i \, = \, w_\mu^i \, r^\mu
\end{equation}
are a new set of Euclidian coordinates obtained from the original ones $r^\mu$ by means  of the components $w_\mu^i$
of the basis vectors $\vec{\mathbf{w}}_\mu$ of the space lattice $\Lambda$. Recalling that:
\begin{equation}\label{derivosugiu}
    \partial_\mu \, =\, \frac{\partial}{\partial r^\mu} \, = \, w_\mu^i \,\partial_i \quad ; \quad \partial_i \, = \,  \frac{\partial}{\partial x^i}
\end{equation}
with a little bit of straightforward algebra we can rewrite eq.(\ref{formaduale}) in the equivalent universal way:
\begin{equation}\label{tensoBeltra2}
    \frac{1}{2} \,  \epsilon_{ijk} \partial_j \Omega_k \, = \, \mu  \, \Omega_i \quad ; \quad \mu \, = \, m
\end{equation}
The next task is that of constructing an ansatz for the vector harmonics $Y_i(\mathbf{x})$ that are eigenfunctions of  $\star_g \mathrm{d}$. Since such eigenfunctions have to be well defined on $\mathrm{T}^3$, their general form is necessarily the following one:
\begin{eqnarray}\label{harmogen}
    Y_i\left(\mathbf{k}\, | \,\mathbf{x}\right) & = & v_i\left(\mathbf{k}\right) \,\cos\left( 2\,\pi \, \mathbf{k}\cdot \mathbf{x}\right) \, +\, \omega_i\left(\mathbf{k}\right) \,\sin\left( 2\,\pi \, \mathbf{k}\cdot \mathbf{x}\right) \quad ; \quad
    \mathbf{k} \, \in \, \Lambda^\star
\end{eqnarray}
The condition that the momentum $\mathbf{k}$ lies in the dual lattice guarantees that $Y_i(\mathbf{x})$ is periodic with respect to the space lattice $\Lambda$. Indeed, by means of the very definition of dual lattice (\ref{reticoloLastar}) it follows that:
\begin{equation}\label{periodico}
    \forall \, \vec{\mathbf{q}} \, \in \, \Lambda \, : \quad Y_i\left(\mathbf{k}\, | \,\mathbf{x}\, + \, \vec{\mathbf{q}}\right) \, = \, Y_i\left(\mathbf{k}\, | \,\mathbf{x}\right)
\end{equation}
Considering next eq. (\ref{tensoBeltra2}) we immediately see that it implies the further condition $\partial^i \, Y_i \, = \, 0$. Imposing such a condition on the general ansatz (\ref{harmogen}) we obtain:
\begin{equation}\label{transversocondo}
   \mathbf{k}\, \cdot \, \vec{\mathbf{v}}\left(\mathbf{k}\right) \, = \, 0 \quad ; \quad \mathbf{k}\, \cdot \, \vec{\mathbf{\omega}}\left(\mathbf{k}\right) \, = \, 0
\end{equation}
which reduces the 6 parameters contained in the general ansatz (\ref{harmogen}) to 4. Imposing next the very equation (\ref{tensoBeltra2}) we get the following two conditions:
\begin{eqnarray}
  \mu \, v_i \left(\mathbf{k}\right) &=& \pi \, \epsilon_{ij\ell} \, k_j \, \omega_\ell \left(\mathbf{k}\right) \label{curlo1} \\
  \mu \, \omega_i \left(\mathbf{k}\right) &=& -\pi \, \epsilon_{ij\ell} \, k_j \, v_\ell \left(\mathbf{k}\right) \label{curlo2}
\end{eqnarray}
The two equations are self consistent if and only if the following condition is verified:
\begin{equation}\label{spectra1}
    \mu^2 \, = \, \pi^2 \, \langle\mathbf{k}\, , \, \mathbf{k}\rangle
\end{equation}
This trivial elementary calculation completely determines the spectrum of the operator $\star_g \, \mathrm{d}$ on $\mathrm{T}_g^3$ endowed with the metric fixed by the choice of a lattice $\Lambda$. The possible eigenvalues are provided by:
\begin{equation}\label{autovalore}
    m_\mathbf{k} \, = \, \pm \, \pi \,  \sqrt{\langle\mathbf{k}\, , \, \mathbf{k}\rangle}  \quad ,\quad \mathbf{k} \, \in \, \Lambda^\star
\end{equation}
The degeneracy of each eigenvalue is geometrically provided by counting the number of intersection points of the dual lattice $\Lambda^\star$ with a sphere whose center is in the origin and whose radius is:
\begin{equation}\label{radius}
    r \, = \,  \sqrt{\langle\mathbf{k}\, , \, \mathbf{k}\rangle}
\end{equation}
For a generic lattice the number of solutions of equation (\ref{radius}) namely the number of intersection points of the lattice with the sphere is just two: $\pm \mathbf{k}$, so that the typical degeneracy of each eigenvalue is just $2$. If the lattice $\Lambda$ is one of the Bravais lattices admitting a non trivial point group  $\mathrm{G}$, then the number of solutions of eq.(\ref{radius})  increases since all lattice vectors $\mathbf{k}$ that sit  in one orbit of  $\mathrm{G}$ have the same norm and therefore are located on the same spherical surface. The degeneracy of the $\star_g \, \mathrm{d}$ eigenvalue is precisely the order of the corresponding $\mathrm{G}$-orbit in the dual lattice $\Lambda^\star$. It is appropriate to note that the spectrum of the $\star_g \, \mathrm{d}$ operator on vectors is just the square root of the spectrum of the Laplacian operator on the same space. Indeed if we have a scalar function $\Phi(\mathbf{x})$ on $\mathrm{T}_g^3$ it admits the expansion in harmonics of the form:
\begin{eqnarray}\label{harmogen2}
    Y\left(\mathbf{k}\, | \, \mathbf{x}\right) & = & a\left(\mathbf{k}\right) \,\cos\left( 2\,\pi \, \mathbf{k}\cdot \mathbf{x}\right) \, +\, b\left(\mathbf{k}\right) \,\sin\left( 2\,\pi \, \mathbf{k}\cdot \mathbf{x}\right) \quad , \quad
    \mathbf{k} \, \in \, \Lambda^\star
\end{eqnarray}
and calculating the Laplacian we obtain
\begin{equation}\label{laplaccio}
    \Delta_g \, Y\left(\mathbf{k}\, | \,\mathbf{x}\right) \, \equiv  \, \frac{1}{4} \, g^{\mu\nu}\, \partial_\mu \, \partial_\nu \, Y\left(\mathbf{k}\, | \,\mathbf{x}\right) \, = \,  \pi^2 \, \langle\mathbf{k}\, , \, \mathbf{k}\rangle \, Y\left(\mathbf{k}\, | \,\mathbf{x}\right)
\end{equation}
In particular all the three components of the vector harmonic (\ref{harmogen}) satisfy the above equation with the above eigenvalue.
\subsection{The algorithm to construct Arnold Beltrami Flows}
\label{algoritmo}
What we described in the previous subsection provides a well defined algorithm to construct a series of Arnold Beltrami flows
that can be summarized in a few clear-cut steps and it is quite suitable for a systematic  computer aided implementation.
\par
The steps are the following ones:
\begin{description}
  \item[a)] Choose a Bravais Lattice $\Lambda$ with a non trivial proper point group  $\mathfrak{P}_\Lambda$.
  \item[b)] Construct the character table and the irreducible representations of $\mathfrak{P}_\Lambda$.
  \item[c)] Analyze the structure of orbits of $\mathfrak{P}_\Lambda$ on the lattice $\Lambda$ and determine the number  of lattice points  contained in each spherical layer $\mathfrak{S}_n$ of the dual lattice $\Lambda^\star$ of quantized radius $r_n$:
      \begin{eqnarray}\label{spherlayer}
     &&\mathbf{k}_{(n)} \, \in \,  \mathfrak{S}_n \quad \Leftrightarrow \quad  \langle \mathbf{k}_{(n)} \, , \, \mathbf{k}_{(n)}\rangle \, = \, r_n^2\nonumber\\
     && 2\,P_n  \, \equiv \, \left|  \mathfrak{S}_n \right|
      \end{eqnarray}
      The number of lattice points in each spherical layer is always even since if $\mathbf{k} \, \in \, \Lambda^\star$ also $-\mathbf{k} \, \in \, \Lambda^\star$ and obviously any vector and its negative have the same norm.
      The spherical layer $\mathfrak{S}_n$ can be composed of one or of more $\mathfrak{P}_\Lambda$-orbits. In any case it corresponds to a fixed eigenvalue $m_n \, = \, \pi \, r_n$ of the $\star \, \mathrm{d}$-operator.
  \item[d)] Construct the most general solution of the Beltrami equation with eigenvalue $m_n$ by using the individual harmonics
  constructed in eq. (\ref{harmogen}):
  \begin{equation}\label{beltramusgen}
    V_i\left(\mathbf{x}\right) \, = \, \sum_{\mathbf{x}\, \in \,\mathfrak{S}_n } \,Y_i\left(\mathbf{k}\, | \,\mathbf{x}\right)
  \end{equation}
  Hidden in each harmonic $Y_i\left(\mathbf{k}\, | \,\mathbf{x}\right)$ there are two parameters that are the remainder of the six parameters $v_i\left(\mathbf{k}\right)$ and $\omega_i\left(\mathbf{k}\right)$ after conditions (\ref{transversocondo},\ref{curlo1},\ref{curlo2}) have been imposed. This would amount to a total of $4\, P_n$ parameters, yet, since the trigonometric functions $\cos(\theta)$ and $\sin(\theta)$ are mapped into plus or minus themselves under change of sign of their argument and since each spherical layer $\mathfrak{S}_n $ contains lattice vectors in pairs $\pm \mathbf{k}$ , it follows that the number of independent parameters is always reduced to $2 P_n$. Hence, at the end of the construction encoded in eq. (\ref{beltramusgen}), we have a Beltrami vector depending on a set of $2 P_n$ parameters that we can call $F_I$ and consider as the components of $2 \, P_n$-component vector $\mathbf{F}$. Ultimately we have an object of the following form:
  \begin{equation}\label{gomez}
    \mathbf{V}\left(\mathbf{x}\, | \, \mathbf{F}\right)
  \end{equation}
 which under the point group   $\mathfrak{P}_\Lambda$ necessarily transforms in the following way:
 \begin{equation}\label{Rtrasformogen}
    \forall \, \gamma \, \in \, \mathfrak{P}_\Lambda \, : \quad \gamma^{-1} \,\cdot \, \mathbf{V}\left(\gamma \,\cdot \,\mathbf{x}\, | \, \mathbf{F}\right) \, = \, \mathbf{V}\left(\mathbf{x}\, | \, \mathfrak{R}[\gamma] \,\cdot \, \mathbf{F}\right)
 \end{equation}
 where $\mathfrak{R}[\gamma]$ are $2\, P_n \times 2\, P_n$ matrices that form a representation of $\mathfrak{P}_\Lambda$. Eq.(\ref{Rtrasformogen}) is necessarily true because any rotation $\gamma \, \in \, \mathfrak{P}_\Lambda$ permutes the elements of $\mathfrak{S}_n$ among themselves.
   \item[e)] Decompose the representation $\mathfrak{R}[\gamma]$ into irreducible representations of $\mathfrak{P}_\Lambda$. Each irreducible subspace $\mathbf{f}_{p}$ of the $2 P_n$ parameter space $\mathbf{F}$ defines a streamline of an Arnold--Beltrami Flow:
       \begin{equation}\label{arnoldosim}
        \frac{\mathrm{d}}{\mathrm{d}t} \, \mathbf{x}(t) \, = \, \mathbf{V}\left(\mathbf{x}(t)\, | \, \mathbf{f}_p\right)
       \end{equation}
       which is worth to analyze.
\end{description}
An obvious question which arises in connection with such a constructive algorithm is the following: how many Arnold--Beltrami flows are there? At first sight it seems that there is an infinite number of such systems since we can arbitrarily increase the radius of the spherical layer and on each new layer it seems that we have new models. Let us however observe that if on two different   spherical layers $\mathfrak{S}_{n_1}$ and $\mathfrak{S}_{n_2}$ there are two orbits of lattice vectors $\mathcal{O}_1$ and $\mathcal{O}_2$ that have the same order
\begin{equation}\label{parorder}
   \ell \, = \,  \left| \mathcal{O}_1\right | \, = \, \left| \mathcal{O}_2\right |
\end{equation}
and furthermore all vectors ${\mathbf{k}}_{(n_2)} \, \in \,\mathcal{O}_2$ are simply proportional to their analogues in orbit $\mathcal{O}_1$:
\begin{equation}\label{proporzio}
    {\mathbf{k}}_{(n_2)} \, = \, \lambda \, {\mathbf{k}}_{(n_1)} \quad ; \quad \lambda \, \in \, \mathbb{Z}
\end{equation}
then we can conclude that:
\begin{equation}\label{riduzione}
    \mathbf{V}_{(n_2)}\left(\mathbf{x}\, | \, \mathbf{f}_p\right) \, = \,  \mathbf{V}_{(n_1)}\left(\, \lambda \,\mathbf{x}\, | \, \mathbf{f}_p\right)
\end{equation}
By redefining  the coordinate fields $\lambda \,\mathbf{x} \, = \, \mathbf{x}^\prime$ and rescaling time $t$ the two differential systems (\ref{arnoldosim}) respectively  constructed from  layer $n_1$ and layer $n_2$ can be identified.
\par
As we shall demonstrate analyzing the case of the cubic lattice and the orbits of the octahedral  group there is always a finite number of $\mathfrak{P}_\Lambda$-orbit type on each lattice $\Lambda$. There is a  maximal orbit $\mathcal{O}_{max}$ that has order equal to the order of the point group:
\begin{equation}\label{orbitmax}
    \left| \mathcal{O}_{max}\right| \, = \, \left| \mathfrak{P}_\Lambda\right|
\end{equation}
and there are a few shortened orbits $\mathcal{O}_{i}$ ($i=\,1,\dots ,\, s$) that have a smaller order:
\begin{equation}\label{carisma}
    \ell_i \, = \, \left| \mathcal{O}_{i}\right| \, < \, \left| \mathfrak{P}_\Lambda\right|
\end{equation}
The fascinating property is that for the shortened orbits which seem to play an analogue role in this context to that of BPS states in another context, property (\ref{proporzio}) is always true. The vectors pertaining to the same orbit $\mathcal{O}_i$ in different spherical layers are always the same up to a multiplicative factor. Hence from the shortened orbits we obtain always a finite number of Arnold--Beltrami flows. It remains the case of the maximal orbit for which property (\ref{proporzio}) is not necessarily imposed. How many independent flows do we obtain considering all the layers? The answer to the posed question is hidden in number theory.
Indeed we have to analyze how many different types of triplets of integer numbers satisfy Diophantine equations of the Fermat type. In section \ref{triplettoni} we provide a systematic classification of such triplets for the cubic lattice showing that there is a finite number of Arnold--Beltrami flows.
\section{The Cubic Lattice and the Octahedral Point Group}
\label{reticolocubico}
Let us now consider, within the general frame presented above the  cubic lattice.
\par
\begin{figure}[!hbt]
\begin{center}
\iffigs
\includegraphics[height=70mm]{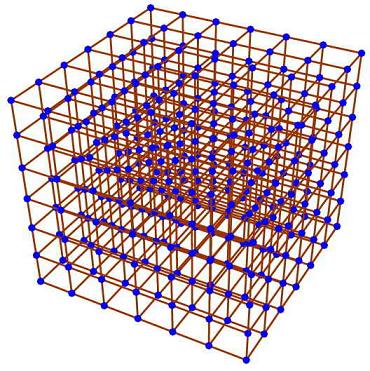}
\else
\end{center}
 \fi
\caption{\it  A view of the self-dual cubic lattice.}
\label{cubicogenerale}
 \iffigs
 \hskip 1cm \unitlength=1.1mm
 \end{center}
  \fi
\end{figure}
The self-dual cubic lattice (momentum and space lattice at the same time) is displayed in fig.\ref{cubicogenerale}.
\par
The basis vectors of the cubic lattice $\Lambda_{cubic}$ are :
\begin{equation}\label{cubicobase}
    \vec{\mathbf{w}}_1 \, = \, \{1,0,0\} \quad ; \quad \vec{\mathbf{w}}_2 \, = \, \{0,1,0\} \quad ; \quad \vec{\mathbf{w}}_3 \, = \, \{0,0,1\}
\end{equation}
which implies that the metric is just the Kronecker delta:
\begin{equation}\label{Kronecca}
    g_{\mu\nu} \, = \, \delta_{\mu\nu}
\end{equation}
and the basis vectors $\vec{\mathbf{e}}^\mu$ of the dual lattice $\Lambda_{cubic}^\star$ coincide with those of the lattice $\Lambda$. Hence the cubic lattice is self-dual:
\begin{equation}\label{autoduali}
    \vec{\mathbf{w}}_\mu \, = \, \vec{\mathbf{e}}^\mu \quad \Rightarrow \quad \Lambda_{cubic} \, = \, \Lambda^\star_{cubic}
\end{equation}
The subgroup of the proper rotation group which maps the cubic lattice into itself is the octahedral  group $\mathrm{O}$ whose order is 24. In the next subsection we recall its structure.
\subsection{Structure of the  Octahedral Group $\mathrm{O_{24}}\sim \mathrm{S_4}$}
\label{ottostruttura}
Abstractly the octahedral Group $\mathrm{O_{24}}\sim \mathrm{S_{4}}$ is isomorphic to the symmetric group of permutations of 4 objects. It is defined by the following generators and relations:
\begin{equation}\label{octapresa}
 T, \, S \quad : \quad    T^3 \, = \, \mbox{\bf e} \quad ; \quad S^2 \, = \, \mbox{\bf e} \quad ; \quad (S\,T)^4 \, = \, \mbox{\bf e}
\end{equation}
On the other hand $\mathrm{O_{24}}$ is a finite, discrete subgroup of the three-dimensional rotation group and any $\gamma \, \in \, \mathrm{O_{24}}\, \subset \, \mathrm{SO(3)}$ of its 24 elements can be uniquely identified by its action on the coordinates $x,y,z$,  as it is displayed below:
\begin{equation}\label{nomiOelemen}
\begin{array}{cc}
\begin{array}{|c|rcl|}
\hline
\mbox{\bf e} & 1_1 & = & \{x,y,z\} \\
 \hline
\null & 2_1 & = & \{-y,-z,x\} \\
\null &  2_2 & = & \{-y,z,-x\} \\
\null & 2_3 & = & \{-z,-x,y\} \\
C_3 & 2_4 & = & \{-z,x,-y\} \\
 \null &2_5 & = & \{z,-x,-y\} \\
\null & 2_6 & = & \{z,x,y\} \\
\null & 2_7 & = & \{y,-z,-x\} \\
\null & 2_8 & = & \{y,z,x\} \\
 \hline
\null & 3_1 & = & \{-x,-y,z\} \\
C_4^2 & 3_2 & = & \{-x,y,-z\} \\
\null & 3_3 & = & \{x,-y,-z\} \\
 \hline
\end{array} & \begin{array}{|c|rcl|}
\hline
\null & 4_1 & = & \{-x,-z,-y\} \\
\null & 4_2 & = & \{-x,z,y\} \\
C_2 &4_3 & = & \{-y,-x,-z\} \\
\null & 4_4 & = & \{-z,-y,-x\} \\
\null & 4_5 & = & \{z,-y,x\} \\
\null & 4_6 & = & \{y,x,-z\} \\
 \hline
\null & 5_1 & = & \{-y,x,z\} \\
\null & 5_2 & = & \{-z,y,x\} \\
C_4 & 5_3 & = & \{z,y,-x\} \\
\null & 5_4 & = & \{y,-x,z\} \\
\null & 5_5 & = & \{x,-z,y\} \\
\null & 5_6 & = & \{x,z,-y\}\\
 \hline
\end{array} \\
\end{array}
\end{equation}
As one sees from the above list the 24 elements are distributed into 5 conjugacy classes mentioned in the first column of the table, according to a nomenclature which is standard in the chemical literature on crystallography. The relation between the abstract and concrete presentation of the octahedral  group is obtained by identifying in the list (\ref{nomiOelemen}) the generators $T$ and $S$ mentioned in eq. (\ref{octapresa}).
Explicitly we have:
\begin{equation}\label{generatiTS}
    T \, = \, 2_8 \, = \, \left(
\begin{array}{lll}
 0 & 1 & 0 \\
 0 & 0 & 1 \\
 1 & 0 & 0
\end{array}
\right)\quad ; \quad S \, = \, 4_6 \, =\left(
\begin{array}{lll}
 0 & 1 & 0 \\
 1 & 0 & 0 \\
 0 & 0 & -1
\end{array}
\right)
\end{equation}
All other elements are reconstructed from the above two using the multiplication table of the group which is displayed below:
{\small
\begin{equation}\label{Omultipla}
\begin{array}{|l|llllllllllllllllllllllll|}
\hline
 \null &   1_1 & 2_1 & 2_2 & 2_3 & 2_4 & 2_5 & 2_6 & 2_7 & 2_8 & 3_1 & 3_2 & 3_3 & 4_1 & 4_2 & 4_3 & 4_4 & 4_5 & 4_6 & 5_1 & 5_2 & 5_3 & 5_4 & 5_5 & 5_6 \\
 \hline
 1_1 &   1_1 & 2_1 & 2_2 & 2_3 & 2_4 & 2_5 & 2_6 & 2_7 & 2_8 & 3_1 & 3_2 & 3_3 & 4_1 & 4_2 & 4_3 & 4_4 & 4_5 & 4_6 & 5_1 & 5_2 & 5_3 & 5_4 & 5_5 & 5_6 \\
 2_1 &   2_1 & 2_5 & 2_4 & 3_3 & 3_2 & 1_1 & 3_1 & 2_6 & 2_3 & 2_7 & 2_2 & 2_8 & 5_3 & 4_4 & 5_6 & 4_6 & 5_4 & 4_2 & 4_1 & 4_3 & 5_1 & 5_5 & 4_5 & 5_2 \\
 2_2 &   2_2 & 2_6 & 2_3 & 1_1 & 3_1 & 3_3 & 3_2 & 2_5 & 2_4 & 2_8 & 2_1 & 2_7 & 4_5 & 5_2 & 5_5 & 5_4 & 4_6 & 4_1 & 4_2 & 5_1 & 4_3 & 5_6 & 5_3 & 4_4 \\
 2_3 &   2_3 & 3_2 & 1_1 & 2_2 & 2_8 & 2_7 & 2_1 & 3_3 & 3_1 & 2_4 & 2_6 & 2_5 & 4_6 & 5_1 & 5_3 & 5_6 & 4_1 & 4_5 & 5_2 & 4_2 & 5_5 & 4_4 & 4_3 & 5_4 \\
 2_4 &   2_4 & 3_1 & 3_3 & 2_1 & 2_7 & 2_8 & 2_2 & 1_1 & 3_2 & 2_3 & 2_5 & 2_6 & 5_4 & 4_3 & 4_5 & 5_5 & 4_2 & 5_3 & 4_4 & 4_1 & 5_6 & 5_2 & 5_1 & 4_6 \\
 2_5 &   2_5 & 1_1 & 3_2 & 2_8 & 2_2 & 2_1 & 2_7 & 3_1 & 3_3 & 2_6 & 2_4 & 2_3 & 5_1 & 4_6 & 5_2 & 4_2 & 5_5 & 4_4 & 5_3 & 5_6 & 4_1 & 4_5 & 5_4 & 4_3 \\
 2_6 &   2_6 & 3_3 & 3_1 & 2_7 & 2_1 & 2_2 & 2_8 & 3_2 & 1_1 & 2_5 & 2_3 & 2_4 & 4_3 & 5_4 & 4_4 & 4_1 & 5_6 & 5_2 & 4_5 & 5_5 & 4_2 & 5_3 & 4_6 & 5_1 \\
 2_7 &   2_7 & 2_3 & 2_6 & 3_1 & 1_1 & 3_2 & 3_3 & 2_4 & 2_5 & 2_1 & 2_8 & 2_2 & 5_2 & 4_5 & 4_2 & 5_1 & 4_3 & 5_6 & 5_5 & 5_4 & 4_6 & 4_1 & 4_4 & 5_3 \\
 2_8 &   2_8 & 2_4 & 2_5 & 3_2 & 3_3 & 3_1 & 1_1 & 2_3 & 2_6 & 2_2 & 2_7 & 2_1 & 4_4 & 5_3 & 4_1 & 4_3 & 5_1 & 5_5 & 5_6 & 4_6 & 5_4 & 4_2 & 5_2 & 4_5 \\
 3_1 &   3_1 & 2_8 & 2_7 & 2_6 & 2_5 & 2_4 & 2_3 & 2_2 & 2_1 & 1_1 & 3_3 & 3_2 & 5_6 & 5_5 & 4_6 & 5_3 & 5_2 & 4_3 & 5_4 & 4_5 & 4_4 & 5_1 & 4_2 & 4_1 \\
 3_2 &   3_2 & 2_7 & 2_8 & 2_5 & 2_6 & 2_3 & 2_4 & 2_1 & 2_2 & 3_3 & 1_1 & 3_1 & 5_5 & 5_6 & 5_4 & 4_5 & 4_4 & 5_1 & 4_6 & 5_3 & 5_2 & 4_3 & 4_1 & 4_2 \\
 3_3 &   3_3 & 2_2 & 2_1 & 2_4 & 2_3 & 2_6 & 2_5 & 2_8 & 2_7 & 3_2 & 3_1 & 1_1 & 4_2 & 4_1 & 5_1 & 5_2 & 5_3 & 5_4 & 4_3 & 4_4 & 4_5 & 4_6 & 5_6 & 5_5 \\
 4_1 &   4_1 & 5_4 & 4_6 & 4_5 & 5_3 & 5_2 & 4_4 & 5_1 & 4_3 & 5_5 & 5_6 & 4_2 & 1_1 & 3_3 & 2_8 & 2_6 & 2_3 & 2_2 & 2_7 & 2_5 & 2_4 & 2_1 & 3_1 & 3_2 \\
 4_2 &   4_2 & 4_6 & 5_4 & 5_3 & 4_5 & 4_4 & 5_2 & 4_3 & 5_1 & 5_6 & 5_5 & 4_1 & 3_3 & 1_1 & 2_7 & 2_5 & 2_4 & 2_1 & 2_8 & 2_6 & 2_3 & 2_2 & 3_2 & 3_1 \\
 4_3 &   4_3 & 5_3 & 5_2 & 5_6 & 4_2 & 5_5 & 4_1 & 4_5 & 4_4 & 4_6 & 5_1 & 5_4 & 2_6 & 2_4 & 1_1 & 2_8 & 2_7 & 3_1 & 3_2 & 2_2 & 2_1 & 3_3 & 2_5 & 2_3 \\
 4_4 &   4_4 & 4_2 & 5_5 & 5_1 & 5_4 & 4_6 & 4_3 & 5_6 & 4_1 & 5_2 & 4_5 & 5_3 & 2_8 & 2_1 & 2_6 & 1_1 & 3_2 & 2_5 & 2_3 & 3_1 & 3_3 & 2_4 & 2_2 & 2_7 \\
 4_5 &   4_5 & 5_6 & 4_1 & 4_6 & 4_3 & 5_1 & 5_4 & 4_2 & 5_5 & 5_3 & 4_4 & 5_2 & 2_2 & 2_7 & 2_4 & 3_2 & 1_1 & 2_3 & 2_5 & 3_3 & 3_1 & 2_6 & 2_8 & 2_1 \\
 4_6 &   4_6 & 4_4 & 4_5 & 4_1 & 5_5 & 4_2 & 5_6 & 5_2 & 5_3 & 4_3 & 5_4 & 5_1 & 2_3 & 2_5 & 3_1 & 2_1 & 2_2 & 1_1 & 3_3 & 2_7 & 2_8 & 3_2 & 2_4 & 2_6 \\
 5_1 &   5_1 & 4_5 & 4_4 & 5_5 & 4_1 & 5_6 & 4_2 & 5_3 & 5_2 & 5_4 & 4_3 & 4_6 & 2_5 & 2_3 & 3_3 & 2_7 & 2_8 & 3_2 & 3_1 & 2_1 & 2_2 & 1_1 & 2_6 & 2_4 \\
 5_2 &   5_2 & 4_1 & 5_6 & 4_3 & 4_6 & 5_4 & 5_1 & 5_5 & 4_2 & 4_4 & 5_3 & 4_5 & 2_7 & 2_2 & 2_5 & 3_3 & 3_1 & 2_6 & 2_4 & 3_2 & 1_1 & 2_3 & 2_1 & 2_8 \\
 5_3 &   5_3 & 5_5 & 4_2 & 5_4 & 5_1 & 4_3 & 4_6 & 4_1 & 5_6 & 4_5 & 5_2 & 4_4 & 2_1 & 2_8 & 2_3 & 3_1 & 3_3 & 2_4 & 2_6 & 1_1 & 3_2 & 2_5 & 2_7 & 2_2 \\
 5_4 &   5_4 & 5_2 & 5_3 & 4_2 & 5_6 & 4_1 & 5_5 & 4_4 & 4_5 & 5_1 & 4_6 & 4_3 & 2_4 & 2_6 & 3_2 & 2_2 & 2_1 & 3_3 & 1_1 & 2_8 & 2_7 & 3_1 & 2_3 & 2_5 \\
 5_5 &   5_5 & 4_3 & 5_1 & 4_4 & 5_2 & 5_3 & 4_5 & 4_6 & 5_4 & 4_1 & 4_2 & 5_6 & 3_2 & 3_1 & 2_2 & 2_4 & 2_5 & 2_8 & 2_1 & 2_3 & 2_6 & 2_7 & 3_3 & 1_1 \\
 5_6 &   5_6 & 5_1 & 4_3 & 5_2 & 4_4 & 4_5 & 5_3 & 5_4 & 4_6 & 4_2 & 4_1 & 5_5 & 3_1 & 3_2 & 2_1 & 2_3 & 2_6 & 2_7 & 2_2 & 2_4 & 2_5 & 2_8 & 1_1 & 3_3\\
 \hline
\end{array}
\end{equation}
}
This observation is important in relation with representation theory. Any linear representation of the group is uniquely specified by giving the matrix representation of the two generators $T=2_8$ and $S=4_6$. In the sequel this will be extensively utilized in the compact codification of the reducible representations that emerge in our calculations.
\subsection{Irreducible representations of the Octahedral Group}
\label{ottoirreppi}
There are five conjugacy classes in $\mathrm{O}_{24}$ and therefore according to theory there are five irreducible representations of the same group, that we name $D_i$, $i\, =\, 1,\dots, 5$. Let us briefly describe them.
\subsubsection{$D_1$ : the identity representation}
The identity representation which exists for all groups is that one where to each element of $\mathrm{O}$ we associate the number $1$
\begin{equation}\label{identD1}
    \forall \, \gamma \, \in \, \mathrm{O}_{24} \,\, : \quad D_1(\gamma) \, = \, 1
\end{equation}
Obviously the character of such a representation is:
\begin{equation}\label{caretterusOD1}
    \chi_1 \, = \, \{1,1,1,1,1\}
\end{equation}
\subsubsection{$D_2$ : the quadratic Vandermonde representation}
The representation $D_2$ is also one-dimensional. It is constructed as follows. Consider the following polynomial of order six in the coordinates of a point in $\mathbb{R}^3$ or $\mathrm{T}^3$:
\begin{equation}\label{vPol}
    \mathfrak{V}(x,y,z) \, = \, (x^2 - y^2) \, (x^2 - z^2) \, (y^2 - z^2)
\end{equation}
As one can explicitly check under the transformations of the octahedral  group listed in eq.(\ref{nomiOelemen}) the polynomial $\mathfrak{V}(x,y,z)$ is always mapped into itself modulo an overall sign. Keeping track of such a sign provides the form of the second one-dimensional representation whose character is explicitly calculated to be the following one:
\begin{equation}\label{caretterusOD2}
    \chi_1 \, = \, \{1,1,1,-1,-1\}
\end{equation}
\subsubsection{$D_3$ : the two-dimensional representation}
The representation $D_3$ is two-dimensional and it corresponds to a homomorphism:
\begin{equation}\label{D3map1}
    D_3 \, : \quad \mathrm{O}_{24} \, \rightarrow \, \mathrm{SL(2,\mathbb{Z})}
\end{equation}
which associates to each element of the octahedral group a $2 \times 2$ integer valued matrix of determinant one.
The homomorphism is completely specified by giving the two matrices representing the two generators:
\begin{equation}\label{D3map2}
    D_3(T) \, = \,
\left(
\begin{array}{ll}
 0 & 1 \\
 -1 & -1
\end{array}
\right) \quad ; \quad D_3(S) \, = \, \left(
\begin{array}{ll}
 0 & 1 \\
 1 & 0
\end{array}
\right)
\end{equation}
The character vector of $D_3$ is easily calculated from the above information and we have:
\begin{equation}\label{caretterusOD3}
    \chi_3 \, = \, \{2,-1,2,0,0\}
\end{equation}
\subsubsection{$D_4$ : the three-dimensional defining representation}
The three dimensional representation $D_4$ is simply the defining representation, where the generators $T$ and $S$ are given by the matrices in eq.(\ref{generatiTS}).
\begin{equation}\label{D4map}
    D_4(T)\, = \, T \quad ; \quad D_4(S) \, = \, S
\end{equation}
From this information the characters are immediately calculated and we get:
\begin{equation}\label{caretterusOD4}
    \chi_3 \, = \, \{3,0,-1,-1,1\}
\end{equation}
\subsubsection{$D_5$ : the three-dimensional unoriented representation}
The three dimensional representation $D_5$ is simply that  where the generators $T$ and $S$ are given by the following  matrices:
\begin{equation}
D_5(T) \, =  \, \left(
\begin{array}{lll}
 0 & 1 & 0 \\
 0 & 0 & 1 \\
 1 & 0 & 0
\end{array}
\right)\quad ; \quad D_5(S) \, = \, \left(
\begin{array}{lll}
 0 & 1 & 0 \\
 1 & 0 & 0 \\
 0 & 0 & 1
\end{array}
\right)
\end{equation}
From this information the characters are immediately calculated and we get:
\begin{equation}\label{caretterusOD4bis}
    \chi_5 \, = \, \{3,0,-1,1,-1\}
\end{equation}
\begin{table}[!hbt]
  \centering
  \begin{eqnarray*}
   \begin{array}{||l||ccccc||}
    \hline
    \hline
{\begin{array}{cc} \null &\mbox{Class}\\
\mbox{Irrep} & \null\\
\end{array}} & \{\mbox{\bf e},1\} & \left\{C_3,8\right\} & \left\{C_4^2,3\right\} & \left\{C_2,6\right\} & \left\{C_4,6\right\} \\
\hline
 \hline
 D_1 \, , \quad \chi_1 \, = \,& 1 & 1 & 1 & 1 & 1 \\
D_2 \, , \quad \chi_2 \, = \,& 1 & 1 & 1 & -1 & -1 \\
 D_3 \, , \quad \chi_3 \, = \, & 2 & -1 & 2 & 0 & 0 \\
 D_4 \, , \quad \chi_4 \, = \, & 3 & 0 & -1 & -1 & 1 \\
D_5 \, , \quad \chi_5 \, = \, & 3 & 0 & -1 & 1 & -1\\
 \hline
 \hline
\end{array}
\end{eqnarray*}
  \caption{Character Table of the proper Octahedral Group}\label{caratteriO}
\end{table}
The table of characters is summarized in Table \ref{caratteriO}.
\section{Extension of the Point Group with Translations and more Group Theory}
\label{maingruppo}
We come now to what constitutes the main mathematical point of the present paper, namely the extension of the point group  with appropriate discrete subgroups of the compactified translation group $\mathrm{U(1)}^3$. This issue bears on a classical topic dating back to the XIX century, which was developed by crystallographers and in particular by the great russian mathematician Fyodorov \cite{fyodorovcryst}. We refer here to the issue of space groups which historically resulted into the classification of the $230$ crystallographic groups, well known in the chemical literature, for which an international system of notations and conventions has been  established that is available in  numerous  encyclopedic tables and books \cite{bravaislat}. Although we will utilize one key-point of the logic that leads to  the classification of space groups, yet our goal happens to be slightly different and what we aim at is not the identification of space groups, rather the construction of what we name a \textit{Universal Classifying Group}, namely of a group which  contains all existing space groups as subgroups. We advocate that such a Universal Classifying Group is the one appropriate to organize the eigenfunctions of the $\star_g \mathrm{d}$-operator into irreducible representations  and eventually to uncover the available hidden symmetries of all Arnold-Beltrami flows.
\subsection{The idea of Space Groups and Frobenius congruences}
\label{spaziogruppi}
The idea of space groups arises naturally in the following way. The covering manifold of the $T^3$ torus is $\mathbb{R}^3$ which can be regarded as the following coset manifold:
\begin{equation}\label{fantacosetto}
    \mathbb{R}^3 \, \simeq \, \frac{\mathbb{E}^3}{\mathrm{SO(3)}} \quad ; \quad  \, \mathbb{E}^3 \, \equiv \, \mathrm{ISO(3)} \, \doteq \, \mathrm{SO(3)} \ltimes \mathcal{T}^3
\end{equation}
where $\mathcal{T}^3$  is the three dimensional translation group acting on $\mathbb{R}^3$ in the standard way:
\begin{equation}\label{trasluco}
    \forall \mathbf{t} \, \in \, \mathcal{T}^3 \, ,\, \forall \mathbf{x} \, \in \, \mathbb{R}^3\quad | \quad\mathbf{t} \, : \,   \mathbf{x} \, \rightarrow \, \mathbf{x}\, + \, \mathbf{t}
\end{equation}
and the Euclidian group $\mathbb{E}^3$ is the semi-direct product  of the proper rotation group $\mathrm{SO(3)}$ with the translation group $\mathcal{T}^3$. Harmonic analysis on $\mathbb{R}^3$ is a complicated matter of functional analysis since $\mathcal{T}^3$ is a non-compact group and its unitary irreducible representations are infinite-dimensional. The landscape changes drastically when we compactify our manifold from $\mathbb{R}^3$ to the three torus $T^3$. Compactification is obtained taking the  quotient of $\mathbb{R}^3$ with respect to the lattice $\Lambda \, \subset \, \mathcal{T}^3$. As a result of this quotient the manifold becomes $\mathbb{S}^1 \times \mathbb{S}^1 \times \mathbb{S}^1$ but also the isometry group is reduced. Instead of $\mathrm{SO(3)}$ as rotation group we are left with its discrete subgroup $\mathfrak{P}_\Lambda \, \subset \, \mathrm{SO(3)}$  which maps the lattice $\Lambda$ into itself (the point group ) and instead of the translation subgroup $\mathcal{T}^3$ we are left with the quotient group:
\begin{equation}\label{quozienTra}
    \mathfrak{T}^3_\Lambda \, \equiv \, \frac{\mathcal{T}^3}{\Lambda} \, \simeq \, \mathrm{U(1)} \times \mathrm{U(1) }\times \mathrm{U(1)}
\end{equation}
In this way we obtain a new group which replaces the Euclidian group and which is the semidirect product of the point group  $\mathfrak{P}_\Lambda$ with $\mathfrak{T}^3_\Lambda $:
\begin{equation}\label{goticoG}
    \mathfrak{G}_{\Lambda}  \, \equiv \, \mathfrak{P}_\Lambda \, \ltimes \,  \mathfrak{T}^3_\Lambda
\end{equation}
The group $\mathfrak{G}_{\Lambda} $  is an exact symmetry of Beltrami equation (\ref{formaduale}) and its action is naturally defined on the parameter space of any of its solutions $\mathbf{V}\left(\mathbf{x} | \mathbf{F}\right)$  that we can obtain by means of the algorithm  described in section \ref{algoritmo}.
To appreciate  this point let us recall that every component of the vector field $\mathbf{V}\left(\mathbf{x} | \mathbf{F}\right)$ associated with a $\mathfrak{P}_\Lambda$ point--orbit $\mathcal{O}$ is a linear combinations of the functions $\cos\left[2\pi \, \mathbf{k}_i \cdot \mathbf{x}\right]$ and
$\sin\left[2\pi \, \mathbf{k}_i \cdot \mathbf{x}\right]$, where $\mathbf{k}_i \, \in \, \mathcal{O}$  are all the momentum vectors contained in the orbit.
Consider next the same functions in a translated point of the three torus $\mathbf{x}^\prime \, = \,\mathbf{x}\,  + \, \mathbf{c}$ where $\mathbf{c} \, = \, \{\xi_1\,,\,\xi_2\, , \, \xi_3\,\}$ is a representative of an equivalence class $\mathfrak{c}$ of constant vectors defined modulo the lattice:
\begin{equation}\label{cribulla}
\mathfrak{c} \, = \, \mathbf{c} \, + \, \mathbf{y} \quad ; \quad \forall \mathbf{y}\, \in \, \Lambda
\end{equation}
The above equivalence classes are the elements  of the quotient group $\mathfrak{T}^3_\Lambda $.
Using standard trigonometric identities $\cos\left[2\pi \, \mathbf{k}_i \cdot \mathbf{x}\, + \, 2\pi \, \mathbf{k}_i \cdot \mathbf{c} \right]$  can be reexpressed as a linear combination of the $\cos\left[2\pi \, \mathbf{k}_i \cdot \mathbf{x}\right]$ and $\sin\left[2\pi \, \mathbf{k}_i \cdot \mathbf{x}\right]$ functions with coefficients that depend on trigonometric functions of $c$. The same is true of $\sin\left[2\pi \, \mathbf{k}_i \cdot \mathbf{x}\, + \, 2\pi \, \mathbf{k}_i \cdot \mathbf{c} \right]$. Note also that because of the periodicity of the trigonometric functions, the shift in their argument by a lattice translation is not-effective so that one deals only with the equivalence classes (\ref{cribulla}).  It follows that for each element $\mathfrak{c}\in \mathfrak{T}^3_\Lambda$  we obtain a matrix representation $\mathcal{M}_\mathfrak{c}$ realized on the $F$ parameters and defined by the following equation:
\begin{eqnarray}\label{traslazionaBuh}
   & \mathbf{V}\left(\mathbf{x}+\mathbf{c} |\mathbf{F}\right)\, =\, \mathbf{V}\left(\mathbf{x} |\mathcal{M}_\mathfrak{c} \mathbf{F}\right) &
\end{eqnarray}
As we already noted in eq.(\ref{Rtrasformogen}), for any group element $\gamma \, \in \, \mathfrak{P}_\Lambda$ we also have a matrix representation induced on the parameter space by the same mechanism:
\begin{equation}\label{RotazioneBuh}
    \forall \, \gamma \, \in \, \mathfrak{P}_\Lambda \, : \quad \gamma^{-1} \,\cdot \, \mathbf{V}\left(\gamma \,\cdot \,\mathbf{x}\, | \, \mathbf{F}\right) \, = \, \mathbf{V}\left(\mathbf{x}\, | \, \mathfrak{R}[\gamma] \,\cdot \, \mathbf{F}\right)
 \end{equation}
 Combining eq.s(\ref{traslazionaBuh}) and (\ref{RotazioneBuh}) we  obtain a matrix realization of the entire group $\mathfrak{G}_{\Lambda} $ in the following way:
 \begin{eqnarray}
   \mathbf{V}\left(\gamma\, \cdot \, \mathbf{x}+\mathbf{c} |\mathbf{F}\right)&=&  \gamma \, \cdot \, \mathbf{V}\left(\mathbf{x} \,  |\, \mathfrak{R}[\gamma] \cdot \mathcal{M}_\mathfrak{c} \,\cdot \,\mathbf{F}\right) \\
   &\Downarrow&\nonumber\\
 \forall \left(\gamma \, , \, \mathfrak{c}\right) \, \in \,  \mathfrak{G}_{\Lambda} &\rightarrow& D\left[ \left(\gamma \, , \, \mathfrak{c}\right) \right] \, = \, \mathfrak{R}[\gamma] \cdot \mathcal{M}_\mathfrak{c}\label{direttocoriandolo}
 \end{eqnarray}
 Actually the construction of Beltrami vector fields in the lowest lying point-orbit, which usually yields a faithful matrix representation of all group elements, can be regarded as an automatic way of taking  the quotient (\ref{quozienTra}) and the resulting representation can be regarded as the defining representation of the group $\mathfrak{G}_{\Lambda}$.
 \par
 The next point in the logic which leads to space groups is the following observation. $\mathfrak{G}_{\Lambda}$ is an unusual mixture of a discrete group (the point group ) with a continuous one (the translation subgroup $\mathfrak{T}^3_\Lambda $). This latter is rather trivial, since its action corresponds to shifting the origin of coordinates in three-dimensional space and, from the point of view of the first order differential system that defines trajectories (see eq.(\ref{streamlines})), it simply corresponds to varying the integration constants. Yet there are in $\mathfrak{G}_{\Lambda}$ some discrete subgroups which can be isomorphic to the point group  $\mathfrak{P}_{\Lambda}$, or to one of its subgroups $H_\Lambda \, \subset \, \mathfrak{P}_{\Lambda}$, without being their conjugate. Such subgroups cannot be disposed of by shifting the origin of coordinates and consequently they can encode non-trivial hidden symmetries of the dynamical system (\ref{streamlines}). The search of such non trivial discrete subgroups of $\mathfrak{G}_{\Lambda}$ is the mission accomplished by crystallographers the result of the mission being the classification of space groups.
 \par
 The simplest and most intuitive way of constructing space-groups relies on the so named Frobenius congruences
 \cite{Aroyo}\cite{Souvignier}. Let us outline this construction.
 Following classical approaches we use a $4\times 4$ matrix representation of the group $\mathfrak{G}_{\Lambda}$:
 \begin{equation}\label{quattroruote}
    \forall \left(\gamma \, , \, \mathfrak{c}\right) \, \in \,  \mathfrak{G}_{\Lambda} \, \rightarrow\,  \hat{D}\left[ \left(\gamma \, , \, \mathfrak{c}\right) \right] \, = \, \left( \begin{array}{c|c}
                             \gamma & \mathfrak{c} \\
                             \hline
                             0& 1
                           \end{array}
    \right)
 \end{equation}
 Performing the matrix product of two elements, in the translation block one has to take into account equivalence modulo lattice $\Lambda$, namely
 \begin{equation}\label{moduloLatte}
    \left( \begin{array}{c|c}
                             \gamma_1 & \mathfrak{c_1} \\
                             \hline
                             0& 1
                           \end{array}
    \right) \, \cdot \, \left( \begin{array}{c|c}
                             \gamma_2 & \mathfrak{c_2} \\
                             \hline
                             0& 1
                           \end{array}
    \right) \, = \,  \left( \begin{array}{c|c}
                             \gamma_1\, \cdot \, \gamma_2 & \gamma_1\,  \mathfrak{c_2} \, + \, \mathfrak{c_1}\, + \, \Lambda \\
                             \hline
                             0& 1
                           \end{array}
    \right)
 \end{equation}
 Utilizing this notation the next step consists of introducing translation deformations of the generators of the point group  searching for deformations that cannot be eliminated by conjugation. We illustrate the steps of such a construction utilizing the example of the cubic lattice and of the octahedral point group.
\subsubsection{Frobenius congruences for the Octahedral Group $\mathrm{O_{24}}$}
The octahedral  group is abstractly defined by the presentation displayed in eq.(\ref{octapresa}). As a first step we parameterize the candidate deformations of the two generators $T$  and $S$ in the following way:
\begin{equation}\label{deformuccia}
    \hat{T} \, = \, \left(
\begin{array}{lll|l}
 0 & 1 & 0 & \tau _1 \\
 0 & 0 & 1 & \tau _2 \\
 1 & 0 & 0 & \tau _3 \\
 \hline
 0 & 0 & 0 & 1
\end{array}
\right) \quad ; \quad \hat{S} \, = \, \left(
\begin{array}{lll|l}
 0 & 0 & 1 & \sigma _1 \\
 0 & -1 & 0 & \sigma _2 \\
 1 & 0 & 0 & \sigma _3 \\
 \hline
 0 & 0 & 0 & 1
\end{array}
\right)
\end{equation}
which should be compared with eq.(\ref{generatiTS}). Next we try impose on the deformed generators the defining relations of $\mathrm{O}_{24}$. By explicit calculation we find:
\begin{eqnarray}\label{finocchiona}
 \hat{T}^3 & = &    \left(
\begin{array}{lll|l}
 1 & 0 & 0 & \tau _1+\tau _2+\tau _3 \\
 0 & 1 & 0 & \tau _1+\tau _2+\tau _3 \\
 0 & 0 & 1 & \tau _1+\tau _2+\tau _3 \\
 \hline
 0 & 0 & 0 & 1
\end{array}
\right) \quad ; \quad \hat{S}^2 \, = \, \left(
\begin{array}{lll|l}
 1 & 0 & 0 & \sigma _1+\sigma _3 \\
 0 & 1 & 0 & 0 \\
 0 & 0 & 1 & \sigma _1+\sigma _3 \\
 \hline
 0 & 0 & 0 & 1
\end{array}
\right) \nonumber\\
\left(\hat{S}\hat{T}\right)^4 & = & \left(
\begin{array}{lll|l}
 1 & 0 & 0 & 4 \sigma _1+4 \tau _3 \\
 0 & 1 & 0 & 0 \\
 0 & 0 & 1 & 0 \\
 \hline
 0 & 0 & 0 & 1
\end{array}
\right)
\end{eqnarray}
so that we obtain the conditions:
\begin{equation}\label{frobeniale}
    \tau _1+\tau _2+\tau _3  \, \in \, \mathbb{Z}\quad ; \quad \sigma _1+\sigma _3 \, \in \, \mathbb{Z} \quad ; \quad 4 \sigma _1+4 \tau _3 \, \in \, \mathbb{Z}
\end{equation}
which are the Frobenius congruences for the present case. Next we consider the effect of conjugation with the most general translation element of the group $\mathfrak{G}_{cubic}$. Just for convenience we parameterize the translation subgroup as follows:
\begin{equation}\label{tmatto}
    \mathfrak{t} \, = \, \left(
\begin{array}{lll|l}
 1 & 0 & 0 & a+c \\
 0 & 1 & 0 & b \\
 0 & 0 & 1 & a-c \\
 \hline
 0 & 0 & 0 & 1
\end{array}
\right)
\end{equation}
and we get:
\begin{equation}\label{giambatto}
    \mathfrak{t} \, \hat{T} \, \mathfrak{t}^{-1} \, = \, \left(
\begin{array}{lll|l}
 0 & 1 & 0 & a-b+c+\tau _1 \\
 0 & 0 & 1 & -a+b+c+\tau _2 \\
 1 & 0 & 0 & \tau _3-2 c \\
 \hline
 0 & 0 & 0 & 1
\end{array}
\right)\quad ; \quad \mathfrak{t} \, \hat{S} \, \mathfrak{t}^{-1} \, = \,\left(
\begin{array}{lll|l}
 0 & 0 & 1 & 2 c+\sigma _1 \\
 0 & -1 & 0 & 2 b+\sigma _2 \\
 1 & 0 & 0 & \sigma _3-2 c \\
 \hline
 0 & 0 & 0 & 1
\end{array}
\right)
\end{equation}
This shows that by using the parameters $b,c$ we can always put $\sigma_1 \, = \, \sigma_2 \, = \,0$, while using the parameter $a$ we can put $\tau_1 \, = \,0$ (this is obviously, not the only possible gauge choice, but it is the most convenient) so that Frobenius congruences reduce to:
\begin{equation}\label{calugone}
     \tau _2+\tau _3  \, \in \, \mathbb{Z}\quad ; \quad \sigma _3 \, \in \, \mathbb{Z} \quad ; \quad 4 \tau _3 \, \in \, \mathbb{Z}
\end{equation}
Eq.(\ref{calugone}) is of great momentum. It tells us that any non trivial subgroup of  $\mathfrak{G}_{cubic}$ which is not conjugate to the point group  contains point group  elements extended with rational translations of the form $\mathfrak{c} \, = \, \left\{ \ft{n_1}{4}\,  , \, \ft{n_2}{4}\,  , \, \ft{n_3}{4}\right\}$. Up to this point our way and that of crystallographers was the same: hereafter our paths separate. The crystallographers classify all possible non trivial groups that extend the point group  with such translation deformations: indeed looking at the crystallographic tables one realizes that  all known space groups for the cubic lattice have translation components of the form $\mathfrak{c} \, = \, \left\{ \ft{n_1}{4}\,  , \, \ft{n_2}{4}\,  , \, \ft{n_3}{4}\right\}$. On the other hand, we do something much simpler which leads to a quite big group containing all possible Space-Groups as subgroups, together with other subgroups that are not space groups in the crystallographic sense.
\subsection{The Universal Classifying Group for the cubic lattice: $\mathrm{\mathrm{G_{1536}}}$}
\label{universalone}
Inspired by the space group  construction and by Frobenius congruences we just consider the subgroup of $\mathfrak{G}_{cubic}$ where translations are quantized in units of $\frac{1}{4}$. In each direction and modulo integers there are just four translations $0, \, \ft 14, \, \ft 12, \, \ft 34$ so that the translation subgroup reduces to $\mathbb{Z}_4 \, \otimes\,\mathbb{Z}_4\, \otimes \, \mathbb{Z}_4$  that has a total of $64$ elements.
In this way we single out a discrete subgroup  $\mathrm{G_{1536}} \, \subset \, \mathfrak{G}_{cubic}$ of order $24 \times 64 \,  =  \, 1536$,  which is simply the semidirect product of the point group $\mathrm{O_{24}}$ with $\mathbb{Z}_4 \, \otimes\,\mathbb{Z}_4\, \otimes \, \mathbb{Z}_4$:
\begin{equation}\label{1536defino}
    \mathfrak{G}_{cubic} \, \supset \, \mathrm{\mathrm{G_{1536}}} \, \simeq \, \mathrm{O_{24}} \, \ltimes \, \left (\mathbb{Z}_4 \, \otimes\,\mathbb{Z}_4\, \otimes \, \mathbb{Z}_4\right)
\end{equation}
We name $\mathrm{\mathrm{G_{1536}}}$ the universal classifying group of the cubic lattice, and its elements can be labeled as follows:
\begin{equation}\label{elementando1536}
    \mathrm{\mathrm{G_{1536}}} \, \in \, \left\{ p_q \, , \, \ft{2 n_1}{4} \, , \, \ft{2 n_2}{4} \, , \, \ft{2 n_3}{4}\right\} \quad \Rightarrow\quad \left\{ \begin{array}{rcl}
   p_q & \in & \mathrm{O_{24}}\\
   \left\{ \ft{n_1}{4}\,  , \, \ft{n_2}{4}\,  , \, \ft{n_3}{4}\right\} & \in & \mathbb{Z}_4 \, \otimes\,\mathbb{Z}_4\, \otimes \, \mathbb{Z}_4\end{array}\right.
\end{equation}
where for the elements of the point group we use the labels $p_q$ established in eq.(\ref{nomiOelemen}) while for the translation part our notation encodes an equivalence class of  translation vectors $\mathfrak{c} \, = \,  \left\{ \ft{n_1}{4}\,  , \, \ft{n_2}{4}\,  , \, \ft{n_3}{4}\right\}$. The reason why we use $\left \{\ft{2 n_1}{4} \, , \, \ft{2 n_2}{4} \, , \, \ft{2 n_3}{4}\right\}$ is simply due to computer convenience. In the quite elaborate  MATHEMATICA codes that we have utilized to derive all our results we internally used such a notation and the automatic LaTeX Export of the outputs is  provided in this way.
In view of eq.(\ref{direttocoriandolo}) we can associate an explicit matrix to each group element of $\mathrm{\mathrm{G_{1536}}}$, starting from the construction of the Beltrami vector field associated with one point orbit of the octahedral  group. We can consider such matrices the defining representation of the group if the representation is faithful. We used the lowest lying $6$-dimensional orbit to be discussed in  section \ref{beltramotti} which we verified to be indeed faithful. Three matrices are sufficient to characterize completely the defining representation just as any other representation: the matrix representing the generator $T$, the matrix representing the generator $S$ and the matrix representing the translation $\left\{ \ft{n_1}{4}\,  , \, \ft{n_2}{4}\,  , \, \ft{n_3}{4}\right\}$. We have found:
\begin{equation}\label{TSindefi1536}
    \mathfrak{R}^{\mbox{defi}}[T] \, = \, \left(
\begin{array}{llllll}
 0 & 0 & 0 & 0 & 1 & 0 \\
 0 & 0 & 0 & 0 & 0 & 1 \\
 0 & 1 & 0 & 0 & 0 & 0 \\
 1 & 0 & 0 & 0 & 0 & 0 \\
 0 & 0 & 0 & 1 & 0 & 0 \\
 0 & 0 & 1 & 0 & 0 & 0
\end{array}
\right)\quad ; \quad \mathfrak{R}^{\mbox{defi}}[S]\, = \, \left(
\begin{array}{llllll}
 0 & 0 & 1 & 0 & 0 & 0 \\
 0 & 0 & 0 & 1 & 0 & 0 \\
 1 & 0 & 0 & 0 & 0 & 0 \\
 0 & 1 & 0 & 0 & 0 & 0 \\
 0 & 0 & 0 & 0 & 0 & -1 \\
 0 & 0 & 0 & 0 & -1 & 0
\end{array}
\right)
\end{equation}
\begin{eqnarray}
   \mathcal{M}_{\{\ft{2 n_1}{4},\ft{2 n_2}{4},\ft{2 n_3}{4}\}}^{\mbox{defi}} &= & \left(
\begin{array}{llllll}
 \cos \left(\ft{\pi}{2}  n _3\right) & 0 & \sin \left(\ft{\pi}{2}  n _3\right) & 0 & 0 & 0 \\
 0 & \cos \left(\ft{\pi}{2}  n _2\right) & 0 & 0 &
   -\sin \left(\ft{\pi}{2}  n _2\right) & 0 \\
 -\sin \left(\ft{\pi}{2}  n _3\right) & 0 & \cos
   \left(\ft{\pi}{2}  n _3\right) & 0 & 0 & 0 \\
 0 & 0 & 0 & \cos \left(\ft{\pi}{2}  n _1\right) & 0 &
   \sin \left(\ft{\pi}{2}  n _1\right) \\
 0 & \sin \left(\ft{\pi}{2}  n _2\right) & 0 & 0 & \cos
   \left(\ft{\pi}{2}  n _2\right) & 0 \\
 0 & 0 & 0 & -\sin \left(\ft{\pi}{2}  n _1\right) & 0 &
   \cos \left(\ft{\pi}{2}  n _1\right)
\end{array}
\right)\nonumber\\
\label{trasladefi1536}
\end{eqnarray}
Relying on the above, matrices any of the 1536 group elements obtains an explicit $6\times 6$ matrix representation upon use of formula
(\ref{direttocoriandolo}). As already stressed we can regard that above as the actual definition of the group $\mathrm{\mathrm{G_{1536}}}$ which from this point on can be studied intrinsically in terms of pure group theory without any further reference to lattices, Beltrami flows or dynamical systems.
\subsection{Structure of the $\mathrm{\mathrm{G_{1536}}}$ group and derivation of its irreps}
The identity card of a finite group is given by the organization of its elements into conjugacy classes, the list of its irreducible representation and finally its character table. Since ours is not any of the crystallographic groups, no explicit information is available in the literature about its conjugacy classes, its irreps and its character table. We were forced to do everything from scratch by ourselves and we could accomplish the task by means   of purposely written MATHEMATICA codes.
Most of our results are presented in the form of tables in the appendices. Since this is a purely mathematical information, we think that it might be useful also in other contexts different from the present context that has motivated their derivation.
\paragraph{Conjugacy Classes}
The conjugacy classes of $\mathrm{\mathrm{G_{1536}}}$ are presented in appendix \ref{coniugato1536}. There are 37 conjugacy classes whose populations is distributed as follows:
\begin{description}
  \item[1)]  2 classes of length 1
  \item[2)]  2  classes of length 3
  \item[3)]  2 classes of  length 6
  \item[4)] 1 class of length 8
  \item[5)] 7 classes of length 12
  \item[6)] 4 classes of length 24
  \item[7)] 13 classes of length 48
  \item[8)] 2 classes of length 96
  \item[9)] 4 classes of length 128
\end{description}
It follows that there must be $37$  irreducible representations whose construction is a task to be solved.
\subsubsection{Strategy to construct the irreducible representations of a solvable group}
\label{startegos}
In general, the derivation of the irreps and of the ensuing character table of a finite group $G$ is  a quite hard task. Yet  a definite constructive algorithm can be devised if $G$ is solvable and if one can establish a chain of normal subgroups ending with an abelian one, whose index is, at each step,  a prime number $q_i$, namely if we have the following situation:
\begin{eqnarray}\label{parrucchiere}
  G &= &   \mathrm{G_{N_p}} \, \rhd \, \mathrm{G_{N_{p-1}}}  \, \rhd \, \dots \, \rhd \, \mathrm{G_{N_1}}\rhd \, \mathrm{G_{N_0}} \, = \, \mbox{abelian group} \nonumber\\
 && \left| \frac{\mathrm{G_{N_i}}}{\mathrm{G_{N_{i-1}}}}\right| \, = \, \frac{\mathrm{N_i}}{\mathrm{N_{i-1}}}\, \equiv \, q_i \, = \, \mbox{prime integer number}
\end{eqnarray}
The algorithm for the construction of the irreducible representations is based on an inductive procedure \cite{Aroyo} that allows to derive the irreps of the group $\mathrm{G_{N_i}}$ if we know those of the group $\mathrm{G_{N_{i-1}}}$ and if the index $q_i$ is a prime number. The first step of the induction is immediately solved because any abelian finite group is necessarily a direct product of cyclic groups $\mathbb{Z}_k$, whose irreps are all one dimensional and obtained by assigning to their generator one of the $k$-th roots of unity. In our case the index $q_i$ is always either $2$ or $3$  which, to none's  wonder,  is the same situation met in the construction of  crystallographic group irreps. Hence we sketch the inductive algorithms with particular reference to the two cases of $q=2$ and $q=3$.
\subsubsection{The inductive algorithm for irreps}
\label{agoinduzio}
To simplify notation we name $\mathcal{G}\, = \, \mathrm{G_{N_i}}$  and $\mathcal{H}\, = \, \mathrm{G_{N_{i-1}}}$. By hypothesis $\mathcal{H} \, \lhd \, \mathcal{G}$ is a normal subgroup. Furthermore $q \, \equiv \, \left |\frac{\mathcal{G}}{\mathcal{H}} \right| \, = \, $ \textit{prime number} (in particular $q\, = \, 2, \mbox{or} \, 3$).
Let us name $D_\alpha\left[\mathcal{H},d_\alpha\right]$ the irreducible representations of the subgroup. The index $\alpha$ (with $\alpha\, = \, 1,\dots, r_H \, \equiv \,  \# \mbox{ of conj. classes of $\mathcal{H}$ }$) enumerates them. In each case $d_\alpha$ denotes the dimension of the corresponding carrying vector space or, in mathematical jargon, of the corresponding module.
\par
The first step to be taken is to distribute the $\mathcal{H}$ irreps into conjugation classes with respect to the bigger group. Conjugation classes of irreps are defined as follows. First one observes that, given an irreducible representation $D_\alpha\left[\mathcal{H},d_\alpha\right]$, for every $g \, \in \, \mathcal{G}$ we can create another irreducible representation  $D^{(g)}_\alpha\left[\mathcal{H},d_\alpha\right]$, named the conjugate of $D_\alpha\left[\mathcal{H},d_\alpha\right]$  with respect to $g$. The new representation is as follows:
\begin{equation}\label{Dconiugo}
    \forall \, h \, \in \, \mathcal{H} \, \quad : \quad D^{(g)}_\alpha\left[\mathcal{H},d_\alpha\right](h) \, = \, D_\alpha\left[\mathcal{H},d_\alpha\right](g^{-1} \, h \, g)
\end{equation}
That the one defined above is a homomorphism of  $\mathcal{H}$ onto  $\mathrm{GL(d_\alpha,\mathbb{R})}$ is obvious and, as a consequence, it is also obvious that the new representation has the same dimension as the first. Secondly if $g = \tilde{h} \, \in \, \mathcal{H}$ is an element of the subgroup we get:
\begin{equation}\label{maialeH}
    \quad D^{(\tilde{h})}_\alpha\left[\mathcal{H},d_\alpha\right](h) \, = \, A^{-1} \, D_\alpha\left[\mathcal{H},d_\alpha\right]( h) \, A \quad \mbox{where} \quad A \, = \, D_\alpha\left[\mathcal{H},d_\alpha\right](\tilde{ h})
\end{equation}
so that conjugation amounts simply to a change of basis (a similarity transformation) inside the same representation. This does not alter the character vector and the new representation is equivalent to the old one.  Hence the only non trivial conjugations to be considered are those with respect to  representatives of the different equivalence classes in $\frac{\mathcal{G}}{\mathcal{H}}$. Let us name $\gamma_i $, $(i=0,\dots ,\,q-1)$ a set of representatives of such equivalence classes and define the orbit of each irrep $D_\alpha\left[\mathcal{H},d_\alpha\right]$ as it follows:
\begin{equation}\label{orbetellosulmare}
    \mbox{Orbit}_\alpha \, \equiv \, \left\{ D^{(\gamma_0)}_\alpha\left[\mathcal{H},d_\alpha\right] \, , \, D^{(\gamma_1)}_\alpha\left[\mathcal{H},d_\alpha\right] \, , \, \dots \, , \,D^{(\gamma_{q-1})}_\alpha\left[\mathcal{H},d_\alpha\right]\right\}
\end{equation}
Since the available irreducible representations are a finite set, every $D^{(\gamma_i)}_\alpha\left[\mathcal{H},d_\alpha\right] $ necessarily is  identified with one of the existing $D_\beta\left[\mathcal{H},d_\beta\right] $. Furthermore, since conjugation preserves the dimension, it follows that $d_\alpha \, = \, d_\beta$. It follows that $\mathcal{H}$-irreps of the same dimensions $d$  arrange themselves into $\mathcal{G}$-orbits:
\begin{equation}\label{orbetellosulmarebis}
    \mbox{Orbit}_\alpha[d] \, = \, \left\{ D_{\alpha_1}\left[\mathcal{H},d\right] \, , \, D_{\alpha_2}\left[\mathcal{H},d\right] \, , \, \dots \, , \,D_{\alpha_q}\left[\mathcal{H},d\right]\right\}
\end{equation}
and there are only two possibilities, either all $\alpha_i \, = \, \alpha$ are equal (self-conjugate representations) or they are all different (non conjugate representations).
\par
Once the irreps of $\mathcal{H}$ have been organized into conjugation orbits, we can proceed to promote them to irreps of the big group $\mathcal{G}$ according to the following scheme:
\begin{description}
  \item[A)] Each self-conjugate $\mathcal{H}$-irrep $D_\alpha\left[\mathcal{H},d\right]$ is uplifted to $q$ distinct irreducible $\mathcal{G}$-representations of the same dimension $d$, namely $D_{\alpha_{i}}\left[\mathcal{G},d\right]$ where $i\, = \, 1,\dots,q$.
  \item[B)] From each orbit $\beta$ of $q$ distinct but conjugate $\mathcal{H}$-irreps $\left\{ D_{\alpha_1}\left[\mathcal{H},d\right] \, , \, D_{\alpha_2}\left[\mathcal{H},d\right] \, , \, \dots \, , \,D_{\alpha_q}\left[\mathcal{H},d\right]\right\}$ one extracts a single ($q\times d$)-dimensional $\mathcal{G}$-representation.
\end{description}
\paragraph{A) Uplifting of self--conjugate representations.} Let $D_{\alpha}\left[\mathcal{H},d\right]$ be a self conjugate irrep. If the index $q$ of the normal subgroup is a prime number,  this means that $\frac{\mathcal{G}}{\mathcal{H}}\, \simeq \, \mathbb{Z}_q$. In this case the representatives $\gamma_{j}$ of the $q$ equivalences classes that form the quotient group can be chosen in the following way:
\begin{equation}\label{gambaletti}
    \gamma_1 \, = \, \mathbf{e} \, , \, \gamma_2 \, = \, g \, , \, \gamma_3 \, = \, g^2 , \, \dots , \, \gamma_q  \, = \, g^{q-1}
\end{equation}
where $g \, \in \,  \mathcal{G}$ is a single group element satisfying $g^q \, = \, \mathbf{e}$. The key-point in up-lifting the representation $D_{\alpha}\left[\mathcal{H},d\right]$ to the bigger group resides in the determination of a $d\times d$ matrix $U$ that should satisfy the following constraints:
\begin{eqnarray}
  U^q &=& \mathbf{1} \\
 \forall \, h \, \in \, \mathcal{H} & : &\quad D_{\alpha}\left[\mathcal{H},d\right]\left(g^{-1}\, h \, g\right) \,=\, U^{-1}\, D_{\alpha}\left[\mathcal{H},d\right]\left(h\right) \, U
\end{eqnarray}
These algebraic equations have exactly $q$ distinct solutions $U_{[j]}$   and each of the solutions leads to one of the irreducible $\mathcal{G}$-representations induced by $D_{\alpha}\left[\mathcal{H},d\right]$. Any element $\gamma \, \in \, \mathcal{G}$ can be written
as $\gamma \, = \, g^p \, h $ with $p=0,1,\dots ,\, q-1$ and $h \, \in \, \mathcal{H}$.
Then it suffices to write:
\begin{equation}\label{corbezzoli}
    D_{a_j}\left[\mathcal{G},d\right](\gamma) \, = \,  D_{a_j}\left[\mathcal{G},d\right]\left (g^p \, h\right) \, = \, U_{[j]}^p \, D_{\alpha}\left[\mathcal{H},d\right] (h)
\end{equation}
\paragraph{B) Uplifting of not self--conjugate representations.} In the case of not self-conjugate representations the induced representation of dimensions $q\times d$ is constructed relying once again on the possibility to write all group elements in the form $\gamma \, = \, g^p \, h $ with $p=0,1,\dots ,\, q-1$ and $h \, \in \, \mathcal{H}$.  Furthermore chosen one representation $D_{\alpha}\left[\mathcal{H},d\right]$ in the $q$-orbit
(\ref{orbetellosulmare}), the other members of the orbit can be represented as $D^{(g^{j})}_\alpha\left[\mathcal{H},d_\alpha\right]$ with $j=1,\,\dots ,\, q-1$. In view of this one writes:
\begin{eqnarray}\label{ciccolone}
     \forall \, h \, \in \, \mathcal{H} &:&\quad D_{\alpha}\left[\mathcal{G},d\right]\left( h \right) \,=\,  \left(\begin{array}{c|c|c|c|c}
                                                                                                                       D_{\alpha}\left[\mathcal{H},d\right](h)& 0& 0 &\dots & 0 \\
                                                                                                                       \hline
                                                                                                                       0& D^{(g)}_{\alpha}\left[\mathcal{H},d\right](h) &0& \dots & 0 \\
                                                                                                                       \hline
                                                                                                                       0 & 0& D^{(g^2)}_{\alpha}\left[\mathcal{H},d\right](h) & \dots & 0 \\
                                                                                                                       \hline
                                                                                                                       \vdots & \vdots & \vdots & \vdots &\vdots \\
                                                                                                                       \hline
                                                                                                                       0& 0 & \dots & 0 & D^{(g^{q-1})}_{\alpha}\left[\mathcal{H},d\right](h)
                                                                                                                     \end{array}
     \right)\nonumber\\
     \null &\null & \null \nonumber\\
     g &:&\quad D_{\alpha}\left[\mathcal{G},d\right]\left( g \right) \,=\,  \left(\begin{array}{c|c|c|c|c}
                                                                                                                       0& 0& 0 &\dots & \mathbf{1} \\
                                                                                                                       \hline
                                                                                                                       \mathbf{1}& 0 &0& \dots & 0 \\
                                                                                                                       \hline
                                                                                                                       0 & \mathbf{1}& 0 & \dots & 0 \\
                                                                                                                       \hline
                                                                                                                       \vdots & \vdots & \vdots & \vdots &\vdots \\
                                                                                                                       \hline
                                                                                                                       0& 0 & \dots &\mathbf{1} & 0
                                                                                                                     \end{array}
     \right)\nonumber\\
     \null &\null & \null \nonumber\\
     \gamma \, = \, g^p \, h  &:&\quad D_{\alpha}\left[\mathcal{G},d\right]\left( g \right) \,=\, \left(D_{\alpha}\left[\mathcal{G},d\right]\left( g \right) \right)^p \, D_{\alpha}\left[\mathcal{G},d\right]\left( h \right) \nonumber\\
   \end{eqnarray}
\subsubsection{Derivation of $\mathrm{\mathrm{G_{1536}}}$ irreps}
\label{ciurlacca}
Utilizing the above described algorithm, implemented by means of purposely written MATHEMATICA codes, we were able to derive the explicit form of the $37$  irreducible representations of $\mathrm{\mathrm{G_{1536}}}$ and its character table. The essential  tool is the following chain of normal subgroups:
\begin{equation}\label{pernicinormaliText}
    \mathrm{\mathrm{G_{1536}} }\, \rhd \, \mathrm{G_{768}} \, \rhd \, \mathrm{G_{256}} \, \rhd \, \mathrm{G_{128}} \, \rhd \, \mathrm{G_{64}}
  \end{equation}
  where $\mathrm{G_{64}} \,  \sim \, \mathbb{Z}_4 \, \times \, \mathbb{Z}_4 \, \times \, \mathbb{Z}_4$ is abelian and corresponds to the compactified translation group. The above chain leads to the following quotient groups:
  \begin{equation}\label{fagianirossiText}
    \frac{\mathrm{\mathrm{G_{1536}} }}{\mathrm{G_{768}}} \, \sim \, \mathbb{Z}_2 \quad ; \quad \frac{\mathrm{G_{768} }}{\mathrm{G_{256}}} \, \sim \, \mathbb{Z}_3 \quad ; \quad \frac{\mathrm{G_{256} }}{\mathrm{G_{128}}} \, \sim \, \mathbb{Z}_2 \quad ; \quad \frac{\mathrm{G_{128} }}{\mathrm{G_{64}}} \, \sim \, \mathbb{Z}_2
  \end{equation}
  The description of the normal subgroups is given in the appendix in sections \ref{coniugato768}, \ref{coniugato256}, \ref{coniugato128}, \ref{coniugato64}.
\par
The result for the irreducible representations, thoroughly described in the appendix \ref{caratterbrut1536}, is summarized here. The $37$ irreps are distributed according to the following pattern:
\begin{description}
  \item[a)] 4 irreps of dimension $1$, namely $D_1,\dots,D_4$
  \item[b)] 2 irreps of dimension $2$, namely $D_5,\dots,D_6$
  \item[c)] 12 irreps of dimension $3$, namely $D_6,\dots,D_{18}$
  \item[d)] 10 irreps of dimension $6$, namely $D_7,\dots,D_{28}$
  \item[e)] 3 irreps of dimension $8$, namely $D_{29},\dots,D_{31}$
  \item[f)] 6 irreps of dimension $12$, namely $D_{32},\dots,D_{37}$
\end{description}
The  character table is displayed in eq.s (\ref{1charto1536}, \ref{2charto1536}).
\par
The irreducible representations of the universal classifying group are a fundamental tool in our classification of Arnold-Beltrami vector fields. Indeed by choosing the various  point group  Orbits of momentum vectors in the cubic lattice, according to their classification presented in the next section \ref{triplettoni}, and constructing the corresponding Arnold-Beltrami fields we obtain all of the $37$ irreducible representations of $\mathrm{\mathrm{G_{1536}}}$. Each representation appears at least once and some of them appear several times. Considering next the subgroups $\mathcal{H}_i$ of $\mathrm{\mathrm{G_{1536}}}$ and the branching rules of  $\mathrm{\mathrm{G_{1536}}}$ irreps with respect to $H_i$ we obtain an explicit algorithm to construct Arnold-Beltrami vector fields with prescribed invariance groups $H_i$. It suffices to select the identity representation of the subgroup in the branching rules. \textit{These are the hidden symmetries advocated in our title.} We come back to the issue of subgroups in the next and following sections.
\section{The spherical layers and the octahedral  lattice orbits}
\label{triplettoni}
\par
Let us now analyze the action of the octahedral  group on the cubic lattice. We define the orbits as the sets of vectors $\mathbf{k}\, \in \, \Lambda$ that can be mapped one into the other by the action of some element of the point group, namely of $\mathrm{O_{24}}$, in the case of the cubic lattice:
\begin{equation}\label{orbitadefi}
    \mathbf{k}_1 \, \in \, \mathcal{O} \quad \mbox{and} \quad \mathbf{k}_2 \, \in \, \mathcal{O} \quad \Rightarrow \quad \exists \, \gamma \, \in \, \mathrm{O_{24}} \; / \; \gamma\,\cdot \, \mathbf{k}_1 \, = \, \mathbf{k}_2
\end{equation}
There are four types of orbits on the cubic lattice:
\paragraph{Orbits of length 6}
Each of these orbits is of the following form:
\begin{equation}\label{orbita6}
\mathcal{O}_6 \, = \,     \left\{
\begin{array}{llllll}
 \{0,0,-n\}, & \{0,0,n\}, & \{0,-n,0\}, & \{0,n,0\}, &
   \{-n,0,0\}, & \{n,0,0\}
\end{array}
\right\}
\end{equation}
where $n \, \in \, \mathbb{Z}$ is any integer number.
\paragraph{Orbits of length 8}
\begin{equation}\label{orbita8}
\mathcal{O}_8 \, = \,     \left\{
\begin{array}{llll}
 \{-n,-n,-n\}, & \{-n,-n,n\}, & \{-n,n,-n\}, & \{-n,n,n\}, \\
   \{n,-n,-n\}, & \{n,-n,n\}, & \{n,n,-n\}, & \{n,n,n\}\\
\end{array}
\right\}
\end{equation}
where $n \, \in \, \mathbb{Z}$ is any integer number.
\paragraph{Orbits of length 12}
\begin{equation}\label{orbita12}
\mathcal{O}_{12} \, = \,     \left\{
\begin{array}{llll}
 \{0,-n,-n\}, & \{0,-n,n\}, & \{0,n,-n\}, & \{0,n,n\}, \\
   \{-n,0,-n\}, & \{-n,0,n\}, & \{-n,-n,0\}, & \{-n,n,0\}, \\
   \{n,0,-n\}, & \{n,0,n\}, & \{n,-n,0\}, & \{n,n,0\}
\end{array}
\right\}
\end{equation}
where $n \, \in \, \mathbb{Z}$ is any integer number.
\begin{table}[!hbt]
  \centering
  \begin{eqnarray*}
\begin{array}{|c|c|l|}
\hline
r^2 &\mbox{Number of Points} &\mbox{Octahedral  Point Group Orbits}\\
\hline
 0 & 1 & \{\{1,1\}\} \\
 1 & 6 & \left\{\left\{O_1,6\right\}\right\} \\
 2 & 12 & \left\{\left\{O_1^2,12\right\}\right\} \\
 3 & 8 & \left\{\left\{O_1^3,8\right\}\right\} \\
 4 & 6 & \left\{\left\{O_1^4,6\right\}\right\} \\
 5 & 24 & \left\{\left\{O_1^5,24\right\}\right\} \\
 6 & 24 & \left\{\left\{O_1^6,24\right\}\right\} \\
 8 & 12 & \left\{\left\{O_1^8,12\right\}\right\} \\
 9 & 30 &
   \left\{\left\{O_1^9,6\right\}\oplus\left\{O_2^9,24\right\}\right
   \} \\
 10 & 24 & \left\{\left\{O_1^{10},24\right\}\right\} \\
 11 & 24 & \left\{\left\{O_1^{11},24\right\}\right\} \\
 12 & 8 & \left\{\left\{O_1^{12},8\right\}\right\} \\
 13 & 24 & \left\{\left\{O_1^{13},24\right\}\right\} \\
 14 & 48 &
   \left\{\left\{O_1^{14},24\right\}\oplus\left\{O_2^{14},24\right\}\right\} \\
 16 & 6 & \left\{\left\{O_1^{16},6\right\}\right\} \\
 17 & 48 &
   \left\{\left\{O_1^{17},24\right\}\oplus\left\{O_2^{17},24\right\}\right\} \\
 18 & 36 &
   \left\{\left\{O_1^{18},24\right\}\oplus\left\{O_2^{18},12\right\}\right\} \\
 19 & 24 & \left\{\left\{O_1^{19},24\right\}\right\} \\
 20 & 24 & \left\{\left\{O_1^{20},24\right\}\right\} \\
 21 & 48 &
   \left\{\left\{O_1^{21},24\right\}\oplus\left\{O_2^{21},24\right\}\right\} \\
 22 & 24 & \left\{\left\{O_1^{22},24\right\}\right\} \\
 24 & 24 & \left\{\left\{O_1^{24},24\right\}\right\} \\
 25 & 30 &
   \left\{\left\{O_1^{25},6\right\}\oplus\left\{O_2^{25},24\right\}
   \right\} \\
 26 & 72 &
   \left\{\left\{O_1^{26},24\right\}\oplus\left\{O_2^{26},24\right\}\oplus\left\{O_3^{26},24\right\}\right\} \\
 27 & 32 &
   \left\{\left\{O_1^{27},24\right\}\oplus\left\{O_2^{27},8\right\}
   \right\} \\
 29 & 72 &
   \left\{\left\{O_1^{29},24\right\}\oplus\left\{O_2^{29},24\right\}\oplus\left\{O_3^{29},24\right\}\right\} \\
 30 & 48 &
   \left\{\left\{O_1^{30},24\right\}\oplus\left\{O_2^{30},24\right\}\right\} \\
 32 & 12 & \left\{\left\{O_1^{32},12\right\}\right\} \\
 33 & 48 &
   \left\{\left\{O_1^{33},24\right\}\oplus\left\{O_2^{33},24\right\}\right\} \\
 34 & 48 &
   \left\{\left\{O_1^{34},24\right\}\oplus\left\{O_2^{34},24\right\}\right\} \\
 35 & 48 &
   \left\{\left\{O_1^{35},24\right\}\oplus\left\{O_2^{35},24\right\}\right\} \\
 36 & 30 &
   \left\{\left\{O_1^{36},6\right\}\oplus\left\{O_2^{36},24\right\}
   \right\} \\
\hline
   \end{array}
\end{eqnarray*}
  \caption{Table of the first spherical layers of momentum vectors in the cubic self--dual lattice}\label{spherdecompo1}
\end{table}
\paragraph{Orbits of length 24}
\begin{equation}\label{orbita24}
\mathcal{O}_{24} \, = \,     \left\{
\begin{array}{llll}
 \{-p,-q,r\}, & \{-p,q,-r\}, & \{-p,-r,-q\}, & \{-p,r,q\}, \\
   \{p,-q,-r\}, & \{p,q,r\}, & \{p,-r,q\}, & \{p,r,-q\}, \\
   \{-q,-p,-r\}, & \{-q,p,r\}, & \{-q,-r,p\}, & \{-q,r,-p\},\\
   \{q,-p,r\}, & \{q,p,-r\}, & \{q,-r,-p\}, & \{q,r,p\}, \\
   \{-r,-p,q\}, & \{-r,p,-q\}, & \{-r,-q,-p\}, & \{-r,q,p\},\\
    \{r,-p,-q\}, & \{r,p,q\}, & \{r,-q,p\}, & \{r,q,-p\},\\
\end{array}
\right\}
\end{equation}
where $\{p, \, q, \, r \}\, \in \, \mathbb{Z}$ is any triplet of integer numbers that are not all three equal in absolute value.
\begin{figure}[!hbt]
\begin{center}
\iffigs
\includegraphics[height=80mm]{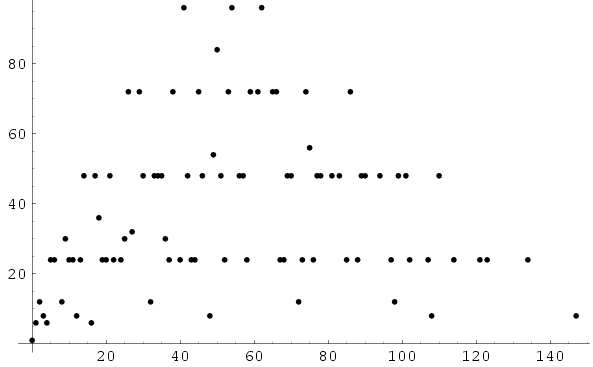}
\else
\end{center}
 \fi
\caption{\it  A view of the distribution of the number of lattice point lying on a surface of squared radius $r^2$. On the horizontal axis we have $r^2$, on the vertical axis we have the number of points lying on that sphere. As one sees the distribution is very much irregular and follows a complicated pattern.}
\label{scaterpoint}
 \iffigs
 \hskip 1cm \unitlength=1.1mm
 \end{center}
  \fi
\end{figure}
Considering the spherical layers of increasing quantized squared radius $r^2$ we discover that for the first low lying layers the points lying on the surface arrange themselves into just one orbit. At $r^2\, = \, 9$ we observe the first splitting of the layer  into two orbits, one of length 6, the other of length 24. Such splittings occur again and again with more and more orbits populating the same spherical surface. Yet single orbits appear also at higher values of $r^2$ as it is shown in Table \ref{spherdecompo1}. The notation adopted in the figure is the following one $\{O^{r^2}_i,\ell\}$ denotes the orbit of length $\ell \, = \, 6, 8, 12, \mbox{or}\, 24$, of  momentum lattice points whose norm is $\mathbf{k}^2 \,= \,r^2$. The index $i$ enumerates the individual orbits placed on the sphere of radius $r$. Predictions of splittings and of orbit degeneracies requires investigations in number theory and diophantine equations that we have not addressed within the scope of the present paper. A visual image of the complicated pattern produced by the distribution of orbits with respect to the quantized radius is provided in fig.\ref{scaterpoint}.
\subsection{Classification of the $48$ types of orbits}
Notwithstanding the number theory complicacies mentioned above the notion of Universal Classifying Group introduces a very effective guide-line to tame the zoo of point orbits displayed in fig. \ref{scaterpoint}. The first observation is that the  group $\mathrm{\mathrm{G_{1536}}}$ has a finite number of irreducible representations so that, irrespectively of the above complicated pattern, the number of different types of Arnold-Beltrami vector fields has also got to be finite,  namely as many as the 37 irreps, times the number of different ways to obtain them from orbits of length 6,8,12 or 24. The second observation is the key role of the number $4$ introduced by Frobenius congruences which was already the clue to the definition of $\mathrm{\mathrm{G_{1536}}}$. What we should expect is that the various orbits should be defined with integers modulo $4$ in other words that we should just consider the possible octahedral  orbits on a lattice with coefficients in $\mathbb{Z}_4$ rather than $\mathbb{Z}$. The easy guess, which is confirmed by computer calculations, is that the pattern of $\mathrm{\mathrm{G_{1536}}}$ representations obtained from the construction of Arnold-Beltrami vector fields according to the algorithm of section \ref{algoritmo} depends only on the equivalence classes of momentum orbits modulo $4$. Hence we have a finite number of such orbits and a finite number of Arnold-Beltrami vector fields which we presently describe.
Let us stress that an embryo of the exhaustive classification of orbits we are going to present was introduced by Arnold in his paper \cite{arnoldorussopapero}. Arnold's  was only an embryo of the complete classification for the following two reasons:
\begin{enumerate}
  \item The type of momenta orbits were partitioned  according to $odd$ and $even$ (namely according to $\mathbb{Z}_2$, rather than $\mathbb{Z}_4$)
  \item The classifying group was taken to be the crystallographic $GS_{24}$, as defined  by us in the appendices (see sect. \ref{coniugatoGS24}), which is too small in comparison with the universal classifying group identified by us in $\mathrm{\mathrm{G_{1536}}}$.
\end{enumerate}
Let us then present the complete classification of point orbits in the momentum lattice.
First we subdivide the momenta into five groups:
\begin{description}
  \item[A)] Momenta of type $\{a,0,0\}$ which generate $\mathrm{O_{24}}$ orbits of length 6 and representations of the universal group $\mathrm{G}_{1536}$ also of dimensions 6.
  \item[B)]Momenta of type $\{a,a,a\}$ which generate $\mathrm{O_{24}}$ orbits of length 8 and representations of the universal group $\mathrm{G}_{1536}$ also of dimensions 8.
  \item[C)] Momenta of type $\{a,a,0\}$ which generate $\mathrm{O_{24}}$ orbits of length 12 and representations of the universal group $\mathrm{G}_{1536}$ also of dimensions 12.
  \item[D)] Momenta of type $\{a,a,b\}$ which generate $\mathrm{O_{24}}$ orbits of length 24 and representations of the universal group $\mathrm{G}_{1536}$ also of dimensions 24.
  \item[E)] Momenta of type $\{a,b,c\}$ which generate $\mathrm{O_{24}}$ orbits of length 24 and representations of the universal group $\mathrm{G}_{1536}$ of dimensions 48.
\end{description}
The reason why in the cases A)\dots D) the dimension of the representation $\mathfrak{R}\left(\mathrm{\mathrm{G_{1536}}}\right)$ coincides with the dimension $|\mathcal{O}|$ of the orbit  is simple. For each momentum in the orbit ($\forall\mathbf{k}_i\, \in \, \mathcal{O}$) also its negative is in the same orbit ($- \, \mathbf{k}_i\, \in \, \mathcal{O}$), hence the number of arguments $\Theta_i \, \equiv \, 2\pi \,\mathbf{k}_i \cdot \mathbf{x}$ of the independent trigonometric functions $\sin \left( \Theta_i\right)$ and $\cos \left(\Theta_i\right)$ is $\ft 12 \times 2 |\mathcal{O}| \, = \, |\mathcal{O}|$ since $\sin \left( \pm\Theta_i\right) \, = \, \pm  \sin \left( \Theta_i\right)$ and $\cos \left( \pm\Theta_i\right) \, = \, \cos \left( \Theta_i\right)$. \par
In case E), instead, the negatives of all the members of the orbit $\mathcal{O}$ are not in $\mathcal{O}$. The number of independent trigonometric functions is therefore 48 and such is the dimension of the representation $\mathfrak{R}\left(\mathrm{\mathrm{G_{1536}}}\right)$.
\par
In each of the five groups one still has  to reduce the entries to $\mathbb{Z}_4$, namely to consider their equivalence class $\mathrm{mod}\,4$. Each different choice of the pattern of  $\mathbb{Z}_4$ classes appearing in an orbit leads to a different decomposition of the representation into irreducible representation of $\mathrm{G}_{1536}$. A simple consideration of the combinatorics leads to the conclusion that there are in total $48$ cases to be considered. The very significant result is that all of the $37$ irreducible representations of $\mathrm{G}_{1536}$ appear at least once in the list of these decompositions. Hence for all the \textit{irrepses} of this group one can find a corresponding Beltrami field and for some \textit{irrepses} such a Beltrami field admits a few inequivalent realizations. The list of the $48$ distinct type of momenta is the following one:
\begin{enumerate}
\item $\mathbf{k} \, = \,  \{0,0, 1+4 \rho \}$
\item $\mathbf{k} \, = \,  \{0,0, 2+4 \rho \}$
\item $\mathbf{k} \, = \,  \{0,0, 3+4 \rho \}$
\item $\mathbf{k} \, = \,  \{0,0, 4+4 \rho \}$
\item $\mathbf{k} \, = \,  \left\{1+4\mu ,1+4 \mu ,1+4 \mu \right\}$
\item $\mathbf{k} \, = \,  \left\{2+4 \mu ,2+4 \mu ,2+4 \mu \right\}$
\item $\mathbf{k} \, = \,  \left\{3+4 \mu ,3+4 \mu ,3+4 \mu \right\}$
\item $\mathbf{k} \, = \,  \left\{4+4 \mu ,4+4 \mu ,4+4 \mu \right\}$
\item $\mathbf{k} \, = \,  \left\{0 ,1+4 \nu ,1+4 \nu \right\}$
\item $\mathbf{k} \, = \,  \left\{0 ,2+4 \nu ,2+4 \nu \right\}$
\item $\mathbf{k} \, = \,  \left\{0 ,3+4 \nu ,3+4 \nu \right\}$
\item $\mathbf{k} \, = \,  \left\{0 ,4+4 \nu ,4+4 \nu \right\}$
\item $\mathbf{k} \, = \,  \left\{1+4  \mu ,1+4 \mu ,2+4 \rho \right\}$
\item $\mathbf{k} \, = \,  \left\{1+4 \mu ,1+4 \mu ,3+4 \rho \right\}$
\item $\mathbf{k} \, = \,  \left\{1+4 \mu ,1+4 \mu ,4+4 \rho \right\}$
\item $\mathbf{k} \, = \,  \left\{1+4\mu ,1+4 \mu ,5+4 \rho \right\}$
\item $\mathbf{k} \, = \,  \{1+4 \mu ,2+4 \mu ,2+4 \rho \}$
\item $\mathbf{k} \, = \,  \left\{2+4\mu ,2+4 \mu ,6+4 \rho \right\}$
\item $\mathbf{k} \, = \,  \left\{2+4  \mu ,2+4 \mu ,3+4 \rho \right\}$
\item $\mathbf{k} \, = \,  \left\{2+4\mu ,2+4 \mu ,4+4 \rho \right\}$
\item $\mathbf{k} \, = \,  \left\{1+4 \mu ,3+4 \mu ,3+4 \rho \right\}$
\item $\mathbf{k} \, = \,  \left\{2+4 \mu ,3+4 \mu ,3+4 \rho \right\}$
\item $\mathbf{k} \, = \,  \left\{3+4  \mu ,3+4 \mu ,7+4 \rho \right\}$
\item $\mathbf{k} \, = \,  \left\{1+4  \mu ,4+4 \mu ,4+4 \rho \right\}$
\item $\mathbf{k} \, = \,  \{2+4 \mu ,4+4 \mu ,4+4 \rho \}$
\item $\mathbf{k} \, = \,  \left\{3+4  \mu ,4+4 \mu ,4+4 \rho \right\}$
\item $\mathbf{k} \, = \,  \left\{4+4 \mu ,4+4 \mu ,8+4 \rho \right\}$
\item $\mathbf{k} \, = \,  \left\{3+4  \mu ,3+4 \mu ,4+4 \rho \right\}$
\item $\mathbf{k} \, = \,  \left\{4+4 \mu ,8+4 \nu ,12+4 \rho \right\}$
\item $\mathbf{k} \, = \,  \{1+4 \mu ,4+4 \nu ,8+4 \rho \}$
\item $\mathbf{k} \, = \,  \left\{2+4 \mu ,4+4 \nu ,8+4 \rho \right\}$
\item $\mathbf{k} \, = \,  \left\{3+4  \mu ,4+4 \nu ,8+4 \rho \right\}$
\item $\mathbf{k} \, = \,  \left\{1+4 \mu ,2+4 \nu ,4+4 \rho \right\}$
\item $\mathbf{k} \, = \,  \left\{1+4 \mu ,3+4 \nu ,4+4 \rho \right\}$
\item $\mathbf{k} \, = \,  \left\{2+4 \mu ,4+4 \nu ,6+4 \rho \right\}$
\item $\mathbf{k} \, = \,  \left\{2+4  \mu ,3+4 \nu ,4+4 \rho \right\}$
\item $\mathbf{k} \, = \,  \left\{1+4 \mu ,5+4 \nu ,9+4 \rho \right\}$
\item $\mathbf{k} \, = \,  \left\{1+4  \mu ,2+4 \nu ,5+4 \rho \right\}$
\item $\mathbf{k} \, = \,  \left\{1+4  \mu ,3+4 \nu ,5+4 \rho \right\}$
\item $\mathbf{k} \, = \,  \left\{1+4  \mu ,2+4 \nu ,6+4 \rho \right\}$
\item $\mathbf{k} \, = \,  \left\{1+4 \mu ,2+4 \nu ,3+4 \rho \right\}$
\item $\mathbf{k} \, = \,  \left\{1+4 \mu ,3+4 \nu ,7+4 \rho \right\}$
\item $\mathbf{k} \, = \,  \left\{2+4 \mu ,6+4 \nu ,10+4 \rho \right\}$
\item $\mathbf{k} \, = \,  \{2+4 \mu ,3+4 \nu ,6+4 \rho \}$
\item $\mathbf{k} \, = \,  \left\{2+4 \mu ,3+4 \nu ,7+4 \rho \right\}$
\item $\mathbf{k} \, = \,  \left\{3+4  \mu ,7+4 \nu ,11+4 \rho \right\}$
\item $\mathbf{k} \, = \,  \left\{1+4 \mu ,4+4 \nu ,5+4 \rho \right\}$
\item $\mathbf{k} \, = \,  \left\{3+4  \mu ,4+4 \nu ,7+4 \rho \right\}$
\end{enumerate}
where $\mu,\nu,\rho \, \in \, \mathbb{Z}$. The simplest and lowest lying representative of each of the $48$ classes of equivalent momenta is obtained choosing $\mu \, = \, \nu \, = \, 0$.
In appendix \ref{salamicrudi},  for each of the $48$ classes enumerated above we provide the decomposition of the corresponding Beltrami vector field parameter space into
$\mathrm{\mathrm{G_{1536}}}$ irreducible representations . These results are the outcome of extensive MATHEMATICA calculations that were performed with purposely written codes. As already stressed the most relevant point is that all the $37$ irreps of the Classifying Group are reproduced: this is the main reason for its name.
\section{Discussion of explicit examples of Arnold-Beltrami Flows from Octahedral  Point Orbits}
\label{beltramotti}
In this section, utilizing the algorithm outlined in section \ref{algoritmo}, we construct the Arnold-Beltrami Flows associated with some of the 48 types of  octahedral Point Orbits in the cubic lattice that have been classified in section \ref{triplettoni}. We consider examples with orbits of length $6, 8, 12$ and $24$. In all cases we devote attention to the groups tructure and we exhibit the explicit form of the Arnold-Beltrami Flows that have the largest possible group of symmetries available in that orbit. For some of these examples we also exhibit computer generated plots of the vector field and of its associated streamlines.
\par
In view of the popularity of the $\mathrm{ABC}$-flows in the  hydrodynamical literature, a particularly in depth analysis is presented of the lowest lying octahedral  orbit of length 6 in which  these models are embedded. Our main concern is to unveil the group theoretical structure behind the celebrated simple form of  eq.(\ref{bagcigaluppi}) which so far seemed to be a sort of miraculous arbitrary invention. In particular we  spot the subgroup of the Universal Classifying Group with respect to which the three parameters $(A,B,C)$ form an irreducible representation. Similarly we exhibit the groups and subgroups associated with the various popularly studied subcases $(A,A,A)$, $(A,B,0)$, $(A,A,0)$ and $(A,0,0)$.
\par
For  the other considered  orbits of length 8, 12, 24 we provide a similar group theoretical analysis although less detailed since, as we already stated above, we mainly confine ourselves to the construction of the Arnold-Beltrami Flows with the largest group of hidden symmetries.
\subsection{The lowest lying octahedral  orbit of length 6 in the cubic lattice and the $\mathrm{ABC}$-flows}
\begin{figure}[!hbt]
\begin{center}
\iffigs
\includegraphics[height=70mm]{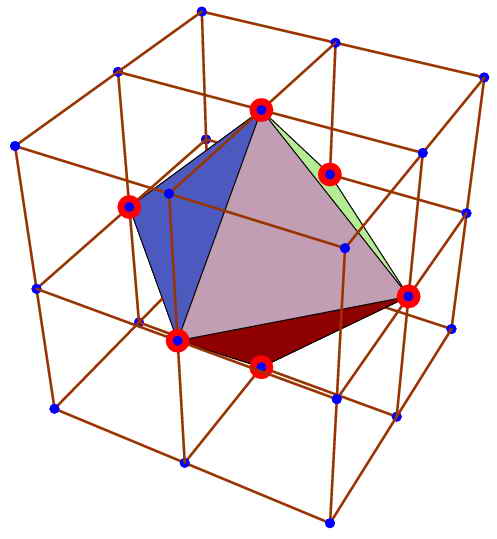}
\else
\end{center}
 \fi
\caption{\it  A view of the octahedral  6-orbit in the cubic lattice, corresponding to the vertices of a regular octahedron.}
\label{ottavertici}
 \iffigs
 \hskip 1cm \unitlength=1.1mm
 \end{center}
  \fi
\end{figure}
Let us now consider case 1) in our list of 48 classes of momentum vectors. If we take the representative $\rho \, = \, 0$ of the class, we obtain the lowest lying orbit of length $6$ of the cubic lattice. Under the action of the octahedral point group  $\mathrm{O_{24}}$, the vector $\mathbf{k}\, = \, \{0,0,1\}$ is mapped into its other five copies that together with it constitute the vertices
\begin{equation}\label{punti6}
    \begin{array}{lllclll}
 p_1 & = & \{-1,0,0\}& ; & p_4 & = & \{0,0,1\} \\
 p_2 & = & \{0,-1,0\}& ; &  p_5 & = & \{0,1,0\} \\
 p_3 & = & \{0,0,-1\}& ; & p_6 & = & \{1,0,0\} \\
 \end{array}
\end{equation}
of a regular octahedron inscribed in the sphere of radius $r^2 \, = \, 1$, as it is depicted in fig.\ref{ottavertici}.
\par
Applying the construction algorithm (see sect. \ref{algoritmo}) of sum over lattice points that belong to a point group  orbit, we obtain the following six parameter vector field:
\begin{equation}\label{orbit6vector}
{\mathbf{V}^{(6)}}({\mathbf{r}}|\mathbf{F})\,  = \,\left(
\begin{array}{l}
 2 \cos \left(2 \pi  \Theta _3\right) F_1+2 \cos \left(2 \pi  \Theta _2\right)
   F_2+2 \sin \left(2 \pi  \Theta _3\right) F_3-2 \sin \left(2 \pi  \Theta
   _2\right) F_5 \\
 -2 \sin \left(2 \pi  \Theta _3\right) F_1+2 \cos \left(2 \pi  \Theta _3\right)
   F_3+2 \cos \left(2 \pi  \Theta _1\right) F_4+2 \sin \left(2 \pi  \Theta
   _1\right) F_6 \\
 2 \sin \left(2 \pi  \Theta _2\right) F_2-2 \sin \left(2 \pi  \Theta _1\right)
   F_4+2 \cos \left(2 \pi  \Theta _2\right) F_5+2 \cos \left(2 \pi  \Theta
   _1\right) F_6
\end{array}
\right)
\end{equation}
where $F_i$ ($i\,=\,1,\dots\, ,6$) are real numbers and where the arguments of the trigonometric functions are the following ones
\begin{equation}\label{angolini1}
  \Theta _1 \, = \,   x,\,\,\Theta _2 \, = \,   y,\,\,\Theta_3 \, = \,  z
\end{equation}
 The action of the octahedral  group $\mathrm{O}_{24}$ on this Beltrami vector field is that presented in eq.(\ref{Rtrasformogen})
where $\gamma$ are the $3 \times 3$ matrices of the fundamental defining representation, while the $6 \times 6$ matrices  $\mathfrak{R}^{(6)}[\gamma ]$ acting on the parameter vector $\mathbf{F}$ and defining a reducible representation of $\mathrm{O}_{24}$ are those determined from the explicit form of the two generators $T,S$ (see eq.(\ref{generatiTS})), displayed in eq. (\ref{TSindefi1536}).
Retrieving from the above generators all the group elements and in particular a representative for each of the conjugacy classes, we can easily compute their traces and in this way establish the character vector of such a representation. We get:
\begin{equation}\label{caratter6}
    \chi\left[\mathbf{6}\right]\, = \, \{6,0,-2,0,0\}
\end{equation}
The multiplicity vector is:
\begin{equation}\label{caratter6bis}
    \mathrm{m}\left[\mathbf{6}\right]\, = \, \{0,0,0,1,1\}
\end{equation}
implying that the six dimensional parameter space decomposes into a $D_4\left[\mathrm{O_{24}},3\right]$ plus a $D_5\left[\mathrm{O_{24}},3\right]$ representations. Utilizing eq. (\ref{proiettori}) we find:
\begin{eqnarray}\label{proiet6}
    \Pi^{4}\left[\mathrm{O_{24}},3\right]\,\mathbf{F} & = &\, \frac{1}{2} \,\left\{  F_1+  F_2,  F_1+  F_2,  F_3+  F_4 ,  F_3+
   F_4 ,  F_5+  F_6,  F_5+  F_6\right\} \nonumber\\
   \Pi^{5}\left[\mathrm{O_{24}},3\right]\,\mathbf{F}& = & \frac{1}{2} \,\left\{  F_1-  F_2,  F_2-  F_1,  F_3-  F_4 ,  F_4 -
   F_3,  F_5-  F_6,  F_6-  F_5\right\}
\end{eqnarray}
which tells us that the two irreducible three-dimensional representations are obtained by identifying pairwise or anti pairwise the six coefficients.
\par
Let us now uplift the point group  representation to a representation of the Universal Classifying Group $\mathrm{\mathrm{G_{1536}}}$. As explained in section \ref{universalone}, this is done by including the elements of the quantized translation group $\mathbb{Z}_4 \times \mathbb{Z}_4 \times \mathbb{Z}_4$ via their $6$-dimensional representation anticipated in eq.(\ref{trasladefi1536}). Indeed, as we already stressed the representation of $\mathrm{\mathrm{G_{1536}}}$ obtained from this fundamental orbit can be regarded as the very definition of the Universal Classifying Group.
\par
What is then this fundamental representation of $\mathrm{\mathrm{G_{1536}}}$ that we have explicitly derived from the constructed Beltrami vector field? It is  an \textit{irreducible, faithful representation} and, from the point of view of the abstract irreps defined by the method of induction (see sections \ref{agoinduzio} and \ref{ciurlacca}), it is the representation $D_{23}\left[\mathrm{\mathrm{G_{1536}}},6\right]$. This is the result mentioned in  appendix \ref{mortadella6} at the top of the list.
In other words, under the action of quantized translations, the two irreducible representations of the point group  described in eq. (\ref{proiet6})  mix up and coalesce into an irreducible $6$-dimensional representation of  $\mathrm{\mathrm{G_{1536}}}$. Our result can be summarized as the branching rule of the representation $D_{23}\left[\mathrm{\mathrm{G_{1536}}},6\right]$ with respect to the point group :
\begin{equation}\label{23intod4ed5}
    D_{23}\left[\mathrm{\mathrm{G_{1536}}},6\right] \, = \, D_4\left[\mathrm{O_{24}},3\right] \, \oplus \, D_5\left[\mathrm{O_{24}},3\right]
\end{equation}
There are however other subgroups of the Universal Classifying Group, different from the point group, with respect to which we can branch the fundamental $D_{23}\left[\mathrm{\mathrm{G_{1536}}},6\right] $ representation. Some special choices of such subgroups are at the origin of the ABC-flows. In the spirit of Frobenius congruence classes the most intriguing subgroups $\mathrm{H} \subset \mathrm{\mathrm{G_{1536}}}$ are those of the space-group-type namely those that contain isomorphic but not conjugate copies of the point group. We focus in particular on the group $\mathrm{GF_{192}}$, described in appendix \ref{coniugatoGF192}, which contains the subgroup
$\mathrm{GS_{24}}$, described in appendix \ref{coniugatoGS24}. This latter  is isomorphic to the point group:  $\mathrm{GS_{24}}\sim\mathrm{O_{24}}$, yet it is not conjugate to it inside $\mathrm{\mathrm{G_{1536}}}$. This means that ${\not\exists} \, \gamma \, \in \, \mathrm{\mathrm{G_{1536}}}$ such that $ \mathrm{GS_{24}}\, = \, \gamma^{-1} \, \mathrm{O_{24}}\, \gamma$.
We have the following chain of inclusions:
\begin{equation}\label{baldacchinoA}
    \mathrm{\mathrm{G_{1536}}} \, \supset \, \mathrm{GF_{192}} \, \supset \, \mathrm{GS_{24}}
\end{equation}
that is parallel to the other one:
\begin{equation}\label{baldacchinoB}
    \mathrm{\mathrm{G_{1536}}} \, \supset \, \mathrm{G_{192}} \, \supset \, \mathrm{O_{24}}
\end{equation}
$\mathrm{G_{192}}$ being another subgroup, isomorphic to $\mathrm{GF_{192}}$, but not conjugate to it in $\mathrm{\mathrm{G_{1536}}}$: it is true that $\mathrm{G_{192}}\sim\mathrm{GF_{192}}$, yet
${\not\exists} \, \gamma \, \in \, \mathrm{\mathrm{G_{1536}}}$ such that $ \mathrm{GF_{192}}\, = \, \gamma^{-1} \, \mathrm{G_{192}}\, \gamma$ (see appendix \ref{coniugatoG192} for the description of $\mathrm{G_{192}}$). Since $\mathrm{G_{192}}$ and $\mathrm{GF_{192}}$ are isomorphic they have the same irreps and the same character table. Yet, since they are not conjugate, the branching rules of the same $\mathrm{\mathrm{G_{1536}}}$ irrep with respect to the former or the latter subgroup can be different. In the case of the representation $D_{23}\left[\mathrm{\mathrm{G_{1536}}},6\right]$, which is that produced by the fundamental  orbit of order six, we have
(see appendix \ref{brancicardo}):
\begin{equation}\label{passerotto}
    D_{23}\left[\mathrm{\mathrm{G_{1536}}},6\right] \, = \, \left\{\begin{array}{rcl}
                                                            D_{20}\left[\mathrm{G_{192}},6 \right] &=&D_4\left[\mathrm{O_{24}},3\right]  \oplus D_5\left[\mathrm{O_{24}},3\right]\\
                                                            D_{12}\left[\mathrm{GF_{192}},3 \right]  \oplus  D_{15}\left[\mathrm{GF_{192}},3 \right]&=&D_1\left[\mathrm{GS_{24}},1\right] \oplus  D_3\left[\mathrm{GS_{24}},2\right] \oplus  D_4\left[\mathrm{GS_{24}},3\right]\\
                                                          \end{array}
     \right.
\end{equation}
where in the second line we have used the branching rules:
\begin{eqnarray}
 D_{12}\left[\mathrm{GF_{192}},3 \right]  &=& D_1\left[\mathrm{GS_{24}},1\right] \oplus  D_3\left[\mathrm{GS_{24}},2\right]  \\
  D_{15}\left[\mathrm{GF_{192}},3 \right] &=&  D_4\left[\mathrm{GS_{24}},3\right] \label{passamontagna}
\end{eqnarray}
that, in view of the isomorphism, are identical with:
\begin{eqnarray}
 D_{12}\left[\mathrm{G_{192}},3 \right]  &=& D_1\left[\mathrm{O_{24}},1\right] \oplus  D_3\left[\mathrm{O_{24}},2\right]  \\
  D_{15}\left[\mathrm{G_{192}},3 \right] &=&  D_4\left[\mathrm{O_{24}},3\right] \label{passamontagnabis}
\end{eqnarray}
Eq.(\ref{passerotto}) has far reaching consequences. While there are no Beltrami vector fields obtained from this orbit that are invariant with respect to the octahedral point group  $\mathrm{O_{24}}$, there exists such an invariant Beltrami flow with respect to the isomorphic $\mathrm{GS_{24}}$: it corresponds to the $D_1\left[\mathrm{GS_{24}},1\right]$ irrep in the second line of (\ref{passerotto}). Furthermore while the six parameter space $\mathbf{F}$ is irreducible with respect to the action of the group $\mathrm{G_{192}}$ (the irrep $D_{20}\left[\mathrm{G_{192}},6 \right]$) it splits into two three-dimensional subspaces with respect to $\mathrm{GF_{192}}$. This is the origin of the ABC-flows. Indeed the ABC Beltrami flows can be identified with the irreducible representation $ D_{12}\left[\mathrm{GF_{192}},3 \right] $. Let us see how. Explicitly we have the following projection operators on the two irreducible representations, $D_{12}$ and $D_{15}$:
\begin{eqnarray}
  \Pi^{(12)}\left[\mathrm{GF_{192}},3\right]\,\mathbf{F}  &=& \left\{F_1,F_2,0,F_4,0,0\right\} \\
  \Pi^{(15)}\left[\mathrm{GF_{192}},3\right]\,\mathbf{F}&=& \left\{0,0,F_3,0,F_5,F_6\right\}
\end{eqnarray}
If we set $F_3\,=\,F_5\, = \, F_6\,=\,0$, we kill the irreducible representation $ D_{15}\left[\mathrm{GF_{192}},3 \right] $ and the residual Beltrami vector field, upon the following identifications:
\begin{equation}\label{agnorizo}
   A\, = \, F_1 \quad ; \quad B\, = \, F_4 \quad ; \quad C \, = \, F_2
\end{equation}
coincides with the time honored ABC flow of eq.(\ref{bagcigaluppi}).
Indeed inserting  the special parameter vector $\mathbf{F}\, = \, \{A,C,0,B,0,0\}$ in eq.(\ref{orbit6vector})  we obtain:
\begin{equation}\label{sicumerulo}
    \mathbf{V}^{(6)}\left(\left\{x,y,z\right\}\, +\,\{ \ft 34 , 0,-\ft 14\}\mid\{A,C,0,B,0,0\}\right) \, = \, \mathbf{V}^{(ABC)}(x,y,z)
\end{equation}
the vector field $\mathbf{V}^{(ABC)}(x,y,z)$ being that defined by eq.(\ref{bagcigaluppi}).
\par
The next step is provided by considering the explicit form of the decomposition of the $ D_{12}\left[\mathrm{GF_{192}},3 \right] $ irrep, \textit{i.e.} the ABC flow, into irreducible representations of the subgroup $\mathrm{GS_{24}}$. The two invariant subspaces are immediately characterized in terms of the parameters $A,B,C$, as it follows:
\begin{eqnarray}
 D_1\left[\mathrm{GS_{24}},1\right]  &\Leftrightarrow& A\, = \, B \, = \, C \, \ne \, 0 \\
 D_3\left[\mathrm{GS_{24}},2\right] &\Leftrightarrow& A\, + \, B \, + \, C \, = \, 0 \label{ergastolo}
\end{eqnarray}
\subsubsection{The $\mathrm{(A,A,A)}$-flow invariant under $\mathrm{GS_{24}}$}
\begin{figure}[!hbt]
\begin{center}
\iffigs
\includegraphics[height=70mm]{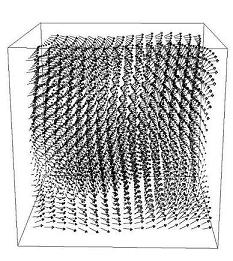}
\includegraphics[height=70mm]{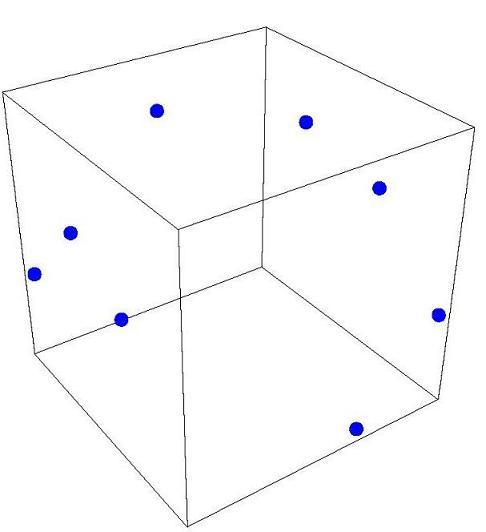}
\else
\end{center}
 \fi
\caption{\it  A plot of the $A:A:A=1$ Beltrami vector field invariant under the group $\mathrm{GS_{24}}$ and a view of its eight stagnation points of eq.(\ref{gommaliqua}).}
\label{AAAcamporella}
 \iffigs
 \hskip 1cm \unitlength=1.1mm
 \end{center}
  \fi
\end{figure}

This information suffices to understand the role of the $A:A:A\, =\,1$ Beltrami vector field often considered in the literature. It is the unique one invariant under the order 24 group $\mathrm{GS_{24}}$ isomorphic to the octahedral point group . Explicitly, in our notations, it takes the following form\footnote{Observe that here and in the sequel we stick to our conventions for $x,y,z$, which differ from those of eq.(\ref{bagcigaluppi}) by the already mentioned shift $\{ \ft 34 , 0,-\ft 14\}$).}:
\begin{equation}\label{AAAfildo}
  \mathbf{V}^{(A,A,A)}(\mathbf{r})\, = \,   \mathbf{V}^{(A,A,A)}(x,y,z)\, \equiv \, 2 A \, \left(
\begin{array}{l}
  (\cos (2 \pi  y)+\cos (2 \pi  z)) \\
  (\cos (2 \pi  x)-\sin (2 \pi  z)) \\
 (\sin (2 \pi  y)-\sin (2 \pi  x))
\end{array}
\right)
\end{equation}
This vector field $\mathbf{V}^{(A,A,A)}(x,y,z)$ is everywhere non singular in the fundamental unit cube (the torus $T^3$) apart from eight isolated   \textit{stagnation points} where it vanishes. They are listed below.
\begin{equation}\label{gommaliqua}
\begin{array}{rclcrcl}
s_1 &= & \left\{\frac{1}{8},\frac{1}{8},\frac{3}{8}\right\}
&; &
s_2 &= &\left\{\frac{1}{8},\frac{3}{8},\frac{1}{8}\right\}\\
s_3 &= & \left\{\frac{3}{8},\frac{1}{8},\frac{5}{8}\right\}
&;&
s_4 &= &\left\{\frac{3}{8},\frac{3}{8},\frac{7}{8}\right\}\\
s_5 &= & \left\{\frac{5}{8},\frac{5}{8},\frac{7}{8}\right\}
&;&
s_6&=&\left\{\frac{5}{8},\frac{7}{8},\frac{5}{8}\right\}\\
s_7&= &\left\{\frac{7}{8},\frac{5}{8},\frac{1}{8}\right\}
&;&
s_8 & =& \left\{\frac{7}{8},\frac{7}{8},\frac{3}{8}\right\}
\end{array}
\end{equation}
A numerical plot of this vector field is displayed in fig. \ref{AAAcamporella}.
\par
\begin{figure}[!hbt]
\begin{center}
\iffigs
\includegraphics[height=80mm]{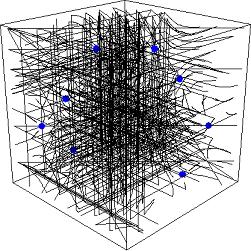}
\else
\end{center}
 \fi
\caption{\it  A plot of $125$ streamlines of the $A:A:A=1$ Beltrami vector field with equally spaced initial conditions. The numerical solutions are smooth in $\mathbb{R}^3$. When a branch reaches a boundary of the unit cube it is continued with its image in the cube modulo the appropriate lattice translation. The circles in this figure are the eight stagnation points. }
\label{AAAfiliderba}
 \iffigs
 \hskip 1cm \unitlength=1.1mm
 \end{center}
  \fi
\end{figure}

In order to provide the reader with a visual impression of the dynamics of this flow, in fig.\ref{AAAfiliderba} we display a set of $5\times 5 \times 5 = 125$ streamlines, namely of numerical integrations of the differential system:
\begin{equation}\label{babbione}
    \frac{dr}{dt} \, = \, \mathbf{V}^{(A,A,A)}(\mathbf{r})
\end{equation}
with initial conditions:
\begin{equation}\label{iniconda}
    \mathbf{r}(0) \, = \, \mathbf{r}_0 \, = \, \left\{\frac{n_1}{5},\, \frac{n_2}{5}, \,\frac{n_3}{5}\right\} \quad ; \quad n_{1,2,3} \, =\, 0,1,2,3,4
\end{equation}

\subsubsection{Chains of subgroups and the flows $(A,B,0)$, $(A,A,0)$ and $(A,0,0)$}
In the literature a lot of attention has been given to the special subcases of the $\mathrm{ABC}$-flow where one or two of the parameters vanish or two are equal among themselves and one vanishes. Also these cases can be thoroughly characterized in group theoretical terms and their special features can be traced back to the hidden subgroups tructure associated with them.
\paragraph{The $(A,B,0)$ case and its associated chain of subgroups}
When we put $C=0$ we define a two dimensional subspace of the representation $D_{12}\left[\mathrm{GF_{192}},3\right]$ which is invariant under some proper subgroup $\mathrm{H}^{\mathrm{(A,B,0)}} \subset \mathrm{GF_{192}}$. This group $\mathrm{H}^{\mathrm{(A,B,0)}}$ can be calculated and found to be of order 64, yet we do not dwell on it because the subgroup of the classifying group $\mathrm{\mathrm{G_{1536}}}$ which leaves the subspace $(A,0,0,B,0,0)$ invariant is larger than $\mathrm{H}^{\mathrm{(A,B,0)}}$ and it is not contained in $\mathrm{GF_{192}}$. It has order 128 and we name it $\mathrm{G_{128}^{(A,B,0)}}$. This short discussion is important because it implies the following: the flows $\mathrm{(A,B,0})$ should not be considered just as a particular case of the $ABC$-flows rather as a different set of flows, whose properties are encoded in the group $\mathrm{G_{128}^{(A,B,0)}}$.
\par
The group $\mathrm{G_{128}^{(A,B,0)}}$ is solvable and a chain of normal subgroups can be found, all of index 2 which ends with the abelian $\mathrm{G_{4}^{(A,B,0)}}$ isomorphic to $\mathbb{Z}_4$. This latter is nothing else than the group of quantized translation in the $y$-direction and its inclusion in the group leaving the space $(A,0,0,B,0,0)$ invariant actually means that the differential system must be  $y$-independent and hence two dimensional. The chain of normal subgroups is displayed here below:
\begin{center}
\begin{picture}(200,100)
\put(-90,45){$\mathbb{Z}_4$}
\put(-75,45){$\sim$}
\put (-60,45){$\mathrm{G_4^{(A,B,0)}}$}
\put (-20,45){$\vartriangleleft $}
\put (-5,45){$\mathrm{G_8^{(A,B,0)}}$}
\put (35,45){$\vartriangleleft $}
\put (50,45){$\mathrm{G_{16}^{(A,B,0)}}$}
\put (90,45){$\vartriangleleft $}
\put (103,47){\line (1,1){20}}
\put (103,47){\line (1,-1){20}}
\put (127,65.5){$\vartriangleleft $}
\put (127,24){$\vartriangleleft $}
\put (142,24){$\mathrm{G_{32}^{(A,A,0)}}$}
\put (142,65.5){$\mathrm{G_{32}^{(A,B,0)}}$}
\put (182,65.5){$\vartriangleleft $}
\put (197,65.5){$\mathrm{G_{64}^{(A,B,0)}}$}
\put (237,65.5){$\vartriangleleft $}
\put (252,65.5){$\mathrm{G_{128}^{(A,B,0)}}$}
\end{picture}
\end{center}
\begin{equation}
\null
\label{goffo}
\end{equation}
and it allows for the construction of irreducible representations of $\mathrm{G_{128}^{(A,B,0)}}$ and all other members of the chain, by means of the induction algorithm. Such a construction we have not done, but all the groups of the chain are listed, with their conjugacy classes in appendix \ref{ABCsubgroups}. The group $\mathrm{G_{128}^{(A,B,0)}}$ leaves the subspace $(A,0,0,B,0,0)$ invariant but still mixes the parameters $A$ and $B$ among themselves. The subgroup $\mathrm{G_{16}^{(A,B,0)}}\lhd \mathrm{G_{128}^{(A,B,0)}} $ instead stabilizes the very vector $(A,0,0,B,0,0)$. This means that any ${\mathrm{(A,B,0)}}$-flow has a hidden symmetry of order 16 provided by the group $\mathrm{G_{16}^{(A,B,0)}}$. The general form of these Beltrami fields is the following one:
\begin{equation}\label{AB0campus}
\mathbf{V}^{(A,B,0)}(\mathbf{r})\, = \,   \mathbf{V}^{(A,B,0)}(x,y,z)\, \equiv \,     \left(
\begin{array}{l}
 A \cos (2 \pi  z) \\
 B \cos (2 \pi  x)-A \sin (2 \pi  z) \\
 -B \sin (2 \pi  x)
\end{array}
\right)
\end{equation}
\par
\begin{figure}[!hbt]
\begin{center}
\iffigs
\includegraphics[height=80mm]{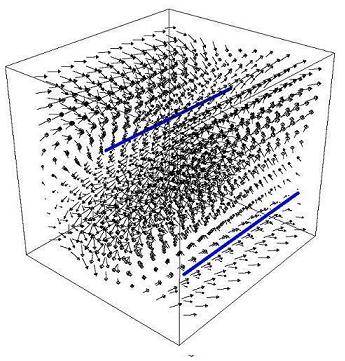}
\includegraphics[height=80mm]{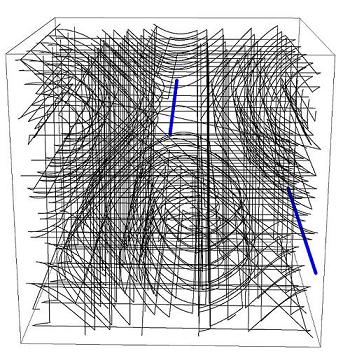}
\else
\end{center}
 \fi
\caption{\it A plot of the Betrami vector field $\mathbf{V}^{(A,A,0)}(\mathbf{r})$ (on the left) where  the two stagnation lines of the flow are visible (fat lines. On the right a family of streamlines with equally spaced initial conditions is displayed.}
\label{AAfilotti}
 \iffigs
 \hskip 1cm \unitlength=1.1mm
 \end{center}
  \fi
\end{figure}

Looking at  eq.(\ref{goffo}) we notice that there is another group of order 32, namely  $\mathrm{G_{32}^{(A,A,0)}}$ which contains $\mathrm{G_{16}^{(A,B,0)}}$ but it is not contained neither in $\mathrm{G_{128}^{(A,B,0)}} $ nor in $\mathrm{GF_{192}}$. This group is the stabilizer of the vector $(A,0,0,A,0,0)$ and hence it is the hidden symmetry group of the flows of type $(A,A,0)$. Once again the very fact that $\mathrm{G_{32}^{(A,A,0)}}$ is not contained in $\mathrm{G_{128}^{(A,B,0)}}$ shows that the $(A,A,0)$ flow should not be considered as a particular case of the $(A,B,0)$-flows rather as a new type of its own. Let us also stress the difference with the case of the $(A,A,A)$-flow. Here the hidden symmetry group $\mathrm{GS_{24}}$ is contained in $\mathrm{GF_{192}}$ and the interpretation of the $(A,A,A)$-flow as a particular case of the $(A,B,C)$-flows is permitted.
Having set:
\begin{equation}\label{AA0campus}
\mathbf{V}^{(A,A,0)}(\mathbf{r})\, = \,   \mathbf{V}^{(A,A,0)}(x,y,z)\, \equiv \,   A \,  \left(
\begin{array}{l}
 \cos (2 \pi  z) \\
 \cos (2 \pi  x)- \sin (2 \pi  z) \\
 - \sin (2 \pi  x)
\end{array}
\right)
\end{equation}
in fig.\ref{AAfilotti} we display a plot of the vector field $\mathbf{V}^{(A,A,0)}(\mathbf{r})$ and a family of its streamlines.
In the case of this flow there are not isolated stagnation points, rather, because of the $y$-independence of the Beltrami vector field, there are two entire stagnation lines explicitly given below:
\begin{equation}\label{cannastagnante}
   \mathrm{sl}_1\, = \, \left\{\frac{1}{2},y,\frac{3}{4}\right\}\quad ; \quad \mathrm{sl}_2\  = \, \left\{1,y,\frac{1}{4
   }\right\}
\end{equation}
Let us finally come to the case of the flow $(A,0,0)$. The one-dimensional subspace of vectors of the form $(A,0,0,0,0,0)$ is left invariant by a rather big subgroup of the classifying group which is of order $256$. We name it  $\mathrm{G_{256}^{(A,0,0)}}$
and its description is given in appendix \ref{ABCsubgroups}. It is a solvable group with a chain of normal subgroups of index $2$ which ends into a subgroup of order $16$ isomorphic to $\mathbb{Z}_4 \times \mathbb{Z}_4$. This information is summarized in the equation below:
\begin{center}
\begin{picture}(60,90)
\put(-120,45){$\mathbb{Z}_4 \times \mathbb{Z}_4$}
\put(-75,45){$\sim$}
\put (-60,45){$\mathrm{G_{16}^{(A,0,0)}}$}
\put (-20,45){$\vartriangleleft $}
\put (-5,45){$\mathrm{G_{32}^{(A,0,0)}}$}
\put (35,45){$\vartriangleleft $}
\put (50,45){$\mathrm{G_{64}^{(A,0,0)}}$}
\put (90,45){$\vartriangleleft $}
\put (103,47){\line (1,-1){20}}
\put (126,27){\circle{3}}
\put (135,24){$\mathrm{G_{128}^{(A,0,0)}}$}
\put (175,24){$\vartriangleleft $}
\put (190,24){$\mathrm{G_{256}^{(A,0,0)}}$}
\put (103,7){\line (1,1){20}}
\put (90,6){$\subset $}
\put (-60,6){$\mathrm{G_{16}^{(A,B,0)}}$}
\put (-20,6){$\vartriangleleft $}
\put (-5,6){$\mathrm{G_{32}^{(A,B,0)}}$}
\put (35,6){$\vartriangleleft $}
\put (50,6){$\mathrm{G_{64}^{(A,B,0)}}$}
\put (-75,6){$\vartriangleleft $}
\put (-115,6){$\mathrm{G_{8}^{(A,B,0)}}$}
\put (-130,6){$\vartriangleleft $}
\put (-170,6){$\mathrm{G_{4}^{(A,B,0)}}$}
\put(-200,6){$\mathbb{Z}_4 $}
\put(-185,6){$\sim$}
\end{picture}
\begin{equation}
\label{goffo2}
\null
\end{equation}
\end{center}
The group $\mathrm{G_{256}^{(A,0,0)}}$ leaves the subspace $(A,0,0,0,0,0)$ invariant but occasionally changes the sign of $A$. The subgroup $\mathrm{G_{128}^{(A,0,0)}} \, \subset \, \mathrm{G_{256}^{(A,0,0)}}$ stabilizes the very vector $(A,0,0,0,0,0)$ and therefore it is the hidden symmetry of the $(A,0,0)$ flows encoded in the planar vector field:
\begin{equation}\label{A00campus}
\mathbf{V}^{(A,0,0)}(\mathbf{r})\, = \,   \mathbf{V}^{(A,0,0)}(x,y,z)\, \equiv \,   A \,
\left(
\begin{array}{l}
 \cos (2 \pi  z) \\
 -\sin (2 \pi  z) \\
 0
\end{array}
\right)
\end{equation}
Looking back at equation (\ref{goffo2}) it is important to note that the group $\mathrm{G_{128}^{(A,0,0)}} \ne \mathrm{G_{128}^{(A,B,0)}}$ is different from the homologous group appearing in the group-chain of the $\mathrm{(A,B,0)}$-flows. So once again the $\mathrm{(A,0,0)}$-flows cannot be regarded as particular cases of the $\mathrm{(A,B,0)}$-flows. Yet the group
$\mathrm{G_{128}^{(A,0,0)}}$ contains the entire chain of normal subgroups $\mathrm{G_{128}^{(A,B,0)}}$ starting from
$\mathrm{G_{64}^{(A,B,0)}}$. There is however a very relevant proviso $\mathrm{G_{64}^{(A,B,0)}}$ is a subgroup of
$\mathrm{G_{128}^{(A,0,0)}}$ but it is not normal.
\begin{figure}[!hbt]
\begin{center}
\iffigs
\includegraphics[height=70mm]{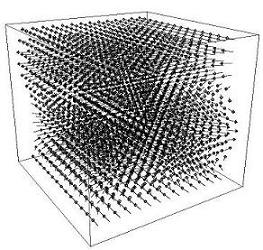}
\includegraphics[height=70mm]{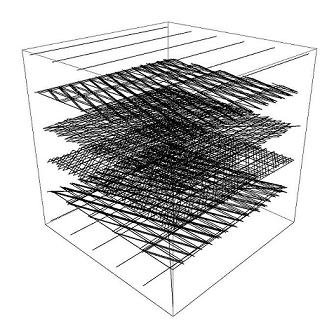}
\else
\end{center}
 \fi
\caption{\it A plot of the Betrami vector field $\mathbf{V}^{(A,0,0)}(\mathbf{r})$ (on the left). On the right a family of streamlines with equally spaced initial conditions is displayed. The planar structure of the streamlines is quite visible.}
\label{A00filucci}
 \iffigs
 \hskip 1cm \unitlength=1.1mm
 \end{center}
  \fi
\end{figure}
In fig. \ref{A00filucci} we show a plot the vector field $\mathbf{V}^{(A,0,0)}(\mathbf{r})$ and a family of its streamlines.
\subsection{The lowest lying octahedral  orbit of length 12 in the cubic lattice and the Beltrami flows respectively invariant under $\mathrm{GP_{24}}$ and $\mathrm{GK_{24}}$}
\label{12orbitsezia}
The next example that we consider corresponds to the length 12 octahedral  orbit of momentum vectors in the class numbered 9) in our list of 48 classes, namely:
\begin{equation}\label{sigfrido12}
    \mathbf{k} \, = \, \{0,1+4\nu,1+4\nu\}
\end{equation}
Choosing the representative $\nu \, = \, 0$, we obtain the following lowest lying orbit of 12 points:
\begin{equation}\label{point12}
    \begin{array}{lllclll}
 p_1 & = & \{-1,-1,0\} & ; & p_7 & = & \{0,1,-1\} \\
 p_2 & = & \{-1,0,-1\}  & ; & p_8 & = & \{0,1,1\} \\
 p_3 & = & \{-1,0,1\} & ; & p_9 & = & \{1,-1,0\} \\
 p_4 & = & \{-1,1,0\} & ; &  p_{10} & = & \{1,0,-1\} \\
 p_5 & = & \{0,-1,-1\} & ; & p_{11} & = & \{1,0,1\} \\
 p_6 & = & \{0,-1,1\} & ; & p_{12} & = & \{1,1,0\}
\end{array}
\end{equation}
\begin{figure}[!hbt]
\begin{center}
\iffigs
\includegraphics[height=80mm]{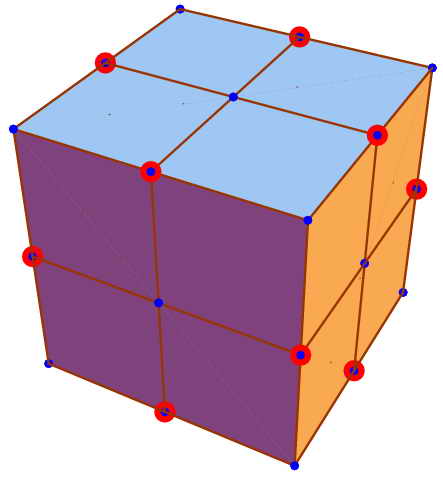}
\else
\end{center}
 \fi
\caption{\it  A view of the octahedral  12-orbit in the cubic lattice, corresponding to the midpoints of the edges of a regular cube.}
\label{cubomezzi}
 \iffigs
 \hskip 1cm \unitlength=1.1mm
 \end{center}
  \fi
\end{figure}
These are the lattice points displayed in fig.\ref{cubomezzi}.
\par
Applying the strategy of sum over lattice points that belong to a point group  orbit, we obtain the following 12 parameter vector field:
\begin{eqnarray}\label{orbit12vector}
{\mathbf{V}^{(12)}}({\mathbf{r}}|\mathbf{F}) & = &\left\{ V_x\, , \, V_y \, , \, V_z\right\} \nonumber\\
V_x & = & 2 \cos \left(2 \pi  \Theta _6\right) F_1+2 \cos \left(2 \pi
    \Theta _5\right) F_2+2 \cos \left(2 \pi  \Theta
   _4\right) F_3-\sqrt{2} \sin \left(2 \pi  \Theta
   _3\right) F_4\nonumber\\
   &&+\sqrt{2} \sin \left(2 \pi  \Theta
   _2\right) F_5-2 \cos \left(2 \pi  \Theta _1\right) F_6-2
   \sqrt{2} \sin \left(2 \pi  \Theta _6\right) F_7-2
   \sqrt{2} \sin \left(2 \pi  \Theta _5\right) F_8\nonumber\\
   &&+\sqrt{2}
   \sin \left(2 \pi  \Theta _4\right) F_9+2 \cos \left(2
   \pi  \Theta _3\right) F_{10}-2 \cos \left(2 \pi  \Theta
   _2\right) F_{11}-\sqrt{2} \sin \left(2 \pi  \Theta
   _1\right) F_{12} \nonumber\\
V_y & = &\sqrt{2} \sin \left(2 \pi  \Theta _6\right) F_1-\sqrt{2}
   \sin \left(2 \pi  \Theta _5\right) F_2+2 \cos \left(2
   \pi  \Theta _4\right) F_3+2 \cos \left(2 \pi  \Theta
   _3\right) F_4\nonumber\\
   &&+2 \cos \left(2 \pi  \Theta _2\right) F_5+2
   \cos \left(2 \pi  \Theta _1\right) F_6+2 \cos \left(2
   \pi  \Theta _6\right) F_7-2 \cos \left(2 \pi  \Theta
   _5\right) F_8\nonumber\\
   &&+\sqrt{2} \sin \left(2 \pi  \Theta
   _4\right) F_9+2 \sqrt{2} \sin \left(2 \pi  \Theta
   _3\right) F_{10}\nonumber\\
   &&+2 \sqrt{2} \sin \left(2 \pi  \Theta
   _2\right) F_{11}+\sqrt{2} \sin \left(2 \pi  \Theta
   _1\right) F_{12}\nonumber\\
V_z & = & \sqrt{2} \sin \left(2 \pi  \Theta _6\right) F_1+\sqrt{2}
   \sin \left(2 \pi  \Theta _5\right) F_2-2 \sqrt{2} \sin
   \left(2 \pi  \Theta _4\right) F_3-\sqrt{2} \sin \left(2
   \pi  \Theta _3\right) F_4\nonumber\\
   &&-\sqrt{2} \sin \left(2 \pi
   \Theta _2\right) F_5-2 \sqrt{2} \sin \left(2 \pi  \Theta
   _1\right) F_6+2 \cos \left(2 \pi  \Theta _6\right) F_7+2
   \cos \left(2 \pi  \Theta _5\right) F_8\nonumber\\
   &&+2 \cos \left(2
   \pi  \Theta _4\right) F_9+2 \cos \left(2 \pi  \Theta
   _3\right) F_{10}+2 \cos \left(2 \pi  \Theta _2\right)
   F_{11}+2 \cos \left(2 \pi  \Theta _1\right) F_{12}\nonumber\\
\end{eqnarray}
where $F_i$ ($i\,=\,1,\dots\, ,12$) are real numbers and the angles $\Theta_i$ are defined as follows:
\begin{equation}\label{angoli12}
   \Theta _1\, = \,  x+y\quad ;\quad\Theta _2\, = \,  x+z\quad ;\quad\Theta _3\, = \,
   x-z\quad ;\quad\Theta _4\, = \,  x-y\quad ;\quad\Theta _5\, = \,  y+z\quad ;\quad\Theta _6\, = \,
   y-z
\end{equation}
\paragraph{Decomposition of this orbit with respect to the Point Group $\mathrm{O_{24}}$}
The action of the octahedral point group  $\mathrm{O_{24}}$  is easily determined on such a vector field by the standard procedure illustrated above. The form of the two $\mathrm{O_{24}}$ generators is displayed below
\begin{eqnarray}\label{TSin12orbitO}
    \mathfrak{R}^{(12)}[T] \, = \, \left(
\begin{array}{llllllllllll}
 0 & 0 & 0 & 0 & 0 & 0 & 0 & 0 & 1 & 0 & 0 & 0 \\
 0 & 0 & 0 & 0 & 0 & 0 & 0 & 0 & 0 & 0 & 0 & 1 \\
 0 & 0 & 0 & 0 & 0 & 0 & 0 & 0 & 0 & 1 & 0 & 0 \\
 1 & 0 & 0 & 0 & 0 & 0 & 0 & 0 & 0 & 0 & 0 & 0 \\
 0 & 1 & 0 & 0 & 0 & 0 & 0 & 0 & 0 & 0 & 0 & 0 \\
 0 & 0 & 0 & 0 & 0 & 0 & 0 & 0 & 0 & 0 & -1 & 0 \\
 0 & 0 & 1 & 0 & 0 & 0 & 0 & 0 & 0 & 0 & 0 & 0 \\
 0 & 0 & 0 & 0 & 0 & 1 & 0 & 0 & 0 & 0 & 0 & 0 \\
 0 & 0 & 0 & 1 & 0 & 0 & 0 & 0 & 0 & 0 & 0 & 0 \\
 0 & 0 & 0 & 0 & 0 & 0 & 1 & 0 & 0 & 0 & 0 & 0 \\
 0 & 0 & 0 & 0 & 0 & 0 & 0 & -1 & 0 & 0 & 0 & 0 \\
 0 & 0 & 0 & 0 & 1 & 0 & 0 & 0 & 0 & 0 & 0 & 0
\end{array}
\right)
\nonumber\\
\mathfrak{R}^{(12)}[S]\, = \,\left(
\begin{array}{llllllllllll}
 0 & 0 & 0 & 0 & 1 & 0 & 0 & 0 & 0 & 0 & 0 & 0 \\
 0 & 0 & 0 & 1 & 0 & 0 & 0 & 0 & 0 & 0 & 0 & 0 \\
 0 & 0 & 1 & 0 & 0 & 0 & 0 & 0 & 0 & 0 & 0 & 0 \\
 0 & 1 & 0 & 0 & 0 & 0 & 0 & 0 & 0 & 0 & 0 & 0 \\
 1 & 0 & 0 & 0 & 0 & 0 & 0 & 0 & 0 & 0 & 0 & 0 \\
 0 & 0 & 0 & 0 & 0 & -1 & 0 & 0 & 0 & 0 & 0 & 0 \\
 0 & 0 & 0 & 0 & 0 & 0 & 0 & 0 & 0 & 0 & -1 & 0 \\
 0 & 0 & 0 & 0 & 0 & 0 & 0 & 0 & 0 & -1 & 0 & 0 \\
 0 & 0 & 0 & 0 & 0 & 0 & 0 & 0 & -1 & 0 & 0 & 0 \\
 0 & 0 & 0 & 0 & 0 & 0 & 0 & -1 & 0 & 0 & 0 & 0 \\
 0 & 0 & 0 & 0 & 0 & 0 & -1 & 0 & 0 & 0 & 0 & 0 \\
 0 & 0 & 0 & 0 & 0 & 0 & 0 & 0 & 0 & 0 & 0 & -1
\end{array}
\right)
\end{eqnarray}
Retrieving from the above generators all the group elements and in particular a representative for each of the conjugacy classes, we  easily compute the character vector of this representation. Explicitly we get:
\begin{equation}\label{caratter12}
    \chi\left[\mathbf{12}\right]\, = \,\{12,0,0,-2,0\}
\end{equation}
The multiplicity vector is:
\begin{equation}\label{multiply8bis}
   \mathrm{m}\left[\mathbf{12}\right]\, = \,\{0,1,1,2,1\}
\end{equation}
This implies that we have a one-dimensional non singlet representation $ D_2\left[\mathrm{O_{24}},1\right]$, a two-dimensional representation $D_3\left[\mathrm{O_{24}},2\right]$, two three dimensional representations $D_4\left[\mathrm{O_{24}},3\right]$ and one three-dimensional representation $D_5\left[\mathrm{O_{24}},3\right]$.
\par
\par
The splitting of the parameter space into these five invariant subspaces requires a little bit of algebraic work that we omit, quoting only the result.
\par
We name $A$ the coefficient corresponding to the one-dimensional $D_2$ representation, $\Gamma_{1,2}$ the two coefficients corresponding to the two-dimensional  $D_3$ representation, $K_{1,2,3}$ the three coefficients corresponding to the three dimensional $D_{4a}$ representation, $\Delta_{1,2,3}$, the three coefficients corresponding to the three dimensional $D_{4b}$ representation and
$\Lambda_{1,2,3}$ the three coefficients corresponding to the three dimensional $D_5$ representation. Expressed in terms of these
variables the parameter vector $\mathbf{F}$ has the following form:
\begin{equation}\label{F12split}
 \mathbf{F}  \, = \,  \left(
\begin{array}{l}
 -A+K_2+\Gamma _1+\Gamma _2 \\
 A+K_2-\Gamma _1-\Gamma _2 \\
 -\Delta _1+\Delta _3+\Lambda _1+\Lambda _2 \\
 -A+K_3-\Gamma _1 \\
 A+K_3+\Gamma _1 \\
 -\Delta _1+\Delta _2+\Lambda _1+\Lambda _3 \\
 -\Delta _1+\Delta _2+\Delta _3-\Lambda _1-\Lambda
   _2-\Lambda _3 \\
 \Delta _1+\Lambda _1 \\
 -A+K_1-\Gamma _2 \\
 \Delta _3+\Lambda _3 \\
 \Delta _2+\Lambda _2 \\
 A+K_1+\Gamma _2
\end{array}
\right)
\end{equation}
In eq.(\ref{F12split}) the decomposition into $\mathrm{O_{24}}$ irreducible representations is fully encoded.
\paragraph{Uplifting to the Universal Classifying Group $\mathrm{\mathrm{G_{1536}}}$}
As in all other cases this reducible representation of the point group   can be uplifted to a representation of the entire classifying group $\mathrm{\mathrm{G_{1536}}}$ including the representation of the translation generators which is calculated to be that displayed in eq.(\ref{12Mtraslo}) (see in appendix \ref{12Mtraslo}). The result of this uplifting is that the $12$-parameter vector field (\ref{orbit12vector}) corresponds to an irreducible $12$-dimensional representation of the classifying group, precisely to $D_{32}\left[\mathrm{\mathrm{G_{1536}}},12\right]$ and the previous results correspond to the following branching rule:
\begin{equation}\label{024diG32}
    D_{32}\left[\mathrm{\mathrm{G_{1536}}},12\right]\, = \,
   D_2\left[\mathrm{O_{24}},1\right]+D_3\left[\mathrm{O_{24}},2\right]+2
    D_4\left[\mathrm{O_{24}},3\right]+D_5\left[\mathrm{O_{24}},3\right]
\end{equation}
It is interesting to consider the branching rule of the same representation with respect to the two subgroups $\mathrm{G_{192}} \, \subset \, \mathrm{\mathrm{G_{1536}}}$ and $\mathrm{GF_{192}} \, \subset \, \mathrm{\mathrm{G_{1536}}}$ that, as we have shown in the previous section, play such an important role in understanding the $\mathrm{ABC}$-flows and  their hidden symmetries.
\paragraph{Decomposition of the orbit with respect to the Groups $\mathrm{G_{192}}$ and $\mathrm{GF_{192}}$}
With the help of the character tables we easily obtain:
\begin{equation}\label{G192diG32}
    D_{32}\left[\mathrm{\mathrm{G_{1536}}},12\right]\, = \,
   D_9\left[\mathrm{G_{192}},3\right]+D_{13}\left[\mathrm{G_{192}},3\right]+
    D_{19}\left[\mathrm{G_{192}},6\right]
\end{equation}
and
\begin{equation}\label{GF192diG32}
    D_{32}\left[\mathrm{\mathrm{G_{1536}}},12\right]\, = \,
   D_9\left[\mathrm{GF_{192}},3\right]+D_{13}\left[\mathrm{GF_{192}},3\right]+
    D_{19}\left[\mathrm{GF_{192}},6\right]
\end{equation}
This result is very interesting. The decomposition of the irreducible representation $D_{32}\left[\mathrm{\mathrm{G_{1536}}},12\right]$ with respect to the two isomorphic but not conjugate subgroups $\mathrm{G_{192}}$ and $\mathrm{GF_{192}}$ is identical. It follows that it is identical also with respect to any of their homologous subgroups.  However the $\mathrm{G_{192}}$ and  $\mathrm{GF_{192}}$ invariant subspaces are far from being the same and the corresponding Beltrami vector fields are different. In the case of the group $\mathrm{G_{192}}$, which contains the point group   as a subgroup $\mathrm{O_{24}}\, \subset  \mathrm{G_{192}}$, the three representations $D_9\left[\mathrm{G_{192}},3\right]$, $D_{13}\left[\mathrm{G_{192}},3\right]$ and $D_{19}\left[\mathrm{G_{192}},6\right]$ simply join together the point group representations in the following way:
\begin{eqnarray}
  D_9\left[\mathrm{G_{192}},3\right] &=& D_{4a}\left[\mathrm{O_{24}},3\right]  \nonumber\\
  D_{13}\left[\mathrm{G_{192}},3\right] &=& D_{2}\left[\mathrm{O_{24}},1\right] \oplus  D_{3}\left[\mathrm{O_{24}},2\right]\nonumber\\
  D_{19}\left[\mathrm{G_{192}},6\right] &=& D_{4b}\left[\mathrm{O_{24}},3\right] \oplus  D_{5}\left[\mathrm{O_{24}},3\right] \label{spittaG192inO24}
\end{eqnarray}
In the case of the group $\mathrm{GF_{192}}$ the invariant subspaces mix   the point group representations in a capricious way and the right hand side of eq.s (\ref{spittaG192inO24}) is not true if in the left hand side we replace $\mathrm{G_{192}}$ with $\mathrm{GF_{192}}$. Yet if we consistently replace $\mathrm{O_{24}}$ with its homologous
$\mathrm{GS_{24}}$ also in the right hand side, then we obtain a true set of equations:
\begin{eqnarray}
  D_9\left[\mathrm{GF_{192}},3\right] &=& D_{4a}\left[\mathrm{GS_{24}},3\right]  \nonumber\\
  D_{13}\left[\mathrm{GF_{192}},3\right] &=& D_{2}\left[\mathrm{GS_{24}},1\right] \oplus  D_{3}\left[\mathrm{GS_{24}},2\right]\nonumber\\
  D_{19}\left[\mathrm{GF_{192}},6\right] &=& D_{4b}\left[\mathrm{GS_{24}},3\right] \oplus  D_{5}\left[\mathrm{GS_{24}},3\right] \label{spittaGF192inGS24}
\end{eqnarray}
Observing eq.s(\ref{spittaG192inO24}) and (\ref{spittaGF192inGS24}) we conclude that from the $12$-parameter vector field
(\ref{orbit12vector}) we cannot extract any one that is invariant under either $\mathrm{O_{24}}$ or $\mathrm{GS_{24}}$ since no $D_1$ representation emerges. This implies that from this orbit we cannot extract any Beltrami vector field with a hidden symmetry isomorphic to that of the proper octahedral  group $\mathrm{O_{24}} \sim \mathrm{S_4}$. Yet there is another abstract group of order $24$ which has isomorphic subgroup copies in   $\mathrm{G_{192}}$  and $\mathrm{GF_{192}}$. This is the abstarct group $\mathrm{A_4} \otimes \mathbb{Z}_2$ and happens to be the group number 13 in the list of the 15 groups of order 24 (see \cite{gappo}). It is the unique group of such an order that has 8 conjugacy classes: its not conjugate copies in the classifying group were respectively named by us $\mathrm{GP_{24} }\subset \mathrm{G_{192}}$ and $\mathrm{GK_{24} }\subset \mathrm{GF_{192}}$ (see appendices \ref{coniugatoGP24} and  \ref{coniugatoGK24} for their description). As we show in the next subsection, there are Beltrami vector fields that have  hidden symmetry either $\mathrm{GP_{24} }$ or $\mathrm{GK_{24} }$.
\subsubsection{Beltrami Flows invariant with respect to the subgroups $\mathrm{GP_{24} }$ and $\mathrm{GK_{24} }$}
As stated above the subgroups $\mathrm{GP_{24} }\subset \mathrm{G_{192}} \subset \mathrm{\mathrm{G_{1536}}}$ and
$\mathrm{GK_{24} }\subset \mathrm{GF_{192}} \subset \mathrm{\mathrm{G_{1536}}}$ are isomorphic among themselves and to the abstract group $\mathrm{A_4} \otimes \mathbb{Z}_2$. This latter, which has order 24, can be defined by the following generators and relations:
\begin{equation}\label{boardone}
 \mathrm{A_4} \otimes \mathbb{Z}_2 \, \equiv \,   \left(\mathcal{T},\,\mathcal{P}\,\mid \, \mathcal{T}^6\, = \, \mathrm{e} \, , \,\mathcal{P}^2 \, = \, \mathrm{e} \, ,\, \left( \mathcal{P}\cdot \mathcal{T}\right)^2 \, = \, \mathrm{e}\right)
\end{equation}
\par
In the case of $\mathrm{GP_{24} }$ we have:
\begin{equation}\label{generiGP24}
  \mathcal{T} \, = \,   \left\{2_8,1,1,1\right\}\quad ; \quad \mathcal{P}\, = \, \left\{3_3,1,0,0\right\}
\end{equation}
and the resulting group is that described in section \ref{coniugatoGP24}.
\par
In the case of $\mathrm{GK_{24} }$ we have instead:
\begin{equation}\label{generiGP24bis}
  \mathcal{T} \, = \,   \left\{2_1,\frac{1}{2},\frac{3}{2},0\right\}\quad ; \quad \mathcal{P}\, = \, \left\{3_3,1,1,0\right\}
\end{equation}
and the resulting group is that described in section \ref{coniugatoGK24}.
\par
We consider next the decomposition of the irreducible representations $D_9$, $D_{13}$ and $D_{19}$ of either $\mathrm{G_{192}}$ or $\mathrm{GF_{192}}$ with respect to either $\mathrm{GP_{24} }$ or $\mathrm{GK_{24} }$
and we get:
\begin{eqnarray}
  D_9\left[\mathrm{G_{192}},3\right] &=& D_{1}\left[\mathrm{GP_{24}},1\right] \oplus D_{2}\left[\mathrm{GP_{24}},1\right] \oplus D_{3}\left[\mathrm{GP_{24}},1\right]  \nonumber\\
  D_{13}\left[\mathrm{G_{192}},3\right] &=& D_{7a}\left[\mathrm{GP_{24}},3\right] \nonumber\\
  D_{19}\left[\mathrm{G_{192}},6\right] &=& D_{7b}\left[\mathrm{GP_{24}},3\right] \oplus  D_{7c}\left[\mathrm{GP_{24}},3\right] \label{spittaG192inGP24}
\end{eqnarray}
and
\begin{eqnarray}
  D_9\left[\mathrm{GF_{192}},3\right] &=& D_{1}\left[\mathrm{GK_{24}},1\right] \oplus D_{2}\left[\mathrm{GK_{24}},1\right] \oplus D_{3}\left[\mathrm{GK_{24}},1\right]  \nonumber\\
  D_{13}\left[\mathrm{GF_{192}},3\right] &=& D_{7a}\left[\mathrm{GK_{24}},3\right] \nonumber\\
  D_{19}\left[\mathrm{GF_{192}},6\right] &=& D_{7b}\left[\mathrm{GK_{24}},3\right] \oplus  D_{7c}\left[\mathrm{GK_{24}},3\right] \label{spittaGF192inGK24}
\end{eqnarray}
This implies that there is both a Beltrami vector field invariant under $\mathrm{GP_{24}}$ and another one invariant under
$\mathrm{GK_{24}}$.
\paragraph{The Beltrami vector field invariant under $\mathrm{GP_{24}}$}
\begin{figure}[!hbt]
\begin{center}
\iffigs
\includegraphics[height=70mm]{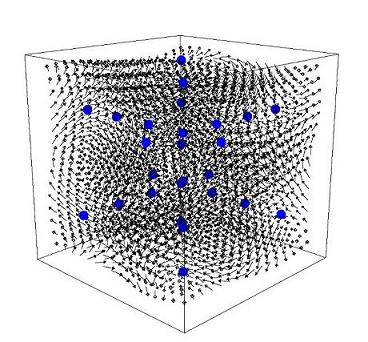}
\includegraphics[height=70mm]{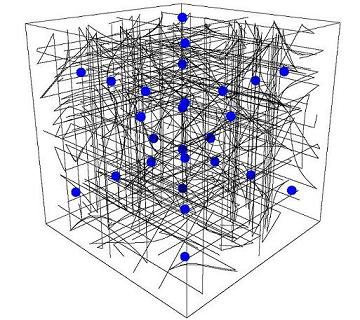}
\else
\end{center}
 \fi
\caption{\it  A plot of the $\mathrm{GP_{24}}$ invariant Beltrami vector field ${\mathbf{V}^{\left(\mathrm{GP_{24}}|D_1\right)}}({\mathbf{r}})$ obtained from  the octahedral  12 point orbit in the cubic lattice (midpoints of the edges of a regular cube). The field is analytically defined in eq.(\ref{mortadellasecca}). On the right we display a family of streamlines of this vector field with equally spaced initial conditions. In both pictures the circles denote the 26 isolated stagnation points of this flow.}
\label{gp24fildati}
 \iffigs
 \hskip 1cm \unitlength=1.1mm
 \end{center}
  \fi
\end{figure}
Applying to the vector field of eq.(\ref{orbit12vector}) the projector onto the singlet representation  $D_{1}\left[\mathrm{GP_{24}},1\right] $, we obtain the following Beltrami vector field
\begin{eqnarray}\label{mortadellasecca}
{\mathbf{V}^{\left(\mathrm{GP_{24}}|D_1\right)}}({\mathbf{r}}) & = &\left\{ V_x\, , \, V_y \, , \, V_z\right\} \nonumber\\
V_x & = &  \cos (2 \pi  (y-z))+2 \cos (2 \pi  (y+z))+\sqrt{2} \sin
   (2 \pi  (x-y))\nonumber\\
   &&-\sqrt{2} \sin (2 \pi  (x+y))-\sqrt{2}
   \sin (2 \pi  (x-z))+\sqrt{2} \sin (2 \pi  (x+z))\nonumber\\
V_y & = &2 \cos (2 \pi  (x-z))+2 \cos (2 \pi  (x+z))+\sqrt{2} \sin
   (2 \pi  (x-y))\nonumber\\
   &&+\sqrt{2} \sin (2 \pi  (x+y))+\sqrt{2}
   \sin (2 \pi  (y-z))-\sqrt{2} \sin (2 \pi  (y+z))\nonumber\\
V_z & = & 2 \cos (2 \pi  (x-y))+2 \cos (2 \pi  (x+y))-\sqrt{2} \sin
   (2 \pi  (x-z))\nonumber\\
   &&+\sqrt{2} \sin (2 \pi  (y-z))-\sqrt{2}
   \sin (2 \pi  (x+z))+\sqrt{2} \sin (2 \pi  (y+z))\nonumber\\
\end{eqnarray}
In the unit cube this vector field has 26 stagnation points:
\begin{equation}\label{stagnonePbis}
    {\mathbf{V}^{\left(\mathrm{GP_{24}}|D_1\right)}}({\mathbf{s}}_i) \quad i=1,\dots, 26
\end{equation}
whose coordinates are explicitly given below:
\begin{equation}\label{comancho}
    \begin{array}{lllclllclll}
 s_1 & = & \left\{0,\frac{1}{4},\frac{1}{4}\right\}
  &; & s_2 & = &
   \left\{0,\frac{3}{4},\frac{3}{4}\right\} &; & s_3 &
   = & \left\{\frac{1}{4},0,\frac{1}{4}\right\} \\
 s_4 & = & \left\{\frac{1}{4},\frac{1}{4},0\right\}
   &; & s_5 & = &
   \left\{\frac{1}{4},\frac{1}{4},\frac{1}{4}\right\} &; & s_6
   & = &
   \left\{\frac{1}{4},\frac{1}{4},\frac{1}{2}\right\} \\
 s_7 & = &
   \left\{\frac{1}{4},\frac{1}{4},\frac{3}{4}\right\} &; & s_8
   & = & \left\{\frac{1}{4},\frac{1}{4},1\right\} &; &
   s_9 & = &
   \left\{\frac{1}{4},\frac{1}{2},\frac{1}{4}\right\} \\
 s_{10} & = &
   \left\{\frac{1}{4},\frac{3}{4},\frac{1}{4}\right\} &; &
   s_{11} & = &
   \left\{\frac{1}{4},\frac{3}{4},\frac{3}{4}\right\} &; &
   s_{12} & = &
   \left\{\frac{1}{4},1,\frac{1}{4}\right\} \\
 s_{13} & = &
   \left\{\frac{1}{2},\frac{1}{4},\frac{1}{4}\right\} &; &
   s_{14} & = &
   \left\{\frac{1}{2},\frac{3}{4},\frac{3}{4}\right\} &; &
   s_{15} & = &
   \left\{\frac{3}{4},0,\frac{3}{4}\right\} \\
 s_{16} & = &
   \left\{\frac{3}{4},\frac{1}{4},\frac{1}{4}\right\} &; &
   s_{17} & = &
   \left\{\frac{3}{4},\frac{1}{4},\frac{3}{4}\right\} &; &
   s_{18} & = &
   \left\{\frac{3}{4},\frac{1}{2},\frac{3}{4}\right\} \\
 s_{19} & = &
   \left\{\frac{3}{4},\frac{3}{4},0\right\} &; & s_{20} &
   = &
   \left\{\frac{3}{4},\frac{3}{4},\frac{1}{4}\right\} &; &
   s_{21} & = &
   \left\{\frac{3}{4},\frac{3}{4},\frac{1}{2}\right\} \\
 s_{22} & = &
   \left\{\frac{3}{4},\frac{3}{4},\frac{3}{4}\right\} &; &
   s_{23} & = &
   \left\{\frac{3}{4},\frac{3}{4},1\right\} &; & s_{24} &
   = & \left\{\frac{3}{4},1,\frac{3}{4}\right\}\\
   s_{25} & = &\left\{1,\frac{1}{4},\frac{1}{4}\right\} &;&
   s_{26} & = &\left\{1,\frac{3}{4},\frac{3}{4}\right\} &;&
\end{array}
\end{equation}
A plot of this vector field and of a family of its streamlines is shown in fig.\ref{gp24fildati}.
\paragraph{The Beltrami vector field invariant under $\mathrm{GK_{24}}$}
Applying to the vector field of eq.(\ref{orbit12vector}) the projector onto the singlet representation  $D_{1}\left[\mathrm{GK_{24}},1\right] $, we obtain the following Beltrami vector field
\begin{eqnarray}\label{mortadellafresca}
{\mathbf{V}^{\left(\mathrm{GK_{24}}|D_1\right)}}({\mathbf{r}}) & = &\left\{ V_x \, , \, V_y \, , \, V_z\right\}\nonumber\\
V_x & = & 2 \cos (2 \pi  (x-y))+2 \cos (2 \pi  (x+y))\nonumber\\
   &&+\sin (2 \pi
   (x-z))-2 \sqrt{2} \sin (2 \pi  (y-z))-\sin (2 \pi
   (x+z))-2 \sqrt{2} \sin (2 \pi  (y+z))  \nonumber\\
V_y & = &2 \cos (2 \pi  (x-y))-2 \cos (2 \pi  (x+y))-\sqrt{2} \cos
   (2 \pi  (x-z))\nonumber\\
   &&-\sqrt{2} \cos (2 \pi  (x+z))+4 \sin (2
   \pi  y) \sin (2 \pi  z)  \nonumber\\
V_z & = & 2 \cos (2 \pi  (y-z))+2 \cos (2 \pi  (y+z))-2 \sqrt{2} \sin
   (2 \pi  (x-y))\nonumber\\
   &&+2 \sqrt{2} \sin (2 \pi  (x+y))+\sin (2
   \pi  (x-z))+\sin (2 \pi  (x+z))
\end{eqnarray}
This vector field has 16 stagnation points:
\begin{equation}\label{stagnonePtris}
    {\mathbf{V}^{\left(\mathrm{GK_{24}}|D_1\right)}}({\mathbf{s}}_i) \quad i=1,\dots, 16
\end{equation}
whose coordinates are explicitly given below:
\begin{equation}\label{pozzanghera}
    \begin{array}{lllllllllll}
 s_1 & = & \left\{\frac{1}{4},0,\frac{1}{4}\right\}
   & ; & s_2 & = &
   \left\{\frac{1}{4},0,\frac{3}{4}\right\} & ; &
   s_3 & = &
   \left\{\frac{3}{4},\frac{1}{4},\frac{1}{4}\right\} \\
 s_4 & = &
   \left\{\frac{1}{4},\frac{1}{2},\frac{1}{4}\right\} &
   ; & s_5 & = &
   \left\{\frac{1}{4},\frac{1}{2},\frac{3}{4}\right\} &
   ; & s_6 & = &
   \left\{\frac{3}{4},\frac{1}{4},\frac{1}{4}\right\} \\
 s_7 & = & \left\{\frac{1}{4},1,\frac{1}{4}\right\}
   & ; & s_8 & = &
   \left\{\frac{1}{4},1,\frac{3}{4}\right\} & ; &
   s_9 & = &
   \left\{\frac{3}{4},\frac{1}{4},\frac{1}{4}\right\} \\
 s_{10} & = &
   \left\{\frac{3}{4},0,\frac{3}{4}\right\} & ; &
   s_{11} & = &
   \left\{\frac{3}{4},\frac{1}{4},\frac{1}{4}\right\} &
   ; & s_{12} & = &
   \left\{\frac{3}{4},\frac{1}{4},\frac{1}{4}\right\} \\
 s_{13} & = &
   \left\{\frac{3}{4},\frac{1}{2},\frac{3}{4}\right\} &
   ; & s_{14} & = &
   \left\{\frac{3}{4},\frac{3}{4},\frac{1}{4}\right\} &
   ; & s_{15} & = &
   \left\{\frac{3}{4},\frac{1}{4},\frac{1}{4}\right\}\\
   s_{16} & = &
   \left\{\frac{3}{4},1,\frac{3}{4}\right\} &\null &
\end{array}
\end{equation}
\begin{figure}[!hbt]
\begin{center}
\iffigs
\includegraphics[height=70mm]{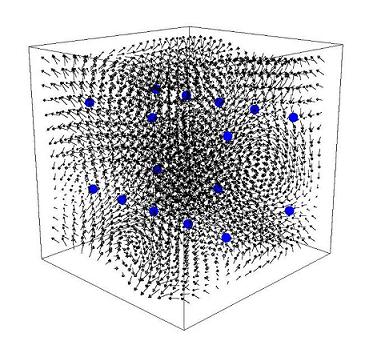}
\includegraphics[height=70mm]{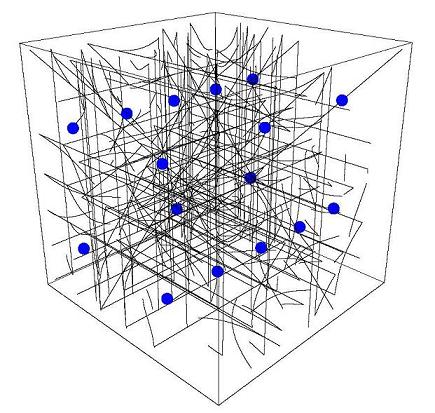}
\else
\end{center}
 \fi
\caption{\it  A plot of the $\mathrm{GK_{24}}$ invariant Beltrami vector field ${\mathbf{V}^{\left(\mathrm{GK_{24}}|D_1\right)}}({\mathbf{r}})$ obtained from  the octahedral  12 point orbit in the cubic lattice (midpoints of the edges of a regular cube). The field is analytically defined in eq.(\ref{mortadellafresca}). On the right we display a family of streamlines of this vector field with equally spaced initial conditions. In both pictures the circles denote the 16 isolated stagnation points of this flow.}
\label{gk24fildati}
 \iffigs
 \hskip 1cm \unitlength=1.1mm
 \end{center}
  \fi
\end{figure}
A plot of this vector field and of a family of its streamlines is shown in fig.\ref{gk24fildati}.
\par
As it is made manifest by the difference in the number of stagnation points and as it can be visually appreciated by comparing
fig.s \ref{gp24fildati} and \ref{gk24fildati}, although their invariance groups are isomorphic, the Beltrami fields ${\mathbf{V}^{\left(\mathrm{GP_{24}}|D_1\right)}}({\mathbf{r}})$ and ${\mathbf{V}^{\left(\mathrm{GK_{24}}|D_1\right)}}({\mathbf{r}})$ are genuinely different. This is an important lesson to be remembered. The complete classification of all invariant Beltrami vector fields requires a complete classification of all subgroups of $\mathrm{\mathrm{\mathrm{G_{1536}}}}$ up to conjugation and not simply up to isomorphism. Furthermore for all such subgroups one needs to find all singlet representations $D_1$.
\subsection{The lowest lying octahedral  orbit of length 8 in the cubic lattice}
\label{orbit8sezia}
\begin{figure}[!hbt]
\begin{center}
\iffigs
\includegraphics[height=70mm]{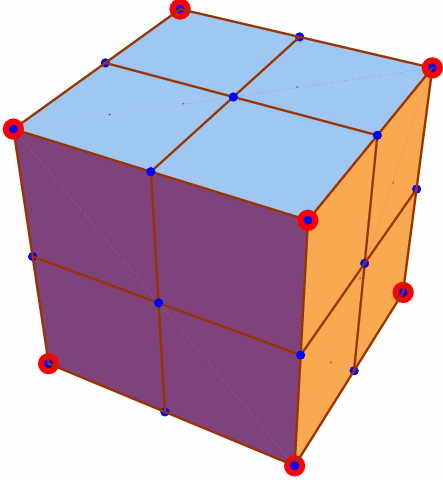}
\else
\end{center}
 \fi
\caption{\it  A view of the octahedral  8-orbit in the cubic lattice, corresponding to the vertices of a regular cube.}
\label{cubovertici}
 \iffigs
 \hskip 1cm \unitlength=1.1mm
 \end{center}
  \fi
\end{figure}

The next case we consider is the class of momentum vectors number 5) in our list of 48, namely:
\begin{equation}\label{sindromeperuviana}
    \mathbf{k} \, = \, \left\{1+4\mu, \,1+4\mu, \,1+4\mu\right\}
\end{equation}
If we choose the lowest lying representative of the class ($\mu\, = \,0$) we obtain the following octahedral  orbit of 8 points:
\begin{equation}\label{punti8}
    \begin{array}{lllclll}
 p_1 & = & \{-1,-1,-1\}& ; & p_5 & = & \{1,-1,-1\}  \\
 p_2 & = & \{-1,-1,1\} & ; & p_6 & = & \{1,-1,1\} \\
 p_3 & = & \{-1,1,-1\} & ; & p_7 & = & \{1,1,-1\} \\
 p_4 & = & \{-1,1,1\} & ; &  p_8 & = & \{1,1,1\} \\
 \end{array}
\end{equation}
These 8 points are the vertices of a regular cube inscribed in the sphere of radius $r^2 \, = \, 3$, as it is displayed in fig.\ref{cubovertici}.
\par
Applying the strategy of sum over lattice points that belong to a point group  orbit, we obtain the following 8 parameter vector field:
\begin{eqnarray}\label{orbit8vector}
{\mathbf{V}^{(8)}}({\mathbf{r}}|\mathbf{F}) & = &\left\{ V_x\, , \, V_y \, , \, V_z\right\} \nonumber\\
V_x & = &  \frac{2}{3} \left(\sqrt{3} \sin \left(2 \pi  \Theta
   _4\right) \left(F_5-F_1\right)+3 \cos \left(2 \pi
   \Theta _4\right) \left(F_1+F_5\right)\right.\nonumber\\
   &&\left.+3 \cos \left(2 \pi
    \Theta _3\right) \left(F_2-F_6\right)+\sqrt{3} \sin
   \left(2 \pi  \Theta _3\right) \left(F_2+F_6\right)+3
   \cos \left(2 \pi  \Theta _2\right)
   \left(F_7-F_3\right)\right.\nonumber\\
    &&\left.-\sqrt{3} \sin \left(2 \pi  \Theta
   _2\right) \left(F_3+F_7\right)+\sqrt{3} \sin \left(2 \pi
    \Theta _1\right) \left(F_4-F_8\right)-3 \cos \left(2
   \pi  \Theta _1\right) \left(F_4+F_8\right)\right) \nonumber\\
V_y & = & \frac{2}{3} \left(3 \cos \left(2 \pi  \Theta _4\right)
   F_1+3 \cos \left(2 \pi  \Theta _3\right) F_2+3 \cos
   \left(2 \pi  \Theta _2\right) F_3\right.\nonumber\\
    &&\left.+3 \cos \left(2 \pi
   \Theta _1\right) F_4+\sqrt{3} \sin \left(2 \pi  \Theta
   _4\right) \left(F_1+2 F_5\right)-\sqrt{3} \sin \left(2
   \pi  \Theta _3\right) \left(F_2-2 F_6\right)\right.\nonumber\\
    &&\left.-\sqrt{3}
   \sin \left(2 \pi  \Theta _2\right) \left(F_3-2
   F_7\right)+\sqrt{3} \sin \left(2 \pi  \Theta _1\right)
   \left(F_4+2 F_8\right)\right) \nonumber\\
V_z & = &\frac{2}{3} \left(-\sqrt{3} \sin \left(2 \pi  \Theta
   _4\right) \left(2 F_1+F_5\right)+\sqrt{3} \sin \left(2
   \pi  \Theta _3\right) \left(F_6-2 F_2\right)\right.\nonumber\\
    &&\left.+\sqrt{3}
   \sin \left(2 \pi  \Theta _2\right) \left(F_7-2
   F_3\right)-\sqrt{3} \sin \left(2 \pi  \Theta _1\right)
   \left(2 F_4+F_8\right)\right.\nonumber\\
    &&\left.+3 \left(\cos \left(2 \pi  \Theta
   _4\right) F_5+\cos \left(2 \pi  \Theta _3\right)
   F_6+\cos \left(2 \pi  \Theta _2\right) F_7+\cos \left(2
   \pi  \Theta _1\right) F_8\right)\right)
\end{eqnarray}
where $F_i$ ($i\,=\,1,\dots\, ,8$) are real numbers and the angles $\Theta_i$ are the following ones:
  \begin{equation}\label{angoli8}
    \begin{array}{rclcrcl}
 \Theta _1& = & x+y+z  &;&
 \Theta _2& = & x+y-z \\
 \Theta _3& = & x-y+z  &;&
 \Theta _4& = & x-y-z
\end{array}
  \end{equation}
  The action of the octahedral  group $\mathrm{O}_{24}$  is easily determined on such a vector field. We have
\begin{equation}\label{gulag}
    \forall \, \gamma \, \in \, \mathrm{O}_{24} \quad  : \quad \gamma^{-1} \cdot \mathbf{V}^{(8)}\left (\gamma \cdot \mathbf{r}|\mathbf{F}\right) \, = \, {\mathbf{V}^{(8)}}\left (\mathbf{r}| \, \mathfrak{R}^{(8)}[\gamma ]\cdot \mathbf{F}\right )
\end{equation}
where, as before, $\gamma$ are the $3 \times 3$ matrices of the fundamental defining representation, while $\mathfrak{R}^{(8)}[\gamma ]$ are $8 \times 8$ matrices acting on the parameter vector $\mathbf{F}$ that defines a reducible representation of $\mathrm{O}_{24}$. The matrix representation of the two generators (\ref{generatiTS}) is explicitly given by:
\begin{equation}\label{TSin8orbitO}
    \mathfrak{R}^{(8)}[T] \, = \,
\left(
\begin{array}{llllllll}
 0 & 0 & -1 & 0 & 0 & 0 & 1 & 0 \\
 1 & 0 & 0 & 0 & 1 & 0 & 0 & 0 \\
 0 & 1 & 0 & 0 & 0 & -1 & 0 & 0 \\
 0 & 0 & 0 & -1 & 0 & 0 & 0 & -1 \\
 0 & 0 & 1 & 0 & 0 & 0 & 0 & 0 \\
 1 & 0 & 0 & 0 & 0 & 0 & 0 & 0 \\
 0 & 1 & 0 & 0 & 0 & 0 & 0 & 0 \\
 0 & 0 & 0 & 1 & 0 & 0 & 0 & 0
\end{array}
\right)
\; ; \; \mathfrak{R}^{(8)}[S]\, = \,\left(
\begin{array}{llllllll}
 1 & 0 & 0 & 0 & 1 & 0 & 0 & 0 \\
 0 & 1 & 0 & 0 & 0 & -1 & 0 & 0 \\
 0 & 0 & 0 & -1 & 0 & 0 & 0 & -1 \\
 0 & 0 & -1 & 0 & 0 & 0 & 1 & 0 \\
 0 & 0 & 0 & 0 & -1 & 0 & 0 & 0 \\
 0 & 0 & 0 & 0 & 0 & -1 & 0 & 0 \\
 0 & 0 & 0 & 0 & 0 & 0 & 0 & -1 \\
 0 & 0 & 0 & 0 & 0 & 0 & -1 & 0
\end{array}
\right)
\end{equation}
Retrieving from the above generators all the group elements and in particular a representative for each of the conjugacy classes, we  easily compute the character vector of this representation. Explicitly we get:
\begin{equation}\label{caratter8}
    \chi\left[\mathbf{8}\right]\, = \, \{8,-1,0,0,0\}
\end{equation}
The multiplicity vector is:
\begin{equation}\label{multiply8}
    \mathrm{m}\left[\mathbf{8}\right]\, = \, \{0,0,1,1,1\}
\end{equation}
implying that the eight dimensional parameter space decomposes into a $D_3\left[\mathrm{O_{24}},2\right]$ plus a $D_4\left[\mathrm{O_{24}},3\right]$ plus a $D_5\left[\mathrm{O_{24}},3\right]$ representation.
\paragraph{Uplifting to the Universal Classifying Group}
 As in all the other cases, shifting the coordinate vector $\mathbf{r}$ by means of the constant vector $\ft 14 \, \mathbf{n} \, = \, \ft 14 \{n_1\,,\,n_2\, , \, n_3\,\}$ induces a rotation on the parameter-vector:
\begin{eqnarray}\label{traslaziona8}
    \mathbf{V}^{(8)}\left(\mathbf{r}+\ft 14 \, \mathbf{n} |\mathbf{F}\right)& =& \mathbf{V}^{(8)}\left(\mathbf{r} |\mathcal{M}_\mathbf{n} \mathbf{F}\right)
\end{eqnarray}
The explicit form of the matrix $\mathcal{M}_\mathbf{n}$ is given in appendix \ref{largone} (see eq.(\ref{8traslatore})).
This information is sufficient to complete the uplifting of the $8$-dimensional representation of the point group  to a representation of the Universal Classifying Group $\mathrm{\mathrm{G_{1536}}}$. It turns out that with respect to this latter the $8$-dimensional representation is an irreducible one, precisely the $D_{30}\left[\mathrm{\mathrm{G_{1536}}},8\right]$. The previous results are summarized in the following branching rule:
\begin{equation}\label{splitta8inO24}
    D_{30}\left[\mathrm{\mathrm{G_{1536}}},8\right] \, = \, D_3\left[\mathrm{O_{24}},2\right] \oplus D_4\left[\mathrm{O_{24}},3\right]\oplus D_5\left[\mathrm{O_{24}},3\right]
\end{equation}
As it happened for the orbit of length 12, the branching rule (\ref{splitta8inO24}) is  uplifted to the subgroup $\mathrm{G_{192}}$:
\begin{equation}\label{splitta8inG192}
    D_{30}\left[\mathrm{\mathrm{G_{1536}}},8\right] \, = \, D_{18}\left[\mathrm{\mathrm{G_{192}}},2\right] \oplus D_5\left[\mathrm{\mathrm{G_{192}}},3\right]\oplus D_6\left[\mathrm{\mathrm{G_{192}}},3\right]
\end{equation}
since we have:
\begin{eqnarray}
  D_{18}\left[\mathrm{G_{192}},2\right]&=& D_3\left[\mathrm{O_{24}},2\right]\nonumber \\
  D_5\left[\mathrm{G_{192}},3\right]&=& D_4\left[\mathrm{O_{24}},3\right] \nonumber\\
  D_6\left[\mathrm{G_{192}},3\right] &=& D_5\left[\mathrm{O_{24}},3\right] \label{caramella192da8}
\end{eqnarray}
Utilizing our character tables we also verify that the decomposition of the representation $D_{30}\left[\mathrm{\mathrm{G_{1536}}},8\right]$ with respect to the group $\mathrm{GF_{192}}$ is identical to its decomposition with respect to the isomorphic (but not conjugate) group $\mathrm{G_{192}}$, namely:
\begin{equation}\label{splitta8inGF192}
    D_{30}\left[\mathrm{\mathrm{G_{1536}}},8\right] \, = \, D_{18}\left[\mathrm{GF_{192}},2\right] \oplus D_5\left[\mathrm{GF_{192}},3\right]\oplus D_6\left[\mathrm{GF_{192}},3\right]
\end{equation}
and we obviously have:
\begin{eqnarray}
  D_{18}\left[\mathrm{GF_{192}},2\right]&=& D_3\left[\mathrm{GS_{24}},2\right]\nonumber \\
  D_5\left[\mathrm{GF_{192}},3\right]&=& D_4\left[\mathrm{GS_{24}},3\right] \nonumber\\
  D_6\left[\mathrm{GF_{192}},3\right] &=& D_5\left[\mathrm{GS_{24}},3\right] \label{caramellaF192da8}
\end{eqnarray}
From the above decompositions it appears that from the 8-parameter vector field (\ref{orbit8vector}) no instance can be extracted of a Beltrami vector field that is invariant  either under $\mathrm{O_{24}}$ or under its homologous $\mathrm{GS_{24}}$. Similarly by explicit decomposition (see the branching rules in appendix \ref{8brancione}) one reaches the conclusion that no singlet $D_1$ representation emerges with respect to the groups $\mathrm{GP_{24}}$ or $\mathrm{GK_{24}}$. There is however another pair of isomorphic (but not conjugate) subgroups $\mathrm{GS_{32}} \, \subset \, \mathrm{G_{192}}\, \subset \, \mathrm{\mathrm{G_{1536}}}$ and $\mathrm{GK_{32}} \, \subset \, \mathrm{GF_{192}}\, \subset \, \mathrm{\mathrm{G_{1536}}}$, both of order $32$, with respect to which singlet invariant vector fields do exist. They are described in the next section.
\subsubsection{Beltrami vector fields with  hidden symmetry $\mathrm{GS_{32}}$ and $\mathrm{GK_{32}}$, respectively}
\begin{figure}[!hbt]
\begin{center}
\iffigs
\includegraphics[height=70mm]{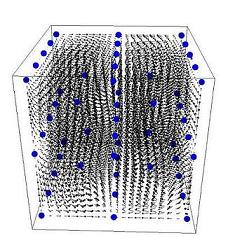}
\includegraphics[height=70mm]{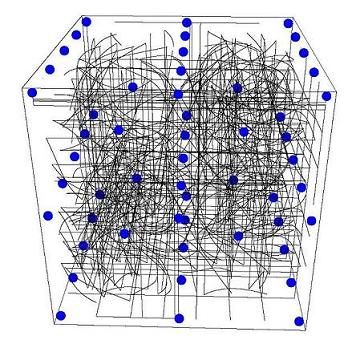}
\else
\end{center}
 \fi
\caption{\it  On the left a plot  of the vector field ${\mathbf{V}^{(\mathrm{GS_{32}})}}({\mathbf{r}}) $ defined in eq.(\ref{gs32flow}) which is invariant under the group $\mathrm{GS_{32}}$. On the right we display a family of streamlines of such a flow. As in previous figures the circles denote the 53 stagnation points.}
\label{gs32fluidos}
 \iffigs
 \hskip 1cm \unitlength=1.1mm
 \end{center}
  \fi
\end{figure}

\begin{figure}[!hbt]
\begin{center}
\iffigs
\includegraphics[height=70mm]{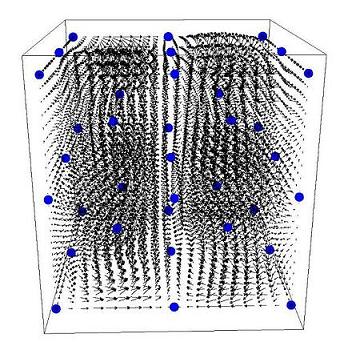}
\includegraphics[height=70mm]{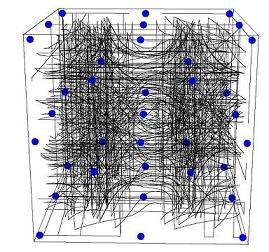}
\else
\end{center}
 \fi
\caption{\it  On the left a plot  of the vector field ${\mathbf{V}^{(\mathrm{GK_{32}})}}({\mathbf{r}}) $ defined in eq.(\ref{gk32flow}) which is invariant under the group $\mathrm{GK_{32}}$. On the right we display a family of streamlines of such a flow. The circles denote the 35 stagnation points of this flow.}
\label{gk32fluidos}
 \iffigs
 \hskip 1cm \unitlength=1.1mm
 \end{center}
  \fi
\end{figure}

The subgroups $\mathrm{GS_{32}}$ and $\mathrm{GK_{32}}$ are explicitly described in sections \ref{coniugatoGS32} and \ref{coniugatoGK32}. Their structure is very simple. They share a normal abelian subgroup of order 16, named $\mathrm{G_{16}}$, which  is isomorphic to $\mathbb{Z}_2^4$:
\begin{center}
\begin{picture}(200,100)
\put (-45,45){$\mathbb{Z}_2\times \mathbb{Z}_2\times \mathbb{Z}_2\times \mathbb{Z}_2$}
\put (50,45){$\sim$}
\put (65,45){$\mathrm{G_{16}}$}
\put (90,45){$\vartriangleleft $}
\put (103,47){\line (1,1){20}}
\put (103,47){\line (1,-1){20}}
\put (127,65.5){$\vartriangleleft $}
\put (127,24){$\vartriangleleft $}
\put (142,24){$\mathrm{GK_{32}}$}
\put (142,65.5){$\mathrm{GS_{32}}$}
\put (180,65.5){$\subset $}
\put (180,24){$\subset $}
\put (197,65.5){$\mathrm{G_{192}}$}
\put (197,24){$\mathrm{GF_{192}}$}
\end{picture}
\end{center}
\begin{equation}
\null
\label{GSGK32}
\end{equation}
The branching rules of the $\mathrm{G_{192}}$ representations (\ref{caramella192da8}) with respect to $\mathrm{GS_{32}}$ are the following ones:
\begin{eqnarray}
  D_{18}\left[\mathrm{G_{192}},2\right]&=& D_1\left[\mathrm{GS_{32}},1\right]\oplus D_2\left[\mathrm{GS_{32}},1\right] \nonumber \\
  D_5\left[\mathrm{G_{192}},3\right]&=& D_3\left[\mathrm{GS_{32}},1\right]\oplus D_9\left[\mathrm{GS_{32}},2\right] \nonumber\\
  D_6\left[\mathrm{G_{192}},3\right] &=&  D_4\left[\mathrm{GS_{32}},1\right]\oplus D_9\left[\mathrm{GS_{32}},2\right] \label{caramellaGS32da8}
\end{eqnarray}
and we similarly have:
\begin{eqnarray}
  D_{18}\left[\mathrm{GF_{192}},2\right]&=& D_1\left[\mathrm{GK_{32}},1\right]\oplus D_2\left[\mathrm{GK_{32}},1\right] \nonumber \\
  D_5\left[\mathrm{GF_{192}},3\right]&=& D_3\left[\mathrm{GK_{32}},1\right]\oplus D_9\left[\mathrm{GK_{32}},2\right] \nonumber\\
  D_6\left[\mathrm{GF_{192}},3\right] &=&  D_4\left[\mathrm{GK_{32}},1\right]\oplus D_9\left[\mathrm{GK_{32}},2\right] \label{caramellaGK32da8}
\end{eqnarray}
The two identity $D_1$ representations appearing in eq.s (\ref{caramellaGS32da8}) and (\ref{caramellaGK32da8}) signalize that from this orbit we can construct Beltrami vector fields invariant with respect either to $\mathrm{GS_{32}}$ or to $\mathrm{GK_{32}}$. Previous experience tells us they should be physically different flows although they have isomorphic hidden symmetries. We construct them in the next two subsections.
\paragraph{The Beltrami vector field invariant under $\mathrm{GS_{32}}$}
Performing the projection onto the $D_1\left[\mathrm{GS_{32}},1\right]$ representation we get the following Beltrami vector field
\begin{eqnarray}
{\mathbf{V}^{(\mathrm{GS_{32}})}}({\mathbf{r}}) & = &\left\{ V_x\, , \, V_y \, , \, V_z\right\} \nonumber\\
 V_x &=& 8 \cos (2 \pi  x) \sin (2 \pi  y) \sin (2 \pi  z)\nonumber \\
  V_y &=& -\cos (2 \pi  (x-y-z))-\cos (2 \pi  (x+y-z))+\cos (2 \pi  (x-y+z))+\cos (2
   \pi  (x+y+z))\nonumber\\
   &&+\sqrt{3} (-\sin (2 \pi  (x-y-z))+\sin (2 \pi  (x+y-z))-\sin
   (2 \pi  (x-y+z))+\sin (2 \pi  (x+y+z))) \nonumber\\
 V_z &=& -\cos (2 \pi  (x-y-z))+\cos (2 \pi  (x+y-z))-\cos (2 \pi  (x-y+z))+\cos (2
   \pi  (x+y+z))\nonumber\\
   &&+\sqrt{3} (\sin (2 \pi  (x-y-z))+\sin (2 \pi  (x+y-z))-\sin
   (2 \pi  (x-y+z))-\sin (2 \pi  (x+y+z)))\nonumber\\
   \label{gs32flow}
\end{eqnarray}
The vector field (\ref{gs32flow}) has 53 stagnation points in the unit cube:
\begin{equation}\label{stagnoneP}
    {\mathbf{V}^{\left(\mathrm{GS_{32}}|D_1\right)}}({\mathbf{s}}_i) \quad i=1,\dots, 53
\end{equation}
whose explicit form is given here below:
\begin{equation}\label{acquaputrida}
    \begin{array}{lllllllllll}
 s_1 & = & \{0,0,0\} & ; & s_2 & =
   & \left\{0,0,\frac{1}{2}\right\} & ; & s_3 &
   = & \left\{\frac{1}{4},0,\frac{1}{2}\right\} \\
 s_4 & = & \left\{0,\frac{1}{2},0\right\} &
   ; & s_5 & = &
   \left\{0,\frac{1}{2},\frac{1}{2}\right\} & ; &
   s_6 & = &
   \left\{\frac{1}{4},0,\frac{1}{2}\right\} \\
 s_7 & = & \{0,1,0\} & ; & s_8 & =
   & \left\{0,1,\frac{1}{2}\right\} & ; & s_9 &
   = & \left\{\frac{1}{4},0,\frac{1}{2}\right\} \\
 s_{10} & = & \left\{\frac{1}{4},0,0\right\} &
   ; & s_{11} & = &
   \left\{\frac{1}{4},0,\frac{1}{2}\right\} & ; &
   s_{12} & = &
   \left\{\frac{1}{4},0,\frac{1}{2}\right\} \\
 s_{13} & = &
   \left\{\frac{1}{4},\frac{1}{4},\frac{1}{4}\right\} &
   ; & s_{14} & = &
   \left\{\frac{1}{4},\frac{1}{4},\frac{3}{4}\right\} &
   ; & s_{15} & = &
   \left\{\frac{1}{4},0,\frac{1}{2}\right\} \\
 s_{16} & = &
   \left\{\frac{1}{4},\frac{1}{2},\frac{1}{2}\right\} &
   ; & s_{17} & = &
   \left\{\frac{1}{4},\frac{1}{2},1\right\} & ; &
   s_{18} & = &
   \left\{\frac{1}{4},0,\frac{1}{2}\right\} \\
 s_{19} & = &
   \left\{\frac{1}{4},\frac{3}{4},\frac{3}{4}\right\} &
   ; & s_{20} & = &
   \left\{\frac{1}{4},1,0\right\} & ; & s_{21} &
   = & \left\{\frac{1}{4},0,\frac{1}{2}\right\} \\
 s_{22} & = & \left\{\frac{1}{4},1,1\right\} &
   ; & s_{23} & = &
   \left\{\frac{1}{2},0,0\right\} & ; & s_{24} &
   = & \left\{\frac{1}{4},0,\frac{1}{2}\right\} \\
 s_{25} & = & \left\{\frac{1}{2},0,1\right\} &
   ; & s_{26} & = &
   \left\{\frac{1}{2},\frac{1}{2},0\right\} & ; &
   s_{27} & = &
   \left\{\frac{1}{4},0,\frac{1}{2}\right\} \\
 s_{28} & = &
   \left\{\frac{1}{2},\frac{1}{2},1\right\} & ; &
   s_{29} & = & \left\{\frac{1}{2},1,0\right\} &
   ; & s_{30} & = &
   \left\{\frac{1}{4},0,\frac{1}{2}\right\} \\
 s_{31} & = & \left\{\frac{1}{2},1,1\right\} &
   ; & s_{32} & = &
   \left\{\frac{3}{4},0,0\right\} & ; & s_{33} &
   = & \left\{\frac{1}{4},0,\frac{1}{2}\right\} \\
 s_{34} & = & \left\{\frac{3}{4},0,1\right\} &
   ; & s_{35} & = &
   \left\{\frac{3}{4},\frac{1}{4},\frac{1}{4}\right\} &
   ; & s_{36} & = &
   \left\{\frac{1}{4},0,\frac{1}{2}\right\} \\
 s_{37} & = &
   \left\{\frac{3}{4},\frac{1}{2},0\right\} & ; &
   s_{38} & = &
   \left\{\frac{3}{4},\frac{1}{2},\frac{1}{2}\right\} &
   ; & s_{39} & = &
   \left\{\frac{1}{4},0,\frac{1}{2}\right\} \\
 s_{40} & = &
   \left\{\frac{3}{4},\frac{3}{4},\frac{1}{4}\right\} &
   ; & s_{41} & = &
   \left\{\frac{3}{4},\frac{3}{4},\frac{3}{4}\right\} &
   ; & s_{42} & = &
   \left\{\frac{1}{4},0,\frac{1}{2}\right\} \\
 s_{43} & = &
   \left\{\frac{3}{4},1,\frac{1}{2}\right\} & ; &
   s_{44} & = & \left\{\frac{3}{4},1,1\right\} &
   ; & s_{45} & = &
   \left\{\frac{1}{4},0,\frac{1}{2}\right\} \\
 s_{46} & = & \left\{1,0,\frac{1}{2}\right\} &
   ; & s_{47} & = & \{1,0,1\} & ;
   & s_{48} & = &
   \left\{\frac{1}{4},0,\frac{1}{2}\right\} \\
 s_{49} & = &
   \left\{1,\frac{1}{2},\frac{1}{2}\right\} & ; &
   s_{50} & = & \left\{1,\frac{1}{2},1\right\} &
   ; & s_{51} & = &
   \left\{\frac{1}{4},0,\frac{1}{2}\right\}
\end{array}
\end{equation}
A plot of this vector field with a family of its streamlines is displayed in fig.\ref{gs32fluidos}.
\paragraph{The Beltrami vector field invariant under $\mathrm{GK_{32}}$}
Performing the projection onto the $D_1\left[\mathrm{GK_{32}},1\right]$ representation we get the following Beltrami vector field
\begin{eqnarray}
{\mathbf{V}^{(\mathrm{GK_{32}})}}({\mathbf{r}}) & = &\left\{ V_x\, , \, V_y \, , \, V_z\right\} \nonumber\\
 V_x &=&\left(-9+\sqrt{3}\right) \cos (2 \pi
   (x-y-z))-\left(-9+\sqrt{3}\right) \cos (2 \pi
   (x+y-z))\nonumber\\
   &&-\left(-9+\sqrt{3}\right) \cos (2 \pi
   (x-y+z))\nonumber \\
   &&+\left(-9+\sqrt{3}\right) \cos (2 \pi
   (x+y+z))-\frac{1}{3} \left(15+7 \sqrt{3}\right) (\sin (2
   \pi  (x-y-z))\nonumber\\
   &&+\sin (2 \pi  (x+y-z))+\sin (2 \pi
   (x-y+z))+\sin (2 \pi  (x+y+z)))\nonumber\\
  V_y &=& \left(-1+3 \sqrt{3}\right) \cos (2 \pi  (x-y-z))+\left(-1+3
   \sqrt{3}\right) \cos (2 \pi  (x+y-z))\nonumber\\
   &&+\left(1-3
   \sqrt{3}\right) \cos (2 \pi  (x-y+z))+\left(1-3
   \sqrt{3}\right) \cos (2 \pi  (x+y+z))\nonumber\\
   &&+\frac{1}{3}
   \left(3+17 \sqrt{3}\right) (-\sin (2 \pi  (x-y-z))\nonumber\\
   &&+\sin
   (2 \pi  (x+y-z))-\sin (2 \pi  (x-y+z))+\sin (2 \pi
   (x+y+z)))\nonumber\\
 V_z &=& \frac{2}{3} \left(-3 \left(4+\sqrt{3}\right) \cos (2 \pi
   (x-y-z))+3 \left(4+\sqrt{3}\right) \cos (2 \pi
   (x+y-z))\right.\nonumber\\
   &&\left.-3 \left(4+\sqrt{3}\right) \cos (2 \pi
   (x-y+z))+3 \left(4+\sqrt{3}\right) \cos (2 \pi
   (x+y+z))\right.\nonumber\\
   &&\left.+\left(-6+5 \sqrt{3}\right) \left(\sin (2 \pi
   (x-y-z))+\sin (2 \pi  (x+y-z))-\sin (2 \pi  (x-y+z))\right.\right.\nonumber\\
   &&\left.\left.-\sin
   (2 \pi  (x+y+z))\right)\right)\nonumber\\
   \label{gk32flow}
\end{eqnarray}
This vector field has 35 stagnation points, namely all those listed in eq.( \ref{acquaputrida}) with the exception the 18 listed below:
\begin{equation}\label{grimaldus}
  \mbox{not stagnation}\, = \, \left\{s_{10},s_{11},s_{12},s_{15},s_{16},s_{17},s_{20},s_{2
   1},s_{22},s_{32},s_{33},s_{34},s_{37},s_{38},s_{39},s_{42
   },s_{43},s_{44}\right\}
\end{equation}
A plot of this vector field with a family of its streamlines is displayed in fig.\ref{gk32fluidos}
\subsection{Example of an octahedral  orbit of length 24 in the cubic lattice}
\begin{figure}[!hbt]
\begin{center}
\iffigs
\includegraphics[height=70mm]{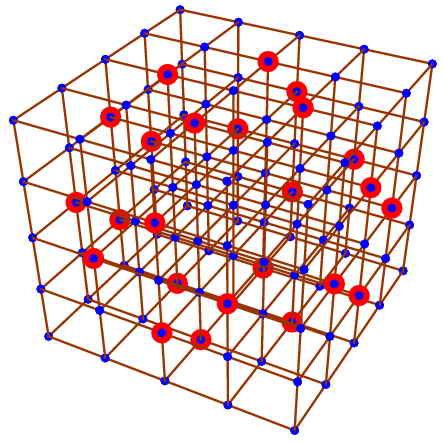}
\includegraphics[height=70mm]{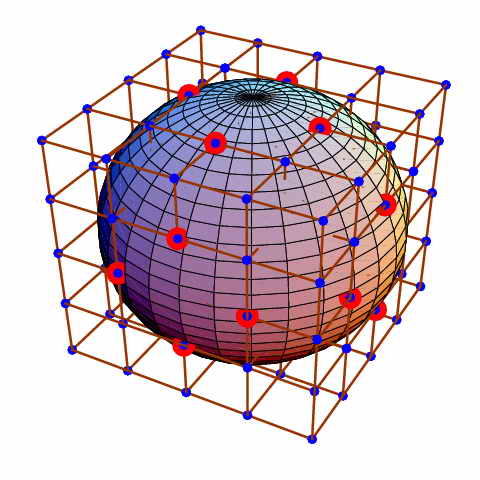}
\else
\end{center}
 \fi
\caption{\it  A view of the considered orbit of length 24 in the cubic lattice: the lattice points intersect the sphere of radius $r^2 \, = \, 6$.}
\label{orbita24PS}
 \iffigs
 \hskip 1cm \unitlength=1.1mm
 \end{center}
  \fi
\end{figure}

As a last example in the present discussion we consider the case numbered  13) in our list of 48 momentum vector classes, namely:
\begin{equation}\label{orbilla24}
  \mathbf{k}\, = \, \{1+4\,\mu, \,1+4\,\mu,2+4\,\rho\}
\end{equation}
Choosing the lowest lying representative in the class ($\mu\, = \, \rho \, = \, 0$) we obtain the octahedral  point orbit of order 24 listed below:
\begin{equation}\label{punti24}
\mathcal{O}^{(24)}_{1,1,2} \, = \, \left\{
\begin{array}{cccc}
\{-2,-1,-1\}, &\{-2,-1,1\}, &\{-2,1,-1\}, &\{-2,1,1\}, \\
\{-1,-2,-1\}, &\{-1,-2,1\}, &\{-1,-1,-2\}, &\{-1,-1,2\}, \\
\{-1,1,-2\}, &\{-1,1,2\}, &\{-1,2,-1\}, &\{-1 ,2,1\}, \\
\{1,-2,-1\}, &\{1,-2,1\}, &\{1,-1,-2\}, &\{1,-1,2\}, \\
\{1,1,-2\}, &\{1,1,2\}, &\{1,2,-1\}, &\{1,2,1\}, \\
\{2,-1,-1\}, &\{2,-1,1\}, &\{2,1,-1\}, &\{2,1,1\}
\end{array}
\right\}
\end{equation}
A geometrical picture of these points in the lattice is displayed in fig.\ref{orbita24PS}.
\par
Starting from the above point orbit, the construction algorithm \ref{algoritmo} produces a vector field containing 24 parameters which has the following form:
\begin{eqnarray}\label{orbit24vector}
{\mathbf{V}^{(24)}}({\mathbf{r}}|\mathbf{F}) & = &\bigoplus_{i=1}^{6} \, \Delta V^{(24)}_{i}(\mathbf{r}|\mathbf{F})
\end{eqnarray}
where:
\begin{equation}\label{deltav1}
  \Delta V^{(24)}_{1}(\mathbf{r}|\mathbf{F}) \, = \,\left(
\begin{array}{l}
 -\left(F_{12}+F_{24}\right) \Omega _1+\left(F_{23}-F_{11}\right)
   \Omega _2+\left(F_{10}-F_{22}\right) \Omega
   _3+\left(F_9+F_{21}\right) \Omega _4 \\
 2 \left(F_{12} \Omega _1+F_{11} \Omega _2+F_{10} \Omega _3+F_9 \Omega
   _4\right) \\
 2 \left(F_{24} \Omega _1+F_{23} \Omega _2+F_{22} \Omega _3+F_{21}
   \Omega _4\right)
\end{array}
\right)
\end{equation}
\begin{equation}\label{deltav2}
  \Delta V^{(24)}_{2}(\mathbf{r}|\mathbf{F}) \, = \,\left(
\begin{array}{l}
 -2 \left(2 F_8+F_{20}\right) \Omega _5+\left(2 F_{19}-4 F_7\right)
   \Omega _6-2 \left(F_6+2 F_{18}\right) \Omega _7-2 \left(F_5-2
   F_{17}\right) \Omega _8 \\
 2 \left(F_8 \Omega _5+F_7 \Omega _6+F_6 \Omega _7+F_5 \Omega _8\right)
   \\
 2 \left(F_{20} \Omega _5+F_{19} \Omega _6+F_{18} \Omega _7+F_{17}
   \Omega _8\right)
\end{array}
\right)
\end{equation}
\begin{equation}\label{deltav3}
  \Delta V^{(24)}_{3}(\mathbf{r}|\mathbf{F}) \, = \, \left(
\begin{array}{l}
 2 \left(F_4-2 F_{16}\right) \Omega _9+2 \left(F_3+2 F_{15}\right)
   \Omega _{10}+\left(4 F_2-2 F_{14}\right) \Omega _{11}+2 \left(2
   F_1+F_{13}\right) \Omega _{12} \\
 2 \left(F_4 \Omega _9+F_3 \Omega _{10}+F_2 \Omega _{11}+F_1 \Omega
   _{12}\right) \\
 2 \left(F_{16} \Omega _9+F_{15} \Omega _{10}+F_{14} \Omega
   _{11}+F_{13} \Omega _{12}\right)
\end{array}
\right)
  \end{equation}
\begin{equation}\label{deltav4}
  \Delta V^{(24)}_{4}(\mathbf{r}|\mathbf{F}) \, = \, \left(
\begin{array}{l}
 \sqrt{\frac{2}{3}} \left(\left(F_{12}-F_{24}\right) \Omega
   _{13}-\left(F_{11}+F_{23}\right) \Omega
   _{14}+\left(F_{10}+F_{22}\right) \Omega
   _{15}+\left(F_{21}-F_9\right) \Omega _{16}\right) \\
 \frac{\left(F_{12}+5 F_{24}\right) \Omega _{13}-\left(F_{11}-5
   F_{23}\right) \Omega _{14}-\left(F_{10}-5 F_{22}\right) \Omega
   _{15}+\left(F_9+5 F_{21}\right) \Omega _{16}}{\sqrt{6}} \\
 \frac{-\left(5 F_{12}+F_{24}\right) \Omega _{13}+\left(F_{23}-5
   F_{11}\right) \Omega _{14}+\left(F_{22}-5 F_{10}\right) \Omega
   _{15}-\left(5 F_9+F_{21}\right) \Omega _{16}}{\sqrt{6}}
\end{array}
\right)
  \end{equation}
\begin{equation}\label{deltav5}
  \Delta V^{(24)}_{5}(\mathbf{r}|\mathbf{F}) \, = \, \left(
\begin{array}{l}
 \sqrt{\frac{2}{3}} \left(\left(F_8-2 F_{20}\right) \Omega
   _{17}-\left(F_7+2 F_{19}\right) \Omega _{18}+\left(2
   F_6-F_{18}\right) \Omega _{19}-\left(2 F_5+F_{17}\right) \Omega
   _{20}\right) \\
 \sqrt{\frac{2}{3}} \left(2 \left(F_8+F_{20}\right) \Omega _{17}+2
   \left(F_{19}-F_7\right) \Omega _{18}+\left(2 F_6+5 F_{18}\right)
   \Omega _{19}+\left(5 F_{17}-2 F_5\right) \Omega _{20}\right) \\
 \sqrt{\frac{2}{3}} \left(-\left(5 F_8+2 F_{20}\right) \Omega
   _{17}+\left(2 F_{19}-5 F_7\right) \Omega _{18}-2
   \left(F_6+F_{18}\right) \Omega _{19}+2 \left(F_{17}-F_5\right)
   \Omega _{20}\right)
\end{array}
\right)
  \end{equation}
\begin{equation}\label{deltav6}
  \Delta V^{(24)}_{6}(\mathbf{r}|\mathbf{F}) \, = \, \left(
\begin{array}{l}
 \sqrt{\frac{2}{3}} \left(\left(2 F_4+F_{16}\right) \Omega
   _{21}+\left(F_{15}-2 F_3\right) \Omega _{22}+\left(F_2+2
   F_{14}\right) \Omega _{23}-\left(F_1-2 F_{13}\right) \Omega
   _{24}\right) \\
 \sqrt{\frac{2}{3}} \left(\left(5 F_{16}-2 F_4\right) \Omega
   _{21}+\left(2 F_3+5 F_{15}\right) \Omega _{22}+2
   \left(F_{14}-F_2\right) \Omega _{23}+2 \left(F_1+F_{13}\right)
   \Omega _{24}\right) \\
 \sqrt{\frac{2}{3}} \left(2 \left(F_{16}-F_4\right) \Omega _{21}-2
   \left(F_3+F_{15}\right) \Omega _{22}+\left(2 F_{14}-5 F_2\right)
   \Omega _{23}-\left(5 F_1+2 F_{13}\right) \Omega _{24}\right)
\end{array}
\right)
  \end{equation}
$F_i$ being the 24 real parameters, $\Omega_i$ denoting the 24 periodic basis functions:
\begin{equation}
\begin{array}{lllllllllllllll}
 \Omega _1 & = & \cos \left(2 \pi  \Theta _1\right) &
   ; & \Omega _2 & = & \cos \left(2 \pi
   \Theta _2\right) & ; & \Omega _3 & = & \cos
   \left(2 \pi  \Theta _3\right) & ; & \Omega _4 &
   = & \cos \left(2 \pi  \Theta _4\right) \\
 \Omega _5 & = & \cos \left(2 \pi  \Theta _5\right) &
   ; & \Omega _6 & = & \cos \left(2 \pi
   \Theta _6\right) & ; & \Omega _7 & = & \cos
   \left(2 \pi  \Theta _7\right) & ; & \Omega _8 &
   = & \cos \left(2 \pi  \Theta _8\right) \\
 \Omega _9 & = & \cos \left(2 \pi  \Theta _9\right) &
   ; & \Omega _{10} & = & \cos \left(2 \pi
   \Theta _{10}\right) & ; & \Omega _{11} & =
   & \cos \left(2 \pi  \Theta _{11}\right) & ; & \Omega
   _{12} & = & \cos \left(2 \pi  \Theta _{12}\right) \\
 \Omega _{13} & = & \sin \left(2 \pi  \Theta _1\right)
   & ; & \Omega _{14} & = & \sin \left(2 \pi
   \Theta _2\right) & ; & \Omega _{15} & = &
   \sin \left(2 \pi  \Theta _3\right) & ; & \Omega
   _{16} & = & \sin \left(2 \pi  \Theta _4\right) \\
 \Omega _{17} & = & \sin \left(2 \pi  \Theta _5\right)
   & ; & \Omega _{18} & = & \sin \left(2 \pi
   \Theta _6\right) & ; & \Omega _{19} & = &
   \sin \left(2 \pi  \Theta _7\right) & ; & \Omega
   _{20} & = & \sin \left(2 \pi  \Theta _8\right) \\
 \Omega _{21} & = & \sin \left(2 \pi  \Theta _9\right)
   & ; & \Omega _{22} & = & \sin \left(2 \pi
   \Theta _{10}\right) & ; & \Omega _{23} & =
   & \sin \left(2 \pi  \Theta _{11}\right) & ; & \Omega
   _{24} & = & \sin \left(2 \pi  \Theta _{12}\right)
\end{array}
\end{equation}
and the 12 independent arguments of the trigonometric functions being those listed below:
\begin{equation}\label{tettoni24}
  \begin{array}{lll}
 \Theta _1 & = & 2 x+y+z \\
 \Theta _2 & = & 2 x+y-z \\
 \Theta _3 & = & 2 x-y+z \\
 \Theta _4 & = & 2 x-y-z \\
 \Theta _5 & = & x+2 y+z \\
 \Theta _6 & = & x+2 y-z \\
 \Theta _7 & = & x+y+2 z \\
 \Theta _8 & = & x+y-2 z \\
 \Theta _9 & = & x-y+2 z \\
 \Theta _{10} & = & x-y-2 z \\
 \Theta _{11} & = & x-2 y+z \\
 \Theta _{12} & = & x-2 y-z
\end{array}
\end{equation}
\subsubsection{Point Group irreps and Uplifting to the Universal Classifying Group}
As in all previous cases we can easily derive the representation of the point group  $\mathrm{O_{24}}$ and of the quantized translations on the parameter space provided by $\mathrm{24} \times \mathrm{24}$-matrices. We do not display the explicit form of the generators of $\mathrm{\mathrm{G_{1536}}}$ since they are too large to fit on paper. We just encode the relevant information in the form of the splitting of the $24$-parameter space in irreducible representation of $\mathrm{\mathrm{G_{1536}}}$ and of its relevant subgroups.
\par
Firstly we find that the 24-dimensional representation of $\mathrm{\mathrm{G_{1536}}}$ is reducible and splits as follows:
\begin{equation}\label{gavrilonka}
    \mathfrak{R}_{(1,1,2)}\left[\mathrm{\mathrm{G_{1536}}},24\right]\, = \, D_{34}\left[\mathrm{\mathrm{G_{1536}}},12\right] \oplus D_{35}\left[\mathrm{\mathrm{G_{1536}}},12\right]
\end{equation}
Secondly we recall from appendix \ref{brancicardo} the following branching rules with respect to the subgroups $\mathrm{G_{192}}$ and $\mathrm{GF_{192}}$:
\begin{eqnarray}
  D_{34}\left[\mathrm{\mathrm{G_{1536}}},12\right]  &=&\left\{  \begin{array}{ccccc}
                                                         D_{10}\left[\mathrm{G_{192}},3\right]  & \oplus & D_{14}\left[\mathrm{G_{192}},3\right] & \oplus & D_{19}\left[\mathrm{G_{192}},6\right] \\
                                                          D_{9}\left[\mathrm{GF_{192}},3\right]  & \oplus & D_{13}\left[\mathrm{GF_{192}},3\right] & \oplus & D_{19}\left[\mathrm{GF_{192}},6\right] \\
                                                       \end{array}
  \right. \nonumber\\
  D_{35}\left[\mathrm{\mathrm{G_{1536}}},12\right] &=& \left\{ \begin{array}{ccccc}
                                                         D_{9}\left[\mathrm{G_{192}},3\right]  & \oplus & D_{13}\left[\mathrm{G_{192}},3\right] & \oplus & D_{19}\left[\mathrm{G_{192}},6\right] \\
                                                          D_{10}\left[\mathrm{GF_{192}},3\right]  & \oplus & D_{14}\left[\mathrm{GF_{192}},3\right] & \oplus & D_{19}\left[\mathrm{GF_{192}},6\right] \\
                                                       \end{array} \right.
                                                       \label{sakuraiOne}
\end{eqnarray}
Thirdly we consider the following branching rules of the considered irreps of the groups $\mathrm{G_{192}}$ and $\mathrm{GF_{192}}$ with respect to theirs subgroups $\mathrm{O_{24}}$ and $\mathrm{GS_{24}}$:
\begin{eqnarray}
  D_{19}\left[\mathrm{G_{192}},6\right] &=&D_{4}\left[\mathrm{O_{24}},3\right] \oplus D_{5}\left[\mathrm{O_{24}},3\right]\nonumber\\
   D_{14}\left[\mathrm{G_{192}},3\right] &=& D_{1}\left[\mathrm{O_{24}},1\right] \oplus D_{3}\left[\mathrm{O_{24}},2\right]\nonumber\\
   D_{13}\left[\mathrm{G_{192}},3\right] &=& D_{2}\left[\mathrm{O_{24}},1\right] \oplus D_{3}\left[\mathrm{O_{24}},2\right] \nonumber\\
   D_{10}\left[\mathrm{G_{192}},3\right] &=&  D_{5}\left[\mathrm{O_{24}},3\right] \nonumber\\
   D_{9}\left[\mathrm{G_{192}},3\right] &=& D_{5}\left[\mathrm{O_{24}},3\right]  \label{filosofobrutto}
  \end{eqnarray}
  \begin{eqnarray}
  D_{19}\left[\mathrm{GF_{192}},6\right] &=&D_{4}\left[\mathrm{GS_{24}},3\right] \oplus D_{5}\left[\mathrm{GS_{24}},3\right]\nonumber\\
   D_{14}\left[\mathrm{GF_{192}},3\right] &=& D_{1}\left[\mathrm{GS_{24}},1\right] \oplus D_{3}\left[\mathrm{GS_{24}},2\right]\nonumber\\
   D_{13}\left[\mathrm{GF_{192}},3\right] &=& D_{2}\left[\mathrm{GS_{24}},1\right] \oplus D_{3}\left[\mathrm{GS_{24}},2\right] \nonumber\\
   D_{10}\left[\mathrm{GF_{192}},3\right] &=&  D_{5}\left[\mathrm{GS_{24}},3\right] \nonumber\\
   D_{9}\left[\mathrm{GF_{192}},3\right] &=& D_{5}\left[\mathrm{GS_{24}},3\right]  \label{filosofobello}
  \end{eqnarray}
  From inspection of eq.s (\ref{filosofobrutto}) and (\ref{filosofobello}) we conclude that there is a singlet representation both of the point group  $\mathrm{O_{24}}$ and of its homologous non conjugate copy $\mathrm{GS_{24}}$. Hence from this orbit we can construct Beltrami vector fields with $\mathrm{O_{24}}$ or $\mathrm{GS_{24}}$ hidden symmetry. This is certainly true, but the situation is even better. There exists a subgroup named by us $\mathrm{Oh_{48}}$ (see appendix \ref{coniugatoOh48} for its description), which is isomorphic to the extended octahedral  group and it is embedded in $\mathrm{\mathrm{G_{1536}}}$ in the following way:
 \begin{equation}\label{filibilatotris}
    \mathrm{O_{24}} \, \subset \, \mathrm{Oh_{48}} \, \subset \, \mathrm{G_{192}}\, \subset \, \mathrm{\mathrm{G_{1536}}}
 \end{equation}
 The branching rule of the entire 24-dimensional representation of the classifying group with respect to $\mathrm{Oh_{48}}$ is the following one:
 \begin{equation}\label{gavrilonkabis}
    \mathfrak{R}_{(1,1,2)}\left[\mathrm{\mathrm{G_{1536}}},24\right]\, = \, D_{1}\left[\mathrm{Oh_{48}},1\right] \oplus D_{3}\left[\mathrm{Oh_{48}},1\right]\oplus \,2 \, D_{5}\left[\mathrm{Oh_{48}},2\right]\oplus \, 3 \, D_{7}\left[\mathrm{Oh_{48}},3\right]\oplus \, 3 \, D_{9}\left[\mathrm{Oh_{48}},3\right]
\end{equation}
 Hence there exists an invariant vector field with respect to the order 48 subgroup $\mathrm{Oh_{48}}$. This is certainly  invariant with respect to all subgroups of the same, in particular $\mathrm{O_{24}}$. As we have only one $\mathrm{O_{24}}$ singlet it means that the unique vector field invariant under $\mathrm{O_{24}}$ has actually an enhanced symmetry $\mathrm{Oh_{48}}$. Furthermore the isomorphism of $\mathrm{G_{192}}$ and $\mathrm{GF_{192}}$ implies that there must exist another subgroup $\mathrm{OKh_{48}}\sim \mathrm{Oh_{48}}$ also isomorphic to the extended octahedral  group and satisfying the inclusion relations homologous to those displayed in eq.(\ref{filibilatotris}), namely:
 \begin{equation}\label{filibilatobis}
    \mathrm{GS_{24}} \, \subset \, \mathrm{OKh_{48}} \, \subset \, \mathrm{GF_{192}}\, \subset \, \mathrm{\mathrm{G_{1536}}}
 \end{equation}
 The branching rules in eq.s (\ref{filosofobrutto}) and (\ref{filosofobello}) imply that the singlet vector field with respect to $\mathrm{GS_{24}}$ is actually invariant with respect to the order 48 group $\mathrm{OKh_{48}}$.
 We have not constructed the vector field invariant with respect to $\mathrm{OKh_{48}}$ and we just constructed the $\mathrm{Oh_{48}}$-invariant one.
\subsubsection{The Beltrami flow invariant under $\mathrm{Oh_{48}}$}
\begin{figure}[!hbt]
\begin{center}
\iffigs
\includegraphics[height=70mm]{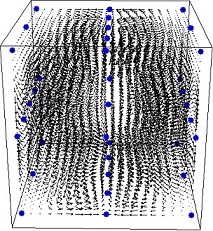}
\includegraphics[height=70mm]{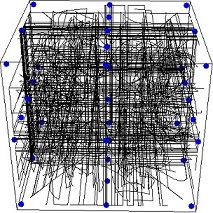}
\else
\end{center}
 \fi
\caption{\it  On the left we exhibit a plot of the Beltrami vector field ${\mathbf{V}^{\left(\mathrm{Oh_{48}}|D_1\right)}}({\mathbf{r}})$ defined in eq.(\ref{salamecottodinizza}), which is invariant under the group $\mathrm{Oh_{48}}$ isomorphic to the extended octahedral  group. On the right we display a family of streamlines of such a flow. }
\label{oh48plottone}
 \iffigs
 \hskip 1cm \unitlength=1.1mm
 \end{center}
  \fi
\end{figure}

Applying the projector onto the irrep $D_{1}\left[\mathrm{Oh_{48}},1\right]$, we obtain the following Beltrami vector field:
\begin{eqnarray}\label{salamecottodinizza}
{\mathbf{V}^{\left(\mathrm{Oh_{48}}|D_1\right)}}({\mathbf{r}}) & = &\left\{ V_x\, , \, V_y \, , \, V_z\right\} \nonumber\\
V_x & = & 18 \Omega _5-18 \Omega _6-18 \Omega _7+18 \Omega _8+18 \Omega _9-18
   \Omega _{10}-18 \Omega _{11}+18 \Omega _{12}\nonumber\\
   &&+\sqrt{6} \left(-\Omega
   _{13}-\Omega _{14}-\Omega _{15}-\Omega _{16}+\Omega _{17}+\Omega
   _{18}+\Omega _{19}+\Omega _{20}+\Omega _{21}+\Omega _{22}+\Omega
   _{23}+\Omega _{24}\right) \nonumber\\
V_y & = & -3 \Omega _1+3 \Omega _2-3 \Omega _3+3 \Omega _4-6 \Omega _5+6 \Omega
   _6+6 \Omega _7-6 \Omega _8+6 \Omega _9-6 \Omega _{10}-6 \Omega
   _{11}+6 \Omega _{12}\nonumber\\
   &&+\sqrt{6} \left(\Omega _{13}+\Omega _{14}-\Omega
   _{15}-\Omega _{16}-4 \Omega _{17}-4 \Omega _{18}+7 \Omega _{19}+7
   \Omega _{20}-7 \Omega _{21}-7 \Omega _{22}+4 \Omega _{23}+4 \Omega
   _{24}\right) \nonumber\\
V_z & = &3 \Omega _1+3 \Omega _2-3 \Omega _3-3 \Omega _4-6 \Omega _5-6 \Omega
   _6+6 \Omega _7+6 \Omega _8-6 \Omega _9-6 \Omega _{10}+6 \Omega
   _{11}+6 \Omega _{12}\nonumber\\
   &&+\sqrt{6} \left(\Omega _{13}-\Omega _{14}+\Omega
   _{15}-\Omega _{16}+7 \Omega _{17}-7 \Omega _{18}-4 \Omega _{19}+4
   \Omega _{20}-4 \Omega _{21}+4 \Omega _{22}+7 \Omega _{23}-7 \Omega
   _{24}\right) \nonumber\\
\end{eqnarray}
A plot of the vector field and a family of its streamlines are displayed in fig.\ref{oh48plottone}. Note that this vector field has 35 stagnation points whose coordinates we do not display for brevity. The stagnation points are as usual denoted by circles in the figure.
\section{The Hexagonal Lattice and the Dihedral Group $\mathrm{{\cal D}_6}$}
\label{hexareticolo}
We come next to a quite short discussion of the hexagonal lattice. In this case we do not construct the Universal Classifying Group and we limit ourselves to display some solutions of the Beltrami Equation corresponding to the lowest lying orbits of the point group  of this lattice which is the dihedral group ${\cal D}_6$.
\par
\begin{figure}[!hbt]
\begin{center}
\iffigs
\includegraphics[height=70mm]{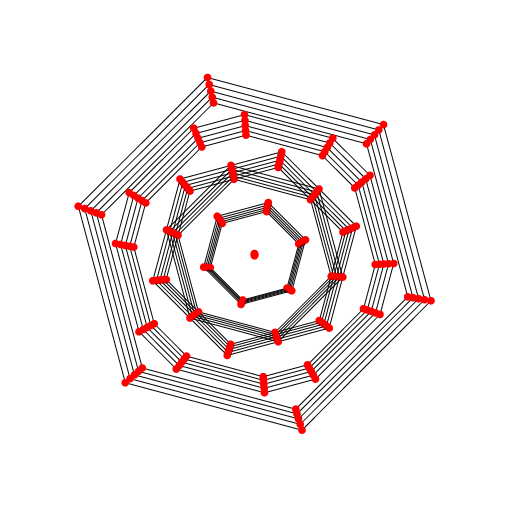}
\includegraphics[height=70mm]{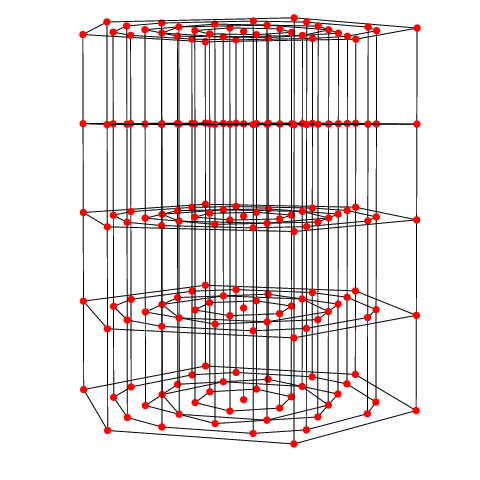}
\else
\end{center}
 \fi
\caption{\it  A view of the hexagonal space lattice $\Lambda_{Hex}$, seen from above and in a front view.}
\label{HexagLatS}
 \iffigs
 \hskip 1cm \unitlength=1.1mm
 \end{center}
  \fi
\end{figure}
\begin{figure}[!hbt]
\begin{center}
\iffigs
\includegraphics[height=70mm]{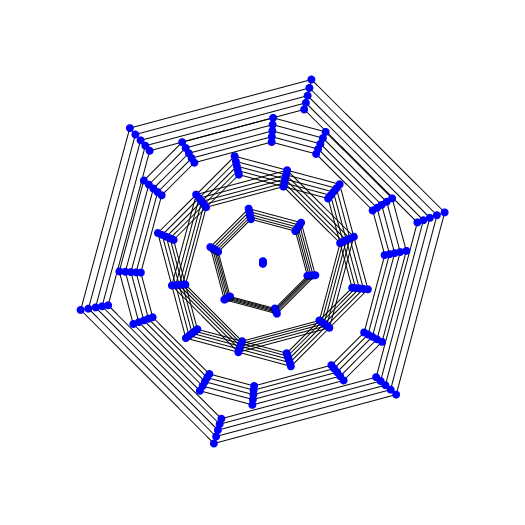}
\includegraphics[height=70mm]{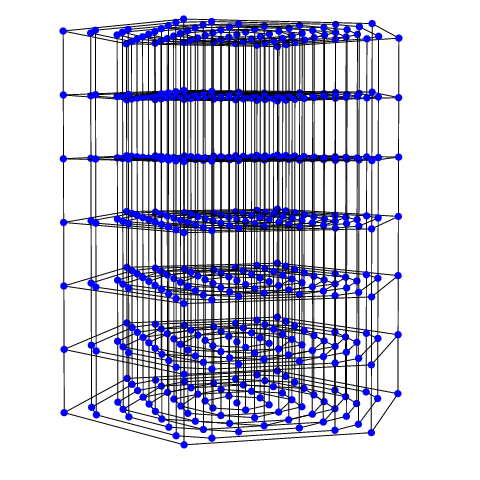}
\else
\end{center}
 \fi
\caption{\it  A view of the hexagonal momentum lattice $\Lambda_{Hex}^\star$, seen from above and in a front view.}
\label{HexagLatP}
 \iffigs
 \hskip 1cm \unitlength=1.1mm
 \end{center}
  \fi
\end{figure}
\begin{figure}[!hbt]
\begin{center}
\iffigs
\includegraphics[height=70mm]{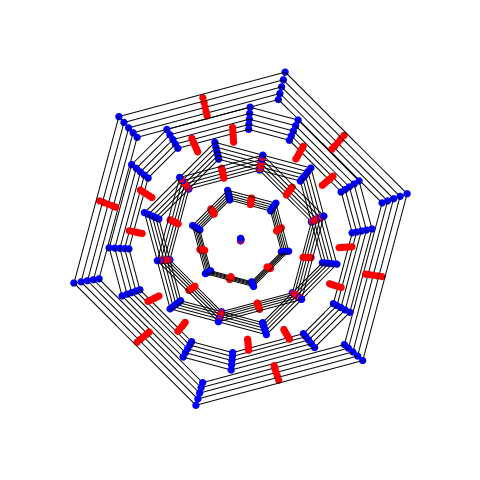}
\includegraphics[height=70mm]{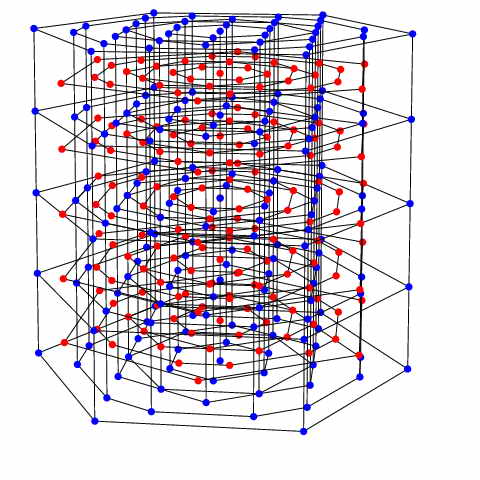}
\else
\end{center}
 \fi
\caption{\it  A comparative view of the hexagonal space and momentum lattices, seen from above and in a front view. The blue points are momentum lattice points ($\in \,\Lambda_{Hex}^\star$), while the red points are space lattice points ($\in \, \Lambda_{Hex}$).}
\label{HexagLatSP}
 \iffigs
 \hskip 1cm \unitlength=1.1mm
 \end{center}
  \fi
\end{figure}

Our main purpose is to illustrate, by means of this example, the new features that appear when the lattice is not self-dual. Since in this section all considered representations are relative to the point group  we simplify the notation mentioning the irreps only as $D_1, \dots, D_6$ without writing in square brackets the group.
\paragraph{The hexagonal lattice}
 $\Lambda_{Hex}$ and its dual $\Lambda^\star_{Hex}$  are displayed in fig.s \ref{HexagLatS},
\ref{HexagLatP} and \ref{HexagLatSP}. These lattices are not self-dual and there is a constant metric which is not diagonal.
\par
The basis vectors of the hexagonal space lattice $\Lambda_{Hex}$ are the following ones:
\begin{equation}\label{Hexagbase}
    \vec{\mathbf{w}}_1 \, = \, \{1,0,0\} \quad ; \quad \vec{\mathbf{w}}_2 \, = \, \left\{\frac{1}{2},-\frac{\sqrt{3}}{2},0\right\} \quad ; \quad \vec{\mathbf{w}}_3 \, = \, \{0,0,1\}
\end{equation}
which implies that the metric is the following non diagonal one:
\begin{equation}\label{nonkronecca}
    g_{\mu\nu} \, = \, \left(
\begin{array}{lll}
 1 & \frac{1}{2} & 0 \\
 \frac{1}{2} & 1 & 0 \\
 0 & 0 & \frac{3}{4}
\end{array}
\right)
\end{equation}
The basis vectors $\vec{\mathbf{e}}^\mu$ of the dual momentum lattice $\Lambda_{Hex}^\star$ do not coincide with those of the lattice $\Lambda_{Hex}$. They are the following ones:
\begin{equation}\label{Hexagdualbas}
    \vec{\mathbf{e}}^1 \, = \, \left\{1,\frac{1}{\sqrt{3}},0\right\} \quad ; \quad \vec{\mathbf{e}}^2 \, = \,\left\{0,-\frac{2}{\sqrt{3}},0\right\} \quad ; \quad \vec{\mathbf{e}}^3 \, = \,\left\{0,0,\frac{2}{\sqrt{3}}\right\}
\end{equation}
The subgroup of the proper rotation group which maps the cubic lattice into itself is the dihedral group $\mathrm{D}_6$ whose order is 12. In the next subsection we recall its structure.
\subsection{The dihedral group $\mathrm{{\cal D}_6}$}
Abstractly the dihedral $\mathrm{{\cal D}_6}$ group is defined by the following generators and relations:
\begin{equation}\label{presentaD6}
    A \; , \; B \quad : \quad A^6 \, = \, \mbox{\bf e} \quad ; \quad B^2 \, = \, \mbox{\bf e} \quad ; \quad \left( BA\right)^2 \, = \, \mbox{\bf e}
\end{equation}
Explicitly in three dimensions we can take the following matrix representation for the generators of $\mathrm{\mathrm{{\cal D}_6}}$:
\begin{equation}\label{d3repD6}
    A \, = \, \left(
\begin{array}{lll}
 \frac{1}{2} & \frac{\sqrt{3}}{2} & 0 \\
 -\frac{\sqrt{3}}{2} & \frac{1}{2} & 0 \\
 0 & 0 & 1
\end{array}
\right) \quad ; \quad B \, = \, \left(
\begin{array}{lll}
 -1 & 0 & 0 \\
 0 & 1 & 0 \\
 0 & 0 & -1
\end{array}
\right)
\end{equation}
The group generated by the above generators has 12 elements that can be arranged into 6 conjugacy classes, as it it is displayed in  Table \ref{nomiD6elem}.
\begin{table}[!hbt]
  \centering
  \begin{eqnarray*}
  \begin{array}{|c|rcl|}
\hline
\mbox{\bf e} & 1_1 &  = & \{x,y,z\} \\
\hline
\null & 2_1 &  = & \left\{\frac{1}{2} \left(x+\sqrt{3}
   y\right),\frac{1}{2} \left(y-\sqrt{3} x\right),z\right\} \\
A & 2_2 &  = & \left\{\frac{1}{2} \left(x-\sqrt{3}
   y\right),\frac{1}{2} \left(\sqrt{3} x+y\right),z\right\} \\
\hline
\null & 3_1 &  = & \left\{\frac{1}{2} \left(\sqrt{3}
   y-x\right),\frac{1}{2} \left(-\sqrt{3} x-y\right),z\right\}
   \\
A^2 & 3_2 &  = & \left\{\frac{1}{2} \left(-x-\sqrt{3}
   y\right),\frac{1}{2} \left(\sqrt{3} x-y\right),z\right\} \\
 \hline
A^3 & 4_1 &  = & \{-x,-y,z\} \\
 \hline
\null & 5_1 &  = & \{-x,y,-z\} \\
B & 5_2 &  = & \left\{\frac{1}{2} \left(x-\sqrt{3}
   y\right),\frac{1}{2} \left(-\sqrt{3} x-y\right),-z\right\}
   \\
\null & 5_3 &  = & \left\{\frac{1}{2} \left(x+\sqrt{3}
   y\right),\frac{1}{2} \left(\sqrt{3} x-y\right),-z\right\} \\
 \hline
\null & 6_1 &  = & \left\{\frac{1}{2} \left(-x-\sqrt{3}
   y\right),\frac{1}{2} \left(y-\sqrt{3} x\right),-z\right\} \\
BA & 6_2 &  = & \{x,-y,-z\} \\
\null & 6_3 &  = & \left\{\frac{1}{2} \left(\sqrt{3}
   y-x\right),\frac{1}{2} \left(\sqrt{3} x+y\right),-z\right\}\\
   \hline
\end{array}
\end{eqnarray*}
  \caption{Conjugacy Classes of the Dihedral Group $\mathrm{{\cal D}_6}$.}\label{nomiD6elem}
\end{table}
In such a table every group element is uniquely identified by its action on the three-dimensional vector $\left\{x,y,z\right\}$.
The multiplication table of the group $\mathrm{{\cal D}_6}$ is shown in Table \ref{Dmultab}.
\begin{table}[!hbt]
  \centering
  \begin{eqnarray*}
    \begin{array}{|l|llllllllllll|}
    \hline
 \null &   1_1 & 2_1 & 2_2 & 3_1 & 3_2 & 4_1 & 5_1 & 5_2 & 5_3 & 6_1 &
   6_2 & 6_3 \\
   \hline
 1_1 &   1_1 & 2_1 & 2_2 & 3_1 & 3_2 & 4_1 & 5_1 & 5_2 & 5_3 & 6_1
   & 6_2 & 6_3 \\
 2_1 &   2_1 & 3_1 & 1_1 & 4_1 & 2_2 & 3_2 & 6_3 & 6_1 & 6_2 & 5_1
   & 5_2 & 5_3 \\
 2_2 &   2_2 & 1_1 & 3_2 & 2_1 & 4_1 & 3_1 & 6_1 & 6_2 & 6_3 & 5_2
   & 5_3 & 5_1 \\
 3_1 &   3_1 & 4_1 & 2_1 & 3_2 & 1_1 & 2_2 & 5_3 & 5_1 & 5_2 & 6_3
   & 6_1 & 6_2 \\
 3_2 &   3_2 & 2_2 & 4_1 & 1_1 & 3_1 & 2_1 & 5_2 & 5_3 & 5_1 & 6_2
   & 6_3 & 6_1 \\
 4_1 &   4_1 & 3_2 & 3_1 & 2_2 & 2_1 & 1_1 & 6_2 & 6_3 & 6_1 & 5_3
   & 5_1 & 5_2 \\
 5_1 &   5_1 & 6_1 & 6_3 & 5_2 & 5_3 & 6_2 & 1_1 & 3_1 & 3_2 & 2_1
   & 4_1 & 2_2 \\
 5_2 &   5_2 & 6_2 & 6_1 & 5_3 & 5_1 & 6_3 & 3_2 & 1_1 & 3_1 & 2_2
   & 2_1 & 4_1 \\
 5_3 &   5_3 & 6_3 & 6_2 & 5_1 & 5_2 & 6_1 & 3_1 & 3_2 & 1_1 & 4_1
   & 2_2 & 2_1 \\
 6_1 &   6_1 & 5_2 & 5_1 & 6_2 & 6_3 & 5_3 & 2_2 & 2_1 & 4_1 & 1_1
   & 3_1 & 3_2 \\
 6_2 &   6_2 & 5_3 & 5_2 & 6_3 & 6_1 & 5_1 & 4_1 & 2_2 & 2_1 & 3_2
   & 1_1 & 3_1 \\
 6_3 &   6_3 & 5_1 & 5_3 & 6_1 & 6_2 & 5_2 & 2_1 & 4_1 & 2_2 & 3_1
   & 3_2 & 1_1\\
   \hline
\end{array}
\end{eqnarray*}
  \caption{Multiplication table of the Dihedral Group $\mathrm{{\cal D}_6}$.}\label{Dmultab}
\end{table}
\subsection{Irreducible representations of the dihedral group $\mathrm{{\cal D}_6}$ and the character table}
The group $\mathrm{{\cal D}_6}$ has six conjugacy classes. Therefore according to theory we expect six irreducible representations
that we name $D_i$, $i\, =\, 1,\dots, 6$. Let us briefly describe them.
The first four representations are one-dimensional.
\subsubsection{$D_1$: the identity representation}
The identity representation which exists for all groups is that one where to each element of $\mathrm{{\cal D}_6}$ we associate the number $1$
\begin{equation}\label{D6identD1}
    \forall \, \gamma \, \in \, \mathrm{O} \,\, : \quad D_1(\gamma) \, = \, 1
\end{equation}
Obviously the character of such a representation is:
\begin{equation}\label{caretterusD6D1}
    \chi_1 \, = \, \{1,1,1,1,1\}
\end{equation}
\subsubsection{$D_2$: the second one-dimensional representation}
The representation $D_2$ is also one-dimensional. It is constructed as follows.
\begin{equation}
\begin{array}{ccccccc}
 \forall \, \gamma & \in & \{\mbox{\bf e}\} & : & D_2(\gamma) & = & 1 \\
  \forall \, \gamma & \in & \{A\} & : & D_2(\gamma) & = & -1 \\
   \forall \, \gamma & \in &\left\{A^2\right\}& : & D_2(\gamma) & = & 1 \\
 \forall \, \gamma & \in  & \left\{A^3\right\}& : & D_2(\gamma) & = & -1 \\
   \forall \, \gamma & \in & \{B\}& : & D_2(\gamma) & = & 1 \\
  \forall \, \gamma & \in  & \{{BA}\} & : & D_2(\gamma) & = & -1 \\
\end{array}
\end{equation}
Clearly the corresponding character vector is the following one:
\begin{equation}\label{caretterusD6D2}
    \chi_2 \, = \, \{1,-1,1,-1,1,-1\}
\end{equation}
Said in another way, this is the representation where $A \, = \, -1$ and $B\, = \,1$.
\subsubsection{$D_3$: the third one-dimensional representation}
The representation $D_3$ is also one-dimensional. It is constructed as follows.
\begin{equation}
\begin{array}{ccccccc}
 \forall \, \gamma & \in & \{\mbox{\bf e}\} & : & D_2(\gamma) & = & 1 \\
  \forall \, \gamma & \in & \{A\} & : & D_2(\gamma) & = & -1 \\
   \forall \, \gamma & \in &\left\{A^2\right\}& : & D_2(\gamma) & = & 1 \\
 \forall \, \gamma & \in  & \left\{A^3\right\}& : & D_2(\gamma) & = & -1 \\
   \forall \, \gamma & \in & \{B\}& : & D_2(\gamma) & = & -1 \\
  \forall \, \gamma & \in  & \{{BA}\} & : & D_2(\gamma) & = & 1 \\
\end{array}
\end{equation}
Clearly the corresponding character vector is the following one:
\begin{equation}\label{caretterusD6D3}
    \chi_3 \, = \, \{1,-1,1,-1,-1,1\}
\end{equation}
Said in another way, this is the representation where $A \, = \, -1$ and $B\, = \, -1$.
\subsubsection{$D_4$: the fourth one-dimensional representation}
The representation $D_4$ is also one-dimensional. It is constructed as follows.
\begin{equation}
\begin{array}{ccccccc}
 \forall \, \gamma & \in & \{\mbox{\bf e}\} & : & D_2(\gamma) & = & 1 \\
  \forall \, \gamma & \in & \{A\} & : & D_2(\gamma) & = & 1 \\
   \forall \, \gamma & \in &\left\{A^2\right\}& : & D_2(\gamma) & = & 1 \\
 \forall \, \gamma & \in  & \left\{A^3\right\}& : & D_2(\gamma) & = & 1 \\
   \forall \, \gamma & \in & \{B\}& : & D_2(\gamma) & = & -1 \\
  \forall \, \gamma & \in  & \{{BA}\} & : & D_2(\gamma) & = & -1 \\
\end{array}
\end{equation}
Clearly the corresponding character vector is the following one:
\begin{equation}\label{caretterusD6D2bis}
    \chi_4 \, = \, \{1,1,1,1,-1,-1\}
\end{equation}
Said in another way, this is the representation where $A \, = \, 1$ and $B\, = \, -1$.
\subsubsection{$D_5$: the first two-dimensional representation}
The representation $D_5$ is two-dimensional and it corresponds to a homomorphism:
\begin{equation}\label{D6mapD5}
    D_5 \, : \quad \mathrm{D_6} \, \rightarrow \, \mathrm{SL(2,\mathbb{C})}
\end{equation}
which associates to each element of the dihedral group a $2 \times 2$ complex valued matrix of determinant one.
The homomorphism is completely specified by giving the two matrices representing the two generators:
\begin{equation}\label{D6mapD5bis}
    D_5(A) \, = \,
\left(
\begin{array}{ll}
 e^{\frac{i \pi }{3}} & 0 \\
 0 & e^{-\frac{i \pi }{3}}
\end{array}
\right)\quad ; \quad D_5(B) \, = \, \left(
\begin{array}{ll}
 0 & 1 \\
 1 & 0
\end{array}
\right)
\end{equation}
The character vector of $D_5$ is easily calculated from the above information and we have:
\begin{equation}\label{caretterusD6D5}
    \chi_5 \, = \, \{2 , \, 1 ,\, -1 ,\, -2,\, 0,\,  0\}
\end{equation}
\subsubsection{$D_6$: the second two-dimensional representation}
The representation $D_6$ is also two-dimensional and it corresponds to a homomorphism:
\begin{equation}\label{D6mapD5tris}
    D_6 \, : \quad \mathrm{D_6} \, \rightarrow \, \mathrm{SL(2,\mathbb{C})}
\end{equation}
which associates to each element of the dihedral group a $2 \times 2$ complex valued matrix of determinant one.
The homomorphism is completely specified by giving the two matrices representing the two generators:
\begin{equation}\label{D6mapD6}
    D_6(A) \, = \,
\left(
\begin{array}{ll}
 e^{\frac{2 i \pi }{3}} & 0 \\
 0 & e^{-\frac{2 i \pi }{3}}
\end{array}
\right)\quad ; \quad D_6(B) \, = \, \left(
\begin{array}{ll}
 0 & 1 \\
 1 & 0
\end{array}
\right)
\end{equation}
The character vector of $D_6$ is easily calculated from the above information and we have:
\begin{equation}\label{caretterusD6D6}
    \chi_6 \, = \, \{2 , \, -1 ,\, -1 ,\, 2,\, 0,\,  0\}
\end{equation}
\begin{table}[!hbt]
  \centering
  \begin{eqnarray*}
    \begin{array}{||l||cccccc||}
    \hline
    \hline
{\begin{array}{cc} \null &\mbox{Class}\\
\mbox{Irrep} & \null\\
\end{array}}& \{\mbox{\bf e},1\} & \{A,2\} & \left\{A^2,2\right\} & \left\{A^3,1\right\} & \{B,3\} &\{{BA},3\} \\
\hline
\hline
 D_1 \, , \quad \chi_1 \, = \,& 1 & 1 & 1 & 1 & 1 & 1 \\
 \hline
 D_2 \, , \quad \chi_2 \, = \,& 1 & -1 & 1 & -1 & 1 & -1 \\
 \hline
 D_3 \, , \quad \chi_3 \, = \,& 1 & -1 & 1 & -1 & -1 & 1 \\
 \hline
 D_4 \, , \quad \chi_4 \, = \,& 1 & 1 & 1 & 1 & -1 & -1 \\
 \hline
 D_5 \, , \quad \chi_5 \, = \,& 2 & 1 & -1 & -2 & 0 & 0 \\
 \hline
 D_6\, , \quad \chi_6 \, = \,& 2 & -1 & -1 & 2 & 0 & 0\\
 \hline
\end{array}
\end{eqnarray*}
  \caption{The character table of the dihedral group $\mathrm{{\cal D}_6}$}\label{D6caratter}
\end{table}
The character table of the $\mathrm{{\cal D}_6}$ group is summarized in Table \ref{D6caratter}.
\subsection{Spherical layers in the Hexagonal lattice and $\mathrm{{\cal D}_6}$ orbits}
Let us now analyze the action of the dihedral group $\mathrm{{\cal D}_6}$ on the hexagonal lattice. Just as in the case of the cubic lattice, we define the orbits as the sets of vectors $\mathbf{k}\, \in \, \Lambda_{Hex}^{\star}$ that can be mapped one into the other by the action of some element of the group $\mathrm{{\cal D}_6}$:
\begin{equation}\label{orbitadefiD6}
    \mathbf{k}_1 \, \in \, \mathcal{O} \quad \mbox{and} \quad \mathbf{k}_2 \, \in \, \mathcal{O} \quad \Rightarrow \quad \exists \, \gamma \, \in \, \mathrm{{\cal D}_6} \; / \; \gamma\,\cdot \, \mathbf{k}_1 \, = \, \mathbf{k}_2
\end{equation}
The hexagonal lattice displays a more variegated bestiary of orbit types with respect to the case of the cubic lattice. There are orbits of length $2$, $6$ and $12$, but those of length $6$ and $12$ appear in a few different types.
\paragraph{Orbits of length 2 }
Each of these orbits is of the following form:
\begin{equation}\label{orbita2}
\mathcal{O}_2 \, = \,     \left\{
\begin{array}{ll}
 \left\{0,0,-\frac{2 r}{\sqrt{3}}\right\}, &
   \left\{0,0,\frac{2 r}{\sqrt{3}}\right\}
\end{array}
\right\}
\end{equation}
where $r \, \in \, \mathbb{Z}$ is any integer number.
\paragraph{Type One Orbits of length 6 }
Each of these orbits is of the following form:
\begin{equation}\label{orbita6One}
\mathcal{O}^{(1)}_6 \, = \,     \left\{
\begin{array}{llllll}
 \left\{0,-\frac{2 p}{\sqrt{3}},0\right\}, &
   \left\{0,\frac{2 p}{\sqrt{3}},0\right\}, &
   \left\{-p,-\frac{p}{\sqrt{3}},0\right\}, &
   \left\{-p,\frac{p}{\sqrt{3}},0\right\}, &
   \left\{p,-\frac{p}{\sqrt{3}},0\right\}, &
   \left\{p,\frac{p}{\sqrt{3}},0\right\}
\end{array}
\right\}
\end{equation}
where $p \, \in \, \mathbb{Z}$ is any integer number.
\paragraph{Type Two Orbits of length 6 }
Each of these orbits is of the following form:
\begin{equation}\label{orbita6Two}
\mathcal{O}^{(2)}_6 \, = \,     \left\{
\begin{array}{llllll}
 \{-2 p,0,0\}, & \left\{-p,-\sqrt{3} p,0\right\}, &
   \left\{-p,\sqrt{3} p,0\right\}, & \left\{p,-\sqrt{3}
   p,0\right\}, & \left\{p,\sqrt{3} p,0\right\}, & \{2
   p,0,0\}
\end{array}
\right\}
\end{equation}
where $p \, \in \, \mathbb{Z}$ is any integer number.
\paragraph{Type One Orbits of length 12 }
Each of these orbits is of the following form:
\begin{equation}\label{orbita12One}
\mathcal{O}^{(1)}_{12} \, = \,     \left\{
\begin{array}{llll}
 \left\{-p,-\frac{p-2 q}{\sqrt{3}},0\right\}, &
   \left\{-p,\frac{p-2 q}{\sqrt{3}},0\right\}, &
   \left\{p,-\frac{p-2 q}{\sqrt{3}},0\right\}, &
   \left\{p,\frac{p-2 q}{\sqrt{3}},0\right\}, \\
   \left\{p-q,-\frac{p+q}{\sqrt{3}},0\right\}, &
   \left\{p-q,\frac{p+q}{\sqrt{3}},0\right\}, &
   \left\{-q,\frac{2 p-q}{\sqrt{3}},0\right\}, &
   \left\{-q,\frac{q-2 p}{\sqrt{3}},0\right\}, \\
   \left\{q,\frac{2 p-q}{\sqrt{3}},0\right\}, &
   \left\{q,\frac{q-2 p}{\sqrt{3}},0\right\}, &
   \left\{q-p,-\frac{p+q}{\sqrt{3}},0\right\}, &
   \left\{q-p,\frac{p+q}{\sqrt{3}},0\right\}\\
\end{array}
\right\}
\end{equation}
where $p,q \, \in \, \mathbb{Z}$ and $q\ne \pm p$ and $q\ne 2\,p$
\paragraph{Type Two Orbits of length 12 }
Each of these orbits is of the following form:
\begin{equation}\label{orbita12Two}
\mathcal{O}^{(2)}_{12} \, = \,     \left\{
\begin{array}{llll}
 \left\{-p,-\frac{p-2 q}{\sqrt{3}},\frac{2
   r}{\sqrt{3}}\right\},
   & \left\{-p,\frac{p-2
   q}{\sqrt{3}},-\frac{2 r}{\sqrt{3}}\right\},
   & \left\{p,-\frac{p-2 q}{\sqrt{3}},-\frac{2
   r}{\sqrt{3}}\right\},
   & \left\{p,\frac{p-2
   q}{\sqrt{3}},\frac{2 r}{\sqrt{3}}\right\},
   \\ \left\{p-q,-\frac{p+q}{\sqrt{3}},\frac{2
   r}{\sqrt{3}}\right\},
   &\left\{p-q,\frac{p+q}{\sqrt{3}},-\frac{2
   r}{\sqrt{3}}\right\},
   & \left\{-q,\frac{2
   p-q}{\sqrt{3}},-\frac{2 r}{\sqrt{3}}\right\}
   &\left\{-q,\frac{q-2 p}{\sqrt{3}},\frac{2
   r}{\sqrt{3}}\right\},
   \\ \left\{q,\frac{2
   p-q}{\sqrt{3}},\frac{2 r}{\sqrt{3}}\right\},
   &\left\{q,\frac{q-2 p}{\sqrt{3}},-\frac{2
   r}{\sqrt{3}}\right\},
   &
   \left\{q-p,-\frac{p+q}{\sqrt{3}},-\frac{2
   r}{\sqrt{3}}\right\},
   &\left\{q-p,\frac{p+q}{\sqrt{3}},\frac{2
   r}{\sqrt{3}}\right\}\\
\end{array}
\right\}
\end{equation}
where $p,q ,r\, \in \, \mathbb{Z}$ and $q\ne \pm p$ and $q\ne 2\,p$
\paragraph{Type Three Orbits of length 12 }
Each of these orbits is of the following form:
\begin{equation}\label{orbita12Three}
\mathcal{O}^{(3)}_{12} \, = \,     \left\{
\begin{array}{llll}
 \left\{0,-\frac{2 p}{\sqrt{3}},-\frac{2
   r}{\sqrt{3}}\right\},
   &
   \left\{0,-\frac{2
   p}{\sqrt{3}},\frac{2 r}{\sqrt{3}}\right\},
   &
   \left\{0,\frac{2 p}{\sqrt{3}},-\frac{2
   r}{\sqrt{3}}\right\},
   &
   \left\{0,\frac{2
   p}{\sqrt{3}},\frac{2 r}{\sqrt{3}}\right\},
   \\
   \left\{-p,-\frac{p}{\sqrt{3}},-\frac{2
   r}{\sqrt{3}}\right\},
   &
   \left\{-p,-\frac{p}{\sqrt{3}},\frac{2
   r}{\sqrt{3}}\right\},
   &
   \left\{-p,\frac{p}{\sqrt{3}},-\frac{2
   r}{\sqrt{3}}\right\},
   &
   \left\{-p,\frac{p}{\sqrt{3}},\frac{2
   r}{\sqrt{3}}\right\},
   \\
   \left\{p,-\frac{p}{\sqrt{3}},-\frac{2
   r}{\sqrt{3}}\right\},
   &
   \left\{p,-\frac{p}{\sqrt{3}},\frac{2
   r}{\sqrt{3}}\right\},
   &
   \left\{p,\frac{p}{\sqrt{3}},-\frac{2
   r}{\sqrt{3}}\right\},
   &
   \left\{p,\frac{p}{\sqrt{3}},\frac{2 r}{\sqrt{3}}\right\}
   \\
\end{array}
\right\}
\end{equation}
where $p,q \, \in \, \mathbb{Z}$ are two arbitrary integer numbers.
\paragraph{Type Four Orbits of length 12 }
Each of these orbits is of the following form:
\begin{equation}\label{orbita12Four}
\mathcal{O}^{(4)}_{12} \, = \,     \left\{
\begin{array}{llll}
 \left\{-2 p,0,-\frac{2 r}{\sqrt{3}}\right\},
 & \left\{-2
   p,0,\frac{2 r}{\sqrt{3}}\right\},
 &
 \left\{-p,-\sqrt{3}
   p,-\frac{2 r}{\sqrt{3}}\right\},
 &
 \left\{-p,-\sqrt{3}
   p,\frac{2 r}{\sqrt{3}}\right\},
 \\
 \left\{-p,\sqrt{3}
   p,-\frac{2 r}{\sqrt{3}}\right\},
 &
 \left\{-p,\sqrt{3}
   p,\frac{2 r}{\sqrt{3}}\right\},
 &
 \left\{p,-\sqrt{3}
   p,-\frac{2 r}{\sqrt{3}}\right\},
 & \left\{p,-\sqrt{3}
   p,\frac{2 r}{\sqrt{3}}\right\},
\\
 \left\{p,\sqrt{3}
   p,-\frac{2 r}{\sqrt{3}}\right\},
 &
 \left\{p,\sqrt{3}
   p,\frac{2 r}{\sqrt{3}}\right\},
 &
 \left\{2 p,0,-\frac{2
   r}{\sqrt{3}}\right\},
 &
 \left\{2 p,0,\frac{2
   r}{\sqrt{3}}\right\},
\end{array}
\right\}
\end{equation}
where $p,q \, \in \, \mathbb{Z}$ are two arbitrary integer numbers.
\begin{table}
  \centering
  \begin{eqnarray*}
\begin{array}{|c|c|c|}
\hline
r^2 &\mbox{Number of Points} &\mbox{Dihedral $\mathrm{{\cal D}_6}$ Group Point Orbits}\\
\hline
0 & 1 & \{1\} \\
 \frac{4}{3} & 8 & \{6\oplus2\} \\
 \frac{8}{3} & 12 & \{12\} \\
 4 & 6 & \{6\} \\
 \frac{16}{3} & 20 & \{12\oplus6\oplus2\} \\
 \frac{20}{3} & 24 & \{12\oplus12\} \\
 \frac{28}{3} & 24 & \{12\oplus12\} \\
 \frac{32}{3} & 36 & \{12\oplus12\oplus12\} \\
 12 & 8 & \{6\oplus2\} \\
 \frac{40}{3} & 24 & \{12\oplus12\} \\
 \frac{44}{3} & 24 & \{12\oplus12\} \\
 16 & 18 & \{6\oplus12\} \\
 \frac{52}{3} & 48 & \{12\oplus12\oplus12\oplus12\} \\
 \frac{56}{3} & 24 & \{12\oplus12\} \\
 \frac{64}{3} & 36 & \{12\oplus12\oplus12\} \\
 \frac{68}{3} & 24 & \{12\oplus12\} \\
 24 & 12 & \{12\} \\
\hline
\end{array}
\end{eqnarray*}
  \caption{Spherical layers in the hexagonal momentum lattice}\label{spherdecompoD6}
\end{table}
As we see the shorter orbit of length $2$ is actually vertical, namely the associated Beltrami Flows correspond to decoupled systems where only the coordinate $z(t)$ obeys a non linear differential equation. The other two coordinates form a free system. Similarly the orbits of length $6$ and the first orbit of length $12$ are all planar. In the corresponding Beltrami Flow there is no dependence on the coordinate $z$ which forms a free system. Presumably all the Beltrami Flows of this type are integrable. Only the maximal orbits of length $12$  of type two, three and four are truly three-dimensional and give rise to systems that might develop a chaos.
\par
In Table \ref{spherdecompoD6} we have displayed the counting of lattice points on the first lying spherical layers of the hexagonal lattice and their splitting into orbits of the dihedral group $\mathrm{{\cal D}_6}$.
In the next section we consider the construction of the Beltrami Flows associated with the first few of such layers.
\section{Beltrami Fields from spherical layers in the Hexagonal Lattice}
\label{HexaFildi}
In this section, as announced, by utilizing the algorithm outlined in section \ref{algoritmo} we construct the Arnold-like Beltrami Flows associated with the first low lying spherical layers of the hexagonal lattice.
\subsection{The lowest lying layer of length 8 in the hexagonal lattice $\Lambda^\star_{Hex}$}
\begin{figure}[!hbt]
\begin{center}
\iffigs
\includegraphics[height=70mm]{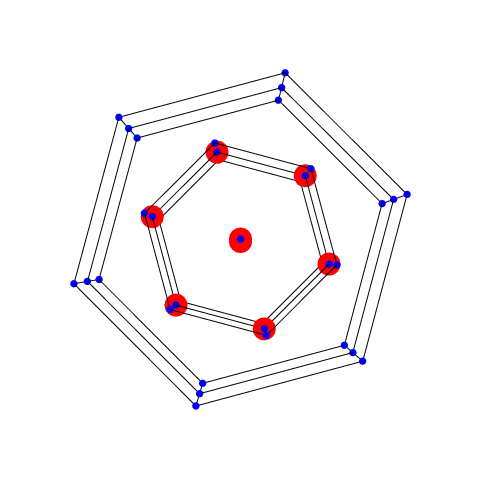}
\includegraphics[height=60mm]{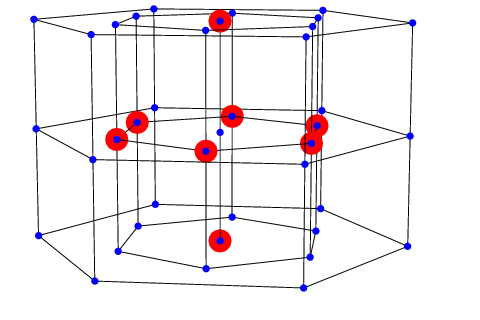}
\else
\end{center}
 \fi
\caption{\it  In this picture we show the location of the 8 points forming the lowest spherical layer in the momentum lattice ($\in \,\Lambda_{Hex}^\star$) whose squared radius is $r^2 \, = \, \frac{4}{3}$. These points split into two orbits under the $\mathrm{{\cal D}_6}$ group. The two points corresponding to the North and South Pole of the sphere form a $2$ orbit. The six points on the equator of the sphere, that are the vertices of an hexagon, form a $6$-orbit.}
\label{HexOrbita8}
 \iffigs
 \hskip 1cm \unitlength=1.1mm
 \end{center}
  \fi
\end{figure}

In fig.\ref{HexOrbita8} we show the location of the momentum lattice points forming the lowest lying spherical layer of length 8.
Under the action of the dihedral group $\mathrm{{\cal D}_6}$ these eight points are split into a $6$ orbit of type one and a $2$-orbit:
\begin{equation}\label{o6p1}
  \mathcal{O}^{(1)}_6 \, = \,  \left\{\begin{array}{llllll}
 \left\{-1,-\frac{1}{\sqrt{3}},0\right\} &
   \left\{-1,\frac{1}{\sqrt{3}},0\right\} &
   \left\{0,-\frac{2}{\sqrt{3}},0\right\} &
   \left\{0,\frac{2}{\sqrt{3}},0\right\} &
   \left\{1,-\frac{1}{\sqrt{3}},0\right\} &
   \left\{1,\frac{1}{\sqrt{3}},0\right\}
\end{array}
\right\}
\end{equation}
\begin{equation}\label{o6p1bis}
  \mathcal{O}_2 \, = \, \left\{ \begin{array}{ll}
 \left\{0,0,-\frac{2}{\sqrt{3}}\right\} &
   \left\{0,0,\frac{2}{\sqrt{3}}\right\}
\end{array}\right\}
\end{equation}
Implementing the construction algorithm we find the following vector field:
\begin{eqnarray}\label{orbit8vectorD6}
{\mathbf{V}^{(8)}}({\mathbf{r}}|\mathbf{F}) & = &\left\{ V_x\, , \, V_y \, , \, V_z\right\} \nonumber\\
V_x & = &  2 \cos \left(\Theta _3\right) F_1+\cos \left(\Theta
   _4\right) F_2-2 \sin \left(\Theta _3\right)
   F_3+\frac{\cos \left(\Theta _2\right)
   F_4}{\sqrt{3}}\nonumber\\
   &&-\frac{\cos \left(\Theta _1\right)
   F_5}{\sqrt{3}}+\sin \left(\Theta _4\right)
   F_6-\frac{1}{2} \sin \left(\Theta _2\right)
   F_7+\frac{1}{2} \sin \left(\Theta _1\right) F_8\nonumber\\
V_y & = &2 \sin \left(\Theta _3\right) F_1+2 \cos \left(\Theta
   _3\right) F_3+\cos \left(\Theta _2\right) F_4+\cos
   \left(\Theta _1\right) F_5\nonumber\\
   &&-\frac{1}{2} \sqrt{3} \sin
   \left(\Theta _2\right) F_7-\frac{1}{2} \sqrt{3} \sin
   \left(\Theta _1\right) F_8  \nonumber\\
V_z & = & -\sin \left(\Theta _4\right) F_2+\frac{2 \sin \left(\Theta
   _2\right) F_4}{\sqrt{3}}+\frac{2 \sin \left(\Theta
   _1\right) F_5}{\sqrt{3}}\nonumber\\
   &&+\cos \left(\Theta _4\right)
   F_6+\cos \left(\Theta _2\right) F_7+\cos \left(\Theta
   _1\right) F_8 \nonumber\\
\end{eqnarray}
where $F_i$ ($i\,=\,1,\dots\, ,8$) are real numbers and the angles $\Theta_i$ are the following ones:
  \begin{equation}\label{angoli8D6}
  \begin{array}{rcl}
 \Theta _1& = & -\frac{2}{3} \pi  \left(3 x+\sqrt{3} y\right)
   \\
 \Theta _2& = & 2 \pi  \left(\frac{y}{\sqrt{3}}-x\right) \\
 \Theta _3& = & -\frac{4 \pi  z}{\sqrt{3}} \\
 \Theta _4& = & -\frac{4 \pi  y}{\sqrt{3}}
\end{array}
  \end{equation}
From the explicit expression of the vector field in full analogy with eq.(\ref{gulag}) we construct the action of the dihedral group $\mathrm{{\cal D}_6}$ on the parameter vector $\mathbf{F}$. Writing:
\begin{equation}\label{gulagD6}
    \forall \, \gamma \, \in \, \mathrm{{\cal D}_6} \quad  : \quad \gamma^{-1} \cdot \mathbf{V}^{(8)}\left (\gamma \cdot \mathbf{r}|\mathbf{F}\right) \, = \, {\mathbf{V}^{(8)}}\left (\mathbf{r}| \, \mathfrak{R}^{(8)}[\gamma ]\cdot \mathbf{F}\right )
\end{equation}
we obtain the form of the reducible representation $\mathfrak{R}^{(8)}[\gamma ]$ which is completely specified giving the images $\mathfrak{R}^{(8)}[A]$, and $\mathfrak{R}^{(8)}[B]$ of the two group generators.
Explicitly we find:
\begin{equation}\label{ABin8orbitD6}
    \mathfrak{R}^{(8)}[A] \, = \, \left(
\begin{array}{llllllll}
 \frac{1}{2} & 0 & -\frac{\sqrt{3}}{2} & 0 & 0 & 0 & 0 & 0
   \\
 0 & 0 & 0 & 0 & -\frac{2}{\sqrt{3}} & 0 & 0 & 0 \\
 \frac{\sqrt{3}}{2} & 0 & \frac{1}{2} & 0 & 0 & 0 & 0 & 0 \\
 0 & \frac{\sqrt{3}}{2} & 0 & 0 & 0 & 0 & 0 & 0 \\
 0 & 0 & 0 & 1 & 0 & 0 & 0 & 0 \\
 0 & 0 & 0 & 0 & 0 & 0 & 0 & 1 \\
 0 & 0 & 0 & 0 & 0 & 1 & 0 & 0 \\
 0 & 0 & 0 & 0 & 0 & 0 & 1 & 0
\end{array}
\right)\; ; \; \mathfrak{R}^{(8)}[B]\, = \,
\left(
\begin{array}{llllllll}
 -1 & 0 & 0 & 0 & 0 & 0 & 0 & 0 \\
 0 & -1 & 0 & 0 & 0 & 0 & 0 & 0 \\
 0 & 0 & 1 & 0 & 0 & 0 & 0 & 0 \\
 0 & 0 & 0 & 0 & 1 & 0 & 0 & 0 \\
 0 & 0 & 0 & 1 & 0 & 0 & 0 & 0 \\
 0 & 0 & 0 & 0 & 0 & -1 & 0 & 0 \\
 0 & 0 & 0 & 0 & 0 & 0 & 0 & -1 \\
 0 & 0 & 0 & 0 & 0 & 0 & -1 & 0
\end{array}
\right)
\end{equation}
Retrieving from the above generators all the group elements and in particular a representative for each of the conjugacy classes, we can easily compute their traces and in this way establish the character vector of this representation. We get:
\begin{equation}\label{caratter8D6}
    \chi\left[\mathbf{8}\right]\, = \, \{8, 1, -1, -2, -2, 0\}
\end{equation}
The multiplicity vector is:
\begin{equation}\label{caratter8D6bis}
    \mathrm{m}\left[\mathbf{8}\right]\, = \, \{0, 0, 1, 1, 2, 1\}
\end{equation}
implying that the 8 dimensional parameter space decomposes into a $D_3$ plus a $D_4$ plus two $D_5$ and one $D_6$ representations.
The corresponding irreducible Beltrami Fields are easily constructed.
\paragraph{Irreducible Beltrami field in the $D_3$ representation}
\begin{eqnarray}\label{orbit8vectorD3}
{\mathbf{V}^{(8|D_3)}}({\mathbf{r}}|\mathbf{F}) & = &\left\{ V_x\, , \, V_y \, , \, V_z\right\} \nonumber\\
& = &\left(
\begin{array}{l}
 \frac{1}{2} \left(-\cos \left(\Theta _1\right)-\cos
   \left(\Theta _2\right)+2 \cos \left(\Theta
   _4\right)\right) \\
 \frac{1}{2} \sqrt{3} \left(\cos \left(\Theta _1\right)-\cos
   \left(\Theta _2\right)\right) \\
 \sin \left(\Theta _1\right)-\sin \left(\Theta
   _2\right)-\sin \left(\Theta _4\right)
\end{array}
\right)
\end{eqnarray}
\paragraph{Irreducible Beltrami field in the $D_4$ representation}
\begin{eqnarray}\label{orbit8vectorD4}
{\mathbf{V}^{(8|D_4)}}({\mathbf{r}}|\mathbf{F}) & = &\left\{ V_x\, , \, V_y \, , \, V_z\right\} \nonumber\\
& = &\left(
\begin{array}{l}
 \frac{1}{2} \left(\sin \left(\Theta _1\right)-\sin
   \left(\Theta _2\right)+2 \sin \left(\Theta
   _4\right)\right) \\
 -\frac{1}{2} \sqrt{3} \left(\sin \left(\Theta
   _1\right)+\sin \left(\Theta _2\right)\right) \\
 \cos \left(\Theta _1\right)+\cos \left(\Theta
   _2\right)+\cos \left(\Theta _4\right)
\end{array}
\right)
\end{eqnarray}
\paragraph{Irreducible Beltrami field in the $D_{5a}$ representation}
\begin{eqnarray}\label{orbit8vectorD5a}
{\mathbf{V}^{(8|D_{5a})}}({\mathbf{r}}|\{A,B\}) & = &\left\{ V_x\, , \, V_y \, , \, V_z\right\} \nonumber\\
& = &\left(
\begin{array}{l}
 2 A \cos \left(\Theta _3\right)-2 B \sin \left(\Theta
   _3\right) \\
 2 \left(B \cos \left(\Theta _3\right)+A \sin \left(\Theta
   _3\right)\right) \\
 0
\end{array}
\right)
\end{eqnarray}
\paragraph{Irreducible Beltrami field in the $D_{5b}$ representation}
\begin{eqnarray}\label{orbit8vectorD5b}
{\mathbf{V}^{(8|D_{5b})}}({\mathbf{r}}|\{A,B\}) & = &\left\{ V_x\, , \, V_y \, , \, V_z\right\} \nonumber\\
& = &\left(
\begin{array}{l}
 -B \cos \left(\Theta _1\right)+A \cos \left(\Theta
   _2\right)+2 (A-B) \cos \left(\Theta _4\right) \\
 \sqrt{3} \left(B \cos \left(\Theta _1\right)+A \cos
   \left(\Theta _2\right)\right) \\
 2 \left(B \sin \left(\Theta _1\right)+A \sin \left(\Theta
   _2\right)+(B-A) \sin \left(\Theta _4\right)\right)
\end{array}
\right)
\end{eqnarray}
\paragraph{Irreducible Beltrami field in the $D_6$ representation}
\begin{eqnarray}\label{orbit8vectorD6bis}
{\mathbf{V}^{(8|D_6)}}({\mathbf{r}}|\{A,B\}) & = &\left\{ V_x\, , \, V_y \, , \, V_z\right\} \nonumber\\
& = &\left(
\begin{array}{l}
 \frac{1}{2} \sqrt{3} B \left(\sin \left(\Theta
   _1\right)-\sin \left(\Theta _2\right)-4 \sin \left(\Theta
   _4\right)\right) \\
 -\frac{3}{2} B \left(\sin \left(\Theta _1\right)+\sin
   \left(\Theta _2\right)\right) \\
 \sqrt{3} B \left(\cos \left(\Theta _1\right)+\cos
   \left(\Theta _2\right)-2 \cos \left(\Theta
   _4\right)\right)
\end{array}
\right)
\end{eqnarray}
Let us now observe that the only angle dependent on the vertical coordinate $z$ is $\Theta_3$ and that all the above irreducible Beltrami fields, except, $D_{4a}$ are independent from $\Theta_3$. Hence all these irreducible Beltrami fields reduce to two dimensional planar systems that are presumably integrable systems.  This is geometrically understandable since the direct sum of all these representations has dimension 6 and reproduces the contribution of the points in orbit $\mathcal{O}_6^{(1)}$ which is just planar. Hence we can easily conjecture
that for all orbits of type $\mathcal{O}_6^{(1)}$ we have the decomposition:
\begin{equation}\label{decompus6one}
\mathcal{O}_6^{(1)} \, \simeq \,   D_2 \oplus   D_3 \oplus D_{5b} \oplus D_5
\end{equation}
each irreducible component being a planar system.
\par
The representation $D_{5a}$ on the other hand corresponds to the contribution of the points in the $2$ orbit and leads to a trivial differential system which is immediately integrated. The coordinate $z=z_0$ is constant in time and the coordinates $x,y$ are linear functions of time.
\par
This example strongly indicates that in the hexagonal lattices the only non trivial systems are those related to orbits of length $12$ and type two, three and four as we have already advocated.
\subsection{The lowest lying orbit of length 12 in the hexagonal lattice $\Lambda^\star_{Hex}$}
\begin{figure}[!hbt]
\begin{center}
\iffigs
\includegraphics[height=70mm]{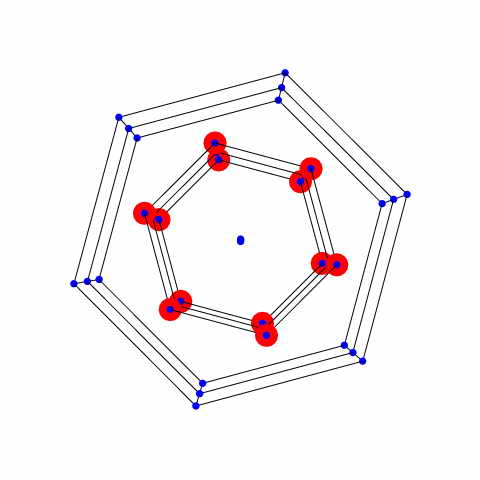}
\includegraphics[height=60mm]{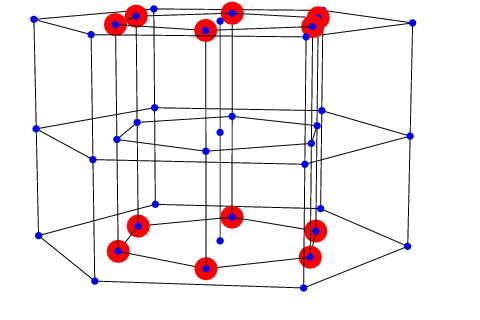}
\else
\end{center}
 \fi
\caption{\it  In this picture we show the location of the 12 points forming the lowest lying $\mathrm{{\cal D}_6}$ orbit in the momentum lattice ($\in \,\Lambda_{Hex}^\star$). They are located on a sphere of squared radius $r^2 \, = \, \frac{8}{3}$.}
\label{HexOrbita12}
 \iffigs
 \hskip 1cm \unitlength=1.1mm
 \end{center}
  \fi
\end{figure}

In fig.\ref{HexOrbita12} we show the location of the momentum lattice points forming the lowest lying spherical layer of length 12 under the action of the dihedral group $\mathrm{{\cal D}_6}$, this layer is irreducible.
Using the standard method we obtain the following general solution of the Beltrami equation depending on $12$ parameters:
\begin{eqnarray}\label{orbit12vectorD6}
{\mathbf{V}^{(12)}}({\mathbf{r}}|\mathbf{F}) & = &\left\{ V_x\, , \, V_y \, , \, V_z\right\} \nonumber\\
V_x & = &  \cos \left(\Theta _6\right) F_1+\cos \left(\Theta _5\right)
   F_2+\sqrt{2} \sin \left(\Theta _6\right) F_7+\sqrt{2}
   \sin \left(\Theta _5\right) F_8+\frac{1}{12} \sin
   \left(\Theta _4\right) \left(6 \sqrt{2} F_3-3 \sqrt{2}
   F_9\right)\nonumber\\
   &&+\frac{1}{12} \cos \left(\Theta _4\right)
   \left(4 \sqrt{3} F_3+8 \sqrt{3} F_9\right)\nonumber\\
   &&+\frac{1}{12}
   \sin \left(\Theta _3\right) \left(-6 \sqrt{2} F_4-3
   \sqrt{2} F_{10}\right)+\frac{1}{12} \cos \left(\Theta
   _3\right) \left(4 \sqrt{3} F_4-8 \sqrt{3}
   F_{10}\right)\nonumber\\
   &&+\frac{1}{12} \sin \left(\Theta _2\right)
   \left(6 \sqrt{2} F_5+3 \sqrt{2}
   F_{11}\right)+\frac{1}{12} \cos \left(\Theta _2\right)
   \left(8 \sqrt{3} F_{11}-4 \sqrt{3}
   F_5\right)\nonumber\\
   &&+\frac{1}{12} \sin \left(\Theta _1\right)
   \left(3 \sqrt{2} F_{12}-6 \sqrt{2}
   F_6\right)+\frac{1}{12} \cos \left(\Theta _1\right)
   \left(-4 \sqrt{3} F_6-8 \sqrt{3} F_{12}\right)\nonumber\\
V_y & = &-\frac{\sin \left(\Theta _6\right) F_1}{\sqrt{2}}+\frac{\sin
   \left(\Theta _5\right) F_2}{\sqrt{2}}\nonumber\\
   &&+\cos \left(\Theta
   _4\right) F_3+\cos \left(\Theta _3\right) F_4+\cos
   \left(\Theta _2\right) F_5+\cos \left(\Theta _1\right)
   F_6\nonumber\\
   &&+\cos \left(\Theta _6\right) F_7-\cos \left(\Theta
   _5\right) F_8+\sin \left(\Theta _4\right)
   \left(-\frac{F_3}{\sqrt{6}}-\frac{7 F_9}{2
   \sqrt{6}}\right)\nonumber\\
   &&+\sin \left(\Theta _3\right)
   \left(\frac{F_4}{\sqrt{6}}-\frac{7 F_{10}}{2
   \sqrt{6}}\right)+\sin \left(\Theta _2\right)
   \left(\frac{F_5}{\sqrt{6}}-\frac{7 F_{11}}{2
   \sqrt{6}}\right)+\sin \left(\Theta _1\right)
   \left(-\frac{F_6}{\sqrt{6}}-\frac{7 F_{12}}{2
   \sqrt{6}}\right)\nonumber\\
V_z & = & -\frac{\sin \left(\Theta _6\right) F_1}{\sqrt{2}}-\frac{\sin
   \left(\Theta _5\right) F_2}{\sqrt{2}}+\cos \left(\Theta
   _6\right) F_7\nonumber\\
   &&+\cos \left(\Theta _5\right) F_8+\cos
   \left(\Theta _4\right) F_9+\frac{1}{6} \sin \left(\Theta
   _4\right) \left(2 \sqrt{6} F_3+\sqrt{6} F_9\right)\nonumber\\
   &&+\cos
   \left(\Theta _3\right) F_{10}+\frac{1}{6} \sin
   \left(\Theta _3\right) \left(2 \sqrt{6} F_4-\sqrt{6}
   F_{10}\right)\nonumber\\
   &&+\cos \left(\Theta _2\right)
   F_{11}+\frac{1}{6} \sin \left(\Theta _2\right) \left(2
   \sqrt{6} F_5-\sqrt{6} F_{11}\right)\nonumber\\
   &&+\cos \left(\Theta
   _1\right) F_{12}+\frac{1}{6} \sin \left(\Theta _1\right)
   \left(2 \sqrt{6} F_6+\sqrt{6} F_{12}\right) \nonumber\\
\end{eqnarray}
where $F_i$ ($i\,=\,1,\dots\, ,8$) are real numbers and the angles $\Theta_i$ are the following ones:
\begin{equation}\label{thettiD6}
    \begin{array}{lll}
 \Theta _1\, = \,  -\frac{2}{3} \pi  \left(3 x+\sqrt{3} (y+2
   z)\right)
   &
   \Theta _2\, = \,  -\frac{2}{3} \pi  \left(3
   x+\sqrt{3} (y-2 z)\right)
   & \Theta _3\, = \,  -\frac{2}{3} \pi
    \left(3 x-\sqrt{3} (y-2 z)\right)
    \\
     \Theta _4\, = \,
   \frac{2}{3} \pi  \left(\sqrt{3} (y+2 z)-3 x\right)
   &
   \Theta _5\, = \,  -\frac{4 \pi  (y+z)}{\sqrt{3}}
   & \Theta_6\, = \,  \frac{4 \pi  (z-y)}{\sqrt{3}}
\end{array}
\end{equation}
The $12$-dimensional representation of $\mathrm{{\cal D}_6}$ in parameter space is also easily constructed and we find:
\begin{equation}\label{Ain12orbitD6}
    \mathfrak{R}^{(12)}[A] \, = \, \left(
\begin{array}{llllllllllll}
 0 & 0 & 0 & 0 & -\frac{2}{\sqrt{3}} & 0 & 0 & 0 & 0 & 0 &
   \frac{1}{\sqrt{3}} & 0 \\
 0 & 0 & 0 & 0 & 0 & -\frac{2}{\sqrt{3}} & 0 & 0 & 0 & 0 & 0
   & -\frac{1}{\sqrt{3}} \\
 0 & \frac{\sqrt{3}}{2} & 0 & 0 & 0 & 0 & 0 & -\frac{1}{2} &
   0 & 0 & 0 & 0 \\
 \frac{\sqrt{3}}{2} & 0 & 0 & 0 & 0 & 0 & \frac{1}{2} & 0 &
   0 & 0 & 0 & 0 \\
 0 & 0 & 1 & 0 & 0 & 0 & 0 & 0 & 1 & 0 & 0 & 0 \\
 0 & 0 & 0 & 1 & 0 & 0 & 0 & 0 & 0 & -1 & 0 & 0 \\
 0 & 0 & 0 & 0 & 0 & 0 & 0 & 0 & 0 & 0 & 1 & 0 \\
 0 & 0 & 0 & 0 & 0 & 0 & 0 & 0 & 0 & 0 & 0 & 1 \\
 0 & 0 & 0 & 0 & 0 & 0 & 0 & 1 & 0 & 0 & 0 & 0 \\
 0 & 0 & 0 & 0 & 0 & 0 & 1 & 0 & 0 & 0 & 0 & 0 \\
 0 & 0 & 0 & 0 & 0 & 0 & 0 & 0 & 1 & 0 & 0 & 0 \\
 0 & 0 & 0 & 0 & 0 & 0 & 0 & 0 & 0 & 1 & 0 & 0
\end{array}
\right)
\end{equation}
\begin{equation}
\label{Bin12orbitD6}
 \mathfrak{R}^{(12)}[B]\, = \, \left(
\begin{array}{llllllllllll}
 0 & -1 & 0 & 0 & 0 & 0 & 0 & 0 & 0 & 0 & 0 & 0 \\
 -1 & 0 & 0 & 0 & 0 & 0 & 0 & 0 & 0 & 0 & 0 & 0 \\
 0 & 0 & 0 & 0 & 1 & 0 & 0 & 0 & 0 & 0 & 0 & 0 \\
 0 & 0 & 0 & 0 & 0 & 1 & 0 & 0 & 0 & 0 & 0 & 0 \\
 0 & 0 & 1 & 0 & 0 & 0 & 0 & 0 & 0 & 0 & 0 & 0 \\
 0 & 0 & 0 & 1 & 0 & 0 & 0 & 0 & 0 & 0 & 0 & 0 \\
 0 & 0 & 0 & 0 & 0 & 0 & 0 & -1 & 0 & 0 & 0 & 0 \\
 0 & 0 & 0 & 0 & 0 & 0 & -1 & 0 & 0 & 0 & 0 & 0 \\
 0 & 0 & 0 & 0 & 0 & 0 & 0 & 0 & 0 & 0 & -1 & 0 \\
 0 & 0 & 0 & 0 & 0 & 0 & 0 & 0 & 0 & 0 & 0 & -1 \\
 0 & 0 & 0 & 0 & 0 & 0 & 0 & 0 & -1 & 0 & 0 & 0 \\
 0 & 0 & 0 & 0 & 0 & 0 & 0 & 0 & 0 & -1 & 0 & 0
\end{array}
\right)
\end{equation}
Retrieving from the above generators all the group elements and in particular a representative for each of the conjugacy classes, we can easily compute their traces and in this way establish the character vector of this representation. We get:
\begin{equation}\label{caratter8D6tris}
    \chi\left[\mathbf{12}\right]\, = \, \{12, 0, 0, 0, 0, 0\}
\end{equation}
The multiplicity vector is:
\begin{equation}\label{caratter8D6quadris}
    \mathrm{m}\left[\mathbf{12}\right]\, = \, \{1, 1, 1, 1, 2, 2\}
\end{equation}
implying that the $12$-dimensional parameter space decomposes into a $D_1$ (invariant Beltrami vector field) plus a $D_2$ plus a $D_3$ plus two $D_5$ and two $D_6$ representations. We present the form of the Beltrami vector fields in the various representations:
\paragraph{Irreducible Beltrami field in the $D_1$ representation}
\begin{eqnarray}\label{orbit12vectorD1}
{\mathbf{V}^{(12|D_1)}}({\mathbf{r}}|\mathbf{F}) & = &\left\{ V_x\, , \, V_y \, , \, V_z\right\} \nonumber\\
V_x & = &\frac{1}{4} \left(-2 \cos \left(\Theta _1\right)+2 \cos
   \left(\Theta _2\right)+2 \cos \left(\Theta _3\right)-2
   \cos \left(\Theta _4\right)\right.\nonumber\\
   &&\left.-4 \cos \left(\Theta
   _5\right)+4 \cos \left(\Theta _6\right)-\sqrt{6} \sin
   \left(\Theta _1\right)-\sqrt{6} \sin \left(\Theta
   _2\right)-\sqrt{6} \sin \left(\Theta _3\right)-\sqrt{6}
   \sin \left(\Theta _4\right)\right) \nonumber\\
V_y & = &\frac{1}{4} \left(2 \sqrt{3} \cos \left(\Theta _1\right)-2
   \sqrt{3} \cos \left(\Theta _2\right)+2 \sqrt{3} \cos
   \left(\Theta _3\right)-2 \sqrt{3} \cos \left(\Theta
   _4\right)\right.\nonumber\\
   &&\left.-\sqrt{2} \sin \left(\Theta _1\right)-\sqrt{2}
   \sin \left(\Theta _2\right)+\sqrt{2} \sin \left(\Theta
   _3\right)+\sqrt{2} \sin \left(\Theta _4\right)-2 \sqrt{2}
   \sin \left(\Theta _5\right)-2 \sqrt{2} \sin \left(\Theta
   _6\right)\right)\nonumber\\
V_z & = &\frac{\sin \left(\Theta _1\right)-\sin \left(\Theta
   _2\right)+\sin \left(\Theta _3\right)-\sin \left(\Theta
   _4\right)+\sin \left(\Theta _5\right)-\sin \left(\Theta
   _6\right)}{\sqrt{2}}\nonumber\\
\end{eqnarray}
\paragraph{Irreducible Beltrami field in the $D_2$ representation}
\begin{eqnarray}\label{orbit12vectorD2}
{\mathbf{V}^{(12|D_2)}}({\mathbf{r}}|\mathbf{F}) & = &\left\{ V_x\, , \, V_y \, , \, V_z\right\} \nonumber\\
V_x & = &\frac{1}{2} \left(-\sqrt{3} \cos \left(\Theta
   _1\right)-\sqrt{3} \cos \left(\Theta _2\right)+\sqrt{3}
   \cos \left(\Theta _3\right)\right.\nonumber\\
   &&\left.+\sqrt{3} \cos \left(\Theta
   _4\right)+\sqrt{2} \sin \left(\Theta _1\right)-\sqrt{2}
   \sin \left(\Theta _2\right)\right.\nonumber\\
   &&\left.+\sqrt{2} \sin \left(\Theta
   _3\right)-\sqrt{2} \sin \left(\Theta _4\right)-2 \sqrt{2}
   \sin \left(\Theta _5\right)+2 \sqrt{2} \sin \left(\Theta
   _6\right)\right) \nonumber\\
V_y & = &\frac{1}{2} \left(-\cos \left(\Theta _1\right)-\cos
   \left(\Theta _2\right)-\cos \left(\Theta _3\right)-\cos
   \left(\Theta _4\right)\right.\nonumber\\
   &&\left.+2 \cos \left(\Theta _5\right)+2
   \cos \left(\Theta _6\right)-\sqrt{6} \sin \left(\Theta
   _1\right)\right.\nonumber\\
   &&\left.+\sqrt{6} \sin \left(\Theta _2\right)+\sqrt{6}
   \sin \left(\Theta _3\right)-\sqrt{6} \sin \left(\Theta
   _4\right)\right)\nonumber\\
V_z & = &\cos \left(\Theta _1\right)-\cos \left(\Theta _2\right)-\cos
   \left(\Theta _3\right)+\cos \left(\Theta _4\right)-\cos
   \left(\Theta _5\right)+\cos \left(\Theta _6\right)\nonumber\\
\end{eqnarray}
\paragraph{Irreducible Beltrami field in the $D_3$ representation}
\begin{eqnarray}\label{orbit12vectorD3}
{\mathbf{V}^{(12|D_3)}}({\mathbf{r}}|\mathbf{F}) & = &\left\{ V_x\, , \, V_y \, , \, V_z\right\} \nonumber\\
V_x & = &\frac{1}{4} \left(-2 \cos \left(\Theta _1\right)-2 \cos
   \left(\Theta _2\right)-2 \cos \left(\Theta _3\right)-2
   \cos \left(\Theta _4\right)\right.\nonumber\\
   &&\left.+4 \cos \left(\Theta
   _5\right)+4 \cos \left(\Theta _6\right)-\sqrt{6} \sin
   \left(\Theta _1\right)+\sqrt{6} \sin \left(\Theta
   _2\right)+\sqrt{6} \sin \left(\Theta _3\right)-\sqrt{6}
   \sin \left(\Theta _4\right)\right)\nonumber\\
V_y & = &\frac{1}{4} \left(2 \sqrt{3} \cos \left(\Theta _1\right)+2
   \sqrt{3} \cos \left(\Theta _2\right)-2 \sqrt{3} \cos
   \left(\Theta _3\right)-2 \sqrt{3} \cos \left(\Theta
   _4\right)\right.\nonumber\\
   &&\left.-\sqrt{2} \sin \left(\Theta _1\right)+\sqrt{2}
   \sin \left(\Theta _2\right)-\sqrt{2} \sin \left(\Theta
   _3\right)+\sqrt{2} \sin \left(\Theta _4\right)+2 \sqrt{2}
   \sin \left(\Theta _5\right)-2 \sqrt{2} \sin \left(\Theta
   _6\right)\right)\nonumber\\
V_z & = &\frac{\sin \left(\Theta _1\right)+\sin \left(\Theta
   _2\right)-\sin \left(\Theta _3\right)-\sin \left(\Theta
   _4\right)-\sin \left(\Theta _5\right)-\sin \left(\Theta
   _6\right)}{\sqrt{2}}\nonumber\\
\end{eqnarray}
\paragraph{Irreducible Beltrami field in the $D_4$ representation}
\begin{eqnarray}\label{orbit12vectorD4}
{\mathbf{V}^{(12|D_4)}}({\mathbf{r}}|\mathbf{F}) & = &\left\{ V_x\, , \, V_y \, , \, V_z\right\} \nonumber\\
V_x & = &\frac{1}{2} \left(-\sqrt{3} \cos \left(\Theta
   _1\right)+\sqrt{3} \cos \left(\Theta _2\right)-\sqrt{3}
   \cos \left(\Theta _3\right)\right.\nonumber\\
   &&\left.+\sqrt{3} \cos \left(\Theta
   _4\right)+\sqrt{2} \sin \left(\Theta _1\right)+\sqrt{2}
   \sin \left(\Theta _2\right)\right.\nonumber\\
   &&\left.-\sqrt{2} \sin \left(\Theta
   _3\right)-\sqrt{2} \sin \left(\Theta _4\right)+2 \sqrt{2}
   \sin \left(\Theta _5\right)+2 \sqrt{2} \sin \left(\Theta
   _6\right)\right)\nonumber\\
V_y & = &\frac{1}{2} \left(-\cos \left(\Theta _1\right)+\cos
   \left(\Theta _2\right)+\cos \left(\Theta _3\right)-\cos
   \left(\Theta _4\right)\right.\nonumber\\
   &&\left.-2 \cos \left(\Theta _5\right)+2
   \cos \left(\Theta _6\right)-\sqrt{6} \sin \left(\Theta
   _1\right)-\sqrt{6} \sin \left(\Theta _2\right)-\sqrt{6}
   \sin \left(\Theta _3\right)-\sqrt{6} \sin \left(\Theta
   _4\right)\right)\nonumber\\
V_z & = &\cos \left(\Theta _1\right)+\cos \left(\Theta _2\right)+\cos
   \left(\Theta _3\right)+\cos \left(\Theta _4\right)+\cos
   \left(\Theta _5\right)+\cos \left(\Theta _6\right)\nonumber\\
\end{eqnarray}
For the representations $D_5$ and $D_6$ we observe that the $2$ irreducible representations of $\mathrm{{\cal D}_6}$ are complex
so that in the real field, not surprisingly can happen that we cannot separate the $4$-dimensional $D_5$ or $D_6$ space into two orthogonal irreducible subspaces. What actually happens is that in these spaces there is an invariant real two dimensional subspace but its orthogonal complement is not invariant. So it is better to keep four parameters in each of these spaces.
\paragraph{Beltrami field in the $D_5$ representation}
Since the formulae in this case become very big we just present the projection on the $D_5$ representation as a substitution rule of the $12$ parameters $F_i$ in equation (\ref{orbit12vectorD6}) in terms of four independent parameters $(A,B,C;D)$. Explicitly we have:
\begin{equation}\label{D5progetta}
   \begin{array}{l}
 F_1\to \frac{2 A-2 B-C+2 D}{\sqrt{3}} \\
 F_2\to \frac{2 A-2 B-C+2 D}{\sqrt{3}} \\
 F_3\to A \\
 F_4\to A \\
 F_5\to B \\
 F_6\to B \\
 F_7\to C \\
 F_8\to -C \\
 F_9\to D-C \\
 F_{10}\to C-D \\
 F_{11}\to D \\
 F_{12}\to -D
\end{array}
\end{equation}
\paragraph{Beltrami field in the $D_6$ representation}
In the same way as above the $D_6$ Beltrami field can be obtained from eq.(\ref{orbit12vectorD6}) with the following substitution:
\begin{equation}\label{D6progetta}
   \left(
\begin{array}{l}
 F_1\to \frac{2 A+2 B-C-2 D}{\sqrt{3}} \\
 F_2\to \frac{-2 A-2 B+C+2 D}{\sqrt{3}} \\
 F_3\to A \\
 F_4\to -A \\
 F_5\to B \\
 F_6\to -B \\
 F_7\to C \\
 F_8\to C \\
 F_9\to -C-D \\
 F_{10}\to -C-D \\
 F_{11}\to D \\
 F_{12}\to D
\end{array}
\right)
\end{equation}
This concludes our short discussion of the hexagonal lattice case. By now it should be clear that the algorithm works for any lattice and that the properties of the corresponding Beltrami fields can be analyzed  analogously to what we did in depth for those living on the cubic lattice. This being emphasized it is time to turn to conclusions.
\section{Conclusions and Outlook}
\label{soklucio}
In this section we plan to summarize the results of this paper and put  them into a perspective for future work and deeper understanding.
\par
Yet, in order to do this some general considerations are necessary which might be particularly useful for those readers, we hope to have some of them, that do not belong to the community of scientists familiar with the ABC-flows and working on such topics. One of the two authors of this paper was in such a position when he was encouraged by the other author to get involved in the matters dealt with here. This author believes that the same shock that he experienced while entering this research field will probably affect any reader  coming from the community of theoretical theorists and that the same fundamental philosophical questions that confronted him at the beginning, remaining so far unanswered, will equally bewilder such a reader. We want  to emphasize that the roots of such a situation are in the very different mentality, we venture to say \textit{weltaunschaung}, characterizing the community of \textit{theoretical theorists} (in this category we include classical and quantum field theorists, supergravity and superstring theorists, general relativists and the like) and the community of mathematical physicists dealing with dynamical systems and related topics, ABC-flows being one prominent province in that realm. For the sake of shortness we refer hereafter  to such a community as to that of the \textit{dynamical theorists}.
\par
So let us first of all single out this difference in \textit{weltanschauung}.
\par
Both scientific communities share a high level of \textit{mathematization} and rely on sophisticated mathematical structures to formulate their constructions, often they use the very same ones, yet there is a drastic difference in their attitude toward Mathematics.
\par
In the community of \textit{theoretical theorists}, Mathematics is scanned in the quest for uniqueness, looking for such a priori conceptual choices that might single out a unique and distinctive mathematical framework able to capture the essence of a Physical Law  and reduce its understanding to First Principles. Theoretical theorists are reductionists: what is most important is not the mathematical structure \textit{per se}, that is utilized to describe a physical process, or its simplified idealized model, rather the conceptual category that singles out such a mathematical structure, in other words its Principle.
\par
In the community of \textit{dynamical theorists} Mathematics is mostly appreciated as a source of diversity, praise being obtained for each new solution of any given equation and a bottle of champagne being opened to welcome the discovery of any new model, for instance of any new Integrable Dynamical System. In the issues addressed in this paper the main goal is the opposite, namely to obtain \textit{the most non integrable systems}, yet the attitude does not change and any new mathematical example of such a type is equally appreciated and welcome.
\par
Two other essential differences in the weltanschauung of the two communities concern the delimitation of the \textit{metatheories} forming the play-ground of model-builders and the role, use and implementation of \textit{symmetries}. The differences in both these issues are quite relevant while trying to assess the character, relevance and perspective of the results we have achieved in this paper. So let us dwell on this point.
\par
In \textit{theoretical theory}, the relevant \textit{metatheories} have been established since long time. Up to the seventies of the last century  they have been \textit{Classical Lagrangian Field Theories} and their quantum descendants, namely the corresponding \textit{Quantum Field Theories} obtained from canonical quantization of the former. Principal distinction in the vast container of \textit{Lagrangian Field Theories} is the space-time symmetry: Lorentz invariance for all models of Particle Physics, Galilei invariance for some field-theoretical description of Newtonian or Statistical Mechanical Systems, the latter in many instances being interpreted as Wick rotations of the former. The efforts of constructive quantum field theorists allowed to establish on a firm basis some general results for these metatheories, the most important among which is the Spin-Statistics theorem. From the middle seventies of the last century, both experiments and theory developments pointed to a further restriction of the \textit{metatheory category} from generic \textit{Relativistic Lagrangian Field Theories} to \textit{Matter Coupled Gauge Theories}. Furthermore a deep conceptual revolution led to replace the concept of fundamental \textit{point-particles} with that of \textit{fundamental extended objects},  \textit{strings} and \textit{superstrings} being the first focus of attention, enlarged, after \textit{the second string revolution} in 1996, to \textit{$p$-branes}. In these developments symmetry played a fundamental conceptual role: bosonic space-time symmetry was enlarged to \textit{supersymmetry}, a necessary step in order to mix space-time and internal symmetries as proved by Coleman and Mandula \cite{colemanmandula}.
\par
Hence, on the basis of this definition of the \textit{metatheory play-ground}, whose change is possible, but only at the price of a conscious, detailed and motivated philosophical revision of some Fundamental Principles (a scientific revolution in the Kuhn sense), the model building work of \textit{theoretical theorists} is aimed  either at the construction of new instances of Lagrangian Field Theories in diverse dimensions, displaying various types of catalogued symmetries as  supersymmetry, conformal symmetry, special bosonic gauge-symmetries, or at the understanding  of general properties of the former, or, still further, at the derivation of new exact solutions of either new or old Field Theories. In all these procedures the role and the use of symmetry is well codified and falls into one of the following cases:
\begin{description}
  \item[A)] Symmetry $\mathrm{G}[\mathcal{L}]$ of the Lagrangian $\mathcal{L}$ and hence of the theory. The classification of possible symmetries of the Lagrangian typically amounts to a classification of theories inside the metatheory. The reduction is almost obtained. We have to invent the Principle that favors one symmetry more than another one.
  \item[B)] Symmetry $\mathrm{G}[\mathcal{S}]$ of a solution $\mathcal{S}$ of the lagrangian field equations which is a subgroup of the lagrangian symmetries: $\mathrm{G}[\mathcal{S}] \subset \mathrm{G}[\mathcal{L}]$. Classification of $\mathrm{G}[\mathcal{S}]$ typically amounts to a classification of solutions $\mathcal{S}$  and this becomes particularly relevant if solutions can be interpreted as vacua of the theory. An extra functional like the energy functional can be typically advocated to select one vacuum symmetry more than another. Here the primary example, with all its far reaching consequences, is the spontaneous symmetry breaking mechanism.
  \item[C)] Hidden symmetry which is  a special declination of case A) or B), when the symmetry of a lagrangian $\mathcal{L}$  or a solution $\mathcal{S}$  is significantly enhanced to a larger one with respect to the obvious one that dictates the construction principles of $\mathcal{L}$.
  \item[D)] Dynamical symmetry: when the irreducible unitary representations of some finite or infinite (super) Lie algebra $\mathbb{G}_D$ constitute the quantum states of a theory (inside the metatheory). In this case dynamics is completely reduced to algebra. Typical instance of this are the two-dimensional conformal field theories where the (super)-Virasoro algebra plays the role of $\mathbb{G}_D$.
\end{description}
\par
In the community of \textit{dynamical theorists} the landscape is much less defined and clarified. The metatheory reference frame is just the vast and almost all containing setup of \textit{dynamical systems}, \textit{i.e.} of first order differential equations, and the main conceptual categories are just those of Poissonian or generalized Poissonian structures. Integrability and non-integrability are the main pursued issues with almost no emphasis on the reduction of choices to First Principles. A vast and mathematically sophisticated literature deals with the construction of a plethora of models each of which mostly plays to role of a Leibniz monad. The role of symmetries in this vast literature is also episodical and not clearly categorized as done in the above list. Indeed what is missing is the attempt to link symmetry to a philosophically motivated selection principle.
\par
The typical attitude of \textit{dynamical theorists} is that any good \textit{mathematical result} will sooner or later find its place in physical theory and therefore it is worth pursuing. This is superficially very similar to the common belief of \textit{theoretical theorists} that all sound and elegant \textit{mathematical architectures}, including in this category all symmetries, have to be realized in the fabrics of Natural Law. The conceptual difference, however, is enormous and it is hidden in the semantic difference between \textit{result} and \textit{architecture}. What is missing in the attitude of \textit{dynamical theorists} is the reductionist tension toward a small set of Economic First Principles rich of consequences but also strongly selective in the sense that they encompass vast landscapes yet rule out many possibilities.
\par
Having spelled out these weltanschauung differences let us come to the case of the ABC-flows which, with their generalizations, constitute the main topics of the present paper.
\par
Leaving aside the issue of periodic boundary conditions, already addressed in the introduction, the main source of discomfort for one of the two authors and, possibly, for the theoretical theorist reader of this paper is that the starting point of the whole thing, namely Euler Equation (\ref{EulerusEqua}), is not a Lagrangian, it is just an equation. This means that the velocity field $\mathrm{U}(t)$ is not uniquely identified as a Lagrangian field and that any of its possible symmetries do not fall in category B) of the above list. This is not too much surprising for any relativist who remembers Einstein's words about the two sides of his field equations. The \textit{left hand side}, said Einstein, meaning the Einstein tensor formed from the metric, \textit{is written on pure marble}. The \textit{right hand side}, meaning the stress energy tensor, \textit{is written on deteriorable wood}. Indeed the stress-energy tensor $T_{\mu\nu}$ of perfect fluids, which contains the velocity field, the energy-density and the pressure, was, according to Einstein, just a modeling of our ignorance, being subject only to the kinematical and almost empty constraint of conservation $\nabla^\mu \, T_{\mu\nu} \, = \, 0$. The mission of theoretical physics was, according to Einstein, to transform the wood into marble by bringing the right hand side to the left, namely by geometrizing it. This is what is currently done in unified field theory models and in particular in models of inflation where the stress energy tensor is derived from a field-theoretical lagrangian. Now what is Euler equation if not the non-relativistic analogue of the stress-energy tensor conservation law? So we can similarly say that the fundamental equation of hydrodynamics (\ref{EulerusEqua}) is a modeling of our ignorance and a priority mission would be that of deriving its main ingredient, namely the velocity field  $\mathrm{U}(t)$ from some field-theoretical lagrangian. This is an extremely urgent question, specially in view of the role that symmetries have in this business.
\par
The relation between point A) and B) in the consideration of symmetries can be  provisionally amended by substituting  $\mathrm{G}[\mathcal{L}]$ of the missing lagrangian $\mathcal{L}$ with the group of symmetries of  the Euler equation (\ref{EulerusEqua}). On $\mathbb{R}^3$ this is just the full Euclidian group $\mathbb{E}^3$. In the case of a $T^3$ torus with a polarized constant metric (\ref{gmunu}) related with some chosen lattice $\Lambda$, it becomes the group defined in eq.(\ref{goticoG}), namely the semidirect product of the lattice point group  with the continuous translation group modulo the lattice $\Lambda$. Thus by setting:
\begin{equation}\label{Agruppo}
    \mathrm{G}[\mathcal{L}] \, \equiv \, \mathfrak{G}_\Lambda
\end{equation}
we can be in business with point A) of the above list.
\par
This point in the discussion offers to us the opportunity to stress the first and probably the \textbf{main of our new results}, namely the concept of  a \textit{Universal Classifying Group} $\mathfrak{GU}_{\Lambda}$. Relying on the suggestion of crystallographers who seek modifications of the lattice point groups, named Space-Groups,  by the inclusion of quantized translations of (\ref{quozienTra}) that cannot be eliminated by conjugation with elements of   $\mathfrak{T}^3$, we have advocated the following three points:
\begin{enumerate}
  \item Frobenius congruence classes define for all lattice $\Lambda$ an abelian subgroup
  \begin{equation}\label{quantotraslo}
     \mathbb{Z}_{k_1} \times \mathbb{Z}_{k_2} \times \mathbb{Z}_{k_3} \, \sim \, \mathfrak{TU}_\Lambda \, \subset \, \mathfrak{T}^3
  \end{equation}
 of quantized translations.
  \item The semidirect product:
  \begin{equation}\label{universopuntuto}
    \mathfrak{GU}_{\Lambda} \, = \, \mathfrak{P}_\Lambda \, \ltimes \, \mathfrak{TU}_\Lambda \,
  \end{equation}
  of the lattice point group  $\mathfrak{P}_\Lambda$ with the discretized translation group mentioned in eq.(\ref{quantotraslo}) constitutes a large discrete group that contains as proper subgroups all possible space groups of crystallography.
  \item The above defined group $ \mathfrak{GU}_{\Lambda}$,  named by us the Universal Classifying Group, is that apt to organize all solutions of the Beltrami equation (\ref{Beltrami}) into irreducible representations and by this token to classify them.
\end{enumerate}
Here comes the second and most severe point of discomfort both for one of the authors and for our hypothetical theoretical theorist reader but, at the same time, here comes also the opportunity to stress the \textbf{second main result of the present paper}.
\par
As emphasized in the introduction the main idea in the whole scientific landscape around ABC-flows  is given by Arnold theorem stating that the only velocity fields able to give rise to chaotic streamlines are those that satisfy Beltrami equation (\ref{Beltrami}). The effort to substantiate this result in a more topological and abstract way leads to the conception of \textit{contact structures}, \textit{contact one-forms} and their associated \textit{Reeb (like) fields}.
Hence the construction and classification of Beltrami vector fields is a main priority in this arena of mathematical hydrodynamics.
\par
Our main contribution, thoroughly developed in this paper, is the recognition that Beltrami equation (\ref{Beltrami}) is nothing else but the eigenvalue equation for a first order Laplace-Beltrami operator, the $\star_g \, \mathrm{d}$ operator that, on a compact Riemannian manifold, as the $T^3$ torus happens to be, has a discrete spectrum, the eigenfunctions being harmonics of the Universal Classifying Group $ \mathfrak{GU}_{\Lambda}$, that can be systematically constructed with a rather simple algorithm and fall into irreps of the same. Special solutions $\mathcal{S}$ can be characterized by invariance under subgroups:
\begin{equation}\label{cosmato}
    \mathrm{G}\left[ \mathcal{S} \right] \, \subset \, \mathfrak{GU}_{\Lambda}
\end{equation}
In this way we  are in business also with point B) of the above list of symmetry conceptions, yet the discomfort is related with the eigenvalue $\lambda$ in eq. (\ref{Beltrami}). Who is going to tell us the value of $\lambda$? If we do not have a theory from which Beltrami equation emerges as a field equation, then we do not have any reason to choose one or another of the possible eigenvalues. We are just in the Leibniz monad world. Every solution of the equation is equally admissible and we can just open as many bottles of champagne as there are eigenvectors and eigenfunctions. These are all the available generalizations of the ABC-flows. A priori it might seem that, since we have infinitely many eigenvalues, there are infinitely many eigenfunctions (with their moduli space) and this implies that we will get fully drunk. Yet the number of irreducible representations of a finite group, like $\mathfrak{GU}_{\Lambda}$, is finite, which sounds a warning that the collection of sparkling wine bottles should also be finite. Indeed, and this is \textbf{third of our main results}, we have proven in this paper that for the cubic lattice there are actually only $48$ different types of eigenfunctions which repeat themselves periodically. $48$ is a large number of bottles, but with some attention we can survive!
\par
Coming back to the issue of a theory able to produce Beltrami equation we observe the following two possibilities:
\begin{description}
  \item[Possibility One) ] Being a first order equation, Beltrami equation might be interpreted in the context of a field-theory as a sort of instantonic equation all of whose solutions are also solutions of the standard second order equations, although the reverse is not true. If we adopt such an ideology  we can easily single out a simple candidate for such a field theory in Euclidian three-dimensions. Let us identify the one-form $\Omega^{[\mathrm{U}]}$ of eq.(\ref{Omfildo}) with a gauge one-form:
      \begin{equation}\label{vanesso}
        \mathbf{A} \, \equiv \, \Omega^{[\mathrm{U}]}
      \end{equation}
 and let us consider the following action functional:
     \begin{eqnarray}\label{cosroe}
        \mathcal{A}& = & \int \, \mathcal{L} \quad ; \quad  \mathcal{L} \, = \, \alpha \, \mathbf{F} \wedge \star_g \, \mathbf{F} \, + \, \beta \, \mathbf{F}\wedge \mathbf{A}\nonumber\\
        \mathbf{F} & = & \mathrm{d} \mathbf{A}
     \end{eqnarray}
which describes a standard $\mathrm{U(1)}$  gauge theory with the addition of a Chern Simons term.  The second order field equation of this theory is:
\begin{equation}\label{barlume}
   \mathrm{ d}\,\star_g \, \mathbf{F} \, + \, \frac{\beta}{\alpha} \, \mathbf{F} \, = \, 0 \quad \Leftrightarrow \quad \mathrm{ d} \left( \star_g \, \mathrm{ d}\mathbf{A}  \, + \,  \frac{\beta}{\alpha} \, \mathbf{A} \right) \, = \, 0
\end{equation}
which is certainly satisfied if:
\begin{equation}\label{BeltramusTwo}
   \star_g \, \mathrm{ d}\mathbf{A}  \, + \,  \frac{\beta}{2 \alpha} \, \mathbf{A}  \, = \, 0
\end{equation}
 That above is indeed Beltrami equation with an eigenvalue $\lambda \, = \, - \,\frac{\beta}{\alpha}$  dictated by the ratio of the two coupling constant in the Lagrangian.
  \item[Possibility Two) ] Beltrami equation is just the field equation of the field theory. In this case we obtain the candidate Lagrangian by making the same identification as in eq.(\ref{vanesso}) and then writing:
      \begin{eqnarray}\label{artasersedue}
        \mathcal{A}& = & \int \, \mathcal{L} \quad ; \quad  \mathcal{L} \, = \, \beta \, \mathbf{A} \wedge \star_g \, \mathbf{A} \, + \, \alpha \, \mathbf{F}\wedge \mathbf{A}\nonumber\\
        \mathbf{F} & = & \mathrm{d} \mathbf{A}
     \end{eqnarray}
The above action describes a $\mathrm{U(1)}$ Chern Simons  gauge theory with the addition of a mass term for the gauge field. Such a mass term might be induced by a Brout-Englert-Higgs  mechanism of spontaneous symmetry breaking. In this case equation (\ref{BeltramusTwo}) is just the complete field equation.
\end{description}
The two mentioned possibilities are quite interesting and challenging in view of the $\mathrm{AdS_4}/\mathrm{CFT_3}$ correspondence which relates a supersymmetric Chern-Simons gauge theory on the boundary with a supergravity theory in the bulk of an anti-de Sitter space  $\mathrm{AdS_4}$ (for the most general form of supersymmetric Chern Simons theories in $d=3$ see \cite{Fabbri:1999ay}  and for several examples of  such theories derived from $\mathrm{AdS_4}/\mathrm{CFT_3}$ correspondence see \cite{Fre':1999xp},\cite{Fabbri:1999hw},\cite{Billo:2000zs},\cite{Fabbri:1999mk}). We plan to come back to such an issue in a forthcoming publication \cite{SCSArnold}. What we want to stress here is that adopting such a point of view one sticks to a Principle that rules out the plethora of Beltrami flows and the feasting with flows of champagne. The eigenvalue in Beltrami equation is a ratio of coupling constants appearing in the Lagrangian and one has to consider only those flows that fit to it. Furthermore it is to be hoped that the ratio $\lambda \, = \,- \,\frac{\beta}{2 \alpha}$ is fixed from other elements of the construction, for instance from the supergravity solution on $\mathrm{AdS_4} \times \text{something}$. In this case it would  obtain an interpretation in terms of First Principles.
\par
Let us now come to a further reason of discomfort in relation with symmetries. In the presentation of our results we have observed that several interesting Arnold-Beltrami Flows are obtained by decomposing the  irreducible representations of the Universal Classifying Group  into irreps of some of its subgroups $\mathrm{H}_i  \, \subset \, \mathfrak{GU}_{\Lambda}$. When an identity representation is available in the branching rules, we obtain an Arnold-Beltrami Flow invariant under the group $\mathrm{H}_i$.
In the existing literature on ABC-flows there is a general feeling  that flows with symmetries have a distinguished and more important role to play than other non invariant ones, yet nowhere such a role is spelled out in a clear way and there is no well established hierarchy of concepts for the interpretation of such symmetries. This is still another manifestation of the problems that the lack of a field theoretical basis for the flows does create.  The streamlines are solutions of non-linear equations and for this reason there is no superposition principle. On the other hand the equation that defines the flows, namely Beltrami equation, is linear and, as such, it leads to a superposition principle. If one is able to anchor such an equation to the firm ground of a field theory, then the moduli space of the solutions, namely the parameters that fill the irreducible representations of the Classifying Group $\mathfrak{GU}_{\Lambda}$, can be explored with standard techniques: as it happens in many  similar situations we should expect that the moduli-space points that correspond to an enhanced symmetry are in some sense singular points and they might dominate some appropriate path-integral.
\par
Thus we believe that all the conceptual problems we have pointed out can be addressed and solved only in the framework of some reasonable field theory for the velocity field $\mathrm{U}$. This is, in our opinion, the priority one issue of this research field.
\par
Finally in the vein of the above remarks let us come back to the discussion of eq.s (\ref{choucrut}) and (\ref{salsicciafresca}) of the introduction. We propose that in higher odd--dimensions $d \, = \,(2\,p \, +\,1)$, instead that by equation (\ref{salsicciafresca}), a \textit{generalized contact structure} be defined by a $p$-contact form $\alpha^{(p)}$ satisfying the condition:
\begin{equation}\label{palettus}
    \alpha^{(p)} \, \wedge \, \mathrm{d}\alpha^{(p)} \, \ne \, 0
\end{equation}
At every point of the manifold $\mathcal{M}_{2p+1}$  the kernel of the contact form is a subspace of the tangent space $\mathrm{T}_p\mathcal{M}$ of codimension $p$  rather than of codimension one. So, as in the classical three dimensional case, also in $2p+1$ dimensions the contact form defines a sub-bundle of the tangent bundle, yet with dimension of the fibre equal to $p+1$ rather than $2p$. The same ideas about maximal non integrability of such a bundle can apply, in the sense that its fibres can be prevented from being the tangent spaces of  any embedded hypersurface $\Sigma_{p+1}\subset M_{2p+1}$ of dimension $p+1$. We stress that the \textit{linear} Beltrami equation  for such generalized contact forms can be linked to supergravity field equations in particular for $p=3$ to the field equations of M-theory when one compactifies $\mathcal{M}_{11} \, = \, \mathcal{M}_{4} \times \mathcal{M}_{7}$. This is another issue that we plan to investigate.
\section*{Acknowledgements}
We would like to thank O.V. Teryaev for the collaboration
at an earlier stage of this investigation. The work of A.S. was partially supported by
the RFBR Grants No. 13-02-91330-NNIO-a and No. 13-02-90602-Arm-a.
\newpage
\part{The Appendices: tables of conjugacy classes, characters, branching rules and orbit splittings}
\appendix
\section{Description of the Universal Classifying Group $\mathrm{\mathrm{G_{1536}}}$ and of its subgroups}
\label{descriptio}
In this appendix we provide the list of elements of the Universal Classifying Group $\mathrm{\mathrm{G_{1536}}}$ and of all its subgroups that happen to be relevant in the constructions discussed in the main text. The most relevant piece of information provided here is the assembling of the group elements into conjugacy classes. This is done both for the Universal Classifying Group $\mathrm{\mathrm{G_{1536}}}$ and for each of its subgroups relevant to us. This arrangement is essential for the calculation of characters and for the decomposition of any representation into irreducible ones. The group elements are uniquely identified by their code:
\begin{equation}\label{garessio}
    \left \{\gamma, \ft{n_1}{2} , \, \ft{n_2}{2} , \, \ft{n_3}{2} \right \} \quad ; \quad \gamma \, \in \, \mathrm{O}_{24}
\end{equation}
where $\gamma $ is an element of the proper octahedral  group labeled according to the notation of  eq.(\ref{nomiOelemen}), while $n_1,n_2,n_3 \, \in \, \left\{0,1,2,3\right\}$ specify a translation. The action of the group element $ \left \{\gamma, \ft{n_1}{2} , \, \ft{n_2}{2} , \, \ft{n_3}{2} \right \}$ on the three Euclidian coordinates $\{x,y,z\}$ is:
\begin{equation}\label{ormea}
    \left \{\gamma, \ft{n_1}{2} , \, \ft{n_2}{2} , \, \ft{n_3}{2} \right \} \, : \, \{x,y,z\} \, \rightarrow \, \gamma \, \cdot \, \left\{x\, ,y\, z \right\}\, +\,
    \left\{ \ft{n_1}{4}, \ft{n_2}{2},  \ft{n_3}{4}\right\}
\end{equation}
The subgroups we explicitly describe in the present section are the following ones:
\begin{description}
  \item[A)] The chain of normal subgroups:
  \begin{equation}\label{pernicinormali}
    \mathrm{\mathrm{G_{1536}} }\, \rhd \, \mathrm{G_{768}} \, \rhd \, \mathrm{G_{256}} \, \rhd \, \mathrm{G_{128}} \, \rhd \, \mathrm{G_{64}}
  \end{equation}
  where $\mathrm{G_{64}} \,  \sim \, \mathbb{Z}_4 \, \times \, \mathbb{Z}_4 \, \times \, \mathbb{Z}_4$ is abelian and corresponds to the compactified translation group. The above chain leads to the following quotient groups:
  \begin{equation}\label{fagianirossi}
    \frac{\mathrm{\mathrm{G_{1536}} }}{\mathrm{G_{768}}} \, \sim \, \mathbb{Z}_2 \quad ; \quad \frac{\mathrm{G_{768} }}{\mathrm{G_{256}}} \, \sim \, \mathbb{Z}_3 \quad ; \quad \frac{\mathrm{G_{256} }}{\mathrm{G_{128}}} \, \sim \, \mathbb{Z}_2 \quad ; \quad \frac{\mathrm{G_{128} }}{\mathrm{G_{64}}} \, \sim \, \mathbb{Z}_2
  \end{equation}
  \item[B)] The  subgroup $\mathrm{G_{192}} \, \subset \, \mathrm{\mathrm{G_{1536}}}$, with respect to which the $6$-dimensional point orbit remains irreducible. $\mathrm{G_{192}}$ is not a normal subgroup but possesses a chain of normal subgroups that make it solvable and allow for the complete calculation of its irreps and character table:
       \begin{equation}\label{pernicirossine}
    \mathrm{G_{192} }\, \rhd \, \mathrm{G_{96}} \, \rhd \, \mathrm{G_{48}} \, \rhd \, \mathrm{G_{16}}
    \end{equation}
  \item[C)] The subgroup $\mathrm{GF_{192}} \, \subset \, \mathrm{\mathrm{G_{1536}}}$, with respect to which the $6$-dimensional point orbit splits into a pair $\mathbf{3} \oplus \mathbf{3}$ of irreducible  representation. The subgroups $\mathrm{G_{192}}$ and $\mathrm{GF_{192}}$ are isomorphic: $\mathrm{G_{192}}\, \sim \, \mathrm{GF_{192}}$, yet they are not conjugate to each other. Indeed the branching rules of $ \mathrm{\mathrm{G_{1536}} }$-irreps with respect to either $\mathrm{G_{192}}$ or $\mathrm{GF_{192}}$ are sometimes different. Just as $\mathrm{G_{192}} $ also $\mathrm{GF_{192}}$ is not a normal subgroup.
  \item[D)] The subgroup $\mathrm{Oh_{48}} \subset \mathrm{G_{192}} $ which is isomorphic to the extended octahedral  group $\mathrm{O_h}$ of crystallography.
  \item[E)] The subgroup $\mathrm{GS_{24}} \subset \mathrm{GF_{192}} $  which is isomorphic but not conjugate to the proper octahedral  group $\mathrm{O_{24}}$ (\textit{i.e.} the point group) and appears as the group of hidden symmetries of a parameterless Arnold-Beltrami flow derived both from the $6$-point orbit and from the lowest lying $24$-point orbit.
  \item[F)] The subgroup $\mathrm{GP_{24}} \subset \mathrm{G_{192}} $ which is not isomorphic to the  proper octahedral  group $\mathrm{O_{24}}$, having a different structure of conjugacy classes, and appears as a group of hidden symmetries of a parameterless Arnold-Beltrami flow derived from the $12$-point orbit.
   \item[G)] The subgroup $\mathrm{GK_{24}} \subset \mathrm{GF_{192}} $ which is isomorphic but not conjugate  to the  group $\mathrm{GP_{24}}$,  and also appears as a group of hidden symmetries of a parameterless Arnold-Beltrami flow derived from the $12$-point orbit. Both $\mathrm{GP_{24}}$  and $\mathrm{GK_{24}} $ are isomorphic to the abstract group $\mathrm{A}_2 \times \mathbb{Z}_2$
   \item[H)] The subgroup $\mathrm{GS_{32}} \subset \mathrm{G_{192}} $ which  appears as a group of hidden symmetries of a parameterless Arnold-Beltrami flow derived from the $8$-point orbit.
       \item[I)] The subgroup $\mathrm{GK_{32}} \subset \mathrm{GF_{192}} $ which is isomorphic but not conjugate to $\mathrm{GS_{32}}$ and   also  appears as a group of hidden symmetries of a parameterless Arnold-Beltrami flow derived from the $8$-point orbit.
\end{description}
\subsection{The Group $\mathrm{\mathrm{G_{1536}}}$}
\label{coniugato1536}
In this section we list all the elements of the space group $\mathrm{\mathrm{G_{1536}}}$, organized into their $37$ conjugacy classes.
\paragraph{Conjugacy class $\mathcal{C}_{1}\left(\mathrm{\mathrm{G_{1536}}}\right)$: $\#$ of elements = $1$}
\begin{equation}\label{G1536classe1}
\left \{
\begin{array}{llll}
 1_1 & 0 & 0 & 0
\end{array}\right\}
\end{equation}
\paragraph{Conjugacy class $\mathcal{C}_{2}\left(\mathrm{\mathrm{G_{1536}}}\right)$: $\#$ of elements = $1$}
\begin{equation}\label{G1536classe2}
\left\{\begin{array}{llll}
 1_1 & 1 & 1 & 1
\end{array}\right\}
\end{equation}
\paragraph{Conjugacy class $\mathcal{C}_{3}\left(\mathrm{\mathrm{G_{1536}}}\right)$: $\#$ of elements = $3$}
\begin{equation}\label{G1536classe3}
\begin{array}{lll}
 \left\{1_1,0,0,1\right\} & \left\{1_1,0,1,0\right\} &
   \left\{1_1,1,0,0\right\}
\end{array}
\end{equation}
\paragraph{Conjugacy class $\mathcal{C}_{4}\left(\mathrm{\mathrm{G_{1536}}}\right)$: $\#$ of elements = $3$}
\begin{equation}\label{G1536classe4}
\begin{array}{lll}
 \left\{1_1,0,1,1\right\} & \left\{1_1,1,0,1\right\} &
   \left\{1_1,1,1,0\right\}
\end{array}
\end{equation}
\paragraph{Conjugacy class $\mathcal{C}_{5}\left(\mathrm{\mathrm{G_{1536}}}\right)$: $\#$ of elements = $6$}
\begin{equation}\label{G1536classe5}
\begin{array}{lll}
 \left\{1_1,0,0,\frac{1}{2}\right\} &
   \left\{1_1,0,0,\frac{3}{2}\right\} &
   \left\{1_1,0,\frac{1}{2},0\right\} \\
 \left\{1_1,0,\frac{3}{2},0\right\} &
   \left\{1_1,\frac{1}{2},0,0\right\} &
   \left\{1_1,\frac{3}{2},0,0\right\}
\end{array}
\end{equation}
\paragraph{Conjugacy class $\mathcal{C}_{6}\left(\mathrm{\mathrm{G_{1536}}}\right)$: $\#$ of elements = $6$}
\begin{equation}\label{G1536classe6}
\begin{array}{lll}
 \left\{1_1,\frac{1}{2},1,1\right\} &
   \left\{1_1,1,\frac{1}{2},1\right\} &
   \left\{1_1,1,1,\frac{1}{2}\right\} \\
 \left\{1_1,1,1,\frac{3}{2}\right\} &
   \left\{1_1,1,\frac{3}{2},1\right\} &
   \left\{1_1,\frac{3}{2},1,1\right\}
\end{array}
\end{equation}
\paragraph{Conjugacy class $\mathcal{C}_{7}\left(\mathrm{\mathrm{G_{1536}}}\right)$: $\#$ of elements = $8$}
{\scriptsize
\begin{equation}\label{G1536classe7}
\begin{array}{llllllll}
 \left\{1_1,\frac{1}{2},\frac{1}{2},\frac{1}{2}\right\} &
   \left\{1_1,\frac{1}{2},\frac{1}{2},\frac{3}{2}\right\}
   &
   \left\{1_1,\frac{1}{2},\frac{3}{2},\frac{1}{2}\right\}
   &
   \left\{1_1,\frac{1}{2},\frac{3}{2},\frac{3}{2}\right\}
   &
   \left\{1_1,\frac{3}{2},\frac{1}{2},\frac{1}{2}\right\}
   &
   \left\{1_1,\frac{3}{2},\frac{1}{2},\frac{3}{2}\right\}
   &
   \left\{1_1,\frac{3}{2},\frac{3}{2},\frac{1}{2}\right\}
   &
   \left\{1_1,\frac{3}{2},\frac{3}{2},\frac{3}{2}\right\}
\end{array}
\end{equation}
}
\paragraph{Conjugacy class $\mathcal{C}_{8}\left(\mathrm{\mathrm{G_{1536}}}\right)$: $\#$ of elements = $12$}
\begin{equation}\label{G1536classe8}
\begin{array}{llll}
 \left\{1_1,0,\frac{1}{2},\frac{1}{2}\right\} &
   \left\{1_1,0,\frac{1}{2},\frac{3}{2}\right\} &
   \left\{1_1,0,\frac{3}{2},\frac{1}{2}\right\} &
   \left\{1_1,0,\frac{3}{2},\frac{3}{2}\right\} \\
 \left\{1_1,\frac{1}{2},0,\frac{1}{2}\right\} &
   \left\{1_1,\frac{1}{2},0,\frac{3}{2}\right\} &
   \left\{1_1,\frac{1}{2},\frac{1}{2},0\right\} &
   \left\{1_1,\frac{1}{2},\frac{3}{2},0\right\} \\
 \left\{1_1,\frac{3}{2},0,\frac{1}{2}\right\} &
   \left\{1_1,\frac{3}{2},0,\frac{3}{2}\right\} &
   \left\{1_1,\frac{3}{2},\frac{1}{2},0\right\} &
   \left\{1_1,\frac{3}{2},\frac{3}{2},0\right\}
\end{array}
\end{equation}
\paragraph{Conjugacy class $\mathcal{C}_{9}\left(\mathrm{\mathrm{G_{1536}}}\right)$: $\#$ of elements = $12$}
\begin{equation}\label{G1536classe9}
\begin{array}{llll}
 \left\{1_1,0,\frac{1}{2},1\right\} &
   \left\{1_1,0,1,\frac{1}{2}\right\} &
   \left\{1_1,0,1,\frac{3}{2}\right\} &
   \left\{1_1,0,\frac{3}{2},1\right\} \\
 \left\{1_1,\frac{1}{2},0,1\right\} &
   \left\{1_1,\frac{1}{2},1,0\right\} &
   \left\{1_1,1,0,\frac{1}{2}\right\} &
   \left\{1_1,1,0,\frac{3}{2}\right\} \\
 \left\{1_1,1,\frac{1}{2},0\right\} &
   \left\{1_1,1,\frac{3}{2},0\right\} &
   \left\{1_1,\frac{3}{2},0,1\right\} &
   \left\{1_1,\frac{3}{2},1,0\right\}
\end{array}
\end{equation}
\paragraph{Conjugacy class $\mathcal{C}_{10}\left(\mathrm{\mathrm{G_{1536}}}\right)$: $\#$ of elements = $12$}
\begin{equation}\label{G1536classe10}
\begin{array}{llll}
 \left\{1_1,\frac{1}{2},\frac{1}{2},1\right\} &
   \left\{1_1,\frac{1}{2},1,\frac{1}{2}\right\} &
   \left\{1_1,\frac{1}{2},1,\frac{3}{2}\right\} &
   \left\{1_1,\frac{1}{2},\frac{3}{2},1\right\} \\
 \left\{1_1,1,\frac{1}{2},\frac{1}{2}\right\} &
   \left\{1_1,1,\frac{1}{2},\frac{3}{2}\right\} &
   \left\{1_1,1,\frac{3}{2},\frac{1}{2}\right\} &
   \left\{1_1,1,\frac{3}{2},\frac{3}{2}\right\} \\
 \left\{1_1,\frac{3}{2},\frac{1}{2},1\right\} &
   \left\{1_1,\frac{3}{2},1,\frac{1}{2}\right\} &
   \left\{1_1,\frac{3}{2},1,\frac{3}{2}\right\} &
   \left\{1_1,\frac{3}{2},\frac{3}{2},1\right\}
\end{array}
\end{equation}
\paragraph{Conjugacy class $\mathcal{C}_{11}\left(\mathrm{\mathrm{G_{1536}}}\right)$: $\#$ of elements = $12$}
\begin{equation}\label{G1536classe11}
\begin{array}{llll}
 \left\{3_1,0,0,0\right\} & \left\{3_1,0,1,0\right\} &
   \left\{3_1,1,0,0\right\} & \left\{3_1,1,1,0\right\} \\
 \left\{3_2,0,0,0\right\} & \left\{3_2,0,0,1\right\} &
   \left\{3_2,1,0,0\right\} & \left\{3_2,1,0,1\right\} \\
 \left\{3_3,0,0,0\right\} & \left\{3_3,0,0,1\right\} &
   \left\{3_3,0,1,0\right\} & \left\{3_3,0,1,1\right\}
\end{array}
\end{equation}
\paragraph{Conjugacy class $\mathcal{C}_{12}\left(\mathrm{\mathrm{G_{1536}}}\right)$: $\#$ of elements = $12$}
\begin{equation}\label{G1536classe12}
\begin{array}{llll}
 \left\{3_1,0,0,1\right\} & \left\{3_1,0,1,1\right\} &
   \left\{3_1,1,0,1\right\} & \left\{3_1,1,1,1\right\} \\
 \left\{3_2,0,1,0\right\} & \left\{3_2,0,1,1\right\} &
   \left\{3_2,1,1,0\right\} & \left\{3_2,1,1,1\right\} \\
 \left\{3_3,1,0,0\right\} & \left\{3_3,1,0,1\right\} &
   \left\{3_3,1,1,0\right\} & \left\{3_3,1,1,1\right\}
\end{array}
\end{equation}
\paragraph{Conjugacy class $\mathcal{C}_{13}\left(\mathrm{\mathrm{G_{1536}}}\right)$: $\#$ of elements = $12$}
\begin{equation}\label{G1536classe13}
\begin{array}{llll}
 \left\{3_1,\frac{1}{2},\frac{1}{2},0\right\} &
   \left\{3_1,\frac{1}{2},\frac{3}{2},0\right\} &
   \left\{3_1,\frac{3}{2},\frac{1}{2},0\right\} &
   \left\{3_1,\frac{3}{2},\frac{3}{2},0\right\} \\
 \left\{3_2,\frac{1}{2},0,\frac{1}{2}\right\} &
   \left\{3_2,\frac{1}{2},0,\frac{3}{2}\right\} &
   \left\{3_2,\frac{3}{2},0,\frac{1}{2}\right\} &
   \left\{3_2,\frac{3}{2},0,\frac{3}{2}\right\} \\
 \left\{3_3,0,\frac{1}{2},\frac{1}{2}\right\} &
   \left\{3_3,0,\frac{1}{2},\frac{3}{2}\right\} &
   \left\{3_3,0,\frac{3}{2},\frac{1}{2}\right\} &
   \left\{3_3,0,\frac{3}{2},\frac{3}{2}\right\}
\end{array}
\end{equation}
\paragraph{Conjugacy class $\mathcal{C}_{14}\left(\mathrm{\mathrm{G_{1536}}}\right)$: $\#$ of elements = $12$}
\begin{equation}\label{G1536classe14}
\begin{array}{llll}
 \left\{3_1,\frac{1}{2},\frac{1}{2},1\right\} &
   \left\{3_1,\frac{1}{2},\frac{3}{2},1\right\} &
   \left\{3_1,\frac{3}{2},\frac{1}{2},1\right\} &
   \left\{3_1,\frac{3}{2},\frac{3}{2},1\right\} \\
 \left\{3_2,\frac{1}{2},1,\frac{1}{2}\right\} &
   \left\{3_2,\frac{1}{2},1,\frac{3}{2}\right\} &
   \left\{3_2,\frac{3}{2},1,\frac{1}{2}\right\} &
   \left\{3_2,\frac{3}{2},1,\frac{3}{2}\right\} \\
 \left\{3_3,1,\frac{1}{2},\frac{1}{2}\right\} &
   \left\{3_3,1,\frac{1}{2},\frac{3}{2}\right\} &
   \left\{3_3,1,\frac{3}{2},\frac{1}{2}\right\} &
   \left\{3_3,1,\frac{3}{2},\frac{3}{2}\right\}
\end{array}
\end{equation}
\paragraph{Conjugacy class $\mathcal{C}_{15}\left(\mathrm{\mathrm{G_{1536}}}\right)$: $\#$ of elements = $24$}
{\scriptsize
\begin{equation}\label{G1536classe15}
\begin{array}{llllllll}
 \left\{3_1,0,0,\frac{1}{2}\right\} &
   \left\{3_1,0,0,\frac{3}{2}\right\} &
   \left\{3_1,0,1,\frac{1}{2}\right\} &
   \left\{3_1,0,1,\frac{3}{2}\right\} &
   \left\{3_1,1,0,\frac{1}{2}\right\} &
   \left\{3_1,1,0,\frac{3}{2}\right\} &
   \left\{3_1,1,1,\frac{1}{2}\right\} &
   \left\{3_1,1,1,\frac{3}{2}\right\} \\
 \left\{3_2,0,\frac{1}{2},0\right\} &
   \left\{3_2,0,\frac{1}{2},1\right\} &
   \left\{3_2,0,\frac{3}{2},0\right\} &
   \left\{3_2,0,\frac{3}{2},1\right\} &
   \left\{3_2,1,\frac{1}{2},0\right\} &
   \left\{3_2,1,\frac{1}{2},1\right\} &
   \left\{3_2,1,\frac{3}{2},0\right\} &
   \left\{3_2,1,\frac{3}{2},1\right\} \\
 \left\{3_3,\frac{1}{2},0,0\right\} &
   \left\{3_3,\frac{1}{2},0,1\right\} &
   \left\{3_3,\frac{1}{2},1,0\right\} &
   \left\{3_3,\frac{1}{2},1,1\right\} &
   \left\{3_3,\frac{3}{2},0,0\right\} &
   \left\{3_3,\frac{3}{2},0,1\right\} &
   \left\{3_3,\frac{3}{2},1,0\right\} &
   \left\{3_3,\frac{3}{2},1,1\right\}
\end{array}
\end{equation}
}
\paragraph{Conjugacy class $\mathcal{C}_{16}\left(\mathrm{\mathrm{G_{1536}}}\right)$: $\#$ of elements = $24$}
{\scriptsize
\begin{equation}\label{G1536classe16}
\begin{array}{llllllll}
 \left\{3_1,0,\frac{1}{2},0\right\} &
   \left\{3_1,0,\frac{3}{2},0\right\} &
   \left\{3_1,\frac{1}{2},0,0\right\} &
   \left\{3_1,\frac{1}{2},1,0\right\} &
   \left\{3_1,1,\frac{1}{2},0\right\} &
   \left\{3_1,1,\frac{3}{2},0\right\} &
   \left\{3_1,\frac{3}{2},0,0\right\} &
   \left\{3_1,\frac{3}{2},1,0\right\} \\
 \left\{3_2,0,0,\frac{1}{2}\right\} &
   \left\{3_2,0,0,\frac{3}{2}\right\} &
   \left\{3_2,\frac{1}{2},0,0\right\} &
   \left\{3_2,\frac{1}{2},0,1\right\} &
   \left\{3_2,1,0,\frac{1}{2}\right\} &
   \left\{3_2,1,0,\frac{3}{2}\right\} &
   \left\{3_2,\frac{3}{2},0,0\right\} &
   \left\{3_2,\frac{3}{2},0,1\right\} \\
 \left\{3_3,0,0,\frac{1}{2}\right\} &
   \left\{3_3,0,0,\frac{3}{2}\right\} &
   \left\{3_3,0,\frac{1}{2},0\right\} &
   \left\{3_3,0,\frac{1}{2},1\right\} &
   \left\{3_3,0,1,\frac{1}{2}\right\} &
   \left\{3_3,0,1,\frac{3}{2}\right\} &
   \left\{3_3,0,\frac{3}{2},0\right\} &
   \left\{3_3,0,\frac{3}{2},1\right\}
\end{array}
\end{equation}
}
\paragraph{Conjugacy class $\mathcal{C}_{17}\left(\mathrm{\mathrm{G_{1536}}}\right)$: $\#$ of elements = $24$}
{\scriptsize
\begin{equation}\label{G1536classe17}
\begin{array}{llllllll}
 \left\{3_1,0,\frac{1}{2},1\right\} &
   \left\{3_1,0,\frac{3}{2},1\right\} &
   \left\{3_1,\frac{1}{2},0,1\right\} &
   \left\{3_1,\frac{1}{2},1,1\right\} &
   \left\{3_1,1,\frac{1}{2},1\right\} &
   \left\{3_1,1,\frac{3}{2},1\right\} &
   \left\{3_1,\frac{3}{2},0,1\right\} &
   \left\{3_1,\frac{3}{2},1,1\right\} \\
 \left\{3_2,0,1,\frac{1}{2}\right\} &
   \left\{3_2,0,1,\frac{3}{2}\right\} &
   \left\{3_2,\frac{1}{2},1,0\right\} &
   \left\{3_2,\frac{1}{2},1,1\right\} &
   \left\{3_2,1,1,\frac{1}{2}\right\} &
   \left\{3_2,1,1,\frac{3}{2}\right\} &
   \left\{3_2,\frac{3}{2},1,0\right\} &
   \left\{3_2,\frac{3}{2},1,1\right\} \\
 \left\{3_3,1,0,\frac{1}{2}\right\} &
   \left\{3_3,1,0,\frac{3}{2}\right\} &
   \left\{3_3,1,\frac{1}{2},0\right\} &
   \left\{3_3,1,\frac{1}{2},1\right\} &
   \left\{3_3,1,1,\frac{1}{2}\right\} &
   \left\{3_3,1,1,\frac{3}{2}\right\} &
   \left\{3_3,1,\frac{3}{2},0\right\} &
   \left\{3_3,1,\frac{3}{2},1\right\}
\end{array}
\end{equation}
}
\paragraph{Conjugacy class $\mathcal{C}_{18}\left(\mathrm{\mathrm{G_{1536}}}\right)$: $\#$ of elements = $24$}
{\scriptsize
\begin{equation}\label{G1536classe18}
\begin{array}{llllllll}
 \left\{3_1,\frac{1}{2},\frac{1}{2},\frac{1}{2}\right\} &
   \left\{3_1,\frac{1}{2},\frac{1}{2},\frac{3}{2}\right\}
   &
   \left\{3_1,\frac{1}{2},\frac{3}{2},\frac{1}{2}\right\}
   &
   \left\{3_1,\frac{1}{2},\frac{3}{2},\frac{3}{2}\right\}
   &
   \left\{3_1,\frac{3}{2},\frac{1}{2},\frac{1}{2}\right\}
   &
   \left\{3_1,\frac{3}{2},\frac{1}{2},\frac{3}{2}\right\}
   &
   \left\{3_1,\frac{3}{2},\frac{3}{2},\frac{1}{2}\right\}
   &
   \left\{3_1,\frac{3}{2},\frac{3}{2},\frac{3}{2}\right\}
   \\
 \left\{3_2,\frac{1}{2},\frac{1}{2},\frac{1}{2}\right\} &
   \left\{3_2,\frac{1}{2},\frac{1}{2},\frac{3}{2}\right\}
   &
   \left\{3_2,\frac{1}{2},\frac{3}{2},\frac{1}{2}\right\}
   &
   \left\{3_2,\frac{1}{2},\frac{3}{2},\frac{3}{2}\right\}
   &
   \left\{3_2,\frac{3}{2},\frac{1}{2},\frac{1}{2}\right\}
   &
   \left\{3_2,\frac{3}{2},\frac{1}{2},\frac{3}{2}\right\}
   &
   \left\{3_2,\frac{3}{2},\frac{3}{2},\frac{1}{2}\right\}
   &
   \left\{3_2,\frac{3}{2},\frac{3}{2},\frac{3}{2}\right\}
   \\
 \left\{3_3,\frac{1}{2},\frac{1}{2},\frac{1}{2}\right\} &
   \left\{3_3,\frac{1}{2},\frac{1}{2},\frac{3}{2}\right\}
   &
   \left\{3_3,\frac{1}{2},\frac{3}{2},\frac{1}{2}\right\}
   &
   \left\{3_3,\frac{1}{2},\frac{3}{2},\frac{3}{2}\right\}
   &
   \left\{3_3,\frac{3}{2},\frac{1}{2},\frac{1}{2}\right\}
   &
   \left\{3_3,\frac{3}{2},\frac{1}{2},\frac{3}{2}\right\}
   &
   \left\{3_3,\frac{3}{2},\frac{3}{2},\frac{1}{2}\right\}
   &
   \left\{3_3,\frac{3}{2},\frac{3}{2},\frac{3}{2}\right\}
\end{array}
\end{equation}
}
\paragraph{Conjugacy class $\mathcal{C}_{19}\left(\mathrm{\mathrm{G_{1536}}}\right)$: $\#$ of elements = $48$}
{\scriptsize
\begin{equation}\label{G1536classe19}
\begin{array}{llllllll}
 \left\{3_1,0,\frac{1}{2},\frac{1}{2}\right\} &
   \left\{3_1,0,\frac{1}{2},\frac{3}{2}\right\} &
   \left\{3_1,0,\frac{3}{2},\frac{1}{2}\right\} &
   \left\{3_1,0,\frac{3}{2},\frac{3}{2}\right\} &
   \left\{3_1,\frac{1}{2},0,\frac{1}{2}\right\} &
   \left\{3_1,\frac{1}{2},0,\frac{3}{2}\right\} &
   \left\{3_1,\frac{1}{2},1,\frac{1}{2}\right\} &
   \left\{3_1,\frac{1}{2},1,\frac{3}{2}\right\} \\
 \left\{3_1,1,\frac{1}{2},\frac{1}{2}\right\} &
   \left\{3_1,1,\frac{1}{2},\frac{3}{2}\right\} &
   \left\{3_1,1,\frac{3}{2},\frac{1}{2}\right\} &
   \left\{3_1,1,\frac{3}{2},\frac{3}{2}\right\} &
   \left\{3_1,\frac{3}{2},0,\frac{1}{2}\right\} &
   \left\{3_1,\frac{3}{2},0,\frac{3}{2}\right\} &
   \left\{3_1,\frac{3}{2},1,\frac{1}{2}\right\} &
   \left\{3_1,\frac{3}{2},1,\frac{3}{2}\right\} \\
 \left\{3_2,0,\frac{1}{2},\frac{1}{2}\right\} &
   \left\{3_2,0,\frac{1}{2},\frac{3}{2}\right\} &
   \left\{3_2,0,\frac{3}{2},\frac{1}{2}\right\} &
   \left\{3_2,0,\frac{3}{2},\frac{3}{2}\right\} &
   \left\{3_2,\frac{1}{2},\frac{1}{2},0\right\} &
   \left\{3_2,\frac{1}{2},\frac{1}{2},1\right\} &
   \left\{3_2,\frac{1}{2},\frac{3}{2},0\right\} &
   \left\{3_2,\frac{1}{2},\frac{3}{2},1\right\} \\
 \left\{3_2,1,\frac{1}{2},\frac{1}{2}\right\} &
   \left\{3_2,1,\frac{1}{2},\frac{3}{2}\right\} &
   \left\{3_2,1,\frac{3}{2},\frac{1}{2}\right\} &
   \left\{3_2,1,\frac{3}{2},\frac{3}{2}\right\} &
   \left\{3_2,\frac{3}{2},\frac{1}{2},0\right\} &
   \left\{3_2,\frac{3}{2},\frac{1}{2},1\right\} &
   \left\{3_2,\frac{3}{2},\frac{3}{2},0\right\} &
   \left\{3_2,\frac{3}{2},\frac{3}{2},1\right\} \\
\end{array}
\end{equation}
}
\paragraph{Conjugacy class $\mathcal{C}_{20}\left(\mathrm{\mathrm{G_{1536}}}\right)$: $\#$ of elements = $48$}
{\scriptsize
\begin{equation}\label{G1536classe20}
\begin{array}{llllllll}
 \left\{4_1,0,0,0\right\} &
   \left\{4_1,0,\frac{1}{2},\frac{1}{2}\right\} &
   \left\{4_1,0,1,1\right\} &
   \left\{4_1,0,\frac{3}{2},\frac{3}{2}\right\} &
   \left\{4_1,1,0,0\right\} &
   \left\{4_1,1,\frac{1}{2},\frac{1}{2}\right\} &
   \left\{4_1,1,1,1\right\} &
   \left\{4_1,1,\frac{3}{2},\frac{3}{2}\right\} \\
 \left\{4_2,0,0,0\right\} &
   \left\{4_2,0,\frac{1}{2},\frac{3}{2}\right\} &
   \left\{4_2,0,1,1\right\} &
   \left\{4_2,0,\frac{3}{2},\frac{1}{2}\right\} &
   \left\{4_2,1,0,0\right\} &
   \left\{4_2,1,\frac{1}{2},\frac{3}{2}\right\} &
   \left\{4_2,1,1,1\right\} &
   \left\{4_2,1,\frac{3}{2},\frac{1}{2}\right\} \\
 \left\{4_3,0,0,0\right\} & \left\{4_3,0,0,1\right\} &
   \left\{4_3,\frac{1}{2},\frac{1}{2},0\right\} &
   \left\{4_3,\frac{1}{2},\frac{1}{2},1\right\} &
   \left\{4_3,1,1,0\right\} & \left\{4_3,1,1,1\right\} &
   \left\{4_3,\frac{3}{2},\frac{3}{2},0\right\} &
   \left\{4_3,\frac{3}{2},\frac{3}{2},1\right\} \\
 \left\{4_4,0,0,0\right\} & \left\{4_4,0,1,0\right\} &
   \left\{4_4,\frac{1}{2},0,\frac{1}{2}\right\} &
   \left\{4_4,\frac{1}{2},1,\frac{1}{2}\right\} &
   \left\{4_4,1,0,1\right\} & \left\{4_4,1,1,1\right\} &
   \left\{4_4,\frac{3}{2},0,\frac{3}{2}\right\} &
   \left\{4_4,\frac{3}{2},1,\frac{3}{2}\right\} \\
 \left\{4_5,0,0,0\right\} & \left\{4_5,0,1,0\right\} &
   \left\{4_5,\frac{1}{2},0,\frac{3}{2}\right\} &
   \left\{4_5,\frac{1}{2},1,\frac{3}{2}\right\} &
   \left\{4_5,1,0,1\right\} & \left\{4_5,1,1,1\right\} &
   \left\{4_5,\frac{3}{2},0,\frac{1}{2}\right\} &
   \left\{4_5,\frac{3}{2},1,\frac{1}{2}\right\} \\
 \left\{4_6,0,0,0\right\} & \left\{4_6,0,0,1\right\} &
   \left\{4_6,\frac{1}{2},\frac{3}{2},0\right\} &
   \left\{4_6,\frac{1}{2},\frac{3}{2},1\right\} &
   \left\{4_6,1,1,0\right\} & \left\{4_6,1,1,1\right\} &
   \left\{4_6,\frac{3}{2},\frac{1}{2},0\right\} &
   \left\{4_6,\frac{3}{2},\frac{1}{2},1\right\}
\end{array}
\end{equation}
}
\paragraph{Conjugacy class $\mathcal{C}_{21}\left(\mathrm{\mathrm{G_{1536}}}\right)$: $\#$ of elements = $48$}
{\scriptsize
\begin{equation}\label{G1536classe21}
\begin{array}{llllllll}
 \left\{4_1,0,0,1\right\} &
   \left\{4_1,0,\frac{1}{2},\frac{3}{2}\right\} &
   \left\{4_1,0,1,0\right\} &
   \left\{4_1,0,\frac{3}{2},\frac{1}{2}\right\} &
   \left\{4_1,1,0,1\right\} &
   \left\{4_1,1,\frac{1}{2},\frac{3}{2}\right\} &
   \left\{4_1,1,1,0\right\} &
   \left\{4_1,1,\frac{3}{2},\frac{1}{2}\right\} \\
 \left\{4_2,0,0,1\right\} &
   \left\{4_2,0,\frac{1}{2},\frac{1}{2}\right\} &
   \left\{4_2,0,1,0\right\} &
   \left\{4_2,0,\frac{3}{2},\frac{3}{2}\right\} &
   \left\{4_2,1,0,1\right\} &
   \left\{4_2,1,\frac{1}{2},\frac{1}{2}\right\} &
   \left\{4_2,1,1,0\right\} &
   \left\{4_2,1,\frac{3}{2},\frac{3}{2}\right\} \\
 \left\{4_3,0,1,0\right\} & \left\{4_3,0,1,1\right\} &
   \left\{4_3,\frac{1}{2},\frac{3}{2},0\right\} &
   \left\{4_3,\frac{1}{2},\frac{3}{2},1\right\} &
   \left\{4_3,1,0,0\right\} & \left\{4_3,1,0,1\right\} &
   \left\{4_3,\frac{3}{2},\frac{1}{2},0\right\} &
   \left\{4_3,\frac{3}{2},\frac{1}{2},1\right\} \\
 \left\{4_4,0,0,1\right\} & \left\{4_4,0,1,1\right\} &
   \left\{4_4,\frac{1}{2},0,\frac{3}{2}\right\} &
   \left\{4_4,\frac{1}{2},1,\frac{3}{2}\right\} &
   \left\{4_4,1,0,0\right\} & \left\{4_4,1,1,0\right\} &
   \left\{4_4,\frac{3}{2},0,\frac{1}{2}\right\} &
   \left\{4_4,\frac{3}{2},1,\frac{1}{2}\right\} \\
 \left\{4_5,0,0,1\right\} & \left\{4_5,0,1,1\right\} &
   \left\{4_5,\frac{1}{2},0,\frac{1}{2}\right\} &
   \left\{4_5,\frac{1}{2},1,\frac{1}{2}\right\} &
   \left\{4_5,1,0,0\right\} & \left\{4_5,1,1,0\right\} &
   \left\{4_5,\frac{3}{2},0,\frac{3}{2}\right\} &
   \left\{4_5,\frac{3}{2},1,\frac{3}{2}\right\} \\
 \left\{4_6,0,1,0\right\} & \left\{4_6,0,1,1\right\} &
   \left\{4_6,\frac{1}{2},\frac{1}{2},0\right\} &
   \left\{4_6,\frac{1}{2},\frac{1}{2},1\right\} &
   \left\{4_6,1,0,0\right\} & \left\{4_6,1,0,1\right\} &
   \left\{4_6,\frac{3}{2},\frac{3}{2},0\right\} &
   \left\{4_6,\frac{3}{2},\frac{3}{2},1\right\}
\end{array}
\end{equation}
}
\paragraph{Conjugacy class $\mathcal{C}_{22}\left(\mathrm{\mathrm{G_{1536}}}\right)$: $\#$ of elements = $48$}
{\scriptsize
\begin{equation}\label{G1536classe22}
\begin{array}{llllllll}
 \left\{4_1,\frac{1}{2},0,0\right\} &
   \left\{4_1,\frac{1}{2},\frac{1}{2},\frac{1}{2}\right\}
   & \left\{4_1,\frac{1}{2},1,1\right\} &
   \left\{4_1,\frac{1}{2},\frac{3}{2},\frac{3}{2}\right\}
   & \left\{4_1,\frac{3}{2},0,0\right\} &
   \left\{4_1,\frac{3}{2},\frac{1}{2},\frac{1}{2}\right\}
   & \left\{4_1,\frac{3}{2},1,1\right\} &
   \left\{4_1,\frac{3}{2},\frac{3}{2},\frac{3}{2}\right\}
   \\
 \left\{4_2,\frac{1}{2},0,0\right\} &
   \left\{4_2,\frac{1}{2},\frac{1}{2},\frac{3}{2}\right\}
   & \left\{4_2,\frac{1}{2},1,1\right\} &
   \left\{4_2,\frac{1}{2},\frac{3}{2},\frac{1}{2}\right\}
   & \left\{4_2,\frac{3}{2},0,0\right\} &
   \left\{4_2,\frac{3}{2},\frac{1}{2},\frac{3}{2}\right\}
   & \left\{4_2,\frac{3}{2},1,1\right\} &
   \left\{4_2,\frac{3}{2},\frac{3}{2},\frac{1}{2}\right\}
   \\
 \left\{4_3,0,0,\frac{1}{2}\right\} &
   \left\{4_3,0,0,\frac{3}{2}\right\} &
   \left\{4_3,\frac{1}{2},\frac{1}{2},\frac{1}{2}\right\}
   &
   \left\{4_3,\frac{1}{2},\frac{1}{2},\frac{3}{2}\right\}
   & \left\{4_3,1,1,\frac{1}{2}\right\} &
   \left\{4_3,1,1,\frac{3}{2}\right\} &
   \left\{4_3,\frac{3}{2},\frac{3}{2},\frac{1}{2}\right\}
   &
   \left\{4_3,\frac{3}{2},\frac{3}{2},\frac{3}{2}\right\}
   \\
 \left\{4_4,0,\frac{1}{2},0\right\} &
   \left\{4_4,0,\frac{3}{2},0\right\} &
   \left\{4_4,\frac{1}{2},\frac{1}{2},\frac{1}{2}\right\}
   &
   \left\{4_4,\frac{1}{2},\frac{3}{2},\frac{1}{2}\right\}
   & \left\{4_4,1,\frac{1}{2},1\right\} &
   \left\{4_4,1,\frac{3}{2},1\right\} &
   \left\{4_4,\frac{3}{2},\frac{1}{2},\frac{3}{2}\right\}
   &
   \left\{4_4,\frac{3}{2},\frac{3}{2},\frac{3}{2}\right\}
   \\
\end{array}
\end{equation}
}
\paragraph{Conjugacy class $\mathcal{C}_{23}\left(\mathrm{\mathrm{G_{1536}}}\right)$: $\#$ of elements = $48$}
{\scriptsize
\begin{equation}\label{G1536classe23}
\begin{array}{llllllll}
 \left\{4_1,\frac{1}{2},0,1\right\} &
   \left\{4_1,\frac{1}{2},\frac{1}{2},\frac{3}{2}\right\}
   & \left\{4_1,\frac{1}{2},1,0\right\} &
   \left\{4_1,\frac{1}{2},\frac{3}{2},\frac{1}{2}\right\}
   & \left\{4_1,\frac{3}{2},0,1\right\} &
   \left\{4_1,\frac{3}{2},\frac{1}{2},\frac{3}{2}\right\}
   & \left\{4_1,\frac{3}{2},1,0\right\} &
   \left\{4_1,\frac{3}{2},\frac{3}{2},\frac{1}{2}\right\}
   \\
 \left\{4_2,\frac{1}{2},0,1\right\} &
   \left\{4_2,\frac{1}{2},\frac{1}{2},\frac{1}{2}\right\}
   & \left\{4_2,\frac{1}{2},1,0\right\} &
   \left\{4_2,\frac{1}{2},\frac{3}{2},\frac{3}{2}\right\}
   & \left\{4_2,\frac{3}{2},0,1\right\} &
   \left\{4_2,\frac{3}{2},\frac{1}{2},\frac{1}{2}\right\}
   & \left\{4_2,\frac{3}{2},1,0\right\} &
   \left\{4_2,\frac{3}{2},\frac{3}{2},\frac{3}{2}\right\}
   \\
 \left\{4_3,0,1,\frac{1}{2}\right\} &
   \left\{4_3,0,1,\frac{3}{2}\right\} &
   \left\{4_3,\frac{1}{2},\frac{3}{2},\frac{1}{2}\right\}
   &
   \left\{4_3,\frac{1}{2},\frac{3}{2},\frac{3}{2}\right\}
   & \left\{4_3,1,0,\frac{1}{2}\right\} &
   \left\{4_3,1,0,\frac{3}{2}\right\} &
   \left\{4_3,\frac{3}{2},\frac{1}{2},\frac{1}{2}\right\}
   &
   \left\{4_3,\frac{3}{2},\frac{1}{2},\frac{3}{2}\right\}
   \\
 \left\{4_4,0,\frac{1}{2},1\right\} &
   \left\{4_4,0,\frac{3}{2},1\right\} &
   \left\{4_4,\frac{1}{2},\frac{1}{2},\frac{3}{2}\right\}
   &
   \left\{4_4,\frac{1}{2},\frac{3}{2},\frac{3}{2}\right\}
   & \left\{4_4,1,\frac{1}{2},0\right\} &
   \left\{4_4,1,\frac{3}{2},0\right\} &
   \left\{4_4,\frac{3}{2},\frac{1}{2},\frac{1}{2}\right\}
   &
   \left\{4_4,\frac{3}{2},\frac{3}{2},\frac{1}{2}\right\}
   \\
\end{array}
\end{equation}
}
\paragraph{Conjugacy class $\mathcal{C}_{24}\left(\mathrm{\mathrm{G_{1536}}}\right)$: $\#$ of elements = $48$}
{\scriptsize
\begin{equation}\label{G1536classe24}
\begin{array}{llllllll}
 \left\{5_1,0,0,0\right\} & \left\{5_1,0,1,0\right\} &
   \left\{5_1,\frac{1}{2},\frac{1}{2},0\right\} &
   \left\{5_1,\frac{1}{2},\frac{3}{2},0\right\} &
   \left\{5_1,1,0,0\right\} & \left\{5_1,1,1,0\right\} &
   \left\{5_1,\frac{3}{2},\frac{1}{2},0\right\} &
   \left\{5_1,\frac{3}{2},\frac{3}{2},0\right\} \\
 \left\{5_2,0,0,0\right\} & \left\{5_2,0,0,1\right\} &
   \left\{5_2,\frac{1}{2},0,\frac{1}{2}\right\} &
   \left\{5_2,\frac{1}{2},0,\frac{3}{2}\right\} &
   \left\{5_2,1,0,0\right\} & \left\{5_2,1,0,1\right\} &
   \left\{5_2,\frac{3}{2},0,\frac{1}{2}\right\} &
   \left\{5_2,\frac{3}{2},0,\frac{3}{2}\right\} \\
 \left\{5_3,0,0,0\right\} & \left\{5_3,0,0,1\right\} &
   \left\{5_3,\frac{1}{2},0,\frac{1}{2}\right\} &
   \left\{5_3,\frac{1}{2},0,\frac{3}{2}\right\} &
   \left\{5_3,1,0,0\right\} & \left\{5_3,1,0,1\right\} &
   \left\{5_3,\frac{3}{2},0,\frac{1}{2}\right\} &
   \left\{5_3,\frac{3}{2},0,\frac{3}{2}\right\} \\
 \left\{5_4,0,0,0\right\} & \left\{5_4,0,1,0\right\} &
   \left\{5_4,\frac{1}{2},\frac{1}{2},0\right\} &
   \left\{5_4,\frac{1}{2},\frac{3}{2},0\right\} &
   \left\{5_4,1,0,0\right\} & \left\{5_4,1,1,0\right\} &
   \left\{5_4,\frac{3}{2},\frac{1}{2},0\right\} &
   \left\{5_4,\frac{3}{2},\frac{3}{2},0\right\} \\
 \left\{5_5,0,0,0\right\} & \left\{5_5,0,0,1\right\} &
   \left\{5_5,0,\frac{1}{2},\frac{1}{2}\right\} &
   \left\{5_5,0,\frac{1}{2},\frac{3}{2}\right\} &
   \left\{5_5,0,1,0\right\} & \left\{5_5,0,1,1\right\} &
   \left\{5_5,0,\frac{3}{2},\frac{1}{2}\right\} &
   \left\{5_5,0,\frac{3}{2},\frac{3}{2}\right\} \\
 \left\{5_6,0,0,0\right\} & \left\{5_6,0,0,1\right\} &
   \left\{5_6,0,\frac{1}{2},\frac{1}{2}\right\} &
   \left\{5_6,0,\frac{1}{2},\frac{3}{2}\right\} &
   \left\{5_6,0,1,0\right\} & \left\{5_6,0,1,1\right\} &
   \left\{5_6,0,\frac{3}{2},\frac{1}{2}\right\} &
   \left\{5_6,0,\frac{3}{2},\frac{3}{2}\right\}
\end{array}
\end{equation}
}
\paragraph{Conjugacy class $\mathcal{C}_{25}\left(\mathrm{\mathrm{G_{1536}}}\right)$: $\#$ of elements = $48$}
{\scriptsize
\begin{equation}\label{G1536classe25}
\begin{array}{llllllll}
 \left\{5_1,0,0,\frac{1}{2}\right\} &
   \left\{5_1,0,1,\frac{1}{2}\right\} &
   \left\{5_1,\frac{1}{2},\frac{1}{2},\frac{1}{2}\right\}
   &
   \left\{5_1,\frac{1}{2},\frac{3}{2},\frac{1}{2}\right\}
   & \left\{5_1,1,0,\frac{1}{2}\right\} &
   \left\{5_1,1,1,\frac{1}{2}\right\} &
   \left\{5_1,\frac{3}{2},\frac{1}{2},\frac{1}{2}\right\}
   &
   \left\{5_1,\frac{3}{2},\frac{3}{2},\frac{1}{2}\right\}
   \\
 \left\{5_2,0,\frac{3}{2},0\right\} &
   \left\{5_2,0,\frac{3}{2},1\right\} &
   \left\{5_2,\frac{1}{2},\frac{3}{2},\frac{1}{2}\right\}
   &
   \left\{5_2,\frac{1}{2},\frac{3}{2},\frac{3}{2}\right\}
   & \left\{5_2,1,\frac{3}{2},0\right\} &
   \left\{5_2,1,\frac{3}{2},1\right\} &
   \left\{5_2,\frac{3}{2},\frac{3}{2},\frac{1}{2}\right\}
   &
   \left\{5_2,\frac{3}{2},\frac{3}{2},\frac{3}{2}\right\}
   \\
 \left\{5_3,0,\frac{1}{2},0\right\} &
   \left\{5_3,0,\frac{1}{2},1\right\} &
   \left\{5_3,\frac{1}{2},\frac{1}{2},\frac{1}{2}\right\}
   &
   \left\{5_3,\frac{1}{2},\frac{1}{2},\frac{3}{2}\right\}
   & \left\{5_3,1,\frac{1}{2},0\right\} &
   \left\{5_3,1,\frac{1}{2},1\right\} &
   \left\{5_3,\frac{3}{2},\frac{1}{2},\frac{1}{2}\right\}
   &
   \left\{5_3,\frac{3}{2},\frac{1}{2},\frac{3}{2}\right\}
   \\
 \left\{5_4,0,0,\frac{3}{2}\right\} &
   \left\{5_4,0,1,\frac{3}{2}\right\} &
   \left\{5_4,\frac{1}{2},\frac{1}{2},\frac{3}{2}\right\}
   &
   \left\{5_4,\frac{1}{2},\frac{3}{2},\frac{3}{2}\right\}
   & \left\{5_4,1,0,\frac{3}{2}\right\} &
   \left\{5_4,1,1,\frac{3}{2}\right\} &
   \left\{5_4,\frac{3}{2},\frac{1}{2},\frac{3}{2}\right\}
   &
   \left\{5_4,\frac{3}{2},\frac{3}{2},\frac{3}{2}\right\}
   \\
\end{array}
\end{equation}
}
\paragraph{Conjugacy class $\mathcal{C}_{26}\left(\mathrm{\mathrm{G_{1536}}}\right)$: $\#$ of elements = $48$}
{\scriptsize
\begin{equation}\label{G1536classe26}
\begin{array}{llllllll}
 \left\{5_1,0,0,1\right\} & \left\{5_1,0,1,1\right\} &
   \left\{5_1,\frac{1}{2},\frac{1}{2},1\right\} &
   \left\{5_1,\frac{1}{2},\frac{3}{2},1\right\} &
   \left\{5_1,1,0,1\right\} & \left\{5_1,1,1,1\right\} &
   \left\{5_1,\frac{3}{2},\frac{1}{2},1\right\} &
   \left\{5_1,\frac{3}{2},\frac{3}{2},1\right\} \\
 \left\{5_2,0,1,0\right\} & \left\{5_2,0,1,1\right\} &
   \left\{5_2,\frac{1}{2},1,\frac{1}{2}\right\} &
   \left\{5_2,\frac{1}{2},1,\frac{3}{2}\right\} &
   \left\{5_2,1,1,0\right\} & \left\{5_2,1,1,1\right\} &
   \left\{5_2,\frac{3}{2},1,\frac{1}{2}\right\} &
   \left\{5_2,\frac{3}{2},1,\frac{3}{2}\right\} \\
 \left\{5_3,0,1,0\right\} & \left\{5_3,0,1,1\right\} &
   \left\{5_3,\frac{1}{2},1,\frac{1}{2}\right\} &
   \left\{5_3,\frac{1}{2},1,\frac{3}{2}\right\} &
   \left\{5_3,1,1,0\right\} & \left\{5_3,1,1,1\right\} &
   \left\{5_3,\frac{3}{2},1,\frac{1}{2}\right\} &
   \left\{5_3,\frac{3}{2},1,\frac{3}{2}\right\} \\
 \left\{5_4,0,0,1\right\} & \left\{5_4,0,1,1\right\} &
   \left\{5_4,\frac{1}{2},\frac{1}{2},1\right\} &
   \left\{5_4,\frac{1}{2},\frac{3}{2},1\right\} &
   \left\{5_4,1,0,1\right\} & \left\{5_4,1,1,1\right\} &
   \left\{5_4,\frac{3}{2},\frac{1}{2},1\right\} &
   \left\{5_4,\frac{3}{2},\frac{3}{2},1\right\} \\
 \left\{5_5,1,0,0\right\} & \left\{5_5,1,0,1\right\} &
   \left\{5_5,1,\frac{1}{2},\frac{1}{2}\right\} &
   \left\{5_5,1,\frac{1}{2},\frac{3}{2}\right\} &
   \left\{5_5,1,1,0\right\} & \left\{5_5,1,1,1\right\} &
   \left\{5_5,1,\frac{3}{2},\frac{1}{2}\right\} &
   \left\{5_5,1,\frac{3}{2},\frac{3}{2}\right\} \\
 \left\{5_6,1,0,0\right\} & \left\{5_6,1,0,1\right\} &
   \left\{5_6,1,\frac{1}{2},\frac{1}{2}\right\} &
   \left\{5_6,1,\frac{1}{2},\frac{3}{2}\right\} &
   \left\{5_6,1,1,0\right\} & \left\{5_6,1,1,1\right\} &
   \left\{5_6,1,\frac{3}{2},\frac{1}{2}\right\} &
   \left\{5_6,1,\frac{3}{2},\frac{3}{2}\right\}
\end{array}
\end{equation}
}
\paragraph{Conjugacy class $\mathcal{C}_{27}\left(\mathrm{\mathrm{G_{1536}}}\right)$: $\#$ of elements = $48$}
{\scriptsize
\begin{equation}\label{G1536classe27}
\begin{array}{llllllll}
 \left\{5_1,0,0,\frac{3}{2}\right\} &
   \left\{5_1,0,1,\frac{3}{2}\right\} &
   \left\{5_1,\frac{1}{2},\frac{1}{2},\frac{3}{2}\right\}
   &
   \left\{5_1,\frac{1}{2},\frac{3}{2},\frac{3}{2}\right\}
   & \left\{5_1,1,0,\frac{3}{2}\right\} &
   \left\{5_1,1,1,\frac{3}{2}\right\} &
   \left\{5_1,\frac{3}{2},\frac{1}{2},\frac{3}{2}\right\}
   &
   \left\{5_1,\frac{3}{2},\frac{3}{2},\frac{3}{2}\right\}
   \\
 \left\{5_2,0,\frac{1}{2},0\right\} &
   \left\{5_2,0,\frac{1}{2},1\right\} &
   \left\{5_2,\frac{1}{2},\frac{1}{2},\frac{1}{2}\right\}
   &
   \left\{5_2,\frac{1}{2},\frac{1}{2},\frac{3}{2}\right\}
   & \left\{5_2,1,\frac{1}{2},0\right\} &
   \left\{5_2,1,\frac{1}{2},1\right\} &
   \left\{5_2,\frac{3}{2},\frac{1}{2},\frac{1}{2}\right\}
   &
   \left\{5_2,\frac{3}{2},\frac{1}{2},\frac{3}{2}\right\}
   \\
 \left\{5_3,0,\frac{3}{2},0\right\} &
   \left\{5_3,0,\frac{3}{2},1\right\} &
   \left\{5_3,\frac{1}{2},\frac{3}{2},\frac{1}{2}\right\}
   &
   \left\{5_3,\frac{1}{2},\frac{3}{2},\frac{3}{2}\right\}
   & \left\{5_3,1,\frac{3}{2},0\right\} &
   \left\{5_3,1,\frac{3}{2},1\right\} &
   \left\{5_3,\frac{3}{2},\frac{3}{2},\frac{1}{2}\right\}
   &
   \left\{5_3,\frac{3}{2},\frac{3}{2},\frac{3}{2}\right\}
   \\
 \left\{5_4,0,0,\frac{1}{2}\right\} &
   \left\{5_4,0,1,\frac{1}{2}\right\} &
   \left\{5_4,\frac{1}{2},\frac{1}{2},\frac{1}{2}\right\}
   &
   \left\{5_4,\frac{1}{2},\frac{3}{2},\frac{1}{2}\right\}
   & \left\{5_4,1,0,\frac{1}{2}\right\} &
   \left\{5_4,1,1,\frac{1}{2}\right\} &
   \left\{5_4,\frac{3}{2},\frac{1}{2},\frac{1}{2}\right\}
   &
   \left\{5_4,\frac{3}{2},\frac{3}{2},\frac{1}{2}\right\}
   \\
\end{array}
\end{equation}
}
\paragraph{Conjugacy class $\mathcal{C}_{28}\left(\mathrm{\mathrm{G_{1536}}}\right)$: $\#$ of elements = $48$}
{\scriptsize
\begin{equation}\label{G1536classe28}
\begin{array}{llllllll}
 \left\{5_1,0,\frac{1}{2},0\right\} &
   \left\{5_1,0,\frac{3}{2},0\right\} &
   \left\{5_1,\frac{1}{2},0,0\right\} &
   \left\{5_1,\frac{1}{2},1,0\right\} &
   \left\{5_1,1,\frac{1}{2},0\right\} &
   \left\{5_1,1,\frac{3}{2},0\right\} &
   \left\{5_1,\frac{3}{2},0,0\right\} &
   \left\{5_1,\frac{3}{2},1,0\right\} \\
 \left\{5_2,0,0,\frac{1}{2}\right\} &
   \left\{5_2,0,0,\frac{3}{2}\right\} &
   \left\{5_2,\frac{1}{2},0,0\right\} &
   \left\{5_2,\frac{1}{2},0,1\right\} &
   \left\{5_2,1,0,\frac{1}{2}\right\} &
   \left\{5_2,1,0,\frac{3}{2}\right\} &
   \left\{5_2,\frac{3}{2},0,0\right\} &
   \left\{5_2,\frac{3}{2},0,1\right\} \\
 \left\{5_3,0,0,\frac{1}{2}\right\} &
   \left\{5_3,0,0,\frac{3}{2}\right\} &
   \left\{5_3,\frac{1}{2},0,0\right\} &
   \left\{5_3,\frac{1}{2},0,1\right\} &
   \left\{5_3,1,0,\frac{1}{2}\right\} &
   \left\{5_3,1,0,\frac{3}{2}\right\} &
   \left\{5_3,\frac{3}{2},0,0\right\} &
   \left\{5_3,\frac{3}{2},0,1\right\} \\
 \left\{5_4,0,\frac{1}{2},0\right\} &
   \left\{5_4,0,\frac{3}{2},0\right\} &
   \left\{5_4,\frac{1}{2},0,0\right\} &
   \left\{5_4,\frac{1}{2},1,0\right\} &
   \left\{5_4,1,\frac{1}{2},0\right\} &
   \left\{5_4,1,\frac{3}{2},0\right\} &
   \left\{5_4,\frac{3}{2},0,0\right\} &
   \left\{5_4,\frac{3}{2},1,0\right\} \\
 \left\{5_5,0,0,\frac{1}{2}\right\} &
   \left\{5_5,0,0,\frac{3}{2}\right\} &
   \left\{5_5,0,\frac{1}{2},0\right\} &
   \left\{5_5,0,\frac{1}{2},1\right\} &
   \left\{5_5,0,1,\frac{1}{2}\right\} &
   \left\{5_5,0,1,\frac{3}{2}\right\} &
   \left\{5_5,0,\frac{3}{2},0\right\} &
   \left\{5_5,0,\frac{3}{2},1\right\} \\
 \left\{5_6,0,0,\frac{1}{2}\right\} &
   \left\{5_6,0,0,\frac{3}{2}\right\} &
   \left\{5_6,0,\frac{1}{2},0\right\} &
   \left\{5_6,0,\frac{1}{2},1\right\} &
   \left\{5_6,0,1,\frac{1}{2}\right\} &
   \left\{5_6,0,1,\frac{3}{2}\right\} &
   \left\{5_6,0,\frac{3}{2},0\right\} &
   \left\{5_6,0,\frac{3}{2},1\right\}
\end{array}
\end{equation}
}
\paragraph{Conjugacy class $\mathcal{C}_{29}\left(\mathrm{\mathrm{G_{1536}}}\right)$: $\#$ of elements = $48$}
{\scriptsize
\begin{equation}\label{G1536classe29}
\begin{array}{llllllll}
 \left\{5_1,0,\frac{1}{2},\frac{1}{2}\right\} &
   \left\{5_1,0,\frac{3}{2},\frac{1}{2}\right\} &
   \left\{5_1,\frac{1}{2},0,\frac{1}{2}\right\} &
   \left\{5_1,\frac{1}{2},1,\frac{1}{2}\right\} &
   \left\{5_1,1,\frac{1}{2},\frac{1}{2}\right\} &
   \left\{5_1,1,\frac{3}{2},\frac{1}{2}\right\} &
   \left\{5_1,\frac{3}{2},0,\frac{1}{2}\right\} &
   \left\{5_1,\frac{3}{2},1,\frac{1}{2}\right\} \\
 \left\{5_2,0,\frac{3}{2},\frac{1}{2}\right\} &
   \left\{5_2,0,\frac{3}{2},\frac{3}{2}\right\} &
   \left\{5_2,\frac{1}{2},\frac{3}{2},0\right\} &
   \left\{5_2,\frac{1}{2},\frac{3}{2},1\right\} &
   \left\{5_2,1,\frac{3}{2},\frac{1}{2}\right\} &
   \left\{5_2,1,\frac{3}{2},\frac{3}{2}\right\} &
   \left\{5_2,\frac{3}{2},\frac{3}{2},0\right\} &
   \left\{5_2,\frac{3}{2},\frac{3}{2},1\right\} \\
 \left\{5_3,0,\frac{1}{2},\frac{1}{2}\right\} &
   \left\{5_3,0,\frac{1}{2},\frac{3}{2}\right\} &
   \left\{5_3,\frac{1}{2},\frac{1}{2},0\right\} &
   \left\{5_3,\frac{1}{2},\frac{1}{2},1\right\} &
   \left\{5_3,1,\frac{1}{2},\frac{1}{2}\right\} &
   \left\{5_3,1,\frac{1}{2},\frac{3}{2}\right\} &
   \left\{5_3,\frac{3}{2},\frac{1}{2},0\right\} &
   \left\{5_3,\frac{3}{2},\frac{1}{2},1\right\} \\
 \left\{5_4,0,\frac{1}{2},\frac{3}{2}\right\} &
   \left\{5_4,0,\frac{3}{2},\frac{3}{2}\right\} &
   \left\{5_4,\frac{1}{2},0,\frac{3}{2}\right\} &
   \left\{5_4,\frac{1}{2},1,\frac{3}{2}\right\} &
   \left\{5_4,1,\frac{1}{2},\frac{3}{2}\right\} &
   \left\{5_4,1,\frac{3}{2},\frac{3}{2}\right\} &
   \left\{5_4,\frac{3}{2},0,\frac{3}{2}\right\} &
   \left\{5_4,\frac{3}{2},1,\frac{3}{2}\right\} \\
\end{array}
\end{equation}
}
\paragraph{Conjugacy class $\mathcal{C}_{30}\left(\mathrm{\mathrm{G_{1536}}}\right)$: $\#$ of elements = $48$}
{\scriptsize
\begin{equation}\label{G1536classe30}
\begin{array}{llllllll}
 \left\{5_1,0,\frac{1}{2},1\right\} &
   \left\{5_1,0,\frac{3}{2},1\right\} &
   \left\{5_1,\frac{1}{2},0,1\right\} &
   \left\{5_1,\frac{1}{2},1,1\right\} &
   \left\{5_1,1,\frac{1}{2},1\right\} &
   \left\{5_1,1,\frac{3}{2},1\right\} &
   \left\{5_1,\frac{3}{2},0,1\right\} &
   \left\{5_1,\frac{3}{2},1,1\right\} \\
 \left\{5_2,0,1,\frac{1}{2}\right\} &
   \left\{5_2,0,1,\frac{3}{2}\right\} &
   \left\{5_2,\frac{1}{2},1,0\right\} &
   \left\{5_2,\frac{1}{2},1,1\right\} &
   \left\{5_2,1,1,\frac{1}{2}\right\} &
   \left\{5_2,1,1,\frac{3}{2}\right\} &
   \left\{5_2,\frac{3}{2},1,0\right\} &
   \left\{5_2,\frac{3}{2},1,1\right\} \\
 \left\{5_3,0,1,\frac{1}{2}\right\} &
   \left\{5_3,0,1,\frac{3}{2}\right\} &
   \left\{5_3,\frac{1}{2},1,0\right\} &
   \left\{5_3,\frac{1}{2},1,1\right\} &
   \left\{5_3,1,1,\frac{1}{2}\right\} &
   \left\{5_3,1,1,\frac{3}{2}\right\} &
   \left\{5_3,\frac{3}{2},1,0\right\} &
   \left\{5_3,\frac{3}{2},1,1\right\} \\
 \left\{5_4,0,\frac{1}{2},1\right\} &
   \left\{5_4,0,\frac{3}{2},1\right\} &
   \left\{5_4,\frac{1}{2},0,1\right\} &
   \left\{5_4,\frac{1}{2},1,1\right\} &
   \left\{5_4,1,\frac{1}{2},1\right\} &
   \left\{5_4,1,\frac{3}{2},1\right\} &
   \left\{5_4,\frac{3}{2},0,1\right\} &
   \left\{5_4,\frac{3}{2},1,1\right\} \\
 \left\{5_5,1,0,\frac{1}{2}\right\} &
   \left\{5_5,1,0,\frac{3}{2}\right\} &
   \left\{5_5,1,\frac{1}{2},0\right\} &
   \left\{5_5,1,\frac{1}{2},1\right\} &
   \left\{5_5,1,1,\frac{1}{2}\right\} &
   \left\{5_5,1,1,\frac{3}{2}\right\} &
   \left\{5_5,1,\frac{3}{2},0\right\} &
   \left\{5_5,1,\frac{3}{2},1\right\} \\
 \left\{5_6,1,0,\frac{1}{2}\right\} &
   \left\{5_6,1,0,\frac{3}{2}\right\} &
   \left\{5_6,1,\frac{1}{2},0\right\} &
   \left\{5_6,1,\frac{1}{2},1\right\} &
   \left\{5_6,1,1,\frac{1}{2}\right\} &
   \left\{5_6,1,1,\frac{3}{2}\right\} &
   \left\{5_6,1,\frac{3}{2},0\right\} &
   \left\{5_6,1,\frac{3}{2},1\right\}
\end{array}
\end{equation}
}
\paragraph{Conjugacy class $\mathcal{C}_{31}\left(\mathrm{\mathrm{G_{1536}}}\right)$: $\#$ of elements = $48$}
{\scriptsize
\begin{equation}\label{G1536classe31}
\begin{array}{llllllll}
 \left\{5_1,0,\frac{1}{2},\frac{3}{2}\right\} &
   \left\{5_1,0,\frac{3}{2},\frac{3}{2}\right\} &
   \left\{5_1,\frac{1}{2},0,\frac{3}{2}\right\} &
   \left\{5_1,\frac{1}{2},1,\frac{3}{2}\right\} &
   \left\{5_1,1,\frac{1}{2},\frac{3}{2}\right\} &
   \left\{5_1,1,\frac{3}{2},\frac{3}{2}\right\} &
   \left\{5_1,\frac{3}{2},0,\frac{3}{2}\right\} &
   \left\{5_1,\frac{3}{2},1,\frac{3}{2}\right\} \\
 \left\{5_2,0,\frac{1}{2},\frac{1}{2}\right\} &
   \left\{5_2,0,\frac{1}{2},\frac{3}{2}\right\} &
   \left\{5_2,\frac{1}{2},\frac{1}{2},0\right\} &
   \left\{5_2,\frac{1}{2},\frac{1}{2},1\right\} &
   \left\{5_2,1,\frac{1}{2},\frac{1}{2}\right\} &
   \left\{5_2,1,\frac{1}{2},\frac{3}{2}\right\} &
   \left\{5_2,\frac{3}{2},\frac{1}{2},0\right\} &
   \left\{5_2,\frac{3}{2},\frac{1}{2},1\right\} \\
 \left\{5_3,0,\frac{3}{2},\frac{1}{2}\right\} &
   \left\{5_3,0,\frac{3}{2},\frac{3}{2}\right\} &
   \left\{5_3,\frac{1}{2},\frac{3}{2},0\right\} &
   \left\{5_3,\frac{1}{2},\frac{3}{2},1\right\} &
   \left\{5_3,1,\frac{3}{2},\frac{1}{2}\right\} &
   \left\{5_3,1,\frac{3}{2},\frac{3}{2}\right\} &
   \left\{5_3,\frac{3}{2},\frac{3}{2},0\right\} &
   \left\{5_3,\frac{3}{2},\frac{3}{2},1\right\} \\
 \left\{5_4,0,\frac{1}{2},\frac{1}{2}\right\} &
   \left\{5_4,0,\frac{3}{2},\frac{1}{2}\right\} &
   \left\{5_4,\frac{1}{2},0,\frac{1}{2}\right\} &
   \left\{5_4,\frac{1}{2},1,\frac{1}{2}\right\} &
   \left\{5_4,1,\frac{1}{2},\frac{1}{2}\right\} &
   \left\{5_4,1,\frac{3}{2},\frac{1}{2}\right\} &
   \left\{5_4,\frac{3}{2},0,\frac{1}{2}\right\} &
   \left\{5_4,\frac{3}{2},1,\frac{1}{2}\right\} \\
\end{array}
\end{equation}
}
\paragraph{Conjugacy class $\mathcal{C}_{32}\left(\mathrm{\mathrm{G_{1536}}}\right)$: $\#$ of elements = $96$}
{\scriptsize
\begin{equation}\label{G1536classe32}
\begin{array}{llllllll}
 \left\{4_1,0,0,\frac{1}{2}\right\} &
   \left\{4_1,0,0,\frac{3}{2}\right\} &
   \left\{4_1,0,\frac{1}{2},0\right\} &
   \left\{4_1,0,\frac{1}{2},1\right\} &
   \left\{4_1,0,1,\frac{1}{2}\right\} &
   \left\{4_1,0,1,\frac{3}{2}\right\} &
   \left\{4_1,0,\frac{3}{2},0\right\} &
   \left\{4_1,0,\frac{3}{2},1\right\} \\
 \left\{4_1,1,0,\frac{1}{2}\right\} &
   \left\{4_1,1,0,\frac{3}{2}\right\} &
   \left\{4_1,1,\frac{1}{2},0\right\} &
   \left\{4_1,1,\frac{1}{2},1\right\} &
   \left\{4_1,1,1,\frac{1}{2}\right\} &
   \left\{4_1,1,1,\frac{3}{2}\right\} &
   \left\{4_1,1,\frac{3}{2},0\right\} &
   \left\{4_1,1,\frac{3}{2},1\right\} \\
 \left\{4_2,0,0,\frac{1}{2}\right\} &
   \left\{4_2,0,0,\frac{3}{2}\right\} &
   \left\{4_2,0,\frac{1}{2},0\right\} &
   \left\{4_2,0,\frac{1}{2},1\right\} &
   \left\{4_2,0,1,\frac{1}{2}\right\} &
   \left\{4_2,0,1,\frac{3}{2}\right\} &
   \left\{4_2,0,\frac{3}{2},0\right\} &
   \left\{4_2,0,\frac{3}{2},1\right\} \\
 \left\{4_2,1,0,\frac{1}{2}\right\} &
   \left\{4_2,1,0,\frac{3}{2}\right\} &
   \left\{4_2,1,\frac{1}{2},0\right\} &
   \left\{4_2,1,\frac{1}{2},1\right\} &
   \left\{4_2,1,1,\frac{1}{2}\right\} &
   \left\{4_2,1,1,\frac{3}{2}\right\} &
   \left\{4_2,1,\frac{3}{2},0\right\} &
   \left\{4_2,1,\frac{3}{2},1\right\} \\
 \left\{4_3,0,\frac{1}{2},0\right\} &
   \left\{4_3,0,\frac{1}{2},1\right\} &
   \left\{4_3,0,\frac{3}{2},0\right\} &
   \left\{4_3,0,\frac{3}{2},1\right\} &
   \left\{4_3,\frac{1}{2},0,0\right\} &
   \left\{4_3,\frac{1}{2},0,1\right\} &
   \left\{4_3,\frac{1}{2},1,0\right\} &
   \left\{4_3,\frac{1}{2},1,1\right\} \\
 \left\{4_3,1,\frac{1}{2},0\right\} &
   \left\{4_3,1,\frac{1}{2},1\right\} &
   \left\{4_3,1,\frac{3}{2},0\right\} &
   \left\{4_3,1,\frac{3}{2},1\right\} &
   \left\{4_3,\frac{3}{2},0,0\right\} &
   \left\{4_3,\frac{3}{2},0,1\right\} &
   \left\{4_3,\frac{3}{2},1,0\right\} &
   \left\{4_3,\frac{3}{2},1,1\right\} \\
\end{array}
\end{equation}
}
\paragraph{Conjugacy class $\mathcal{C}_{33}\left(\mathrm{\mathrm{G_{1536}}}\right)$: $\#$ of elements = $96$}
{\scriptsize
\begin{equation}\label{G1536classe33}
\begin{array}{llllllll}
 \left\{4_1,\frac{1}{2},0,\frac{1}{2}\right\} &
   \left\{4_1,\frac{1}{2},0,\frac{3}{2}\right\} &
   \left\{4_1,\frac{1}{2},\frac{1}{2},0\right\} &
   \left\{4_1,\frac{1}{2},\frac{1}{2},1\right\} &
   \left\{4_1,\frac{1}{2},1,\frac{1}{2}\right\} &
   \left\{4_1,\frac{1}{2},1,\frac{3}{2}\right\} &
   \left\{4_1,\frac{1}{2},\frac{3}{2},0\right\} &
   \left\{4_1,\frac{1}{2},\frac{3}{2},1\right\} \\
 \left\{4_1,\frac{3}{2},0,\frac{1}{2}\right\} &
   \left\{4_1,\frac{3}{2},0,\frac{3}{2}\right\} &
   \left\{4_1,\frac{3}{2},\frac{1}{2},0\right\} &
   \left\{4_1,\frac{3}{2},\frac{1}{2},1\right\} &
   \left\{4_1,\frac{3}{2},1,\frac{1}{2}\right\} &
   \left\{4_1,\frac{3}{2},1,\frac{3}{2}\right\} &
   \left\{4_1,\frac{3}{2},\frac{3}{2},0\right\} &
   \left\{4_1,\frac{3}{2},\frac{3}{2},1\right\} \\
 \left\{4_2,\frac{1}{2},0,\frac{1}{2}\right\} &
   \left\{4_2,\frac{1}{2},0,\frac{3}{2}\right\} &
   \left\{4_2,\frac{1}{2},\frac{1}{2},0\right\} &
   \left\{4_2,\frac{1}{2},\frac{1}{2},1\right\} &
   \left\{4_2,\frac{1}{2},1,\frac{1}{2}\right\} &
   \left\{4_2,\frac{1}{2},1,\frac{3}{2}\right\} &
   \left\{4_2,\frac{1}{2},\frac{3}{2},0\right\} &
   \left\{4_2,\frac{1}{2},\frac{3}{2},1\right\} \\
 \left\{4_2,\frac{3}{2},0,\frac{1}{2}\right\} &
   \left\{4_2,\frac{3}{2},0,\frac{3}{2}\right\} &
   \left\{4_2,\frac{3}{2},\frac{1}{2},0\right\} &
   \left\{4_2,\frac{3}{2},\frac{1}{2},1\right\} &
   \left\{4_2,\frac{3}{2},1,\frac{1}{2}\right\} &
   \left\{4_2,\frac{3}{2},1,\frac{3}{2}\right\} &
   \left\{4_2,\frac{3}{2},\frac{3}{2},0\right\} &
   \left\{4_2,\frac{3}{2},\frac{3}{2},1\right\} \\
\end{array}
\end{equation}
}
\paragraph{Conjugacy class $\mathcal{C}_{34}\left(\mathrm{\mathrm{G_{1536}}}\right)$: $\#$ of elements = $128$}
{\scriptsize
\begin{equation}\label{G1536classe34}
\begin{array}{llllllll}
 \left\{2_1,0,0,0\right\} &
   \left\{2_1,0,\frac{1}{2},\frac{1}{2}\right\} &
   \left\{2_1,0,1,1\right\} &
   \left\{2_1,0,\frac{3}{2},\frac{3}{2}\right\} &
   \left\{2_1,\frac{1}{2},0,\frac{3}{2}\right\} &
   \left\{2_1,\frac{1}{2},\frac{1}{2},0\right\} &
   \left\{2_1,\frac{1}{2},1,\frac{1}{2}\right\} &
   \left\{2_1,\frac{1}{2},\frac{3}{2},1\right\} \\
 \left\{2_1,1,0,1\right\} &
   \left\{2_1,1,\frac{1}{2},\frac{3}{2}\right\} &
   \left\{2_1,1,1,0\right\} &
   \left\{2_1,1,\frac{3}{2},\frac{1}{2}\right\} &
   \left\{2_1,\frac{3}{2},0,\frac{1}{2}\right\} &
   \left\{2_1,\frac{3}{2},\frac{1}{2},1\right\} &
   \left\{2_1,\frac{3}{2},1,\frac{3}{2}\right\} &
   \left\{2_1,\frac{3}{2},\frac{3}{2},0\right\} \\
 \left\{2_2,0,0,0\right\} &
   \left\{2_2,0,\frac{1}{2},\frac{3}{2}\right\} &
   \left\{2_2,0,1,1\right\} &
   \left\{2_2,0,\frac{3}{2},\frac{1}{2}\right\} &
   \left\{2_2,\frac{1}{2},0,\frac{1}{2}\right\} &
   \left\{2_2,\frac{1}{2},\frac{1}{2},0\right\} &
   \left\{2_2,\frac{1}{2},1,\frac{3}{2}\right\} &
   \left\{2_2,\frac{1}{2},\frac{3}{2},1\right\} \\
 \left\{2_2,1,0,1\right\} &
   \left\{2_2,1,\frac{1}{2},\frac{1}{2}\right\} &
   \left\{2_2,1,1,0\right\} &
   \left\{2_2,1,\frac{3}{2},\frac{3}{2}\right\} &
   \left\{2_2,\frac{3}{2},0,\frac{3}{2}\right\} &
   \left\{2_2,\frac{3}{2},\frac{1}{2},1\right\} &
   \left\{2_2,\frac{3}{2},1,\frac{1}{2}\right\} &
   \left\{2_2,\frac{3}{2},\frac{3}{2},0\right\} \\
 \left\{2_3,0,0,0\right\} &
   \left\{2_3,0,\frac{1}{2},\frac{3}{2}\right\} &
   \left\{2_3,0,1,1\right\} &
   \left\{2_3,0,\frac{3}{2},\frac{1}{2}\right\} &
   \left\{2_3,\frac{1}{2},0,\frac{1}{2}\right\} &
   \left\{2_3,\frac{1}{2},\frac{1}{2},0\right\} &
   \left\{2_3,\frac{1}{2},1,\frac{3}{2}\right\} &
   \left\{2_3,\frac{1}{2},\frac{3}{2},1\right\} \\
\end{array}
\end{equation}
}
\paragraph{Conjugacy class $\mathcal{C}_{35}\left(\mathrm{\mathrm{G_{1536}}}\right)$: $\#$ of elements = $128$}
{\scriptsize
\begin{equation}\label{G1536classe35}
\begin{array}{llllllll}
 \left\{2_1,0,0,\frac{1}{2}\right\} &
   \left\{2_1,0,\frac{1}{2},1\right\} &
   \left\{2_1,0,1,\frac{3}{2}\right\} &
   \left\{2_1,0,\frac{3}{2},0\right\} &
   \left\{2_1,\frac{1}{2},0,0\right\} &
   \left\{2_1,\frac{1}{2},\frac{1}{2},\frac{1}{2}\right\}
   & \left\{2_1,\frac{1}{2},1,1\right\} &
   \left\{2_1,\frac{1}{2},\frac{3}{2},\frac{3}{2}\right\}
   \\
 \left\{2_1,1,0,\frac{3}{2}\right\} &
   \left\{2_1,1,\frac{1}{2},0\right\} &
   \left\{2_1,1,1,\frac{1}{2}\right\} &
   \left\{2_1,1,\frac{3}{2},1\right\} &
   \left\{2_1,\frac{3}{2},0,1\right\} &
   \left\{2_1,\frac{3}{2},\frac{1}{2},\frac{3}{2}\right\}
   & \left\{2_1,\frac{3}{2},1,0\right\} &
   \left\{2_1,\frac{3}{2},\frac{3}{2},\frac{1}{2}\right\}
   \\
 \left\{2_2,0,0,\frac{1}{2}\right\} &
   \left\{2_2,0,\frac{1}{2},0\right\} &
   \left\{2_2,0,1,\frac{3}{2}\right\} &
   \left\{2_2,0,\frac{3}{2},1\right\} &
   \left\{2_2,\frac{1}{2},0,1\right\} &
   \left\{2_2,\frac{1}{2},\frac{1}{2},\frac{1}{2}\right\}
   & \left\{2_2,\frac{1}{2},1,0\right\} &
   \left\{2_2,\frac{1}{2},\frac{3}{2},\frac{3}{2}\right\}
   \\
 \left\{2_2,1,0,\frac{3}{2}\right\} &
   \left\{2_2,1,\frac{1}{2},1\right\} &
   \left\{2_2,1,1,\frac{1}{2}\right\} &
   \left\{2_2,1,\frac{3}{2},0\right\} &
   \left\{2_2,\frac{3}{2},0,0\right\} &
   \left\{2_2,\frac{3}{2},\frac{1}{2},\frac{3}{2}\right\}
   & \left\{2_2,\frac{3}{2},1,1\right\} &
   \left\{2_2,\frac{3}{2},\frac{3}{2},\frac{1}{2}\right\}
   \\
 \left\{2_3,0,0,\frac{3}{2}\right\} &
   \left\{2_3,0,\frac{1}{2},1\right\} &
   \left\{2_3,0,1,\frac{1}{2}\right\} &
   \left\{2_3,0,\frac{3}{2},0\right\} &
   \left\{2_3,\frac{1}{2},0,0\right\} &
   \left\{2_3,\frac{1}{2},\frac{1}{2},\frac{3}{2}\right\}
   & \left\{2_3,\frac{1}{2},1,1\right\} &
   \left\{2_3,\frac{1}{2},\frac{3}{2},\frac{1}{2}\right\}
   \\
\end{array}
\end{equation}
}
\paragraph{Conjugacy class $\mathcal{C}_{36}\left(\mathrm{\mathrm{G_{1536}}}\right)$: $\#$-elements = $128$}
{\scriptsize
\begin{equation}\label{G1536classe36}
\begin{array}{llllllll}
 \left\{2_1,0,0,1\right\} &
   \left\{2_1,0,\frac{1}{2},\frac{3}{2}\right\} &
   \left\{2_1,0,1,0\right\} &
   \left\{2_1,0,\frac{3}{2},\frac{1}{2}\right\} &
   \left\{2_1,\frac{1}{2},0,\frac{1}{2}\right\} &
   \left\{2_1,\frac{1}{2},\frac{1}{2},1\right\} &
   \left\{2_1,\frac{1}{2},1,\frac{3}{2}\right\} &
   \left\{2_1,\frac{1}{2},\frac{3}{2},0\right\} \\
 \left\{2_1,1,0,0\right\} &
   \left\{2_1,1,\frac{1}{2},\frac{1}{2}\right\} &
   \left\{2_1,1,1,1\right\} &
   \left\{2_1,1,\frac{3}{2},\frac{3}{2}\right\} &
   \left\{2_1,\frac{3}{2},0,\frac{3}{2}\right\} &
   \left\{2_1,\frac{3}{2},\frac{1}{2},0\right\} &
   \left\{2_1,\frac{3}{2},1,\frac{1}{2}\right\} &
   \left\{2_1,\frac{3}{2},\frac{3}{2},1\right\} \\
 \left\{2_2,0,0,1\right\} &
   \left\{2_2,0,\frac{1}{2},\frac{1}{2}\right\} &
   \left\{2_2,0,1,0\right\} &
   \left\{2_2,0,\frac{3}{2},\frac{3}{2}\right\} &
   \left\{2_2,\frac{1}{2},0,\frac{3}{2}\right\} &
   \left\{2_2,\frac{1}{2},\frac{1}{2},1\right\} &
   \left\{2_2,\frac{1}{2},1,\frac{1}{2}\right\} &
   \left\{2_2,\frac{1}{2},\frac{3}{2},0\right\} \\
 \left\{2_2,1,0,0\right\} &
   \left\{2_2,1,\frac{1}{2},\frac{3}{2}\right\} &
   \left\{2_2,1,1,1\right\} &
   \left\{2_2,1,\frac{3}{2},\frac{1}{2}\right\} &
   \left\{2_2,\frac{3}{2},0,\frac{1}{2}\right\} &
   \left\{2_2,\frac{3}{2},\frac{1}{2},0\right\} &
   \left\{2_2,\frac{3}{2},1,\frac{3}{2}\right\} &
   \left\{2_2,\frac{3}{2},\frac{3}{2},1\right\} \\
 \left\{2_3,0,0,1\right\} &
   \left\{2_3,0,\frac{1}{2},\frac{1}{2}\right\} &
   \left\{2_3,0,1,0\right\} &
   \left\{2_3,0,\frac{3}{2},\frac{3}{2}\right\} &
   \left\{2_3,\frac{1}{2},0,\frac{3}{2}\right\} &
   \left\{2_3,\frac{1}{2},\frac{1}{2},1\right\} &
   \left\{2_3,\frac{1}{2},1,\frac{1}{2}\right\} &
   \left\{2_3,\frac{1}{2},\frac{3}{2},0\right\} \\
\end{array}
\end{equation}
}
\paragraph{Conjugacy class $\mathcal{C}_{37}\left(\mathrm{\mathrm{G_{1536}}}\right)$: $\#$-elements = $128$}
{\scriptsize
\begin{equation}\label{G1536classe37}
\begin{array}{llllllll}
 \left\{2_1,0,0,\frac{3}{2}\right\} &
   \left\{2_1,0,\frac{1}{2},0\right\} &
   \left\{2_1,0,1,\frac{1}{2}\right\} &
   \left\{2_1,0,\frac{3}{2},1\right\} &
   \left\{2_1,\frac{1}{2},0,1\right\} &
   \left\{2_1,\frac{1}{2},\frac{1}{2},\frac{3}{2}\right\}
   & \left\{2_1,\frac{1}{2},1,0\right\} &
   \left\{2_1,\frac{1}{2},\frac{3}{2},\frac{1}{2}\right\}
   \\
 \left\{2_1,1,0,\frac{1}{2}\right\} &
   \left\{2_1,1,\frac{1}{2},1\right\} &
   \left\{2_1,1,1,\frac{3}{2}\right\} &
   \left\{2_1,1,\frac{3}{2},0\right\} &
   \left\{2_1,\frac{3}{2},0,0\right\} &
   \left\{2_1,\frac{3}{2},\frac{1}{2},\frac{1}{2}\right\}
   & \left\{2_1,\frac{3}{2},1,1\right\} &
   \left\{2_1,\frac{3}{2},\frac{3}{2},\frac{3}{2}\right\}
   \\
 \left\{2_2,0,0,\frac{3}{2}\right\} &
   \left\{2_2,0,\frac{1}{2},1\right\} &
   \left\{2_2,0,1,\frac{1}{2}\right\} &
   \left\{2_2,0,\frac{3}{2},0\right\} &
   \left\{2_2,\frac{1}{2},0,0\right\} &
   \left\{2_2,\frac{1}{2},\frac{1}{2},\frac{3}{2}\right\}
   & \left\{2_2,\frac{1}{2},1,1\right\} &
   \left\{2_2,\frac{1}{2},\frac{3}{2},\frac{1}{2}\right\}
   \\
 \left\{2_2,1,0,\frac{1}{2}\right\} &
   \left\{2_2,1,\frac{1}{2},0\right\} &
   \left\{2_2,1,1,\frac{3}{2}\right\} &
   \left\{2_2,1,\frac{3}{2},1\right\} &
   \left\{2_2,\frac{3}{2},0,1\right\} &
   \left\{2_2,\frac{3}{2},\frac{1}{2},\frac{1}{2}\right\}
   & \left\{2_2,\frac{3}{2},1,0\right\} &
   \left\{2_2,\frac{3}{2},\frac{3}{2},\frac{3}{2}\right\}
   \\
 \left\{2_3,0,0,\frac{1}{2}\right\} &
   \left\{2_3,0,\frac{1}{2},0\right\} &
   \left\{2_3,0,1,\frac{3}{2}\right\} &
   \left\{2_3,0,\frac{3}{2},1\right\} &
   \left\{2_3,\frac{1}{2},0,1\right\} &
   \left\{2_3,\frac{1}{2},\frac{1}{2},\frac{1}{2}\right\}
   & \left\{2_3,\frac{1}{2},1,0\right\} &
   \left\{2_3,\frac{1}{2},\frac{3}{2},\frac{3}{2}\right\}
   \\
\end{array}
\end{equation}
}
\newpage
\subsection{The Group $\mathrm{G_{768}}$}
\label{coniugato768}
In this section we list all the elements of the space group $\mathrm{G_{768}}$, organized into their $32$ conjugacy classes.
\paragraph{Conjugacy class $\mathcal{C}_{1}\left(\mathrm{G_{768}}\right)$: $\#$ of elements = $1$}
\begin{equation}\label{G768classe1}
\begin{array}{l}
 \left\{1_1,0,0,0\right\}
\end{array}
\end{equation}
\paragraph{Conjugacy class $\mathcal{C}_{2}\left(\mathrm{G_{768}}\right)$: $\#$ of elements = $1$}
\begin{equation}\label{G768classe2}
\begin{array}{l}
 \left\{1_1,1,1,1\right\}
\end{array}
\end{equation}
\paragraph{Conjugacy class $\mathcal{C}_{3}\left(\mathrm{G_{768}}\right)$: $\#$ of elements = $3$}
\begin{equation}\label{G768classe3}
\begin{array}{lll}
 \left\{1_1,0,0,1\right\} & \left\{1_1,0,1,0\right\} &
   \left\{1_1,1,0,0\right\}
\end{array}
\end{equation}
\paragraph{Conjugacy class $\mathcal{C}_{4}\left(\mathrm{G_{768}}\right)$: $\#$ of elements = $3$}
\begin{equation}\label{G768classe4}
\begin{array}{lll}
 \left\{1_1,0,1,1\right\} & \left\{1_1,1,0,1\right\} &
   \left\{1_1,1,1,0\right\}
\end{array}
\end{equation}
\paragraph{Conjugacy class $\mathcal{C}_{5}\left(\mathrm{G_{768}}\right)$: $\#$ of elements = $4$}
\begin{equation}\label{G768classe5}
\begin{array}{llll}
 \left\{1_1,\frac{1}{2},\frac{1}{2},\frac{1}{2}\right\} &
   \left\{1_1,\frac{1}{2},\frac{3}{2},\frac{3}{2}\right\}
   &
   \left\{1_1,\frac{3}{2},\frac{1}{2},\frac{3}{2}\right\}
   &
   \left\{1_1,\frac{3}{2},\frac{3}{2},\frac{1}{2}\right\}
\end{array}
\end{equation}
\paragraph{Conjugacy class $\mathcal{C}_{6}\left(\mathrm{G_{768}}\right)$: $\#$ of elements = $4$}
\begin{equation}\label{G768classe6}
\begin{array}{llll}
 \left\{1_1,\frac{1}{2},\frac{1}{2},\frac{3}{2}\right\} &
   \left\{1_1,\frac{1}{2},\frac{3}{2},\frac{1}{2}\right\}
   &
   \left\{1_1,\frac{3}{2},\frac{1}{2},\frac{1}{2}\right\}
   &
   \left\{1_1,\frac{3}{2},\frac{3}{2},\frac{3}{2}\right\}
\end{array}
\end{equation}
\paragraph{Conjugacy class $\mathcal{C}_{7}\left(\mathrm{G_{768}}\right)$: $\#$ of elements = $6$}
\begin{equation}\label{G768classe7}
\begin{array}{lll}
 \left\{1_1,0,0,\frac{1}{2}\right\} &
   \left\{1_1,0,0,\frac{3}{2}\right\} &
   \left\{1_1,0,\frac{1}{2},0\right\} \\
 \left\{1_1,0,\frac{3}{2},0\right\} &
   \left\{1_1,\frac{1}{2},0,0\right\} &
   \left\{1_1,\frac{3}{2},0,0\right\}
\end{array}
\end{equation}
\paragraph{Conjugacy class $\mathcal{C}_{8}\left(\mathrm{G_{768}}\right)$: $\#$ of elements = $6$}
\begin{equation}\label{G768classe8}
\begin{array}{lll}
 \left\{1_1,0,\frac{1}{2},1\right\} &
   \left\{1_1,0,\frac{3}{2},1\right\} &
   \left\{1_1,\frac{1}{2},1,0\right\} \\
 \left\{1_1,1,0,\frac{1}{2}\right\} &
   \left\{1_1,1,0,\frac{3}{2}\right\} &
   \left\{1_1,\frac{3}{2},1,0\right\}
\end{array}
\end{equation}
\paragraph{Conjugacy class $\mathcal{C}_{9}\left(\mathrm{G_{768}}\right)$: $\#$ of elements = $6$}
\begin{equation}\label{G768classe9}
\begin{array}{lll}
 \left\{1_1,0,1,\frac{1}{2}\right\} &
   \left\{1_1,0,1,\frac{3}{2}\right\} &
   \left\{1_1,\frac{1}{2},0,1\right\} \\
 \left\{1_1,1,\frac{1}{2},0\right\} &
   \left\{1_1,1,\frac{3}{2},0\right\} &
   \left\{1_1,\frac{3}{2},0,1\right\}
\end{array}
\end{equation}
\paragraph{Conjugacy class $\mathcal{C}_{10}\left(\mathrm{G_{768}}\right)$: $\#$ of elements = $6$}
\begin{equation}\label{G768classe10}
\begin{array}{lll}
 \left\{1_1,\frac{1}{2},1,1\right\} &
   \left\{1_1,1,\frac{1}{2},1\right\} &
   \left\{1_1,1,1,\frac{1}{2}\right\} \\
 \left\{1_1,1,1,\frac{3}{2}\right\} &
   \left\{1_1,1,\frac{3}{2},1\right\} &
   \left\{1_1,\frac{3}{2},1,1\right\}
\end{array}
\end{equation}
\paragraph{Conjugacy class $\mathcal{C}_{11}\left(\mathrm{G_{768}}\right)$: $\#$ of elements = $12$}
{\scriptsize
\begin{equation}\label{G768classe11}
\begin{array}{llllll}
 \left\{1_1,0,\frac{1}{2},\frac{1}{2}\right\} &
   \left\{1_1,0,\frac{1}{2},\frac{3}{2}\right\} &
   \left\{1_1,0,\frac{3}{2},\frac{1}{2}\right\} &
   \left\{1_1,0,\frac{3}{2},\frac{3}{2}\right\} &
   \left\{1_1,\frac{1}{2},0,\frac{1}{2}\right\} &
   \left\{1_1,\frac{1}{2},0,\frac{3}{2}\right\} \\
 \left\{1_1,\frac{1}{2},\frac{1}{2},0\right\} &
   \left\{1_1,\frac{1}{2},\frac{3}{2},0\right\} &
   \left\{1_1,\frac{3}{2},0,\frac{1}{2}\right\} &
   \left\{1_1,\frac{3}{2},0,\frac{3}{2}\right\} &
   \left\{1_1,\frac{3}{2},\frac{1}{2},0\right\} &
   \left\{1_1,\frac{3}{2},\frac{3}{2},0\right\}
\end{array}
\end{equation}
}
\paragraph{Conjugacy class $\mathcal{C}_{12}\left(\mathrm{G_{768}}\right)$: $\#$ of elements = $12$}
{\scriptsize
\begin{equation}\label{G768classe12}
\begin{array}{llllll}
 \left\{1_1,\frac{1}{2},\frac{1}{2},1\right\} &
   \left\{1_1,\frac{1}{2},1,\frac{1}{2}\right\} &
   \left\{1_1,\frac{1}{2},1,\frac{3}{2}\right\} &
   \left\{1_1,\frac{1}{2},\frac{3}{2},1\right\} &
   \left\{1_1,1,\frac{1}{2},\frac{1}{2}\right\} &
   \left\{1_1,1,\frac{1}{2},\frac{3}{2}\right\} \\
 \left\{1_1,1,\frac{3}{2},\frac{1}{2}\right\} &
   \left\{1_1,1,\frac{3}{2},\frac{3}{2}\right\} &
   \left\{1_1,\frac{3}{2},\frac{1}{2},1\right\} &
   \left\{1_1,\frac{3}{2},1,\frac{1}{2}\right\} &
   \left\{1_1,\frac{3}{2},1,\frac{3}{2}\right\} &
   \left\{1_1,\frac{3}{2},\frac{3}{2},1\right\}
\end{array}
\end{equation}
}
\paragraph{Conjugacy class $\mathcal{C}_{13}\left(\mathrm{G_{768}}\right)$: $\#$ of elements = $12$}
{\scriptsize
\begin{equation}\label{G768classe13}
\begin{array}{llllll}
 \left\{3_1,0,0,0\right\} & \left\{3_1,0,1,0\right\} &
   \left\{3_1,1,0,0\right\} & \left\{3_1,1,1,0\right\} &
   \left\{3_2,0,0,0\right\} & \left\{3_2,0,0,1\right\} \\
 \left\{3_2,1,0,0\right\} & \left\{3_2,1,0,1\right\} &
   \left\{3_3,0,0,0\right\} & \left\{3_3,0,0,1\right\} &
   \left\{3_3,0,1,0\right\} & \left\{3_3,0,1,1\right\}
\end{array}
\end{equation}
}
\paragraph{Conjugacy class $\mathcal{C}_{14}\left(\mathrm{G_{768}}\right)$: $\#$ of elements = $12$}
{\scriptsize
\begin{equation}\label{G768classe14}
\begin{array}{llllll}
 \left\{3_1,0,0,1\right\} & \left\{3_1,0,1,1\right\} &
   \left\{3_1,1,0,1\right\} & \left\{3_1,1,1,1\right\} &
   \left\{3_2,0,1,0\right\} & \left\{3_2,0,1,1\right\} \\
 \left\{3_2,1,1,0\right\} & \left\{3_2,1,1,1\right\} &
   \left\{3_3,1,0,0\right\} & \left\{3_3,1,0,1\right\} &
   \left\{3_3,1,1,0\right\} & \left\{3_3,1,1,1\right\}
\end{array}
\end{equation}
}
\paragraph{Conjugacy class $\mathcal{C}_{15}\left(\mathrm{G_{768}}\right)$: $\#$ of elements = $12$}
{\scriptsize
\begin{equation}\label{G768classe15}
\begin{array}{llllll}
 \left\{3_1,0,\frac{1}{2},0\right\} &
   \left\{3_1,0,\frac{3}{2},0\right\} &
   \left\{3_1,1,\frac{1}{2},0\right\} &
   \left\{3_1,1,\frac{3}{2},0\right\} &
   \left\{3_2,\frac{1}{2},0,0\right\} &
   \left\{3_2,\frac{1}{2},0,1\right\} \\
 \left\{3_2,\frac{3}{2},0,0\right\} &
   \left\{3_2,\frac{3}{2},0,1\right\} &
   \left\{3_3,0,0,\frac{1}{2}\right\} &
   \left\{3_3,0,0,\frac{3}{2}\right\} &
   \left\{3_3,0,1,\frac{1}{2}\right\} &
   \left\{3_3,0,1,\frac{3}{2}\right\}
\end{array}
\end{equation}
}
\paragraph{Conjugacy class $\mathcal{C}_{16}\left(\mathrm{G_{768}}\right)$: $\#$ of elements = $12$}
{\scriptsize
\begin{equation}\label{G768classe16}
\begin{array}{llllll}
 \left\{3_1,0,\frac{1}{2},1\right\} &
   \left\{3_1,0,\frac{3}{2},1\right\} &
   \left\{3_1,1,\frac{1}{2},1\right\} &
   \left\{3_1,1,\frac{3}{2},1\right\} &
   \left\{3_2,\frac{1}{2},1,0\right\} &
   \left\{3_2,\frac{1}{2},1,1\right\} \\
 \left\{3_2,\frac{3}{2},1,0\right\} &
   \left\{3_2,\frac{3}{2},1,1\right\} &
   \left\{3_3,1,0,\frac{1}{2}\right\} &
   \left\{3_3,1,0,\frac{3}{2}\right\} &
   \left\{3_3,1,1,\frac{1}{2}\right\} &
   \left\{3_3,1,1,\frac{3}{2}\right\}
\end{array}
\end{equation}
}
\paragraph{Conjugacy class $\mathcal{C}_{17}\left(\mathrm{G_{768}}\right)$: $\#$ of elements = $12$}
{\scriptsize
\begin{equation}\label{G768classe17}
\begin{array}{llllll}
 \left\{3_1,\frac{1}{2},0,0\right\} &
   \left\{3_1,\frac{1}{2},1,0\right\} &
   \left\{3_1,\frac{3}{2},0,0\right\} &
   \left\{3_1,\frac{3}{2},1,0\right\} &
   \left\{3_2,0,0,\frac{1}{2}\right\} &
   \left\{3_2,0,0,\frac{3}{2}\right\} \\
 \left\{3_2,1,0,\frac{1}{2}\right\} &
   \left\{3_2,1,0,\frac{3}{2}\right\} &
   \left\{3_3,0,\frac{1}{2},0\right\} &
   \left\{3_3,0,\frac{1}{2},1\right\} &
   \left\{3_3,0,\frac{3}{2},0\right\} &
   \left\{3_3,0,\frac{3}{2},1\right\}
\end{array}
\end{equation}
}
\paragraph{Conjugacy class $\mathcal{C}_{18}\left(\mathrm{G_{768}}\right)$: $\#$ of elements = $12$}
{\scriptsize
\begin{equation}\label{G768classe18}
\begin{array}{llllll}
 \left\{3_1,\frac{1}{2},0,1\right\} &
   \left\{3_1,\frac{1}{2},1,1\right\} &
   \left\{3_1,\frac{3}{2},0,1\right\} &
   \left\{3_1,\frac{3}{2},1,1\right\} &
   \left\{3_2,0,1,\frac{1}{2}\right\} &
   \left\{3_2,0,1,\frac{3}{2}\right\} \\
 \left\{3_2,1,1,\frac{1}{2}\right\} &
   \left\{3_2,1,1,\frac{3}{2}\right\} &
   \left\{3_3,1,\frac{1}{2},0\right\} &
   \left\{3_3,1,\frac{1}{2},1\right\} &
   \left\{3_3,1,\frac{3}{2},0\right\} &
   \left\{3_3,1,\frac{3}{2},1\right\}
\end{array}
\end{equation}
}
\paragraph{Conjugacy class $\mathcal{C}_{19}\left(\mathrm{G_{768}}\right)$: $\#$ of elements = $12$}
{\scriptsize
\begin{equation}\label{G768classe19}
%
\end{equation}
}
\paragraph{Conjugacy class $\mathcal{C}_{20}\left(\mathrm{G_{768}}\right)$: $\#$ of elements = $12$}
{\scriptsize
\begin{equation}\label{G768classe20}
%
\end{equation}
}
\paragraph{Conjugacy class $\mathcal{C}_{21}\left(\mathrm{G_{768}}\right)$: $\#$ of elements = $24$}
{\scriptsize
\begin{equation}\label{G768classe21}
%
\end{equation}
}
\paragraph{Conjugacy class $\mathcal{C}_{22}\left(\mathrm{G_{768}}\right)$: $\#$ of elements = $24$}
{\scriptsize
\begin{equation}\label{G768classe22}
%
\end{equation}
}
\paragraph{Conjugacy class $\mathcal{C}_{23}\left(\mathrm{G_{768}}\right)$: $\#$ of elements = $24$}
{\scriptsize
\begin{equation}\label{G768classe23}
%
\end{equation}
}
\paragraph{Conjugacy class $\mathcal{C}_{24}\left(\mathrm{G_{768}}\right)$: $\#$ of elements = $24$}
{\scriptsize
\begin{equation}\label{G768classe24}
%
\end{equation}
}
\paragraph{Conjugacy class $\mathcal{C}_{25}\left(\mathrm{G_{768}}\right)$: $\#$ of elements = $64$}
{\scriptsize
\begin{equation}\label{G768classe25}
%
\end{equation}
}
\paragraph{Conjugacy class $\mathcal{C}_{26}\left(\mathrm{G_{768}}\right)$: $\#$ of elements = $64$}
{\scriptsize
\begin{equation}\label{G768classe26}
%
\end{equation}
}
\paragraph{Conjugacy class $\mathcal{C}_{27}\left(\mathrm{G_{768}}\right)$: $\#$ of elements = $64$}
{\scriptsize
\begin{equation}\label{G768classe27}
%
\end{equation}
}
\paragraph{Conjugacy class $\mathcal{C}_{28}\left(\mathrm{G_{768}}\right)$: $\#$ of elements = $64$}
{\scriptsize
\begin{equation}\label{G768classe28}
%
\end{equation}
}
\paragraph{Conjugacy class $\mathcal{C}_{29}\left(\mathrm{G_{768}}\right)$: $\#$ of elements = $64$}
{\scriptsize
\begin{equation}\label{G768classe29}
%
\end{equation}
}
\paragraph{Conjugacy class $\mathcal{C}_{30}\left(\mathrm{G_{768}}\right)$: $\#$ of elements = $64$}
{\scriptsize
\begin{equation}\label{G768classe30}
%
\end{equation}
}
\paragraph{Conjugacy class $\mathcal{C}_{31}\left(\mathrm{G_{768}}\right)$: $\#$ of elements = $64$}
{\scriptsize
\begin{equation}\label{G768classe31}
%
\end{equation}
}
\paragraph{Conjugacy class $\mathcal{C}_{32}\left(\mathrm{G_{768}}\right)$: $\#$ of elements = $64$}
{\scriptsize
\begin{equation}\label{G768classe32}
%
\end{equation}
}
\subsection{The Group $\mathrm{G_{256}}$}
\label{coniugato256}
In this section we list all the elements of the space group $\mathrm{G_{256}}$, organized into their $64$ conjugacy classes.
\paragraph{Conjugacy class $\mathcal{C}_{1}\left(\mathrm{G_{256}}\right)$: $\#$ of elements = $1$}
\begin{equation}\label{G256classe1}
\begin{array}{l}
 \left\{1_1,0,0,0\right\}
\end{array}
\end{equation}
\paragraph{Conjugacy class $\mathcal{C}_{2}\left(\mathrm{G_{256}}\right)$: $\#$ of elements = $1$}
\begin{equation}\label{G256classe2}
\begin{array}{l}
 \left\{1_1,0,0,1\right\}
\end{array}
\end{equation}
\paragraph{Conjugacy class $\mathcal{C}_{3}\left(\mathrm{G_{256}}\right)$: $\#$ of elements = $1$}
\begin{equation}\label{G256classe3}
\begin{array}{l}
 \left\{1_1,0,1,0\right\}
\end{array}
\end{equation}
\paragraph{Conjugacy class $\mathcal{C}_{4}\left(\mathrm{G_{256}}\right)$: $\#$ of elements = $1$}
\begin{equation}\label{G256classe4}
\begin{array}{l}
 \left\{1_1,0,1,1\right\}
\end{array}
\end{equation}
\paragraph{Conjugacy class $\mathcal{C}_{5}\left(\mathrm{G_{256}}\right)$: $\#$ of elements = $1$}
\begin{equation}\label{G256classe5}
\begin{array}{l}
 \left\{1_1,1,0,0\right\}
\end{array}
\end{equation}
\paragraph{Conjugacy class $\mathcal{C}_{6}\left(\mathrm{G_{256}}\right)$: $\#$ of elements = $1$}
\begin{equation}\label{G256classe6}
\begin{array}{l}
 \left\{1_1,1,0,1\right\}
\end{array}
\end{equation}
\paragraph{Conjugacy class $\mathcal{C}_{7}\left(\mathrm{G_{256}}\right)$: $\#$ of elements = $1$}
\begin{equation}\label{G256classe7}
\begin{array}{l}
 \left\{1_1,1,1,0\right\}
\end{array}
\end{equation}
\paragraph{Conjugacy class $\mathcal{C}_{8}\left(\mathrm{G_{256}}\right)$: $\#$ of elements = $1$}
\begin{equation}\label{G256classe8}
%
\end{equation}
\paragraph{Conjugacy class $\mathcal{C}_{9}\left(\mathrm{G_{256}}\right)$: $\#$ of elements = $2$}
\begin{equation}\label{G256classe9}
%
\end{equation}
\paragraph{Conjugacy class $\mathcal{C}_{10}\left(\mathrm{G_{256}}\right)$: $\#$ of elements = $2$}
\begin{equation}\label{G256classe10}
%
\end{equation}
\paragraph{Conjugacy class $\mathcal{C}_{11}\left(\mathrm{G_{256}}\right)$: $\#$ of elements = $2$}
\begin{equation}\label{G256classe11}
%
\end{equation}
\paragraph{Conjugacy class $\mathcal{C}_{12}\left(\mathrm{G_{256}}\right)$: $\#$ of elements = $2$}
\begin{equation}\label{G256classe12}
%
\end{equation}
\paragraph{Conjugacy class $\mathcal{C}_{13}\left(\mathrm{G_{256}}\right)$: $\#$ of elements = $2$}
\begin{equation}\label{G256classe13}
%
\end{equation}
\paragraph{Conjugacy class $\mathcal{C}_{14}\left(\mathrm{G_{256}}\right)$: $\#$ of elements = $2$}
\begin{equation}\label{G256classe14}
%
\end{equation}
\paragraph{Conjugacy class $\mathcal{C}_{15}\left(\mathrm{G_{256}}\right)$: $\#$ of elements = $2$}
\begin{equation}\label{G256classe15}
%
\end{equation}
\paragraph{Conjugacy class $\mathcal{C}_{16}\left(\mathrm{G_{256}}\right)$: $\#$ of elements = $2$}
\begin{equation}\label{G256classe16}
%
\end{equation}
\paragraph{Conjugacy class $\mathcal{C}_{17}\left(\mathrm{G_{256}}\right)$: $\#$ of elements = $2$}
\begin{equation}\label{G256classe17}
%
\end{equation}
\paragraph{Conjugacy class $\mathcal{C}_{18}\left(\mathrm{G_{256}}\right)$: $\#$ of elements = $2$}
\begin{equation}\label{G256classe18}
%
\end{equation}
\paragraph{Conjugacy class $\mathcal{C}_{19}\left(\mathrm{G_{256}}\right)$: $\#$ of elements = $2$}
\begin{equation}\label{G256classe19}
%
\end{equation}
\paragraph{Conjugacy class $\mathcal{C}_{20}\left(\mathrm{G_{256}}\right)$: $\#$ of elements = $2$}
\begin{equation}\label{G256classe20}
%
\end{equation}
\paragraph{Conjugacy class $\mathcal{C}_{21}\left(\mathrm{G_{256}}\right)$: $\#$ of elements = $4$}
\begin{equation}\label{G256classe21}
%
\end{equation}
\paragraph{Conjugacy class $\mathcal{C}_{22}\left(\mathrm{G_{256}}\right)$: $\#$ of elements = $4$}
\begin{equation}\label{G256classe22}
%
\end{equation}
\paragraph{Conjugacy class $\mathcal{C}_{23}\left(\mathrm{G_{256}}\right)$: $\#$ of elements = $4$}
\begin{equation}\label{G256classe23}
%
\end{equation}
\paragraph{Conjugacy class $\mathcal{C}_{24}\left(\mathrm{G_{256}}\right)$: $\#$ of elements = $4$}
\begin{equation}\label{G256classe24}
%
\end{equation}
\paragraph{Conjugacy class $\mathcal{C}_{25}\left(\mathrm{G_{256}}\right)$: $\#$ of elements = $4$}
\begin{equation}\label{G256classe25}
%
\end{equation}
\paragraph{Conjugacy class $\mathcal{C}_{26}\left(\mathrm{G_{256}}\right)$: $\#$ of elements = $4$}
\begin{equation}\label{G256classe26}
%
\end{equation}
\paragraph{Conjugacy class $\mathcal{C}_{27}\left(\mathrm{G_{256}}\right)$: $\#$ of elements = $4$}
\begin{equation}\label{G256classe27}
%
\end{equation}
\paragraph{Conjugacy class $\mathcal{C}_{28}\left(\mathrm{G_{256}}\right)$: $\#$ of elements = $4$}
\begin{equation}\label{G256classe28}
%
\end{equation}
\paragraph{Conjugacy class $\mathcal{C}_{29}\left(\mathrm{G_{256}}\right)$: $\#$ of elements = $4$}
\begin{equation}\label{G256classe29}
%
\end{equation}
\paragraph{Conjugacy class $\mathcal{C}_{30}\left(\mathrm{G_{256}}\right)$: $\#$ of elements = $4$}
\begin{equation}\label{G256classe30}
%
\end{equation}
\paragraph{Conjugacy class $\mathcal{C}_{31}\left(\mathrm{G_{256}}\right)$: $\#$ of elements = $4$}
\begin{equation}\label{G256classe31}
%
\end{equation}
\paragraph{Conjugacy class $\mathcal{C}_{32}\left(\mathrm{G_{256}}\right)$: $\#$ of elements = $4$}
\begin{equation}\label{G256classe32}
%
\end{equation}
\paragraph{Conjugacy class $\mathcal{C}_{33}\left(\mathrm{G_{256}}\right)$: $\#$ of elements = $4$}
\begin{equation}\label{G256classe33}
%
\end{equation}
\paragraph{Conjugacy class $\mathcal{C}_{34}\left(\mathrm{G_{256}}\right)$: $\#$ of elements = $4$}
\begin{equation}\label{G256classe34}
%
\end{equation}
\paragraph{Conjugacy class $\mathcal{C}_{35}\left(\mathrm{G_{256}}\right)$: $\#$ of elements = $4$}
\begin{equation}\label{G256classe35}
%
\end{equation}
\paragraph{Conjugacy class $\mathcal{C}_{36}\left(\mathrm{G_{256}}\right)$: $\#$ of elements = $4$}
\begin{equation}\label{G256classe36}
%
\end{equation}
\paragraph{Conjugacy class $\mathcal{C}_{37}\left(\mathrm{G_{256}}\right)$: $\#$ of elements = $4$}
\begin{equation}\label{G256classe37}
%
\end{equation}
\paragraph{Conjugacy class $\mathcal{C}_{38}\left(\mathrm{G_{256}}\right)$: $\#$ of elements = $4$}
\begin{equation}\label{G256classe38}
%
\end{equation}
\paragraph{Conjugacy class $\mathcal{C}_{39}\left(\mathrm{G_{256}}\right)$: $\#$ of elements = $4$}
\begin{equation}\label{G256classe39}
%
\end{equation}
\paragraph{Conjugacy class $\mathcal{C}_{40}\left(\mathrm{G_{256}}\right)$: $\#$ of elements = $4$}
\begin{equation}\label{G256classe40}
%
\end{equation}
\paragraph{Conjugacy class $\mathcal{C}_{41}\left(\mathrm{G_{256}}\right)$: $\#$ of elements = $4$}
\begin{equation}\label{G256classe41}
%
\end{equation}
\paragraph{Conjugacy class $\mathcal{C}_{42}\left(\mathrm{G_{256}}\right)$: $\#$ of elements = $4$}
\begin{equation}\label{G256classe42}
%
\end{equation}
\paragraph{Conjugacy class $\mathcal{C}_{43}\left(\mathrm{G_{256}}\right)$: $\#$ of elements = $4$}
\begin{equation}\label{G256classe43}
%
\end{equation}
\paragraph{Conjugacy class $\mathcal{C}_{44}\left(\mathrm{G_{256}}\right)$: $\#$ of elements = $4$}
\begin{equation}\label{G256classe44}
%
\end{equation}
\paragraph{Conjugacy class $\mathcal{C}_{45}\left(\mathrm{G_{256}}\right)$: $\#$ of elements = $4$}
\begin{equation}\label{G256classe45}
%
\end{equation}
\paragraph{Conjugacy class $\mathcal{C}_{46}\left(\mathrm{G_{256}}\right)$: $\#$ of elements = $4$}
\begin{equation}\label{G256classe46}
%
\end{equation}
\paragraph{Conjugacy class $\mathcal{C}_{47}\left(\mathrm{G_{256}}\right)$: $\#$ of elements = $4$}
\begin{equation}\label{G256classe47}
%
\end{equation}
\paragraph{Conjugacy class $\mathcal{C}_{48}\left(\mathrm{G_{256}}\right)$: $\#$ of elements = $4$}
\begin{equation}\label{G256classe48}
%
\end{equation}
\paragraph{Conjugacy class $\mathcal{C}_{49}\left(\mathrm{G_{256}}\right)$: $\#$ of elements = $4$}
\begin{equation}\label{G256classe49}
%
\end{equation}
\paragraph{Conjugacy class $\mathcal{C}_{50}\left(\mathrm{G_{256}}\right)$: $\#$ of elements = $4$}
\begin{equation}\label{G256classe50}
%
\end{equation}
\paragraph{Conjugacy class $\mathcal{C}_{51}\left(\mathrm{G_{256}}\right)$: $\#$ of elements = $4$}
\begin{equation}\label{G256classe51}
%
\end{equation}
\paragraph{Conjugacy class $\mathcal{C}_{52}\left(\mathrm{G_{256}}\right)$: $\#$ of elements = $4$}
\begin{equation}\label{G256classe52}
%
\end{equation}
\paragraph{Conjugacy class $\mathcal{C}_{53}\left(\mathrm{G_{256}}\right)$: $\#$ of elements = $8$}
{\scriptsize
\begin{equation}\label{G256classe53}
%
\end{equation}
}
\paragraph{Conjugacy class $\mathcal{C}_{54}\left(\mathrm{G_{256}}\right)$: $\#$ of elements = $8$}
{\scriptsize
\begin{equation}\label{G256classe54}
%
\end{equation}
}
\paragraph{Conjugacy class $\mathcal{C}_{55}\left(\mathrm{G_{256}}\right)$: $\#$ of elements = $8$}
{\scriptsize
\begin{equation}\label{G256classe55}
%
\end{equation}
}
\paragraph{Conjugacy class $\mathcal{C}_{56}\left(\mathrm{G_{256}}\right)$: $\#$ of elements = $8$}
{\scriptsize
\begin{equation}\label{G256classe56}
%
\end{equation}
}
\paragraph{Conjugacy class $\mathcal{C}_{57}\left(\mathrm{G_{256}}\right)$: $\#$ of elements = $8$}
{\scriptsize
\begin{equation}\label{G256classe57}
%
\end{equation}
}
\paragraph{Conjugacy class $\mathcal{C}_{58}\left(\mathrm{G_{256}}\right)$: $\#$ of elements = $8$}
{\scriptsize
\begin{equation}\label{G256classe58}
%
\end{equation}
}
\paragraph{Conjugacy class $\mathcal{C}_{59}\left(\mathrm{G_{256}}\right)$: $\#$ of elements = $8$}
{\scriptsize
\begin{equation}\label{G256classe59}
%
\end{equation}
}
\paragraph{Conjugacy class $\mathcal{C}_{60}\left(\mathrm{G_{256}}\right)$: $\#$ of elements = $8$}
{\scriptsize
\begin{equation}\label{G256classe60}
%
\end{equation}
}
\paragraph{Conjugacy class $\mathcal{C}_{61}\left(\mathrm{G_{256}}\right)$: $\#$ of elements = $8$}
{\scriptsize
\begin{equation}\label{G256classe61}
%
\end{equation}
}
\paragraph{Conjugacy class $\mathcal{C}_{62}\left(\mathrm{G_{256}}\right)$: $\#$ of elements = $8$}
{\scriptsize
\begin{equation}\label{G256classe62}
%
\end{equation}
}
\paragraph{Conjugacy class $\mathcal{C}_{63}\left(\mathrm{G_{256}}\right)$: $\#$ of elements = $8$}
{\scriptsize
\begin{equation}\label{G256classe63}
%
\end{equation}
}
\paragraph{Conjugacy class $\mathcal{C}_{64}\left(\mathrm{G_{256}}\right)$: $\#$ of elements = $8$}
{\scriptsize
\begin{equation}\label{G256classe64}
%
\end{equation}
}
\subsection{The Group $\mathrm{G_{128}}$}
\label{coniugato128}
In this section we list all the elements of the space group $\mathrm{G_{128}}$, organized into their $56$ conjugacy classes.
\paragraph{Conjugacy class $\mathcal{C}_{1}\left(\mathrm{G_{128}}\right)$: $\#$ of elements = $1$}
\begin{equation}\label{G128classe1}
%
\end{equation}
\paragraph{Conjugacy class $\mathcal{C}_{2}\left(\mathrm{G_{128}}\right)$: $\#$ of elements = $1$}
\begin{equation}\label{G128classe2}
%
\end{equation}
\paragraph{Conjugacy class $\mathcal{C}_{3}\left(\mathrm{G_{128}}\right)$: $\#$ of elements = $1$}
\begin{equation}\label{G128classe3}
%
\end{equation}
\paragraph{Conjugacy class $\mathcal{C}_{4}\left(\mathrm{G_{128}}\right)$: $\#$ of elements = $1$}
\begin{equation}\label{G128classe4}
%
\end{equation}
\paragraph{Conjugacy class $\mathcal{C}_{5}\left(\mathrm{G_{128}}\right)$: $\#$ of elements = $1$}
\begin{equation}\label{G128classe5}
%
\end{equation}
\paragraph{Conjugacy class $\mathcal{C}_{6}\left(\mathrm{G_{128}}\right)$: $\#$ of elements = $1$}
\begin{equation}\label{G128classe6}
%
\end{equation}
\paragraph{Conjugacy class $\mathcal{C}_{7}\left(\mathrm{G_{128}}\right)$: $\#$ of elements = $1$}
\begin{equation}\label{G128classe7}
%
\end{equation}
\paragraph{Conjugacy class $\mathcal{C}_{8}\left(\mathrm{G_{128}}\right)$: $\#$ of elements = $1$}
\begin{equation}\label{G128classe8}
%
\end{equation}
\paragraph{Conjugacy class $\mathcal{C}_{9}\left(\mathrm{G_{128}}\right)$: $\#$ of elements = $1$}
\begin{equation}\label{G128classe9}
%
\end{equation}
\paragraph{Conjugacy class $\mathcal{C}_{10}\left(\mathrm{G_{128}}\right)$: $\#$ of elements = $1$}
\begin{equation}\label{G128classe10}
%
\end{equation}
\paragraph{Conjugacy class $\mathcal{C}_{11}\left(\mathrm{G_{128}}\right)$: $\#$ of elements = $1$}
\begin{equation}\label{G128classe11}
%
\end{equation}
\paragraph{Conjugacy class $\mathcal{C}_{12}\left(\mathrm{G_{128}}\right)$: $\#$ of elements = $1$}
\begin{equation}\label{G128classe12}
%
\end{equation}
\paragraph{Conjugacy class $\mathcal{C}_{13}\left(\mathrm{G_{128}}\right)$: $\#$ of elements = $1$}
\begin{equation}\label{G128classe13}
%
\end{equation}
\paragraph{Conjugacy class $\mathcal{C}_{14}\left(\mathrm{G_{128}}\right)$: $\#$ of elements = $1$}
\begin{equation}\label{G128classe14}
%
\end{equation}
\paragraph{Conjugacy class $\mathcal{C}_{15}\left(\mathrm{G_{128}}\right)$: $\#$ of elements = $1$}
\begin{equation}\label{G128classe15}
%
\end{equation}
\paragraph{Conjugacy class $\mathcal{C}_{16}\left(\mathrm{G_{128}}\right)$: $\#$ of elements = $1$}
\begin{equation}\label{G128classe16}
%
\end{equation}
\paragraph{Conjugacy class $\mathcal{C}_{17}\left(\mathrm{G_{128}}\right)$: $\#$ of elements = $2$}
\begin{equation}\label{G128classe17}
%
\end{equation}
\paragraph{Conjugacy class $\mathcal{C}_{18}\left(\mathrm{G_{128}}\right)$: $\#$ of elements = $2$}
\begin{equation}\label{G128classe18}
%
\end{equation}
\paragraph{Conjugacy class $\mathcal{C}_{19}\left(\mathrm{G_{128}}\right)$: $\#$ of elements = $2$}
\begin{equation}\label{G128classe19}
%
\end{equation}
\paragraph{Conjugacy class $\mathcal{C}_{20}\left(\mathrm{G_{128}}\right)$: $\#$ of elements = $2$}
\begin{equation}\label{G128classe20}
%
\end{equation}
\paragraph{Conjugacy class $\mathcal{C}_{21}\left(\mathrm{G_{128}}\right)$: $\#$ of elements = $2$}
\begin{equation}\label{G128classe21}
%
\end{equation}
\paragraph{Conjugacy class $\mathcal{C}_{22}\left(\mathrm{G_{128}}\right)$: $\#$ of elements = $2$}
\begin{equation}\label{G128classe22}
%
\end{equation}
\paragraph{Conjugacy class $\mathcal{C}_{23}\left(\mathrm{G_{128}}\right)$: $\#$ of elements = $2$}
\begin{equation}\label{G128classe23}
%
\end{equation}
\paragraph{Conjugacy class $\mathcal{C}_{24}\left(\mathrm{G_{128}}\right)$: $\#$ of elements = $2$}
\begin{equation}\label{G128classe24}
%
\end{equation}
\paragraph{Conjugacy class $\mathcal{C}_{25}\left(\mathrm{G_{128}}\right)$: $\#$ of elements = $2$}
\begin{equation}\label{G128classe25}
%
\end{equation}
\paragraph{Conjugacy class $\mathcal{C}_{26}\left(\mathrm{G_{128}}\right)$: $\#$ of elements = $2$}
\begin{equation}\label{G128classe26}
%
\end{equation}
\paragraph{Conjugacy class $\mathcal{C}_{27}\left(\mathrm{G_{128}}\right)$: $\#$ of elements = $2$}
\begin{equation}\label{G128classe27}
%
\end{equation}
\paragraph{Conjugacy class $\mathcal{C}_{28}\left(\mathrm{G_{128}}\right)$: $\#$ of elements = $2$}
\begin{equation}\label{G128classe28}
%
\end{equation}
\paragraph{Conjugacy class $\mathcal{C}_{29}\left(\mathrm{G_{128}}\right)$: $\#$ of elements = $2$}
\begin{equation}\label{G128classe29}
%
\end{equation}
\paragraph{Conjugacy class $\mathcal{C}_{30}\left(\mathrm{G_{128}}\right)$: $\#$ of elements = $2$}
\begin{equation}\label{G128classe30}
%
\end{equation}
\paragraph{Conjugacy class $\mathcal{C}_{31}\left(\mathrm{G_{128}}\right)$: $\#$ of elements = $2$}
\begin{equation}\label{G128classe31}
%
\end{equation}
\paragraph{Conjugacy class $\mathcal{C}_{32}\left(\mathrm{G_{128}}\right)$: $\#$ of elements = $2$}
\begin{equation}\label{G128classe32}
%
\end{equation}
\paragraph{Conjugacy class $\mathcal{C}_{33}\left(\mathrm{G_{128}}\right)$: $\#$ of elements = $2$}
\begin{equation}\label{G128classe33}
%
\end{equation}
\paragraph{Conjugacy class $\mathcal{C}_{34}\left(\mathrm{G_{128}}\right)$: $\#$ of elements = $2$}
\begin{equation}\label{G128classe34}
%
\end{equation}
\paragraph{Conjugacy class $\mathcal{C}_{35}\left(\mathrm{G_{128}}\right)$: $\#$ of elements = $2$}
\begin{equation}\label{G128classe35}
%
\end{equation}
\paragraph{Conjugacy class $\mathcal{C}_{36}\left(\mathrm{G_{128}}\right)$: $\#$ of elements = $2$}
\begin{equation}\label{G128classe36}
%
\end{equation}
\paragraph{Conjugacy class $\mathcal{C}_{37}\left(\mathrm{G_{128}}\right)$: $\#$ of elements = $2$}
\begin{equation}\label{G128classe37}
%
\end{equation}
\paragraph{Conjugacy class $\mathcal{C}_{38}\left(\mathrm{G_{128}}\right)$: $\#$ of elements = $2$}
\begin{equation}\label{G128classe38}
%
\end{equation}
\paragraph{Conjugacy class $\mathcal{C}_{39}\left(\mathrm{G_{128}}\right)$: $\#$ of elements = $2$}
\begin{equation}\label{G128classe39}
%
\end{equation}
\paragraph{Conjugacy class $\mathcal{C}_{40}\left(\mathrm{G_{128}}\right)$: $\#$ of elements = $2$}
\begin{equation}\label{G128classe40}
%
\end{equation}
\paragraph{Conjugacy class $\mathcal{C}_{41}\left(\mathrm{G_{128}}\right)$: $\#$ of elements = $4$}
\begin{equation}\label{G128classe41}
%
\end{equation}
\paragraph{Conjugacy class $\mathcal{C}_{42}\left(\mathrm{G_{128}}\right)$: $\#$ of elements = $4$}
\begin{equation}\label{G128classe42}
%
\end{equation}
\paragraph{Conjugacy class $\mathcal{C}_{43}\left(\mathrm{G_{128}}\right)$: $\#$ of elements = $4$}
\begin{equation}\label{G128classe43}
%
\end{equation}
\paragraph{Conjugacy class $\mathcal{C}_{44}\left(\mathrm{G_{128}}\right)$: $\#$ of elements = $4$}
\begin{equation}\label{G128classe44}
%
\end{equation}
\paragraph{Conjugacy class $\mathcal{C}_{45}\left(\mathrm{G_{128}}\right)$: $\#$ of elements = $4$}
\begin{equation}\label{G128classe45}
%
\end{equation}
\paragraph{Conjugacy class $\mathcal{C}_{46}\left(\mathrm{G_{128}}\right)$: $\#$ of elements = $4$}
\begin{equation}\label{G128classe46}
%
\end{equation}
\paragraph{Conjugacy class $\mathcal{C}_{47}\left(\mathrm{G_{128}}\right)$: $\#$ of elements = $4$}
\begin{equation}\label{G128classe47}
%
\end{equation}
\paragraph{Conjugacy class $\mathcal{C}_{48}\left(\mathrm{G_{128}}\right)$: $\#$ of elements = $4$}
\begin{equation}\label{G128classe48}
%
\end{equation}
\paragraph{Conjugacy class $\mathcal{C}_{49}\left(\mathrm{G_{128}}\right)$: $\#$ of elements = $4$}
\begin{equation}\label{G128classe49}
%
\end{equation}
\paragraph{Conjugacy class $\mathcal{C}_{50}\left(\mathrm{G_{128}}\right)$: $\#$ of elements = $4$}
\begin{equation}\label{G128classe50}
%
\end{equation}
\paragraph{Conjugacy class $\mathcal{C}_{51}\left(\mathrm{G_{128}}\right)$: $\#$ of elements = $4$}
\begin{equation}\label{G128classe51}
%
\end{equation}
\paragraph{Conjugacy class $\mathcal{C}_{52}\left(\mathrm{G_{128}}\right)$: $\#$ of elements = $4$}
\begin{equation}\label{G128classe52}
%
\end{equation}
\paragraph{Conjugacy class $\mathcal{C}_{53}\left(\mathrm{G_{128}}\right)$: $\#$ of elements = $4$}
\begin{equation}\label{G128classe53}
%
\end{equation}
\paragraph{Conjugacy class $\mathcal{C}_{54}\left(\mathrm{G_{128}}\right)$: $\#$ of elements = $4$}
\begin{equation}\label{G128classe54}
%
\end{equation}
\paragraph{Conjugacy class $\mathcal{C}_{55}\left(\mathrm{G_{128}}\right)$: $\#$ of elements = $4$}
\begin{equation}\label{G128classe55}
%
\end{equation}
\paragraph{Conjugacy class $\mathcal{C}_{56}\left(\mathrm{G_{128}}\right)$: $\#$ of elements = $4$}
\begin{equation}\label{G128classe56}
%
\end{equation}
\subsection{The group $\mathrm{G_{64}}$}
\label{coniugato64}
The group $\mathrm{G_{64}}$ is abelian. Hence there are just as many conjugacy classes as there are elements, every conjugacy class containing just one element. For this reason it suffices to list the $64$ elements, which are displayed below:
{\scriptsize
\begin{equation}\label{G64elementi}
    %
\end{equation}
}
Abstractly the group $\mathrm{G_{64}}$ is isomorphic to $\mathbb{Z}_4 \times \mathbb{Z}_4 \times \mathbb{Z}_4$.
\subsection{The Group $\mathrm{G_{192}}$}
\label{coniugatoG192}
In this section we list all the elements of the space group $\mathrm{G_{192}}$, organized into their $20$ conjugacy classes.
\paragraph{Conjugacy class $\mathcal{C}_{1}\left(\mathrm{G_{192}}\right)$: $\#$ of elements = $1$}
\begin{equation}\label{G192classe1}
%
\end{equation}
\paragraph{Conjugacy class $\mathcal{C}_{2}\left(\mathrm{G_{192}}\right)$: $\#$ of elements = $1$}
\begin{equation}\label{G192classe2}
%
\end{equation}
\paragraph{Conjugacy class $\mathcal{C}_{3}\left(\mathrm{G_{192}}\right)$: $\#$ of elements = $3$}
\begin{equation}\label{G192classe3}
%
\end{equation}
\paragraph{Conjugacy class $\mathcal{C}_{4}\left(\mathrm{G_{192}}\right)$: $\#$ of elements = $3$}
\begin{equation}\label{G192classe4}
%
\end{equation}
\paragraph{Conjugacy class $\mathcal{C}_{5}\left(\mathrm{G_{192}}\right)$: $\#$ of elements = $3$}
\begin{equation}\label{G192classe5}
%
\end{equation}
\paragraph{Conjugacy class $\mathcal{C}_{6}\left(\mathrm{G_{192}}\right)$: $\#$ of elements = $3$}
\begin{equation}\label{G192classe6}
%
\end{equation}
\paragraph{Conjugacy class $\mathcal{C}_{7}\left(\mathrm{G_{192}}\right)$: $\#$ of elements = $3$}
\begin{equation}\label{G192classe7}
%
\end{equation}
\paragraph{Conjugacy class $\mathcal{C}_{8}\left(\mathrm{G_{192}}\right)$: $\#$ of elements = $3$}
\begin{equation}\label{G192classe8}
%
\end{equation}
\paragraph{Conjugacy class $\mathcal{C}_{9}\left(\mathrm{G_{192}}\right)$: $\#$ of elements = $6$}
\begin{equation}\label{G192classe9}
%
\end{equation}
\paragraph{Conjugacy class $\mathcal{C}_{10}\left(\mathrm{G_{192}}\right)$: $\#$ of elements = $6$}
\begin{equation}\label{G192classe10}
%
\end{equation}
\paragraph{Conjugacy class $\mathcal{C}_{11}\left(\mathrm{G_{192}}\right)$: $\#$ of elements = $12$}
\begin{equation}\label{G192classe11}
%
\end{equation}
\paragraph{Conjugacy class $\mathcal{C}_{12}\left(\mathrm{G_{192}}\right)$: $\#$ of elements = $12$}
\begin{equation}\label{G192classe12}
%
\end{equation}
\paragraph{Conjugacy class $\mathcal{C}_{13}\left(\mathrm{G_{192}}\right)$: $\#$ of elements = $12$}
\begin{equation}\label{G192classe13}
%
\end{equation}
\paragraph{Conjugacy class $\mathcal{C}_{14}\left(\mathrm{G_{192}}\right)$: $\#$ of elements = $12$}
\begin{equation}\label{G192classe14}
%
\end{equation}
\paragraph{Conjugacy class $\mathcal{C}_{15}\left(\mathrm{G_{192}}\right)$: $\#$ of elements = $12$}
\begin{equation}\label{G192classe15}
%
\end{equation}
\paragraph{Conjugacy class $\mathcal{C}_{16}\left(\mathrm{G_{192}}\right)$: $\#$ of elements = $12$}
\begin{equation}\label{G192classe16}
%
\end{equation}
\paragraph{Conjugacy class $\mathcal{C}_{17}\left(\mathrm{G_{192}}\right)$: $\#$ of elements = $12$}
\begin{equation}\label{G192classe17}
%
\end{equation}
\paragraph{Conjugacy class $\mathcal{C}_{18}\left(\mathrm{G_{192}}\right)$: $\#$ of elements = $12$}
\begin{equation}\label{G192classe18}
%
\end{equation}
\paragraph{Conjugacy class $\mathcal{C}_{19}\left(\mathrm{G_{192}}\right)$: $\#$ of elements = $32$}
{\scriptsize
\begin{equation}\label{G192classe19}
%
\end{equation}
}
\paragraph{Conjugacy class $\mathcal{C}_{20}\left(\mathrm{G_{192}}\right)$: $\#$ of elements = $32$}
{\scriptsize
\begin{equation}\label{G192classe20}
%
\end{equation}
}
\subsection{The Group $\mathrm{GF_{192}}$}
\label{coniugatoGF192}
In this section we list all the elements of the space group $\mathrm{GF_{192}}$, organized into their $20$ conjugacy classes.
\paragraph{Conjugacy class $\mathcal{C}_{1}\left(\mathrm{GF_{192}}\right)$: $\#$ of elements = $1$}
\begin{equation}\label{GF192classe1}
%
\end{equation}
\paragraph{Conjugacy class $\mathcal{C}_{2}\left(\mathrm{GF_{192}}\right)$: $\#$ of elements = $1$}
\begin{equation}\label{GF192classe2}
%
\end{equation}
\paragraph{Conjugacy class $\mathcal{C}_{3}\left(\mathrm{GF_{192}}\right)$: $\#$ of elements = $3$}
\begin{equation}\label{GF192classe3}
%
\end{equation}
\paragraph{Conjugacy class $\mathcal{C}_{4}\left(\mathrm{GF_{192}}\right)$: $\#$ of elements = $3$}
\begin{equation}\label{GF192classe4}
%
\end{equation}
\paragraph{Conjugacy class $\mathcal{C}_{5}\left(\mathrm{GF_{192}}\right)$: $\#$ of elements = $3$}
\begin{equation}\label{GF192classe5}
%
\end{equation}
\paragraph{Conjugacy class $\mathcal{C}_{6}\left(\mathrm{GF_{192}}\right)$: $\#$ of elements = $3$}
\begin{equation}\label{GF192classe6}
%
\end{equation}
\paragraph{Conjugacy class $\mathcal{C}_{7}\left(\mathrm{GF_{192}}\right)$: $\#$ of elements = $3$}
\begin{equation}\label{GF192classe7}
%
\end{equation}
\paragraph{Conjugacy class $\mathcal{C}_{8}\left(\mathrm{GF_{192}}\right)$: $\#$ of elements = $3$}
\begin{equation}\label{GF192classe8}
%
\end{equation}
\paragraph{Conjugacy class $\mathcal{C}_{9}\left(\mathrm{GF_{192}}\right)$: $\#$ of elements = $6$}
\begin{equation}\label{GF192classe9}
%
\end{equation}
\paragraph{Conjugacy class $\mathcal{C}_{10}\left(\mathrm{GF_{192}}\right)$: $\#$ of elements = $6$}
\begin{equation}\label{GF192classe10}
%
\end{equation}
\paragraph{Conjugacy class $\mathcal{C}_{11}\left(\mathrm{GF_{192}}\right)$: $\#$ of elements = $12$}
\begin{equation}\label{GF192classe11}
%
\end{equation}
\paragraph{Conjugacy class $\mathcal{C}_{12}\left(\mathrm{GF_{192}}\right)$: $\#$ of elements = $12$}
\begin{equation}\label{GF192classe12}
%
\end{equation}
\paragraph{Conjugacy class $\mathcal{C}_{13}\left(\mathrm{GF_{192}}\right)$: $\#$ of elements = $12$}
\begin{equation}\label{GF192classe13}
%
\end{equation}
\paragraph{Conjugacy class $\mathcal{C}_{14}\left(\mathrm{GF_{192}}\right)$: $\#$ of elements = $12$}
\begin{equation}\label{GF192classe14}
%
\end{equation}
\paragraph{Conjugacy class $\mathcal{C}_{15}\left(\mathrm{GF_{192}}\right)$: $\#$ of elements = $12$}
\begin{equation}\label{GF192classe15}
%
\end{equation}
\paragraph{Conjugacy class $\mathcal{C}_{16}\left(\mathrm{GF_{192}}\right)$: $\#$ of elements = $12$}
\begin{equation}\label{GF192classe16}
%
\end{equation}
\paragraph{Conjugacy class $\mathcal{C}_{17}\left(\mathrm{GF_{192}}\right)$: $\#$ of elements = $12$}
\begin{equation}\label{GF192classe17}
%
\end{equation}
\paragraph{Conjugacy class $\mathcal{C}_{18}\left(\mathrm{GF_{192}}\right)$: $\#$ of elements = $12$}
\begin{equation}\label{GF192classe18}
%
\end{equation}
\paragraph{Conjugacy class $\mathcal{C}_{19}\left(\mathrm{GF_{192}}\right)$: $\#$ of elements = $32$}
{\scriptsize
\begin{equation}\label{GF192classe19}
%
\end{equation}
}
\paragraph{Conjugacy class $\mathcal{C}_{20}\left(\mathrm{GF_{192}}\right)$: $\#$ of elements = $32$}
{\scriptsize
\begin{equation}\label{GF192classe20}
%
\end{equation}
}
\subsection{The Group $\mathrm{Oh_{48}}$}
\label{coniugatoOh48}
In this section we list all the elements of the space group $\mathrm{Oh_{48}}$, organized into their $10$ conjugacy classes that, in this case, are arranged according to the order which is customary in crystallography for the extended octahedral  group.
\paragraph{Conjugacy class $\mathcal{C}_{1}\left(\mathrm{Oh_{48}}\right)$: $\#$ of elements = $1$}
\begin{equation}\label{Oh48classe1}
%
\end{equation}
\paragraph{Conjugacy class $\mathcal{C}_{2}\left(\mathrm{Oh_{48}}\right)$: $\#$ of elements = $8$}
\begin{equation}\label{Oh48classe2}
%
\end{equation}
\paragraph{Conjugacy class $\mathcal{C}_{3}\left(\mathrm{Oh_{48}}\right)$: $\#$ of elements = $3$}
\begin{equation}\label{Oh48classe3}
%
\end{equation}
\paragraph{Conjugacy class $\mathcal{C}_{4}\left(\mathrm{Oh_{48}}\right)$: $\#$ of elements = $6$}
\begin{equation}\label{Oh48classe4}
%
\end{equation}
\paragraph{Conjugacy class $\mathcal{C}_{5}\left(\mathrm{Oh_{48}}\right)$: $\#$ of elements = $6$}
\begin{equation}\label{Oh48classe5}
%
\end{equation}
\paragraph{Conjugacy class $\mathcal{C}_{6}\left(\mathrm{Oh_{48}}\right)$: $\#$ of elements = $1$}
\begin{equation}\label{Oh48classe6}
%
\end{equation}
\paragraph{Conjugacy class $\mathcal{C}_{7}\left(\mathrm{Oh_{48}}\right)$: $\#$ of elements = $8$}
\begin{equation}\label{Oh48classe7}
%
\end{equation}
\paragraph{Conjugacy class $\mathcal{C}_{8}\left(\mathrm{Oh_{48}}\right)$: $\#$ of elements = $3$}
\begin{equation}\label{Oh48classe8}
%
\end{equation}
\paragraph{Conjugacy class $\mathcal{C}_{9}\left(\mathrm{Oh_{48}}\right)$: $\#$ of elements = $6$}
\begin{equation}\label{Oh48classe9}
%
\end{equation}
\paragraph{Conjugacy class $\mathcal{C}_{10}\left(\mathrm{Oh_{48}}\right)$: $\#$ of elements = $6$}
\begin{equation}\label{Oh48classe10}
%
\end{equation}
\subsection{The Group $\mathrm{GS_{24}}$}
\label{coniugatoGS24}
In this section we list all the elements of the space group $\mathrm{GS_{24}}$, organized into their $5$ conjugacy classes that, in this case are arranged according to the order which is customary in crystallography for the proper octahedral  group.
\paragraph{Conjugacy class $\mathcal{C}_{1}\left(\mathrm{GS_{24}}\right)$: $\#$ of elements = $1$}
\begin{equation}\label{GS24classe1}
%
\end{equation}
\paragraph{Conjugacy class $\mathcal{C}_{2}\left(\mathrm{GS_{24}}\right)$: $\#$ of elements = $8$}
\begin{equation}\label{GS24classe2}
%
\end{equation}
\paragraph{Conjugacy class $\mathcal{C}_{3}\left(\mathrm{GS_{24}}\right)$: $\#$ of elements = $3$}
\begin{equation}\label{GS24classe3}
%
\end{equation}
\paragraph{Conjugacy class $\mathcal{C}_{4}\left(\mathrm{GS_{24}}\right)$: $\#$ of elements = $6$}
\begin{equation}\label{GS24classe4}
%
\end{equation}
\paragraph{Conjugacy class $\mathcal{C}_{5}\left(\mathrm{GS_{24}}\right)$: $\#$ of elements = $6$}
\begin{equation}\label{GS24classe5}
%
\end{equation}
\subsection{The Group $\mathrm{GP_{24}}$}
\label{coniugatoGP24}
In this section we list all the elements of the space group $\mathrm{GP_{24}}$, organized into their $8$ conjugacy classes.
\paragraph{Conjugacy class $\mathcal{C}_{1}\left(\mathrm{GP_{24}}\right)$: $\#$ of elements = $1$}
\begin{equation}\label{GP24classe1}
%
\end{equation}
\paragraph{Conjugacy class $\mathcal{C}_{2}\left(\mathrm{GP_{24}}\right)$: $\#$ of elements = $1$}
\begin{equation}\label{GP24classe2}
%
\end{equation}
\paragraph{Conjugacy class $\mathcal{C}_{3}\left(\mathrm{GP_{24}}\right)$: $\#$ of elements = $3$}
\begin{equation}\label{GP24classe3}
%
\end{equation}
\paragraph{Conjugacy class $\mathcal{C}_{4}\left(\mathrm{GP_{24}}\right)$: $\#$ of elements = $3$}
\begin{equation}\label{GP24classe4}
%
\end{equation}
\paragraph{Conjugacy class $\mathcal{C}_{5}\left(\mathrm{GP_{24}}\right)$: $\#$ of elements = $4$}
\begin{equation}\label{GP24classe5}
%
\end{equation}
\paragraph{Conjugacy class $\mathcal{C}_{6}\left(\mathrm{GP_{24}}\right)$: $\#$ of elements = $4$}
\begin{equation}\label{GP24classe6}
%
\end{equation}
\paragraph{Conjugacy class $\mathcal{C}_{7}\left(\mathrm{GP_{24}}\right)$: $\#$ of elements = $4$}
\begin{equation}\label{GP24classe7}
%
\end{equation}
\paragraph{Conjugacy class $\mathcal{C}_{8}\left(\mathrm{GP_{24}}\right)$: $\#$ of elements = $4$}
\begin{equation}\label{GP24classe8}
%
\end{equation}
\subsection{The Group $\mathrm{GK_{24}}$}
\label{coniugatoGK24}
In this section we list all the elements of the space group $\mathrm{GK_{24}}$, organized into their $8$ conjugacy classes.
\paragraph{Conjugacy class $\mathcal{C}_{1}\left(\mathrm{GK_{24}}\right)$: $\#$ of elements = $1$}
\begin{equation}\label{GK24classe1}
%
\end{equation}
\paragraph{Conjugacy class $\mathcal{C}_{2}\left(\mathrm{GK_{24}}\right)$: $\#$ of elements = $1$}
\begin{equation}\label{GK24classe2}
%
\end{equation}
\paragraph{Conjugacy class $\mathcal{C}_{3}\left(\mathrm{GK_{24}}\right)$: $\#$ of elements = $3$}
\begin{equation}\label{GK24classe3}
%
\end{equation}
\paragraph{Conjugacy class $\mathcal{C}_{4}\left(\mathrm{GK_{24}}\right)$: $\#$ of elements = $3$}
\begin{equation}\label{GK24classe4}
%
\end{equation}
\paragraph{Conjugacy class $\mathcal{C}_{5}\left(\mathrm{GK_{24}}\right)$: $\#$ of elements = $4$}
\begin{equation}\label{GK24classe5}
%
\end{equation}
\paragraph{Conjugacy class $\mathcal{C}_{6}\left(\mathrm{GK_{24}}\right)$: $\#$ of elements = $4$}
\begin{equation}\label{GK24classe6}
%
\end{equation}
\paragraph{Conjugacy class $\mathcal{C}_{7}\left(\mathrm{GK_{24}}\right)$: $\#$ of elements = $4$}
\begin{equation}\label{GK24classe7}
%
\end{equation}
\paragraph{Conjugacy class $\mathcal{C}_{8}\left(\mathrm{GK_{24}}\right)$: $\#$ of elements = $4$}
\begin{equation}\label{GK24classe8}
%
\end{equation}
\subsection{The Group $\mathrm{GS_{32}}$}
\label{coniugatoGS32}
In this section we list all the elements of the space group $\mathrm{GS_{32}}$, organized into their $14$ conjugacy classes.
\paragraph{Conjugacy class $\mathcal{C}_{1}\left(\mathrm{GS_{32}}\right)$: $\#$ of elements = $1$}
\begin{equation}\label{GS32classe1}
%
\end{equation}
\paragraph{Conjugacy class $\mathcal{C}_{2}\left(\mathrm{GS_{32}}\right)$: $\#$ of elements = $1$}
\begin{equation}\label{GS32classe2}
%
\end{equation}
\paragraph{Conjugacy class $\mathcal{C}_{3}\left(\mathrm{GS_{32}}\right)$: $\#$ of elements = $1$}
\begin{equation}\label{GS32classe3}
%
\end{equation}
\paragraph{Conjugacy class $\mathcal{C}_{4}\left(\mathrm{GS_{32}}\right)$: $\#$ of elements = $1$}
\begin{equation}\label{GS32classe4}
%
\end{equation}
\paragraph{Conjugacy class $\mathcal{C}_{5}\left(\mathrm{GS_{32}}\right)$: $\#$ of elements = $2$}
\begin{equation}\label{GS32classe5}
%
\end{equation}
\paragraph{Conjugacy class $\mathcal{C}_{6}\left(\mathrm{GS_{32}}\right)$: $\#$ of elements = $2$}
\begin{equation}\label{GS32classe6}
%
\end{equation}
\paragraph{Conjugacy class $\mathcal{C}_{7}\left(\mathrm{GS_{32}}\right)$: $\#$ of elements = $2$}
\begin{equation}\label{GS32classe7}
%
\end{equation}
\paragraph{Conjugacy class $\mathcal{C}_{8}\left(\mathrm{GS_{32}}\right)$: $\#$ of elements = $2$}
\begin{equation}\label{GS32classe8}
%
\end{equation}
\paragraph{Conjugacy class $\mathcal{C}_{9}\left(\mathrm{GS_{32}}\right)$: $\#$ of elements = $2$}
\begin{equation}\label{GS32classe9}
%
\end{equation}
\paragraph{Conjugacy class $\mathcal{C}_{10}\left(\mathrm{GS_{32}}\right)$: $\#$ of elements = $2$}
\begin{equation}\label{GS32classe10}
%
\end{equation}
\paragraph{Conjugacy class $\mathcal{C}_{11}\left(\mathrm{GS_{32}}\right)$: $\#$ of elements = $4$}
\begin{equation}\label{GS32classe11}
%
\end{equation}
\paragraph{Conjugacy class $\mathcal{C}_{12}\left(\mathrm{GS_{32}}\right)$: $\#$ of elements = $4$}
\begin{equation}\label{GS32classe12}
%
\end{equation}
\paragraph{Conjugacy class $\mathcal{C}_{13}\left(\mathrm{GS_{32}}\right)$: $\#$ of elements = $4$}
\begin{equation}\label{GS32classe13}
%
\end{equation}
\paragraph{Conjugacy class $\mathcal{C}_{14}\left(\mathrm{GS_{32}}\right)$: $\#$ of elements = $4$}
\begin{equation}\label{GS32classe14}
%
\end{equation}
\subsection{The Group $\mathrm{GK_{32}}$}
\label{coniugatoGK32}
In this section we list all the elements of the space group $\mathrm{GK_{32}}$, organized into their $14$ conjugacy classes.
\paragraph{Conjugacy class $\mathcal{C}_{1}\left(\mathrm{GK_{32}}\right)$: $\#$ of elements = $1$}
\begin{equation}\label{GK32classe1}
%
\end{equation}
\paragraph{Conjugacy class $\mathcal{C}_{2}\left(\mathrm{GK_{32}}\right)$: $\#$ of elements = $1$}
\begin{equation}\label{GK32classe2}
%
\end{equation}
\paragraph{Conjugacy class $\mathcal{C}_{3}\left(\mathrm{GK_{32}}\right)$: $\#$ of elements = $1$}
\begin{equation}\label{GK32classe3}
%
\end{equation}
\paragraph{Conjugacy class $\mathcal{C}_{4}\left(\mathrm{GK_{32}}\right)$: $\#$ of elements = $1$}
\begin{equation}\label{GK32classe4}
%
\end{equation}
\paragraph{Conjugacy class $\mathcal{C}_{5}\left(\mathrm{GK_{32}}\right)$: $\#$ of elements = $2$}
\begin{equation}\label{GK32classe5}
%
\end{equation}
\paragraph{Conjugacy class $\mathcal{C}_{6}\left(\mathrm{GK_{32}}\right)$: $\#$ of elements = $2$}
\begin{equation}\label{GK32classe6}
%
\end{equation}
\paragraph{Conjugacy class $\mathcal{C}_{7}\left(\mathrm{GK_{32}}\right)$: $\#$ of elements = $2$}
\begin{equation}\label{GK32classe7}
%
\end{equation}
\paragraph{Conjugacy class $\mathcal{C}_{8}\left(\mathrm{GK_{32}}\right)$: $\#$ of elements = $2$}
\begin{equation}\label{GK32classe8}
%
\end{equation}
\paragraph{Conjugacy class $\mathcal{C}_{9}\left(\mathrm{GK_{32}}\right)$: $\#$ of elements = $2$}
\begin{equation}\label{GK32classe9}
%
\end{equation}
\paragraph{Conjugacy class $\mathcal{C}_{10}\left(\mathrm{GK_{32}}\right)$: $\#$ of elements = $2$}
\begin{equation}\label{GK32classe10}
%
\end{equation}
\paragraph{Conjugacy class $\mathcal{C}_{11}\left(\mathrm{GK_{32}}\right)$: $\#$ of elements = $4$}
\begin{equation}\label{GK32classe11}
%
\end{equation}
\paragraph{Conjugacy class $\mathcal{C}_{12}\left(\mathrm{GK_{32}}\right)$: $\#$ of elements = $4$}
\begin{equation}\label{GK32classe12}
%
\end{equation}
\paragraph{Conjugacy class $\mathcal{C}_{13}\left(\mathrm{GK_{32}}\right)$: $\#$ of elements = $4$}
\begin{equation}\label{GK32classe13}
%
\end{equation}
\paragraph{Conjugacy class $\mathcal{C}_{14}\left(\mathrm{GK_{32}}\right)$: $\#$ of elements = $4$}
\begin{equation}\label{GK32classe14}
%
\end{equation}
\section{Character tables of the considered discrete groups}
\label{carateropoli}
In this section we present the results for the irreducible representations of the various groups listed in section \ref{descriptio} and we display the character tables of each of them. As explained in the main text the basis to obtain such results has been the implementation in a series of purposely written MATHEMATICA codes of the algorithm described in sections \ref{startegos} and \ref{agoinduzio}.
\subsection{Character Table of the Group $\mathrm{\mathrm{G_{1536}}}$}
\label{caratterbrut1536}
The big ambient group $\mathrm{\mathrm{G_{1536}}}$ has $37$ conjugacy classes and therefore $37$ irreducible representations that are distributed according to the following pattern:
\begin{description}
  \item[a)] 4 irreps of dimension $1$, namely $D_1,\dots,D_4$
  \item[b)] 2 irreps of dimension $2$, namely $D_5,\dots,D_6$
  \item[c)] 12 irreps of dimension $3$, namely $D_6,\dots,D_{18}$
  \item[d)] 10 irreps of dimension $6$, namely $D_7,\dots,D_{28}$
  \item[e)] 3 irreps of dimension $8$, namely $D_{29},\dots,D_{31}$
  \item[f)] 6 irreps of dimension $12$, namely $D_{32},\dots,D_{37}$
\end{description}
The corresponding character table that we have calculated with the procedures described in the main text is displayed below. For pure typographical reasons we were forced to split the character table in two parts in order to fit it into the page.
{\scriptsize \begin{equation}\label{1charto1536}
\begin{array}{|l|llllllllllllllllll|}
\hline
 0 & C_1 & C_2 & C_3 & C_4 & C_5 & C_6 & C_7 & C_8 & C_9
   & C_{10} & C_{11} & C_{12} & C_{13} & C_{14} & C_{15}
   & C_{16} & C_{17} & C_{18} \\
   \hline
 D_1 & 1 & 1 & 1 & 1 & 1 & 1 & 1 & 1 & 1 & 1 & 1 & 1 & 1
   & 1 & 1 & 1 & 1 & 1 \\
 D_2 & 1 & 1 & 1 & 1 & 1 & 1 & 1 & 1 & 1 & 1 & 1 & 1 & 1
   & 1 & 1 & 1 & 1 & 1 \\
 D_3 & 1 & 1 & 1 & 1 & -1 & -1 & -1 & 1 & -1 & 1 & 1 & 1
   & 1 & 1 & -1 & -1 & -1 & -1 \\
 D_4 & 1 & 1 & 1 & 1 & -1 & -1 & -1 & 1 & -1 & 1 & 1 & 1
   & 1 & 1 & -1 & -1 & -1 & -1 \\
 D_5 & 2 & 2 & 2 & 2 & 2 & 2 & 2 & 2 & 2 & 2 & 2 & 2 & 2
   & 2 & 2 & 2 & 2 & 2 \\
 D_6 & 2 & 2 & 2 & 2 & -2 & -2 & -2 & 2 & -2 & 2 & 2 & 2
   & 2 & 2 & -2 & -2 & -2 & -2 \\
 D_7 & 3 & 3 & 3 & 3 & 3 & 3 & 3 & 3 & 3 & 3 & -1 & -1 &
   -1 & -1 & -1 & -1 & -1 & -1 \\
 D_8 & 3 & 3 & 3 & 3 & 3 & 3 & 3 & 3 & 3 & 3 & -1 & -1 &
   -1 & -1 & -1 & -1 & -1 & -1 \\
 D_9 & 3 & 3 & 3 & 3 & 1 & 1 & -3 & -1 & 1 & -1 & 3 & 3 &
   -1 & -1 & 1 & 1 & 1 & -3 \\
 D_{10} & 3 & 3 & 3 & 3 & 1 & 1 & -3 & -1 & 1 & -1 & 3 &
   3 & -1 & -1 & 1 & 1 & 1 & -3 \\
 D_{11} & 3 & 3 & 3 & 3 & 1 & 1 & -3 & -1 & 1 & -1 & -1 &
   -1 & 3 & 3 & -3 & 1 & 1 & 1 \\
 D_{12} & 3 & 3 & 3 & 3 & 1 & 1 & -3 & -1 & 1 & -1 & -1 &
   -1 & 3 & 3 & -3 & 1 & 1 & 1 \\
 D_{13} & 3 & 3 & 3 & 3 & -1 & -1 & 3 & -1 & -1 & -1 & 3
   & 3 & -1 & -1 & -1 & -1 & -1 & 3 \\
 D_{14} & 3 & 3 & 3 & 3 & -1 & -1 & 3 & -1 & -1 & -1 & 3
   & 3 & -1 & -1 & -1 & -1 & -1 & 3 \\
 D_{15} & 3 & 3 & 3 & 3 & -1 & -1 & 3 & -1 & -1 & -1 & -1
   & -1 & 3 & 3 & 3 & -1 & -1 & -1 \\
 D_{16} & 3 & 3 & 3 & 3 & -1 & -1 & 3 & -1 & -1 & -1 & -1
   & -1 & 3 & 3 & 3 & -1 & -1 & -1 \\
 D_{17} & 3 & 3 & 3 & 3 & -3 & -3 & -3 & 3 & -3 & 3 & -1
   & -1 & -1 & -1 & 1 & 1 & 1 & 1 \\
 D_{18} & 3 & 3 & 3 & 3 & -3 & -3 & -3 & 3 & -3 & 3 & -1
   & -1 & -1 & -1 & 1 & 1 & 1 & 1 \\
 D_{19} & 6 & 6 & 6 & 6 & 2 & 2 & -6 & -2 & 2 & -2 & -2 &
   -2 & -2 & -2 & 2 & -2 & -2 & 2 \\
 D_{20} & 6 & 6 & 6 & 6 & -2 & -2 & 6 & -2 & -2 & -2 & -2
   & -2 & -2 & -2 & -2 & 2 & 2 & -2 \\
 D_{21} & 6 & -6 & 2 & -2 & 4 & -4 & 0 & 2 & 0 & -2 & 2 &
   -2 & 2 & -2 & 0 & 2 & -2 & 0 \\
 D_{22} & 6 & -6 & 2 & -2 & 4 & -4 & 0 & 2 & 0 & -2 & 2 &
   -2 & 2 & -2 & 0 & 2 & -2 & 0 \\
 D_{23} & 6 & -6 & 2 & -2 & 4 & -4 & 0 & 2 & 0 & -2 & -2
   & 2 & -2 & 2 & 0 & -2 & 2 & 0 \\
 D_{24} & 6 & -6 & 2 & -2 & 4 & -4 & 0 & 2 & 0 & -2 & -2
   & 2 & -2 & 2 & 0 & -2 & 2 & 0 \\
 D_{25} & 6 & -6 & 2 & -2 & -4 & 4 & 0 & 2 & 0 & -2 & 2 &
   -2 & 2 & -2 & 0 & -2 & 2 & 0 \\
 D_{26} & 6 & -6 & 2 & -2 & -4 & 4 & 0 & 2 & 0 & -2 & 2 &
   -2 & 2 & -2 & 0 & -2 & 2 & 0 \\
 D_{27} & 6 & -6 & 2 & -2 & -4 & 4 & 0 & 2 & 0 & -2 & -2
   & 2 & -2 & 2 & 0 & 2 & -2 & 0 \\
 D_{28} & 6 & -6 & 2 & -2 & -4 & 4 & 0 & 2 & 0 & -2 & -2
   & 2 & -2 & 2 & 0 & 2 & -2 & 0 \\
 D_{29} & 8 & -8 & -8 & 8 & 0 & 0 & 0 & 0 & 0 & 0 & 0 & 0
   & 0 & 0 & 0 & 0 & 0 & 0 \\
 D_{30} & 8 & -8 & -8 & 8 & 0 & 0 & 0 & 0 & 0 & 0 & 0 & 0
   & 0 & 0 & 0 & 0 & 0 & 0 \\
 D_{31} & 8 & -8 & -8 & 8 & 0 & 0 & 0 & 0 & 0 & 0 & 0 & 0
   & 0 & 0 & 0 & 0 & 0 & 0 \\
 D_{32} & 12 & 12 & -4 & -4 & 4 & 4 & 0 & 0 & -4 & 0 & 0
   & 0 & 0 & 0 & 0 & 0 & 0 & 0 \\
 D_{33} & 12 & 12 & -4 & -4 & 4 & 4 & 0 & 0 & -4 & 0 & 0
   & 0 & 0 & 0 & 0 & 0 & 0 & 0 \\
 D_{34} & 12 & 12 & -4 & -4 & -4 & -4 & 0 & 0 & 4 & 0 & 0
   & 0 & 0 & 0 & 0 & 0 & 0 & 0 \\
 D_{35} & 12 & 12 & -4 & -4 & -4 & -4 & 0 & 0 & 4 & 0 & 0
   & 0 & 0 & 0 & 0 & 0 & 0 & 0 \\
 D_{36} & 12 & -12 & 4 & -4 & 0 & 0 & 0 & -4 & 0 & 4 & 4
   & -4 & -4 & 4 & 0 & 0 & 0 & 0 \\
 D_{37} & 12 & -12 & 4 & -4 & 0 & 0 & 0 & -4 & 0 & 4 & -4
   & 4 & 4 & -4 & 0 & 0 & 0 & 0\\
   \hline
\end{array}
\end{equation}}
{\scriptsize
\begin{equation}\label{2charto1536}
    \begin{array}{|l|lllllllllllllllllll|}
 \hline
 0 & C_{19} & C_{20} & C_{21} & C_{22} & C_{23} & C_{24}
   & C_{25} & C_{26} & C_{27} & C_{28} & C_{29} & C_{30}
   & C_{31} & C_{32} & C_{33} & C_{34} & C_{35} & C_{36}
   & C_{37} \\
 \hline
 D_1 & 1 & 1 & 1 & 1 & 1 & 1 & 1 & 1 & 1 & 1 & 1 & 1 & 1
   & 1 & 1 & 1 & 1 & 1 & 1 \\
 D_2 & 1 & -1 & -1 & -1 & -1 & -1 & -1 & -1 & -1 & -1 &
   -1 & -1 & -1 & -1 & -1 & 1 & 1 & 1 & 1 \\
 D_3 & 1 & -1 & -1 & 1 & 1 & -1 & 1 & -1 & 1 & 1 & -1 & 1
   & -1 & 1 & -1 & 1 & -1 & 1 & -1 \\
 D_4 & 1 & 1 & 1 & -1 & -1 & 1 & -1 & 1 & -1 & -1 & 1 &
   -1 & 1 & -1 & 1 & 1 & -1 & 1 & -1 \\
 D_5 & 2 & 0 & 0 & 0 & 0 & 0 & 0 & 0 & 0 & 0 & 0 & 0 & 0
   & 0 & 0 & -1 & -1 & -1 & -1 \\
 D_6 & 2 & 0 & 0 & 0 & 0 & 0 & 0 & 0 & 0 & 0 & 0 & 0 & 0
   & 0 & 0 & -1 & 1 & -1 & 1 \\
 D_7 & -1 & 1 & 1 & 1 & 1 & -1 & -1 & -1 & -1 & -1 & -1 &
   -1 & -1 & 1 & 1 & 0 & 0 & 0 & 0 \\
 D_8 & -1 & -1 & -1 & -1 & -1 & 1 & 1 & 1 & 1 & 1 & 1 & 1
   & 1 & -1 & -1 & 0 & 0 & 0 & 0 \\
 D_9 & -1 & 1 & 1 & -1 & -1 & 1 & -1 & 1 & -1 & 1 & -1 &
   1 & -1 & 1 & -1 & 0 & 0 & 0 & 0 \\
 D_{10} & -1 & -1 & -1 & 1 & 1 & -1 & 1 & -1 & 1 & -1 & 1
   & -1 & 1 & -1 & 1 & 0 & 0 & 0 & 0 \\
 D_{11} & -1 & -1 & -1 & 1 & 1 & 1 & -1 & 1 & -1 & 1 & -1
   & 1 & -1 & -1 & 1 & 0 & 0 & 0 & 0 \\
 D_{12} & -1 & 1 & 1 & -1 & -1 & -1 & 1 & -1 & 1 & -1 & 1
   & -1 & 1 & 1 & -1 & 0 & 0 & 0 & 0 \\
 D_{13} & -1 & -1 & -1 & -1 & -1 & -1 & -1 & -1 & -1 & 1
   & 1 & 1 & 1 & 1 & 1 & 0 & 0 & 0 & 0 \\
 D_{14} & -1 & 1 & 1 & 1 & 1 & 1 & 1 & 1 & 1 & -1 & -1 &
   -1 & -1 & -1 & -1 & 0 & 0 & 0 & 0 \\
 D_{15} & -1 & -1 & -1 & -1 & -1 & 1 & 1 & 1 & 1 & -1 &
   -1 & -1 & -1 & 1 & 1 & 0 & 0 & 0 & 0 \\
 D_{16} & -1 & 1 & 1 & 1 & 1 & -1 & -1 & -1 & -1 & 1 & 1
   & 1 & 1 & -1 & -1 & 0 & 0 & 0 & 0 \\
 D_{17} & -1 & -1 & -1 & 1 & 1 & 1 & -1 & 1 & -1 & -1 & 1
   & -1 & 1 & 1 & -1 & 0 & 0 & 0 & 0 \\
 D_{18} & -1 & 1 & 1 & -1 & -1 & -1 & 1 & -1 & 1 & 1 & -1
   & 1 & -1 & -1 & 1 & 0 & 0 & 0 & 0 \\
 D_{19} & 2 & 0 & 0 & 0 & 0 & 0 & 0 & 0 & 0 & 0 & 0 & 0 &
   0 & 0 & 0 & 0 & 0 & 0 & 0 \\
 D_{20} & 2 & 0 & 0 & 0 & 0 & 0 & 0 & 0 & 0 & 0 & 0 & 0 &
   0 & 0 & 0 & 0 & 0 & 0 & 0 \\
 D_{21} & 0 & 0 & 0 & 0 & 0 & -2 & 0 & 2 & 0 & -2 & 0 & 2
   & 0 & 0 & 0 & 0 & 0 & 0 & 0 \\
 D_{22} & 0 & 0 & 0 & 0 & 0 & 2 & 0 & -2 & 0 & 2 & 0 & -2
   & 0 & 0 & 0 & 0 & 0 & 0 & 0 \\
 D_{23} & 0 & 0 & 0 & 0 & 0 & 0 & -2 & 0 & 2 & 0 & -2 & 0
   & 2 & 0 & 0 & 0 & 0 & 0 & 0 \\
 D_{24} & 0 & 0 & 0 & 0 & 0 & 0 & 2 & 0 & -2 & 0 & 2 & 0
   & -2 & 0 & 0 & 0 & 0 & 0 & 0 \\
 D_{25} & 0 & 0 & 0 & 0 & 0 & -2 & 0 & 2 & 0 & 2 & 0 & -2
   & 0 & 0 & 0 & 0 & 0 & 0 & 0 \\
 D_{26} & 0 & 0 & 0 & 0 & 0 & 2 & 0 & -2 & 0 & -2 & 0 & 2
   & 0 & 0 & 0 & 0 & 0 & 0 & 0 \\
 D_{27} & 0 & 0 & 0 & 0 & 0 & 0 & -2 & 0 & 2 & 0 & 2 & 0
   & -2 & 0 & 0 & 0 & 0 & 0 & 0 \\
 D_{28} & 0 & 0 & 0 & 0 & 0 & 0 & 2 & 0 & -2 & 0 & -2 & 0
   & 2 & 0 & 0 & 0 & 0 & 0 & 0 \\
 D_{29} & 0 & 0 & 0 & 0 & 0 & 0 & 0 & 0 & 0 & 0 & 0 & 0 &
   0 & 0 & 0 & 2 & 0 & -2 & 0 \\
 D_{30} & 0 & 0 & 0 & 0 & 0 & 0 & 0 & 0 & 0 & 0 & 0 & 0 &
   0 & 0 & 0 & -1 & -\sqrt{3} & 1 & \sqrt{3} \\
 D_{31} & 0 & 0 & 0 & 0 & 0 & 0 & 0 & 0 & 0 & 0 & 0 & 0 &
   0 & 0 & 0 & -1 & \sqrt{3} & 1 & -\sqrt{3} \\
 D_{32} & 0 & -2 & 2 & -2 & 2 & 0 & 0 & 0 & 0 & 0 & 0 & 0
   & 0 & 0 & 0 & 0 & 0 & 0 & 0 \\
 D_{33} & 0 & 2 & -2 & 2 & -2 & 0 & 0 & 0 & 0 & 0 & 0 & 0
   & 0 & 0 & 0 & 0 & 0 & 0 & 0 \\
 D_{34} & 0 & 2 & -2 & -2 & 2 & 0 & 0 & 0 & 0 & 0 & 0 & 0
   & 0 & 0 & 0 & 0 & 0 & 0 & 0 \\
 D_{35} & 0 & -2 & 2 & 2 & -2 & 0 & 0 & 0 & 0 & 0 & 0 & 0
   & 0 & 0 & 0 & 0 & 0 & 0 & 0 \\
 D_{36} & 0 & 0 & 0 & 0 & 0 & 0 & 0 & 0 & 0 & 0 & 0 & 0 &
   0 & 0 & 0 & 0 & 0 & 0 & 0 \\
 D_{37} & 0 & 0 & 0 & 0 & 0 & 0 & 0 & 0 & 0 & 0 & 0 & 0 &
   0 & 0 & 0 & 0 & 0 & 0 & 0\\
 \hline
\end{array}
\end{equation}
}
\subsection{Character Table of the Group $\mathrm{G_{768}}$}
The group $\mathrm{G_{768}}$ has $32$ conjugacy classes and therefore $32$ irreducible representations that are distributed according to the following pattern:
\begin{description}
  \item[a)] 6 irreps of dimension $1$, namely $D_1,\dots,D_6$
  \item[b)] 10 irreps of dimension $3$, namely $D_6,\dots,D_{16}$
  \item[c)] 6 irreps of dimension $4$, namely $D_{16},\dots,D_{22}$
  \item[d)] 8 irreps of dimension $6$, namely $D_{23},\dots,D_{30}$
  \item[e)] 2 irreps of dimension $12$, namely $D_{31}, D_{32}$
\end{description}
The corresponding character table that we have calculated with the procedures described in the main text is displayed below. For pure typographical reasons we were forced to split the character table in two parts in order to fit it into the page.
{\scriptsize \begin{equation}\label{1charto768}
\begin{array}{|l|llllllllllllllll|}
\hline
 0 & C_1 & C_2 & C_3 & C_4 & C_5 & C_6 & C_7 & C_8 & C_9
   & C_{10} & C_{11} & C_{12} & C_{13} & C_{14} & C_{15}
   & C_{16} \\
 \hline
 D_1 & 1 & 1 & 1 & 1 & 1 & 1 & 1 & 1 & 1 & 1 & 1 & 1 & 1
   & 1 & 1 & 1 \\
 D_2 & 1 & 1 & 1 & 1 & 1 & 1 & 1 & 1 & 1 & 1 & 1 & 1 & 1
   & 1 & 1 & 1 \\
 D_3 & 1 & 1 & 1 & 1 & 1 & 1 & 1 & 1 & 1 & 1 & 1 & 1 & 1
   & 1 & 1 & 1 \\
 D_4 & 1 & 1 & 1 & 1 & -1 & -1 & -1 & -1 & -1 & -1 & 1 &
   1 & 1 & 1 & -1 & -1 \\
 D_5 & 1 & 1 & 1 & 1 & -1 & -1 & -1 & -1 & -1 & -1 & 1 &
   1 & 1 & 1 & -1 & -1 \\
 D_6 & 1 & 1 & 1 & 1 & -1 & -1 & -1 & -1 & -1 & -1 & 1 &
   1 & 1 & 1 & -1 & -1 \\
 D_7 & 3 & 3 & 3 & 3 & 3 & 3 & 3 & 3 & 3 & 3 & 3 & 3 & -1
   & -1 & -1 & -1 \\
 D_8 & 3 & 3 & 3 & 3 & -3 & -3 & 1 & 1 & 1 & 1 & -1 & -1
   & 3 & 3 & 1 & 1 \\
 D_9 & 3 & 3 & 3 & 3 & -3 & -3 & 1 & 1 & 1 & 1 & -1 & -1
   & -1 & -1 & 1 & 1 \\
 D_{10} & 3 & 3 & 3 & 3 & -3 & -3 & 1 & 1 & 1 & 1 & -1 &
   -1 & -1 & -1 & 1 & 1 \\
 D_{11} & 3 & 3 & 3 & 3 & -3 & -3 & 1 & 1 & 1 & 1 & -1 &
   -1 & -1 & -1 & -3 & -3 \\
 D_{12} & 3 & 3 & 3 & 3 & 3 & 3 & -1 & -1 & -1 & -1 & -1
   & -1 & 3 & 3 & -1 & -1 \\
 D_{13} & 3 & 3 & 3 & 3 & 3 & 3 & -1 & -1 & -1 & -1 & -1
   & -1 & -1 & -1 & -1 & -1 \\
 D_{14} & 3 & 3 & 3 & 3 & 3 & 3 & -1 & -1 & -1 & -1 & -1
   & -1 & -1 & -1 & 3 & 3 \\
 D_{15} & 3 & 3 & 3 & 3 & 3 & 3 & -1 & -1 & -1 & -1 & -1
   & -1 & -1 & -1 & -1 & -1 \\
 D_{16} & 3 & 3 & 3 & 3 & -3 & -3 & -3 & -3 & -3 & -3 & 3
   & 3 & -1 & -1 & 1 & 1 \\
 D_{17} & 4 & -4 & -4 & 4 & -4 i & 4 i & 0 & 0 & 0 & 0 &
   0 & 0 & 0 & 0 & 0 & 0 \\
 D_{18} & 4 & -4 & -4 & 4 & -4 i & 4 i & 0 & 0 & 0 & 0 &
   0 & 0 & 0 & 0 & 0 & 0 \\
 D_{19} & 4 & -4 & -4 & 4 & -4 i & 4 i & 0 & 0 & 0 & 0 &
   0 & 0 & 0 & 0 & 0 & 0 \\
 D_{20} & 4 & -4 & -4 & 4 & 4 i & -4 i & 0 & 0 & 0 & 0 &
   0 & 0 & 0 & 0 & 0 & 0 \\
 D_{21} & 4 & -4 & -4 & 4 & 4 i & -4 i & 0 & 0 & 0 & 0 &
   0 & 0 & 0 & 0 & 0 & 0 \\
 D_{22} & 4 & -4 & -4 & 4 & 4 i & -4 i & 0 & 0 & 0 & 0 &
   0 & 0 & 0 & 0 & 0 & 0 \\
 D_{23} & 6 & -6 & 2 & -2 & 0 & 0 & 4 & 0 & 0 & -4 & 2 &
   -2 & 2 & -2 & 2 & -2 \\
 D_{24} & 6 & -6 & 2 & -2 & 0 & 0 & 4 & 0 & 0 & -4 & 2 &
   -2 & -2 & 2 & -2 & 2 \\
 D_{25} & 6 & -6 & 2 & -2 & 0 & 0 & 0 & 4 & -4 & 0 & -2 &
   2 & 2 & -2 & -2 & 2 \\
 D_{26} & 6 & -6 & 2 & -2 & 0 & 0 & 0 & 4 & -4 & 0 & -2 &
   2 & -2 & 2 & 2 & -2 \\
 D_{27} & 6 & -6 & 2 & -2 & 0 & 0 & 0 & -4 & 4 & 0 & -2 &
   2 & 2 & -2 & 2 & -2 \\
 D_{28} & 6 & -6 & 2 & -2 & 0 & 0 & 0 & -4 & 4 & 0 & -2 &
   2 & -2 & 2 & -2 & 2 \\
 D_{29} & 6 & -6 & 2 & -2 & 0 & 0 & -4 & 0 & 0 & 4 & 2 &
   -2 & 2 & -2 & -2 & 2 \\
 D_{30} & 6 & -6 & 2 & -2 & 0 & 0 & -4 & 0 & 0 & 4 & 2 &
   -2 & -2 & 2 & 2 & -2 \\
 D_{31} & 12 & 12 & -4 & -4 & 0 & 0 & 4 & -4 & -4 & 4 & 0
   & 0 & 0 & 0 & 0 & 0 \\
 D_{32} & 12 & 12 & -4 & -4 & 0 & 0 & -4 & 4 & 4 & -4 & 0
   & 0 & 0 & 0 & 0 & 0\\
   \hline
\end{array}
\end{equation}}
{\tiny
\begin{equation}\label{2charto768}
\begin{array}{lllllllllllllllll}
 0 & C_{17} & C_{18} & C_{19} & C_{20} & C_{21} & C_{22}
   & C_{23} & C_{24} & C_{25} & C_{26} & C_{27} & C_{28}
   & C_{29} & C_{30} & C_{31} & C_{32} \\
 D_1 & 1 & 1 & 1 & 1 & 1 & 1 & 1 & 1 & 1 & 1 & 1 & 1 & 1
   & 1 & 1 & 1 \\
 D_2 & 1 & 1 & 1 & 1 & 1 & 1 & 1 & 1 & -\sqrt[3]{-1} &
   -\sqrt[3]{-1} & -\sqrt[3]{-1} & -\sqrt[3]{-1} &
   (-1)^{2/3} & (-1)^{2/3} & (-1)^{2/3} & (-1)^{2/3} \\
 D_3 & 1 & 1 & 1 & 1 & 1 & 1 & 1 & 1 & (-1)^{2/3} &
   (-1)^{2/3} & (-1)^{2/3} & (-1)^{2/3} & -\sqrt[3]{-1} &
   -\sqrt[3]{-1} & -\sqrt[3]{-1} & -\sqrt[3]{-1} \\
 D_4 & -1 & -1 & 1 & 1 & -1 & 1 & 1 & -1 & 1 & -1 & 1 &
   -1 & 1 & -1 & 1 & -1 \\
 D_5 & -1 & -1 & 1 & 1 & -1 & 1 & 1 & -1 & -\sqrt[3]{-1}
   & \sqrt[3]{-1} & -\sqrt[3]{-1} & \sqrt[3]{-1} &
   (-1)^{2/3} & -(-1)^{2/3} & (-1)^{2/3} & -(-1)^{2/3} \\
 D_6 & -1 & -1 & 1 & 1 & -1 & 1 & 1 & -1 & (-1)^{2/3} &
   -(-1)^{2/3} & (-1)^{2/3} & -(-1)^{2/3} & -\sqrt[3]{-1}
   & \sqrt[3]{-1} & -\sqrt[3]{-1} & \sqrt[3]{-1} \\
 D_7 & -1 & -1 & -1 & -1 & -1 & -1 & -1 & -1 & 0 & 0 & 0
   & 0 & 0 & 0 & 0 & 0 \\
 D_8 & 1 & 1 & -1 & -1 & 1 & -1 & -1 & -3 & 0 & 0 & 0 & 0
   & 0 & 0 & 0 & 0 \\
 D_9 & 1 & 1 & 3 & 3 & -3 & -1 & -1 & 1 & 0 & 0 & 0 & 0 &
   0 & 0 & 0 & 0 \\
 D_{10} & -3 & -3 & -1 & -1 & 1 & 3 & -1 & 1 & 0 & 0 & 0
   & 0 & 0 & 0 & 0 & 0 \\
 D_{11} & 1 & 1 & -1 & -1 & 1 & -1 & 3 & 1 & 0 & 0 & 0 &
   0 & 0 & 0 & 0 & 0 \\
 D_{12} & -1 & -1 & -1 & -1 & -1 & -1 & -1 & 3 & 0 & 0 &
   0 & 0 & 0 & 0 & 0 & 0 \\
 D_{13} & 3 & 3 & -1 & -1 & -1 & 3 & -1 & -1 & 0 & 0 & 0
   & 0 & 0 & 0 & 0 & 0 \\
 D_{14} & -1 & -1 & -1 & -1 & -1 & -1 & 3 & -1 & 0 & 0 &
   0 & 0 & 0 & 0 & 0 & 0 \\
 D_{15} & -1 & -1 & 3 & 3 & 3 & -1 & -1 & -1 & 0 & 0 & 0
   & 0 & 0 & 0 & 0 & 0 \\
 D_{16} & 1 & 1 & -1 & -1 & 1 & -1 & -1 & 1 & 0 & 0 & 0 &
   0 & 0 & 0 & 0 & 0 \\
 D_{17} & 0 & 0 & 0 & 0 & 0 & 0 & 0 & 0 & 1 & -i & -1 & i
   & 1 & -i & -1 & i \\
 D_{18} & 0 & 0 & 0 & 0 & 0 & 0 & 0 & 0 & -\sqrt[3]{-1} &
   (-1)^{5/6} & \sqrt[3]{-1} & -(-1)^{5/6} & (-1)^{2/3} &
   \sqrt[6]{-1} & -(-1)^{2/3} & -\sqrt[6]{-1} \\
 D_{19} & 0 & 0 & 0 & 0 & 0 & 0 & 0 & 0 & (-1)^{2/3} &
   \sqrt[6]{-1} & -(-1)^{2/3} & -\sqrt[6]{-1} &
   -\sqrt[3]{-1} & (-1)^{5/6} & \sqrt[3]{-1} &
   -(-1)^{5/6} \\
 D_{20} & 0 & 0 & 0 & 0 & 0 & 0 & 0 & 0 & 1 & i & -1 & -i
   & 1 & i & -1 & -i \\
 D_{21} & 0 & 0 & 0 & 0 & 0 & 0 & 0 & 0 & -\sqrt[3]{-1} &
   -(-1)^{5/6} & \sqrt[3]{-1} & (-1)^{5/6} & (-1)^{2/3} &
   -\sqrt[6]{-1} & -(-1)^{2/3} & \sqrt[6]{-1} \\
 D_{22} & 0 & 0 & 0 & 0 & 0 & 0 & 0 & 0 & (-1)^{2/3} &
   -\sqrt[6]{-1} & -(-1)^{2/3} & \sqrt[6]{-1} &
   -\sqrt[3]{-1} & -(-1)^{5/6} & \sqrt[3]{-1} &
   (-1)^{5/6} \\
 D_{23} & 2 & -2 & 2 & -2 & 0 & 0 & 0 & 0 & 0 & 0 & 0 & 0
   & 0 & 0 & 0 & 0 \\
 D_{24} & -2 & 2 & -2 & 2 & 0 & 0 & 0 & 0 & 0 & 0 & 0 & 0
   & 0 & 0 & 0 & 0 \\
 D_{25} & 2 & -2 & -2 & 2 & 0 & 0 & 0 & 0 & 0 & 0 & 0 & 0
   & 0 & 0 & 0 & 0 \\
 D_{26} & -2 & 2 & 2 & -2 & 0 & 0 & 0 & 0 & 0 & 0 & 0 & 0
   & 0 & 0 & 0 & 0 \\
 D_{27} & -2 & 2 & -2 & 2 & 0 & 0 & 0 & 0 & 0 & 0 & 0 & 0
   & 0 & 0 & 0 & 0 \\
 D_{28} & 2 & -2 & 2 & -2 & 0 & 0 & 0 & 0 & 0 & 0 & 0 & 0
   & 0 & 0 & 0 & 0 \\
 D_{29} & -2 & 2 & 2 & -2 & 0 & 0 & 0 & 0 & 0 & 0 & 0 & 0
   & 0 & 0 & 0 & 0 \\
 D_{30} & 2 & -2 & -2 & 2 & 0 & 0 & 0 & 0 & 0 & 0 & 0 & 0
   & 0 & 0 & 0 & 0 \\
 D_{31} & 0 & 0 & 0 & 0 & 0 & 0 & 0 & 0 & 0 & 0 & 0 & 0 &
   0 & 0 & 0 & 0 \\
 D_{32} & 0 & 0 & 0 & 0 & 0 & 0 & 0 & 0 & 0 & 0 & 0 & 0 &
   0 & 0 & 0 & 0
\end{array}
\end{equation}
}
\subsection{Character Table of the Group $\mathrm{G_{256}}$}
The group $\mathrm{G_{256}}$ has $64$ conjugacy classes and therefore $64$ irreducible representations that are distributed according to the following pattern:
\begin{description}
  \item[a)] 32 irreps of dimension $1$, namely $D_1,\dots,D_{32}$
  \item[b)] 24 irreps of dimension $2$, namely $D_{33},\dots,D_{56}$
  \item[c)] 8 irreps of dimension $4$, namely $D_{57},\dots,D_{64}$
\end{description}
The corresponding character table that we have calculated with the procedures described in the main text is displayed below. For pure typographical reasons we were forced to split the character table in three parts in order to fit it into the page.
{\tiny
\begin{equation}\label{1charto256}
\begin{array}{|l|llllllllllllllllllllll|}
\hline
0 & C_1 & C_2 & C_3 & C_4 & C_5 & C_6 & C_7 & C_8 & C_9
   & C_{10} & C_{11} & C_{12} & C_{13} & C_{14} & C_{15}
   & C_{16} & C_{17} & C_{18} & C_{19} & C_{20} & C_{21}
   & C_{22} \\
 \hline
 D_1 & 1 & 1 & 1 & 1 & 1 & 1 & 1 & 1 & 1 & 1 & 1 & 1 & 1
   & 1 & 1 & 1 & 1 & 1 & 1 & 1 & 1 & 1 \\
 D_2 & 1 & 1 & 1 & 1 & 1 & 1 & 1 & 1 & 1 & 1 & 1 & 1 & 1
   & 1 & 1 & 1 & 1 & 1 & 1 & 1 & 1 & 1 \\
 D_3 & 1 & 1 & 1 & 1 & 1 & 1 & 1 & 1 & 1 & 1 & 1 & 1 & 1
   & 1 & 1 & 1 & 1 & 1 & 1 & 1 & 1 & 1 \\
 D_4 & 1 & 1 & 1 & 1 & 1 & 1 & 1 & 1 & 1 & 1 & 1 & 1 & 1
   & 1 & 1 & 1 & 1 & 1 & 1 & 1 & 1 & 1 \\
 D_5 & 1 & 1 & 1 & 1 & 1 & 1 & 1 & 1 & -1 & 1 & 1 & -1 &
   1 & 1 & 1 & 1 & -1 & 1 & 1 & -1 & -1 & -1 \\
 D_6 & 1 & 1 & 1 & 1 & 1 & 1 & 1 & 1 & -1 & 1 & 1 & -1 &
   1 & 1 & 1 & 1 & -1 & 1 & 1 & -1 & -1 & -1 \\
 D_7 & 1 & 1 & 1 & 1 & 1 & 1 & 1 & 1 & -1 & 1 & 1 & -1 &
   1 & 1 & 1 & 1 & -1 & 1 & 1 & -1 & -1 & -1 \\
 D_8 & 1 & 1 & 1 & 1 & 1 & 1 & 1 & 1 & -1 & 1 & 1 & -1 &
   1 & 1 & 1 & 1 & -1 & 1 & 1 & -1 & -1 & -1 \\
 D_9 & 1 & 1 & 1 & 1 & 1 & 1 & 1 & 1 & 1 & -1 & -1 & 1 &
   1 & 1 & 1 & 1 & 1 & -1 & -1 & 1 & -1 & 1 \\
 D_{10} & 1 & 1 & 1 & 1 & 1 & 1 & 1 & 1 & 1 & -1 & -1 & 1
   & 1 & 1 & 1 & 1 & 1 & -1 & -1 & 1 & -1 & 1 \\
 D_{11} & 1 & 1 & 1 & 1 & 1 & 1 & 1 & 1 & 1 & -1 & -1 & 1
   & 1 & 1 & 1 & 1 & 1 & -1 & -1 & 1 & -1 & 1 \\
 D_{12} & 1 & 1 & 1 & 1 & 1 & 1 & 1 & 1 & 1 & -1 & -1 & 1
   & 1 & 1 & 1 & 1 & 1 & -1 & -1 & 1 & -1 & 1 \\
 D_{13} & 1 & 1 & 1 & 1 & 1 & 1 & 1 & 1 & -1 & -1 & -1 &
   -1 & 1 & 1 & 1 & 1 & -1 & -1 & -1 & -1 & 1 & -1 \\
 D_{14} & 1 & 1 & 1 & 1 & 1 & 1 & 1 & 1 & -1 & -1 & -1 &
   -1 & 1 & 1 & 1 & 1 & -1 & -1 & -1 & -1 & 1 & -1 \\
 D_{15} & 1 & 1 & 1 & 1 & 1 & 1 & 1 & 1 & -1 & -1 & -1 &
   -1 & 1 & 1 & 1 & 1 & -1 & -1 & -1 & -1 & 1 & -1 \\
 D_{16} & 1 & 1 & 1 & 1 & 1 & 1 & 1 & 1 & -1 & -1 & -1 &
   -1 & 1 & 1 & 1 & 1 & -1 & -1 & -1 & -1 & 1 & -1 \\
 D_{17} & 1 & 1 & 1 & 1 & 1 & 1 & 1 & 1 & 1 & 1 & 1 & 1 &
   -1 & -1 & -1 & -1 & 1 & 1 & 1 & 1 & 1 & -1 \\
 D_{18} & 1 & 1 & 1 & 1 & 1 & 1 & 1 & 1 & 1 & 1 & 1 & 1 &
   -1 & -1 & -1 & -1 & 1 & 1 & 1 & 1 & 1 & -1 \\
 D_{19} & 1 & 1 & 1 & 1 & 1 & 1 & 1 & 1 & 1 & 1 & 1 & 1 &
   -1 & -1 & -1 & -1 & 1 & 1 & 1 & 1 & 1 & -1 \\
 D_{20} & 1 & 1 & 1 & 1 & 1 & 1 & 1 & 1 & 1 & 1 & 1 & 1 &
   -1 & -1 & -1 & -1 & 1 & 1 & 1 & 1 & 1 & -1 \\
 D_{21} & 1 & 1 & 1 & 1 & 1 & 1 & 1 & 1 & -1 & 1 & 1 & -1
   & -1 & -1 & -1 & -1 & -1 & 1 & 1 & -1 & -1 & 1 \\
 D_{22} & 1 & 1 & 1 & 1 & 1 & 1 & 1 & 1 & -1 & 1 & 1 & -1
   & -1 & -1 & -1 & -1 & -1 & 1 & 1 & -1 & -1 & 1 \\
 D_{23} & 1 & 1 & 1 & 1 & 1 & 1 & 1 & 1 & -1 & 1 & 1 & -1
   & -1 & -1 & -1 & -1 & -1 & 1 & 1 & -1 & -1 & 1 \\
 D_{24} & 1 & 1 & 1 & 1 & 1 & 1 & 1 & 1 & -1 & 1 & 1 & -1
   & -1 & -1 & -1 & -1 & -1 & 1 & 1 & -1 & -1 & 1 \\
 D_{25} & 1 & 1 & 1 & 1 & 1 & 1 & 1 & 1 & 1 & -1 & -1 & 1
   & -1 & -1 & -1 & -1 & 1 & -1 & -1 & 1 & -1 & -1 \\
 D_{26} & 1 & 1 & 1 & 1 & 1 & 1 & 1 & 1 & 1 & -1 & -1 & 1
   & -1 & -1 & -1 & -1 & 1 & -1 & -1 & 1 & -1 & -1 \\
 D_{27} & 1 & 1 & 1 & 1 & 1 & 1 & 1 & 1 & 1 & -1 & -1 & 1
   & -1 & -1 & -1 & -1 & 1 & -1 & -1 & 1 & -1 & -1 \\
 D_{28} & 1 & 1 & 1 & 1 & 1 & 1 & 1 & 1 & 1 & -1 & -1 & 1
   & -1 & -1 & -1 & -1 & 1 & -1 & -1 & 1 & -1 & -1 \\
 D_{29} & 1 & 1 & 1 & 1 & 1 & 1 & 1 & 1 & -1 & -1 & -1 &
   -1 & -1 & -1 & -1 & -1 & -1 & -1 & -1 & -1 & 1 & 1 \\
 D_{30} & 1 & 1 & 1 & 1 & 1 & 1 & 1 & 1 & -1 & -1 & -1 &
   -1 & -1 & -1 & -1 & -1 & -1 & -1 & -1 & -1 & 1 & 1 \\
 D_{31} & 1 & 1 & 1 & 1 & 1 & 1 & 1 & 1 & -1 & -1 & -1 &
   -1 & -1 & -1 & -1 & -1 & -1 & -1 & -1 & -1 & 1 & 1 \\
 D_{32} & 1 & 1 & 1 & 1 & 1 & 1 & 1 & 1 & -1 & -1 & -1 &
   -1 & -1 & -1 & -1 & -1 & -1 & -1 & -1 & -1 & 1 & 1 \\
 D_{33} & 2 & -2 & 2 & -2 & 2 & -2 & 2 & -2 & 0 & 2 & -2
   & 0 & 2 & -2 & 2 & -2 & 0 & 2 & -2 & 0 & 0 & 0 \\
 D_{34} & 2 & -2 & 2 & -2 & 2 & -2 & 2 & -2 & 0 & 2 & -2
   & 0 & 2 & -2 & 2 & -2 & 0 & 2 & -2 & 0 & 0 & 0 \\
 D_{35} & 2 & -2 & 2 & -2 & 2 & -2 & 2 & -2 & 0 & -2 & 2
   & 0 & 2 & -2 & 2 & -2 & 0 & -2 & 2 & 0 & 0 & 0 \\
 D_{36} & 2 & -2 & 2 & -2 & 2 & -2 & 2 & -2 & 0 & -2 & 2
   & 0 & 2 & -2 & 2 & -2 & 0 & -2 & 2 & 0 & 0 & 0 \\
 D_{37} & 2 & -2 & 2 & -2 & 2 & -2 & 2 & -2 & 0 & 2 & -2
   & 0 & -2 & 2 & -2 & 2 & 0 & 2 & -2 & 0 & 0 & 0 \\
 D_{38} & 2 & -2 & 2 & -2 & 2 & -2 & 2 & -2 & 0 & 2 & -2
   & 0 & -2 & 2 & -2 & 2 & 0 & 2 & -2 & 0 & 0 & 0 \\
 D_{39} & 2 & -2 & 2 & -2 & 2 & -2 & 2 & -2 & 0 & -2 & 2
   & 0 & -2 & 2 & -2 & 2 & 0 & -2 & 2 & 0 & 0 & 0 \\
 D_{40} & 2 & -2 & 2 & -2 & 2 & -2 & 2 & -2 & 0 & -2 & 2
   & 0 & -2 & 2 & -2 & 2 & 0 & -2 & 2 & 0 & 0 & 0 \\
 D_{41} & 2 & 2 & -2 & -2 & 2 & 2 & -2 & -2 & 2 & 0 & 0 &
   -2 & 2 & 2 & -2 & -2 & 2 & 0 & 0 & -2 & 0 & 2 \\
 D_{42} & 2 & 2 & -2 & -2 & 2 & 2 & -2 & -2 & 2 & 0 & 0 &
   -2 & 2 & 2 & -2 & -2 & 2 & 0 & 0 & -2 & 0 & 2 \\
 D_{43} & 2 & 2 & -2 & -2 & 2 & 2 & -2 & -2 & -2 & 0 & 0
   & 2 & 2 & 2 & -2 & -2 & -2 & 0 & 0 & 2 & 0 & -2 \\
 D_{44} & 2 & 2 & -2 & -2 & 2 & 2 & -2 & -2 & -2 & 0 & 0
   & 2 & 2 & 2 & -2 & -2 & -2 & 0 & 0 & 2 & 0 & -2 \\
 D_{45} & 2 & 2 & 2 & 2 & -2 & -2 & -2 & -2 & 2 & 2 & 2 &
   2 & 0 & 0 & 0 & 0 & -2 & -2 & -2 & -2 & 2 & 0 \\
 D_{46} & 2 & 2 & 2 & 2 & -2 & -2 & -2 & -2 & 2 & 2 & 2 &
   2 & 0 & 0 & 0 & 0 & -2 & -2 & -2 & -2 & 2 & 0 \\
 D_{47} & 2 & 2 & 2 & 2 & -2 & -2 & -2 & -2 & -2 & 2 & 2
   & -2 & 0 & 0 & 0 & 0 & 2 & -2 & -2 & 2 & -2 & 0 \\
 D_{48} & 2 & 2 & 2 & 2 & -2 & -2 & -2 & -2 & -2 & 2 & 2
   & -2 & 0 & 0 & 0 & 0 & 2 & -2 & -2 & 2 & -2 & 0 \\
 D_{49} & 2 & 2 & 2 & 2 & -2 & -2 & -2 & -2 & 2 & -2 & -2
   & 2 & 0 & 0 & 0 & 0 & -2 & 2 & 2 & -2 & -2 & 0 \\
 D_{50} & 2 & 2 & 2 & 2 & -2 & -2 & -2 & -2 & 2 & -2 & -2
   & 2 & 0 & 0 & 0 & 0 & -2 & 2 & 2 & -2 & -2 & 0 \\
 D_{51} & 2 & 2 & 2 & 2 & -2 & -2 & -2 & -2 & -2 & -2 &
   -2 & -2 & 0 & 0 & 0 & 0 & 2 & 2 & 2 & 2 & 2 & 0 \\
 D_{52} & 2 & 2 & 2 & 2 & -2 & -2 & -2 & -2 & -2 & -2 &
   -2 & -2 & 0 & 0 & 0 & 0 & 2 & 2 & 2 & 2 & 2 & 0 \\
 D_{53} & 2 & 2 & -2 & -2 & 2 & 2 & -2 & -2 & 2 & 0 & 0 &
   -2 & -2 & -2 & 2 & 2 & 2 & 0 & 0 & -2 & 0 & -2 \\
 D_{54} & 2 & 2 & -2 & -2 & 2 & 2 & -2 & -2 & 2 & 0 & 0 &
   -2 & -2 & -2 & 2 & 2 & 2 & 0 & 0 & -2 & 0 & -2 \\
 D_{55} & 2 & 2 & -2 & -2 & 2 & 2 & -2 & -2 & -2 & 0 & 0
   & 2 & -2 & -2 & 2 & 2 & -2 & 0 & 0 & 2 & 0 & 2 \\
 D_{56} & 2 & 2 & -2 & -2 & 2 & 2 & -2 & -2 & -2 & 0 & 0
   & 2 & -2 & -2 & 2 & 2 & -2 & 0 & 0 & 2 & 0 & 2 \\
 D_{57} & 4 & -4 & -4 & 4 & 4 & -4 & -4 & 4 & 0 & 0 & 0 &
   0 & 4 & -4 & -4 & 4 & 0 & 0 & 0 & 0 & 0 & 0 \\
 D_{58} & 4 & -4 & 4 & -4 & -4 & 4 & -4 & 4 & 0 & 4 & -4
   & 0 & 0 & 0 & 0 & 0 & 0 & -4 & 4 & 0 & 0 & 0 \\
 D_{59} & 4 & 4 & -4 & -4 & -4 & -4 & 4 & 4 & 4 & 0 & 0 &
   -4 & 0 & 0 & 0 & 0 & -4 & 0 & 0 & 4 & 0 & 0 \\
 D_{60} & 4 & -4 & -4 & 4 & -4 & 4 & 4 & -4 & 0 & 0 & 0 &
   0 & 0 & 0 & 0 & 0 & 0 & 0 & 0 & 0 & 0 & 0 \\
 D_{61} & 4 & 4 & -4 & -4 & -4 & -4 & 4 & 4 & -4 & 0 & 0
   & 4 & 0 & 0 & 0 & 0 & 4 & 0 & 0 & -4 & 0 & 0 \\
 D_{62} & 4 & -4 & -4 & 4 & -4 & 4 & 4 & -4 & 0 & 0 & 0 &
   0 & 0 & 0 & 0 & 0 & 0 & 0 & 0 & 0 & 0 & 0 \\
 D_{63} & 4 & -4 & 4 & -4 & -4 & 4 & -4 & 4 & 0 & -4 & 4
   & 0 & 0 & 0 & 0 & 0 & 0 & 4 & -4 & 0 & 0 & 0 \\
 D_{64} & 4 & -4 & -4 & 4 & 4 & -4 & -4 & 4 & 0 & 0 & 0 &
   0 & -4 & 4 & 4 & -4 & 0 & 0 & 0 & 0 & 0 & 0\\
   \hline
\end{array}
\end{equation}}
{\tiny
\begin{equation}\label{2charto256}
\begin{array}{|l|llllllllllllllllllllll|}
\hline
 0 & C_{23} & C_{24} & C_{25} & C_{26} & C_{27} & C_{28}
   & C_{29} & C_{30} & C_{31} & C_{32} & C_{33} & C_{34}
   & C_{35} & C_{36} & C_{37} & C_{38} & C_{39} & C_{40}
   & C_{41} & C_{42} & C_{43} & C_{44} \\
 \hline
 D_1 & 1 & 1 & 1 & 1 & 1 & 1 & 1 & 1 & 1 & 1 & 1 & 1 & 1
   & 1 & 1 & 1 & 1 & 1 & 1 & 1 & 1 & 1 \\
 D_2 & 1 & 1 & 1 & 1 & 1 & 1 & 1 & 1 & 1 & 1 & 1 & 1 & 1
   & 1 & -1 & -1 & -1 & -1 & -1 & -1 & -1 & -1 \\
 D_3 & 1 & 1 & 1 & 1 & 1 & 1 & -1 & -1 & -1 & -1 & -1 &
   -1 & -1 & -1 & 1 & 1 & 1 & 1 & 1 & 1 & 1 & 1 \\
 D_4 & 1 & 1 & 1 & 1 & 1 & 1 & -1 & -1 & -1 & -1 & -1 &
   -1 & -1 & -1 & -1 & -1 & -1 & -1 & -1 & -1 & -1 & -1
   \\
 D_5 & 1 & -1 & 1 & -1 & -1 & -1 & 1 & 1 & 1 & 1 & 1 & 1
   & 1 & 1 & 1 & -1 & 1 & -1 & 1 & -1 & 1 & -1 \\
 D_6 & 1 & -1 & 1 & -1 & -1 & -1 & 1 & 1 & 1 & 1 & 1 & 1
   & 1 & 1 & -1 & 1 & -1 & 1 & -1 & 1 & -1 & 1 \\
 D_7 & 1 & -1 & 1 & -1 & -1 & -1 & -1 & -1 & -1 & -1 & -1
   & -1 & -1 & -1 & 1 & -1 & 1 & -1 & 1 & -1 & 1 & -1 \\
 D_8 & 1 & -1 & 1 & -1 & -1 & -1 & -1 & -1 & -1 & -1 & -1
   & -1 & -1 & -1 & -1 & 1 & -1 & 1 & -1 & 1 & -1 & 1 \\
 D_9 & -1 & -1 & -1 & -1 & 1 & -1 & 1 & 1 & -1 & -1 & 1 &
   1 & -1 & -1 & 1 & 1 & 1 & 1 & 1 & 1 & 1 & 1 \\
 D_{10} & -1 & -1 & -1 & -1 & 1 & -1 & 1 & 1 & -1 & -1 &
   1 & 1 & -1 & -1 & -1 & -1 & -1 & -1 & -1 & -1 & -1 &
   -1 \\
 D_{11} & -1 & -1 & -1 & -1 & 1 & -1 & -1 & -1 & 1 & 1 &
   -1 & -1 & 1 & 1 & 1 & 1 & 1 & 1 & 1 & 1 & 1 & 1 \\
 D_{12} & -1 & -1 & -1 & -1 & 1 & -1 & -1 & -1 & 1 & 1 &
   -1 & -1 & 1 & 1 & -1 & -1 & -1 & -1 & -1 & -1 & -1 &
   -1 \\
 D_{13} & -1 & 1 & -1 & 1 & -1 & 1 & 1 & 1 & -1 & -1 & 1
   & 1 & -1 & -1 & 1 & -1 & 1 & -1 & 1 & -1 & 1 & -1 \\
 D_{14} & -1 & 1 & -1 & 1 & -1 & 1 & 1 & 1 & -1 & -1 & 1
   & 1 & -1 & -1 & -1 & 1 & -1 & 1 & -1 & 1 & -1 & 1 \\
 D_{15} & -1 & 1 & -1 & 1 & -1 & 1 & -1 & -1 & 1 & 1 & -1
   & -1 & 1 & 1 & 1 & -1 & 1 & -1 & 1 & -1 & 1 & -1 \\
 D_{16} & -1 & 1 & -1 & 1 & -1 & 1 & -1 & -1 & 1 & 1 & -1
   & -1 & 1 & 1 & -1 & 1 & -1 & 1 & -1 & 1 & -1 & 1 \\
 D_{17} & -1 & -1 & -1 & -1 & -1 & 1 & 1 & 1 & 1 & 1 & -1
   & -1 & -1 & -1 & 1 & 1 & 1 & 1 & -1 & -1 & -1 & -1 \\
 D_{18} & -1 & -1 & -1 & -1 & -1 & 1 & 1 & 1 & 1 & 1 & -1
   & -1 & -1 & -1 & -1 & -1 & -1 & -1 & 1 & 1 & 1 & 1 \\
 D_{19} & -1 & -1 & -1 & -1 & -1 & 1 & -1 & -1 & -1 & -1
   & 1 & 1 & 1 & 1 & 1 & 1 & 1 & 1 & -1 & -1 & -1 & -1 \\
 D_{20} & -1 & -1 & -1 & -1 & -1 & 1 & -1 & -1 & -1 & -1
   & 1 & 1 & 1 & 1 & -1 & -1 & -1 & -1 & 1 & 1 & 1 & 1 \\
 D_{21} & -1 & 1 & -1 & 1 & 1 & -1 & 1 & 1 & 1 & 1 & -1 &
   -1 & -1 & -1 & 1 & -1 & 1 & -1 & -1 & 1 & -1 & 1 \\
 D_{22} & -1 & 1 & -1 & 1 & 1 & -1 & 1 & 1 & 1 & 1 & -1 &
   -1 & -1 & -1 & -1 & 1 & -1 & 1 & 1 & -1 & 1 & -1 \\
 D_{23} & -1 & 1 & -1 & 1 & 1 & -1 & -1 & -1 & -1 & -1 &
   1 & 1 & 1 & 1 & 1 & -1 & 1 & -1 & -1 & 1 & -1 & 1 \\
 D_{24} & -1 & 1 & -1 & 1 & 1 & -1 & -1 & -1 & -1 & -1 &
   1 & 1 & 1 & 1 & -1 & 1 & -1 & 1 & 1 & -1 & 1 & -1 \\
 D_{25} & 1 & 1 & 1 & 1 & -1 & -1 & 1 & 1 & -1 & -1 & -1
   & -1 & 1 & 1 & 1 & 1 & 1 & 1 & -1 & -1 & -1 & -1 \\
 D_{26} & 1 & 1 & 1 & 1 & -1 & -1 & 1 & 1 & -1 & -1 & -1
   & -1 & 1 & 1 & -1 & -1 & -1 & -1 & 1 & 1 & 1 & 1 \\
 D_{27} & 1 & 1 & 1 & 1 & -1 & -1 & -1 & -1 & 1 & 1 & 1 &
   1 & -1 & -1 & 1 & 1 & 1 & 1 & -1 & -1 & -1 & -1 \\
 D_{28} & 1 & 1 & 1 & 1 & -1 & -1 & -1 & -1 & 1 & 1 & 1 &
   1 & -1 & -1 & -1 & -1 & -1 & -1 & 1 & 1 & 1 & 1 \\
 D_{29} & 1 & -1 & 1 & -1 & 1 & 1 & 1 & 1 & -1 & -1 & -1
   & -1 & 1 & 1 & 1 & -1 & 1 & -1 & -1 & 1 & -1 & 1 \\
 D_{30} & 1 & -1 & 1 & -1 & 1 & 1 & 1 & 1 & -1 & -1 & -1
   & -1 & 1 & 1 & -1 & 1 & -1 & 1 & 1 & -1 & 1 & -1 \\
 D_{31} & 1 & -1 & 1 & -1 & 1 & 1 & -1 & -1 & 1 & 1 & 1 &
   1 & -1 & -1 & 1 & -1 & 1 & -1 & -1 & 1 & -1 & 1 \\
 D_{32} & 1 & -1 & 1 & -1 & 1 & 1 & -1 & -1 & 1 & 1 & 1 &
   1 & -1 & -1 & -1 & 1 & -1 & 1 & 1 & -1 & 1 & -1 \\
 D_{33} & 2 & 0 & -2 & 0 & 0 & 0 & 2 & -2 & 2 & -2 & 2 &
   -2 & 2 & -2 & 0 & 0 & 0 & 0 & 0 & 0 & 0 & 0 \\
 D_{34} & 2 & 0 & -2 & 0 & 0 & 0 & -2 & 2 & -2 & 2 & -2 &
   2 & -2 & 2 & 0 & 0 & 0 & 0 & 0 & 0 & 0 & 0 \\
 D_{35} & -2 & 0 & 2 & 0 & 0 & 0 & 2 & -2 & -2 & 2 & 2 &
   -2 & -2 & 2 & 0 & 0 & 0 & 0 & 0 & 0 & 0 & 0 \\
 D_{36} & -2 & 0 & 2 & 0 & 0 & 0 & -2 & 2 & 2 & -2 & -2 &
   2 & 2 & -2 & 0 & 0 & 0 & 0 & 0 & 0 & 0 & 0 \\
 D_{37} & -2 & 0 & 2 & 0 & 0 & 0 & 2 & -2 & 2 & -2 & -2 &
   2 & -2 & 2 & 0 & 0 & 0 & 0 & 0 & 0 & 0 & 0 \\
 D_{38} & -2 & 0 & 2 & 0 & 0 & 0 & -2 & 2 & -2 & 2 & 2 &
   -2 & 2 & -2 & 0 & 0 & 0 & 0 & 0 & 0 & 0 & 0 \\
 D_{39} & 2 & 0 & -2 & 0 & 0 & 0 & 2 & -2 & -2 & 2 & -2 &
   2 & 2 & -2 & 0 & 0 & 0 & 0 & 0 & 0 & 0 & 0 \\
 D_{40} & 2 & 0 & -2 & 0 & 0 & 0 & -2 & 2 & 2 & -2 & 2 &
   -2 & -2 & 2 & 0 & 0 & 0 & 0 & 0 & 0 & 0 & 0 \\
 D_{41} & 0 & 0 & 0 & 0 & -2 & 0 & 0 & 0 & 0 & 0 & 0 & 0
   & 0 & 0 & -2 & -2 & 2 & 2 & -2 & -2 & 2 & 2 \\
 D_{42} & 0 & 0 & 0 & 0 & -2 & 0 & 0 & 0 & 0 & 0 & 0 & 0
   & 0 & 0 & 2 & 2 & -2 & -2 & 2 & 2 & -2 & -2 \\
 D_{43} & 0 & 0 & 0 & 0 & 2 & 0 & 0 & 0 & 0 & 0 & 0 & 0 &
   0 & 0 & -2 & 2 & 2 & -2 & -2 & 2 & 2 & -2 \\
 D_{44} & 0 & 0 & 0 & 0 & 2 & 0 & 0 & 0 & 0 & 0 & 0 & 0 &
   0 & 0 & 2 & -2 & -2 & 2 & 2 & -2 & -2 & 2 \\
 D_{45} & 0 & 0 & 0 & 0 & 0 & -2 & 0 & 0 & 0 & 0 & 0 & 0
   & 0 & 0 & 0 & 0 & 0 & 0 & 0 & 0 & 0 & 0 \\
 D_{46} & 0 & 0 & 0 & 0 & 0 & -2 & 0 & 0 & 0 & 0 & 0 & 0
   & 0 & 0 & 0 & 0 & 0 & 0 & 0 & 0 & 0 & 0 \\
 D_{47} & 0 & 0 & 0 & 0 & 0 & 2 & 0 & 0 & 0 & 0 & 0 & 0 &
   0 & 0 & 0 & 0 & 0 & 0 & 0 & 0 & 0 & 0 \\
 D_{48} & 0 & 0 & 0 & 0 & 0 & 2 & 0 & 0 & 0 & 0 & 0 & 0 &
   0 & 0 & 0 & 0 & 0 & 0 & 0 & 0 & 0 & 0 \\
 D_{49} & 0 & 0 & 0 & 0 & 0 & 2 & 0 & 0 & 0 & 0 & 0 & 0 &
   0 & 0 & 0 & 0 & 0 & 0 & 0 & 0 & 0 & 0 \\
 D_{50} & 0 & 0 & 0 & 0 & 0 & 2 & 0 & 0 & 0 & 0 & 0 & 0 &
   0 & 0 & 0 & 0 & 0 & 0 & 0 & 0 & 0 & 0 \\
 D_{51} & 0 & 0 & 0 & 0 & 0 & -2 & 0 & 0 & 0 & 0 & 0 & 0
   & 0 & 0 & 0 & 0 & 0 & 0 & 0 & 0 & 0 & 0 \\
 D_{52} & 0 & 0 & 0 & 0 & 0 & -2 & 0 & 0 & 0 & 0 & 0 & 0
   & 0 & 0 & 0 & 0 & 0 & 0 & 0 & 0 & 0 & 0 \\
 D_{53} & 0 & 0 & 0 & 0 & 2 & 0 & 0 & 0 & 0 & 0 & 0 & 0 &
   0 & 0 & -2 & -2 & 2 & 2 & 2 & 2 & -2 & -2 \\
 D_{54} & 0 & 0 & 0 & 0 & 2 & 0 & 0 & 0 & 0 & 0 & 0 & 0 &
   0 & 0 & 2 & 2 & -2 & -2 & -2 & -2 & 2 & 2 \\
 D_{55} & 0 & 0 & 0 & 0 & -2 & 0 & 0 & 0 & 0 & 0 & 0 & 0
   & 0 & 0 & -2 & 2 & 2 & -2 & 2 & -2 & -2 & 2 \\
 D_{56} & 0 & 0 & 0 & 0 & -2 & 0 & 0 & 0 & 0 & 0 & 0 & 0
   & 0 & 0 & 2 & -2 & -2 & 2 & -2 & 2 & 2 & -2 \\
 D_{57} & 0 & 0 & 0 & 0 & 0 & 0 & 0 & 0 & 0 & 0 & 0 & 0 &
   0 & 0 & 0 & 0 & 0 & 0 & 0 & 0 & 0 & 0 \\
 D_{58} & 0 & 0 & 0 & 0 & 0 & 0 & 0 & 0 & 0 & 0 & 0 & 0 &
   0 & 0 & 0 & 0 & 0 & 0 & 0 & 0 & 0 & 0 \\
 D_{59} & 0 & 0 & 0 & 0 & 0 & 0 & 0 & 0 & 0 & 0 & 0 & 0 &
   0 & 0 & 0 & 0 & 0 & 0 & 0 & 0 & 0 & 0 \\
 D_{60} & 0 & -4 i & 0 & 4 i & 0 & 0 & 0 & 0 & 0 & 0 & 0
   & 0 & 0 & 0 & 0 & 0 & 0 & 0 & 0 & 0 & 0 & 0 \\
 D_{61} & 0 & 0 & 0 & 0 & 0 & 0 & 0 & 0 & 0 & 0 & 0 & 0 &
   0 & 0 & 0 & 0 & 0 & 0 & 0 & 0 & 0 & 0 \\
 D_{62} & 0 & 4 i & 0 & -4 i & 0 & 0 & 0 & 0 & 0 & 0 & 0
   & 0 & 0 & 0 & 0 & 0 & 0 & 0 & 0 & 0 & 0 & 0 \\
 D_{63} & 0 & 0 & 0 & 0 & 0 & 0 & 0 & 0 & 0 & 0 & 0 & 0 &
   0 & 0 & 0 & 0 & 0 & 0 & 0 & 0 & 0 & 0 \\
 D_{64} & 0 & 0 & 0 & 0 & 0 & 0 & 0 & 0 & 0 & 0 & 0 & 0 &
   0 & 0 & 0 & 0 & 0 & 0 & 0 & 0 & 0 & 0\\
\hline
\end{array}
\end{equation}
}
{\tiny
\begin{equation}\label{3charto256}
\begin{array}{|l|llllllllllllllllllll|}
\hline
 0 & C_{45} & C_{46} & C_{47} & C_{48} & C_{49} & C_{50}
   & C_{51} & C_{52} & C_{53} & C_{54} & C_{55} & C_{56}
   & C_{57} & C_{58} & C_{59} & C_{60} & C_{61} & C_{62}
   & C_{63} & C_{64} \\
 \hline
 D_1 & 1 & 1 & 1 & 1 & 1 & 1 & 1 & 1 & 1 & 1 & 1 & 1 & 1
   & 1 & 1 & 1 & 1 & 1 & 1 & 1 \\
 D_2 & 1 & 1 & 1 & 1 & 1 & 1 & 1 & 1 & 1 & 1 & 1 & 1 & 1
   & 1 & -1 & -1 & -1 & -1 & -1 & -1 \\
 D_3 & 1 & 1 & 1 & 1 & 1 & 1 & -1 & -1 & -1 & -1 & -1 &
   -1 & -1 & -1 & 1 & 1 & 1 & 1 & 1 & 1 \\
 D_4 & 1 & 1 & 1 & 1 & 1 & 1 & -1 & -1 & -1 & -1 & -1 &
   -1 & -1 & -1 & -1 & -1 & -1 & -1 & -1 & -1 \\
 D_5 & 1 & -1 & 1 & -1 & -1 & -1 & 1 & 1 & 1 & 1 & 1 & 1
   & 1 & 1 & 1 & -1 & 1 & -1 & 1 & -1 \\
 D_6 & 1 & -1 & 1 & -1 & -1 & -1 & 1 & 1 & 1 & 1 & 1 & 1
   & 1 & 1 & -1 & 1 & -1 & 1 & -1 & 1 \\
 D_7 & 1 & -1 & 1 & -1 & -1 & -1 & -1 & -1 & -1 & -1 & -1
   & -1 & -1 & -1 & 1 & -1 & 1 & -1 & 1 & -1 \\
 D_8 & 1 & -1 & 1 & -1 & -1 & -1 & -1 & -1 & -1 & -1 & -1
   & -1 & -1 & -1 & -1 & 1 & -1 & 1 & -1 & 1 \\
 D_9 & -1 & -1 & -1 & -1 & 1 & -1 & 1 & 1 & -1 & -1 & 1 &
   1 & -1 & -1 & 1 & 1 & 1 & 1 & 1 & 1 \\
 D_{10} & -1 & -1 & -1 & -1 & 1 & -1 & 1 & 1 & -1 & -1 &
   1 & 1 & -1 & -1 & -1 & -1 & -1 & -1 & -1 & -1 \\
 D_{11} & -1 & -1 & -1 & -1 & 1 & -1 & -1 & -1 & 1 & 1 &
   -1 & -1 & 1 & 1 & 1 & 1 & 1 & 1 & 1 & 1 \\
 D_{12} & -1 & -1 & -1 & -1 & 1 & -1 & -1 & -1 & 1 & 1 &
   -1 & -1 & 1 & 1 & -1 & -1 & -1 & -1 & -1 & -1 \\
 D_{13} & -1 & 1 & -1 & 1 & -1 & 1 & 1 & 1 & -1 & -1 & 1
   & 1 & -1 & -1 & 1 & -1 & 1 & -1 & 1 & -1 \\
 D_{14} & -1 & 1 & -1 & 1 & -1 & 1 & 1 & 1 & -1 & -1 & 1
   & 1 & -1 & -1 & -1 & 1 & -1 & 1 & -1 & 1 \\
 D_{15} & -1 & 1 & -1 & 1 & -1 & 1 & -1 & -1 & 1 & 1 & -1
   & -1 & 1 & 1 & 1 & -1 & 1 & -1 & 1 & -1 \\
 D_{16} & -1 & 1 & -1 & 1 & -1 & 1 & -1 & -1 & 1 & 1 & -1
   & -1 & 1 & 1 & -1 & 1 & -1 & 1 & -1 & 1 \\
 D_{17} & -1 & -1 & -1 & -1 & -1 & 1 & 1 & 1 & 1 & 1 & -1
   & -1 & -1 & -1 & 1 & 1 & 1 & 1 & -1 & -1 \\
 D_{18} & -1 & -1 & -1 & -1 & -1 & 1 & 1 & 1 & 1 & 1 & -1
   & -1 & -1 & -1 & -1 & -1 & -1 & -1 & 1 & 1 \\
 D_{19} & -1 & -1 & -1 & -1 & -1 & 1 & -1 & -1 & -1 & -1
   & 1 & 1 & 1 & 1 & 1 & 1 & 1 & 1 & -1 & -1 \\
 D_{20} & -1 & -1 & -1 & -1 & -1 & 1 & -1 & -1 & -1 & -1
   & 1 & 1 & 1 & 1 & -1 & -1 & -1 & -1 & 1 & 1 \\
 D_{21} & -1 & 1 & -1 & 1 & 1 & -1 & 1 & 1 & 1 & 1 & -1 &
   -1 & -1 & -1 & 1 & -1 & 1 & -1 & -1 & 1 \\
 D_{22} & -1 & 1 & -1 & 1 & 1 & -1 & 1 & 1 & 1 & 1 & -1 &
   -1 & -1 & -1 & -1 & 1 & -1 & 1 & 1 & -1 \\
 D_{23} & -1 & 1 & -1 & 1 & 1 & -1 & -1 & -1 & -1 & -1 &
   1 & 1 & 1 & 1 & 1 & -1 & 1 & -1 & -1 & 1 \\
 D_{24} & -1 & 1 & -1 & 1 & 1 & -1 & -1 & -1 & -1 & -1 &
   1 & 1 & 1 & 1 & -1 & 1 & -1 & 1 & 1 & -1 \\
 D_{25} & 1 & 1 & 1 & 1 & -1 & -1 & 1 & 1 & -1 & -1 & -1
   & -1 & 1 & 1 & 1 & 1 & 1 & 1 & -1 & -1 \\
 D_{26} & 1 & 1 & 1 & 1 & -1 & -1 & 1 & 1 & -1 & -1 & -1
   & -1 & 1 & 1 & -1 & -1 & -1 & -1 & 1 & 1 \\
 D_{27} & 1 & 1 & 1 & 1 & -1 & -1 & -1 & -1 & 1 & 1 & 1 &
   1 & -1 & -1 & 1 & 1 & 1 & 1 & -1 & -1 \\
 D_{28} & 1 & 1 & 1 & 1 & -1 & -1 & -1 & -1 & 1 & 1 & 1 &
   1 & -1 & -1 & -1 & -1 & -1 & -1 & 1 & 1 \\
 D_{29} & 1 & -1 & 1 & -1 & 1 & 1 & 1 & 1 & -1 & -1 & -1
   & -1 & 1 & 1 & 1 & -1 & 1 & -1 & -1 & 1 \\
 D_{30} & 1 & -1 & 1 & -1 & 1 & 1 & 1 & 1 & -1 & -1 & -1
   & -1 & 1 & 1 & -1 & 1 & -1 & 1 & 1 & -1 \\
 D_{31} & 1 & -1 & 1 & -1 & 1 & 1 & -1 & -1 & 1 & 1 & 1 &
   1 & -1 & -1 & 1 & -1 & 1 & -1 & -1 & 1 \\
 D_{32} & 1 & -1 & 1 & -1 & 1 & 1 & -1 & -1 & 1 & 1 & 1 &
   1 & -1 & -1 & -1 & 1 & -1 & 1 & 1 & -1 \\
 D_{33} & 2 & 0 & -2 & 0 & 0 & 0 & 2 & -2 & 2 & -2 & 2 &
   -2 & 2 & -2 & 0 & 0 & 0 & 0 & 0 & 0 \\
 D_{34} & 2 & 0 & -2 & 0 & 0 & 0 & -2 & 2 & -2 & 2 & -2 &
   2 & -2 & 2 & 0 & 0 & 0 & 0 & 0 & 0 \\
 D_{35} & -2 & 0 & 2 & 0 & 0 & 0 & 2 & -2 & -2 & 2 & 2 &
   -2 & -2 & 2 & 0 & 0 & 0 & 0 & 0 & 0 \\
 D_{36} & -2 & 0 & 2 & 0 & 0 & 0 & -2 & 2 & 2 & -2 & -2 &
   2 & 2 & -2 & 0 & 0 & 0 & 0 & 0 & 0 \\
 D_{37} & -2 & 0 & 2 & 0 & 0 & 0 & 2 & -2 & 2 & -2 & -2 &
   2 & -2 & 2 & 0 & 0 & 0 & 0 & 0 & 0 \\
 D_{38} & -2 & 0 & 2 & 0 & 0 & 0 & -2 & 2 & -2 & 2 & 2 &
   -2 & 2 & -2 & 0 & 0 & 0 & 0 & 0 & 0 \\
 D_{39} & 2 & 0 & -2 & 0 & 0 & 0 & 2 & -2 & -2 & 2 & -2 &
   2 & 2 & -2 & 0 & 0 & 0 & 0 & 0 & 0 \\
 D_{40} & 2 & 0 & -2 & 0 & 0 & 0 & -2 & 2 & 2 & -2 & 2 &
   -2 & -2 & 2 & 0 & 0 & 0 & 0 & 0 & 0 \\
 D_{41} & 0 & 0 & 0 & 0 & -2 & 0 & 0 & 0 & 0 & 0 & 0 & 0
   & 0 & 0 & -2 & -2 & 2 & 2 & -2 & -2 \\
 D_{42} & 0 & 0 & 0 & 0 & -2 & 0 & 0 & 0 & 0 & 0 & 0 & 0
   & 0 & 0 & 2 & 2 & -2 & -2 & 2 & 2 \\
 D_{43} & 0 & 0 & 0 & 0 & 2 & 0 & 0 & 0 & 0 & 0 & 0 & 0 &
   0 & 0 & -2 & 2 & 2 & -2 & -2 & 2 \\
 D_{44} & 0 & 0 & 0 & 0 & 2 & 0 & 0 & 0 & 0 & 0 & 0 & 0 &
   0 & 0 & 2 & -2 & -2 & 2 & 2 & -2 \\
 D_{45} & 0 & 0 & 0 & 0 & 0 & -2 & 0 & 0 & 0 & 0 & 0 & 0
   & 0 & 0 & 0 & 0 & 0 & 0 & 0 & 0 \\
 D_{46} & 0 & 0 & 0 & 0 & 0 & -2 & 0 & 0 & 0 & 0 & 0 & 0
   & 0 & 0 & 0 & 0 & 0 & 0 & 0 & 0 \\
 D_{47} & 0 & 0 & 0 & 0 & 0 & 2 & 0 & 0 & 0 & 0 & 0 & 0 &
   0 & 0 & 0 & 0 & 0 & 0 & 0 & 0 \\
 D_{48} & 0 & 0 & 0 & 0 & 0 & 2 & 0 & 0 & 0 & 0 & 0 & 0 &
   0 & 0 & 0 & 0 & 0 & 0 & 0 & 0 \\
 D_{49} & 0 & 0 & 0 & 0 & 0 & 2 & 0 & 0 & 0 & 0 & 0 & 0 &
   0 & 0 & 0 & 0 & 0 & 0 & 0 & 0 \\
 D_{50} & 0 & 0 & 0 & 0 & 0 & 2 & 0 & 0 & 0 & 0 & 0 & 0 &
   0 & 0 & 0 & 0 & 0 & 0 & 0 & 0 \\
 D_{51} & 0 & 0 & 0 & 0 & 0 & -2 & 0 & 0 & 0 & 0 & 0 & 0
   & 0 & 0 & 0 & 0 & 0 & 0 & 0 & 0 \\
 D_{52} & 0 & 0 & 0 & 0 & 0 & -2 & 0 & 0 & 0 & 0 & 0 & 0
   & 0 & 0 & 0 & 0 & 0 & 0 & 0 & 0 \\
 D_{53} & 0 & 0 & 0 & 0 & 2 & 0 & 0 & 0 & 0 & 0 & 0 & 0 &
   0 & 0 & -2 & -2 & 2 & 2 & 2 & 2 \\
 D_{54} & 0 & 0 & 0 & 0 & 2 & 0 & 0 & 0 & 0 & 0 & 0 & 0 &
   0 & 0 & 2 & 2 & -2 & -2 & -2 & -2 \\
 D_{55} & 0 & 0 & 0 & 0 & -2 & 0 & 0 & 0 & 0 & 0 & 0 & 0
   & 0 & 0 & -2 & 2 & 2 & -2 & 2 & -2 \\
 D_{56} & 0 & 0 & 0 & 0 & -2 & 0 & 0 & 0 & 0 & 0 & 0 & 0
   & 0 & 0 & 2 & -2 & -2 & 2 & -2 & 2 \\
 D_{57} & 0 & 0 & 0 & 0 & 0 & 0 & 0 & 0 & 0 & 0 & 0 & 0 &
   0 & 0 & 0 & 0 & 0 & 0 & 0 & 0 \\
 D_{58} & 0 & 0 & 0 & 0 & 0 & 0 & 0 & 0 & 0 & 0 & 0 & 0 &
   0 & 0 & 0 & 0 & 0 & 0 & 0 & 0 \\
 D_{59} & 0 & 0 & 0 & 0 & 0 & 0 & 0 & 0 & 0 & 0 & 0 & 0 &
   0 & 0 & 0 & 0 & 0 & 0 & 0 & 0 \\
 D_{60} & 0 & -4 i & 0 & 4 i & 0 & 0 & 0 & 0 & 0 & 0 & 0
   & 0 & 0 & 0 & 0 & 0 & 0 & 0 & 0 & 0 \\
 D_{61} & 0 & 0 & 0 & 0 & 0 & 0 & 0 & 0 & 0 & 0 & 0 & 0 &
   0 & 0 & 0 & 0 & 0 & 0 & 0 & 0 \\
 D_{62} & 0 & 4 i & 0 & -4 i & 0 & 0 & 0 & 0 & 0 & 0 & 0
   & 0 & 0 & 0 & 0 & 0 & 0 & 0 & 0 & 0 \\
 D_{63} & 0 & 0 & 0 & 0 & 0 & 0 & 0 & 0 & 0 & 0 & 0 & 0 &
   0 & 0 & 0 & 0 & 0 & 0 & 0 & 0 \\
 D_{64} & 0 & 0 & 0 & 0 & 0 & 0 & 0 & 0 & 0 & 0 & 0 & 0 &
   0 & 0 & 0 & 0 & 0 & 0 & 0 & 0\\
 \hline
\end{array}
\end{equation}
}
\subsection{Character Table of the Group $\mathrm{G_{128}}$}
The group $\mathrm{G_{128}}$ has $56$ conjugacy classes and therefore $56$ irreducible representations that are distributed according to the following pattern:
\begin{description}
  \item[a)] 32 irreps of dimension $1$, namely $D_1,\dots,D_{32}$
  \item[b)] 24 irreps of dimension $2$, namely $D_{33},\dots,D_{56}$
\end{description}
The corresponding character table that we have calculated with the procedures described in the main text is displayed below. For pure typographical reasons we were forced to split the character table in three parts in order to fit it into the page.
{\tiny
\begin{equation}\label{1charto128}
\begin{array}{|l|lllllllllllllllllll|}
\hline
 0 & C_1 & C_2 & C_3 & C_4 & C_5 & C_6 & C_7 & C_8 & C_9
   & C_{10} & C_{11} & C_{12} & C_{13} & C_{14} & C_{15}
   & C_{16} & C_{17} & C_{18} & C_{19} \\
\hline
 D_1 & 1 & 1 & 1 & 1 & 1 & 1 & 1 & 1 & 1 & 1 & 1 & 1 & 1
   & 1 & 1 & 1 & 1 & 1 & 1 \\
 D_2 & 1 & 1 & 1 & 1 & 1 & 1 & 1 & 1 & 1 & 1 & 1 & 1 & 1
   & 1 & 1 & 1 & 1 & 1 & 1 \\
 D_3 & 1 & i & -1 & -i & 1 & i & -1 & -i & 1 & i & -1 &
   -i & 1 & i & -1 & -i & 1 & i & -1 \\
 D_4 & 1 & i & -1 & -i & 1 & i & -1 & -i & 1 & i & -1 &
   -i & 1 & i & -1 & -i & 1 & i & -1 \\
 D_5 & 1 & -1 & 1 & -1 & 1 & -1 & 1 & -1 & 1 & -1 & 1 &
   -1 & 1 & -1 & 1 & -1 & 1 & -1 & 1 \\
 D_6 & 1 & -1 & 1 & -1 & 1 & -1 & 1 & -1 & 1 & -1 & 1 &
   -1 & 1 & -1 & 1 & -1 & 1 & -1 & 1 \\
 D_7 & 1 & -i & -1 & i & 1 & -i & -1 & i & 1 & -i & -1 &
   i & 1 & -i & -1 & i & 1 & -i & -1 \\
 D_8 & 1 & -i & -1 & i & 1 & -i & -1 & i & 1 & -i & -1 &
   i & 1 & -i & -1 & i & 1 & -i & -1 \\
 D_9 & 1 & 1 & 1 & 1 & 1 & 1 & 1 & 1 & 1 & 1 & 1 & 1 & 1
   & 1 & 1 & 1 & -1 & -1 & -1 \\
 D_{10} & 1 & 1 & 1 & 1 & 1 & 1 & 1 & 1 & 1 & 1 & 1 & 1 &
   1 & 1 & 1 & 1 & -1 & -1 & -1 \\
 D_{11} & 1 & i & -1 & -i & 1 & i & -1 & -i & 1 & i & -1
   & -i & 1 & i & -1 & -i & -1 & -i & 1 \\
 D_{12} & 1 & i & -1 & -i & 1 & i & -1 & -i & 1 & i & -1
   & -i & 1 & i & -1 & -i & -1 & -i & 1 \\
 D_{13} & 1 & -1 & 1 & -1 & 1 & -1 & 1 & -1 & 1 & -1 & 1
   & -1 & 1 & -1 & 1 & -1 & -1 & 1 & -1 \\
 D_{14} & 1 & -1 & 1 & -1 & 1 & -1 & 1 & -1 & 1 & -1 & 1
   & -1 & 1 & -1 & 1 & -1 & -1 & 1 & -1 \\
 D_{15} & 1 & -i & -1 & i & 1 & -i & -1 & i & 1 & -i & -1
   & i & 1 & -i & -1 & i & -1 & i & 1 \\
 D_{16} & 1 & -i & -1 & i & 1 & -i & -1 & i & 1 & -i & -1
   & i & 1 & -i & -1 & i & -1 & i & 1 \\
 D_{17} & 1 & 1 & 1 & 1 & 1 & 1 & 1 & 1 & 1 & 1 & 1 & 1 &
   1 & 1 & 1 & 1 & 1 & 1 & 1 \\
 D_{18} & 1 & 1 & 1 & 1 & 1 & 1 & 1 & 1 & 1 & 1 & 1 & 1 &
   1 & 1 & 1 & 1 & 1 & 1 & 1 \\
 D_{19} & 1 & i & -1 & -i & 1 & i & -1 & -i & 1 & i & -1
   & -i & 1 & i & -1 & -i & 1 & i & -1 \\
 D_{20} & 1 & i & -1 & -i & 1 & i & -1 & -i & 1 & i & -1
   & -i & 1 & i & -1 & -i & 1 & i & -1 \\
 D_{21} & 1 & -1 & 1 & -1 & 1 & -1 & 1 & -1 & 1 & -1 & 1
   & -1 & 1 & -1 & 1 & -1 & 1 & -1 & 1 \\
 D_{22} & 1 & -1 & 1 & -1 & 1 & -1 & 1 & -1 & 1 & -1 & 1
   & -1 & 1 & -1 & 1 & -1 & 1 & -1 & 1 \\
 D_{23} & 1 & -i & -1 & i & 1 & -i & -1 & i & 1 & -i & -1
   & i & 1 & -i & -1 & i & 1 & -i & -1 \\
 D_{24} & 1 & -i & -1 & i & 1 & -i & -1 & i & 1 & -i & -1
   & i & 1 & -i & -1 & i & 1 & -i & -1 \\
 D_{25} & 1 & 1 & 1 & 1 & 1 & 1 & 1 & 1 & 1 & 1 & 1 & 1 &
   1 & 1 & 1 & 1 & -1 & -1 & -1 \\
 D_{26} & 1 & 1 & 1 & 1 & 1 & 1 & 1 & 1 & 1 & 1 & 1 & 1 &
   1 & 1 & 1 & 1 & -1 & -1 & -1 \\
 D_{27} & 1 & i & -1 & -i & 1 & i & -1 & -i & 1 & i & -1
   & -i & 1 & i & -1 & -i & -1 & -i & 1 \\
 D_{28} & 1 & i & -1 & -i & 1 & i & -1 & -i & 1 & i & -1
   & -i & 1 & i & -1 & -i & -1 & -i & 1 \\
 D_{29} & 1 & -1 & 1 & -1 & 1 & -1 & 1 & -1 & 1 & -1 & 1
   & -1 & 1 & -1 & 1 & -1 & -1 & 1 & -1 \\
 D_{30} & 1 & -1 & 1 & -1 & 1 & -1 & 1 & -1 & 1 & -1 & 1
   & -1 & 1 & -1 & 1 & -1 & -1 & 1 & -1 \\
 D_{31} & 1 & -i & -1 & i & 1 & -i & -1 & i & 1 & -i & -1
   & i & 1 & -i & -1 & i & -1 & i & 1 \\
 D_{32} & 1 & -i & -1 & i & 1 & -i & -1 & i & 1 & -i & -1
   & i & 1 & -i & -1 & i & -1 & i & 1 \\
 D_{33} & 2 & 2 & 2 & 2 & -2 & -2 & -2 & -2 & 2 & 2 & 2 &
   2 & -2 & -2 & -2 & -2 & 0 & 0 & 0 \\
 D_{34} & 2 & 2 i & -2 & -2 i & -2 & -2 i & 2 & 2 i & 2 &
   2 i & -2 & -2 i & -2 & -2 i & 2 & 2 i & 0 & 0 & 0 \\
 D_{35} & 2 & -2 & 2 & -2 & -2 & 2 & -2 & 2 & 2 & -2 & 2
   & -2 & -2 & 2 & -2 & 2 & 0 & 0 & 0 \\
 D_{36} & 2 & -2 i & -2 & 2 i & -2 & 2 i & 2 & -2 i & 2 &
   -2 i & -2 & 2 i & -2 & 2 i & 2 & -2 i & 0 & 0 & 0 \\
 D_{37} & 2 & 2 & 2 & 2 & 2 & 2 & 2 & 2 & -2 & -2 & -2 &
   -2 & -2 & -2 & -2 & -2 & 2 & 2 & 2 \\
 D_{38} & 2 & 2 i & -2 & -2 i & 2 & 2 i & -2 & -2 i & -2
   & -2 i & 2 & 2 i & -2 & -2 i & 2 & 2 i & 2 & 2 i & -2
   \\
 D_{39} & 2 & -2 & 2 & -2 & 2 & -2 & 2 & -2 & -2 & 2 & -2
   & 2 & -2 & 2 & -2 & 2 & 2 & -2 & 2 \\
 D_{40} & 2 & -2 i & -2 & 2 i & 2 & -2 i & -2 & 2 i & -2
   & 2 i & 2 & -2 i & -2 & 2 i & 2 & -2 i & 2 & -2 i & -2
   \\
 D_{41} & 2 & 2 & 2 & 2 & -2 & -2 & -2 & -2 & -2 & -2 &
   -2 & -2 & 2 & 2 & 2 & 2 & 0 & 0 & 0 \\
 D_{42} & 2 & 2 i & -2 & -2 i & -2 & -2 i & 2 & 2 i & -2
   & -2 i & 2 & 2 i & 2 & 2 i & -2 & -2 i & 0 & 0 & 0 \\
 D_{43} & 2 & -2 & 2 & -2 & -2 & 2 & -2 & 2 & -2 & 2 & -2
   & 2 & 2 & -2 & 2 & -2 & 0 & 0 & 0 \\
 D_{44} & 2 & -2 i & -2 & 2 i & -2 & 2 i & 2 & -2 i & -2
   & 2 i & 2 & -2 i & 2 & -2 i & -2 & 2 i & 0 & 0 & 0 \\
 D_{45} & 2 & 2 & 2 & 2 & 2 & 2 & 2 & 2 & -2 & -2 & -2 &
   -2 & -2 & -2 & -2 & -2 & -2 & -2 & -2 \\
 D_{46} & 2 & 2 i & -2 & -2 i & 2 & 2 i & -2 & -2 i & -2
   & -2 i & 2 & 2 i & -2 & -2 i & 2 & 2 i & -2 & -2 i & 2
   \\
 D_{47} & 2 & -2 & 2 & -2 & 2 & -2 & 2 & -2 & -2 & 2 & -2
   & 2 & -2 & 2 & -2 & 2 & -2 & 2 & -2 \\
 D_{48} & 2 & -2 i & -2 & 2 i & 2 & -2 i & -2 & 2 i & -2
   & 2 i & 2 & -2 i & -2 & 2 i & 2 & -2 i & -2 & 2 i & 2
   \\
 D_{49} & 2 & 2 & 2 & 2 & -2 & -2 & -2 & -2 & -2 & -2 &
   -2 & -2 & 2 & 2 & 2 & 2 & 0 & 0 & 0 \\
 D_{50} & 2 & 2 i & -2 & -2 i & -2 & -2 i & 2 & 2 i & -2
   & -2 i & 2 & 2 i & 2 & 2 i & -2 & -2 i & 0 & 0 & 0 \\
 D_{51} & 2 & -2 & 2 & -2 & -2 & 2 & -2 & 2 & -2 & 2 & -2
   & 2 & 2 & -2 & 2 & -2 & 0 & 0 & 0 \\
 D_{52} & 2 & -2 i & -2 & 2 i & -2 & 2 i & 2 & -2 i & -2
   & 2 i & 2 & -2 i & 2 & -2 i & -2 & 2 i & 0 & 0 & 0 \\
 D_{53} & 2 & 2 & 2 & 2 & -2 & -2 & -2 & -2 & 2 & 2 & 2 &
   2 & -2 & -2 & -2 & -2 & 0 & 0 & 0 \\
 D_{54} & 2 & 2 i & -2 & -2 i & -2 & -2 i & 2 & 2 i & 2 &
   2 i & -2 & -2 i & -2 & -2 i & 2 & 2 i & 0 & 0 & 0 \\
 D_{55} & 2 & -2 & 2 & -2 & -2 & 2 & -2 & 2 & 2 & -2 & 2
   & -2 & -2 & 2 & -2 & 2 & 0 & 0 & 0 \\
 D_{56} & 2 & -2 i & -2 & 2 i & -2 & 2 i & 2 & -2 i & 2 &
   -2 i & -2 & 2 i & -2 & 2 i & 2 & -2 i & 0 & 0 & 0\\
\hline
\end{array}
\end{equation}
}
{\tiny
\begin{equation}\label{2charto128}
\begin{array}{|l|lllllllllllllllllll|}
\hline
 0 & C_{20} & C_{21} & C_{22} & C_{23} & C_{24} & C_{25}
   & C_{26} & C_{27} & C_{28} & C_{29} & C_{30} & C_{31}
   & C_{32} & C_{33} & C_{34} & C_{35} & C_{36} & C_{37}
   & C_{38} \\
 \hline
D_1 & 1 & 1 & 1 & 1 & 1 & 1 & 1 & 1 & 1 & 1 & 1 & 1 & 1
   & 1 & 1 & 1 & 1 & 1 & 1 \\
 D_2 & 1 & 1 & 1 & 1 & 1 & 1 & 1 & 1 & 1 & 1 & 1 & 1 & 1
   & 1 & 1 & 1 & 1 & 1 & 1 \\
 D_3 & -i & 1 & i & -1 & -i & 1 & i & -1 & -i & 1 & i &
   -1 & -i & 1 & i & -1 & -i & 1 & i \\
 D_4 & -i & 1 & i & -1 & -i & 1 & i & -1 & -i & 1 & i &
   -1 & -i & 1 & i & -1 & -i & 1 & i \\
 D_5 & -1 & 1 & -1 & 1 & -1 & 1 & -1 & 1 & -1 & 1 & -1 &
   1 & -1 & 1 & -1 & 1 & -1 & 1 & -1 \\
 D_6 & -1 & 1 & -1 & 1 & -1 & 1 & -1 & 1 & -1 & 1 & -1 &
   1 & -1 & 1 & -1 & 1 & -1 & 1 & -1 \\
 D_7 & i & 1 & -i & -1 & i & 1 & -i & -1 & i & 1 & -i &
   -1 & i & 1 & -i & -1 & i & 1 & -i \\
 D_8 & i & 1 & -i & -1 & i & 1 & -i & -1 & i & 1 & -i &
   -1 & i & 1 & -i & -1 & i & 1 & -i \\
 D_9 & -1 & 1 & 1 & 1 & 1 & -1 & -1 & -1 & -1 & 1 & 1 & 1
   & 1 & -1 & -1 & -1 & -1 & -1 & -1 \\
 D_{10} & -1 & 1 & 1 & 1 & 1 & -1 & -1 & -1 & -1 & 1 & 1
   & 1 & 1 & -1 & -1 & -1 & -1 & -1 & -1 \\
 D_{11} & i & 1 & i & -1 & -i & -1 & -i & 1 & i & 1 & i &
   -1 & -i & -1 & -i & 1 & i & -1 & -i \\
 D_{12} & i & 1 & i & -1 & -i & -1 & -i & 1 & i & 1 & i &
   -1 & -i & -1 & -i & 1 & i & -1 & -i \\
 D_{13} & 1 & 1 & -1 & 1 & -1 & -1 & 1 & -1 & 1 & 1 & -1
   & 1 & -1 & -1 & 1 & -1 & 1 & -1 & 1 \\
 D_{14} & 1 & 1 & -1 & 1 & -1 & -1 & 1 & -1 & 1 & 1 & -1
   & 1 & -1 & -1 & 1 & -1 & 1 & -1 & 1 \\
 D_{15} & -i & 1 & -i & -1 & i & -1 & i & 1 & -i & 1 & -i
   & -1 & i & -1 & i & 1 & -i & -1 & i \\
 D_{16} & -i & 1 & -i & -1 & i & -1 & i & 1 & -i & 1 & -i
   & -1 & i & -1 & i & 1 & -i & -1 & i \\
 D_{17} & 1 & -1 & -1 & -1 & -1 & -1 & -1 & -1 & -1 & -1
   & -1 & -1 & -1 & -1 & -1 & -1 & -1 & 1 & 1 \\
 D_{18} & 1 & -1 & -1 & -1 & -1 & -1 & -1 & -1 & -1 & -1
   & -1 & -1 & -1 & -1 & -1 & -1 & -1 & 1 & 1 \\
 D_{19} & -i & -1 & -i & 1 & i & -1 & -i & 1 & i & -1 &
   -i & 1 & i & -1 & -i & 1 & i & 1 & i \\
 D_{20} & -i & -1 & -i & 1 & i & -1 & -i & 1 & i & -1 &
   -i & 1 & i & -1 & -i & 1 & i & 1 & i \\
 D_{21} & -1 & -1 & 1 & -1 & 1 & -1 & 1 & -1 & 1 & -1 & 1
   & -1 & 1 & -1 & 1 & -1 & 1 & 1 & -1 \\
 D_{22} & -1 & -1 & 1 & -1 & 1 & -1 & 1 & -1 & 1 & -1 & 1
   & -1 & 1 & -1 & 1 & -1 & 1 & 1 & -1 \\
 D_{23} & i & -1 & i & 1 & -i & -1 & i & 1 & -i & -1 & i
   & 1 & -i & -1 & i & 1 & -i & 1 & -i \\
 D_{24} & i & -1 & i & 1 & -i & -1 & i & 1 & -i & -1 & i
   & 1 & -i & -1 & i & 1 & -i & 1 & -i \\
 D_{25} & -1 & -1 & -1 & -1 & -1 & 1 & 1 & 1 & 1 & -1 &
   -1 & -1 & -1 & 1 & 1 & 1 & 1 & -1 & -1 \\
 D_{26} & -1 & -1 & -1 & -1 & -1 & 1 & 1 & 1 & 1 & -1 &
   -1 & -1 & -1 & 1 & 1 & 1 & 1 & -1 & -1 \\
 D_{27} & i & -1 & -i & 1 & i & 1 & i & -1 & -i & -1 & -i
   & 1 & i & 1 & i & -1 & -i & -1 & -i \\
 D_{28} & i & -1 & -i & 1 & i & 1 & i & -1 & -i & -1 & -i
   & 1 & i & 1 & i & -1 & -i & -1 & -i \\
 D_{29} & 1 & -1 & 1 & -1 & 1 & 1 & -1 & 1 & -1 & -1 & 1
   & -1 & 1 & 1 & -1 & 1 & -1 & -1 & 1 \\
 D_{30} & 1 & -1 & 1 & -1 & 1 & 1 & -1 & 1 & -1 & -1 & 1
   & -1 & 1 & 1 & -1 & 1 & -1 & -1 & 1 \\
 D_{31} & -i & -1 & i & 1 & -i & 1 & -i & -1 & i & -1 & i
   & 1 & -i & 1 & -i & -1 & i & -1 & i \\
 D_{32} & -i & -1 & i & 1 & -i & 1 & -i & -1 & i & -1 & i
   & 1 & -i & 1 & -i & -1 & i & -1 & i \\
 D_{33} & 0 & 2 & 2 & 2 & 2 & 0 & 0 & 0 & 0 & -2 & -2 &
   -2 & -2 & 0 & 0 & 0 & 0 & 0 & 0 \\
 D_{34} & 0 & 2 & 2 i & -2 & -2 i & 0 & 0 & 0 & 0 & -2 &
   -2 i & 2 & 2 i & 0 & 0 & 0 & 0 & 0 & 0 \\
 D_{35} & 0 & 2 & -2 & 2 & -2 & 0 & 0 & 0 & 0 & -2 & 2 &
   -2 & 2 & 0 & 0 & 0 & 0 & 0 & 0 \\
 D_{36} & 0 & 2 & -2 i & -2 & 2 i & 0 & 0 & 0 & 0 & -2 &
   2 i & 2 & -2 i & 0 & 0 & 0 & 0 & 0 & 0 \\
 D_{37} & 2 & 0 & 0 & 0 & 0 & 0 & 0 & 0 & 0 & 0 & 0 & 0 &
   0 & 0 & 0 & 0 & 0 & -2 & -2 \\
 D_{38} & -2 i & 0 & 0 & 0 & 0 & 0 & 0 & 0 & 0 & 0 & 0 &
   0 & 0 & 0 & 0 & 0 & 0 & -2 & -2 i \\
 D_{39} & -2 & 0 & 0 & 0 & 0 & 0 & 0 & 0 & 0 & 0 & 0 & 0
   & 0 & 0 & 0 & 0 & 0 & -2 & 2 \\
 D_{40} & 2 i & 0 & 0 & 0 & 0 & 0 & 0 & 0 & 0 & 0 & 0 & 0
   & 0 & 0 & 0 & 0 & 0 & -2 & 2 i \\
 D_{41} & 0 & 0 & 0 & 0 & 0 & -2 & -2 & -2 & -2 & 0 & 0 &
   0 & 0 & 2 & 2 & 2 & 2 & 0 & 0 \\
 D_{42} & 0 & 0 & 0 & 0 & 0 & -2 & -2 i & 2 & 2 i & 0 & 0
   & 0 & 0 & 2 & 2 i & -2 & -2 i & 0 & 0 \\
 D_{43} & 0 & 0 & 0 & 0 & 0 & -2 & 2 & -2 & 2 & 0 & 0 & 0
   & 0 & 2 & -2 & 2 & -2 & 0 & 0 \\
 D_{44} & 0 & 0 & 0 & 0 & 0 & -2 & 2 i & 2 & -2 i & 0 & 0
   & 0 & 0 & 2 & -2 i & -2 & 2 i & 0 & 0 \\
 D_{45} & -2 & 0 & 0 & 0 & 0 & 0 & 0 & 0 & 0 & 0 & 0 & 0
   & 0 & 0 & 0 & 0 & 0 & 2 & 2 \\
 D_{46} & 2 i & 0 & 0 & 0 & 0 & 0 & 0 & 0 & 0 & 0 & 0 & 0
   & 0 & 0 & 0 & 0 & 0 & 2 & 2 i \\
 D_{47} & 2 & 0 & 0 & 0 & 0 & 0 & 0 & 0 & 0 & 0 & 0 & 0 &
   0 & 0 & 0 & 0 & 0 & 2 & -2 \\
 D_{48} & -2 i & 0 & 0 & 0 & 0 & 0 & 0 & 0 & 0 & 0 & 0 &
   0 & 0 & 0 & 0 & 0 & 0 & 2 & -2 i \\
 D_{49} & 0 & 0 & 0 & 0 & 0 & 2 & 2 & 2 & 2 & 0 & 0 & 0 &
   0 & -2 & -2 & -2 & -2 & 0 & 0 \\
 D_{50} & 0 & 0 & 0 & 0 & 0 & 2 & 2 i & -2 & -2 i & 0 & 0
   & 0 & 0 & -2 & -2 i & 2 & 2 i & 0 & 0 \\
 D_{51} & 0 & 0 & 0 & 0 & 0 & 2 & -2 & 2 & -2 & 0 & 0 & 0
   & 0 & -2 & 2 & -2 & 2 & 0 & 0 \\
 D_{52} & 0 & 0 & 0 & 0 & 0 & 2 & -2 i & -2 & 2 i & 0 & 0
   & 0 & 0 & -2 & 2 i & 2 & -2 i & 0 & 0 \\
 D_{53} & 0 & -2 & -2 & -2 & -2 & 0 & 0 & 0 & 0 & 2 & 2 &
   2 & 2 & 0 & 0 & 0 & 0 & 0 & 0 \\
 D_{54} & 0 & -2 & -2 i & 2 & 2 i & 0 & 0 & 0 & 0 & 2 & 2
   i & -2 & -2 i & 0 & 0 & 0 & 0 & 0 & 0 \\
 D_{55} & 0 & -2 & 2 & -2 & 2 & 0 & 0 & 0 & 0 & 2 & -2 &
   2 & -2 & 0 & 0 & 0 & 0 & 0 & 0 \\
 D_{56} & 0 & -2 & 2 i & 2 & -2 i & 0 & 0 & 0 & 0 & 2 &
   -2 i & -2 & 2 i & 0 & 0 & 0 & 0 & 0 & 0\\
\hline
\end{array}
\end{equation}
}
{\tiny
\begin{equation}\label{3charto128}
\begin{array}{|l|lllllllllllllllllll|}
\hline
 0 & C_{39} & C_{40} & C_{41} & C_{42} & C_{43} & C_{44}
   & C_{45} & C_{46} & C_{47} & C_{48} & C_{49} & C_{50}
   & C_{51} & C_{52} & C_{53} & C_{54} & C_{55} & C_{56}
   & 0 \\
\hline
 D_1 & 1 & 1 & 1 & 1 & 1 & 1 & 1 & 1 & 1 & 1 & 1 & 1 & 1
   & 1 & 1 & 1 & 1 & 1 & 0 \\
 D_2 & 1 & 1 & -1 & -1 & -1 & -1 & -1 & -1 & -1 & -1 & -1
   & -1 & -1 & -1 & -1 & -1 & -1 & -1 & 0 \\
 D_3 & -1 & -i & 1 & i & -1 & -i & 1 & i & -1 & -i & 1 &
   i & -1 & -i & 1 & i & -1 & -i & 0 \\
 D_4 & -1 & -i & -1 & -i & 1 & i & -1 & -i & 1 & i & -1 &
   -i & 1 & i & -1 & -i & 1 & i & 0 \\
 D_5 & 1 & -1 & 1 & -1 & 1 & -1 & 1 & -1 & 1 & -1 & 1 &
   -1 & 1 & -1 & 1 & -1 & 1 & -1 & 0 \\
 D_6 & 1 & -1 & -1 & 1 & -1 & 1 & -1 & 1 & -1 & 1 & -1 &
   1 & -1 & 1 & -1 & 1 & -1 & 1 & 0 \\
 D_7 & -1 & i & 1 & -i & -1 & i & 1 & -i & -1 & i & 1 &
   -i & -1 & i & 1 & -i & -1 & i & 0 \\
 D_8 & -1 & i & -1 & i & 1 & -i & -1 & i & 1 & -i & -1 &
   i & 1 & -i & -1 & i & 1 & -i & 0 \\
 D_9 & -1 & -1 & 1 & 1 & 1 & 1 & -1 & -1 & -1 & -1 & 1 &
   1 & 1 & 1 & -1 & -1 & -1 & -1 & 0 \\
 D_{10} & -1 & -1 & -1 & -1 & -1 & -1 & 1 & 1 & 1 & 1 &
   -1 & -1 & -1 & -1 & 1 & 1 & 1 & 1 & 0 \\
 D_{11} & 1 & i & 1 & i & -1 & -i & -1 & -i & 1 & i & 1 &
   i & -1 & -i & -1 & -i & 1 & i & 0 \\
 D_{12} & 1 & i & -1 & -i & 1 & i & 1 & i & -1 & -i & -1
   & -i & 1 & i & 1 & i & -1 & -i & 0 \\
 D_{13} & -1 & 1 & 1 & -1 & 1 & -1 & -1 & 1 & -1 & 1 & 1
   & -1 & 1 & -1 & -1 & 1 & -1 & 1 & 0 \\
 D_{14} & -1 & 1 & -1 & 1 & -1 & 1 & 1 & -1 & 1 & -1 & -1
   & 1 & -1 & 1 & 1 & -1 & 1 & -1 & 0 \\
 D_{15} & 1 & -i & 1 & -i & -1 & i & -1 & i & 1 & -i & 1
   & -i & -1 & i & -1 & i & 1 & -i & 0 \\
 D_{16} & 1 & -i & -1 & i & 1 & -i & 1 & -i & -1 & i & -1
   & i & 1 & -i & 1 & -i & -1 & i & 0 \\
 D_{17} & 1 & 1 & 1 & 1 & 1 & 1 & 1 & 1 & 1 & 1 & -1 & -1
   & -1 & -1 & -1 & -1 & -1 & -1 & 0 \\
 D_{18} & 1 & 1 & -1 & -1 & -1 & -1 & -1 & -1 & -1 & -1 &
   1 & 1 & 1 & 1 & 1 & 1 & 1 & 1 & 0 \\
 D_{19} & -1 & -i & 1 & i & -1 & -i & 1 & i & -1 & -i &
   -1 & -i & 1 & i & -1 & -i & 1 & i & 0 \\
 D_{20} & -1 & -i & -1 & -i & 1 & i & -1 & -i & 1 & i & 1
   & i & -1 & -i & 1 & i & -1 & -i & 0 \\
 D_{21} & 1 & -1 & 1 & -1 & 1 & -1 & 1 & -1 & 1 & -1 & -1
   & 1 & -1 & 1 & -1 & 1 & -1 & 1 & 0 \\
 D_{22} & 1 & -1 & -1 & 1 & -1 & 1 & -1 & 1 & -1 & 1 & 1
   & -1 & 1 & -1 & 1 & -1 & 1 & -1 & 0 \\
 D_{23} & -1 & i & 1 & -i & -1 & i & 1 & -i & -1 & i & -1
   & i & 1 & -i & -1 & i & 1 & -i & 0 \\
 D_{24} & -1 & i & -1 & i & 1 & -i & -1 & i & 1 & -i & 1
   & -i & -1 & i & 1 & -i & -1 & i & 0 \\
 D_{25} & -1 & -1 & 1 & 1 & 1 & 1 & -1 & -1 & -1 & -1 &
   -1 & -1 & -1 & -1 & 1 & 1 & 1 & 1 & 0 \\
 D_{26} & -1 & -1 & -1 & -1 & -1 & -1 & 1 & 1 & 1 & 1 & 1
   & 1 & 1 & 1 & -1 & -1 & -1 & -1 & 0 \\
 D_{27} & 1 & i & 1 & i & -1 & -i & -1 & -i & 1 & i & -1
   & -i & 1 & i & 1 & i & -1 & -i & 0 \\
 D_{28} & 1 & i & -1 & -i & 1 & i & 1 & i & -1 & -i & 1 &
   i & -1 & -i & -1 & -i & 1 & i & 0 \\
 D_{29} & -1 & 1 & 1 & -1 & 1 & -1 & -1 & 1 & -1 & 1 & -1
   & 1 & -1 & 1 & 1 & -1 & 1 & -1 & 0 \\
 D_{30} & -1 & 1 & -1 & 1 & -1 & 1 & 1 & -1 & 1 & -1 & 1
   & -1 & 1 & -1 & -1 & 1 & -1 & 1 & 0 \\
 D_{31} & 1 & -i & 1 & -i & -1 & i & -1 & i & 1 & -i & -1
   & i & 1 & -i & 1 & -i & -1 & i & 0 \\
 D_{32} & 1 & -i & -1 & i & 1 & -i & 1 & -i & -1 & i & 1
   & -i & -1 & i & -1 & i & 1 & -i & 0 \\
 D_{33} & 0 & 0 & 0 & 0 & 0 & 0 & 0 & 0 & 0 & 0 & 0 & 0 &
   0 & 0 & 0 & 0 & 0 & 0 & 0 \\
 D_{34} & 0 & 0 & 0 & 0 & 0 & 0 & 0 & 0 & 0 & 0 & 0 & 0 &
   0 & 0 & 0 & 0 & 0 & 0 & 0 \\
 D_{35} & 0 & 0 & 0 & 0 & 0 & 0 & 0 & 0 & 0 & 0 & 0 & 0 &
   0 & 0 & 0 & 0 & 0 & 0 & 0 \\
 D_{36} & 0 & 0 & 0 & 0 & 0 & 0 & 0 & 0 & 0 & 0 & 0 & 0 &
   0 & 0 & 0 & 0 & 0 & 0 & 0 \\
 D_{37} & -2 & -2 & 0 & 0 & 0 & 0 & 0 & 0 & 0 & 0 & 0 & 0
   & 0 & 0 & 0 & 0 & 0 & 0 & 0 \\
 D_{38} & 2 & 2 i & 0 & 0 & 0 & 0 & 0 & 0 & 0 & 0 & 0 & 0
   & 0 & 0 & 0 & 0 & 0 & 0 & 0 \\
 D_{39} & -2 & 2 & 0 & 0 & 0 & 0 & 0 & 0 & 0 & 0 & 0 & 0
   & 0 & 0 & 0 & 0 & 0 & 0 & 0 \\
 D_{40} & 2 & -2 i & 0 & 0 & 0 & 0 & 0 & 0 & 0 & 0 & 0 &
   0 & 0 & 0 & 0 & 0 & 0 & 0 & 0 \\
 D_{41} & 0 & 0 & 0 & 0 & 0 & 0 & 0 & 0 & 0 & 0 & 0 & 0 &
   0 & 0 & 0 & 0 & 0 & 0 & 0 \\
 D_{42} & 0 & 0 & 0 & 0 & 0 & 0 & 0 & 0 & 0 & 0 & 0 & 0 &
   0 & 0 & 0 & 0 & 0 & 0 & 0 \\
 D_{43} & 0 & 0 & 0 & 0 & 0 & 0 & 0 & 0 & 0 & 0 & 0 & 0 &
   0 & 0 & 0 & 0 & 0 & 0 & 0 \\
 D_{44} & 0 & 0 & 0 & 0 & 0 & 0 & 0 & 0 & 0 & 0 & 0 & 0 &
   0 & 0 & 0 & 0 & 0 & 0 & 0 \\
 D_{45} & 2 & 2 & 0 & 0 & 0 & 0 & 0 & 0 & 0 & 0 & 0 & 0 &
   0 & 0 & 0 & 0 & 0 & 0 & 0 \\
 D_{46} & -2 & -2 i & 0 & 0 & 0 & 0 & 0 & 0 & 0 & 0 & 0 &
   0 & 0 & 0 & 0 & 0 & 0 & 0 & 0 \\
 D_{47} & 2 & -2 & 0 & 0 & 0 & 0 & 0 & 0 & 0 & 0 & 0 & 0
   & 0 & 0 & 0 & 0 & 0 & 0 & 0 \\
 D_{48} & -2 & 2 i & 0 & 0 & 0 & 0 & 0 & 0 & 0 & 0 & 0 &
   0 & 0 & 0 & 0 & 0 & 0 & 0 & 0 \\
 D_{49} & 0 & 0 & 0 & 0 & 0 & 0 & 0 & 0 & 0 & 0 & 0 & 0 &
   0 & 0 & 0 & 0 & 0 & 0 & 0 \\
 D_{50} & 0 & 0 & 0 & 0 & 0 & 0 & 0 & 0 & 0 & 0 & 0 & 0 &
   0 & 0 & 0 & 0 & 0 & 0 & 0 \\
 D_{51} & 0 & 0 & 0 & 0 & 0 & 0 & 0 & 0 & 0 & 0 & 0 & 0 &
   0 & 0 & 0 & 0 & 0 & 0 & 0 \\
 D_{52} & 0 & 0 & 0 & 0 & 0 & 0 & 0 & 0 & 0 & 0 & 0 & 0 &
   0 & 0 & 0 & 0 & 0 & 0 & 0 \\
 D_{53} & 0 & 0 & 0 & 0 & 0 & 0 & 0 & 0 & 0 & 0 & 0 & 0 &
   0 & 0 & 0 & 0 & 0 & 0 & 0 \\
 D_{54} & 0 & 0 & 0 & 0 & 0 & 0 & 0 & 0 & 0 & 0 & 0 & 0 &
   0 & 0 & 0 & 0 & 0 & 0 & 0 \\
 D_{55} & 0 & 0 & 0 & 0 & 0 & 0 & 0 & 0 & 0 & 0 & 0 & 0 &
   0 & 0 & 0 & 0 & 0 & 0 & 0 \\
 D_{56} & 0 & 0 & 0 & 0 & 0 & 0 & 0 & 0 & 0 & 0 & 0 & 0 &
   0 & 0 & 0 & 0 & 0 & 0 & 0\\
\hline
\end{array}
\end{equation}
}
\subsection{Character Table of the Group $\mathrm{G_{64}}$}
The group $\mathrm{G_{64}}$ is abelian and has order $64$. Hence it has exactly  $64$ conjugacy classes and  $64$ irreducible $1$-dimensional representations.
The corresponding character table that we have calculated with the procedures described in the main text is displayed below. For pure typographical reasons we were forced to split the character table in three parts in order to fit it into the page. In the formulae below
$\omega \, = \, \exp\left[\frac{2\, \pi}{3} \, {\rm i} \right]$ is a cubic root of the unity.
{\tiny
\begin{equation}\label{1charto64}
\begin{array}{|l|llllllllllllllllllllll|}
\hline
 0 & C_1 & C_2 & C_3 & C_4 & C_5 & C_6 & C_7 & C_8 & C_9
   & C_{10} & C_{11} & C_{12} & C_{13} & C_{14} & C_{15}
   & C_{16} & C_{17} & C_{18} & C_{19} & C_{20} & C_{21}
   & C_{22} \\
   \hline
 D_1 & 1 & 1 & 1 & 1 & 1 & 1 & 1 & 1 & 1 & 1 & 1 & 1 & 1
   & 1 & 1 & 1 & 1 & 1 & 1 & 0 & 0 & 0 \\
 D_2 & 1 & \omega  & \omega ^2 & \omega ^3 & 1 & \omega
   & \omega ^2 & \omega ^3 & 1 & \omega  & \omega ^2 &
   \omega ^3 & 1 & \omega  & \omega ^2 & \omega ^3 & 1 &
   \omega  & \omega ^2 & 0 & 0 & 0 \\
 D_3 & 1 & \omega ^2 & 1 & \omega ^2 & 1 & \omega ^2 & 1
   & \omega ^2 & 1 & \omega ^2 & 1 & \omega ^2 & 1 &
   \omega ^2 & 1 & \omega ^2 & 1 & \omega ^2 & 1 & 0 & 0
   & 0 \\
 D_4 & 1 & \omega ^3 & \omega ^2 & \omega  & 1 & \omega
   ^3 & \omega ^2 & \omega  & 1 & \omega ^3 & \omega ^2 &
   \omega  & 1 & \omega ^3 & \omega ^2 & \omega  & 1 &
   \omega ^3 & \omega ^2 & 0 & 0 & 0 \\
 D_5 & 1 & 1 & 1 & 1 & \omega  & \omega  & \omega  &
   \omega  & \omega ^2 & \omega ^2 & \omega ^2 & \omega
   ^2 & \omega ^3 & \omega ^3 & \omega ^3 & \omega ^3 & 1
   & 1 & 1 & 0 & 0 & 0 \\
 D_6 & 1 & \omega  & \omega ^2 & \omega ^3 & \omega  &
   \omega ^2 & \omega ^3 & 1 & \omega ^2 & \omega ^3 & 1
   & \omega  & \omega ^3 & 1 & \omega  & \omega ^2 & 1 &
   \omega  & \omega ^2 & 0 & 0 & 0 \\
 D_7 & 1 & \omega ^2 & 1 & \omega ^2 & \omega  & \omega
   ^3 & \omega  & \omega ^3 & \omega ^2 & 1 & \omega ^2 &
   1 & \omega ^3 & \omega  & \omega ^3 & \omega  & 1 &
   \omega ^2 & 1 & 0 & 0 & 0 \\
 D_8 & 1 & \omega ^3 & \omega ^2 & \omega  & \omega  & 1
   & \omega ^3 & \omega ^2 & \omega ^2 & \omega  & 1 &
   \omega ^3 & \omega ^3 & \omega ^2 & \omega  & 1 & 1 &
   \omega ^3 & \omega ^2 & 0 & 0 & 0 \\
 D_9 & 1 & 1 & 1 & 1 & \omega ^2 & \omega ^2 & \omega ^2
   & \omega ^2 & 1 & 1 & 1 & 1 & \omega ^2 & \omega ^2 &
   \omega ^2 & \omega ^2 & 1 & 1 & 1 & 0 & 0 & 0 \\
 D_{10} & 1 & \omega  & \omega ^2 & \omega ^3 & \omega ^2
   & \omega ^3 & 1 & \omega  & 1 & \omega  & \omega ^2 &
   \omega ^3 & \omega ^2 & \omega ^3 & 1 & \omega  & 1 &
   \omega  & \omega ^2 & 0 & 0 & 0 \\
 D_{11} & 1 & \omega ^2 & 1 & \omega ^2 & \omega ^2 & 1 &
   \omega ^2 & 1 & 1 & \omega ^2 & 1 & \omega ^2 & \omega
   ^2 & 1 & \omega ^2 & 1 & 1 & \omega ^2 & 1 & 0 & 0 & 0
   \\
 D_{12} & 1 & \omega ^3 & \omega ^2 & \omega  & \omega ^2
   & \omega  & 1 & \omega ^3 & 1 & \omega ^3 & \omega ^2
   & \omega  & \omega ^2 & \omega  & 1 & \omega ^3 & 1 &
   \omega ^3 & \omega ^2 & 0 & 0 & 0 \\
 D_{13} & 1 & 1 & 1 & 1 & \omega ^3 & \omega ^3 & \omega
   ^3 & \omega ^3 & \omega ^2 & \omega ^2 & \omega ^2 &
   \omega ^2 & \omega  & \omega  & \omega  & \omega  & 1
   & 1 & 1 & 0 & 0 & 0 \\
 D_{14} & 1 & \omega  & \omega ^2 & \omega ^3 & \omega ^3
   & 1 & \omega  & \omega ^2 & \omega ^2 & \omega ^3 & 1
   & \omega  & \omega  & \omega ^2 & \omega ^3 & 1 & 1 &
   \omega  & \omega ^2 & 0 & 0 & 0 \\
 D_{15} & 1 & \omega ^2 & 1 & \omega ^2 & \omega ^3 &
   \omega  & \omega ^3 & \omega  & \omega ^2 & 1 & \omega
   ^2 & 1 & \omega  & \omega ^3 & \omega  & \omega ^3 & 1
   & \omega ^2 & 1 & 0 & 0 & 0 \\
 D_{16} & 1 & \omega ^3 & \omega ^2 & \omega  & \omega ^3
   & \omega ^2 & \omega  & 1 & \omega ^2 & \omega  & 1 &
   \omega ^3 & \omega  & 1 & \omega ^3 & \omega ^2 & 1 &
   \omega ^3 & \omega ^2 & 0 & 0 & 0 \\
 D_{17} & 1 & 1 & 1 & 1 & 1 & 1 & 1 & 1 & 1 & 1 & 1 & 1 &
   1 & 1 & 1 & 1 & \omega  & \omega  & \omega  & 0 & 0 &
   0 \\
 D_{18} & 1 & \omega  & \omega ^2 & \omega ^3 & 1 &
   \omega  & \omega ^2 & \omega ^3 & 1 & \omega  & \omega
   ^2 & \omega ^3 & 1 & \omega  & \omega ^2 & \omega ^3 &
   \omega  & \omega ^2 & \omega ^3 & 0 & 0 & 0 \\
 D_{19} & 1 & \omega ^2 & 1 & \omega ^2 & 1 & \omega ^2 &
   1 & \omega ^2 & 1 & \omega ^2 & 1 & \omega ^2 & 1 &
   \omega ^2 & 1 & \omega ^2 & \omega  & \omega ^3 &
   \omega  & 0 & 0 & 0 \\
 D_{20} & 1 & \omega ^3 & \omega ^2 & \omega  & 1 &
   \omega ^3 & \omega ^2 & \omega  & 1 & \omega ^3 &
   \omega ^2 & \omega  & 1 & \omega ^3 & \omega ^2 &
   \omega  & \omega  & 1 & \omega ^3 & 0 & 0 & 0 \\
 D_{21} & 1 & 1 & 1 & 1 & \omega  & \omega  & \omega  &
   \omega  & \omega ^2 & \omega ^2 & \omega ^2 & \omega
   ^2 & \omega ^3 & \omega ^3 & \omega ^3 & \omega ^3 &
   \omega  & \omega  & \omega  & 0 & 0 & 0 \\
 D_{22} & 1 & \omega  & \omega ^2 & \omega ^3 & \omega  &
   \omega ^2 & \omega ^3 & 1 & \omega ^2 & \omega ^3 & 1
   & \omega  & \omega ^3 & 1 & \omega  & \omega ^2 &
   \omega  & \omega ^2 & \omega ^3 & 0 & 0 & 0 \\
 D_{23} & 1 & \omega ^2 & 1 & \omega ^2 & \omega  &
   \omega ^3 & \omega  & \omega ^3 & \omega ^2 & 1 &
   \omega ^2 & 1 & \omega ^3 & \omega  & \omega ^3 &
   \omega  & \omega  & \omega ^3 & \omega  & 0 & 0 & 0 \\
 D_{24} & 1 & \omega ^3 & \omega ^2 & \omega  & \omega  &
   1 & \omega ^3 & \omega ^2 & \omega ^2 & \omega  & 1 &
   \omega ^3 & \omega ^3 & \omega ^2 & \omega  & 1 &
   \omega  & 1 & \omega ^3 & 0 & 0 & 0 \\
 D_{25} & 1 & 1 & 1 & 1 & \omega ^2 & \omega ^2 & \omega
   ^2 & \omega ^2 & 1 & 1 & 1 & 1 & \omega ^2 & \omega ^2
   & \omega ^2 & \omega ^2 & \omega  & \omega  & \omega
   & 0 & 0 & 0 \\
 D_{26} & 1 & \omega  & \omega ^2 & \omega ^3 & \omega ^2
   & \omega ^3 & 1 & \omega  & 1 & \omega  & \omega ^2 &
   \omega ^3 & \omega ^2 & \omega ^3 & 1 & \omega  &
   \omega  & \omega ^2 & \omega ^3 & 0 & 0 & 0 \\
 D_{27} & 1 & \omega ^2 & 1 & \omega ^2 & \omega ^2 & 1 &
   \omega ^2 & 1 & 1 & \omega ^2 & 1 & \omega ^2 & \omega
   ^2 & 1 & \omega ^2 & 1 & \omega  & \omega ^3 & \omega
   & 0 & 0 & 0 \\
 D_{28} & 1 & \omega ^3 & \omega ^2 & \omega  & \omega ^2
   & \omega  & 1 & \omega ^3 & 1 & \omega ^3 & \omega ^2
   & \omega  & \omega ^2 & \omega  & 1 & \omega ^3 &
   \omega  & 1 & \omega ^3 & 0 & 0 & 0 \\
 D_{29} & 1 & 1 & 1 & 1 & \omega ^3 & \omega ^3 & \omega
   ^3 & \omega ^3 & \omega ^2 & \omega ^2 & \omega ^2 &
   \omega ^2 & \omega  & \omega  & \omega  & \omega  &
   \omega  & \omega  & \omega  & 0 & 0 & 0 \\
 D_{30} & 1 & \omega  & \omega ^2 & \omega ^3 & \omega ^3
   & 1 & \omega  & \omega ^2 & \omega ^2 & \omega ^3 & 1
   & \omega  & \omega  & \omega ^2 & \omega ^3 & 1 &
   \omega  & \omega ^2 & \omega ^3 & 0 & 0 & 0 \\
 D_{31} & 1 & \omega ^2 & 1 & \omega ^2 & \omega ^3 &
   \omega  & \omega ^3 & \omega  & \omega ^2 & 1 & \omega
   ^2 & 1 & \omega  & \omega ^3 & \omega  & \omega ^3 &
   \omega  & \omega ^3 & \omega  & 0 & 0 & 0 \\
 D_{32} & 1 & \omega ^3 & \omega ^2 & \omega  & \omega ^3
   & \omega ^2 & \omega  & 1 & \omega ^2 & \omega  & 1 &
   \omega ^3 & \omega  & 1 & \omega ^3 & \omega ^2 &
   \omega  & 1 & \omega ^3 & 0 & 0 & 0 \\
 D_{33} & 1 & 1 & 1 & 1 & 1 & 1 & 1 & 1 & 1 & 1 & 1 & 1 &
   1 & 1 & 1 & 1 & \omega ^2 & \omega ^2 & \omega ^2 & 0
   & 0 & 0 \\
 D_{34} & 1 & \omega  & \omega ^2 & \omega ^3 & 1 &
   \omega  & \omega ^2 & \omega ^3 & 1 & \omega  & \omega
   ^2 & \omega ^3 & 1 & \omega  & \omega ^2 & \omega ^3 &
   \omega ^2 & \omega ^3 & 1 & 0 & 0 & 0 \\
 D_{35} & 1 & \omega ^2 & 1 & \omega ^2 & 1 & \omega ^2 &
   1 & \omega ^2 & 1 & \omega ^2 & 1 & \omega ^2 & 1 &
   \omega ^2 & 1 & \omega ^2 & \omega ^2 & 1 & \omega ^2
   & 0 & 0 & 0 \\
 D_{36} & 1 & \omega ^3 & \omega ^2 & \omega  & 1 &
   \omega ^3 & \omega ^2 & \omega  & 1 & \omega ^3 &
   \omega ^2 & \omega  & 1 & \omega ^3 & \omega ^2 &
   \omega  & \omega ^2 & \omega  & 1 & 0 & 0 & 0 \\
 D_{37} & 1 & 1 & 1 & 1 & \omega  & \omega  & \omega  &
   \omega  & \omega ^2 & \omega ^2 & \omega ^2 & \omega
   ^2 & \omega ^3 & \omega ^3 & \omega ^3 & \omega ^3 &
   \omega ^2 & \omega ^2 & \omega ^2 & 0 & 0 & 0 \\
 D_{38} & 1 & \omega  & \omega ^2 & \omega ^3 & \omega  &
   \omega ^2 & \omega ^3 & 1 & \omega ^2 & \omega ^3 & 1
   & \omega  & \omega ^3 & 1 & \omega  & \omega ^2 &
   \omega ^2 & \omega ^3 & 1 & 0 & 0 & 0 \\
 D_{39} & 1 & \omega ^2 & 1 & \omega ^2 & \omega  &
   \omega ^3 & \omega  & \omega ^3 & \omega ^2 & 1 &
   \omega ^2 & 1 & \omega ^3 & \omega  & \omega ^3 &
   \omega  & \omega ^2 & 1 & \omega ^2 & 0 & 0 & 0 \\
 D_{40} & 1 & \omega ^3 & \omega ^2 & \omega  & \omega  &
   1 & \omega ^3 & \omega ^2 & \omega ^2 & \omega  & 1 &
   \omega ^3 & \omega ^3 & \omega ^2 & \omega  & 1 &
   \omega ^2 & \omega  & 1 & 0 & 0 & 0 \\
 D_{41} & 1 & 1 & 1 & 1 & \omega ^2 & \omega ^2 & \omega
   ^2 & \omega ^2 & 1 & 1 & 1 & 1 & \omega ^2 & \omega ^2
   & \omega ^2 & \omega ^2 & \omega ^2 & \omega ^2 &
   \omega ^2 & 0 & 0 & 0 \\
 D_{42} & 1 & \omega  & \omega ^2 & \omega ^3 & \omega ^2
   & \omega ^3 & 1 & \omega  & 1 & \omega  & \omega ^2 &
   \omega ^3 & \omega ^2 & \omega ^3 & 1 & \omega  &
   \omega ^2 & \omega ^3 & 1 & 0 & 0 & 0 \\
 D_{43} & 1 & \omega ^2 & 1 & \omega ^2 & \omega ^2 & 1 &
   \omega ^2 & 1 & 1 & \omega ^2 & 1 & \omega ^2 & \omega
   ^2 & 1 & \omega ^2 & 1 & \omega ^2 & 1 & \omega ^2 & 0
   & 0 & 0 \\
 D_{44} & 1 & \omega ^3 & \omega ^2 & \omega  & \omega ^2
   & \omega  & 1 & \omega ^3 & 1 & \omega ^3 & \omega ^2
   & \omega  & \omega ^2 & \omega  & 1 & \omega ^3 &
   \omega ^2 & \omega  & 1 & 0 & 0 & 0 \\
 D_{45} & 1 & 1 & 1 & 1 & \omega ^3 & \omega ^3 & \omega
   ^3 & \omega ^3 & \omega ^2 & \omega ^2 & \omega ^2 &
   \omega ^2 & \omega  & \omega  & \omega  & \omega  &
   \omega ^2 & \omega ^2 & \omega ^2 & 0 & 0 & 0 \\
 D_{46} & 1 & \omega  & \omega ^2 & \omega ^3 & \omega ^3
   & 1 & \omega  & \omega ^2 & \omega ^2 & \omega ^3 & 1
   & \omega  & \omega  & \omega ^2 & \omega ^3 & 1 &
   \omega ^2 & \omega ^3 & 1 & 0 & 0 & 0 \\
 D_{47} & 1 & \omega ^2 & 1 & \omega ^2 & \omega ^3 &
   \omega  & \omega ^3 & \omega  & \omega ^2 & 1 & \omega
   ^2 & 1 & \omega  & \omega ^3 & \omega  & \omega ^3 &
   \omega ^2 & 1 & \omega ^2 & 0 & 0 & 0 \\
 D_{48} & 1 & \omega ^3 & \omega ^2 & \omega  & \omega ^3
   & \omega ^2 & \omega  & 1 & \omega ^2 & \omega  & 1 &
   \omega ^3 & \omega  & 1 & \omega ^3 & \omega ^2 &
   \omega ^2 & \omega  & 1 & 0 & 0 & 0 \\
 D_{49} & 1 & 1 & 1 & 1 & 1 & 1 & 1 & 1 & 1 & 1 & 1 & 1 &
   1 & 1 & 1 & 1 & \omega ^3 & \omega ^3 & \omega ^3 & 0
   & 0 & 0 \\
 D_{50} & 1 & \omega  & \omega ^2 & \omega ^3 & 1 &
   \omega  & \omega ^2 & \omega ^3 & 1 & \omega  & \omega
   ^2 & \omega ^3 & 1 & \omega  & \omega ^2 & \omega ^3 &
   \omega ^3 & 1 & \omega  & 0 & 0 & 0 \\
 D_{51} & 1 & \omega ^2 & 1 & \omega ^2 & 1 & \omega ^2 &
   1 & \omega ^2 & 1 & \omega ^2 & 1 & \omega ^2 & 1 &
   \omega ^2 & 1 & \omega ^2 & \omega ^3 & \omega  &
   \omega ^3 & 0 & 0 & 0 \\
 D_{52} & 1 & \omega ^3 & \omega ^2 & \omega  & 1 &
   \omega ^3 & \omega ^2 & \omega  & 1 & \omega ^3 &
   \omega ^2 & \omega  & 1 & \omega ^3 & \omega ^2 &
   \omega  & \omega ^3 & \omega ^2 & \omega  & 0 & 0 & 0
   \\
 D_{53} & 1 & 1 & 1 & 1 & \omega  & \omega  & \omega  &
   \omega  & \omega ^2 & \omega ^2 & \omega ^2 & \omega
   ^2 & \omega ^3 & \omega ^3 & \omega ^3 & \omega ^3 &
   \omega ^3 & \omega ^3 & \omega ^3 & 0 & 0 & 0 \\
 D_{54} & 1 & \omega  & \omega ^2 & \omega ^3 & \omega  &
   \omega ^2 & \omega ^3 & 1 & \omega ^2 & \omega ^3 & 1
   & \omega  & \omega ^3 & 1 & \omega  & \omega ^2 &
   \omega ^3 & 1 & \omega  & 0 & 0 & 0 \\
 D_{55} & 1 & \omega ^2 & 1 & \omega ^2 & \omega  &
   \omega ^3 & \omega  & \omega ^3 & \omega ^2 & 1 &
   \omega ^2 & 1 & \omega ^3 & \omega  & \omega ^3 &
   \omega  & \omega ^3 & \omega  & \omega ^3 & 0 & 0 & 0
   \\
 D_{56} & 1 & \omega ^3 & \omega ^2 & \omega  & \omega  &
   1 & \omega ^3 & \omega ^2 & \omega ^2 & \omega  & 1 &
   \omega ^3 & \omega ^3 & \omega ^2 & \omega  & 1 &
   \omega ^3 & \omega ^2 & \omega  & 0 & 0 & 0 \\
 D_{57} & 1 & 1 & 1 & 1 & \omega ^2 & \omega ^2 & \omega
   ^2 & \omega ^2 & 1 & 1 & 1 & 1 & \omega ^2 & \omega ^2
   & \omega ^2 & \omega ^2 & \omega ^3 & \omega ^3 &
   \omega ^3 & 0 & 0 & 0 \\
 D_{58} & 1 & \omega  & \omega ^2 & \omega ^3 & \omega ^2
   & \omega ^3 & 1 & \omega  & 1 & \omega  & \omega ^2 &
   \omega ^3 & \omega ^2 & \omega ^3 & 1 & \omega  &
   \omega ^3 & 1 & \omega  & 0 & 0 & 0 \\
 D_{59} & 1 & \omega ^2 & 1 & \omega ^2 & \omega ^2 & 1 &
   \omega ^2 & 1 & 1 & \omega ^2 & 1 & \omega ^2 & \omega
   ^2 & 1 & \omega ^2 & 1 & \omega ^3 & \omega  & \omega
   ^3 & 0 & 0 & 0 \\
 D_{60} & 1 & \omega ^3 & \omega ^2 & \omega  & \omega ^2
   & \omega  & 1 & \omega ^3 & 1 & \omega ^3 & \omega ^2
   & \omega  & \omega ^2 & \omega  & 1 & \omega ^3 &
   \omega ^3 & \omega ^2 & \omega  & 0 & 0 & 0 \\
 D_{61} & 1 & 1 & 1 & 1 & \omega ^3 & \omega ^3 & \omega
   ^3 & \omega ^3 & \omega ^2 & \omega ^2 & \omega ^2 &
   \omega ^2 & \omega  & \omega  & \omega  & \omega  &
   \omega ^3 & \omega ^3 & \omega ^3 & 0 & 0 & 0 \\
 D_{62} & 1 & \omega  & \omega ^2 & \omega ^3 & \omega ^3
   & 1 & \omega  & \omega ^2 & \omega ^2 & \omega ^3 & 1
   & \omega  & \omega  & \omega ^2 & \omega ^3 & 1 &
   \omega ^3 & 1 & \omega  & 0 & 0 & 0 \\
 D_{63} & 1 & \omega ^2 & 1 & \omega ^2 & \omega ^3 &
   \omega  & \omega ^3 & \omega  & \omega ^2 & 1 & \omega
   ^2 & 1 & \omega  & \omega ^3 & \omega  & \omega ^3 &
   \omega ^3 & \omega  & \omega ^3 & 0 & 0 & 0 \\
 D_{64} & 1 & \omega ^3 & \omega ^2 & \omega  & \omega ^3
   & \omega ^2 & \omega  & 1 & \omega ^2 & \omega  & 1 &
   \omega ^3 & \omega  & 1 & \omega ^3 & \omega ^2 &
   \omega ^3 & \omega ^2 & \omega  & 0 & 0 & 0\\
 \hline
\end{array}
\end{equation}
}
{\tiny
\begin{equation}\label{2charto64}
\begin{array}{|l|llllllllllllllllllllll|}
\hline
 0 & C_{23} & C_{24} & C_{25} & C_{26} & C_{27} & C_{28}
   & C_{29} & C_{30} & C_{31} & C_{32} & C_{33} & C_{34}
   & C_{35} & C_{36} & C_{37} & C_{38} & C_{39} & C_{40}
   & C_{41} & C_{42} & C_{43} & C_{44} \\
   \hline
 D_1 & 1 & 1 & 1 & 1 & 1 & 1 & 1 & 1 & 1 & 1 & 1 & 1 & 1
   & 1 & 1 & 1 & 1 & 1 & 1 & 1 & 1 & 1 \\
 D_2 & \omega ^2 & \omega ^3 & 1 & \omega  & \omega ^2 &
   \omega ^3 & 1 & \omega  & \omega ^2 & \omega ^3 & 1 &
   \omega  & \omega ^2 & \omega ^3 & 1 & \omega  & \omega
   ^2 & \omega ^3 & 1 & \omega  & \omega ^2 & \omega ^3
   \\
 D_3 & 1 & \omega ^2 & 1 & \omega ^2 & 1 & \omega ^2 & 1
   & \omega ^2 & 1 & \omega ^2 & 1 & \omega ^2 & 1 &
   \omega ^2 & 1 & \omega ^2 & 1 & \omega ^2 & 1 & \omega
   ^2 & 1 & \omega ^2 \\
 D_4 & \omega ^2 & \omega  & 1 & \omega ^3 & \omega ^2 &
   \omega  & 1 & \omega ^3 & \omega ^2 & \omega  & 1 &
   \omega ^3 & \omega ^2 & \omega  & 1 & \omega ^3 &
   \omega ^2 & \omega  & 1 & \omega ^3 & \omega ^2 &
   \omega  \\
 D_5 & \omega  & \omega  & \omega ^2 & \omega ^2 & \omega
   ^2 & \omega ^2 & \omega ^3 & \omega ^3 & \omega ^3 &
   \omega ^3 & 1 & 1 & 1 & 1 & \omega  & \omega  & \omega
    & \omega  & \omega ^2 & \omega ^2 & \omega ^2 &
   \omega ^2 \\
 D_6 & \omega ^3 & 1 & \omega ^2 & \omega ^3 & 1 & \omega
    & \omega ^3 & 1 & \omega  & \omega ^2 & 1 & \omega  &
   \omega ^2 & \omega ^3 & \omega  & \omega ^2 & \omega
   ^3 & 1 & \omega ^2 & \omega ^3 & 1 & \omega  \\
 D_7 & \omega  & \omega ^3 & \omega ^2 & 1 & \omega ^2 &
   1 & \omega ^3 & \omega  & \omega ^3 & \omega  & 1 &
   \omega ^2 & 1 & \omega ^2 & \omega  & \omega ^3 &
   \omega  & \omega ^3 & \omega ^2 & 1 & \omega ^2 & 1 \\
 D_8 & \omega ^3 & \omega ^2 & \omega ^2 & \omega  & 1 &
   \omega ^3 & \omega ^3 & \omega ^2 & \omega  & 1 & 1 &
   \omega ^3 & \omega ^2 & \omega  & \omega  & 1 & \omega
   ^3 & \omega ^2 & \omega ^2 & \omega  & 1 & \omega ^3
   \\
 D_9 & \omega ^2 & \omega ^2 & 1 & 1 & 1 & 1 & \omega ^2
   & \omega ^2 & \omega ^2 & \omega ^2 & 1 & 1 & 1 & 1 &
   \omega ^2 & \omega ^2 & \omega ^2 & \omega ^2 & 1 & 1
   & 1 & 1 \\
 D_{10} & 1 & \omega  & 1 & \omega  & \omega ^2 & \omega
   ^3 & \omega ^2 & \omega ^3 & 1 & \omega  & 1 & \omega
   & \omega ^2 & \omega ^3 & \omega ^2 & \omega ^3 & 1 &
   \omega  & 1 & \omega  & \omega ^2 & \omega ^3 \\
 D_{11} & \omega ^2 & 1 & 1 & \omega ^2 & 1 & \omega ^2 &
   \omega ^2 & 1 & \omega ^2 & 1 & 1 & \omega ^2 & 1 &
   \omega ^2 & \omega ^2 & 1 & \omega ^2 & 1 & 1 & \omega
   ^2 & 1 & \omega ^2 \\
 D_{12} & 1 & \omega ^3 & 1 & \omega ^3 & \omega ^2 &
   \omega  & \omega ^2 & \omega  & 1 & \omega ^3 & 1 &
   \omega ^3 & \omega ^2 & \omega  & \omega ^2 & \omega
   & 1 & \omega ^3 & 1 & \omega ^3 & \omega ^2 & \omega
   \\
 D_{13} & \omega ^3 & \omega ^3 & \omega ^2 & \omega ^2 &
   \omega ^2 & \omega ^2 & \omega  & \omega  & \omega  &
   \omega  & 1 & 1 & 1 & 1 & \omega ^3 & \omega ^3 &
   \omega ^3 & \omega ^3 & \omega ^2 & \omega ^2 & \omega
   ^2 & \omega ^2 \\
 D_{14} & \omega  & \omega ^2 & \omega ^2 & \omega ^3 & 1
   & \omega  & \omega  & \omega ^2 & \omega ^3 & 1 & 1 &
   \omega  & \omega ^2 & \omega ^3 & \omega ^3 & 1 &
   \omega  & \omega ^2 & \omega ^2 & \omega ^3 & 1 &
   \omega  \\
 D_{15} & \omega ^3 & \omega  & \omega ^2 & 1 & \omega ^2
   & 1 & \omega  & \omega ^3 & \omega  & \omega ^3 & 1 &
   \omega ^2 & 1 & \omega ^2 & \omega ^3 & \omega  &
   \omega ^3 & \omega  & \omega ^2 & 1 & \omega ^2 & 1 \\
 D_{16} & \omega  & 1 & \omega ^2 & \omega  & 1 & \omega
   ^3 & \omega  & 1 & \omega ^3 & \omega ^2 & 1 & \omega
   ^3 & \omega ^2 & \omega  & \omega ^3 & \omega ^2 &
   \omega  & 1 & \omega ^2 & \omega  & 1 & \omega ^3 \\
 D_{17} & \omega  & \omega  & \omega  & \omega  & \omega
   & \omega  & \omega  & \omega  & \omega  & \omega  &
   \omega ^2 & \omega ^2 & \omega ^2 & \omega ^2 & \omega
   ^2 & \omega ^2 & \omega ^2 & \omega ^2 & \omega ^2 &
   \omega ^2 & \omega ^2 & \omega ^2 \\
 D_{18} & \omega ^3 & 1 & \omega  & \omega ^2 & \omega ^3
   & 1 & \omega  & \omega ^2 & \omega ^3 & 1 & \omega ^2
   & \omega ^3 & 1 & \omega  & \omega ^2 & \omega ^3 & 1
   & \omega  & \omega ^2 & \omega ^3 & 1 & \omega  \\
 D_{19} & \omega  & \omega ^3 & \omega  & \omega ^3 &
   \omega  & \omega ^3 & \omega  & \omega ^3 & \omega  &
   \omega ^3 & \omega ^2 & 1 & \omega ^2 & 1 & \omega ^2
   & 1 & \omega ^2 & 1 & \omega ^2 & 1 & \omega ^2 & 1 \\
 D_{20} & \omega ^3 & \omega ^2 & \omega  & 1 & \omega ^3
   & \omega ^2 & \omega  & 1 & \omega ^3 & \omega ^2 &
   \omega ^2 & \omega  & 1 & \omega ^3 & \omega ^2 &
   \omega  & 1 & \omega ^3 & \omega ^2 & \omega  & 1 &
   \omega ^3 \\
 D_{21} & \omega ^2 & \omega ^2 & \omega ^3 & \omega ^3 &
   \omega ^3 & \omega ^3 & 1 & 1 & 1 & 1 & \omega ^2 &
   \omega ^2 & \omega ^2 & \omega ^2 & \omega ^3 & \omega
   ^3 & \omega ^3 & \omega ^3 & 1 & 1 & 1 & 1 \\
 D_{22} & 1 & \omega  & \omega ^3 & 1 & \omega  & \omega
   ^2 & 1 & \omega  & \omega ^2 & \omega ^3 & \omega ^2 &
   \omega ^3 & 1 & \omega  & \omega ^3 & 1 & \omega  &
   \omega ^2 & 1 & \omega  & \omega ^2 & \omega ^3 \\
 D_{23} & \omega ^2 & 1 & \omega ^3 & \omega  & \omega ^3
   & \omega  & 1 & \omega ^2 & 1 & \omega ^2 & \omega ^2
   & 1 & \omega ^2 & 1 & \omega ^3 & \omega  & \omega ^3
   & \omega  & 1 & \omega ^2 & 1 & \omega ^2 \\
 D_{24} & 1 & \omega ^3 & \omega ^3 & \omega ^2 & \omega
   & 1 & 1 & \omega ^3 & \omega ^2 & \omega  & \omega ^2
   & \omega  & 1 & \omega ^3 & \omega ^3 & \omega ^2 &
   \omega  & 1 & 1 & \omega ^3 & \omega ^2 & \omega  \\
 D_{25} & \omega ^3 & \omega ^3 & \omega  & \omega  &
   \omega  & \omega  & \omega ^3 & \omega ^3 & \omega ^3
   & \omega ^3 & \omega ^2 & \omega ^2 & \omega ^2 &
   \omega ^2 & 1 & 1 & 1 & 1 & \omega ^2 & \omega ^2 &
   \omega ^2 & \omega ^2 \\
 D_{26} & \omega  & \omega ^2 & \omega  & \omega ^2 &
   \omega ^3 & 1 & \omega ^3 & 1 & \omega  & \omega ^2 &
   \omega ^2 & \omega ^3 & 1 & \omega  & 1 & \omega  &
   \omega ^2 & \omega ^3 & \omega ^2 & \omega ^3 & 1 &
   \omega  \\
 D_{27} & \omega ^3 & \omega  & \omega  & \omega ^3 &
   \omega  & \omega ^3 & \omega ^3 & \omega  & \omega ^3
   & \omega  & \omega ^2 & 1 & \omega ^2 & 1 & 1 & \omega
   ^2 & 1 & \omega ^2 & \omega ^2 & 1 & \omega ^2 & 1 \\
 D_{28} & \omega  & 1 & \omega  & 1 & \omega ^3 & \omega
   ^2 & \omega ^3 & \omega ^2 & \omega  & 1 & \omega ^2 &
   \omega  & 1 & \omega ^3 & 1 & \omega ^3 & \omega ^2 &
   \omega  & \omega ^2 & \omega  & 1 & \omega ^3 \\
 D_{29} & 1 & 1 & \omega ^3 & \omega ^3 & \omega ^3 &
   \omega ^3 & \omega ^2 & \omega ^2 & \omega ^2 & \omega
   ^2 & \omega ^2 & \omega ^2 & \omega ^2 & \omega ^2 &
   \omega  & \omega  & \omega  & \omega  & 1 & 1 & 1 & 1
   \\
 D_{30} & \omega ^2 & \omega ^3 & \omega ^3 & 1 & \omega
   & \omega ^2 & \omega ^2 & \omega ^3 & 1 & \omega  &
   \omega ^2 & \omega ^3 & 1 & \omega  & \omega  & \omega
   ^2 & \omega ^3 & 1 & 1 & \omega  & \omega ^2 & \omega
   ^3 \\
 D_{31} & 1 & \omega ^2 & \omega ^3 & \omega  & \omega ^3
   & \omega  & \omega ^2 & 1 & \omega ^2 & 1 & \omega ^2
   & 1 & \omega ^2 & 1 & \omega  & \omega ^3 & \omega  &
   \omega ^3 & 1 & \omega ^2 & 1 & \omega ^2 \\
 D_{32} & \omega ^2 & \omega  & \omega ^3 & \omega ^2 &
   \omega  & 1 & \omega ^2 & \omega  & 1 & \omega ^3 &
   \omega ^2 & \omega  & 1 & \omega ^3 & \omega  & 1 &
   \omega ^3 & \omega ^2 & 1 & \omega ^3 & \omega ^2 &
   \omega  \\
 D_{33} & \omega ^2 & \omega ^2 & \omega ^2 & \omega ^2 &
   \omega ^2 & \omega ^2 & \omega ^2 & \omega ^2 & \omega
   ^2 & \omega ^2 & 1 & 1 & 1 & 1 & 1 & 1 & 1 & 1 & 1 & 1
   & 1 & 1 \\
 D_{34} & 1 & \omega  & \omega ^2 & \omega ^3 & 1 &
   \omega  & \omega ^2 & \omega ^3 & 1 & \omega  & 1 &
   \omega  & \omega ^2 & \omega ^3 & 1 & \omega  & \omega
   ^2 & \omega ^3 & 1 & \omega  & \omega ^2 & \omega ^3
   \\
 D_{35} & \omega ^2 & 1 & \omega ^2 & 1 & \omega ^2 & 1 &
   \omega ^2 & 1 & \omega ^2 & 1 & 1 & \omega ^2 & 1 &
   \omega ^2 & 1 & \omega ^2 & 1 & \omega ^2 & 1 & \omega
   ^2 & 1 & \omega ^2 \\
 D_{36} & 1 & \omega ^3 & \omega ^2 & \omega  & 1 &
   \omega ^3 & \omega ^2 & \omega  & 1 & \omega ^3 & 1 &
   \omega ^3 & \omega ^2 & \omega  & 1 & \omega ^3 &
   \omega ^2 & \omega  & 1 & \omega ^3 & \omega ^2 &
   \omega  \\
 D_{37} & \omega ^3 & \omega ^3 & 1 & 1 & 1 & 1 & \omega
   & \omega  & \omega  & \omega  & 1 & 1 & 1 & 1 & \omega
    & \omega  & \omega  & \omega  & \omega ^2 & \omega ^2
   & \omega ^2 & \omega ^2 \\
 D_{38} & \omega  & \omega ^2 & 1 & \omega  & \omega ^2 &
   \omega ^3 & \omega  & \omega ^2 & \omega ^3 & 1 & 1 &
   \omega  & \omega ^2 & \omega ^3 & \omega  & \omega ^2
   & \omega ^3 & 1 & \omega ^2 & \omega ^3 & 1 & \omega
   \\
 D_{39} & \omega ^3 & \omega  & 1 & \omega ^2 & 1 &
   \omega ^2 & \omega  & \omega ^3 & \omega  & \omega ^3
   & 1 & \omega ^2 & 1 & \omega ^2 & \omega  & \omega ^3
   & \omega  & \omega ^3 & \omega ^2 & 1 & \omega ^2 & 1
   \\
 D_{40} & \omega  & 1 & 1 & \omega ^3 & \omega ^2 &
   \omega  & \omega  & 1 & \omega ^3 & \omega ^2 & 1 &
   \omega ^3 & \omega ^2 & \omega  & \omega  & 1 & \omega
   ^3 & \omega ^2 & \omega ^2 & \omega  & 1 & \omega ^3
   \\
 D_{41} & 1 & 1 & \omega ^2 & \omega ^2 & \omega ^2 &
   \omega ^2 & 1 & 1 & 1 & 1 & 1 & 1 & 1 & 1 & \omega ^2
   & \omega ^2 & \omega ^2 & \omega ^2 & 1 & 1 & 1 & 1 \\
 D_{42} & \omega ^2 & \omega ^3 & \omega ^2 & \omega ^3 &
   1 & \omega  & 1 & \omega  & \omega ^2 & \omega ^3 & 1
   & \omega  & \omega ^2 & \omega ^3 & \omega ^2 & \omega
   ^3 & 1 & \omega  & 1 & \omega  & \omega ^2 & \omega ^3
   \\
 D_{43} & 1 & \omega ^2 & \omega ^2 & 1 & \omega ^2 & 1 &
   1 & \omega ^2 & 1 & \omega ^2 & 1 & \omega ^2 & 1 &
   \omega ^2 & \omega ^2 & 1 & \omega ^2 & 1 & 1 & \omega
   ^2 & 1 & \omega ^2 \\
 D_{44} & \omega ^2 & \omega  & \omega ^2 & \omega  & 1 &
   \omega ^3 & 1 & \omega ^3 & \omega ^2 & \omega  & 1 &
   \omega ^3 & \omega ^2 & \omega  & \omega ^2 & \omega
   & 1 & \omega ^3 & 1 & \omega ^3 & \omega ^2 & \omega
   \\
 D_{45} & \omega  & \omega  & 1 & 1 & 1 & 1 & \omega ^3 &
   \omega ^3 & \omega ^3 & \omega ^3 & 1 & 1 & 1 & 1 &
   \omega ^3 & \omega ^3 & \omega ^3 & \omega ^3 & \omega
   ^2 & \omega ^2 & \omega ^2 & \omega ^2 \\
 D_{46} & \omega ^3 & 1 & 1 & \omega  & \omega ^2 &
   \omega ^3 & \omega ^3 & 1 & \omega  & \omega ^2 & 1 &
   \omega  & \omega ^2 & \omega ^3 & \omega ^3 & 1 &
   \omega  & \omega ^2 & \omega ^2 & \omega ^3 & 1 &
   \omega  \\
 D_{47} & \omega  & \omega ^3 & 1 & \omega ^2 & 1 &
   \omega ^2 & \omega ^3 & \omega  & \omega ^3 & \omega
   & 1 & \omega ^2 & 1 & \omega ^2 & \omega ^3 & \omega
   & \omega ^3 & \omega  & \omega ^2 & 1 & \omega ^2 & 1
   \\
 D_{48} & \omega ^3 & \omega ^2 & 1 & \omega ^3 & \omega
   ^2 & \omega  & \omega ^3 & \omega ^2 & \omega  & 1 & 1
   & \omega ^3 & \omega ^2 & \omega  & \omega ^3 & \omega
   ^2 & \omega  & 1 & \omega ^2 & \omega  & 1 & \omega ^3
   \\
 D_{49} & \omega ^3 & \omega ^3 & \omega ^3 & \omega ^3 &
   \omega ^3 & \omega ^3 & \omega ^3 & \omega ^3 & \omega
   ^3 & \omega ^3 & \omega ^2 & \omega ^2 & \omega ^2 &
   \omega ^2 & \omega ^2 & \omega ^2 & \omega ^2 & \omega
   ^2 & \omega ^2 & \omega ^2 & \omega ^2 & \omega ^2 \\
 D_{50} & \omega  & \omega ^2 & \omega ^3 & 1 & \omega  &
   \omega ^2 & \omega ^3 & 1 & \omega  & \omega ^2 &
   \omega ^2 & \omega ^3 & 1 & \omega  & \omega ^2 &
   \omega ^3 & 1 & \omega  & \omega ^2 & \omega ^3 & 1 &
   \omega  \\
 D_{51} & \omega ^3 & \omega  & \omega ^3 & \omega  &
   \omega ^3 & \omega  & \omega ^3 & \omega  & \omega ^3
   & \omega  & \omega ^2 & 1 & \omega ^2 & 1 & \omega ^2
   & 1 & \omega ^2 & 1 & \omega ^2 & 1 & \omega ^2 & 1 \\
 D_{52} & \omega  & 1 & \omega ^3 & \omega ^2 & \omega  &
   1 & \omega ^3 & \omega ^2 & \omega  & 1 & \omega ^2 &
   \omega  & 1 & \omega ^3 & \omega ^2 & \omega  & 1 &
   \omega ^3 & \omega ^2 & \omega  & 1 & \omega ^3 \\
 D_{53} & 1 & 1 & \omega  & \omega  & \omega  & \omega  &
   \omega ^2 & \omega ^2 & \omega ^2 & \omega ^2 & \omega
   ^2 & \omega ^2 & \omega ^2 & \omega ^2 & \omega ^3 &
   \omega ^3 & \omega ^3 & \omega ^3 & 1 & 1 & 1 & 1 \\
 D_{54} & \omega ^2 & \omega ^3 & \omega  & \omega ^2 &
   \omega ^3 & 1 & \omega ^2 & \omega ^3 & 1 & \omega  &
   \omega ^2 & \omega ^3 & 1 & \omega  & \omega ^3 & 1 &
   \omega  & \omega ^2 & 1 & \omega  & \omega ^2 & \omega
   ^3 \\
 D_{55} & 1 & \omega ^2 & \omega  & \omega ^3 & \omega  &
   \omega ^3 & \omega ^2 & 1 & \omega ^2 & 1 & \omega ^2
   & 1 & \omega ^2 & 1 & \omega ^3 & \omega  & \omega ^3
   & \omega  & 1 & \omega ^2 & 1 & \omega ^2 \\
 D_{56} & \omega ^2 & \omega  & \omega  & 1 & \omega ^3 &
   \omega ^2 & \omega ^2 & \omega  & 1 & \omega ^3 &
   \omega ^2 & \omega  & 1 & \omega ^3 & \omega ^3 &
   \omega ^2 & \omega  & 1 & 1 & \omega ^3 & \omega ^2 &
   \omega  \\
 D_{57} & \omega  & \omega  & \omega ^3 & \omega ^3 &
   \omega ^3 & \omega ^3 & \omega  & \omega  & \omega  &
   \omega  & \omega ^2 & \omega ^2 & \omega ^2 & \omega
   ^2 & 1 & 1 & 1 & 1 & \omega ^2 & \omega ^2 & \omega ^2
   & \omega ^2 \\
 D_{58} & \omega ^3 & 1 & \omega ^3 & 1 & \omega  &
   \omega ^2 & \omega  & \omega ^2 & \omega ^3 & 1 &
   \omega ^2 & \omega ^3 & 1 & \omega  & 1 & \omega  &
   \omega ^2 & \omega ^3 & \omega ^2 & \omega ^3 & 1 &
   \omega  \\
 D_{59} & \omega  & \omega ^3 & \omega ^3 & \omega  &
   \omega ^3 & \omega  & \omega  & \omega ^3 & \omega  &
   \omega ^3 & \omega ^2 & 1 & \omega ^2 & 1 & 1 & \omega
   ^2 & 1 & \omega ^2 & \omega ^2 & 1 & \omega ^2 & 1 \\
 D_{60} & \omega ^3 & \omega ^2 & \omega ^3 & \omega ^2 &
   \omega  & 1 & \omega  & 1 & \omega ^3 & \omega ^2 &
   \omega ^2 & \omega  & 1 & \omega ^3 & 1 & \omega ^3 &
   \omega ^2 & \omega  & \omega ^2 & \omega  & 1 & \omega
   ^3 \\
 D_{61} & \omega ^2 & \omega ^2 & \omega  & \omega  &
   \omega  & \omega  & 1 & 1 & 1 & 1 & \omega ^2 & \omega
   ^2 & \omega ^2 & \omega ^2 & \omega  & \omega  &
   \omega  & \omega  & 1 & 1 & 1 & 1 \\
 D_{62} & 1 & \omega  & \omega  & \omega ^2 & \omega ^3 &
   1 & 1 & \omega  & \omega ^2 & \omega ^3 & \omega ^2 &
   \omega ^3 & 1 & \omega  & \omega  & \omega ^2 & \omega
   ^3 & 1 & 1 & \omega  & \omega ^2 & \omega ^3 \\
 D_{63} & \omega ^2 & 1 & \omega  & \omega ^3 & \omega  &
   \omega ^3 & 1 & \omega ^2 & 1 & \omega ^2 & \omega ^2
   & 1 & \omega ^2 & 1 & \omega  & \omega ^3 & \omega  &
   \omega ^3 & 1 & \omega ^2 & 1 & \omega ^2 \\
 D_{64} & 1 & \omega ^3 & \omega  & 1 & \omega ^3 &
   \omega ^2 & 1 & \omega ^3 & \omega ^2 & \omega  &
   \omega ^2 & \omega  & 1 & \omega ^3 & \omega  & 1 &
   \omega ^3 & \omega ^2 & 1 & \omega ^3 & \omega ^2 &
   \omega\\
   \hline
\end{array}
\end{equation}
}
{\tiny
\begin{equation}\label{3charto64}
\begin{array}{|l|llllllllllllllllllll|}
\hline
 0 & C_{45} & C_{46} & C_{47} & C_{48} & C_{49} & C_{50}
   & C_{51} & C_{52} & C_{53} & C_{54} & C_{55} & C_{56}
   & C_{57} & C_{58} & C_{59} & C_{60} & C_{61} & C_{62}
   & C_{63} & C_{64} \\
 \hline
 D_1 & 1 & 1 & 1 & 1 & 1 & 1 & 1 & 1 & 1 & 1 & 1 & 1 & 1
   & 1 & 1 & 1 & 1 & 1 & 1 & 1 \\
 D_2 & 1 & \omega  & \omega ^2 & \omega ^3 & 1 & \omega
   & \omega ^2 & \omega ^3 & 1 & \omega  & \omega ^2 &
   \omega ^3 & 1 & \omega  & \omega ^2 & \omega ^3 & 1 &
   \omega  & \omega ^2 & \omega ^3 \\
 D_3 & 1 & \omega ^2 & 1 & \omega ^2 & 1 & \omega ^2 & 1
   & \omega ^2 & 1 & \omega ^2 & 1 & \omega ^2 & 1 &
   \omega ^2 & 1 & \omega ^2 & 1 & \omega ^2 & 1 & \omega
   ^2 \\
 D_4 & 1 & \omega ^3 & \omega ^2 & \omega  & 1 & \omega
   ^3 & \omega ^2 & \omega  & 1 & \omega ^3 & \omega ^2 &
   \omega  & 1 & \omega ^3 & \omega ^2 & \omega  & 1 &
   \omega ^3 & \omega ^2 & \omega  \\
 D_5 & \omega ^3 & \omega ^3 & \omega ^3 & \omega ^3 & 1
   & 1 & 1 & 1 & \omega  & \omega  & \omega  & \omega  &
   \omega ^2 & \omega ^2 & \omega ^2 & \omega ^2 & \omega
   ^3 & \omega ^3 & \omega ^3 & \omega ^3 \\
 D_6 & \omega ^3 & 1 & \omega  & \omega ^2 & 1 & \omega
   & \omega ^2 & \omega ^3 & \omega  & \omega ^2 & \omega
   ^3 & 1 & \omega ^2 & \omega ^3 & 1 & \omega  & \omega
   ^3 & 1 & \omega  & \omega ^2 \\
 D_7 & \omega ^3 & \omega  & \omega ^3 & \omega  & 1 &
   \omega ^2 & 1 & \omega ^2 & \omega  & \omega ^3 &
   \omega  & \omega ^3 & \omega ^2 & 1 & \omega ^2 & 1 &
   \omega ^3 & \omega  & \omega ^3 & \omega  \\
 D_8 & \omega ^3 & \omega ^2 & \omega  & 1 & 1 & \omega
   ^3 & \omega ^2 & \omega  & \omega  & 1 & \omega ^3 &
   \omega ^2 & \omega ^2 & \omega  & 1 & \omega ^3 &
   \omega ^3 & \omega ^2 & \omega  & 1 \\
 D_9 & \omega ^2 & \omega ^2 & \omega ^2 & \omega ^2 & 1
   & 1 & 1 & 1 & \omega ^2 & \omega ^2 & \omega ^2 &
   \omega ^2 & 1 & 1 & 1 & 1 & \omega ^2 & \omega ^2 &
   \omega ^2 & \omega ^2 \\
 D_{10} & \omega ^2 & \omega ^3 & 1 & \omega  & 1 &
   \omega  & \omega ^2 & \omega ^3 & \omega ^2 & \omega
   ^3 & 1 & \omega  & 1 & \omega  & \omega ^2 & \omega ^3
   & \omega ^2 & \omega ^3 & 1 & \omega  \\
 D_{11} & \omega ^2 & 1 & \omega ^2 & 1 & 1 & \omega ^2 &
   1 & \omega ^2 & \omega ^2 & 1 & \omega ^2 & 1 & 1 &
   \omega ^2 & 1 & \omega ^2 & \omega ^2 & 1 & \omega ^2
   & 1 \\
 D_{12} & \omega ^2 & \omega  & 1 & \omega ^3 & 1 &
   \omega ^3 & \omega ^2 & \omega  & \omega ^2 & \omega
   & 1 & \omega ^3 & 1 & \omega ^3 & \omega ^2 & \omega
   & \omega ^2 & \omega  & 1 & \omega ^3 \\
 D_{13} & \omega  & \omega  & \omega  & \omega  & 1 & 1 &
   1 & 1 & \omega ^3 & \omega ^3 & \omega ^3 & \omega ^3
   & \omega ^2 & \omega ^2 & \omega ^2 & \omega ^2 &
   \omega  & \omega  & \omega  & \omega  \\
 D_{14} & \omega  & \omega ^2 & \omega ^3 & 1 & 1 &
   \omega  & \omega ^2 & \omega ^3 & \omega ^3 & 1 &
   \omega  & \omega ^2 & \omega ^2 & \omega ^3 & 1 &
   \omega  & \omega  & \omega ^2 & \omega ^3 & 1 \\
 D_{15} & \omega  & \omega ^3 & \omega  & \omega ^3 & 1 &
   \omega ^2 & 1 & \omega ^2 & \omega ^3 & \omega  &
   \omega ^3 & \omega  & \omega ^2 & 1 & \omega ^2 & 1 &
   \omega  & \omega ^3 & \omega  & \omega ^3 \\
 D_{16} & \omega  & 1 & \omega ^3 & \omega ^2 & 1 &
   \omega ^3 & \omega ^2 & \omega  & \omega ^3 & \omega
   ^2 & \omega  & 1 & \omega ^2 & \omega  & 1 & \omega ^3
   & \omega  & 1 & \omega ^3 & \omega ^2 \\
 D_{17} & \omega ^2 & \omega ^2 & \omega ^2 & \omega ^2 &
   \omega ^3 & \omega ^3 & \omega ^3 & \omega ^3 & \omega
   ^3 & \omega ^3 & \omega ^3 & \omega ^3 & \omega ^3 &
   \omega ^3 & \omega ^3 & \omega ^3 & \omega ^3 & \omega
   ^3 & \omega ^3 & \omega ^3 \\
 D_{18} & \omega ^2 & \omega ^3 & 1 & \omega  & \omega ^3
   & 1 & \omega  & \omega ^2 & \omega ^3 & 1 & \omega  &
   \omega ^2 & \omega ^3 & 1 & \omega  & \omega ^2 &
   \omega ^3 & 1 & \omega  & \omega ^2 \\
 D_{19} & \omega ^2 & 1 & \omega ^2 & 1 & \omega ^3 &
   \omega  & \omega ^3 & \omega  & \omega ^3 & \omega  &
   \omega ^3 & \omega  & \omega ^3 & \omega  & \omega ^3
   & \omega  & \omega ^3 & \omega  & \omega ^3 & \omega
   \\
 D_{20} & \omega ^2 & \omega  & 1 & \omega ^3 & \omega ^3
   & \omega ^2 & \omega  & 1 & \omega ^3 & \omega ^2 &
   \omega  & 1 & \omega ^3 & \omega ^2 & \omega  & 1 &
   \omega ^3 & \omega ^2 & \omega  & 1 \\
 D_{21} & \omega  & \omega  & \omega  & \omega  & \omega
   ^3 & \omega ^3 & \omega ^3 & \omega ^3 & 1 & 1 & 1 & 1
   & \omega  & \omega  & \omega  & \omega  & \omega ^2 &
   \omega ^2 & \omega ^2 & \omega ^2 \\
 D_{22} & \omega  & \omega ^2 & \omega ^3 & 1 & \omega ^3
   & 1 & \omega  & \omega ^2 & 1 & \omega  & \omega ^2 &
   \omega ^3 & \omega  & \omega ^2 & \omega ^3 & 1 &
   \omega ^2 & \omega ^3 & 1 & \omega  \\
 D_{23} & \omega  & \omega ^3 & \omega  & \omega ^3 &
   \omega ^3 & \omega  & \omega ^3 & \omega  & 1 & \omega
   ^2 & 1 & \omega ^2 & \omega  & \omega ^3 & \omega  &
   \omega ^3 & \omega ^2 & 1 & \omega ^2 & 1 \\
 D_{24} & \omega  & 1 & \omega ^3 & \omega ^2 & \omega ^3
   & \omega ^2 & \omega  & 1 & 1 & \omega ^3 & \omega ^2
   & \omega  & \omega  & 1 & \omega ^3 & \omega ^2 &
   \omega ^2 & \omega  & 1 & \omega ^3 \\
 D_{25} & 1 & 1 & 1 & 1 & \omega ^3 & \omega ^3 & \omega
   ^3 & \omega ^3 & \omega  & \omega  & \omega  & \omega
   & \omega ^3 & \omega ^3 & \omega ^3 & \omega ^3 &
   \omega  & \omega  & \omega  & \omega  \\
 D_{26} & 1 & \omega  & \omega ^2 & \omega ^3 & \omega ^3
   & 1 & \omega  & \omega ^2 & \omega  & \omega ^2 &
   \omega ^3 & 1 & \omega ^3 & 1 & \omega  & \omega ^2 &
   \omega  & \omega ^2 & \omega ^3 & 1 \\
 D_{27} & 1 & \omega ^2 & 1 & \omega ^2 & \omega ^3 &
   \omega  & \omega ^3 & \omega  & \omega  & \omega ^3 &
   \omega  & \omega ^3 & \omega ^3 & \omega  & \omega ^3
   & \omega  & \omega  & \omega ^3 & \omega  & \omega ^3
   \\
 D_{28} & 1 & \omega ^3 & \omega ^2 & \omega  & \omega ^3
   & \omega ^2 & \omega  & 1 & \omega  & 1 & \omega ^3 &
   \omega ^2 & \omega ^3 & \omega ^2 & \omega  & 1 &
   \omega  & 1 & \omega ^3 & \omega ^2 \\
 D_{29} & \omega ^3 & \omega ^3 & \omega ^3 & \omega ^3 &
   \omega ^3 & \omega ^3 & \omega ^3 & \omega ^3 & \omega
   ^2 & \omega ^2 & \omega ^2 & \omega ^2 & \omega  &
   \omega  & \omega  & \omega  & 1 & 1 & 1 & 1 \\
 D_{30} & \omega ^3 & 1 & \omega  & \omega ^2 & \omega ^3
   & 1 & \omega  & \omega ^2 & \omega ^2 & \omega ^3 & 1
   & \omega  & \omega  & \omega ^2 & \omega ^3 & 1 & 1 &
   \omega  & \omega ^2 & \omega ^3 \\
 D_{31} & \omega ^3 & \omega  & \omega ^3 & \omega  &
   \omega ^3 & \omega  & \omega ^3 & \omega  & \omega ^2
   & 1 & \omega ^2 & 1 & \omega  & \omega ^3 & \omega  &
   \omega ^3 & 1 & \omega ^2 & 1 & \omega ^2 \\
 D_{32} & \omega ^3 & \omega ^2 & \omega  & 1 & \omega ^3
   & \omega ^2 & \omega  & 1 & \omega ^2 & \omega  & 1 &
   \omega ^3 & \omega  & 1 & \omega ^3 & \omega ^2 & 1 &
   \omega ^3 & \omega ^2 & \omega  \\
 D_{33} & 1 & 1 & 1 & 1 & \omega ^2 & \omega ^2 & \omega
   ^2 & \omega ^2 & \omega ^2 & \omega ^2 & \omega ^2 &
   \omega ^2 & \omega ^2 & \omega ^2 & \omega ^2 & \omega
   ^2 & \omega ^2 & \omega ^2 & \omega ^2 & \omega ^2 \\
 D_{34} & 1 & \omega  & \omega ^2 & \omega ^3 & \omega ^2
   & \omega ^3 & 1 & \omega  & \omega ^2 & \omega ^3 & 1
   & \omega  & \omega ^2 & \omega ^3 & 1 & \omega  &
   \omega ^2 & \omega ^3 & 1 & \omega  \\
 D_{35} & 1 & \omega ^2 & 1 & \omega ^2 & \omega ^2 & 1 &
   \omega ^2 & 1 & \omega ^2 & 1 & \omega ^2 & 1 & \omega
   ^2 & 1 & \omega ^2 & 1 & \omega ^2 & 1 & \omega ^2 & 1
   \\
 D_{36} & 1 & \omega ^3 & \omega ^2 & \omega  & \omega ^2
   & \omega  & 1 & \omega ^3 & \omega ^2 & \omega  & 1 &
   \omega ^3 & \omega ^2 & \omega  & 1 & \omega ^3 &
   \omega ^2 & \omega  & 1 & \omega ^3 \\
 D_{37} & \omega ^3 & \omega ^3 & \omega ^3 & \omega ^3 &
   \omega ^2 & \omega ^2 & \omega ^2 & \omega ^2 & \omega
   ^3 & \omega ^3 & \omega ^3 & \omega ^3 & 1 & 1 & 1 & 1
   & \omega  & \omega  & \omega  & \omega  \\
 D_{38} & \omega ^3 & 1 & \omega  & \omega ^2 & \omega ^2
   & \omega ^3 & 1 & \omega  & \omega ^3 & 1 & \omega  &
   \omega ^2 & 1 & \omega  & \omega ^2 & \omega ^3 &
   \omega  & \omega ^2 & \omega ^3 & 1 \\
 D_{39} & \omega ^3 & \omega  & \omega ^3 & \omega  &
   \omega ^2 & 1 & \omega ^2 & 1 & \omega ^3 & \omega  &
   \omega ^3 & \omega  & 1 & \omega ^2 & 1 & \omega ^2 &
   \omega  & \omega ^3 & \omega  & \omega ^3 \\
 D_{40} & \omega ^3 & \omega ^2 & \omega  & 1 & \omega ^2
   & \omega  & 1 & \omega ^3 & \omega ^3 & \omega ^2 &
   \omega  & 1 & 1 & \omega ^3 & \omega ^2 & \omega  &
   \omega  & 1 & \omega ^3 & \omega ^2 \\
 D_{41} & \omega ^2 & \omega ^2 & \omega ^2 & \omega ^2 &
   \omega ^2 & \omega ^2 & \omega ^2 & \omega ^2 & 1 & 1
   & 1 & 1 & \omega ^2 & \omega ^2 & \omega ^2 & \omega
   ^2 & 1 & 1 & 1 & 1 \\
 D_{42} & \omega ^2 & \omega ^3 & 1 & \omega  & \omega ^2
   & \omega ^3 & 1 & \omega  & 1 & \omega  & \omega ^2 &
   \omega ^3 & \omega ^2 & \omega ^3 & 1 & \omega  & 1 &
   \omega  & \omega ^2 & \omega ^3 \\
 D_{43} & \omega ^2 & 1 & \omega ^2 & 1 & \omega ^2 & 1 &
   \omega ^2 & 1 & 1 & \omega ^2 & 1 & \omega ^2 & \omega
   ^2 & 1 & \omega ^2 & 1 & 1 & \omega ^2 & 1 & \omega ^2
   \\
 D_{44} & \omega ^2 & \omega  & 1 & \omega ^3 & \omega ^2
   & \omega  & 1 & \omega ^3 & 1 & \omega ^3 & \omega ^2
   & \omega  & \omega ^2 & \omega  & 1 & \omega ^3 & 1 &
   \omega ^3 & \omega ^2 & \omega  \\
 D_{45} & \omega  & \omega  & \omega  & \omega  & \omega
   ^2 & \omega ^2 & \omega ^2 & \omega ^2 & \omega  &
   \omega  & \omega  & \omega  & 1 & 1 & 1 & 1 & \omega
   ^3 & \omega ^3 & \omega ^3 & \omega ^3 \\
 D_{46} & \omega  & \omega ^2 & \omega ^3 & 1 & \omega ^2
   & \omega ^3 & 1 & \omega  & \omega  & \omega ^2 &
   \omega ^3 & 1 & 1 & \omega  & \omega ^2 & \omega ^3 &
   \omega ^3 & 1 & \omega  & \omega ^2 \\
 D_{47} & \omega  & \omega ^3 & \omega  & \omega ^3 &
   \omega ^2 & 1 & \omega ^2 & 1 & \omega  & \omega ^3 &
   \omega  & \omega ^3 & 1 & \omega ^2 & 1 & \omega ^2 &
   \omega ^3 & \omega  & \omega ^3 & \omega  \\
 D_{48} & \omega  & 1 & \omega ^3 & \omega ^2 & \omega ^2
   & \omega  & 1 & \omega ^3 & \omega  & 1 & \omega ^3 &
   \omega ^2 & 1 & \omega ^3 & \omega ^2 & \omega  &
   \omega ^3 & \omega ^2 & \omega  & 1 \\
 D_{49} & \omega ^2 & \omega ^2 & \omega ^2 & \omega ^2 &
   \omega  & \omega  & \omega  & \omega  & \omega  &
   \omega  & \omega  & \omega  & \omega  & \omega  &
   \omega  & \omega  & \omega  & \omega  & \omega  &
   \omega  \\
 D_{50} & \omega ^2 & \omega ^3 & 1 & \omega  & \omega  &
   \omega ^2 & \omega ^3 & 1 & \omega  & \omega ^2 &
   \omega ^3 & 1 & \omega  & \omega ^2 & \omega ^3 & 1 &
   \omega  & \omega ^2 & \omega ^3 & 1 \\
 D_{51} & \omega ^2 & 1 & \omega ^2 & 1 & \omega  &
   \omega ^3 & \omega  & \omega ^3 & \omega  & \omega ^3
   & \omega  & \omega ^3 & \omega  & \omega ^3 & \omega
   & \omega ^3 & \omega  & \omega ^3 & \omega  & \omega
   ^3 \\
 D_{52} & \omega ^2 & \omega  & 1 & \omega ^3 & \omega  &
   1 & \omega ^3 & \omega ^2 & \omega  & 1 & \omega ^3 &
   \omega ^2 & \omega  & 1 & \omega ^3 & \omega ^2 &
   \omega  & 1 & \omega ^3 & \omega ^2 \\
 D_{53} & \omega  & \omega  & \omega  & \omega  & \omega
   & \omega  & \omega  & \omega  & \omega ^2 & \omega ^2
   & \omega ^2 & \omega ^2 & \omega ^3 & \omega ^3 &
   \omega ^3 & \omega ^3 & 1 & 1 & 1 & 1 \\
 D_{54} & \omega  & \omega ^2 & \omega ^3 & 1 & \omega  &
   \omega ^2 & \omega ^3 & 1 & \omega ^2 & \omega ^3 & 1
   & \omega  & \omega ^3 & 1 & \omega  & \omega ^2 & 1 &
   \omega  & \omega ^2 & \omega ^3 \\
 D_{55} & \omega  & \omega ^3 & \omega  & \omega ^3 &
   \omega  & \omega ^3 & \omega  & \omega ^3 & \omega ^2
   & 1 & \omega ^2 & 1 & \omega ^3 & \omega  & \omega ^3
   & \omega  & 1 & \omega ^2 & 1 & \omega ^2 \\
 D_{56} & \omega  & 1 & \omega ^3 & \omega ^2 & \omega  &
   1 & \omega ^3 & \omega ^2 & \omega ^2 & \omega  & 1 &
   \omega ^3 & \omega ^3 & \omega ^2 & \omega  & 1 & 1 &
   \omega ^3 & \omega ^2 & \omega  \\
 D_{57} & 1 & 1 & 1 & 1 & \omega  & \omega  & \omega  &
   \omega  & \omega ^3 & \omega ^3 & \omega ^3 & \omega
   ^3 & \omega  & \omega  & \omega  & \omega  & \omega ^3
   & \omega ^3 & \omega ^3 & \omega ^3 \\
 D_{58} & 1 & \omega  & \omega ^2 & \omega ^3 & \omega  &
   \omega ^2 & \omega ^3 & 1 & \omega ^3 & 1 & \omega  &
   \omega ^2 & \omega  & \omega ^2 & \omega ^3 & 1 &
   \omega ^3 & 1 & \omega  & \omega ^2 \\
 D_{59} & 1 & \omega ^2 & 1 & \omega ^2 & \omega  &
   \omega ^3 & \omega  & \omega ^3 & \omega ^3 & \omega
   & \omega ^3 & \omega  & \omega  & \omega ^3 & \omega
   & \omega ^3 & \omega ^3 & \omega  & \omega ^3 & \omega
    \\
 D_{60} & 1 & \omega ^3 & \omega ^2 & \omega  & \omega  &
   1 & \omega ^3 & \omega ^2 & \omega ^3 & \omega ^2 &
   \omega  & 1 & \omega  & 1 & \omega ^3 & \omega ^2 &
   \omega ^3 & \omega ^2 & \omega  & 1 \\
 D_{61} & \omega ^3 & \omega ^3 & \omega ^3 & \omega ^3 &
   \omega  & \omega  & \omega  & \omega  & 1 & 1 & 1 & 1
   & \omega ^3 & \omega ^3 & \omega ^3 & \omega ^3 &
   \omega ^2 & \omega ^2 & \omega ^2 & \omega ^2 \\
 D_{62} & \omega ^3 & 1 & \omega  & \omega ^2 & \omega  &
   \omega ^2 & \omega ^3 & 1 & 1 & \omega  & \omega ^2 &
   \omega ^3 & \omega ^3 & 1 & \omega  & \omega ^2 &
   \omega ^2 & \omega ^3 & 1 & \omega  \\
 D_{63} & \omega ^3 & \omega  & \omega ^3 & \omega  &
   \omega  & \omega ^3 & \omega  & \omega ^3 & 1 & \omega
   ^2 & 1 & \omega ^2 & \omega ^3 & \omega  & \omega ^3 &
   \omega  & \omega ^2 & 1 & \omega ^2 & 1 \\
 D_{64} & \omega ^3 & \omega ^2 & \omega  & 1 & \omega  &
   1 & \omega ^3 & \omega ^2 & 1 & \omega ^3 & \omega ^2
   & \omega  & \omega ^3 & \omega ^2 & \omega  & 1 &
   \omega ^2 & \omega  & 1 & \omega ^3\\
 \hline
\end{array}
\end{equation}
}
\subsection{Character Table of the Group $\mathrm{G_{192}}$}
The group $\mathrm{G_{192}}$ has 20 conjugacy classes and therefore it has  $20$  irreducible representations that are distributed according to the following pattern:
\begin{description}
  \item[a)] 4 irreps of dimension $1$, namely $D_1,\dots,D_4$
  \item[b)] 12 irreps of dimension $3$, namely $D_5,\dots,D_{16}$
  \item[c)] 2 irreps of dimension $2$, namely $D_{17},D_{18}$
  \item[d)] 2 irreps of dimension $6$, namely $D_{19},D_{20}$
\end{description}
The character table is displayed below, where  by $\epsilon$ we have denoted the cubic root of unity $\epsilon = \exp\left[\frac{2\pi}{3} \, {\rm i}\right]$.
{\scriptsize
\begin{equation}\label{charto192}
\begin{array}{|l|llllllllllllllllllll|}
\hline
 0 & C_1 & C_2 & C_3 & C_4 & C_5 & C_6 & C_7 & C_8 & C_9 & C_{10} & C_{11} &
   C_{12} & C_{13} & C_{14} & C_{15} & C_{16} & C_{17} & C_{18} & C_{19} &
   C_{20} \\
   \hline
 D_1 & 1 & 1 & 1 & 1 & 1 & 1 & 1 & 1 & 1 & 1 & 1 & 1 & 1 & 1 & 1 & 1 & 1 & 1
   & 1 & 1 \\
 D_2 & 1 & 1 & 1 & 1 & 1 & 1 & 1 & 1 & 1 & 1 & -1 & -1 & -1 & -1 & -1 & -1 &
   -1 & -1 & 1 & 1 \\
 D_3 & 1 & -1 & -1 & 1 & 1 & -1 & 1 & -1 & -1 & 1 & 1 & -1 & -1 & 1 & 1 & -1
   & -1 & 1 & 1 & -1 \\
 D_4 & 1 & -1 & -1 & 1 & 1 & -1 & 1 & -1 & -1 & 1 & -1 & 1 & 1 & -1 & -1 & 1
   & 1 & -1 & 1 & -1 \\
 D_5 & 3 & -3 & -3 & 3 & -1 & 1 & -1 & 1 & 1 & -1 & -1 & 1 & 1 & -1 & 1 & -1
   & -1 & 1 & 0 & 0 \\
 D_6 & 3 & -3 & -3 & 3 & -1 & 1 & -1 & 1 & 1 & -1 & 1 & -1 & -1 & 1 & -1 & 1
   & 1 & -1 & 0 & 0 \\
 D_7 & 3 & 3 & 3 & 3 & -1 & -1 & -1 & -1 & -1 & -1 & -1 & -1 & -1 & -1 & 1 &
   1 & 1 & 1 & 0 & 0 \\
 D_8 & 3 & 3 & 3 & 3 & -1 & -1 & -1 & -1 & -1 & -1 & 1 & 1 & 1 & 1 & -1 & -1
   & -1 & -1 & 0 & 0 \\
 D_9 & 3 & 3 & -1 & -1 & -1 & 3 & 3 & -1 & -1 & -1 & -1 & 1 & -1 & 1 & 1 & 1
   & -1 & -1 & 0 & 0 \\
 D_{10} & 3 & 3 & -1 & -1 & -1 & 3 & 3 & -1 & -1 & -1 & 1 & -1 & 1 & -1 & -1
   & -1 & 1 & 1 & 0 & 0 \\
 D_{11} & 3 & -3 & 1 & -1 & -1 & -3 & 3 & 1 & 1 & -1 & -1 & -1 & 1 & 1 & 1 &
   -1 & 1 & -1 & 0 & 0 \\
 D_{12} & 3 & -3 & 1 & -1 & -1 & -3 & 3 & 1 & 1 & -1 & 1 & 1 & -1 & -1 & -1
   & 1 & -1 & 1 & 0 & 0 \\
 D_{13} & 3 & 3 & -1 & -1 & 3 & -1 & -1 & 3 & -1 & -1 & -1 & 1 & -1 & 1 & -1
   & -1 & 1 & 1 & 0 & 0 \\
 D_{14} & 3 & 3 & -1 & -1 & 3 & -1 & -1 & 3 & -1 & -1 & 1 & -1 & 1 & -1 & 1
   & 1 & -1 & -1 & 0 & 0 \\
 D_{15} & 3 & -3 & 1 & -1 & 3 & 1 & -1 & -3 & 1 & -1 & -1 & -1 & 1 & 1 & -1
   & 1 & -1 & 1 & 0 & 0 \\
 D_{16} & 3 & -3 & 1 & -1 & 3 & 1 & -1 & -3 & 1 & -1 & 1 & 1 & -1 & -1 & 1 &
   -1 & 1 & -1 & 0 & 0 \\
 D_{17} & 2 & 2 & 2 & 2 & 2 & 2 & 2 & 2 & 2 & 2 & 0 & 0 & 0 & 0 & 0 & 0 & 0
   & 0 & \epsilon  (\epsilon +1) & \epsilon  (\epsilon +1) \\
 D_{18} & 2 & -2 & -2 & 2 & 2 & -2 & 2 & -2 & -2 & 2 & 0 & 0 & 0 & 0 & 0 & 0
   & 0 & 0 & \epsilon  (\epsilon +1) & -\epsilon  (\epsilon +1) \\
 D_{19} & 6 & 6 & -2 & -2 & -2 & -2 & -2 & -2 & 2 & 2 & 0 & 0 & 0 & 0 & 0 &
   0 & 0 & 0 & 0 & 0 \\
 D_{20} & 6 & -6 & 2 & -2 & -2 & 2 & -2 & 2 & -2 & 2 & 0 & 0 & 0 & 0 & 0 & 0
   & 0 & 0 & 0 & 0\\
   \hline
\end{array}
\end{equation}
}
\subsection{Character Table of the Group $\mathrm{G_{96}}$}
The group $\mathrm{G_{96}}$ has 16 conjugacy classes and therefore it has  $16$  irreducible representations that are distributed according to the following pattern:
\begin{description}
  \item[a)] 6 irreps of dimension $1$, namely $D_1,\dots,D_6$
  \item[b)] 10 irreps of dimension $3$, namely $D_7,\dots,D_{16}$
\end{description}
The character table is displayed below, where  by $\epsilon$ we have denoted the cubic root of unity $\epsilon = \exp\left[\frac{2\pi}{3} \, {\rm i}\right]$.
{\scriptsize
\begin{equation}\label{charto96}
\begin{array}{|l|llllllllllllllll|}
\hline
 0 & C_1 & C_2 & C_3 & C_4 & C_5 & C_6 & C_7 & C_8 & C_9 & C_{10} & C_{11} &
   C_{12} & C_{13} & C_{14} & C_{15} & C_{16} \\
   \hline
 D_1 & 1 & 1 & 1 & 1 & 1 & 1 & 1 & 1 & 1 & 1 & 1 & 1 & 1 & 1 & 1 & 1 \\
 D_2 & 1 & -1 & -1 & 1 & 1 & -1 & -1 & 1 & -1 & 1 & 1 & -1 & 1 & -1 & 1 & -1
   \\
 D_3 & 1 & 1 & 1 & 1 & 1 & 1 & 1 & 1 & 1 & 1 & 1 & 1 & \epsilon  & \epsilon
   & \epsilon ^2 & \epsilon ^2 \\
 D_4 & 1 & -1 & -1 & 1 & 1 & -1 & -1 & 1 & -1 & 1 & 1 & -1 & \epsilon  &
   -\epsilon  & \epsilon ^2 & -\epsilon ^2 \\
 D_5 & 1 & 1 & 1 & 1 & 1 & 1 & 1 & 1 & 1 & 1 & 1 & 1 & \epsilon ^2 &
   \epsilon ^2 & \epsilon  & \epsilon  \\
 D_6 & 1 & -1 & -1 & 1 & 1 & -1 & -1 & 1 & -1 & 1 & 1 & -1 & \epsilon ^2 &
   -\epsilon ^2 & \epsilon  & -\epsilon  \\
 D_7 & 3 & -3 & -3 & 3 & -1 & 1 & 1 & -1 & 1 & -1 & -1 & 1 & 0 & 0 & 0 & 0
   \\
 D_8 & 3 & 3 & 3 & 3 & -1 & -1 & -1 & -1 & -1 & -1 & -1 & -1 & 0 & 0 & 0 & 0
   \\
 D_9 & 3 & 3 & -1 & -1 & -1 & -1 & 3 & -1 & -1 & 3 & -1 & -1 & 0 & 0 & 0 & 0
   \\
 D_{10} & 3 & -3 & 1 & -1 & -1 & 1 & -3 & -1 & 1 & 3 & -1 & 1 & 0 & 0 & 0 &
   0 \\
 D_{11} & 3 & 3 & -1 & -1 & -1 & 3 & -1 & -1 & -1 & -1 & 3 & -1 & 0 & 0 & 0
   & 0 \\
 D_{12} & 3 & -3 & 1 & -1 & -1 & -3 & 1 & -1 & 1 & -1 & 3 & 1 & 0 & 0 & 0 &
   0 \\
 D_{13} & 3 & 3 & -1 & -1 & 3 & -1 & -1 & -1 & -1 & -1 & -1 & 3 & 0 & 0 & 0
   & 0 \\
 D_{14} & 3 & -3 & 1 & -1 & 3 & 1 & 1 & -1 & 1 & -1 & -1 & -3 & 0 & 0 & 0 &
   0 \\
 D_{15} & 3 & 3 & -1 & -1 & -1 & -1 & -1 & 3 & 3 & -1 & -1 & -1 & 0 & 0 & 0
   & 0 \\
 D_{16} & 3 & -3 & 1 & -1 & -1 & 1 & 1 & 3 & -3 & -1 & -1 & 1 & 0 & 0 & 0 &
   0\\
   \hline
\end{array}
\end{equation}
}
\subsection{Character Table of the Group $\mathrm{G_{48}}$}
The group $\mathrm{G_{48}}$ has 8 conjugacy classes and therefore it has  $8$  irreducible representations that are distributed according to the following pattern:
\begin{description}
  \item[a)] 3 irreps of dimension $1$, namely $D_1,\dots,D_3$
  \item[b)] 5 irreps of dimension $3$, namely $D_4,\dots,D_{8}$
\end{description}
The character table is displayed below, where  by $\epsilon$ we have denoted the cubic root of unity $\epsilon = \exp\left[\frac{2\pi}{3} \, {\rm i}\right]$.
{
\begin{equation}\label{chartoG48}
\begin{array}{|l|llllllll|}
\hline
 0 & C_1 & C_2 & C_3 & C_4 & C_5 & C_6 & C_7 & C_8 \\
 \hline
 D_1 & 1 & 1 & 1 & 1 & 1 & 1 & 1 & 1 \\
 D_2 & 1 & 1 & 1 & 1 & 1 & 1 & \epsilon  & \epsilon ^2 \\
 D_3 & 1 & 1 & 1 & 1 & 1 & 1 & \epsilon ^2 & \epsilon  \\
 D_4 & 3 & 3 & -1 & -1 & -1 & -1 & 0 & 0 \\
 D_5 & 3 & -1 & -1 & -1 & 3 & -1 & 0 & 0 \\
 D_6 & 3 & -1 & -1 & -1 & -1 & 3 & 0 & 0 \\
 D_7 & 3 & -1 & 3 & -1 & -1 & -1 & 0 & 0 \\
 D_8 & 3 & -1 & -1 & 3 & -1 & -1 & 0 & 0\\
 \hline
\end{array}
\end{equation}
}
\subsection{Character Table of the Group $\mathrm{G_{16}}$}
The group $\mathrm{G_{16}}$ is abelian with order 16. Therefore it has  has 16  conjugacy classes and  $16$  one-dimensional irreducible representations.
The character table is displayed below.
{
\begin{equation}\label{chartoG16}
\begin{array}{|l|llllllllllllllll|}
\hline
 0 & C_1 & C_2 & C_3 & C_4 & C_5 & C_6 & C_7 & C_8 & C_9 & C_{10} & C_{11} &
   C_{12} & C_{13} & C_{14} & C_{15} & C_{16} \\
   \hline
 D_1 & 1 & 1 & 1 & 1 & 1 & 1 & 1 & 1 & 1 & 1 & 1 & 1 & 1 & 1 & 1 & 1 \\
 D_2 & 1 & 1 & 1 & 1 & 1 & 1 & 1 & 1 & -1 & -1 & -1 & -1 & -1 & -1 & -1 & -1
   \\
 D_3 & 1 & 1 & 1 & 1 & -1 & -1 & -1 & -1 & 1 & 1 & 1 & 1 & -1 & -1 & -1 & -1
   \\
 D_4 & 1 & 1 & 1 & 1 & -1 & -1 & -1 & -1 & -1 & -1 & -1 & -1 & 1 & 1 & 1 & 1
   \\
 D_5 & 1 & 1 & -1 & -1 & -1 & -1 & 1 & 1 & 1 & 1 & -1 & -1 & -1 & -1 & 1 & 1
   \\
 D_6 & 1 & 1 & -1 & -1 & -1 & -1 & 1 & 1 & -1 & -1 & 1 & 1 & 1 & 1 & -1 & -1
   \\
 D_7 & 1 & 1 & -1 & -1 & 1 & 1 & -1 & -1 & 1 & 1 & -1 & -1 & 1 & 1 & -1 & -1
   \\
 D_8 & 1 & 1 & -1 & -1 & 1 & 1 & -1 & -1 & -1 & -1 & 1 & 1 & -1 & -1 & 1 & 1
   \\
 D_9 & 1 & -1 & 1 & -1 & 1 & -1 & 1 & -1 & 1 & -1 & 1 & -1 & 1 & -1 & 1 & -1
   \\
 D_{10} & 1 & -1 & 1 & -1 & 1 & -1 & 1 & -1 & -1 & 1 & -1 & 1 & -1 & 1 & -1
   & 1 \\
 D_{11} & 1 & -1 & 1 & -1 & -1 & 1 & -1 & 1 & 1 & -1 & 1 & -1 & -1 & 1 & -1
   & 1 \\
 D_{12} & 1 & -1 & 1 & -1 & -1 & 1 & -1 & 1 & -1 & 1 & -1 & 1 & 1 & -1 & 1 &
   -1 \\
 D_{13} & 1 & -1 & -1 & 1 & -1 & 1 & 1 & -1 & 1 & -1 & -1 & 1 & -1 & 1 & 1 &
   -1 \\
 D_{14} & 1 & -1 & -1 & 1 & -1 & 1 & 1 & -1 & -1 & 1 & 1 & -1 & 1 & -1 & -1
   & 1 \\
 D_{15} & 1 & -1 & -1 & 1 & 1 & -1 & -1 & 1 & 1 & -1 & -1 & 1 & 1 & -1 & -1
   & 1 \\
 D_{16} & 1 & -1 & -1 & 1 & 1 & -1 & -1 & 1 & -1 & 1 & 1 & -1 & -1 & 1 & 1 &
   -1\\
   \hline
\end{array}
\end{equation}
}
\subsection{Character Table of the Group $\mathrm{GS_{32}}$}
The group $\mathrm{GS_{32}}$ has 14 conjugacy classes and therefore it has  $14$  irreducible representations that are distributed according to the following pattern:
\begin{description}
  \item[a)] 8 irreps of dimension $1$, namely $D_1,\dots,D_8$
  \item[b)] 6 irreps of dimension $2$, namely $D_9,\dots,D_{14}$
\end{description}
The character table is displayed below,
{
\begin{equation}\label{chartoGS32}
\begin{array}{|l|llllllllllllll|}
\hline
 0 & C_1 & C_2 & C_3 & C_4 & C_5 & C_6 & C_7 & C_8 & C_9 & C_{10} & C_{11} &
   C_{12} & C_{13} & C_{14} \\
 \hline
 D_1 & 1 & 1 & 1 & 1 & 1 & 1 & 1 & 1 & 1 & 1 & 1 & 1 & 1 & 1 \\
 D_2 & 1 & 1 & 1 & 1 & 1 & 1 & 1 & 1 & 1 & 1 & -1 & -1 & -1 & -1 \\
 D_3 & 1 & 1 & 1 & 1 & 1 & -1 & -1 & -1 & -1 & 1 & 1 & 1 & -1 & -1 \\
 D_4 & 1 & 1 & 1 & 1 & 1 & -1 & -1 & -1 & -1 & 1 & -1 & -1 & 1 & 1 \\
 D_5 & 1 & 1 & 1 & 1 & -1 & 1 & 1 & -1 & -1 & -1 & -1 & 1 & -1 & 1 \\
 D_6 & 1 & 1 & 1 & 1 & -1 & 1 & 1 & -1 & -1 & -1 & 1 & -1 & 1 & -1 \\
 D_7 & 1 & 1 & 1 & 1 & -1 & -1 & -1 & 1 & 1 & -1 & -1 & 1 & 1 & -1 \\
 D_8 & 1 & 1 & 1 & 1 & -1 & -1 & -1 & 1 & 1 & -1 & 1 & -1 & -1 & 1 \\
 D_9 & 2 & 2 & -2 & -2 & 2 & 0 & 0 & 0 & 0 & -2 & 0 & 0 & 0 & 0 \\
 D_{10} & 2 & 2 & -2 & -2 & -2 & 0 & 0 & 0 & 0 & 2 & 0 & 0 & 0 & 0 \\
 D_{11} & 2 & -2 & -2 & 2 & 0 & 0 & 0 & 2 & -2 & 0 & 0 & 0 & 0 & 0 \\
 D_{12} & 2 & -2 & 2 & -2 & 0 & 2 & -2 & 0 & 0 & 0 & 0 & 0 & 0 & 0 \\
 D_{13} & 2 & -2 & 2 & -2 & 0 & -2 & 2 & 0 & 0 & 0 & 0 & 0 & 0 & 0 \\
 D_{14} & 2 & -2 & -2 & 2 & 0 & 0 & 0 & -2 & 2 & 0 & 0 & 0 & 0 & 0\\
 \hline
\end{array}
\end{equation}
}
\subsection{Character Table of the Group $\mathrm{GP_{24}}$}
The group $\mathrm{GP_{24}}$ has 8 conjugacy classes and therefore it has  $8$  irreducible representations that are distributed according to the following pattern:
\begin{description}
  \item[a)] 6 irreps of dimension $1$, namely $D_1,\dots,D_6$
  \item[b)] 2 irreps of dimension $2$, namely $D_7,D_{8}$
\end{description}
The character table is displayed below,
{
\begin{equation}\label{chartoGP24}
\begin{array}{|l|llllllll|}
\hline
 0 & C_1 & C_2 & C_3 & C_4 & C_5 & C_6 & C_7 & C_8 \\
 \hline
 D_1 & 1 & 1 & 1 & 1 & 1 & 1 & 1 & 1 \\
 D_2 & 1 & 1 & 1 & 1 & (-1)^{2/3} & (-1)^{2/3} & -\sqrt[3]{-1} &
   -\sqrt[3]{-1} \\
 D_3 & 1 & 1 & 1 & 1 & -\sqrt[3]{-1} & -\sqrt[3]{-1} & (-1)^{2/3} &
   (-1)^{2/3} \\
 D_4 & 1 & -1 & -1 & 1 & -1 & 1 & -1 & 1 \\
 D_5 & 1 & -1 & -1 & 1 & -(-1)^{2/3} & (-1)^{2/3} & \sqrt[3]{-1} &
   -\sqrt[3]{-1} \\
 D_6 & 1 & -1 & -1 & 1 & \sqrt[3]{-1} & -\sqrt[3]{-1} & -(-1)^{2/3} &
   (-1)^{2/3} \\
 D_7 & 3 & 3 & -1 & -1 & 0 & 0 & 0 & 0 \\
 D_8 & 3 & -3 & 1 & -1 & 0 & 0 & 0 & 0\\
 \hline
\end{array}
\end{equation}
}
\subsection{Character Table of the Group $\mathrm{Oh_{48}}$}
The group $\mathrm{Oh_{48}}$ is isomorphic to the extended octahedral  group and has 10 conjugacy classes.  Therefore it has  $10$  irreducible representations that are distributed according to the following pattern:
\begin{description}
  \item[a)] 4 irreps of dimension $1$, namely $D_1,\dots,D_4$
  \item[b)] 2 irreps of dimension $2$, namely $D_5,D_{6}$
  \item[c)] 4 irreps of dimension $3$, namely $D_7,\dots D_{10}$
\end{description}
The character table is displayed below,
{
\begin{equation}\label{chartoOh48}
\begin{array}{|l|llllllllll|}
\hline
 0 & C_1 & C_2 & C_3 & C_4 & C_5 & C_6 & C_7 & C_8 & C_9 & C_{10} \\
 \hline
 D_1 & 1 & 1 & 1 & 1 & 1 & 1 & 1 & 1 & 1 & 1 \\
 D_2 & 1 & 1 & 1 & 1 & 1 & -1 & -1 & -1 & -1 & -1 \\
 D_3 & 1 & 1 & 1 & -1 & -1 & 1 & 1 & 1 & -1 & -1 \\
 D_4 & 1 & 1 & 1 & -1 & -1 & -1 & -1 & -1 & 1 & 1 \\
 D_5 & 2 & -1 & 2 & 0 & 0 & 2 & -1 & 2 & 0 & 0 \\
 D_6 & 2 & -1 & 2 & 0 & 0 & -2 & 1 & -2 & 0 & 0 \\
 D_7 & 3 & 0 & -1 & -1 & 1 & 3 & 0 & -1 & -1 & 1 \\
 D_8 & 3 & 0 & -1 & -1 & 1 & -3 & 0 & 1 & 1 & -1 \\
 D_9 & 3 & 0 & -1 & 1 & -1 & 3 & 0 & -1 & 1 & -1 \\
 D_{10} & 3 & 0 & -1 & 1 & -1 & -3 & 0 & 1 & -1 & 1\\
 \hline
\end{array}
\end{equation}
}
\subsection{Character Table of the Group $\mathrm{GS_{24}}$}
The group $\mathrm{GS_{24}}$ is isomorphic to the proper octahedral  group $\mathrm{O_{24}}$ and has 5 conjugacy classes.  Therefore it has  $5$  irreducible representations that are distributed according to the following pattern:
\begin{description}
  \item[a)] 2 irreps of dimension $1$, namely $D_1,\dots,D_2$
  \item[b)] 1 irrep of dimension $2$, namely $D_3$
  \item[c)] 2 irreps of dimension $3$, namely $D_4,D_{5}$
\end{description}
The character table is displayed below,
{
\begin{equation}\label{GS24}
\begin{array}{|l|lllll|}
\hline
 0 & C_1 & C_2 & C_3 & C_4 & C_5 \\
 \hline
 D_1 & 1 & 1 & 1 & 1 & 1 \\
 D_2 & 1 & 1 & 1 & -1 & -1 \\
 D_3 & 2 & -1 & 2 & 0 & 0 \\
 D_4 & 3 & 0 & -1 & -1 & 1 \\
 D_5 & 3 & 0 & -1 & 1 & -1\\
 \hline
\end{array}
\end{equation}
}
\section{Classification of the momentum vectors and of the corresponding $\mathrm{G}_{1536}$ \textit{irreps} }
\label{salamicrudi}
In this section we list the results obtained by means of a MATHEMATICA computer code relative to decomposition into \textit{irreps} of the representations of the group $\mathrm{G}_{1536}$ generated by the various octahedral  group orbits of momentum vectors.
We find that there are five types of momentum vectors on the lattice:
\begin{description}
  \item[A)] Momenta of type $\{a,0,0\}$ which generate representations of the universal group $\mathrm{G}_{1536}$ of dimensions 6.
  \item[B)]Momenta of type $\{a,a,a\}$ which generate representations of the universal group $\mathrm{G}_{1536}$ of dimensions 8.
  \item[C)] Momenta of type $\{a,a,0\}$ which generate representations of the universal group $\mathrm{G}_{1536}$ of dimensions 12
  \item[D)] Momenta of type $\{a,a,b\}$ which generate representations of the universal group $\mathrm{G}_{1536}$ of dimensions 24
  \item[E)] Momenta of type $\{a,b,c\}$ which generate representations of the universal group $\mathrm{G}_{1536}$ of dimensions 48
\end{description}
In each of the five groups one still has  to reduce the entries to $\mathbb{Z}_4$, namely to consider their equivalence class $\mathrm{mod}\,4$. Each different choice of the pattern of  $\mathbb{Z}_4$ classes appearing in an orbit leads to different decomposition of the representation into irreducible representation of $\mathrm{G}_{1536}$. A simple consideration of the combinatorics leads to the conclusion that there are in total $48$ cases to be considered. The very significant result is that all of the $37$ irreducible representations of $\mathrm{G}_{1536}$ appears at least once in the list of these decompositions. Hence for all the \textit{irrepses} of this group one can find a corresponding Beltrami field and for some \textit{irrepses} such a Beltrami field admits a few inequivalent realizations.
\par
In the sections below we list the decompositions of  the representations generated by all of the $48$ classes of momentum vectors.
The numbers $4\mu$, $4\nu$, $4\rho$ with $\mu,\nu,\rho\, = \, 0,\pm 1,\pm 2, \dots$ represent the equivalence class of $\mathbb{Z}_4$ integers.
\subsection{  Classes of momentum vectors yielding orbits of length 6: $\{$a,0,0$\}$}
\label{mortadella6}
\noindent\(
\pmb{}\\
\pmb{\text{ Class of the momentum vector  = } \{0,0, 1+4 \rho \}}\\
\pmb{\text{ Dimension of the $\mathrm{G}_{1536}$  representation  = } 6}\\
\pmb{\text{ Orbit   = } D_{23}[\mathrm{G}_{1536},6]}\\
\pmb{\text{ Class of the momentum vector  = } \{0,0, 2+4 \rho \}}\\
\pmb{\text{ Dimension of the $\mathrm{G}_{1536}$  representation  = } 6}\\
\pmb{\text{ Orbit   = } D_{19}[\mathrm{G}_{1536},6]}\\
\pmb{\text{ Class of the momentum vector  = } \{0,0, 3+4 \rho \}}\\
\pmb{\text{ Dimension of the $\mathrm{G}_{1536}$  representation  = } 6}\\
\pmb{\text{ Orbit   = } D_{24}[\mathrm{G}_{1536},6]}\\
\pmb{\text{ Class of the momentum vector  = } \{0,0, 4+4 \rho \}}\\
\pmb{\text{ Dimension of the $\mathrm{G}_{1536}$  representation  = } 6}\\
\pmb{\text{ Orbit   = } D_7[\mathrm{G}_{1536},3]+D_8[\mathrm{G}_{1536},3]}\\
\pmb{ }\)
\subsection{Classes of momentum vectors yielding orbits of length 8: $\{$a,a,a$\}$}
\noindent\(
\pmb{}\\
\pmb{\text{ Class of the momentum vector  = } \left\{1+4\mu ,1+4 \mu ,1+4 \mu \right\}}\\
\pmb{\text{ Dimension of the $\mathrm{G}_{1536}$  representation  = } 8}\\
\pmb{\text{ Orbit   = } D_{30}[\mathrm{G}_{1536},8]}\\
\pmb{\text{ Class of the momentum vector  = } \left\{2+4 \mu ,2+4 \mu ,2+4 \mu \right\}}\\
\pmb{\text{ Dimension of the $\mathrm{G}_{1536}$  representation  = } 8}\\
\pmb{\text{ Orbit   = } D_6[\mathrm{G}_{1536},2]+D_{17}[\mathrm{G}_{1536},3]+D_{18}[\mathrm{G}_{1536},3]}\\
\pmb{\text{ Class of the momentum vector  = } \left\{3+4 \mu ,3+4 \mu ,3+4 \mu \right\}}\\
\pmb{\text{ Dimension of the $\mathrm{G}_{1536}$  representation  = } 8}\\
\pmb{\text{ Orbit   = } D_{31}[\mathrm{G}_{1536},8]}\\
\pmb{\text{ Class of the momentum vector  = } \left\{4+4 \mu ,4+4 \mu ,4+4 \mu \right\}}\\
\pmb{\text{ Dimension of the $\mathrm{G}_{1536}$  representation  = } 8}\\
\pmb{\text{ Orbit   = } D_5[\mathrm{G}_{1536},2]+D_7[\mathrm{G}_{1536},3]+D_8[\mathrm{G}_{1536},3]}\\
\pmb{ }\)
\subsection{Classes of momentum vectors yielding orbits of length 12: $\{$a,a,0$\}$}
\noindent\(
\pmb{}\\
\pmb{\text{ Class of the momentum vector  = } \left\{0 ,1+4 \nu ,1+4 \nu \right\}}\\
\pmb{\text{ Dimension of the $\mathrm{G}_{1536}$  representation  = } 12}\\
\pmb{\text{ Orbit   = } D_{32}[\mathrm{G}_{1536},12]}\\
\pmb{\text{ Class of the momentum vector  = } \left\{0 ,2+4 \nu ,2+4 \nu \right\}}\\
\pmb{\text{ Dimension of the $\mathrm{G}_{1536}$  representation  = } 12}\\
\pmb{\text{ Orbit   = } D_{13}[\mathrm{G}_{1536},3]+D_{15}[\mathrm{G}_{1536},3]+D_{20}[\mathrm{G}_{1536},6]}\\
\pmb{\text{ Class of the momentum vector  = } \left\{0 ,3+4 \nu ,3+4 \nu \right\}}\\
\pmb{\text{ Dimension of the $\mathrm{G}_{1536}$  representation  = } 12}\\
\pmb{\text{ Orbit   = } D_{32}[\mathrm{G}_{1536},12]}\\
\pmb{\text{ Class of the momentum vector  = } \left\{0 ,4+4 \nu ,4+4 \nu \right\}}\\
\pmb{\text{ Dimension of the $\mathrm{G}_{1536}$  representation  = } 12}\\
\pmb{\text{ Orbit   = } D_2[\mathrm{G}_{1536},1]+D_5[\mathrm{G}_{1536},2]+D_7[\mathrm{G}_{1536},3]+2
D_8[\mathrm{G}_{1536},3]}\\
\pmb{ }\)
\subsection{Classes of momentum vectors yielding orbits of length 24: $\{$a,a,b$\}$}
\noindent\(
\pmb{}\\
\pmb{\text{ Class of the momentum vector  = } \left\{1+4  \mu ,1+4 \mu ,2+4 \rho \right\}}\\
\pmb{\text{ Dimension of the $\mathrm{G}_{1536}$  representation  = } 24}\\
\pmb{\text{ Orbit   = } D_{34}[\mathrm{G}_{1536},12]+D_{35}[\mathrm{G}_{1536},12]}\\
\pmb{\text{ Class of the momentum vector  = } \left\{1+4 \mu ,1+4 \mu ,3+4 \rho \right\}}\\
\pmb{\text{ Dimension of the $\mathrm{G}_{1536}$  representation  = } 24}\\
\pmb{\text{ Orbit   = } D_{29}[\mathrm{G}_{1536},8]+D_{30}[\mathrm{G}_{1536},8]+D_{31}[\mathrm{G}_{1536},8]}\\
\pmb{\text{ Class of the momentum vector  = } \left\{1+4 \mu ,1+4 \mu ,4+4 \rho \right\}}\\
\pmb{\text{ Dimension of the $\mathrm{G}_{1536}$  representation  = } 24}\\
\pmb{\text{ Orbit   = } D_{32}[\mathrm{G}_{1536},12]+D_{33}[\mathrm{G}_{1536},12]}\\
\pmb{\text{ Class of the momentum vector  = } \left\{1+4\mu ,1+4 \mu ,5+4 \rho \right\}}\\
\pmb{\text{ Dimension of the $\mathrm{G}_{1536}$  representation  = } 24}\\
\pmb{\text{ Orbit   = } D_{29}[\mathrm{G}_{1536},8]+D_{30}[\mathrm{G}_{1536},8]+D_{31}[\mathrm{G}_{1536},8]}\\
\pmb{\text{ Class of the momentum vector  = } \{1+4 \mu ,2+4 \mu ,2+4 \rho \}}\\
\pmb{\text{ Dimension of the $\mathrm{G}_{1536}$  representation  = } 24}\\
\pmb{\text{ Orbit   = } D_{25}[\mathrm{G}_{1536},6]+D_{26}[\mathrm{G}_{1536},6]+D_{27}[\mathrm{G}_{1536},6]+D_{28}[\mathrm{G}_{1536},6]}\\
\pmb{\text{ Class of the momentum vector  = } \left\{2+4\mu ,2+4 \mu ,6+4 \rho \right\}}\\
\pmb{\text{ Dimension of the $\mathrm{G}_{1536}$  representation  = } 24}\\
\pmb{\text{ Orbit   = } D_3[\mathrm{G}_{1536},1]+D_4[\mathrm{G}_{1536},1]+2 D_6[\mathrm{G}_{1536},2]+3
D_{17}[\mathrm{G}_{1536},3]+3 D_{18}[\mathrm{G}_{1536},3]}\\
\pmb{\text{ Class of the momentum vector  = } \left\{2+4  \mu ,2+4 \mu ,3+4 \rho \right\}}\\
\pmb{\text{ Dimension of the $\mathrm{G}_{1536}$  representation  = } 24}\\
\pmb{\text{ Orbit   = } D_{25}[\mathrm{G}_{1536},6]+D_{26}[\mathrm{G}_{1536},6]+D_{27}[\mathrm{G}_{1536},6]+D_{28}[\mathrm{G}_{1536},6]}\\
\pmb{\text{ Class of the momentum vector  = } \left\{2+4\mu ,2+4 \mu ,4+4 \rho \right\}}\\
\pmb{\text{ Dimension of the $\mathrm{G}_{1536}$  representation  = } 24}\\
\pmb{\text{ Orbit   = } D_{13}[\mathrm{G}_{1536},3]+D_{14}[\mathrm{G}_{1536},3]+D_{15}[\mathrm{G}_{1536},3]+D_{16}[\mathrm{G}_{1536},3]+2
D_{20}[\mathrm{G}_{1536},6]}\\
\pmb{\text{ Class of the momentum vector  = } \left\{1+4 \mu ,3+4 \mu ,3+4 \rho \right\}}\\
\pmb{\text{ Dimension of the $\mathrm{G}_{1536}$  representation  = } 24}\\
\pmb{\text{ Orbit   = } D_{29}[\mathrm{G}_{1536},8]+D_{30}[\mathrm{G}_{1536},8]+D_{31}[\mathrm{G}_{1536},8]}\\
\pmb{\text{ Class of the momentum vector  = } \left\{2+4 \mu ,3+4 \mu ,3+4 \rho \right\}}\\
\pmb{\text{ Dimension of the $\mathrm{G}_{1536}$  representation  = } 24}\\
\pmb{\text{ Orbit   = } D_{34}[\mathrm{G}_{1536},12]+D_{35}[\mathrm{G}_{1536},12]}\\
\pmb{\text{ Class of the momentum vector  = } \left\{3+4  \mu ,3+4 \mu ,7+4 \rho \right\}}\\
\pmb{\text{ Dimension of the $\mathrm{G}_{1536}$  representation  = } 24}\\
\pmb{\text{ Orbit   = } D_{29}[\mathrm{G}_{1536},8]+D_{30}[\mathrm{G}_{1536},8]+D_{31}[\mathrm{G}_{1536},8]}\\
\pmb{\text{ Class of the momentum vector  = } \left\{1+4  \mu ,4+4 \mu ,4+4 \rho \right\}}\\
\pmb{\text{ Dimension of the $\mathrm{G}_{1536}$  representation  = } 24}\\
\pmb{\text{ Orbit   = } D_{21}[\mathrm{G}_{1536},6]+D_{22}[\mathrm{G}_{1536},6]+D_{23}[\mathrm{G}_{1536},6]+D_{24}[\mathrm{G}_{1536},6]}\\
\pmb{\text{ Class of the momentum vector  = } \{2+4 \mu ,4+4 \mu ,4+4 \rho \}}\\
\pmb{\text{ Dimension of the $\mathrm{G}_{1536}$  representation  = } 24}\\
\pmb{\text{ Orbit   = } D_9[\mathrm{G}_{1536},3]+D_{10}[\mathrm{G}_{1536},3]+D_{11}[\mathrm{G}_{1536},3]+D_{12}[\mathrm{G}_{1536},3]+2
D_{19}[\mathrm{G}_{1536},6]}\\
\pmb{\text{ Class of the momentum vector  = } \left\{3+4  \mu ,4+4 \mu ,4+4 \rho \right\}}\\
\pmb{\text{ Dimension of the $\mathrm{G}_{1536}$  representation  = } 24}\\
\pmb{\text{ Orbit   = } D_{21}[\mathrm{G}_{1536},6]+D_{22}[\mathrm{G}_{1536},6]+D_{23}[\mathrm{G}_{1536},6]+D_{24}[\mathrm{G}_{1536},6]}\\
\pmb{\text{ Class of the momentum vector  = } \left\{4+4 \mu ,4+4 \mu ,8+4 \rho \right\}}\\
\pmb{\text{ Dimension of the $\mathrm{G}_{1536}$  representation  = } 24}\\
\pmb{\text{ Orbit   = } D_1[\mathrm{G}_{1536},1]+D_2[\mathrm{G}_{1536},1]+2 D_5[\mathrm{G}_{1536},2]+3
D_7[\mathrm{G}_{1536},3]+3 D_8[\mathrm{G}_{1536},3]}\\
\pmb{\text{ Class of the momentum vector  = } \left\{3+4  \mu ,3+4 \mu ,4+4 \rho \right\}}\\
\pmb{\text{ Dimension of the $\mathrm{G}_{1536}$  representation  = } 24}\\
\pmb{\text{ Orbit   = } D_{32}[\mathrm{G}_{1536},12]+D_{33}[\mathrm{G}_{1536},12]}\\
\pmb{ }\)
\subsection{Classes of momentum vectors yielding point orbits of length 24 and $\mathrm{G_{1536}}$ representations of dimensions 48: $\{$a,b,c$\}$}
\noindent\(
\pmb{}\\
\pmb{\text{ Class of the momentum vector  = } \left\{4+4 \mu ,8+4 \nu ,12+4 \rho \right\}}\\
\pmb{\text{ Dimension of the $\mathrm{G}_{1536}$  representation  = } 48}\\
\pmb{\text{ Orbit   = } 2 D_1[\mathrm{G}_{1536},1]+2 D_2[\mathrm{G}_{1536},1]+4 D_5[\mathrm{G}_{1536},2]+6
D_7[\mathrm{G}_{1536},3]+6 D_8[\mathrm{G}_{1536},3]}\\
\pmb{\text{ Class of the momentum vector  = } \{1+4 \mu ,4+4 \nu ,8+4 \rho \}}\\
\pmb{\text{ Dimension of the $\mathrm{G}_{1536}$  representation  = } 48}\\
\pmb{\text{ Orbit   = } 2 D_{21}[\mathrm{G}_{1536},6]+2 D_{22}[\mathrm{G}_{1536},6]+2
D_{23}[\mathrm{G}_{1536},6]+2 D_{24}[\mathrm{G}_{1536},6]}\\
\pmb{\text{ Class of the momentum vector  = } \left\{2+4 \mu ,4+4 \nu ,8+4 \rho \right\}}\\
\pmb{\text{ Dimension of the $\mathrm{G}_{1536}$  representation  = } 48}\\
\pmb{\text{ Orbit   = } 2 D_9[\mathrm{G}_{1536},3]+2 D_{10}[\mathrm{G}_{1536},3]+2 D_{11}[\mathrm{G}_{1536},3]+2
D_{12}[\mathrm{G}_{1536},3]+4 D_{19}[\mathrm{G}_{1536},6]}\\
\pmb{\text{ Class of the momentum vector  = } \left\{3+4  \mu ,4+4 \nu ,8+4 \rho \right\}}\\
\pmb{\text{ Dimension of the $\mathrm{G}_{1536}$  representation  = } 48}\\
\pmb{\text{ Orbit   = } 2 D_{21}[\mathrm{G}_{1536},6]+2 D_{22}[\mathrm{G}_{1536},6]+2
D_{23}[\mathrm{G}_{1536},6]+2 D_{24}[\mathrm{G}_{1536},6]}\\
\pmb{\text{ Class of the momentum vector  = } \left\{1+4 \mu ,2+4 \nu ,4+4 \rho \right\}}\\
\pmb{\text{ Dimension of the $\mathrm{G}_{1536}$  representation  = } 48}\\
\pmb{\text{ Orbit   = } 2 D_{36}[\mathrm{G}_{1536},12]+2 D_{37}[\mathrm{G}_{1536},12]}\\
\pmb{\text{ Class of the momentum vector  = } \left\{1+4 \mu ,3+4 \nu ,4+4 \rho \right\}}\\
\pmb{\text{ Dimension of the $\mathrm{G}_{1536}$  representation  = } 48}\\
\pmb{\text{ Orbit   = } 2 D_{32}[\mathrm{G}_{1536},12]+2 D_{33}[\mathrm{G}_{1536},12]}\\
\pmb{\text{ Class of the momentum vector  = } \left\{2+4 \mu ,4+4 \nu ,6+4 \rho \right\}}\\
\pmb{\text{ Dimension of the $\mathrm{G}_{1536}$  representation  = } 48}\\
\pmb{\text{ Orbit   = } 2 D_{13}[\mathrm{G}_{1536},3]+2 D_{14}[\mathrm{G}_{1536},3]+2
D_{15}[\mathrm{G}_{1536},3]+2 D_{16}[\mathrm{G}_{1536},3]+4 D_{20}[\mathrm{G}_{1536},6]}\\
\pmb{\text{ Class of the momentum vector  = } \left\{2+4  \mu ,3+4 \nu ,4+4 \rho \right\}}\\
\pmb{\text{ Dimension of the $\mathrm{G}_{1536}$  representation  = } 48}\\
\pmb{\text{ Orbit   = } 2 D_{36}[\mathrm{G}_{1536},12]+2 D_{37}[\mathrm{G}_{1536},12]}\\
\pmb{\text{ Class of the momentum vector  = } \left\{1+4 \mu ,5+4 \nu ,9+4 \rho \right\}}\\
\pmb{\text{ Dimension of the $\mathrm{G}_{1536}$  representation  = } 48}\\
\pmb{\text{ Orbit   = } 2 D_{29}[\mathrm{G}_{1536},8]+2 D_{30}[\mathrm{G}_{1536},8]+2
D_{31}[\mathrm{G}_{1536},8]}\\
\pmb{\text{ Class of the momentum vector  = } \left\{1+4  \mu ,2+4 \nu ,5+4 \rho \right\}}\\
\pmb{\text{ Dimension of the $\mathrm{G}_{1536}$  representation  = } 48}\\
\pmb{\text{ Orbit   = } 2 D_{34}[\mathrm{G}_{1536},12]+2 D_{35}[\mathrm{G}_{1536},12]}\\
\pmb{\text{ Class of the momentum vector  = } \left\{1+4  \mu ,3+4 \nu ,5+4 \rho \right\}}\\
\pmb{\text{ Dimension of the $\mathrm{G}_{1536}$  representation  = } 48}\\
\pmb{\text{ Orbit   = } 2 D_{29}[\mathrm{G}_{1536},8]+2 D_{30}[\mathrm{G}_{1536},8]+2
D_{31}[\mathrm{G}_{1536},8]}\\
\pmb{\text{ Class of the momentum vector  = } \left\{1+4  \mu ,2+4 \nu ,6+4 \rho \right\}}\\
\pmb{\text{ Dimension of the $\mathrm{G}_{1536}$  representation  = } 48}\\
\pmb{\text{ Orbit   = } 2 D_{25}[\mathrm{G}_{1536},6]+2 D_{26}[\mathrm{G}_{1536},6]+2
D_{27}[\mathrm{G}_{1536},6]+2 D_{28}[\mathrm{G}_{1536},6]}\\
\pmb{\text{ Class of the momentum vector  = } \left\{1+4 \mu ,2+4 \nu ,3+4 \rho \right\}}\\
\pmb{\text{ Dimension of the $\mathrm{G}_{1536}$  representation  = } 48}\\
\pmb{\text{ Orbit   = } 2 D_{34}[\mathrm{G}_{1536},12]+2 D_{35}[\mathrm{G}_{1536},12]}\\
\pmb{\text{ Class of the momentum vector  = } \left\{1+4 \mu ,3+4 \nu ,7+4 \rho \right\}}\\
\pmb{\text{ Dimension of the $\mathrm{G}_{1536}$  representation  = } 48}\\
\pmb{\text{ Orbit   = } 2 D_{29}[\mathrm{G}_{1536},8]+2 D_{30}[\mathrm{G}_{1536},8]+2
D_{31}[\mathrm{G}_{1536},8]}\\
\pmb{\text{ Class of the momentum vector  = } \left\{2+4 \mu ,6+4 \nu ,10+4 \rho \right\}}\\
\pmb{\text{ Dimension of the $\mathrm{G}_{1536}$  representation  = } 48}\\
\pmb{\text{ Orbit   = } 2 D_3[\mathrm{G}_{1536},1]+2 D_4[\mathrm{G}_{1536},1]+4 D_6[\mathrm{G}_{1536},2]+6
D_{17}[\mathrm{G}_{1536},3]+6 D_{18}[\mathrm{G}_{1536},3]}\\
\pmb{\text{ Class of the momentum vector  = } \{2+4 \mu ,3+4 \nu ,6+4 \rho \}}\\
\pmb{\text{ Dimension of the $\mathrm{G}_{1536}$  representation  = } 48}\\
\pmb{\text{ Orbit   = } 2 D_{25}[\mathrm{G}_{1536},6]+2 D_{26}[\mathrm{G}_{1536},6]+2
D_{27}[\mathrm{G}_{1536},6]+2 D_{28}[\mathrm{G}_{1536},6]}\\
\pmb{\text{ Class of the momentum vector  = } \left\{2+4 \mu ,3+4 \nu ,7+4 \rho \right\}}\\
\pmb{\text{ Dimension of the $\mathrm{G}_{1536}$  representation  = } 48}\\
\pmb{\text{ Orbit   = } 2 D_{34}[\mathrm{G}_{1536},12]+2 D_{35}[\mathrm{G}_{1536},12]}\\
\pmb{\text{ Class of the momentum vector  = } \left\{3+4  \mu ,7+4 \nu ,11+4 \rho \right\}}\\
\pmb{\text{ Dimension of the $\mathrm{G}_{1536}$  representation  = } 48}\\
\pmb{\text{ Orbit   = } 2 D_{29}[\mathrm{G}_{1536},8]+2 D_{30}[\mathrm{G}_{1536},8]+2
D_{31}[\mathrm{G}_{1536},8]}\\
\pmb{\text{ Class of the momentum vector  = } \left\{1+4 \mu ,4+4 \nu ,5+4 \rho \right\}}\\
\pmb{\text{ Dimension of the $\mathrm{G}_{1536}$  representation  = } 48}\\
\pmb{\text{ Orbit   = } 2 D_{32}[\mathrm{G}_{1536},12]+2 D_{33}[\mathrm{G}_{1536},12]}\\
\pmb{\text{ Class of the momentum vector  = } \left\{3+4  \mu ,4+4 \nu ,7+4 \rho \right\}}\\
\pmb{\text{ Dimension of the $\mathrm{G}_{1536}$  representation  = } 48}\\
\pmb{\text{ Orbit   = } 2 D_{32}[\mathrm{G}_{1536},12]+2 D_{33}[\mathrm{G}_{1536},12]}\\
\pmb{ }\)
\section{Branching rules of $\mathrm{G}_{1536}$ irreps}
\label{brancicardo}
In this section we present the branching rules of all of the 37 irreducible representations of the Universal Classifying Group $\mathrm{G}_{1536}$ with respect to all of its 16 subgroups mentioned in Appendix \ref{descriptio}, namely:
\begin{equation}\label{pidocchioso}
  \begin{array}{rlcrlcrlcrl}
     1) & \mathrm{G}_{768} & ; & 5) & \mathrm{G}_{192} & ; & 9) &\mathrm{GF}_{192}  & ; & 13) & \mathrm{GS}_{24} \\
     2) & \mathrm{G}_{256} & ; & 6) & \mathrm{G}_{96} & ; & 10) & \mathrm{GF}_{96} & ; & 14) & \mathrm{GP}_{24} \\
     3) & \mathrm{G}_{128} & ; & 7) & \mathrm{G}_{48} & ; & 11) & \mathrm{GF}_{48} & ; & 15) & \mathrm{O}_{24} \\
     4) & \mathrm{G}_{64} & ; & 8) & \mathrm{G}_{16} & ; & 12) & \mathrm{GS}_{32} & ; & 16) & \mathrm{Oh}_{48}
   \end{array}
\end{equation}
The information contained in the following tables adjoined with the information provided in the tables of appendix \ref{salamicrudi} allows to spot all cases of Beltrami vector fields invariant under some group of symmetries including (or not including translations), namely the appearance of a $D_1(H,1)$ representation for some subgroup $H\subset \mathrm{G}_{1536}$ in the branching rule of some $D_x(\mathrm{G}_{1536},y)$ irreducible representation produced in one of the momentum vector orbits classified in appendix \ref{salamicrudi}. Looking at those tables one realizes that all 37 irreps of $\mathrm{G}_{1536}$ appear at least once. Hence any identity representation appearing in any of the following branching rules corresponds to an existing $H$-invariant Beltrami vector field.
\subsection{Branching rules of the irreprs of dimension 1 and 2}
\label{1e2brancione}
\noindent\(D_1\left[\mathrm{G}_{1536},1\right]\text{ = }D_1[\text{$\mathrm{G}_{16}$},1]\)\\
\noindent\(D_1\left[\mathrm{G}_{1536},1\right]\text{ = }D_1[\text{$\mathrm{G}_{48}$},1]\)\\
\noindent\(D_1\left[\mathrm{G}_{1536},1\right]\text{ = }D_1[\text{$\mathrm{G}_{64}$},1]\)\\
\noindent\(D_1\left[\mathrm{G}_{1536},1\right]\text{ = }D_1[\text{$\mathrm{G}_{96}$},1]\)\\
\noindent\(D_1\left[\mathrm{G}_{1536},1\right]\text{ = }D_1[\text{$\mathrm{G}_{128}$},1]\)\\
\noindent\(D_1\left[\mathrm{G}_{1536},1\right]\text{ = }D_1[\text{$\mathrm{G}_{192}$},1]\)\\
\noindent\(D_1\left[\mathrm{G}_{1536},1\right]\text{ = }D_1[\text{$\mathrm{G}_{256}$},1]\)\\
\noindent\(D_1\left[\mathrm{G}_{1536},1\right]\text{ = }D_1[\text{$\mathrm{G}_{768}$},1]\)\\
\noindent\(D_1\left[\mathrm{G}_{1536},1\right]\text{ = }D_1[\text{$\mathrm{GF}_{48}$},1]\)\\
\noindent\(D_1\left[\mathrm{G}_{1536},1\right]\text{ = }D_1[\text{$\mathrm{GF}_{192}$},1]\)\\
\noindent\(D_1\left[\mathrm{G}_{1536},1\right]\text{ = }D_1[\text{$\mathrm{GF}_{96}$},1]\)\\
\noindent\(D_1\left[\mathrm{G}_{1536},1\right]\text{ = }D_1[\text{$\mathrm{GP}_{24}$},1]\)\\
\noindent\(D_1\left[\mathrm{G}_{1536},1\right]\text{ = }D_1[\text{$\mathrm{GS}_{24}$},1]\)\\
\noindent\(D_1\left[\mathrm{G}_{1536},1\right]\text{ = }D_1[\text{$\mathrm{GS}_{32}$},1]\)\\
\noindent\(D_1\left[\mathrm{G}_{1536},1\right]\text{ = }D_1[\text{$\mathrm{O}_{24}$},1]\)\\
\noindent\(D_1\left[\mathrm{G}_{1536},1\right]\text{ = }D_1[\text{$\mathrm{Oh}_{48}$},1]\)\\
\noindent\(D_2\left[\mathrm{G}_{1536},1\right]\text{ = }D_1[\text{$\mathrm{G}_{16}$},1]\)\\
\noindent\(D_2\left[\mathrm{G}_{1536},1\right]\text{ = }D_1[\text{$\mathrm{G}_{48}$},1]\)\\
\noindent\(D_2\left[\mathrm{G}_{1536},1\right]\text{ = }D_1[\text{$\mathrm{G}_{64}$},1]\)\\
\noindent\(D_2\left[\mathrm{G}_{1536},1\right]\text{ = }D_1[\text{$\mathrm{G}_{96}$},1]\)\\
\noindent\(D_2\left[\mathrm{G}_{1536},1\right]\text{ = }D_1[\text{$\mathrm{G}_{128}$},1]\)\\
\noindent\(D_2\left[\mathrm{G}_{1536},1\right]\text{ = }D_2[\text{$\mathrm{G}_{192}$},1]\)\\
\noindent\(D_2\left[\mathrm{G}_{1536},1\right]\text{ = }D_1[\text{$\mathrm{G}_{256}$},1]\)\\
\noindent\(D_2\left[\mathrm{G}_{1536},1\right]\text{ = }D_1[\text{$\mathrm{G}_{768}$},1]\)\\
\noindent\(D_2\left[\mathrm{G}_{1536},1\right]\text{ = }D_1[\text{$\mathrm{GF}_{48}$},1]\)\\
\noindent\(D_2\left[\mathrm{G}_{1536},1\right]\text{ = }D_2[\text{$\mathrm{GF}_{192}$},1]\)\\
\noindent\(D_2\left[\mathrm{G}_{1536},1\right]\text{ = }D_2[\text{$\mathrm{GF}_{96}$},1]\)\\
\noindent\(D_2\left[\mathrm{G}_{1536},1\right]\text{ = }D_1[\text{$\mathrm{GP}_{24}$},1]\)\\
\noindent\(D_2\left[\mathrm{G}_{1536},1\right]\text{ = }D_2[\text{$\mathrm{GS}_{24}$},1]\)\\
\noindent\(D_2\left[\mathrm{G}_{1536},1\right]\text{ = }D_2[\text{$\mathrm{GS}_{32}$},1]\)\\
\noindent\(D_2\left[\mathrm{G}_{1536},1\right]\text{ = }D_2[\text{$\mathrm{O}_{24}$},1]\)\\
\noindent\(D_2\left[\mathrm{G}_{1536},1\right]\text{ = }D_3[\text{$\mathrm{Oh}_{48}$},1]\)\\
\noindent\(D_3\left[\mathrm{G}_{1536},1\right]\text{ = }D_1[\text{$\mathrm{G}_{16}$},1]\)\\
\noindent\(D_3\left[\mathrm{G}_{1536},1\right]\text{ = }D_1[\text{$\mathrm{G}_{48}$},1]\)\\
\noindent\(D_3\left[\mathrm{G}_{1536},1\right]\text{ = }D_{43}[\text{$\mathrm{G}_{64}$},1]\)\\
\noindent\(D_3\left[\mathrm{G}_{1536},1\right]\text{ = }D_1[\text{$\mathrm{G}_{96}$},1]\)\\
\noindent\(D_3\left[\mathrm{G}_{1536},1\right]\text{ = }D_{29}[\text{$\mathrm{G}_{128}$},1]\)\\
\noindent\(D_3\left[\mathrm{G}_{1536},1\right]\text{ = }D_2[\text{$\mathrm{G}_{192}$},1]\)\\
\noindent\(D_3\left[\mathrm{G}_{1536},1\right]\text{ = }D_{29}[\text{$\mathrm{G}_{256}$},1]\)\\
\noindent\(D_3\left[\mathrm{G}_{1536},1\right]\text{ = }D_4[\text{$\mathrm{G}_{768}$},1]\)\\
\noindent\(D_3\left[\mathrm{G}_{1536},1\right]\text{ = }D_1[\text{$\mathrm{GF}_{48}$},1]\)\\
\noindent\(D_3\left[\mathrm{G}_{1536},1\right]\text{ = }D_1[\text{$\mathrm{GF}_{192}$},1]\)\\
\noindent\(D_3\left[\mathrm{G}_{1536},1\right]\text{ = }D_1[\text{$\mathrm{GF}_{96}$},1]\)\\
\noindent\(D_3\left[\mathrm{G}_{1536},1\right]\text{ = }D_1[\text{$\mathrm{GP}_{24}$},1]\)\\
\noindent\(D_3\left[\mathrm{G}_{1536},1\right]\text{ = }D_1[\text{$\mathrm{GS}_{24}$},1]\)\\
\noindent\(D_3\left[\mathrm{G}_{1536},1\right]\text{ = }D_2[\text{$\mathrm{GS}_{32}$},1]\)\\
\noindent\(D_3\left[\mathrm{G}_{1536},1\right]\text{ = }D_2[\text{$\mathrm{O}_{24}$},1]\)\\
\noindent\(D_3\left[\mathrm{G}_{1536},1\right]\text{ = }D_3[\text{$\mathrm{Oh}_{48}$},1]\)\\
\noindent\(D_4\left[\mathrm{G}_{1536},1\right]\text{ = }D_1[\text{$\mathrm{G}_{16}$},1]\)\\
\noindent\(D_4\left[\mathrm{G}_{1536},1\right]\text{ = }D_1[\text{$\mathrm{G}_{48}$},1]\)\\
\noindent\(D_4\left[\mathrm{G}_{1536},1\right]\text{ = }D_{43}[\text{$\mathrm{G}_{64}$},1]\)\\
\noindent\(D_4\left[\mathrm{G}_{1536},1\right]\text{ = }D_1[\text{$\mathrm{G}_{96}$},1]\)\\
\noindent\(D_4\left[\mathrm{G}_{1536},1\right]\text{ = }D_{29}[\text{$\mathrm{G}_{128}$},1]\)\\
\noindent\(D_4\left[\mathrm{G}_{1536},1\right]\text{ = }D_1[\text{$\mathrm{G}_{192}$},1]\)\\
\noindent\(D_4\left[\mathrm{G}_{1536},1\right]\text{ = }D_{29}[\text{$\mathrm{G}_{256}$},1]\)\\
\noindent\(D_4\left[\mathrm{G}_{1536},1\right]\text{ = }D_4[\text{$\mathrm{G}_{768}$},1]\)\\
\noindent\(D_4\left[\mathrm{G}_{1536},1\right]\text{ = }D_1[\text{$\mathrm{GF}_{48}$},1]\)\\
\noindent\(D_4\left[\mathrm{G}_{1536},1\right]\text{ = }D_2[\text{$\mathrm{GF}_{192}$},1]\)\\
\noindent\(D_4\left[\mathrm{G}_{1536},1\right]\text{ = }D_2[\text{$\mathrm{GF}_{96}$},1]\)\\
\noindent\(D_4\left[\mathrm{G}_{1536},1\right]\text{ = }D_1[\text{$\mathrm{GP}_{24}$},1]\)\\
\noindent\(D_4\left[\mathrm{G}_{1536},1\right]\text{ = }D_2[\text{$\mathrm{GS}_{24}$},1]\)\\
\noindent\(D_4\left[\mathrm{G}_{1536},1\right]\text{ = }D_1[\text{$\mathrm{GS}_{32}$},1]\)\\
\noindent\(D_4\left[\mathrm{G}_{1536},1\right]\text{ = }D_1[\text{$\mathrm{O}_{24}$},1]\)\\
\noindent\(D_4\left[\mathrm{G}_{1536},1\right]\text{ = }D_1[\text{$\mathrm{Oh}_{48}$},1]\)\\
\noindent\(D_5\left[\mathrm{G}_{1536},2\right]\text{ = }2 D_1[\text{$\mathrm{G}_{16}$},1]\)\\
\noindent\(D_5\left[\mathrm{G}_{1536},2\right]\text{ = }D_2[\text{$\mathrm{G}_{48}$},1]+D_3[\text{$\mathrm{G}_{48}$},1]\)\\
\noindent\(D_5\left[\mathrm{G}_{1536},2\right]\text{ = }2 D_1[\text{$\mathrm{G}_{64}$},1]\)\\
\noindent\(D_5\left[\mathrm{G}_{1536},2\right]\text{ = }D_3[\text{$\mathrm{G}_{96}$},1]+D_5[\text{$\mathrm{G}_{96}$},1]\)\\
\noindent\(D_5\left[\mathrm{G}_{1536},2\right]\text{ = }2 D_1[\text{$\mathrm{G}_{128}$},1]\)\\
\noindent\(D_5\left[\mathrm{G}_{1536},2\right]\text{ = }D_{17}[\text{$\mathrm{G}_{192}$},2]\)\\
\noindent\(D_5\left[\mathrm{G}_{1536},2\right]\text{ = }2 D_1[\text{$\mathrm{G}_{256}$},1]\)\\
\noindent\(D_5\left[\mathrm{G}_{1536},2\right]\text{ = }D_2[\text{$\mathrm{G}_{768}$},1]+D_3[\text{$\mathrm{G}_{768}$},1]\)\\
\noindent\(D_5\left[\mathrm{G}_{1536},2\right]\text{ = }D_2[\text{$\mathrm{GF}_{48}$},1]+D_3[\text{$\mathrm{GF}_{48}$},1]\)\\
\noindent\(D_5\left[\mathrm{G}_{1536},2\right]\text{ = }D_{17}[\text{$\mathrm{GF}_{192}$},2]\)\\
\noindent\(D_5\left[\mathrm{G}_{1536},2\right]\text{ = }D_3[\text{$\mathrm{GF}_{96}$},2]\)\\
\noindent\(D_5\left[\mathrm{G}_{1536},2\right]\text{ = }D_2[\text{$\mathrm{GP}_{24}$},1]+D_3[\text{$\mathrm{GP}_{24}$},1]\)\\
\noindent\(D_5\left[\mathrm{G}_{1536},2\right]\text{ = }D_3[\text{$\mathrm{GS}_{24}$},2]\)\\
\noindent\(D_5\left[\mathrm{G}_{1536},2\right]\text{ = }D_1[\text{$\mathrm{GS}_{32}$},1]+D_2[\text{$\mathrm{GS}_{32}$},1]\)\\
\noindent\(D_5\left[\mathrm{G}_{1536},2\right]\text{ = }D_3[\text{$\mathrm{O}_{24}$},2]\)\\
\noindent\(D_5\left[\mathrm{G}_{1536},2\right]\text{ = }D_5[\text{$\mathrm{Oh}_{48}$},2]\)\\
\noindent\(D_6\left[\mathrm{G}_{1536},2\right]\text{ = }2 D_1[\text{$\mathrm{G}_{16}$},1]\)\\
\noindent\(D_6\left[\mathrm{G}_{1536},2\right]\text{ = }D_2[\text{$\mathrm{G}_{48}$},1]+D_3[\text{$\mathrm{G}_{48}$},1]\)\\
\noindent\(D_6\left[\mathrm{G}_{1536},2\right]\text{ = }2 D_{43}[\text{$\mathrm{G}_{64}$},1]\)\\
\noindent\(D_6\left[\mathrm{G}_{1536},2\right]\text{ = }D_3[\text{$\mathrm{G}_{96}$},1]+D_5[\text{$\mathrm{G}_{96}$},1]\)\\
\noindent\(D_6\left[\mathrm{G}_{1536},2\right]\text{ = }2 D_{29}[\text{$\mathrm{G}_{128}$},1]\)\\
\noindent\(D_6\left[\mathrm{G}_{1536},2\right]\text{ = }D_{17}[\text{$\mathrm{G}_{192}$},2]\)\\
\noindent\(D_6\left[\mathrm{G}_{1536},2\right]\text{ = }2 D_{29}[\text{$\mathrm{G}_{256}$},1]\)\\
\noindent\(D_6\left[\mathrm{G}_{1536},2\right]\text{ = }D_5[\text{$\mathrm{G}_{768}$},1]+D_6[\text{$\mathrm{G}_{768}$},1]\)\\
\noindent\(D_6\left[\mathrm{G}_{1536},2\right]\text{ = }D_2[\text{$\mathrm{GF}_{48}$},1]+D_3[\text{$\mathrm{GF}_{48}$},1]\)\\
\noindent\(D_6\left[\mathrm{G}_{1536},2\right]\text{ = }D_{17}[\text{$\mathrm{GF}_{192}$},2]\)\\
\noindent\(D_6\left[\mathrm{G}_{1536},2\right]\text{ = }D_3[\text{$\mathrm{GF}_{96}$},2]\)\\
\noindent\(D_6\left[\mathrm{G}_{1536},2\right]\text{ = }D_2[\text{$\mathrm{GP}_{24}$},1]+D_3[\text{$\mathrm{GP}_{24}$},1]\)\\
\noindent\(D_6\left[\mathrm{G}_{1536},2\right]\text{ = }D_3[\text{$\mathrm{GS}_{24}$},2]\)\\
\noindent\(D_6\left[\mathrm{G}_{1536},2\right]\text{ = }D_1[\text{$\mathrm{GS}_{32}$},1]+D_2[\text{$\mathrm{GS}_{32}$},1]\)\\
\noindent\(D_6\left[\mathrm{G}_{1536},2\right]\text{ = }D_3[\text{$\mathrm{O}_{24}$},2]\)\\
\noindent\(D_6\left[\mathrm{G}_{1536},2\right]\text{ = }D_5[\text{$\mathrm{Oh}_{48}$},2]\)\\
\subsection{Branching rules of the irreps of dimensions 3}
\label{3brancione}
\noindent\(D_7\left[\mathrm{G}_{1536},3\right]\text{ = }D_2[\text{$\mathrm{G}_{16}$},1]+D_3[\text{$\mathrm{G}_{16}$},1]+D_4[\text{$\mathrm{G}_{16}$},1]\)\\
\noindent\(D_7\left[\mathrm{G}_{1536},3\right]\text{ = }D_4[\text{$\mathrm{G}_{48}$},3]\)\\
\noindent\(D_7\left[\mathrm{G}_{1536},3\right]\text{ = }3 D_1[\text{$\mathrm{G}_{64}$},1]\)\\
\noindent\(D_7\left[\mathrm{G}_{1536},3\right]\text{ = }D_8[\text{$\mathrm{G}_{96}$},3]\)\\
\noindent\(D_7\left[\mathrm{G}_{1536},3\right]\text{ = }D_1[\text{$\mathrm{G}_{128}$},1]+2 D_2[\text{$\mathrm{G}_{128}$},1]\)\\
\noindent\(D_7\left[\mathrm{G}_{1536},3\right]\text{ = }D_8[\text{$\mathrm{G}_{192}$},3]\)\\
\noindent\(D_7\left[\mathrm{G}_{1536},3\right]\text{ = }D_2[\text{$\mathrm{G}_{256}$},1]+D_3[\text{$\mathrm{G}_{256}$},1]+D_4[\text{$\mathrm{G}_{256}$},1]\)\\
\noindent\(D_7\left[\mathrm{G}_{1536},3\right]\text{ = }D_7[\text{$\mathrm{G}_{768}$},3]\)\\
\noindent\(D_7\left[\mathrm{G}_{1536},3\right]\text{ = }D_4[\text{$\mathrm{GF}_{48}$},3]\)\\
\noindent\(D_7\left[\mathrm{G}_{1536},3\right]\text{ = }D_8[\text{$\mathrm{GF}_{192}$},3]\)\\
\noindent\(D_7\left[\mathrm{G}_{1536},3\right]\text{ = }D_4[\text{$\mathrm{GF}_{96}$},3]\)\\
\noindent\(D_7\left[\mathrm{G}_{1536},3\right]\text{ = }D_7[\text{$\mathrm{GP}_{24}$},3]\)\\
\noindent\(D_7\left[\mathrm{G}_{1536},3\right]\text{ = }D_5[\text{$\mathrm{GS}_{24}$},3]\)\\
\noindent\(D_7\left[\mathrm{G}_{1536},3\right]\text{ = }D_3[\text{$\mathrm{GS}_{32}$},1]+D_9[\text{$\mathrm{GS}_{32}$},2]\)\\
\noindent\(D_7\left[\mathrm{G}_{1536},3\right]\text{ = }D_5[\text{$\mathrm{O}_{24}$},3]\)\\
\noindent\(D_7\left[\mathrm{G}_{1536},3\right]\text{ = }D_9[\text{$\mathrm{Oh}_{48}$},3]\)\\
\noindent\(D_8\left[\mathrm{G}_{1536},3\right]\text{ = }D_2[\text{$\mathrm{G}_{16}$},1]+D_3[\text{$\mathrm{G}_{16}$},1]+D_4[\text{$\mathrm{G}_{16}$},1]\)\\
\noindent\(D_8\left[\mathrm{G}_{1536},3\right]\text{ = }D_4[\text{$\mathrm{G}_{48}$},3]\)\\
\noindent\(D_8\left[\mathrm{G}_{1536},3\right]\text{ = }3 D_1[\text{$\mathrm{G}_{64}$},1]\)\\
\noindent\(D_8\left[\mathrm{G}_{1536},3\right]\text{ = }D_8[\text{$\mathrm{G}_{96}$},3]\)\\
\noindent\(D_8\left[\mathrm{G}_{1536},3\right]\text{ = }D_1[\text{$\mathrm{G}_{128}$},1]+2 D_2[\text{$\mathrm{G}_{128}$},1]\)\\
\noindent\(D_8\left[\mathrm{G}_{1536},3\right]\text{ = }D_7[\text{$\mathrm{G}_{192}$},3]\)\\
\noindent\(D_8\left[\mathrm{G}_{1536},3\right]\text{ = }D_2[\text{$\mathrm{G}_{256}$},1]+D_3[\text{$\mathrm{G}_{256}$},1]+D_4[\text{$\mathrm{G}_{256}$},1]\)\\
\noindent\(D_8\left[\mathrm{G}_{1536},3\right]\text{ = }D_7[\text{$\mathrm{G}_{768}$},3]\)\\
\noindent\(D_8\left[\mathrm{G}_{1536},3\right]\text{ = }D_4[\text{$\mathrm{GF}_{48}$},3]\)\\
\noindent\(D_8\left[\mathrm{G}_{1536},3\right]\text{ = }D_7[\text{$\mathrm{GF}_{192}$},3]\)\\
\noindent\(D_8\left[\mathrm{G}_{1536},3\right]\text{ = }D_5[\text{$\mathrm{GF}_{96}$},3]\)\\
\noindent\(D_8\left[\mathrm{G}_{1536},3\right]\text{ = }D_7[\text{$\mathrm{GP}_{24}$},3]\)\\
\noindent\(D_8\left[\mathrm{G}_{1536},3\right]\text{ = }D_4[\text{$\mathrm{GS}_{24}$},3]\)\\
\noindent\(D_8\left[\mathrm{G}_{1536},3\right]\text{ = }D_4[\text{$\mathrm{GS}_{32}$},1]+D_9[\text{$\mathrm{GS}_{32}$},2]\)\\
\noindent\(D_8\left[\mathrm{G}_{1536},3\right]\text{ = }D_4[\text{$\mathrm{O}_{24}$},3]\)\\
\noindent\(D_8\left[\mathrm{G}_{1536},3\right]\text{ = }D_7[\text{$\mathrm{Oh}_{48}$},3]\)\\
\noindent\(D_9\left[\mathrm{G}_{1536},3\right]\text{ = }3 D_1[\text{$\mathrm{G}_{16}$},1]\)\\
\noindent\(D_9\left[\mathrm{G}_{1536},3\right]\text{ = }D_1[\text{$\mathrm{G}_{48}$},1]+D_2[\text{$\mathrm{G}_{48}$},1]+D_3[\text{$\mathrm{G}_{48}$},1]\)\\
\noindent\(D_9\left[\mathrm{G}_{1536},3\right]\text{ = }D_3[\text{$\mathrm{G}_{64}$},1]+D_9[\text{$\mathrm{G}_{64}$},1]+D_{33}[\text{$\mathrm{G}_{64}$},1]\)\\
\noindent\(D_9\left[\mathrm{G}_{1536},3\right]\text{ = }D_1[\text{$\mathrm{G}_{96}$},1]+D_3[\text{$\mathrm{G}_{96}$},1]+D_5[\text{$\mathrm{G}_{96}$},1]\)\\
\noindent\(D_9\left[\mathrm{G}_{1536},3\right]\text{ = }D_5[\text{$\mathrm{G}_{128}$},1]+D_9[\text{$\mathrm{G}_{128}$},1]+D_{17}[\text{$\mathrm{G}_{128}$},1]\)\\
\noindent\(D_9\left[\mathrm{G}_{1536},3\right]\text{ = }D_1[\text{$\mathrm{G}_{192}$},1]+D_{17}[\text{$\mathrm{G}_{192}$},2]\)\\
\noindent\(D_9\left[\mathrm{G}_{1536},3\right]\text{ = }D_5[\text{$\mathrm{G}_{256}$},1]+D_9[\text{$\mathrm{G}_{256}$},1]+D_{17}[\text{$\mathrm{G}_{256}$},1]\)\\
\noindent\(D_9\left[\mathrm{G}_{1536},3\right]\text{ = }D_8[\text{$\mathrm{G}_{768}$},3]\)\\
\noindent\(D_9\left[\mathrm{G}_{1536},3\right]\text{ = }D_1[\text{$\mathrm{GF}_{48}$},1]+D_2[\text{$\mathrm{GF}_{48}$},1]+D_3[\text{$\mathrm{GF}_{48}$},1]\)\\
\noindent\(D_9\left[\mathrm{G}_{1536},3\right]\text{ = }D_2[\text{$\mathrm{GF}_{192}$},1]+D_{17}[\text{$\mathrm{GF}_{192}$},2]\)\\
\noindent\(D_9\left[\mathrm{G}_{1536},3\right]\text{ = }D_2[\text{$\mathrm{GF}_{96}$},1]+D_3[\text{$\mathrm{GF}_{96}$},2]\)\\
\noindent\(D_9\left[\mathrm{G}_{1536},3\right]\text{ = }D_1[\text{$\mathrm{GP}_{24}$},1]+D_2[\text{$\mathrm{GP}_{24}$},1]+D_3[\text{$\mathrm{GP}_{24}$},1]\)\\
\noindent\(D_9\left[\mathrm{G}_{1536},3\right]\text{ = }D_2[\text{$\mathrm{GS}_{24}$},1]+D_3[\text{$\mathrm{GS}_{24}$},2]\)\\
\noindent\(D_9\left[\mathrm{G}_{1536},3\right]\text{ = }2 D_1[\text{$\mathrm{GS}_{32}$},1]+D_2[\text{$\mathrm{GS}_{32}$},1]\)\\
\noindent\(D_9\left[\mathrm{G}_{1536},3\right]\text{ = }D_1[\text{$\mathrm{O}_{24}$},1]+D_3[\text{$\mathrm{O}_{24}$},2]\)\\
\noindent\(D_9\left[\mathrm{G}_{1536},3\right]\text{ = }D_1[\text{$\mathrm{Oh}_{48}$},1]+D_5[\text{$\mathrm{Oh}_{48}$},2]\)\\
\noindent\(D_{10}\left[\mathrm{G}_{1536},3\right]\text{ = }3 D_1[\text{$\mathrm{G}_{16}$},1]\)\\
\noindent\(D_{10}\left[\mathrm{G}_{1536},3\right]\text{ = }D_1[\text{$\mathrm{G}_{48}$},1]+D_2[\text{$\mathrm{G}_{48}$},1]+D_3[\text{$\mathrm{G}_{48}$},1]\)\\
\noindent\(D_{10}\left[\mathrm{G}_{1536},3\right]\text{ = }D_3[\text{$\mathrm{G}_{64}$},1]+D_9[\text{$\mathrm{G}_{64}$},1]+D_{33}[\text{$\mathrm{G}_{64}$},1]\)\\
\noindent\(D_{10}\left[\mathrm{G}_{1536},3\right]\text{ = }D_1[\text{$\mathrm{G}_{96}$},1]+D_3[\text{$\mathrm{G}_{96}$},1]+D_5[\text{$\mathrm{G}_{96}$},1]\)\\
\noindent\(D_{10}\left[\mathrm{G}_{1536},3\right]\text{ = }D_5[\text{$\mathrm{G}_{128}$},1]+D_9[\text{$\mathrm{G}_{128}$},1]+D_{17}[\text{$\mathrm{G}_{128}$},1]\)\\
\noindent\(D_{10}\left[\mathrm{G}_{1536},3\right]\text{ = }D_2[\text{$\mathrm{G}_{192}$},1]+D_{17}[\text{$\mathrm{G}_{192}$},2]\)\\
\noindent\(D_{10}\left[\mathrm{G}_{1536},3\right]\text{ = }D_5[\text{$\mathrm{G}_{256}$},1]+D_9[\text{$\mathrm{G}_{256}$},1]+D_{17}[\text{$\mathrm{G}_{256}$},1]\)\\
\noindent\(D_{10}\left[\mathrm{G}_{1536},3\right]\text{ = }D_8[\text{$\mathrm{G}_{768}$},3]\)\\
\noindent\(D_{10}\left[\mathrm{G}_{1536},3\right]\text{ = }D_1[\text{$\mathrm{GF}_{48}$},1]+D_2[\text{$\mathrm{GF}_{48}$},1]+D_3[\text{$\mathrm{GF}_{48}$},1]\)\\
\noindent\(D_{10}\left[\mathrm{G}_{1536},3\right]\text{ = }D_1[\text{$\mathrm{GF}_{192}$},1]+D_{17}[\text{$\mathrm{GF}_{192}$},2]\)\\
\noindent\(D_{10}\left[\mathrm{G}_{1536},3\right]\text{ = }D_1[\text{$\mathrm{GF}_{96}$},1]+D_3[\text{$\mathrm{GF}_{96}$},2]\)\\
\noindent\(D_{10}\left[\mathrm{G}_{1536},3\right]\text{ = }D_1[\text{$\mathrm{GP}_{24}$},1]+D_2[\text{$\mathrm{GP}_{24}$},1]+D_3[\text{$\mathrm{GP}_{24}$},1]\)\\
\noindent\(D_{10}\left[\mathrm{G}_{1536},3\right]\text{ = }D_1[\text{$\mathrm{GS}_{24}$},1]+D_3[\text{$\mathrm{GS}_{24}$},2]\)\\
\noindent\(D_{10}\left[\mathrm{G}_{1536},3\right]\text{ = }D_1[\text{$\mathrm{GS}_{32}$},1]+2 D_2[\text{$\mathrm{GS}_{32}$},1]\)\\
\noindent\(D_{10}\left[\mathrm{G}_{1536},3\right]\text{ = }D_2[\text{$\mathrm{O}_{24}$},1]+D_3[\text{$\mathrm{O}_{24}$},2]\)\\
\noindent\(D_{10}\left[\mathrm{G}_{1536},3\right]\text{ = }D_3[\text{$\mathrm{Oh}_{48}$},1]+D_5[\text{$\mathrm{Oh}_{48}$},2]\)\\
\noindent\(D_{11}\left[\mathrm{G}_{1536},3\right]\text{ = }D_2[\text{$\mathrm{G}_{16}$},1]+D_3[\text{$\mathrm{G}_{16}$},1]+D_4[\text{$\mathrm{G}_{16}$},1]\)\\
\noindent\(D_{11}\left[\mathrm{G}_{1536},3\right]\text{ = }D_4[\text{$\mathrm{G}_{48}$},3]\)\\
\noindent\(D_{11}\left[\mathrm{G}_{1536},3\right]\text{ = }D_3[\text{$\mathrm{G}_{64}$},1]+D_9[\text{$\mathrm{G}_{64}$},1]+D_{33}[\text{$\mathrm{G}_{64}$},1]\)\\
\noindent\(D_{11}\left[\mathrm{G}_{1536},3\right]\text{ = }D_8[\text{$\mathrm{G}_{96}$},3]\)\\
\noindent\(D_{11}\left[\mathrm{G}_{1536},3\right]\text{ = }D_5[\text{$\mathrm{G}_{128}$},1]+D_{10}[\text{$\mathrm{G}_{128}$},1]+D_{18}[\text{$\mathrm{G}_{128}$},1]\)\\
\noindent\(D_{11}\left[\mathrm{G}_{1536},3\right]\text{ = }D_7[\text{$\mathrm{G}_{192}$},3]\)\\
\noindent\(D_{11}\left[\mathrm{G}_{1536},3\right]\text{ = }D_6[\text{$\mathrm{G}_{256}$},1]+D_{11}[\text{$\mathrm{G}_{256}$},1]+D_{20}[\text{$\mathrm{G}_{256}$},1]\)\\
\noindent\(D_{11}\left[\mathrm{G}_{1536},3\right]\text{ = }D_9[\text{$\mathrm{G}_{768}$},3]\)\\
\noindent\(D_{11}\left[\mathrm{G}_{1536},3\right]\text{ = }D_4[\text{$\mathrm{GF}_{48}$},3]\)\\
\noindent\(D_{11}\left[\mathrm{G}_{1536},3\right]\text{ = }D_8[\text{$\mathrm{GF}_{192}$},3]\)\\
\noindent\(D_{11}\left[\mathrm{G}_{1536},3\right]\text{ = }D_4[\text{$\mathrm{GF}_{96}$},3]\)\\
\noindent\(D_{11}\left[\mathrm{G}_{1536},3\right]\text{ = }D_7[\text{$\mathrm{GP}_{24}$},3]\)\\
\noindent\(D_{11}\left[\mathrm{G}_{1536},3\right]\text{ = }D_5[\text{$\mathrm{GS}_{24}$},3]\)\\
\noindent\(D_{11}\left[\mathrm{G}_{1536},3\right]\text{ = }D_4[\text{$\mathrm{GS}_{32}$},1]+D_9[\text{$\mathrm{GS}_{32}$},2]\)\\
\noindent\(D_{11}\left[\mathrm{G}_{1536},3\right]\text{ = }D_4[\text{$\mathrm{O}_{24}$},3]\)\\
\noindent\(D_{11}\left[\mathrm{G}_{1536},3\right]\text{ = }D_7[\text{$\mathrm{Oh}_{48}$},3]\)\\
\noindent\(D_{12}\left[\mathrm{G}_{1536},3\right]\text{ = }D_2[\text{$\mathrm{G}_{16}$},1]+D_3[\text{$\mathrm{G}_{16}$},1]+D_4[\text{$\mathrm{G}_{16}$},1]\)\\
\noindent\(D_{12}\left[\mathrm{G}_{1536},3\right]\text{ = }D_4[\text{$\mathrm{G}_{48}$},3]\)\\
\noindent\(D_{12}\left[\mathrm{G}_{1536},3\right]\text{ = }D_3[\text{$\mathrm{G}_{64}$},1]+D_9[\text{$\mathrm{G}_{64}$},1]+D_{33}[\text{$\mathrm{G}_{64}$},1]\)\\
\noindent\(D_{12}\left[\mathrm{G}_{1536},3\right]\text{ = }D_8[\text{$\mathrm{G}_{96}$},3]\)\\
\noindent\(D_{12}\left[\mathrm{G}_{1536},3\right]\text{ = }D_5[\text{$\mathrm{G}_{128}$},1]+D_{10}[\text{$\mathrm{G}_{128}$},1]+D_{18}[\text{$\mathrm{G}_{128}$},1]\)\\
\noindent\(D_{12}\left[\mathrm{G}_{1536},3\right]\text{ = }D_8[\text{$\mathrm{G}_{192}$},3]\)\\
\noindent\(D_{12}\left[\mathrm{G}_{1536},3\right]\text{ = }D_6[\text{$\mathrm{G}_{256}$},1]+D_{11}[\text{$\mathrm{G}_{256}$},1]+D_{20}[\text{$\mathrm{G}_{256}$},1]\)\\
\noindent\(D_{12}\left[\mathrm{G}_{1536},3\right]\text{ = }D_9[\text{$\mathrm{G}_{768}$},3]\)\\
\noindent\(D_{12}\left[\mathrm{G}_{1536},3\right]\text{ = }D_4[\text{$\mathrm{GF}_{48}$},3]\)\\
\noindent\(D_{12}\left[\mathrm{G}_{1536},3\right]\text{ = }D_7[\text{$\mathrm{GF}_{192}$},3]\)\\
\noindent\(D_{12}\left[\mathrm{G}_{1536},3\right]\text{ = }D_5[\text{$\mathrm{GF}_{96}$},3]\)\\
\noindent\(D_{12}\left[\mathrm{G}_{1536},3\right]\text{ = }D_7[\text{$\mathrm{GP}_{24}$},3]\)\\
\noindent\(D_{12}\left[\mathrm{G}_{1536},3\right]\text{ = }D_4[\text{$\mathrm{GS}_{24}$},3]\)\\
\noindent\(D_{12}\left[\mathrm{G}_{1536},3\right]\text{ = }D_3[\text{$\mathrm{GS}_{32}$},1]+D_9[\text{$\mathrm{GS}_{32}$},2]\)\\
\noindent\(D_{12}\left[\mathrm{G}_{1536},3\right]\text{ = }D_5[\text{$\mathrm{O}_{24}$},3]\)\\
\noindent\(D_{12}\left[\mathrm{G}_{1536},3\right]\text{ = }D_9[\text{$\mathrm{Oh}_{48}$},3]\)\\
\noindent\(D_{13}\left[\mathrm{G}_{1536},3\right]\text{ = }3 D_1[\text{$\mathrm{G}_{16}$},1]\)\\
\noindent\(D_{13}\left[\mathrm{G}_{1536},3\right]\text{ = }D_1[\text{$\mathrm{G}_{48}$},1]+D_2[\text{$\mathrm{G}_{48}$},1]+D_3[\text{$\mathrm{G}_{48}$},1]\)\\
\noindent\(D_{13}\left[\mathrm{G}_{1536},3\right]\text{ = }D_{11}[\text{$\mathrm{G}_{64}$},1]+D_{35}[\text{$\mathrm{G}_{64}$},1]+D_{41}[\text{$\mathrm{G}_{64}$},1]\)\\
\noindent\(D_{13}\left[\mathrm{G}_{1536},3\right]\text{ = }D_1[\text{$\mathrm{G}_{96}$},1]+D_3[\text{$\mathrm{G}_{96}$},1]+D_5[\text{$\mathrm{G}_{96}$},1]\)\\
\noindent\(D_{13}\left[\mathrm{G}_{1536},3\right]\text{ = }D_{13}[\text{$\mathrm{G}_{128}$},1]+D_{21}[\text{$\mathrm{G}_{128}$},1]+D_{25}[\text{$\mathrm{G}_{128}$},1]\)\\
\noindent\(D_{13}\left[\mathrm{G}_{1536},3\right]\text{ = }D_2[\text{$\mathrm{G}_{192}$},1]+D_{17}[\text{$\mathrm{G}_{192}$},2]\)\\
\noindent\(D_{13}\left[\mathrm{G}_{1536},3\right]\text{ = }D_{13}[\text{$\mathrm{G}_{256}$},1]+D_{21}[\text{$\mathrm{G}_{256}$},1]+D_{25}[\text{$\mathrm{G}_{256}$},1]\)\\
\noindent\(D_{13}\left[\mathrm{G}_{1536},3\right]\text{ = }D_{12}[\text{$\mathrm{G}_{768}$},3]\)\\
\noindent\(D_{13}\left[\mathrm{G}_{1536},3\right]\text{ = }D_1[\text{$\mathrm{GF}_{48}$},1]+D_2[\text{$\mathrm{GF}_{48}$},1]+D_3[\text{$\mathrm{GF}_{48}$},1]\)\\
\noindent\(D_{13}\left[\mathrm{G}_{1536},3\right]\text{ = }D_2[\text{$\mathrm{GF}_{192}$},1]+D_{17}[\text{$\mathrm{GF}_{192}$},2]\)\\
\noindent\(D_{13}\left[\mathrm{G}_{1536},3\right]\text{ = }D_2[\text{$\mathrm{GF}_{96}$},1]+D_3[\text{$\mathrm{GF}_{96}$},2]\)\\
\noindent\(D_{13}\left[\mathrm{G}_{1536},3\right]\text{ = }D_1[\text{$\mathrm{GP}_{24}$},1]+D_2[\text{$\mathrm{GP}_{24}$},1]+D_3[\text{$\mathrm{GP}_{24}$},1]\)\\
\noindent\(D_{13}\left[\mathrm{G}_{1536},3\right]\text{ = }D_2[\text{$\mathrm{GS}_{24}$},1]+D_3[\text{$\mathrm{GS}_{24}$},2]\)\\
\noindent\(D_{13}\left[\mathrm{G}_{1536},3\right]\text{ = }D_1[\text{$\mathrm{GS}_{32}$},1]+2 D_2[\text{$\mathrm{GS}_{32}$},1]\)\\
\noindent\(D_{13}\left[\mathrm{G}_{1536},3\right]\text{ = }D_2[\text{$\mathrm{O}_{24}$},1]+D_3[\text{$\mathrm{O}_{24}$},2]\)\\
\noindent\(D_{13}\left[\mathrm{G}_{1536},3\right]\text{ = }D_3[\text{$\mathrm{Oh}_{48}$},1]+D_5[\text{$\mathrm{Oh}_{48}$},2]\)\\
\noindent\(D_{14}\left[\mathrm{G}_{1536},3\right]\text{ = }3 D_1[\text{$\mathrm{G}_{16}$},1]\)\\
\noindent\(D_{14}\left[\mathrm{G}_{1536},3\right]\text{ = }D_1[\text{$\mathrm{G}_{48}$},1]+D_2[\text{$\mathrm{G}_{48}$},1]+D_3[\text{$\mathrm{G}_{48}$},1]\)\\
\noindent\(D_{14}\left[\mathrm{G}_{1536},3\right]\text{ = }D_{11}[\text{$\mathrm{G}_{64}$},1]+D_{35}[\text{$\mathrm{G}_{64}$},1]+D_{41}[\text{$\mathrm{G}_{64}$},1]\)\\
\noindent\(D_{14}\left[\mathrm{G}_{1536},3\right]\text{ = }D_1[\text{$\mathrm{G}_{96}$},1]+D_3[\text{$\mathrm{G}_{96}$},1]+D_5[\text{$\mathrm{G}_{96}$},1]\)\\
\noindent\(D_{14}\left[\mathrm{G}_{1536},3\right]\text{ = }D_{13}[\text{$\mathrm{G}_{128}$},1]+D_{21}[\text{$\mathrm{G}_{128}$},1]+D_{25}[\text{$\mathrm{G}_{128}$},1]\)\\
\noindent\(D_{14}\left[\mathrm{G}_{1536},3\right]\text{ = }D_1[\text{$\mathrm{G}_{192}$},1]+D_{17}[\text{$\mathrm{G}_{192}$},2]\)\\
\noindent\(D_{14}\left[\mathrm{G}_{1536},3\right]\text{ = }D_{13}[\text{$\mathrm{G}_{256}$},1]+D_{21}[\text{$\mathrm{G}_{256}$},1]+D_{25}[\text{$\mathrm{G}_{256}$},1]\)\\
\noindent\(D_{14}\left[\mathrm{G}_{1536},3\right]\text{ = }D_{12}[\text{$\mathrm{G}_{768}$},3]\)\\
\noindent\(D_{14}\left[\mathrm{G}_{1536},3\right]\text{ = }D_1[\text{$\mathrm{GF}_{48}$},1]+D_2[\text{$\mathrm{GF}_{48}$},1]+D_3[\text{$\mathrm{GF}_{48}$},1]\)\\
\noindent\(D_{14}\left[\mathrm{G}_{1536},3\right]\text{ = }D_1[\text{$\mathrm{GF}_{192}$},1]+D_{17}[\text{$\mathrm{GF}_{192}$},2]\)\\
\noindent\(D_{14}\left[\mathrm{G}_{1536},3\right]\text{ = }D_1[\text{$\mathrm{GF}_{96}$},1]+D_3[\text{$\mathrm{GF}_{96}$},2]\)\\
\noindent\(D_{14}\left[\mathrm{G}_{1536},3\right]\text{ = }D_1[\text{$\mathrm{GP}_{24}$},1]+D_2[\text{$\mathrm{GP}_{24}$},1]+D_3[\text{$\mathrm{GP}_{24}$},1]\)\\
\noindent\(D_{14}\left[\mathrm{G}_{1536},3\right]\text{ = }D_1[\text{$\mathrm{GS}_{24}$},1]+D_3[\text{$\mathrm{GS}_{24}$},2]\)\\
\noindent\(D_{14}\left[\mathrm{G}_{1536},3\right]\text{ = }2 D_1[\text{$\mathrm{GS}_{32}$},1]+D_2[\text{$\mathrm{GS}_{32}$},1]\)\\
\noindent\(D_{14}\left[\mathrm{G}_{1536},3\right]\text{ = }D_1[\text{$\mathrm{O}_{24}$},1]+D_3[\text{$\mathrm{O}_{24}$},2]\)\\
\noindent\(D_{14}\left[\mathrm{G}_{1536},3\right]\text{ = }D_1[\text{$\mathrm{Oh}_{48}$},1]+D_5[\text{$\mathrm{Oh}_{48}$},2]\)\\
\noindent\(D_{15}\left[\mathrm{G}_{1536},3\right]\text{ = }D_2[\text{$\mathrm{G}_{16}$},1]+D_3[\text{$\mathrm{G}_{16}$},1]+D_4[\text{$\mathrm{G}_{16}$},1]\)\\
\noindent\(D_{15}\left[\mathrm{G}_{1536},3\right]\text{ = }D_4[\text{$\mathrm{G}_{48}$},3]\)\\
\noindent\(D_{15}\left[\mathrm{G}_{1536},3\right]\text{ = }D_{11}[\text{$\mathrm{G}_{64}$},1]+D_{35}[\text{$\mathrm{G}_{64}$},1]+D_{41}[\text{$\mathrm{G}_{64}$},1]\)\\
\noindent\(D_{15}\left[\mathrm{G}_{1536},3\right]\text{ = }D_8[\text{$\mathrm{G}_{96}$},3]\)\\
\noindent\(D_{15}\left[\mathrm{G}_{1536},3\right]\text{ = }D_{14}[\text{$\mathrm{G}_{128}$},1]+D_{22}[\text{$\mathrm{G}_{128}$},1]+D_{25}[\text{$\mathrm{G}_{128}$},1]\)\\
\noindent\(D_{15}\left[\mathrm{G}_{1536},3\right]\text{ = }D_7[\text{$\mathrm{G}_{192}$},3]\)\\
\noindent\(D_{15}\left[\mathrm{G}_{1536},3\right]\text{ = }D_{16}[\text{$\mathrm{G}_{256}$},1]+D_{23}[\text{$\mathrm{G}_{256}$},1]+D_{26}[\text{$\mathrm{G}_{256}$},1]\)\\
\noindent\(D_{15}\left[\mathrm{G}_{1536},3\right]\text{ = }D_{15}[\text{$\mathrm{G}_{768}$},3]\)\\
\noindent\(D_{15}\left[\mathrm{G}_{1536},3\right]\text{ = }D_4[\text{$\mathrm{GF}_{48}$},3]\)\\
\noindent\(D_{15}\left[\mathrm{G}_{1536},3\right]\text{ = }D_7[\text{$\mathrm{GF}_{192}$},3]\)\\
\noindent\(D_{15}\left[\mathrm{G}_{1536},3\right]\text{ = }D_5[\text{$\mathrm{GF}_{96}$},3]\)\\
\noindent\(D_{15}\left[\mathrm{G}_{1536},3\right]\text{ = }D_7[\text{$\mathrm{GP}_{24}$},3]\)\\
\noindent\(D_{15}\left[\mathrm{G}_{1536},3\right]\text{ = }D_4[\text{$\mathrm{GS}_{24}$},3]\)\\
\noindent\(D_{15}\left[\mathrm{G}_{1536},3\right]\text{ = }D_4[\text{$\mathrm{GS}_{32}$},1]+D_9[\text{$\mathrm{GS}_{32}$},2]\)\\
\noindent\(D_{15}\left[\mathrm{G}_{1536},3\right]\text{ = }D_4[\text{$\mathrm{O}_{24}$},3]\)\\
\noindent\(D_{15}\left[\mathrm{G}_{1536},3\right]\text{ = }D_7[\text{$\mathrm{Oh}_{48}$},3]\)\\
\noindent\(D_{16}\left[\mathrm{G}_{1536},3\right]\text{ = }D_2[\text{$\mathrm{G}_{16}$},1]+D_3[\text{$\mathrm{G}_{16}$},1]+D_4[\text{$\mathrm{G}_{16}$},1]\)\\
\noindent\(D_{16}\left[\mathrm{G}_{1536},3\right]\text{ = }D_4[\text{$\mathrm{G}_{48}$},3]\)\\
\noindent\(D_{16}\left[\mathrm{G}_{1536},3\right]\text{ = }D_{11}[\text{$\mathrm{G}_{64}$},1]+D_{35}[\text{$\mathrm{G}_{64}$},1]+D_{41}[\text{$\mathrm{G}_{64}$},1]\)\\
\noindent\(D_{16}\left[\mathrm{G}_{1536},3\right]\text{ = }D_8[\text{$\mathrm{G}_{96}$},3]\)\\
\noindent\(D_{16}\left[\mathrm{G}_{1536},3\right]\text{ = }D_{14}[\text{$\mathrm{G}_{128}$},1]+D_{22}[\text{$\mathrm{G}_{128}$},1]+D_{25}[\text{$\mathrm{G}_{128}$},1]\)\\
\noindent\(D_{16}\left[\mathrm{G}_{1536},3\right]\text{ = }D_8[\text{$\mathrm{G}_{192}$},3]\)\\
\noindent\(D_{16}\left[\mathrm{G}_{1536},3\right]\text{ = }D_{16}[\text{$\mathrm{G}_{256}$},1]+D_{23}[\text{$\mathrm{G}_{256}$},1]+D_{26}[\text{$\mathrm{G}_{256}$},1]\)\\
\noindent\(D_{16}\left[\mathrm{G}_{1536},3\right]\text{ = }D_{15}[\text{$\mathrm{G}_{768}$},3]\)\\
\noindent\(D_{16}\left[\mathrm{G}_{1536},3\right]\text{ = }D_4[\text{$\mathrm{GF}_{48}$},3]\)\\
\noindent\(D_{16}\left[\mathrm{G}_{1536},3\right]\text{ = }D_8[\text{$\mathrm{GF}_{192}$},3]\)\\
\noindent\(D_{16}\left[\mathrm{G}_{1536},3\right]\text{ = }D_4[\text{$\mathrm{GF}_{96}$},3]\)\\
\noindent\(D_{16}\left[\mathrm{G}_{1536},3\right]\text{ = }D_7[\text{$\mathrm{GP}_{24}$},3]\)\\
\noindent\(D_{16}\left[\mathrm{G}_{1536},3\right]\text{ = }D_5[\text{$\mathrm{GS}_{24}$},3]\)\\
\noindent\(D_{16}\left[\mathrm{G}_{1536},3\right]\text{ = }D_3[\text{$\mathrm{GS}_{32}$},1]+D_9[\text{$\mathrm{GS}_{32}$},2]\)\\
\noindent\(D_{16}\left[\mathrm{G}_{1536},3\right]\text{ = }D_5[\text{$\mathrm{O}_{24}$},3]\)\\
\noindent\(D_{16}\left[\mathrm{G}_{1536},3\right]\text{ = }D_9[\text{$\mathrm{Oh}_{48}$},3]\)\\
\noindent\(D_{17}\left[\mathrm{G}_{1536},3\right]\text{ = }D_2[\text{$\mathrm{G}_{16}$},1]+D_3[\text{$\mathrm{G}_{16}$},1]+D_4[\text{$\mathrm{G}_{16}$},1]\)\\
\noindent\(D_{17}\left[\mathrm{G}_{1536},3\right]\text{ = }D_4[\text{$\mathrm{G}_{48}$},3]\)\\
\noindent\(D_{17}\left[\mathrm{G}_{1536},3\right]\text{ = }3 D_{43}[\text{$\mathrm{G}_{64}$},1]\)\\
\noindent\(D_{17}\left[\mathrm{G}_{1536},3\right]\text{ = }D_8[\text{$\mathrm{G}_{96}$},3]\)\\
\noindent\(D_{17}\left[\mathrm{G}_{1536},3\right]\text{ = }D_{29}[\text{$\mathrm{G}_{128}$},1]+2 D_{30}[\text{$\mathrm{G}_{128}$},1]\)\\
\noindent\(D_{17}\left[\mathrm{G}_{1536},3\right]\text{ = }D_7[\text{$\mathrm{G}_{192}$},3]\)\\
\noindent\(D_{17}\left[\mathrm{G}_{1536},3\right]\text{ = }D_{30}[\text{$\mathrm{G}_{256}$},1]+D_{31}[\text{$\mathrm{G}_{256}$},1]+D_{32}[\text{$\mathrm{G}_{256}$},1]\)\\
\noindent\(D_{17}\left[\mathrm{G}_{1536},3\right]\text{ = }D_{16}[\text{$\mathrm{G}_{768}$},3]\)\\
\noindent\(D_{17}\left[\mathrm{G}_{1536},3\right]\text{ = }D_4[\text{$\mathrm{GF}_{48}$},3]\)\\
\noindent\(D_{17}\left[\mathrm{G}_{1536},3\right]\text{ = }D_8[\text{$\mathrm{GF}_{192}$},3]\)\\
\noindent\(D_{17}\left[\mathrm{G}_{1536},3\right]\text{ = }D_4[\text{$\mathrm{GF}_{96}$},3]\)\\
\noindent\(D_{17}\left[\mathrm{G}_{1536},3\right]\text{ = }D_7[\text{$\mathrm{GP}_{24}$},3]\)\\
\noindent\(D_{17}\left[\mathrm{G}_{1536},3\right]\text{ = }D_5[\text{$\mathrm{GS}_{24}$},3]\)\\
\noindent\(D_{17}\left[\mathrm{G}_{1536},3\right]\text{ = }D_4[\text{$\mathrm{GS}_{32}$},1]+D_9[\text{$\mathrm{GS}_{32}$},2]\)\\
\noindent\(D_{17}\left[\mathrm{G}_{1536},3\right]\text{ = }D_4[\text{$\mathrm{O}_{24}$},3]\)\\
\noindent\(D_{17}\left[\mathrm{G}_{1536},3\right]\text{ = }D_7[\text{$\mathrm{Oh}_{48}$},3]\)\\
\noindent\(D_{18}\left[\mathrm{G}_{1536},3\right]\text{ = }D_2[\text{$\mathrm{G}_{16}$},1]+D_3[\text{$\mathrm{G}_{16}$},1]+D_4[\text{$\mathrm{G}_{16}$},1]\)\\
\noindent\(D_{18}\left[\mathrm{G}_{1536},3\right]\text{ = }D_4[\text{$\mathrm{G}_{48}$},3]\)\\
\noindent\(D_{18}\left[\mathrm{G}_{1536},3\right]\text{ = }3 D_{43}[\text{$\mathrm{G}_{64}$},1]\)\\
\noindent\(D_{18}\left[\mathrm{G}_{1536},3\right]\text{ = }D_8[\text{$\mathrm{G}_{96}$},3]\)\\
\noindent\(D_{18}\left[\mathrm{G}_{1536},3\right]\text{ = }D_{29}[\text{$\mathrm{G}_{128}$},1]+2 D_{30}[\text{$\mathrm{G}_{128}$},1]\)\\
\noindent\(D_{18}\left[\mathrm{G}_{1536},3\right]\text{ = }D_8[\text{$\mathrm{G}_{192}$},3]\)\\
\noindent\(D_{18}\left[\mathrm{G}_{1536},3\right]\text{ = }D_{30}[\text{$\mathrm{G}_{256}$},1]+D_{31}[\text{$\mathrm{G}_{256}$},1]+D_{32}[\text{$\mathrm{G}_{256}$},1]\)\\
\noindent\(D_{18}\left[\mathrm{G}_{1536},3\right]\text{ = }D_{16}[\text{$\mathrm{G}_{768}$},3]\)\\
\noindent\(D_{18}\left[\mathrm{G}_{1536},3\right]\text{ = }D_4[\text{$\mathrm{GF}_{48}$},3]\)\\
\noindent\(D_{18}\left[\mathrm{G}_{1536},3\right]\text{ = }D_7[\text{$\mathrm{GF}_{192}$},3]\)\\
\noindent\(D_{18}\left[\mathrm{G}_{1536},3\right]\text{ = }D_5[\text{$\mathrm{GF}_{96}$},3]\)\\
\noindent\(D_{18}\left[\mathrm{G}_{1536},3\right]\text{ = }D_7[\text{$\mathrm{GP}_{24}$},3]\)\\
\noindent\(D_{18}\left[\mathrm{G}_{1536},3\right]\text{ = }D_4[\text{$\mathrm{GS}_{24}$},3]\)\\
\noindent\(D_{18}\left[\mathrm{G}_{1536},3\right]\text{ = }D_3[\text{$\mathrm{GS}_{32}$},1]+D_9[\text{$\mathrm{GS}_{32}$},2]\)\\
\noindent\(D_{18}\left[\mathrm{G}_{1536},3\right]\text{ = }D_5[\text{$\mathrm{O}_{24}$},3]\)\\
\noindent\(D_{18}\left[\mathrm{G}_{1536},3\right]\text{ = }D_9[\text{$\mathrm{Oh}_{48}$},3]\)\\
\subsection{Branching rules of the irreps of dimensions 6}
\label{6brancione}
\noindent\(D_{19}\left[\mathrm{G}_{1536},6\right]\text{ = }2 D_2[\text{$\mathrm{G}_{16}$},1]+2 D_3[\text{$\mathrm{G}_{16}$},1]+2 D_4[\text{$\mathrm{G}_{16}$},1]\)\\
\noindent\(D_{19}\left[\mathrm{G}_{1536},6\right]\text{ = }2 D_4[\text{$\mathrm{G}_{48}$},3]\)\\
\noindent\(D_{19}\left[\mathrm{G}_{1536},6\right]\text{ = }2 D_3[\text{$\mathrm{G}_{64}$},1]+2 D_9[\text{$\mathrm{G}_{64}$},1]+2 D_{33}[\text{$\mathrm{G}_{64}$},1]\)\\
\noindent\(D_{19}\left[\mathrm{G}_{1536},6\right]\text{ = }2 D_8[\text{$\mathrm{G}_{96}$},3]\)\\
\noindent\(D_{19}\left[\mathrm{G}_{1536},6\right]\text{ = }2 D_6[\text{$\mathrm{G}_{128}$},1]+D_9[\text{$\mathrm{G}_{128}$},1]+D_{10}[\text{$\mathrm{G}_{128}$},1]+D_{17}[\text{$\mathrm{G}_{128}$},1]+D_{18}[\text{$\mathrm{G}_{128}$},1]\)\\
\noindent\(D_{19}\left[\mathrm{G}_{1536},6\right]\text{ = }D_7[\text{$\mathrm{G}_{192}$},3]+D_8[\text{$\mathrm{G}_{192}$},3]\)\\
\noindent\(D_{19}\left[\mathrm{G}_{1536},6\right]\text{ = }D_7[\text{$\mathrm{G}_{256}$},1]+D_8[\text{$\mathrm{G}_{256}$},1]+D_{10}[\text{$\mathrm{G}_{256}$},1]+D_{12}[\text{$\mathrm{G}_{256}$},1]+D_{18}[\text{$\mathrm{G}_{256}$},1]+D_{19}[\text{$\mathrm{G}_{256}$},1]\)\\
\noindent\(D_{19}\left[\mathrm{G}_{1536},6\right]\text{ = }D_{10}[\text{$\mathrm{G}_{768}$},3]+D_{11}[\text{$\mathrm{G}_{768}$},3]\)\\
\noindent\(D_{19}\left[\mathrm{G}_{1536},6\right]\text{ = }2 D_4[\text{$\mathrm{GF}_{48}$},3]\)\\
\noindent\(D_{19}\left[\mathrm{G}_{1536},6\right]\text{ = }D_7[\text{$\mathrm{GF}_{192}$},3]+D_8[\text{$\mathrm{GF}_{192}$},3]\)\\
\noindent\(D_{19}\left[\mathrm{G}_{1536},6\right]\text{ = }D_4[\text{$\mathrm{GF}_{96}$},3]+D_5[\text{$\mathrm{GF}_{96}$},3]\)\\
\noindent\(D_{19}\left[\mathrm{G}_{1536},6\right]\text{ = }2 D_7[\text{$\mathrm{GP}_{24}$},3]\)\\
\noindent\(D_{19}\left[\mathrm{G}_{1536},6\right]\text{ = }D_4[\text{$\mathrm{GS}_{24}$},3]+D_5[\text{$\mathrm{GS}_{24}$},3]\)\\
\noindent\(D_{19}\left[\mathrm{G}_{1536},6\right]\text{ = }D_3[\text{$\mathrm{GS}_{32}$},1]+D_4[\text{$\mathrm{GS}_{32}$},1]+2 D_9[\text{$\mathrm{GS}_{32}$},2]\)\\
\noindent\(D_{19}\left[\mathrm{G}_{1536},6\right]\text{ = }D_4[\text{$\mathrm{O}_{24}$},3]+D_5[\text{$\mathrm{O}_{24}$},3]\)\\
\noindent\(D_{19}\left[\mathrm{G}_{1536},6\right]\text{ = }D_7[\text{$\mathrm{Oh}_{48}$},3]+D_9[\text{$\mathrm{Oh}_{48}$},3]\)\\
\noindent\(D_{20}\left[\mathrm{G}_{1536},6\right]\text{ = }2 D_2[\text{$\mathrm{G}_{16}$},1]+2 D_3[\text{$\mathrm{G}_{16}$},1]+2 D_4[\text{$\mathrm{G}_{16}$},1]\)\\
\noindent\(D_{20}\left[\mathrm{G}_{1536},6\right]\text{ = }2 D_4[\text{$\mathrm{G}_{48}$},3]\)\\
\noindent\(D_{20}\left[\mathrm{G}_{1536},6\right]\text{ = }2 D_{11}[\text{$\mathrm{G}_{64}$},1]+2 D_{35}[\text{$\mathrm{G}_{64}$},1]+2 D_{41}[\text{$\mathrm{G}_{64}$},1]\)\\
\noindent\(D_{20}\left[\mathrm{G}_{1536},6\right]\text{ = }2 D_8[\text{$\mathrm{G}_{96}$},3]\)\\
\noindent\(D_{20}\left[\mathrm{G}_{1536},6\right]\text{ = }D_{13}[\text{$\mathrm{G}_{128}$},1]+D_{14}[\text{$\mathrm{G}_{128}$},1]+D_{21}[\text{$\mathrm{G}_{128}$},1]+D_{22}[\text{$\mathrm{G}_{128}$},1]+2 D_{26}[\text{$\mathrm{G}_{128}$},1]\)\\
\noindent\(D_{20}\left[\mathrm{G}_{1536},6\right]\text{ = }D_7[\text{$\mathrm{G}_{192}$},3]+D_8[\text{$\mathrm{G}_{192}$},3]\)\\
\noindent\(D_{20}\left[\mathrm{G}_{1536},6\right]\text{ = }D_{14}[\text{$\mathrm{G}_{256}$},1]+D_{15}[\text{$\mathrm{G}_{256}$},1]+D_{22}[\text{$\mathrm{G}_{256}$},1]+D_{24}[\text{$\mathrm{G}_{256}$},1]+D_{27}[\text{$\mathrm{G}_{256}$},1]+D_{28}[\text{$\mathrm{G}_{256}$},1]\)\\
\noindent\(D_{20}\left[\mathrm{G}_{1536},6\right]\text{ = }D_{13}[\text{$\mathrm{G}_{768}$},3]+D_{14}[\text{$\mathrm{G}_{768}$},3]\)\\
\noindent\(D_{20}\left[\mathrm{G}_{1536},6\right]\text{ = }2 D_4[\text{$\mathrm{GF}_{48}$},3]\)\\
\noindent\(D_{20}\left[\mathrm{G}_{1536},6\right]\text{ = }D_7[\text{$\mathrm{GF}_{192}$},3]+D_8[\text{$\mathrm{GF}_{192}$},3]\)\\
\noindent\(D_{20}\left[\mathrm{G}_{1536},6\right]\text{ = }D_4[\text{$\mathrm{GF}_{96}$},3]+D_5[\text{$\mathrm{GF}_{96}$},3]\)\\
\noindent\(D_{20}\left[\mathrm{G}_{1536},6\right]\text{ = }2 D_7[\text{$\mathrm{GP}_{24}$},3]\)\\
\noindent\(D_{20}\left[\mathrm{G}_{1536},6\right]\text{ = }D_4[\text{$\mathrm{GS}_{24}$},3]+D_5[\text{$\mathrm{GS}_{24}$},3]\)\\
\noindent\(D_{20}\left[\mathrm{G}_{1536},6\right]\text{ = }D_3[\text{$\mathrm{GS}_{32}$},1]+D_4[\text{$\mathrm{GS}_{32}$},1]+2 D_9[\text{$\mathrm{GS}_{32}$},2]\)\\
\noindent\(D_{20}\left[\mathrm{G}_{1536},6\right]\text{ = }D_4[\text{$\mathrm{O}_{24}$},3]+D_5[\text{$\mathrm{O}_{24}$},3]\)\\
\noindent\(D_{20}\left[\mathrm{G}_{1536},6\right]\text{ = }D_7[\text{$\mathrm{Oh}_{48}$},3]+D_9[\text{$\mathrm{Oh}_{48}$},3]\)\\
\noindent\(D_{21}\left[\mathrm{G}_{1536},6\right]\text{ = }D_6[\text{$\mathrm{G}_{16}$},1]+D_7[\text{$\mathrm{G}_{16}$},1]+D_9[\text{$\mathrm{G}_{16}$},1]+D_{11}[\text{$\mathrm{G}_{16}$},1]+D_{15}[\text{$\mathrm{G}_{16}$},1]+D_{16}[\text{$\mathrm{G}_{16}$},1]\)\\
\noindent\(D_{21}\left[\mathrm{G}_{1536},6\right]\text{ = }D_6[\text{$\mathrm{G}_{48}$},3]+D_7[\text{$\mathrm{G}_{48}$},3]\)\\
\noindent\(D_{21}\left[\mathrm{G}_{1536},6\right]\text{ = }D_2[\text{$\mathrm{G}_{64}$},1]+D_4[\text{$\mathrm{G}_{64}$},1]+D_5[\text{$\mathrm{G}_{64}$},1]+D_{13}[\text{$\mathrm{G}_{64}$},1]+D_{17}[\text{$\mathrm{G}_{64}$},1]+D_{49}[\text{$\mathrm{G}_{64}$},1]\)\\
\noindent\(D_{21}\left[\mathrm{G}_{1536},6\right]\text{ = }D_{12}[\text{$\mathrm{G}_{96}$},3]+D_{14}[\text{$\mathrm{G}_{96}$},3]\)\\
\noindent\(D_{21}\left[\mathrm{G}_{1536},6\right]\text{ = }D_3[\text{$\mathrm{G}_{128}$},1]+D_7[\text{$\mathrm{G}_{128}$},1]+D_{33}[\text{$\mathrm{G}_{128}$},2]+D_{37}[\text{$\mathrm{G}_{128}$},2]\)\\
\noindent\(D_{21}\left[\mathrm{G}_{1536},6\right]\text{ = }D_{12}[\text{$\mathrm{G}_{192}$},3]+D_{15}[\text{$\mathrm{G}_{192}$},3]\)\\
\noindent\(D_{21}\left[\mathrm{G}_{1536},6\right]\text{ = }D_{33}[\text{$\mathrm{G}_{256}$},2]+D_{42}[\text{$\mathrm{G}_{256}$},2]+D_{45}[\text{$\mathrm{G}_{256}$},2]\)\\
\noindent\(D_{21}\left[\mathrm{G}_{1536},6\right]\text{ = }D_{23}[\text{$\mathrm{G}_{768}$},6]\)\\
\noindent\(D_{21}\left[\mathrm{G}_{1536},6\right]\text{ = }D_5[\text{$\mathrm{GF}_{48}$},3]+D_8[\text{$\mathrm{GF}_{48}$},3]\)\\
\noindent\(D_{21}\left[\mathrm{G}_{1536},6\right]\text{ = }D_{20}[\text{$\mathrm{GF}_{192}$},6]\)\\
\noindent\(D_{21}\left[\mathrm{G}_{1536},6\right]\text{ = }D_{10}[\text{$\mathrm{GF}_{96}$},6]\)\\
\noindent\(D_{21}\left[\mathrm{G}_{1536},6\right]\text{ = }D_4[\text{$\mathrm{GP}_{24}$},1]+D_5[\text{$\mathrm{GP}_{24}$},1]+D_6[\text{$\mathrm{GP}_{24}$},1]+D_8[\text{$\mathrm{GP}_{24}$},3]\)\\
\noindent\(D_{21}\left[\mathrm{G}_{1536},6\right]\text{ = }D_4[\text{$\mathrm{GS}_{24}$},3]+D_5[\text{$\mathrm{GS}_{24}$},3]\)\\
\noindent\(D_{21}\left[\mathrm{G}_{1536},6\right]\text{ = }D_5[\text{$\mathrm{GS}_{32}$},1]+D_8[\text{$\mathrm{GS}_{32}$},1]+D_{12}[\text{$\mathrm{GS}_{32}$},2]+D_{14}[\text{$\mathrm{GS}_{32}$},2]\)\\
\noindent\(D_{21}\left[\mathrm{G}_{1536},6\right]\text{ = }D_2[\text{$\mathrm{O}_{24}$},1]+D_3[\text{$\mathrm{O}_{24}$},2]+D_5[\text{$\mathrm{O}_{24}$},3]\)\\
\noindent\(D_{21}\left[\mathrm{G}_{1536},6\right]\text{ = }D_4[\text{$\mathrm{Oh}_{48}$},1]+D_6[\text{$\mathrm{Oh}_{48}$},2]+D_{10}[\text{$\mathrm{Oh}_{48}$},3]\)\\
\noindent\(D_{22}\left[\mathrm{G}_{1536},6\right]\text{ = }D_6[\text{$\mathrm{G}_{16}$},1]+D_7[\text{$\mathrm{G}_{16}$},1]+D_9[\text{$\mathrm{G}_{16}$},1]+D_{11}[\text{$\mathrm{G}_{16}$},1]+D_{15}[\text{$\mathrm{G}_{16}$},1]+D_{16}[\text{$\mathrm{G}_{16}$},1]\)\\
\noindent\(D_{22}\left[\mathrm{G}_{1536},6\right]\text{ = }D_6[\text{$\mathrm{G}_{48}$},3]+D_7[\text{$\mathrm{G}_{48}$},3]\)\\
\noindent\(D_{22}\left[\mathrm{G}_{1536},6\right]\text{ = }D_2[\text{$\mathrm{G}_{64}$},1]+D_4[\text{$\mathrm{G}_{64}$},1]+D_5[\text{$\mathrm{G}_{64}$},1]+D_{13}[\text{$\mathrm{G}_{64}$},1]+D_{17}[\text{$\mathrm{G}_{64}$},1]+D_{49}[\text{$\mathrm{G}_{64}$},1]\)\\
\noindent\(D_{22}\left[\mathrm{G}_{1536},6\right]\text{ = }D_{12}[\text{$\mathrm{G}_{96}$},3]+D_{14}[\text{$\mathrm{G}_{96}$},3]\)\\
\noindent\(D_{22}\left[\mathrm{G}_{1536},6\right]\text{ = }D_3[\text{$\mathrm{G}_{128}$},1]+D_7[\text{$\mathrm{G}_{128}$},1]+D_{33}[\text{$\mathrm{G}_{128}$},2]+D_{37}[\text{$\mathrm{G}_{128}$},2]\)\\
\noindent\(D_{22}\left[\mathrm{G}_{1536},6\right]\text{ = }D_{11}[\text{$\mathrm{G}_{192}$},3]+D_{16}[\text{$\mathrm{G}_{192}$},3]\)\\
\noindent\(D_{22}\left[\mathrm{G}_{1536},6\right]\text{ = }D_{33}[\text{$\mathrm{G}_{256}$},2]+D_{42}[\text{$\mathrm{G}_{256}$},2]+D_{45}[\text{$\mathrm{G}_{256}$},2]\)\\
\noindent\(D_{22}\left[\mathrm{G}_{1536},6\right]\text{ = }D_{23}[\text{$\mathrm{G}_{768}$},6]\)\\
\noindent\(D_{22}\left[\mathrm{G}_{1536},6\right]\text{ = }D_5[\text{$\mathrm{GF}_{48}$},3]+D_8[\text{$\mathrm{GF}_{48}$},3]\)\\
\noindent\(D_{22}\left[\mathrm{G}_{1536},6\right]\text{ = }D_{20}[\text{$\mathrm{GF}_{192}$},6]\)\\
\noindent\(D_{22}\left[\mathrm{G}_{1536},6\right]\text{ = }D_{10}[\text{$\mathrm{GF}_{96}$},6]\)\\
\noindent\(D_{22}\left[\mathrm{G}_{1536},6\right]\text{ = }D_4[\text{$\mathrm{GP}_{24}$},1]+D_5[\text{$\mathrm{GP}_{24}$},1]+D_6[\text{$\mathrm{GP}_{24}$},1]+D_8[\text{$\mathrm{GP}_{24}$},3]\)\\
\noindent\(D_{22}\left[\mathrm{G}_{1536},6\right]\text{ = }D_4[\text{$\mathrm{GS}_{24}$},3]+D_5[\text{$\mathrm{GS}_{24}$},3]\)\\
\noindent\(D_{22}\left[\mathrm{G}_{1536},6\right]\text{ = }D_6[\text{$\mathrm{GS}_{32}$},1]+D_7[\text{$\mathrm{GS}_{32}$},1]+D_{12}[\text{$\mathrm{GS}_{32}$},2]+D_{14}[\text{$\mathrm{GS}_{32}$},2]\)\\
\noindent\(D_{22}\left[\mathrm{G}_{1536},6\right]\text{ = }D_1[\text{$\mathrm{O}_{24}$},1]+D_3[\text{$\mathrm{O}_{24}$},2]+D_4[\text{$\mathrm{O}_{24}$},3]\)\\
\noindent\(D_{22}\left[\mathrm{G}_{1536},6\right]\text{ = }D_2[\text{$\mathrm{Oh}_{48}$},1]+D_6[\text{$\mathrm{Oh}_{48}$},2]+D_8[\text{$\mathrm{Oh}_{48}$},3]\)\\
\noindent\(D_{23}\left[\mathrm{G}_{1536},6\right]\text{ = }D_5[\text{$\mathrm{G}_{16}$},1]+D_8[\text{$\mathrm{G}_{16}$},1]+D_{10}[\text{$\mathrm{G}_{16}$},1]+D_{12}[\text{$\mathrm{G}_{16}$},1]+D_{13}[\text{$\mathrm{G}_{16}$},1]+D_{14}[\text{$\mathrm{G}_{16}$},1]\)\\
\noindent\(D_{23}\left[\mathrm{G}_{1536},6\right]\text{ = }D_5[\text{$\mathrm{G}_{48}$},3]+D_8[\text{$\mathrm{G}_{48}$},3]\)\\
\noindent\(D_{23}\left[\mathrm{G}_{1536},6\right]\text{ = }D_2[\text{$\mathrm{G}_{64}$},1]+D_4[\text{$\mathrm{G}_{64}$},1]+D_5[\text{$\mathrm{G}_{64}$},1]+D_{13}[\text{$\mathrm{G}_{64}$},1]+D_{17}[\text{$\mathrm{G}_{64}$},1]+D_{49}[\text{$\mathrm{G}_{64}$},1]\)\\
\noindent\(D_{23}\left[\mathrm{G}_{1536},6\right]\text{ = }D_{10}[\text{$\mathrm{G}_{96}$},3]+D_{16}[\text{$\mathrm{G}_{96}$},3]\)\\
\noindent\(D_{23}\left[\mathrm{G}_{1536},6\right]\text{ = }D_4[\text{$\mathrm{G}_{128}$},1]+D_8[\text{$\mathrm{G}_{128}$},1]+D_{33}[\text{$\mathrm{G}_{128}$},2]+D_{37}[\text{$\mathrm{G}_{128}$},2]\)\\
\noindent\(D_{23}\left[\mathrm{G}_{1536},6\right]\text{ = }D_{20}[\text{$\mathrm{G}_{192}$},6]\)\\
\noindent\(D_{23}\left[\mathrm{G}_{1536},6\right]\text{ = }D_{34}[\text{$\mathrm{G}_{256}$},2]+D_{41}[\text{$\mathrm{G}_{256}$},2]+D_{46}[\text{$\mathrm{G}_{256}$},2]\)\\
\noindent\(D_{23}\left[\mathrm{G}_{1536},6\right]\text{ = }D_{24}[\text{$\mathrm{G}_{768}$},6]\)\\
\noindent\(D_{23}\left[\mathrm{G}_{1536},6\right]\text{ = }D_6[\text{$\mathrm{GF}_{48}$},3]+D_7[\text{$\mathrm{GF}_{48}$},3]\)\\
\noindent\(D_{23}\left[\mathrm{G}_{1536},6\right]\text{ = }D_{12}[\text{$\mathrm{GF}_{192}$},3]+D_{15}[\text{$\mathrm{GF}_{192}$},3]\)\\
\noindent\(D_{23}\left[\mathrm{G}_{1536},6\right]\text{ = }D_6[\text{$\mathrm{GF}_{96}$},3]+D_9[\text{$\mathrm{GF}_{96}$},3]\)\\
\noindent\(D_{23}\left[\mathrm{G}_{1536},6\right]\text{ = }2 D_8[\text{$\mathrm{GP}_{24}$},3]\)\\
\noindent\(D_{23}\left[\mathrm{G}_{1536},6\right]\text{ = }D_1[\text{$\mathrm{GS}_{24}$},1]+D_3[\text{$\mathrm{GS}_{24}$},2]+D_4[\text{$\mathrm{GS}_{24}$},3]\)\\
\noindent\(D_{23}\left[\mathrm{G}_{1536},6\right]\text{ = }D_{10}[\text{$\mathrm{GS}_{32}$},2]+D_{11}[\text{$\mathrm{GS}_{32}$},2]+D_{13}[\text{$\mathrm{GS}_{32}$},2]\)\\
\noindent\(D_{23}\left[\mathrm{G}_{1536},6\right]\text{ = }D_4[\text{$\mathrm{O}_{24}$},3]+D_5[\text{$\mathrm{O}_{24}$},3]\)\\
\noindent\(D_{23}\left[\mathrm{G}_{1536},6\right]\text{ = }D_8[\text{$\mathrm{Oh}_{48}$},3]+D_{10}[\text{$\mathrm{Oh}_{48}$},3]\)\\
\noindent\(D_{24}\left[\mathrm{G}_{1536},6\right]\text{ = }D_5[\text{$\mathrm{G}_{16}$},1]+D_8[\text{$\mathrm{G}_{16}$},1]+D_{10}[\text{$\mathrm{G}_{16}$},1]+D_{12}[\text{$\mathrm{G}_{16}$},1]+D_{13}[\text{$\mathrm{G}_{16}$},1]+D_{14}[\text{$\mathrm{G}_{16}$},1]\)\\
\noindent\(D_{24}\left[\mathrm{G}_{1536},6\right]\text{ = }D_5[\text{$\mathrm{G}_{48}$},3]+D_8[\text{$\mathrm{G}_{48}$},3]\)\\
\noindent\(D_{24}\left[\mathrm{G}_{1536},6\right]\text{ = }D_2[\text{$\mathrm{G}_{64}$},1]+D_4[\text{$\mathrm{G}_{64}$},1]+D_5[\text{$\mathrm{G}_{64}$},1]+D_{13}[\text{$\mathrm{G}_{64}$},1]+D_{17}[\text{$\mathrm{G}_{64}$},1]+D_{49}[\text{$\mathrm{G}_{64}$},1]\)\\
\noindent\(D_{24}\left[\mathrm{G}_{1536},6\right]\text{ = }D_{10}[\text{$\mathrm{G}_{96}$},3]+D_{16}[\text{$\mathrm{G}_{96}$},3]\)\\
\noindent\(D_{24}\left[\mathrm{G}_{1536},6\right]\text{ = }D_4[\text{$\mathrm{G}_{128}$},1]+D_8[\text{$\mathrm{G}_{128}$},1]+D_{33}[\text{$\mathrm{G}_{128}$},2]+D_{37}[\text{$\mathrm{G}_{128}$},2]\)\\
\noindent\(D_{24}\left[\mathrm{G}_{1536},6\right]\text{ = }D_{20}[\text{$\mathrm{G}_{192}$},6]\)\\
\noindent\(D_{24}\left[\mathrm{G}_{1536},6\right]\text{ = }D_{34}[\text{$\mathrm{G}_{256}$},2]+D_{41}[\text{$\mathrm{G}_{256}$},2]+D_{46}[\text{$\mathrm{G}_{256}$},2]\)\\
\noindent\(D_{24}\left[\mathrm{G}_{1536},6\right]\text{ = }D_{24}[\text{$\mathrm{G}_{768}$},6]\)\\
\noindent\(D_{24}\left[\mathrm{G}_{1536},6\right]\text{ = }D_6[\text{$\mathrm{GF}_{48}$},3]+D_7[\text{$\mathrm{GF}_{48}$},3]\)\\
\noindent\(D_{24}\left[\mathrm{G}_{1536},6\right]\text{ = }D_{11}[\text{$\mathrm{GF}_{192}$},3]+D_{16}[\text{$\mathrm{GF}_{192}$},3]\)\\
\noindent\(D_{24}\left[\mathrm{G}_{1536},6\right]\text{ = }D_7[\text{$\mathrm{GF}_{96}$},3]+D_8[\text{$\mathrm{GF}_{96}$},3]\)\\
\noindent\(D_{24}\left[\mathrm{G}_{1536},6\right]\text{ = }2 D_8[\text{$\mathrm{GP}_{24}$},3]\)\\
\noindent\(D_{24}\left[\mathrm{G}_{1536},6\right]\text{ = }D_2[\text{$\mathrm{GS}_{24}$},1]+D_3[\text{$\mathrm{GS}_{24}$},2]+D_5[\text{$\mathrm{GS}_{24}$},3]\)\\
\noindent\(D_{24}\left[\mathrm{G}_{1536},6\right]\text{ = }D_{10}[\text{$\mathrm{GS}_{32}$},2]+D_{11}[\text{$\mathrm{GS}_{32}$},2]+D_{13}[\text{$\mathrm{GS}_{32}$},2]\)\\
\noindent\(D_{24}\left[\mathrm{G}_{1536},6\right]\text{ = }D_4[\text{$\mathrm{O}_{24}$},3]+D_5[\text{$\mathrm{O}_{24}$},3]\)\\
\noindent\(D_{24}\left[\mathrm{G}_{1536},6\right]\text{ = }D_8[\text{$\mathrm{Oh}_{48}$},3]+D_{10}[\text{$\mathrm{Oh}_{48}$},3]\)\\
\noindent\(D_{25}\left[\mathrm{G}_{1536},6\right]\text{ = }D_6[\text{$\mathrm{G}_{16}$},1]+D_7[\text{$\mathrm{G}_{16}$},1]+D_9[\text{$\mathrm{G}_{16}$},1]+D_{11}[\text{$\mathrm{G}_{16}$},1]+D_{15}[\text{$\mathrm{G}_{16}$},1]+D_{16}[\text{$\mathrm{G}_{16}$},1]\)\\
\noindent\(D_{25}\left[\mathrm{G}_{1536},6\right]\text{ = }D_6[\text{$\mathrm{G}_{48}$},3]+D_7[\text{$\mathrm{G}_{48}$},3]\)\\
\noindent\(D_{25}\left[\mathrm{G}_{1536},6\right]\text{ = }D_{27}[\text{$\mathrm{G}_{64}$},1]+D_{39}[\text{$\mathrm{G}_{64}$},1]+D_{42}[\text{$\mathrm{G}_{64}$},1]+D_{44}[\text{$\mathrm{G}_{64}$},1]+D_{47}[\text{$\mathrm{G}_{64}$},1]+D_{59}[\text{$\mathrm{G}_{64}$},1]\)\\
\noindent\(D_{25}\left[\mathrm{G}_{1536},6\right]\text{ = }D_{12}[\text{$\mathrm{G}_{96}$},3]+D_{14}[\text{$\mathrm{G}_{96}$},3]\)\\
\noindent\(D_{25}\left[\mathrm{G}_{1536},6\right]\text{ = }D_{27}[\text{$\mathrm{G}_{128}$},1]+D_{31}[\text{$\mathrm{G}_{128}$},1]+D_{47}[\text{$\mathrm{G}_{128}$},2]+D_{55}[\text{$\mathrm{G}_{128}$},2]\)\\
\noindent\(D_{25}\left[\mathrm{G}_{1536},6\right]\text{ = }D_{12}[\text{$\mathrm{G}_{192}$},3]+D_{15}[\text{$\mathrm{G}_{192}$},3]\)\\
\noindent\(D_{25}\left[\mathrm{G}_{1536},6\right]\text{ = }D_{39}[\text{$\mathrm{G}_{256}$},2]+D_{51}[\text{$\mathrm{G}_{256}$},2]+D_{56}[\text{$\mathrm{G}_{256}$},2]\)\\
\noindent\(D_{25}\left[\mathrm{G}_{1536},6\right]\text{ = }D_{29}[\text{$\mathrm{G}_{768}$},6]\)\\
\noindent\(D_{25}\left[\mathrm{G}_{1536},6\right]\text{ = }D_5[\text{$\mathrm{GF}_{48}$},3]+D_8[\text{$\mathrm{GF}_{48}$},3]\)\\
\noindent\(D_{25}\left[\mathrm{G}_{1536},6\right]\text{ = }D_{20}[\text{$\mathrm{GF}_{192}$},6]\)\\
\noindent\(D_{25}\left[\mathrm{G}_{1536},6\right]\text{ = }D_{10}[\text{$\mathrm{GF}_{96}$},6]\)\\
\noindent\(D_{25}\left[\mathrm{G}_{1536},6\right]\text{ = }D_4[\text{$\mathrm{GP}_{24}$},1]+D_5[\text{$\mathrm{GP}_{24}$},1]+D_6[\text{$\mathrm{GP}_{24}$},1]+D_8[\text{$\mathrm{GP}_{24}$},3]\)\\
\noindent\(D_{25}\left[\mathrm{G}_{1536},6\right]\text{ = }D_4[\text{$\mathrm{GS}_{24}$},3]+D_5[\text{$\mathrm{GS}_{24}$},3]\)\\
\noindent\(D_{25}\left[\mathrm{G}_{1536},6\right]\text{ = }D_5[\text{$\mathrm{GS}_{32}$},1]+D_8[\text{$\mathrm{GS}_{32}$},1]+D_{12}[\text{$\mathrm{GS}_{32}$},2]+D_{14}[\text{$\mathrm{GS}_{32}$},2]\)\\
\noindent\(D_{25}\left[\mathrm{G}_{1536},6\right]\text{ = }D_2[\text{$\mathrm{O}_{24}$},1]+D_3[\text{$\mathrm{O}_{24}$},2]+D_5[\text{$\mathrm{O}_{24}$},3]\)\\
\noindent\(D_{25}\left[\mathrm{G}_{1536},6\right]\text{ = }D_4[\text{$\mathrm{Oh}_{48}$},1]+D_6[\text{$\mathrm{Oh}_{48}$},2]+D_{10}[\text{$\mathrm{Oh}_{48}$},3]\)\\
\noindent\(D_{26}\left[\mathrm{G}_{1536},6\right]\text{ = }D_6[\text{$\mathrm{G}_{16}$},1]+D_7[\text{$\mathrm{G}_{16}$},1]+D_9[\text{$\mathrm{G}_{16}$},1]+D_{11}[\text{$\mathrm{G}_{16}$},1]+D_{15}[\text{$\mathrm{G}_{16}$},1]+D_{16}[\text{$\mathrm{G}_{16}$},1]\)\\
\noindent\(D_{26}\left[\mathrm{G}_{1536},6\right]\text{ = }D_6[\text{$\mathrm{G}_{48}$},3]+D_7[\text{$\mathrm{G}_{48}$},3]\)\\
\noindent\(D_{26}\left[\mathrm{G}_{1536},6\right]\text{ = }D_{27}[\text{$\mathrm{G}_{64}$},1]+D_{39}[\text{$\mathrm{G}_{64}$},1]+D_{42}[\text{$\mathrm{G}_{64}$},1]+D_{44}[\text{$\mathrm{G}_{64}$},1]+D_{47}[\text{$\mathrm{G}_{64}$},1]+D_{59}[\text{$\mathrm{G}_{64}$},1]\)\\
\noindent\(D_{26}\left[\mathrm{G}_{1536},6\right]\text{ = }D_{12}[\text{$\mathrm{G}_{96}$},3]+D_{14}[\text{$\mathrm{G}_{96}$},3]\)\\
\noindent\(D_{26}\left[\mathrm{G}_{1536},6\right]\text{ = }D_{27}[\text{$\mathrm{G}_{128}$},1]+D_{31}[\text{$\mathrm{G}_{128}$},1]+D_{47}[\text{$\mathrm{G}_{128}$},2]+D_{55}[\text{$\mathrm{G}_{128}$},2]\)\\
\noindent\(D_{26}\left[\mathrm{G}_{1536},6\right]\text{ = }D_{11}[\text{$\mathrm{G}_{192}$},3]+D_{16}[\text{$\mathrm{G}_{192}$},3]\)\\
\noindent\(D_{26}\left[\mathrm{G}_{1536},6\right]\text{ = }D_{39}[\text{$\mathrm{G}_{256}$},2]+D_{51}[\text{$\mathrm{G}_{256}$},2]+D_{56}[\text{$\mathrm{G}_{256}$},2]\)\\
\noindent\(D_{26}\left[\mathrm{G}_{1536},6\right]\text{ = }D_{29}[\text{$\mathrm{G}_{768}$},6]\)\\
\noindent\(D_{26}\left[\mathrm{G}_{1536},6\right]\text{ = }D_5[\text{$\mathrm{GF}_{48}$},3]+D_8[\text{$\mathrm{GF}_{48}$},3]\)\\
\noindent\(D_{26}\left[\mathrm{G}_{1536},6\right]\text{ = }D_{20}[\text{$\mathrm{GF}_{192}$},6]\)\\
\noindent\(D_{26}\left[\mathrm{G}_{1536},6\right]\text{ = }D_{10}[\text{$\mathrm{GF}_{96}$},6]\)\\
\noindent\(D_{26}\left[\mathrm{G}_{1536},6\right]\text{ = }D_4[\text{$\mathrm{GP}_{24}$},1]+D_5[\text{$\mathrm{GP}_{24}$},1]+D_6[\text{$\mathrm{GP}_{24}$},1]+D_8[\text{$\mathrm{GP}_{24}$},3]\)\\
\noindent\(D_{26}\left[\mathrm{G}_{1536},6\right]\text{ = }D_4[\text{$\mathrm{GS}_{24}$},3]+D_5[\text{$\mathrm{GS}_{24}$},3]\)\\
\noindent\(D_{26}\left[\mathrm{G}_{1536},6\right]\text{ = }D_6[\text{$\mathrm{GS}_{32}$},1]+D_7[\text{$\mathrm{GS}_{32}$},1]+D_{12}[\text{$\mathrm{GS}_{32}$},2]+D_{14}[\text{$\mathrm{GS}_{32}$},2]\)\\
\noindent\(D_{26}\left[\mathrm{G}_{1536},6\right]\text{ = }D_1[\text{$\mathrm{O}_{24}$},1]+D_3[\text{$\mathrm{O}_{24}$},2]+D_4[\text{$\mathrm{O}_{24}$},3]\)\\
\noindent\(D_{26}\left[\mathrm{G}_{1536},6\right]\text{ = }D_2[\text{$\mathrm{Oh}_{48}$},1]+D_6[\text{$\mathrm{Oh}_{48}$},2]+D_8[\text{$\mathrm{Oh}_{48}$},3]\)\\
\noindent\(D_{27}\left[\mathrm{G}_{1536},6\right]\text{ = }D_5[\text{$\mathrm{G}_{16}$},1]+D_8[\text{$\mathrm{G}_{16}$},1]+D_{10}[\text{$\mathrm{G}_{16}$},1]+D_{12}[\text{$\mathrm{G}_{16}$},1]+D_{13}[\text{$\mathrm{G}_{16}$},1]+D_{14}[\text{$\mathrm{G}_{16}$},1]\)\\
\noindent\(D_{27}\left[\mathrm{G}_{1536},6\right]\text{ = }D_5[\text{$\mathrm{G}_{48}$},3]+D_8[\text{$\mathrm{G}_{48}$},3]\)\\
\noindent\(D_{27}\left[\mathrm{G}_{1536},6\right]\text{ = }D_{27}[\text{$\mathrm{G}_{64}$},1]+D_{39}[\text{$\mathrm{G}_{64}$},1]+D_{42}[\text{$\mathrm{G}_{64}$},1]+D_{44}[\text{$\mathrm{G}_{64}$},1]+D_{47}[\text{$\mathrm{G}_{64}$},1]+D_{59}[\text{$\mathrm{G}_{64}$},1]\)\\
\noindent\(D_{27}\left[\mathrm{G}_{1536},6\right]\text{ = }D_{10}[\text{$\mathrm{G}_{96}$},3]+D_{16}[\text{$\mathrm{G}_{96}$},3]\)\\
\noindent\(D_{27}\left[\mathrm{G}_{1536},6\right]\text{ = }D_{28}[\text{$\mathrm{G}_{128}$},1]+D_{32}[\text{$\mathrm{G}_{128}$},1]+D_{47}[\text{$\mathrm{G}_{128}$},2]+D_{55}[\text{$\mathrm{G}_{128}$},2]\)\\
\noindent\(D_{27}\left[\mathrm{G}_{1536},6\right]\text{ = }D_{20}[\text{$\mathrm{G}_{192}$},6]\)\\
\noindent\(D_{27}\left[\mathrm{G}_{1536},6\right]\text{ = }D_{40}[\text{$\mathrm{G}_{256}$},2]+D_{52}[\text{$\mathrm{G}_{256}$},2]+D_{55}[\text{$\mathrm{G}_{256}$},2]\)\\
\noindent\(D_{27}\left[\mathrm{G}_{1536},6\right]\text{ = }D_{30}[\text{$\mathrm{G}_{768}$},6]\)\\
\noindent\(D_{27}\left[\mathrm{G}_{1536},6\right]\text{ = }D_6[\text{$\mathrm{GF}_{48}$},3]+D_7[\text{$\mathrm{GF}_{48}$},3]\)\\
\noindent\(D_{27}\left[\mathrm{G}_{1536},6\right]\text{ = }D_{12}[\text{$\mathrm{GF}_{192}$},3]+D_{15}[\text{$\mathrm{GF}_{192}$},3]\)\\
\noindent\(D_{27}\left[\mathrm{G}_{1536},6\right]\text{ = }D_6[\text{$\mathrm{GF}_{96}$},3]+D_9[\text{$\mathrm{GF}_{96}$},3]\)\\
\noindent\(D_{27}\left[\mathrm{G}_{1536},6\right]\text{ = }2 D_8[\text{$\mathrm{GP}_{24}$},3]\)\\
\noindent\(D_{27}\left[\mathrm{G}_{1536},6\right]\text{ = }D_1[\text{$\mathrm{GS}_{24}$},1]+D_3[\text{$\mathrm{GS}_{24}$},2]+D_4[\text{$\mathrm{GS}_{24}$},3]\)\\
\noindent\(D_{27}\left[\mathrm{G}_{1536},6\right]\text{ = }D_{10}[\text{$\mathrm{GS}_{32}$},2]+D_{11}[\text{$\mathrm{GS}_{32}$},2]+D_{13}[\text{$\mathrm{GS}_{32}$},2]\)\\
\noindent\(D_{27}\left[\mathrm{G}_{1536},6\right]\text{ = }D_4[\text{$\mathrm{O}_{24}$},3]+D_5[\text{$\mathrm{O}_{24}$},3]\)\\
\noindent\(D_{27}\left[\mathrm{G}_{1536},6\right]\text{ = }D_8[\text{$\mathrm{Oh}_{48}$},3]+D_{10}[\text{$\mathrm{Oh}_{48}$},3]\)\\
\noindent\(D_{28}\left[\mathrm{G}_{1536},6\right]\text{ = }D_5[\text{$\mathrm{G}_{16}$},1]+D_8[\text{$\mathrm{G}_{16}$},1]+D_{10}[\text{$\mathrm{G}_{16}$},1]+D_{12}[\text{$\mathrm{G}_{16}$},1]+D_{13}[\text{$\mathrm{G}_{16}$},1]+D_{14}[\text{$\mathrm{G}_{16}$},1]\)\\
\noindent\(D_{28}\left[\mathrm{G}_{1536},6\right]\text{ = }D_5[\text{$\mathrm{G}_{48}$},3]+D_8[\text{$\mathrm{G}_{48}$},3]\)\\
\noindent\(D_{28}\left[\mathrm{G}_{1536},6\right]\text{ = }D_{27}[\text{$\mathrm{G}_{64}$},1]+D_{39}[\text{$\mathrm{G}_{64}$},1]+D_{42}[\text{$\mathrm{G}_{64}$},1]+D_{44}[\text{$\mathrm{G}_{64}$},1]+D_{47}[\text{$\mathrm{G}_{64}$},1]+D_{59}[\text{$\mathrm{G}_{64}$},1]\)\\
\noindent\(D_{28}\left[\mathrm{G}_{1536},6\right]\text{ = }D_{10}[\text{$\mathrm{G}_{96}$},3]+D_{16}[\text{$\mathrm{G}_{96}$},3]\)\\
\noindent\(D_{28}\left[\mathrm{G}_{1536},6\right]\text{ = }D_{28}[\text{$\mathrm{G}_{128}$},1]+D_{32}[\text{$\mathrm{G}_{128}$},1]+D_{47}[\text{$\mathrm{G}_{128}$},2]+D_{55}[\text{$\mathrm{G}_{128}$},2]\)\\
\noindent\(D_{28}\left[\mathrm{G}_{1536},6\right]\text{ = }D_{20}[\text{$\mathrm{G}_{192}$},6]\)\\
\noindent\(D_{28}\left[\mathrm{G}_{1536},6\right]\text{ = }D_{40}[\text{$\mathrm{G}_{256}$},2]+D_{52}[\text{$\mathrm{G}_{256}$},2]+D_{55}[\text{$\mathrm{G}_{256}$},2]\)\\
\noindent\(D_{28}\left[\mathrm{G}_{1536},6\right]\text{ = }D_{30}[\text{$\mathrm{G}_{768}$},6]\)\\
\noindent\(D_{28}\left[\mathrm{G}_{1536},6\right]\text{ = }D_6[\text{$\mathrm{GF}_{48}$},3]+D_7[\text{$\mathrm{GF}_{48}$},3]\)\\
\noindent\(D_{28}\left[\mathrm{G}_{1536},6\right]\text{ = }D_{11}[\text{$\mathrm{GF}_{192}$},3]+D_{16}[\text{$\mathrm{GF}_{192}$},3]\)\\
\noindent\(D_{28}\left[\mathrm{G}_{1536},6\right]\text{ = }D_7[\text{$\mathrm{GF}_{96}$},3]+D_8[\text{$\mathrm{GF}_{96}$},3]\)\\
\noindent\(D_{28}\left[\mathrm{G}_{1536},6\right]\text{ = }2 D_8[\text{$\mathrm{GP}_{24}$},3]\)\\
\noindent\(D_{28}\left[\mathrm{G}_{1536},6\right]\text{ = }D_2[\text{$\mathrm{GS}_{24}$},1]+D_3[\text{$\mathrm{GS}_{24}$},2]+D_5[\text{$\mathrm{GS}_{24}$},3]\)\\
\noindent\(D_{28}\left[\mathrm{G}_{1536},6\right]\text{ = }D_{10}[\text{$\mathrm{GS}_{32}$},2]+D_{11}[\text{$\mathrm{GS}_{32}$},2]+D_{13}[\text{$\mathrm{GS}_{32}$},2]\)\\
\noindent\(D_{28}\left[\mathrm{G}_{1536},6\right]\text{ = }D_4[\text{$\mathrm{O}_{24}$},3]+D_5[\text{$\mathrm{O}_{24}$},3]\)\\
\noindent\(D_{28}\left[\mathrm{G}_{1536},6\right]\text{ = }D_8[\text{$\mathrm{Oh}_{48}$},3]+D_{10}[\text{$\mathrm{Oh}_{48}$},3]\)\\
\subsection{Branching rules of the irreps of dimension 8}
\label{8brancione}
\noindent\(D_{29}\left[\mathrm{G}_{1536},8\right]\text{ = }2 D_1[\text{$\mathrm{G}_{16}$},1]+2 D_2[\text{$\mathrm{G}_{16}$},1]+2 D_3[\text{$\mathrm{G}_{16}$},1]+2 D_4[\text{$\mathrm{G}_{16}$},1]\)\\
\noindent\(D_{29}\left[\mathrm{G}_{1536},8\right]\text{ = }2 D_1[\text{$\mathrm{G}_{48}$},1]+2 D_4[\text{$\mathrm{G}_{48}$},3]\)\\
\noindent\(D_{29}\left[\mathrm{G}_{1536},8\right]\text{ = }D_{22}[\text{$\mathrm{G}_{64}$},1]+D_{24}[\text{$\mathrm{G}_{64}$},1]+D_{30}[\text{$\mathrm{G}_{64}$},1]+D_{32}[\text{$\mathrm{G}_{64}$},1]+D_{54}[\text{$\mathrm{G}_{64}$},1]+D_{56}[\text{$\mathrm{G}_{64}$},1]+D_{62}[\text{$\mathrm{G}_{64}$},1]+D_{64}[\text{$\mathrm{G}_{64}$},1]\)\\
\noindent\(D_{29}\left[\mathrm{G}_{1536},8\right]\text{ = }2 D_2[\text{$\mathrm{G}_{96}$},1]+2 D_7[\text{$\mathrm{G}_{96}$},3]\)\\
\noindent\(D_{29}\left[\mathrm{G}_{1536},8\right]\text{ = }D_{42}[\text{$\mathrm{G}_{128}$},2]+D_{44}[\text{$\mathrm{G}_{128}$},2]+D_{50}[\text{$\mathrm{G}_{128}$},2]+D_{52}[\text{$\mathrm{G}_{128}$},2]\)\\
\noindent\(D_{29}\left[\mathrm{G}_{1536},8\right]\text{ = }D_3[\text{$\mathrm{G}_{192}$},1]+D_4[\text{$\mathrm{G}_{192}$},1]+D_5[\text{$\mathrm{G}_{192}$},3]+D_6[\text{$\mathrm{G}_{192}$},3]\)\\
\noindent\(D_{29}\left[\mathrm{G}_{1536},8\right]\text{ = }D_{60}[\text{$\mathrm{G}_{256}$},4]+D_{62}[\text{$\mathrm{G}_{256}$},4]\)\\
\noindent\(D_{29}\left[\mathrm{G}_{1536},8\right]\text{ = }D_{17}[\text{$\mathrm{G}_{768}$},4]+D_{20}[\text{$\mathrm{G}_{768}$},4]\)\\
\noindent\(D_{29}\left[\mathrm{G}_{1536},8\right]\text{ = }2 D_1[\text{$\mathrm{GF}_{48}$},1]+2 D_4[\text{$\mathrm{GF}_{48}$},3]\)\\
\noindent\(D_{29}\left[\mathrm{G}_{1536},8\right]\text{ = }D_3[\text{$\mathrm{GF}_{192}$},1]+D_4[\text{$\mathrm{GF}_{192}$},1]+D_5[\text{$\mathrm{GF}_{192}$},3]+D_6[\text{$\mathrm{GF}_{192}$},3]\)\\
\noindent\(D_{29}\left[\mathrm{G}_{1536},8\right]\text{ = }D_1[\text{$\mathrm{GF}_{96}$},1]+D_2[\text{$\mathrm{GF}_{96}$},1]+D_4[\text{$\mathrm{GF}_{96}$},3]+D_5[\text{$\mathrm{GF}_{96}$},3]\)\\
\noindent\(D_{29}\left[\mathrm{G}_{1536},8\right]\text{ = }2 D_4[\text{$\mathrm{GP}_{24}$},1]+2 D_8[\text{$\mathrm{GP}_{24}$},3]\)\\
\noindent\(D_{29}\left[\mathrm{G}_{1536},8\right]\text{ = }D_1[\text{$\mathrm{GS}_{24}$},1]+D_2[\text{$\mathrm{GS}_{24}$},1]+D_4[\text{$\mathrm{GS}_{24}$},3]+D_5[\text{$\mathrm{GS}_{24}$},3]\)\\
\noindent\(D_{29}\left[\mathrm{G}_{1536},8\right]\text{ = }D_1[\text{$\mathrm{GS}_{32}$},1]+D_2[\text{$\mathrm{GS}_{32}$},1]+D_3[\text{$\mathrm{GS}_{32}$},1]+D_4[\text{$\mathrm{GS}_{32}$},1]+2 D_9[\text{$\mathrm{GS}_{32}$},2]\)\\
\noindent\(D_{29}\left[\mathrm{G}_{1536},8\right]\text{ = }D_1[\text{$\mathrm{O}_{24}$},1]+D_2[\text{$\mathrm{O}_{24}$},1]+D_4[\text{$\mathrm{O}_{24}$},3]+D_5[\text{$\mathrm{O}_{24}$},3]\)\\
\noindent\(D_{29}\left[\mathrm{G}_{1536},8\right]\text{ = }D_2[\text{$\mathrm{Oh}_{48}$},1]+D_4[\text{$\mathrm{Oh}_{48}$},1]+D_8[\text{$\mathrm{Oh}_{48}$},3]+D_{10}[\text{$\mathrm{Oh}_{48}$},3]\)\\
\noindent\(D_{30}\left[\mathrm{G}_{1536},8\right]\text{ = }2 D_1[\text{$\mathrm{G}_{16}$},1]+2 D_2[\text{$\mathrm{G}_{16}$},1]+2 D_3[\text{$\mathrm{G}_{16}$},1]+2 D_4[\text{$\mathrm{G}_{16}$},1]\)\\
\noindent\(D_{30}\left[\mathrm{G}_{1536},8\right]\text{ = }D_2[\text{$\mathrm{G}_{48}$},1]+D_3[\text{$\mathrm{G}_{48}$},1]+2 D_4[\text{$\mathrm{G}_{48}$},3]\)\\
\noindent\(D_{30}\left[\mathrm{G}_{1536},8\right]\text{ = }D_{22}[\text{$\mathrm{G}_{64}$},1]+D_{24}[\text{$\mathrm{G}_{64}$},1]+D_{30}[\text{$\mathrm{G}_{64}$},1]+D_{32}[\text{$\mathrm{G}_{64}$},1]+D_{54}[\text{$\mathrm{G}_{64}$},1]+D_{56}[\text{$\mathrm{G}_{64}$},1]+D_{62}[\text{$\mathrm{G}_{64}$},1]+D_{64}[\text{$\mathrm{G}_{64}$},1]\)\\
\noindent\(D_{30}\left[\mathrm{G}_{1536},8\right]\text{ = }D_4[\text{$\mathrm{G}_{96}$},1]+D_6[\text{$\mathrm{G}_{96}$},1]+2 D_7[\text{$\mathrm{G}_{96}$},3]\)\\
\noindent\(D_{30}\left[\mathrm{G}_{1536},8\right]\text{ = }D_{42}[\text{$\mathrm{G}_{128}$},2]+D_{44}[\text{$\mathrm{G}_{128}$},2]+D_{50}[\text{$\mathrm{G}_{128}$},2]+D_{52}[\text{$\mathrm{G}_{128}$},2]\)\\
\noindent\(D_{30}\left[\mathrm{G}_{1536},8\right]\text{ = }D_5[\text{$\mathrm{G}_{192}$},3]+D_6[\text{$\mathrm{G}_{192}$},3]+D_{18}[\text{$\mathrm{G}_{192}$},2]\)\\
\noindent\(D_{30}\left[\mathrm{G}_{1536},8\right]\text{ = }D_{60}[\text{$\mathrm{G}_{256}$},4]+D_{62}[\text{$\mathrm{G}_{256}$},4]\)\\
\noindent\(D_{30}\left[\mathrm{G}_{1536},8\right]\text{ = }D_{18}[\text{$\mathrm{G}_{768}$},4]+D_{22}[\text{$\mathrm{G}_{768}$},4]\)\\
\noindent\(D_{30}\left[\mathrm{G}_{1536},8\right]\text{ = }D_2[\text{$\mathrm{GF}_{48}$},1]+D_3[\text{$\mathrm{GF}_{48}$},1]+2 D_4[\text{$\mathrm{GF}_{48}$},3]\)\\
\noindent\(D_{30}\left[\mathrm{G}_{1536},8\right]\text{ = }D_5[\text{$\mathrm{GF}_{192}$},3]+D_6[\text{$\mathrm{GF}_{192}$},3]+D_{18}[\text{$\mathrm{GF}_{192}$},2]\)\\
\noindent\(D_{30}\left[\mathrm{G}_{1536},8\right]\text{ = }D_3[\text{$\mathrm{GF}_{96}$},2]+D_4[\text{$\mathrm{GF}_{96}$},3]+D_5[\text{$\mathrm{GF}_{96}$},3]\)\\
\noindent\(D_{30}\left[\mathrm{G}_{1536},8\right]\text{ = }D_5[\text{$\mathrm{GP}_{24}$},1]+D_6[\text{$\mathrm{GP}_{24}$},1]+2 D_8[\text{$\mathrm{GP}_{24}$},3]\)\\
\noindent\(D_{30}\left[\mathrm{G}_{1536},8\right]\text{ = }D_3[\text{$\mathrm{GS}_{24}$},2]+D_4[\text{$\mathrm{GS}_{24}$},3]+D_5[\text{$\mathrm{GS}_{24}$},3]\)\\
\noindent\(D_{30}\left[\mathrm{G}_{1536},8\right]\text{ = }D_1[\text{$\mathrm{GS}_{32}$},1]+D_2[\text{$\mathrm{GS}_{32}$},1]+D_3[\text{$\mathrm{GS}_{32}$},1]+D_4[\text{$\mathrm{GS}_{32}$},1]+2 D_9[\text{$\mathrm{GS}_{32}$},2]\)\\
\noindent\(D_{30}\left[\mathrm{G}_{1536},8\right]\text{ = }D_3[\text{$\mathrm{O}_{24}$},2]+D_4[\text{$\mathrm{O}_{24}$},3]+D_5[\text{$\mathrm{O}_{24}$},3]\)\\
\noindent\(D_{30}\left[\mathrm{G}_{1536},8\right]\text{ = }D_6[\text{$\mathrm{Oh}_{48}$},2]+D_8[\text{$\mathrm{Oh}_{48}$},3]+D_{10}[\text{$\mathrm{Oh}_{48}$},3]\)\\
\noindent\(D_{31}\left[\mathrm{G}_{1536},8\right]\text{ = }2 D_1[\text{$\mathrm{G}_{16}$},1]+2 D_2[\text{$\mathrm{G}_{16}$},1]+2 D_3[\text{$\mathrm{G}_{16}$},1]+2 D_4[\text{$\mathrm{G}_{16}$},1]\)\\
\noindent\(D_{31}\left[\mathrm{G}_{1536},8\right]\text{ = }D_2[\text{$\mathrm{G}_{48}$},1]+D_3[\text{$\mathrm{G}_{48}$},1]+2 D_4[\text{$\mathrm{G}_{48}$},3]\)\\
\noindent\(D_{31}\left[\mathrm{G}_{1536},8\right]\text{ = }D_{22}[\text{$\mathrm{G}_{64}$},1]+D_{24}[\text{$\mathrm{G}_{64}$},1]+D_{30}[\text{$\mathrm{G}_{64}$},1]+D_{32}[\text{$\mathrm{G}_{64}$},1]+D_{54}[\text{$\mathrm{G}_{64}$},1]+D_{56}[\text{$\mathrm{G}_{64}$},1]+D_{62}[\text{$\mathrm{G}_{64}$},1]+D_{64}[\text{$\mathrm{G}_{64}$},1]\)\\
\noindent\(D_{31}\left[\mathrm{G}_{1536},8\right]\text{ = }D_4[\text{$\mathrm{G}_{96}$},1]+D_6[\text{$\mathrm{G}_{96}$},1]+2 D_7[\text{$\mathrm{G}_{96}$},3]\)\\
\noindent\(D_{31}\left[\mathrm{G}_{1536},8\right]\text{ = }D_{42}[\text{$\mathrm{G}_{128}$},2]+D_{44}[\text{$\mathrm{G}_{128}$},2]+D_{50}[\text{$\mathrm{G}_{128}$},2]+D_{52}[\text{$\mathrm{G}_{128}$},2]\)\\
\noindent\(D_{31}\left[\mathrm{G}_{1536},8\right]\text{ = }D_5[\text{$\mathrm{G}_{192}$},3]+D_6[\text{$\mathrm{G}_{192}$},3]+D_{18}[\text{$\mathrm{G}_{192}$},2]\)\\
\noindent\(D_{31}\left[\mathrm{G}_{1536},8\right]\text{ = }D_{60}[\text{$\mathrm{G}_{256}$},4]+D_{62}[\text{$\mathrm{G}_{256}$},4]\)\\
\noindent\(D_{31}\left[\mathrm{G}_{1536},8\right]\text{ = }D_{19}[\text{$\mathrm{G}_{768}$},4]+D_{21}[\text{$\mathrm{G}_{768}$},4]\)\\
\noindent\(D_{31}\left[\mathrm{G}_{1536},8\right]\text{ = }D_2[\text{$\mathrm{GF}_{48}$},1]+D_3[\text{$\mathrm{GF}_{48}$},1]+2 D_4[\text{$\mathrm{GF}_{48}$},3]\)\\
\noindent\(D_{31}\left[\mathrm{G}_{1536},8\right]\text{ = }D_5[\text{$\mathrm{GF}_{192}$},3]+D_6[\text{$\mathrm{GF}_{192}$},3]+D_{18}[\text{$\mathrm{GF}_{192}$},2]\)\\
\noindent\(D_{31}\left[\mathrm{G}_{1536},8\right]\text{ = }D_3[\text{$\mathrm{GF}_{96}$},2]+D_4[\text{$\mathrm{GF}_{96}$},3]+D_5[\text{$\mathrm{GF}_{96}$},3]\)\\
\noindent\(D_{31}\left[\mathrm{G}_{1536},8\right]\text{ = }D_5[\text{$\mathrm{GP}_{24}$},1]+D_6[\text{$\mathrm{GP}_{24}$},1]+2 D_8[\text{$\mathrm{GP}_{24}$},3]\)\\
\noindent\(D_{31}\left[\mathrm{G}_{1536},8\right]\text{ = }D_3[\text{$\mathrm{GS}_{24}$},2]+D_4[\text{$\mathrm{GS}_{24}$},3]+D_5[\text{$\mathrm{GS}_{24}$},3]\)\\
\noindent\(D_{31}\left[\mathrm{G}_{1536},8\right]\text{ = }D_1[\text{$\mathrm{GS}_{32}$},1]+D_2[\text{$\mathrm{GS}_{32}$},1]+D_3[\text{$\mathrm{GS}_{32}$},1]+D_4[\text{$\mathrm{GS}_{32}$},1]+2 D_9[\text{$\mathrm{GS}_{32}$},2]\)\\
\noindent\(D_{31}\left[\mathrm{G}_{1536},8\right]\text{ = }D_3[\text{$\mathrm{O}_{24}$},2]+D_4[\text{$\mathrm{O}_{24}$},3]+D_5[\text{$\mathrm{O}_{24}$},3]\)\\
\noindent\(D_{31}\left[\mathrm{G}_{1536},8\right]\text{ = }D_6[\text{$\mathrm{Oh}_{48}$},2]+D_8[\text{$\mathrm{Oh}_{48}$},3]+D_{10}[\text{$\mathrm{Oh}_{48}$},3]\)\\
\subsection{Branching rules of the irreps of dimension 12}
\label{12brancione}
\noindent\(D_{32}\left[\mathrm{G}_{1536},12\right]
\text{ = }D_5[\text{ $\mathrm{G}_{16}$},1]+D_6[\text{ $\mathrm{G}_{16}$},1]+D_7[\text{ $\mathrm{G}_{16}$},1]+D_8[\text{ $\mathrm{G}_{16}$},1]\)\\
\noindent\(+D_9[\text{ $\mathrm{G}_{16}$},1]+D_{10}[\text{ $\mathrm{G}_{16}$},1]\)\\
\noindent\(+D_{10}[\text{ $\mathrm{G}_{16}$},1]+D_{11}[\text{ $\mathrm{G}_{16}$},1]+D_{12}[\text{ $\mathrm{G}_{16}$},1]+D_{13}[\text{ $\mathrm{G}_{16}$},1]+D_{14}[\text{ $\mathrm{G}_{16}$},1]+D_{15}[\text{ $\mathrm{G}_{16}$},1]+D_{16}[\text{ $\mathrm{G}_{16}$},1] \)\\
\noindent\(D_{32}\left[\mathrm{G}_{1536},12\right]
\text{ = } D_5[\text{ $\mathrm{G}_{48}$},3]+D_6[\text{ $\mathrm{G}_{48}$},3]+D_7[\text{ $\mathrm{G}_{48}$},3]+D_8[\text{ $\mathrm{G}_{48}$},3]\)\\
\noindent\(D_{32}\left[\mathrm{G}_{1536},12\right]
\text{ = } D_6[\text{ $\mathrm{G}_{64}$},1]+D_8[\text{ $\mathrm{G}_{64}$},1]+D_{14}[\text{ $\mathrm{G}_{64}$},1]+D_{16}[\text{ $\mathrm{G}_{64}$},1]+D_{18}[\text{ $\mathrm{G}_{64}$},1]\)\\
\noindent\(+D_{20}[\text{ $\mathrm{G}_{64}$},1]+D_{21}[\text{ $\mathrm{G}_{64}$},1]+D_{29}[\text{ $\mathrm{G}_{64}$},1]+D_{50}[\text{ $\mathrm{G}_{64}$},1]+D_{52}[\text{ $\mathrm{G}_{64}$},1]+D_{53}[\text{ $\mathrm{G}_{64}$},1]+D_{61}[\text{ $\mathrm{G}_{64}$},1]\)\\
\noindent\(D_{32}\left[\mathrm{G}_{1536},12\right]
\text{ = } D_9[\text{ $\mathrm{G}_{96}$},3]+D_{11}[\text{ $\mathrm{G}_{96}$},3]+D_{13}[\text{ $\mathrm{G}_{96}$},3]+D_{15}[\text{ $\mathrm{G}_{96}$},3]\)\\
\noindent\(D_{32}\left[\mathrm{G}_{1536},12\right]
\text{ = } D_{34}[\text{ $\mathrm{G}_{128}$},2]+D_{36}[\text{ $\mathrm{G}_{128}$},2]+D_{38}[\text{ $\mathrm{G}_{128}$},2]+D_{40}[\text{ $\mathrm{G}_{128}$},2]+D_{41}[\text{ $\mathrm{G}_{128}$},2]+D_{49}[\text{ $\mathrm{G}_{128}$},2]\)\\
\noindent\(D_{32}\left[\mathrm{G}_{1536},12\right]
\text{ = } D_9[\text{ $\mathrm{G}_{192}$},3]+D_{13}[\text{ $\mathrm{G}_{192}$},3]+D_{19}[\text{ $\mathrm{G}_{192}$},6]\)\\
\noindent\(D_{32}\left[\mathrm{G}_{1536},12\right]
\text{ = } D_{57}[\text{ $\mathrm{G}_{256}$},4]+D_{58}[\text{ $\mathrm{G}_{256}$},4]+D_{59}[\text{ $\mathrm{G}_{256}$},4]\)\\
\noindent\(D_{32}\left[\mathrm{G}_{1536},12\right]
\text{ = } D_{31}[\text{ $\mathrm{G}_{768}$},12]\)\\
\noindent\(D_{32}\left[\mathrm{G}_{1536},12\right]
\text{ = }D_5[\text{ $\mathrm{GF}_{48}$},3]+D_6[\text{ $\mathrm{GF}_{48}$},3]+D_7[\text{ $\mathrm{GF}_{48}$},3]+D_8[\text{ $\mathrm{GF}_{48}$},3]\)\\
\noindent\(D_{32}\left[\mathrm{G}_{1536},12\right]
\text{ = } D_9[\text{ $\mathrm{GF}_{192}$},3]+D_{13}[\text{ $\mathrm{GF}_{192}$},3]+D_{19}[\text{ $\mathrm{GF}_{192}$},6]\)\\
\noindent\(D_{32}\left[\mathrm{G}_{1536},12\right]
\text{ = } D_6[\text{ $\mathrm{GF}_{96}$},3]+D_8[\text{ $\mathrm{GF}_{96}$},3]+D_{10}[\text{ $\mathrm{GF}_{96}$},6]\)\\
\noindent\(D_{32}\left[\mathrm{G}_{1536},12\right]
\text{ = } D_1[\text{ $\mathrm{GP}_{24}$},1]+D_2[\text{ $\mathrm{GP}_{24}$},1]+D_3[\text{ $\mathrm{GP}_{24}$},1]+3 D_7[\text{ $\mathrm{GP}_{24}$},3]\)\\
\noindent\(D_{32}\left[\mathrm{G}_{1536},12\right]
\text{ = } D_2[\text{ $\mathrm{GS}_{24}$},1]+D_3[\text{ $\mathrm{GS}_{24}$},2]+2 D_4[\text{ $\mathrm{GS}_{24}$},3]+D_5[\text{ $\mathrm{GS}_{24}$},3]\)\\
\noindent\(D_{32}\left[\mathrm{G}_{1536},12\right]
\text{ = } D_6[\text{ $\mathrm{GS}_{32}$},1]+D_8[\text{ $\mathrm{GS}_{32}$},1]+D_{10}[\text{ $\mathrm{GS}_{32}$},2]\)\\
\noindent\(+D_{11}[\text{ $\mathrm{GS}_{32}$},2]+D_{12}[\text{ $\mathrm{GS}_{32}$},2]\)\\
\noindent\(+D_{13}[\text{ $\mathrm{GS}_{32}$},2]+D_{14}[\text{ $\mathrm{GS}_{32}$},2]\)\\
\noindent\(D_{32}\left[\mathrm{G}_{1536},12\right]
\text{ = } D_2[\text{ $\mathrm{O}_{24}$},1]+D_3[\text{ $\mathrm{O}_{24}$},2]+2 D_4[\text{ $\mathrm{O}_{24}$},3]+D_5[\text{ $\mathrm{O}_{24}$},3]\)\\
\noindent\(D_{32}\left[\mathrm{G}_{1536},12\right]
\text{ = } D_3[\text{ $\mathrm{Oh}_{48}$},1]+D_5[\text{ $\mathrm{Oh}_{48}$},2]+2 D_7[\text{ $\mathrm{Oh}_{48}$},3]+D_9[\text{ $\mathrm{Oh}_{48}$},3]\)\\
\noindent\(D_{33}\left[\mathrm{G}_{1536},12\right]
\text{ = } D_5[\text{ $\mathrm{G}_{16}$},1]+D_6[\text{ $\mathrm{G}_{16}$},1]+D_7[\text{ $\mathrm{G}_{16}$},1]+D_8[\text{ $\mathrm{G}_{16}$},1]\)\\
\noindent\(+D_9[\text{ $\mathrm{G}_{16}$},1]+D_{10}[\text{ $\mathrm{G}_{16}$},1]+D_{11}[\text{ $\mathrm{G}_{16}$},1]\)\\
\noindent\(+D_{12}[\text{ $\mathrm{G}_{16}$},1]+D_{13}[\text{ $\mathrm{G}_{16}$},1]+D_{14}[\text{ $\mathrm{G}_{16}$},1]+D_{15}[\text{ $\mathrm{G}_{16}$},1]+D_{16}[\text{ $\mathrm{G}_{16}$},1]\)\\
\noindent\(D_{33}\left[\mathrm{G}_{1536},12\right]
\text{ = } D_5[\text{ $\mathrm{G}_{48}$},3]+D_6[\text{ $\mathrm{G}_{48}$},3]+D_7[\text{ $\mathrm{G}_{48}$},3]+D_8[\text{ $\mathrm{G}_{48}$},3]\)\\
\noindent\(D_{33}\left[\mathrm{G}_{1536},12\right]
\text{ = } D_6[\text{ $\mathrm{G}_{64}$},1]+D_8[\text{ $\mathrm{G}_{64}$},1]+D_{14}[\text{ $\mathrm{G}_{64}$},1]+D_{16}[\text{ $\mathrm{G}_{64}$},1]+D_{18}[\text{ $\mathrm{G}_{64}$},1]+D_{20}[\text{ $\mathrm{G}_{64}$},1]\)\\
\noindent\(+D_{21}[\text{ $\mathrm{G}_{64}$},1]+D_{29}[\text{ $\mathrm{G}_{64}$},1]+D_{50}[\text{ $\mathrm{G}_{64}$},1]+D_{52}[\text{ $\mathrm{G}_{64}$},1]+D_{53}[\text{ $\mathrm{G}_{64}$},1]+D_{61}[\text{ $\mathrm{G}_{64}$},1]\)\\
\noindent\(D_{33}\left[\mathrm{G}_{1536},12\right]
\text{ = } D_9[\text{ $\mathrm{G}_{96}$},3]+D_{11}[\text{ $\mathrm{G}_{96}$},3]+D_{13}[\text{ $\mathrm{G}_{96}$},3]+D_{15}[\text{ $\mathrm{G}_{96}$},3]\)\\
\noindent\(D_{33}\left[\mathrm{G}_{1536},12\right]
\text{ = } D_{34}[\text{ $\mathrm{G}_{128}$},2]+D_{36}[\text{ $\mathrm{G}_{128}$},2]+D_{38}[\text{ $\mathrm{G}_{128}$},2]+D_{40}[\text{ $\mathrm{G}_{128}$},2]+D_{41}[\text{ $\mathrm{G}_{128}$},2]+D_{49}[\text{ $\mathrm{G}_{128}$},2]\)\\
\noindent\(D_{33}\left[\mathrm{G}_{1536},12\right]
\text{ = } D_{10}[\text{ $\mathrm{G}_{192}$},3]+D_{14}[\text{ $\mathrm{G}_{192}$},3]+D_{19}[\text{ $\mathrm{G}_{192}$},6]\)\\
\noindent\(D_{33}\left[\mathrm{G}_{1536},12\right]
\text{ = } D_{57}[\text{ $\mathrm{G}_{256}$},4]+D_{58}[\text{ $\mathrm{G}_{256}$},4]+D_{59}[\text{ $\mathrm{G}_{256}$},4]\)\\
\noindent\(D_{33}\left[\mathrm{G}_{1536},12\right]
\text{ = } D_{31}[\text{ $\mathrm{G}_{768}$},12]\)\\
\noindent\(D_{33}\left[\mathrm{G}_{1536},12\right]
\text{ = } D_5[\text{ $\mathrm{GF}_{48}$},3]+D_6[\text{ $\mathrm{GF}_{48}$},3]+D_7[\text{ $\mathrm{GF}_{48}$},3]+D_8[\text{ $\mathrm{GF}_{48}$},3]\)\\
\noindent\(D_{33}\left[\mathrm{G}_{1536},12\right]
\text{ = } D_{10}[\text{ $\mathrm{GF}_{192}$},3]+D_{14}[\text{ $\mathrm{GF}_{192}$},3]+D_{19}[\text{ $\mathrm{GF}_{192}$},6]\)\\
\noindent\(D_{33}\left[\mathrm{G}_{1536},12\right]
\text{ = } D_7[\text{ $\mathrm{GF}_{96}$},3]+D_9[\text{ $\mathrm{GF}_{96}$},3]+D_{10}[\text{ $\mathrm{GF}_{96}$},6]\)\\
\noindent\(D_{33}\left[\mathrm{G}_{1536},12\right]
\text{ = } D_1[\text{ $\mathrm{GP}_{24}$},1]+D_2[\text{ $\mathrm{GP}_{24}$},1]+D_3[\text{ $\mathrm{GP}_{24}$},1]+3 D_7[\text{ $\mathrm{GP}_{24}$},3]\)\\
\noindent\(D_{33}\left[\mathrm{G}_{1536},12\right]
\text{ = } D_1[\text{ $\mathrm{GS}_{24}$},1]+D_3[\text{ $\mathrm{GS}_{24}$},2]+D_4[\text{ $\mathrm{GS}_{24}$},3]+2 D_5[\text{ $\mathrm{GS}_{24}$},3]\)\\
\noindent\(D_{33}\left[\mathrm{G}_{1536},12\right]
\text{ = } D_5[\text{ $\mathrm{GS}_{32}$},1]+D_7[\text{ $\mathrm{GS}_{32}$},1]+D_{10}[\text{ $\mathrm{GS}_{32}$},2]+D_{11}[\text{ $\mathrm{GS}_{32}$},2]\)\\
\noindent\(+D_{12}[\text{ $\mathrm{GS}_{32}$},2]+D_{13}[\text{ $\mathrm{GS}_{32}$},2]+D_{14}[\text{ $\mathrm{GS}_{32}$},2]\)\\
\noindent\(D_{33}\left[\mathrm{G}_{1536},12\right]
\text{ = } D_1[\text{ $\mathrm{O}_{24}$},1]+D_3[\text{ $\mathrm{O}_{24}$},2]+D_4[\text{ $\mathrm{O}_{24}$},3]+2 D_5[\text{ $\mathrm{O}_{24}$},3]\)\\
\noindent\(D_{33}\left[\mathrm{G}_{1536},12\right]
\text{ = } D_1[\text{ $\mathrm{Oh}_{48}$},1]+D_5[\text{ $\mathrm{Oh}_{48}$},2]+D_7[\text{ $\mathrm{Oh}_{48}$},3]+2 D_9[\text{ $\mathrm{Oh}_{48}$},3]\)\\
\noindent\(D_{34}\left[\mathrm{G}_{1536},12\right]
\text{ = } D_5[\text{ $\mathrm{G}_{16}$},1]+D_6[\text{ $\mathrm{G}_{16}$},1]+D_7[\text{ $\mathrm{G}_{16}$},1]+D_8[\text{ $\mathrm{G}_{16}$},1]+D_9[\text{ $\mathrm{G}_{16}$},1]+D_{10}[\text{ $\mathrm{G}_{16}$},1]+D_{11}[\text{ $\mathrm{G}_{16}$},1]\)\\
\noindent\(+D_{12}[\text{ $\mathrm{G}_{16}$},1]+D_{13}[\text{ $\mathrm{G}_{16}$},1]+D_{14}[\text{ $\mathrm{G}_{16}$},1]+D_{15}[\text{ $\mathrm{G}_{16}$},1]+D_{16}[\text{ $\mathrm{G}_{16}$},1]\)\\
\noindent\(D_{34}\left[\mathrm{G}_{1536},12\right]
\text{ = } D_5[\text{ $\mathrm{G}_{48}$},3]+D_6[\text{ $\mathrm{G}_{48}$},3]+D_7[\text{ $\mathrm{G}_{48}$},3]+D_8[\text{ $\mathrm{G}_{48}$},3]\)\\
\noindent\(D_{34}\left[\mathrm{G}_{1536},12\right]
\text{ = } D_{23}[\text{ $\mathrm{G}_{64}$},1]+D_{26}[\text{ $\mathrm{G}_{64}$},1]+D_{28}[\text{ $\mathrm{G}_{64}$},1]+D_{31}[\text{ $\mathrm{G}_{64}$},1]+D_{38}[\text{ $\mathrm{G}_{64}$},1]\)\\
\noindent\(+D_{40}[\text{ $\mathrm{G}_{64}$},1]+D_{46}[\text{ $\mathrm{G}_{64}$},1]\)
\noindent\(+D_{48}[\text{ $\mathrm{G}_{64}$},1]+D_{55}[\text{ $\mathrm{G}_{64}$},1]+D_{58}[\text{ $\mathrm{G}_{64}$},1]+D_{60}[\text{ $\mathrm{G}_{64}$},1]+D_{63}[\text{ $\mathrm{G}_{64}$},1]\)\\
\noindent\(D_{34}\left[\mathrm{G}_{1536},12\right]
\text{ = }D_9[\text{ $\mathrm{G}_{96}$},3]+D_{11}[\text{ $\mathrm{G}_{96}$},3]+D_{13}[\text{ $\mathrm{G}_{96}$},3]+D_{15}[\text{ $\mathrm{G}_{96}$},3]\)\\
\noindent\(D_{34}\left[\mathrm{G}_{1536},12\right]
\text{ = } D_{43}[\text{ $\mathrm{G}_{128}$},2]+D_{46}[\text{ $\mathrm{G}_{128}$},2]+D_{48}[\text{ $\mathrm{G}_{128}$},2]+D_{51}[\text{ $\mathrm{G}_{128}$},2]+D_{54}[\text{ $\mathrm{G}_{128}$},2]+D_{56}[\text{ $\mathrm{G}_{128}$},2]\)\\
\noindent\(D_{34}\left[\mathrm{G}_{1536},12\right]
\text{ = } D_{10}[\text{ $\mathrm{G}_{192}$},3]+D_{14}[\text{ $\mathrm{G}_{192}$},3]+D_{19}[\text{ $\mathrm{G}_{192}$},6]\)\\
\noindent\(D_{34}\left[\mathrm{G}_{1536},12\right]
\text{ = } D_{61}[\text{ $\mathrm{G}_{256}$},4]+D_{63}[\text{ $\mathrm{G}_{256}$},4]+D_{64}[\text{ $\mathrm{G}_{256}$},4]\)\\
\noindent\(D_{34}\left[\mathrm{G}_{1536},12\right]
\text{ = } D_{32}[\text{ $\mathrm{G}_{768}$},12]\)\\
\noindent\(D_{34}\left[\mathrm{G}_{1536},12\right]
\text{ = } D_5[\text{ $\mathrm{GF}_{48}$},3]+D_6[\text{ $\mathrm{GF}_{48}$},3]+D_7[\text{ $\mathrm{GF}_{48}$},3]+D_8[\text{ $\mathrm{GF}_{48}$},3]\)\\
\noindent\(D_{34}\left[\mathrm{G}_{1536},12\right]
\text{ = } D_9[\text{ $\mathrm{GF}_{192}$},3]+D_{13}[\text{ $\mathrm{GF}_{192}$},3]+D_{19}[\text{ $\mathrm{GF}_{192}$},6]\)\\
\noindent\(D_{34}\left[\mathrm{G}_{1536},12\right]
\text{ = } D_6[\text{ $\mathrm{GF}_{96}$},3]+D_8[\text{ $\mathrm{GF}_{96}$},3]+D_{10}[\text{ $\mathrm{GF}_{96}$},6]\)\\
\noindent\(D_{34}\left[\mathrm{G}_{1536},12\right]
\text{ = } D_1[\text{ $\mathrm{GP}_{24}$},1]+D_2[\text{ $\mathrm{GP}_{24}$},1]+D_3[\text{ $\mathrm{GP}_{24}$},1]+3 D_7[\text{ $\mathrm{GP}_{24}$},3]\)\\
\noindent\(D_{34}\left[\mathrm{G}_{1536},12\right]
\text{ = } D_2[\text{ $\mathrm{GS}_{24}$},1]+D_3[\text{ $\mathrm{GS}_{24}$},2]+2 D_4[\text{ $\mathrm{GS}_{24}$},3]+D_5[\text{ $\mathrm{GS}_{24}$},3]\)\\
\noindent\(D_{34}\left[\mathrm{G}_{1536},12\right]
\text{ = } D_5[\text{ $\mathrm{GS}_{32}$},1]+D_7[\text{ $\mathrm{GS}_{32}$},1]+D_{10}[\text{ $\mathrm{GS}_{32}$},2]+D_{11}[\text{ $\mathrm{GS}_{32}$},2]+D_{12}[\text{ $\mathrm{GS}_{32}$},2]\)\\
\noindent\(+D_{13}[\text{ $\mathrm{GS}_{32}$},2]+D_{14}[\text{ $\mathrm{GS}_{32}$},2]\)\\
\noindent\(D_{34}\left[\mathrm{G}_{1536},12\right]
\text{ = } D_1[\text{ $\mathrm{O}_{24}$},1]+D_3[\text{ $\mathrm{O}_{24}$},2]+D_4[\text{ $\mathrm{O}_{24}$},3]+2 D_5[\text{ $\mathrm{O}_{24}$},3]\)\\
\noindent\(D_{34}\left[\mathrm{G}_{1536},12\right]
\text{ = } D_1[\text{ $\mathrm{Oh}_{48}$},1]+D_5[\text{ $\mathrm{Oh}_{48}$},2]+D_7[\text{ $\mathrm{Oh}_{48}$},3]+2 D_9[\text{ $\mathrm{Oh}_{48}$},3]\)\\
\noindent\(D_{35}\left[\mathrm{G}_{1536},12\right]
\text{ = } D_5[\text{ $\mathrm{G}_{16}$},1]+D_6[\text{ $\mathrm{G}_{16}$},1]+D_7[\text{ $\mathrm{G}_{16}$},1]+D_8[\text{ $\mathrm{G}_{16}$},1]+D_9[\text{ $\mathrm{G}_{16}$},1]\)\\
\noindent\(+D_{10}[\text{ $\mathrm{G}_{16}$},1]+D_{11}[\text{ $\mathrm{G}_{16}$},1]+D_{12}[\text{ $\mathrm{G}_{16}$},1]+D_{13}[\text{ $\mathrm{G}_{16}$},1]+D_{14}[\text{ $\mathrm{G}_{16}$},1]+D_{15}[\text{ $\mathrm{G}_{16}$},1]+D_{16}[\text{ $\mathrm{G}_{16}$},1]\)\\
\noindent\(D_{35}\left[\mathrm{G}_{1536},12\right]
\text{ = } D_5[\text{ $\mathrm{G}_{48}$},3]+D_6[\text{ $\mathrm{G}_{48}$},3]+D_7[\text{ $\mathrm{G}_{48}$},3]+D_8[\text{ $\mathrm{G}_{48}$},3]\)\\
\noindent\(D_{35}\left[\mathrm{G}_{1536},12\right]
\text{ = } D_{23}[\text{ $\mathrm{G}_{64}$},1]+D_{26}[\text{ $\mathrm{G}_{64}$},1]+D_{28}[\text{ $\mathrm{G}_{64}$},1]+D_{31}[\text{ $\mathrm{G}_{64}$},1]+D_{38}[\text{ $\mathrm{G}_{64}$},1]+D_{40}[\text{ $\mathrm{G}_{64}$},1]+D_{46}[\text{ $\mathrm{G}_{64}$},1]\)\\
\noindent\(+D_{48}[\text{ $\mathrm{G}_{64}$},1]+D_{55}[\text{ $\mathrm{G}_{64}$},1]+D_{58}[\text{ $\mathrm{G}_{64}$},1]+D_{60}[\text{ $\mathrm{G}_{64}$},1]+D_{63}[\text{ $\mathrm{G}_{64}$},1]\)\\
\noindent\(D_{35}\left[\mathrm{G}_{1536},12\right]
\text{ = } D_9[\text{ $\mathrm{G}_{96}$},3]+D_{11}[\text{ $\mathrm{G}_{96}$},3]+D_{13}[\text{ $\mathrm{G}_{96}$},3]+D_{15}[\text{ $\mathrm{G}_{96}$},3]\)\\
\noindent\(D_{35}\left[\mathrm{G}_{1536},12\right]
\text{ = } D_{43}[\text{ $\mathrm{G}_{128}$},2]+D_{46}[\text{ $\mathrm{G}_{128}$},2]+D_{48}[\text{ $\mathrm{G}_{128}$},2]+D_{51}[\text{ $\mathrm{G}_{128}$},2]+D_{54}[\text{ $\mathrm{G}_{128}$},2]+D_{56}[\text{ $\mathrm{G}_{128}$},2]\)\\
\noindent\(D_{35}\left[\mathrm{G}_{1536},12\right]
\text{ = } D_9[\text{ $\mathrm{G}_{192}$},3]+D_{13}[\text{ $\mathrm{G}_{192}$},3]+D_{19}[\text{ $\mathrm{G}_{192}$},6]\)\\
\noindent\(D_{35}\left[\mathrm{G}_{1536},12\right]
\text{ = } D_{61}[\text{ $\mathrm{G}_{256}$},4]+D_{63}[\text{ $\mathrm{G}_{256}$},4]+D_{64}[\text{ $\mathrm{G}_{256}$},4]\)\\
\noindent\(D_{35}\left[\mathrm{G}_{1536},12\right]
\text{ = } D_{32}[\text{ $\mathrm{G}_{768}$},12]\)\\
\noindent\(D_{35}\left[\mathrm{G}_{1536},12\right]
\text{ = } D_5[\text{ $\mathrm{GF}_{48}$},3]+D_6[\text{ $\mathrm{GF}_{48}$},3]+D_7[\text{ $\mathrm{GF}_{48}$},3]+D_8[\text{ $\mathrm{GF}_{48}$},3]\)\\
\noindent\(D_{35}\left[\mathrm{G}_{1536},12\right]
\text{ = } D_{10}[\text{ $\mathrm{GF}_{192}$},3]+D_{14}[\text{ $\mathrm{GF}_{192}$},3]+D_{19}[\text{ $\mathrm{GF}_{192}$},6]\)\\
\noindent\(D_{35}\left[\mathrm{G}_{1536},12\right]
\text{ = } D_7[\text{ $\mathrm{GF}_{96}$},3]+D_9[\text{ $\mathrm{GF}_{96}$},3]+D_{10}[\text{ $\mathrm{GF}_{96}$},6]\)\\
\noindent\(D_{35}\left[\mathrm{G}_{1536},12\right]
\text{ = } D_1[\text{ $\mathrm{GP}_{24}$},1]+D_2[\text{ $\mathrm{GP}_{24}$},1]+D_3[\text{ $\mathrm{GP}_{24}$},1]+3 D_7[\text{ $\mathrm{GP}_{24}$},3]\)\\
\noindent\(D_{35}\left[\mathrm{G}_{1536},12\right]
\text{ = } D_1[\text{ $\mathrm{GS}_{24}$},1]+D_3[\text{ $\mathrm{GS}_{24}$},2]+D_4[\text{ $\mathrm{GS}_{24}$},3]+2 D_5[\text{ $\mathrm{GS}_{24}$},3]\)\\
\noindent\(D_{35}\left[\mathrm{G}_{1536},12\right]
\text{ = } D_6[\text{ $\mathrm{GS}_{32}$},1]+D_8[\text{ $\mathrm{GS}_{32}$},1]+D_{10}[\text{ $\mathrm{GS}_{32}$},2]+D_{11}[\text{ $\mathrm{GS}_{32}$},2]\)\\
\noindent\(+D_{12}[\text{ $\mathrm{GS}_{32}$},2]+D_{13}[\text{ $\mathrm{GS}_{32}$},2]+D_{14}[\text{ $\mathrm{GS}_{32}$},2]\)\\
\noindent\(D_{35}\left[\mathrm{G}_{1536},12\right]
\text{ = } D_2[\text{ $\mathrm{O}_{24}$},1]+D_3[\text{ $\mathrm{O}_{24}$},2]+2 D_4[\text{ $\mathrm{O}_{24}$},3]+D_5[\text{ $\mathrm{O}_{24}$},3]\)\\
\noindent\(D_{35}\left[\mathrm{G}_{1536},12\right]
\text{ = } D_3[\text{ $\mathrm{Oh}_{48}$},1]+D_5[\text{ $\mathrm{Oh}_{48}$},2]+2 D_7[\text{ $\mathrm{Oh}_{48}$},3]+D_9[\text{ $\mathrm{Oh}_{48}$},3]\)\\
\noindent\(D_{36}\left[\mathrm{G}_{1536},12\right]
\text{ = } 2 D_6[\text{ $\mathrm{G}_{16}$},1]+2 D_7[\text{ $\mathrm{G}_{16}$},1]+2 D_9[\text{ $\mathrm{G}_{16}$},1]+2 D_{11}[\text{ $\mathrm{G}_{16}$},1]+2
D_{15}[\text{ $\mathrm{G}_{16}$},1]+2 D_{16}[\text{ $\mathrm{G}_{16}$},1]\)\\
\noindent\(D_{36}\left[\mathrm{G}_{1536},12\right]
\text{ = } 2 D_6[\text{ $\mathrm{G}_{48}$},3]+2 D_7[\text{ $\mathrm{G}_{48}$},3]\)\\
\noindent\(D_{36}\left[\mathrm{G}_{1536},12\right]
\text{ = } D_7[\text{ $\mathrm{G}_{64}$},1]+D_{10}[\text{ $\mathrm{G}_{64}$},1]+D_{12}[\text{ $\mathrm{G}_{64}$},1]+D_{15}[\text{ $\mathrm{G}_{64}$},1]+D_{19}[\text{ $\mathrm{G}_{64}$},1]+D_{25}[\text{ $\mathrm{G}_{64}$},1]+D_{34}[\text{ $\mathrm{G}_{64}$},1]\)\\
\noindent\(+D_{36}[\text{ $\mathrm{G}_{64}$},1]+D_{37}[\text{ $\mathrm{G}_{64}$},1]+D_{45}[\text{ $\mathrm{G}_{64}$},1]+D_{51}[\text{ $\mathrm{G}_{64}$},1]+D_{57}[\text{ $\mathrm{G}_{64}$},1]\)\\
\noindent\(D_{36}\left[\mathrm{G}_{1536},12\right]
\text{ = } 2 D_{12}[\text{ $\mathrm{G}_{96}$},3]+2 D_{14}[\text{ $\mathrm{G}_{96}$},3]\)\\
\noindent\(D_{36}\left[\mathrm{G}_{1536},12\right]
\text{ = } D_{11}[\text{ $\mathrm{G}_{128}$},1]+D_{15}[\text{ $\mathrm{G}_{128}$},1]+D_{19}[\text{ $\mathrm{G}_{128}$},1]+D_{23}[\text{ $\mathrm{G}_{128}$},1]+D_{35}[\text{ $\mathrm{G}_{128}$},2]\)\\
\noindent\(+D_{39}[\text{ $\mathrm{G}_{128}$},2]+D_{45}[\text{ $\mathrm{G}_{128}$},2]+D_{53}[\text{ $\mathrm{G}_{128}$},2]\)\\
\noindent\(D_{36}\left[\mathrm{G}_{1536},12\right]
\text{ = } D_{11}[\text{ $\mathrm{G}_{192}$},3]+D_{12}[\text{ $\mathrm{G}_{192}$},3]+D_{15}[\text{ $\mathrm{G}_{192}$},3]+D_{16}[\text{ $\mathrm{G}_{192}$},3]\)\\
\noindent\(D_{36}\left[\mathrm{G}_{1536},12\right]
\text{ = } D_{35}[\text{ $\mathrm{G}_{256}$},2]+D_{37}[\text{ $\mathrm{G}_{256}$},2]+D_{44}[\text{ $\mathrm{G}_{256}$},2]+D_{47}[\text{ $\mathrm{G}_{256}$},2]\)\\
\noindent\(+D_{49}[\text{ $\mathrm{G}_{256}$},2]+D_{54}[\text{ $\mathrm{G}_{256}$},2]\)\\
\noindent\(D_{36}\left[\mathrm{G}_{1536},12\right]
\text{ = } D_{25}[\text{ $\mathrm{G}_{768}$},6]+D_{27}[\text{ $\mathrm{G}_{768}$},6]\)\\
\noindent\(D_{36}\left[\mathrm{G}_{1536},12\right]
\text{ = } 2 D_5[\text{ $\mathrm{GF}_{48}$},3]+2 D_8[\text{ $\mathrm{GF}_{48}$},3]\)\\
\noindent\(D_{36}\left[\mathrm{G}_{1536},12\right]
\text{ = } 2 D_{20}[\text{ $\mathrm{GF}_{192}$},6]\)\\
\noindent\(D_{36}\left[\mathrm{G}_{1536},12\right]
\text{ = } 2 D_{10}[\text{ $\mathrm{GF}_{96}$},6]\)\\
\noindent\(D_{36}\left[\mathrm{G}_{1536},12\right]
\text{ = } 2 D_4[\text{ $\mathrm{GP}_{24}$},1]+2 D_5[\text{ $\mathrm{GP}_{24}$},1]+2 D_6[\text{ $\mathrm{GP}_{24}$},1]+2 D_8[\text{ $\mathrm{GP}_{24}$},3]\)\\
\noindent\(D_{36}\left[\mathrm{G}_{1536},12\right]
\text{ = } 2 D_4[\text{ $\mathrm{GS}_{24}$},3]+2 D_5[\text{ $\mathrm{GS}_{24}$},3]\)\\
\noindent\(D_{36}\left[\mathrm{G}_{1536},12\right]
\text{ = } D_5[\text{ $\mathrm{GS}_{32}$},1]+D_6[\text{ $\mathrm{GS}_{32}$},1]+D_7[\text{ $\mathrm{GS}_{32}$},1]+D_8[\text{ $\mathrm{GS}_{32}$},1]+2
D_{12}[\text{ $\mathrm{GS}_{32}$},2]+2 D_{14}[\text{ $\mathrm{GS}_{32}$},2]\)\\
\noindent\(D_{36}\left[\mathrm{G}_{1536},12\right]
\text{ = } D_1[\text{ $\mathrm{O}_{24}$},1]+D_2[\text{ $\mathrm{O}_{24}$},1]+2 D_3[\text{ $\mathrm{O}_{24}$},2]+D_4[\text{ $\mathrm{O}_{24}$},3]+D_5[\text{ $\mathrm{O}_{24}$},3]\)\\
\noindent\(D_{36}\left[\mathrm{G}_{1536},12\right]
\text{ = } D_2[\text{ $\mathrm{Oh}_{48}$},1]+D_4[\text{ $\mathrm{Oh}_{48}$},1]+2 D_6[\text{ $\mathrm{Oh}_{48}$},2]+D_8[\text{ $\mathrm{Oh}_{48}$},3]+D_{10}[\text{ $\mathrm{Oh}_{48}$},3]\)\\
\noindent\(D_{37}\left[\mathrm{G}_{1536},12\right]
\text{ = } 2 D_5[\text{ $\mathrm{G}_{16}$},1]+2 D_8[\text{ $\mathrm{G}_{16}$},1]+2 D_{10}[\text{ $\mathrm{G}_{16}$},1]+2 D_{12}[\text{ $\mathrm{G}_{16}$},1]+2
D_{13}[\text{ $\mathrm{G}_{16}$},1]+2 D_{14}[\text{ $\mathrm{G}_{16}$},1]\)\\
\noindent\(D_{37}\left[\mathrm{G}_{1536},12\right]
\text{ = } 2 D_5[\text{ $\mathrm{G}_{48}$},3]+2 D_8[\text{ $\mathrm{G}_{48}$},3]\)\\
\noindent\(D_{37}\left[\mathrm{G}_{1536},12\right]
\text{ = } D_7[\text{ $\mathrm{G}_{64}$},1]+D_{10}[\text{ $\mathrm{G}_{64}$},1]+D_{12}[\text{ $\mathrm{G}_{64}$},1]+D_{15}[\text{ $\mathrm{G}_{64}$},1]+D_{19}[\text{ $\mathrm{G}_{64}$},1]+D_{25}[\text{ $\mathrm{G}_{64}$},1]\)\\
\noindent\(+D_{34}[\text{ $\mathrm{G}_{64}$},1]+D_{36}[\text{ $\mathrm{G}_{64}$},1]+D_{37}[\text{ $\mathrm{G}_{64}$},1]+D_{45}[\text{ $\mathrm{G}_{64}$},1]+D_{51}[\text{ $\mathrm{G}_{64}$},1]+D_{57}[\text{ $\mathrm{G}_{64}$},1]\)\\
\noindent\(D_{37}\left[\mathrm{G}_{1536},12\right]
\text{ = } 2 D_{10}[\text{ $\mathrm{G}_{96}$},3]+2 D_{16}[\text{ $\mathrm{G}_{96}$},3]\)\\
\noindent\(D_{37}\left[\mathrm{G}_{1536},12\right]
\text{ = } D_{12}[\text{ $\mathrm{G}_{128}$},1]+D_{16}[\text{ $\mathrm{G}_{128}$},1]+D_{20}[\text{ $\mathrm{G}_{128}$},1]+D_{24}[\text{ $\mathrm{G}_{128}$},1]+D_{35}[\text{ $\mathrm{G}_{128}$},2]+D_{39}[\text{ $\mathrm{G}_{128}$},2]\)\\
\noindent\(+D_{45}[\text{ $\mathrm{G}_{128}$},2]+D_{53}[\text{ $\mathrm{G}_{128}$},2]\)
\section{Other relevant subgroups}
\label{ABCsubgroups}
In this appendix we list the additional chains of subgroups of the Universal Classifying Group $\mathrm{\mathrm{G_{1536}}}$ that emerge in the analysis of the classical model of $\mathrm{ABC}$-flows. As in previous cases we just give for each of them the conjugacy classes of which they are composed. The interesting network of interrelation between these subgroups, which explains the various cases and subcases of $\mathrm{ABC}$-flows is thoroughly discussed in the main text.
\subsection{The Group $\mathrm{G}^{\mathrm{(A,B,0)}}_{128}$}
{\scriptsize
\noindent \textit{\small Conjugacy Class $\mathcal{C}_{1}\left(\mathrm{G}^{\mathrm{(A,B,0)}}_{128}\right)$}
\begin{equation}

\end{equation}
}
\subsection{The Group $\mathrm{G}^{\mathrm{(A,B,0)}}_{64}$}
{\scriptsize
\noindent
\textit{\small Conjugacy Class $\mathcal{C}_{1}\left(\mathrm{G}^{\mathrm{(A,B,0)}}_{64}\right)$}\begin{equation}
%
\end{equation}
}
\subsection{The Group $\mathrm{G^{(A,B,0)}_{32}}$}
{\scriptsize
\noindent
\textit{\small Conjugacy Class $\mathcal{C}_{1}\left(\mathrm{G_{32}^{(A,B,0)}}\right)$}\begin{equation}
%
\end{equation}
}
\subsection{The Group $\mathrm{G^{(A,B,0)}_{16}}$}
{\scriptsize
\noindent
\textit{\small Conjugacy Class $\mathcal{C}_{1}\left(\mathrm{G_{16}^{(A,B,0)}}\right)$}\begin{equation}
%
\end{equation}
}
\subsection{The Group $\mathrm{G^{(A,B,0)}_{8}}$}
{\scriptsize
\noindent
\textit{\small Conjugacy Class $\mathcal{C}_{1}\left(\mathrm{G_{8}^{(A,B,0)}}\right)$}\begin{equation}
%
\end{equation}
}
\subsection{The Group $\mathrm{G^{(A,B,0)}_{4}}$}
{\scriptsize
\noindent
\textit{\small Abelian: every element is a conjugacy class}\begin{equation}
%
\end{equation}
}
\subsection{The Group $\mathrm{G^{(A,0,0)}_{256}}$}
{\scriptsize
\noindent
\textit{\small Conjugacy Class $\mathcal{C}_{1}\left(\mathrm{G_{256}^{(A,0,0)}}\right)$}\begin{equation}
%
\end{equation}
}
\subsection{The Group $\mathrm{G^{(A,0,0)}_{128}}$}
{\scriptsize
\noindent
\textit{\small Conjugacy Class $\mathcal{C}_{1}\left(\mathrm{G_{128}^{(A,0,0)}}\right)$}\begin{equation}
%
\end{equation}
}
\subsection{The Group $\mathrm{G^{(A,0,0)}_{64}}$}
{\scriptsize
\noindent
\textit{\small Conjugacy Class $\mathcal{C}_{1}\left(\mathrm{G_{64}^{(A,0,0)}}\right)$}\begin{equation}
%
\end{equation}
}
\subsection{The Group $\mathrm{G^{(A,0,0)}_{32}}$}
{\scriptsize
\noindent
\textit{\small Conjugacy Class $\mathcal{C}_{1}\left(\mathrm{G_{32}^{(A,0,0)}}\right)$}\begin{equation}
%
\end{equation}
}
\subsection{The Group $\mathrm{G^{(A,0,0)}_{16}}$}
{\scriptsize
\noindent
\textit{\small Abelian: every element is a conjugacy class}\begin{equation}
%
\end{equation}
}
\subsection{The Group $\mathrm{G^{(A,A,0)}_{32}}$}
{\scriptsize
\noindent
\textit{\small Conjugacy Class $\mathcal{C}_{1}\left(\mathrm{G_{32}^{(A,A,0)}}\right)$}\begin{equation}
%
\end{equation}
}
\newpage
\begin{landscape}
\section{Some large formulae for the main text}
\label{largone}
In this Appendix we write in landscape format some formulae quoted in the main text that are too large for the standard view
\par
For section \ref{12orbitsezia} we write here the form of the translation generators in the lowest lying octahedral  orbit of length 12:
{\tiny
\begin{eqnarray}
\mathcal{M}^{(12)}_{\{n_1,0,0\}}&=&\left(
%
\right)
\nonumber\\
\label{12Mtraslo}
\end{eqnarray}
}
For section \ref{orbit8sezia} we write here the form of the translation generators in the lowest lying octahedral  orbit of length 8:
{\tiny
\begin{eqnarray}
 &\mathcal{M}^{(8)}_{\{n_1,0,0\}}\, =& \nonumber\\
 &\left(
%
\right)&  \nonumber\\
\label{8traslatore}
\end{eqnarray}
}
\end{landscape}
\newpage
\part{The Bibliography}

\end{document}